\documentclass[a4paper, 11pt, oneside]{book}
\pdfoutput=1
\usepackage{hyperref}
\hypersetup{colorlinks}
\usepackage[a4paper,margin=1in]{geometry}
\usepackage{graphicx}
\usepackage[onehalfspacing]{setspace}
\usepackage{xcolor}
\usepackage{xspace}

\usepackage{physics2}
\usepackage{derivative}
\usepackage{diffcoeff}
\usepackage{amsmath}
\usepackage{amsfonts}
\usepackage{amssymb}
\usepackage[legacycolonsymbols]{mathtools}
\usepackage{mathrsfs}

\usepackage{cleveref}
\usepackage{minitoc}
\usepackage{url}
\usepackage{siunitx}
\usepackage{bm}
\usepackage{cite}
\usepackage{copyrightbox}
\usepackage{lineno}
\usepackage[bottom]{footmisc}
\definecolor{MK_One_One}{RGB}{100,60,10}  
\definecolor{MK_One_Two}{RGB}{186,149,71} 
\definecolor{MK_One_Three}{RGB}{216,202,165}
\definecolor{MK_One_Four}{RGB}{169,204,199} 
\definecolor{MK_One_Five}{RGB}{60,150,142} 
\definecolor{MK_One_Six}{RGB}{1,72,64} 

\hypersetup{
colorlinks=true,
 linkcolor=MK_One_Five
,citecolor=MK_One_Two
,filecolor=MK_One_Three
,urlcolor=MK_One_Five
,menucolor=MK_One_Five
,runcolor=MK_One_Four
,linkbordercolor=MK_One_One
,citebordercolor=MK_One_Two
,filebordercolor=MK_One_Three
,urlbordercolor=MK_One_Six
,menubordercolor=MK_One_Five
,runbordercolor=MK_One_Four
}

\newcommand{\LiteBIRD}{\textit{LiteBIRD}\xspace}
\newcommand{\LB}{\textit{\LiteBIRD}\xspace}
\newcommand{\Planck}{\textit{Planck}\xspace}
\newcommand{\COBE}{\textit{COBE}\xspace}
\newcommand{\WMAP}{\textit{WMAP}\xspace}
\newcommand{\EPIC}{\textit{EPIC}\xspace}

\newcommand{\PICO}{\textit{PICO}\xspace}
\newcommand{\CORE}{\textit{CORE}\xspace}

\newcommand{\healpix}{\texttt{HEALPix}\xspace}
\newcommand{\Falcons}{\texttt{Falcons.jl}\xspace}

\newcommand{\SBM}{\texttt{SBM}\xspace}

\newcommand{\Nhits}{N_{\rm{hits}}}
\newcommand{\Nside}{N_{\rm{side}}}
\newcommand{\Npix}{N_{\rm{pix}}}
\newcommand{\Ndets}{N_{\rm{dets}}}
\newcommand{\Ntot}{\Nhits^{\rm tot}}
\newcommand{\tmu}{_{\texttt{T}}^{(\mu)}}
\newcommand{\bmu}{_{\texttt{B}}^{(\mu)}}
\newcommand{\pmu}{^{(\mu)}}
\newcommand{\ozeta}{\overline{\zeta}}
\newcommand{\oeth}{\overline{\eth}}
\newcommand{\spin}{\textit{spin}\xspace}
\newcommand{\Spin}{\textit{Spin}\xspace}
\newcommand{\h}[1][2]{{{}_{#1}\tilde{h}}}
\newcommand{\htot}[1][2]{\h[#1]^{\rm tot}}
\newcommand{\hmean}[1][2]{{\ab <\ab|\h[#1]|^2>}}
\newcommand{\St}[1][2]{{{}_{#1}\tilde{S}}}
\newcommand{\Sd}[1][2]{{{}_{#1}\tilde{S}^{d}}}
\newcommand{\Dt}[1][2]{{{}_{#1}\tilde{D}}}
\newcommand{\Dd}[1][2]{{{}_{#1}\tilde{D}^{d}}}
\newcommand{\Z}[1][2]{{{}_{#1}Z}}
\newcommand{\hZ}[1][2]{{{}_{#1}\hat{Z}}}
\newcommand{\M}[1][2]{{{}_{#1}\tilde{M}}}
\newcommand{\arcmin}{$^{\prime}$\xspace}
\newcommand{\Ltwo}{$\mathrm{L_{2}}$\xspace}

\usephysicsmodule{ab}
\newcommand{\mqty}[1]{\begin{matrix}
    #1
\end{matrix}}
\difdef { f } {}
{ outer-Ldelim = \left . , outer-Rdelim = \right |, sub-nudge = 0 mu }

\newcommand{\SC}{\textit{standard configuration}\xspace}
\newcommand{\BC}{\textit{balanced configuration}\xspace}
\newcommand{\FC}{\textit{flipped configuration}\xspace}
\newcommand{\moire}{Moir\'e\xspace}
\newcommand{\Tbetalow}{$T_{\beta}^{\mathrm{lower}}$\xspace}
\newcommand{\tbl}{T_{\beta}^{\mathrm{lower}}}
\newcommand{\sigmahits}{\sigma_{\rm{hits}}}

\newcommand{\Nmod}{N_{\rm{mod}}}
\newcommand{\Nmargin}{N_{\rm{margin}}}

\crefname{figure}{figure}{figures} \Crefname{figure}{Figure}{Figures}

\newcommand{\chapabstract}[1]{
\begin{quote}
    \singlespacing
    \rule{14cm}{1pt} #1 \vskip-4mm \rule{14cm}{1pt}
\end{quote}
}

\usephysicsmodule{ab} % 括弧サイズの自動調整
\usephysicsmodule{ab.braket} % braket記法
\usephysicsmodule{diagmat} % 対角行列
\usephysicsmodule[showleft=2,showtop=2]{xmat} % n×m行列

\makeatletter
\newcommand{\vb}{\@ifstar\boldsymbol\mathbf}
\newcommand{\va}[1]{\@ifstar{\vec{#1}}{\vec{\mathrm{#1}}}}
\newcommand{\vu}[1]{%
\@ifstar{\hat{\boldsymbol{#1}}}{\hat{\mathbf{#1}}}}
\makeatother
\DeclareMathOperator{\grad}{\nabla}
\renewcommand{\Re}{\operatorname{Re}}
\renewcommand{\Im}{\operatorname{Im}}
\newcommand\blfootnote[1]{
    \begingroup
    \renewcommand\thefootnote{}\footnote{#1}
    \addtocounter{footnote}{-1}
    \endgroup
}

\newcommand{\apj}{Astrophysical Journal}
\newcommand{\apjl}{Astrophysical Journal, Letters}
\newcommand{\apjs}{Astrophysical Journal, Supplement}

\newcommand{\aap}{Astronomy and Astrophysics}

\newcommand{\jcap}{Journal of Cosmology and Astroparticle Physics}

\newcommand{\mnras}{Monthly Notices of the RAS}
\newcommand{\na}{New Astronomy}

\newcommand{\nat}{Nature}

\newcommand{\prd}{Physical Review D}

\newcommand{\prl}{Physical Review Letters}

\newcommand{\sovast}{Soviet Astronomy}

\newcommand{\ptep}{Progress of Theoretical and Experimental Physics}

\begin{document}
    %\documentclass{book}
%\usepackage{graphicx}
%\usepackage[a4paper,margin=1in]{geometry}
%\usepackage[charter]{mathdesign}
%-----------------------------------------------------------------
%\begin{document}

% ----------------------------------------------------------------
\begin{titlepage}
\begin{center}

\vspace*{3cm}

% タイトルページ内でのみ行間を変更
\renewcommand{\baselinestretch}{2}\selectfont
{\Large Design of the full-sky scanning strategy and systematic effect \\ control in a cosmic microwave background probe}\\
\renewcommand{\baselinestretch}{1}\selectfont

% ----------------------------------------------------------------
\vspace{1.0cm}
{\Large by}\\[5pt]

\vspace{1.0cm}
{\Large Yusuke Takase}\\[5pt]
{\large takase\_y@s.okayama-u.ac.jp}\\[14pt]

\renewcommand{\baselinestretch}{2}\selectfont
\vspace{0.5cm}
{\Large A Doctoral Thesis}\\[5pt]
% ----------------------------------------------------------------

\vspace{2cm}
{\Large Submitted to} \\
{\Large the Graduate School of the Okayama University}\\
{\Large in March, 2025}\\
{\Large in Partial Fulfillment of the Requirements}\\
{\Large for the Degree of Doctor of Philosophy in Science}\\
{\Large in the Division of Mathematics and Physics} \\[1cm]
\renewcommand{\baselinestretch}{1}\selectfont

\vspace{0.5cm}
{\Large Thesis supervisor: Hirokazu Ishino} \\[14pt]
{\Large Professor of Astrophysics} \\[14pt]

\end{center}
\end{titlepage}
% ----------------------------------------------------------------
%\end{document}
    \frontmatter
    \setcounter{tocdepth}{3}

    \dominitoc
    \tableofcontents
    \listoffigures
    \listoftables
    %\blfootnote{Compiled at \today}
    \blfootnote{Compiled at March 28, 2025}
    %-------------------
    \mainmatter

    \chapter*{\center Abstract}

The quest for primordial $B$-mode polarization signatures in the Cosmic
Microwave Background~(CMB) is one of the most ambitious endeavors in
contemporary cosmology. Such a discovery would serve as a smoking gun for primordial
gravitational waves produced by tensor perturbations in the universe's nascent moments,
and would allow the precise determination of the tensor-to-scalar ratio, $r$ ---
a crucial parameter for distinguishing between competing inflationary models.
This in-depth investigation requires unparalleled precision in mapping the large-scale
angular scales of the CMB, necessitating full-sky observations from space-based platforms
free from the distortions of the Earth's atmosphere.

Given that the expected $B$-mode signatures are approximately three to four orders
of magnitude fainter than the CMB temperature anisotropies, the search for their
detection requires the implementation of well-designed in-flight calibration and
systematic effects mitigation strategies. Our investigation begins with a comprehensive
analysis of scanning strategy parameter optimizations, examining their influence
on three critical areas: the efficiency of in-flight calibration procedures, the
suppression of inherent systematic effects, and the development of robust null-test
methods for characterizing systematic effects.

The next generation of space-based observatories, exemplified by \LiteBIRD, which
incorporates Half-Wave Plate~(HWP) modulation technology, heralds a paradigm
shift in polarization measurements. This advanced approach enables single-detector
observations, bypassing the traditional need for differential detection by
orthogonal pairing employed in previous experimental configurations, and thereby
eliminating the systematic complexities associated with it. While the HWP
modulation mechanism is exceptionally effective in suppressing various
systematic effects, residual perturbations remain. Through sophisticated
analytical frameworks for the mapping process, emphasizing signals of \spin corresponding
to specific axiality, we systematically evaluate the suppression of these
systematic effects and elucidate optimal scanning strategy characteristics
within the multi-dimensional parameter space of spacecraft scan configurations, culminating
in an optimized scanning strategy design for comprehensive full-sky polarization
surveys.

In addition, we explore the effectiveness of optimal scanning strategies in
mitigating systematic effects through extensive simulation studies, including benchmark
systematic effects both with and without the implementation of the HWP. Using
the \spin-based mapping formalism, we evaluate the performance of the HWP. The
HWP-enabled configuration emerges as an effective solution in polarization
reconstruction, with negligible residual systematic effects. Conversely, the configuration
without HWP exhibits significant systematic effects that affect the estimation of
$r$, although the application of mitigation techniques we develop effectively
reduces systematic uncertainties and improves the estimation of $r$.

    \chapter{Introduction}
\minitoc

\chapabstract{ This thesis investigates scanning strategy optimization and systematic effect estimation and removal techniques for next-generation Cosmic Microwave Background~(CMB) polarization space missions. We focus on improving $B$-mode polarization measurements to detect primordial gravitational waves --- a key signature of cosmic inflation. The research develops novel map-making methods exploiting rotational symmetries, particularly for missions using continuously rotating Half-Wave Plates~(HWP). Through optimized scanning strategies and systematic effect control, this work aims to enhance the precision of tensor-to-scalar ratio, $r$ measurements, advancing our understanding of the early universe. }

\section{Probing inflationary models with CMB polarization}

How was our universe born and shaped? This remains one of the most profound
mysteries in modern cosmology. The key to unraveling this enigma lies in the Cosmic
Microwave Background~(CMB) --- the oldest light in the universe, still
observable today \cite{penzias1965measurement}.

The CMB, a cornerstone prediction of Big Bang cosmology, represents primordial
photons, emitted when the Universe was hot and dense, that have been redshifted
and cooled by cosmic expansion, and are now primarily observable in the microwave
regime \cite{gamow1946expanding,dicke1965cosmic}. This ancient radiation
exhibits a remarkably uniform $\sim$3\,K blackbody spectrum across all sky directions,
with tiny temperature fluctuations of approximately 10\,\si{\mu K} \cite{mather1990preliminary,bennett1996fouryear}.
These observations confirmed that the early universe existed in thermal
equilibrium with small matter density perturbations, which eventually seeded today's
large-scale structures, galaxies, and stars. However, this very uniformity presents
a paradox: the horizon problem, where causally disconnected regions inexplicably
share nearly identical temperatures, challenging the conventional Big Bang
theory.

Inflation theory emerged as a solution to this contradiction
\cite{sato1981firstorder,guth1981inflationary,kazanas1980dynamics}. By positing a
period of superluminal expansion in the early universe, inflation explains how
quantum fluctuations crossed the horizon scale and became frozen, resolving the horizon
problem. How can we verify that inflation actually occurred?

Inflationary models predict that quantum fluctuations in spacetime, specifically
tensor perturbations, were stretched into primordial gravitational waves
\cite{guzzetti2016gravitational,grishchuk1974amplification}. These waves,
inherently tensor in nature, generate two distinct types of quadrupole anisotropies
in the matter distribution perpendicular to their propagation direction. While
scalar acoustic waves also produce quadrupole anisotropies, they generate only one
type, due to their longitudinal nature. These quadrupole anisotropies, both
tensor and scalar, interact with the CMB through Thomson scattering, producing
polarization on large-angular scales during the last-scattering epoch. This polarization
can be decomposed in Fourier space into even-parity $E$-modes and odd-parity $B$-modes
\cite{seljak1997polarization,zaldarriaga1997allsky}. Importantly, while primordial
gravitational waves generate both $E$- and $B$-modes, acoustic waves produce only
$E$-modes. Thus, the detection of $B$-mode polarization, which is not degenerate
with $E$-modes, would provide compelling evidence for primordial gravitational waves
and allow us to estimate the tensor-to-scalar ratio $r$, which is proportional
to the square amplitude of these waves \cite{seljak1997signature}. The tensor-to-scalar
ratio is crucial for constraining the energy scale and potential shape of inflation,
thereby enabling us to test and refine inflationary models.

Just as previous generations confirmed the Big Bang theory through CMB
temperature anisotropies, we now seek to validate inflationary theory through CMB
polarization measurements, potentially uncovering direct evidence of cosmic
inflation.

\section{Importance of scanning strategy and systematic effects control}

The precision of CMB polarization measurements directly influences the estimation
of $r$. While statistical uncertainties in CMB observations have dramatically
improved through technological advancements in TES~(Transition-Edge Sensor)
bolometers and multiplexed detector arrays \cite{irwin1995TES,polarbear2010new},
with ground-based \textit{BICEP2/Keck} experiments achieving an upper limit of
$r<0.036$ \cite{2021BICEP}. However, three critical challenges remain for
further improvement:

\begin{enumerate}
    \item Ground-based experiments inherently struggle to observe large-angular scales
        where primordial gravitational wave $B$-modes are most prominent because
        its sky coverage is limited.

    \item Ground-based observation limitations:
        \begin{itemize}
            \item Systematic effects from atmospheric emission and fluctuations \cite{errard2015atomospheric}

            \item Restricted frequency bands due to atmospheric windows, leading
                to incomplete foregrounds (galactic synchrotron and dust emission)
                separation \cite{thomas1990atomwindow}
        \end{itemize}

    \item Instrumental systematic effects become increasingly problematic as the
        experiment's sensitivity is improved.
\end{enumerate}
These challenges can be addressed through:
\begin{itemize}
    \item Space-based observations for challenges 1 and 2

    \item A two-pronged approach for challenge 3:
        \begin{enumerate}
            \item Development of observation strategy that physically suppress systematic
                effects

            \item Analytical techniques to isolate systematic effects from observational
                data
        \end{enumerate}
\end{itemize}

For space missions, consideration of a full-sky scanning strategy is essential.
An optimized scanning strategy can average out systematic effects per sky pixel,
thereby suppressing spurious $B$-mode signals at large-angular scales \cite{OptimalScan}.
Additionally, while polarization signals exhibit a $180^{\circ}$ rotational symmetry,
referred to as \spin, most systematic effects display different symmetries. This
distinction allows for the effective separation of polarization signals from
systematics \cite{spin_characterisation}.\footnote{In this thesis we distinguish
the `spin' (normal font), which is the rotation around the maximum inertial axis
of a spacecraft, and `\spin' (italic font), which is an integer characterizing systematic
effects.}

The next-generation CMB polarization probe \LiteBIRD, led by JAXA, uniquely focuses
on polarization measurements \cite{PTEP2023}. It features a continuously
rotating Half-Wave Plate~(HWP)
\cite{sakurai2020breadboard,columbro2020polarization}, which controls incident light
polarization states modulates the incoming polarized light and effectively suppresses
instrumental $1/f$ noise at large-angular scales \cite{kusaka2014modulation}.

The optimization of scanning strategies for a space mission with HWPs that adopts
a polarization modulator remains understudied, as does the development of
analytical methods exploiting systematic effects' \spin with in presence of HWPs.
This research on full-sky scanning strategy optimization and systematic effect
removal techniques is crucial for improving polarization measurement precision
in next-generation space missions like \LiteBIRD.

This thesis addresses these challenges by proposing a scanning strategy that
effectively minimize systematic effects for next-generation CMB polarization
probes aimed at producing high-precision polarization maps. We present methods for
rapid systematic effect estimation and removal in observed polarization maps. By
combining these approaches, we aim to enhance $r$ measurement precision in next-generation
CMB polarization probes.

\section{Contents of this thesis}
The thesis is organized as follows: \cref{chap:CMB_polarization} presents a
comprehensive foundation of CMB observations and its anisotropies, elucidating the
fundamental concepts of $E$-mode and $B$-mode polarization. This chapter
establishes the critical connection between $B$-mode polarization and primordial
gravitational waves, describing their role as potential evidence for cosmic inflation.

\Cref{chap:CMB_space_missions} examines the trajectory of CMB space missions,
with particular attention to their scanning methodologies. The chapter introduces
the \LiteBIRD mission, which serves as the primary motivation for this research,
and elaborates on the theoretical foundations and practical implications of its Half-Wave
Plate (HWP) technology. \Cref{chap:formalism} introduces an innovative map-making
framework that exploits \spin moment decomposition of the signal. We develop a rigorous
mathematical formalism for reconstructing temperature and polarization maps from
time-series data in \spin space, pioneering the first comprehensive treatment of
HWP observations observations with a rotating HWP in multi-detector systems.

\Cref{chap:scanning_strategy_optimization} presents a detailed optimization
analysis of scanning strategies for next-generation CMB polarization probes, with
specific application to \LiteBIRD. This analysis yields essential design
principles for scanning strategies and spacecraft configurations in future CMB
polarimeters. \Cref{chap:systematics} investigates characteristic systematic effects
in CMB polarization measurements, demonstrating how an optimized scanning
strategy effectively suppresses some systematic effects. The chapter also proposes
novel methodologies for enhancing polarization map fidelity through systematic effect
isolation and quantifies their impact on $r$ measurements.

\Cref{chap:conclusion} synthesizes the research findings and explores future directions
in the field. The content of this thesis is based on the following publication
which is published by \textit{\jcap}:
\begin{description}
    \item[\cite{takase2024scan}] \textbf{Y.~Takase}, L.~Vacher, H.~Ishino, G.~Patanchon, L.~Montier, S.~Stever et al., 
    \textit{Multi-dimensional optimisation of the scanning strategy for the \LiteBIRD space mission}, \href{https://iopscience.iop.org/article/10.1088/1475-7516/2024/12/036}{\textit{Journal of Cosmology and Astroparticle Physic} \textbf{2024} (2024) 036} %\bibentry{takase2024scan}
\end{description}

    \chapter{CMB polarization}
\label{chap:CMB_polarization} \minitoc

\chapabstract{ This chapter explores the fundamental aspects of CMB polarization, beginning with its historical detection. We discuss the blackbody nature of CMB radiation and its temperature anisotropies, as observed by major satellite missions including \COBE, \WMAP, and \Planck. The chapter then delves into polarization anisotropies, explaining their quantification through Stokes parameters and their decomposition into $E$-mode and $B$-mode polarizations. We explain how these polarization modes are generated, with $E$-modes arising from scalar perturbations and $B$-modes from tensor perturbations. The chapter concludes with a detailed interpretation of angular power spectra and their implications for understanding primordial gravitational waves and cosmic inflation, highlighting current observational constraints on the tensor-to-scalar ratio. This chapter draws extensively from the comprehensive Japanese textbook on \textit{Cosmic Microwave Background Radiation} by Eiichiro Komatsu \cite{komatsu2019cmb}, which serves as the primary reference source.}

\section{Detection of the CMB}
In 1964, a groundbreaking discovery occurred at Bell Labs in Holmdel, New Jersey.
Arno Penzias and Robert W.~Wilson were conducting radio observations of
Cassiopeia at 7.35\,cm wavelength using a 6-meter horn antenna. While comparing sky
temperatures with a 5\,K calibration source, they detected an unexplained 3.5\,K
excess radiation that remained constant across the sky and seasons, even after accounting
for known atmospheric, ground, and antenna emissions. This mysterious signal was
later identified as the Cosmic Microwave Background~(CMB), whose existence had been
theoretically predicted by Robert H.~Dicke. The discovery was published in 1965,
with Penzias and Wilson reporting the observation ref.~\cite{penzias1965measurement}
and Dicke providing the theoretical framework ref.~\cite{dicke1965cosmic}. This momentous
finding earned Penzias and Wilson the 1978 Nobel Prize in Physics.

The Big Bang theory had predicted that the early universe existed as a hot,
dense plasma where light and matter maintained frequent energy exchanges through
interactions. This would have resulted in thermal equilibrium, producing
blackbody radiation. The spectral radiance of such radiation follows the Planck distribution:
\begin{equation}
    B_{\nu}(T) = \frac{2h \nu^{3}}{c^{2}}\frac{1}{\exp(h \nu /k_{B}T) - 1}. \label{eq:Planck}
\end{equation}
where $B_{\nu}(T)$ represents the spectral radiance (\si{W.\meter^{-2}.sr^{-1} Hz^{-1}}),
$c$ is the vacuum speed of light, $h$ is Planck constant, $k_{B}$ is Boltzmann
constant, $\nu$ denotes frequency, and $T$ is temperature.

\begin{figure}
    \centering
    \copyrightbox{ \includegraphics[width=0.9\columnwidth]{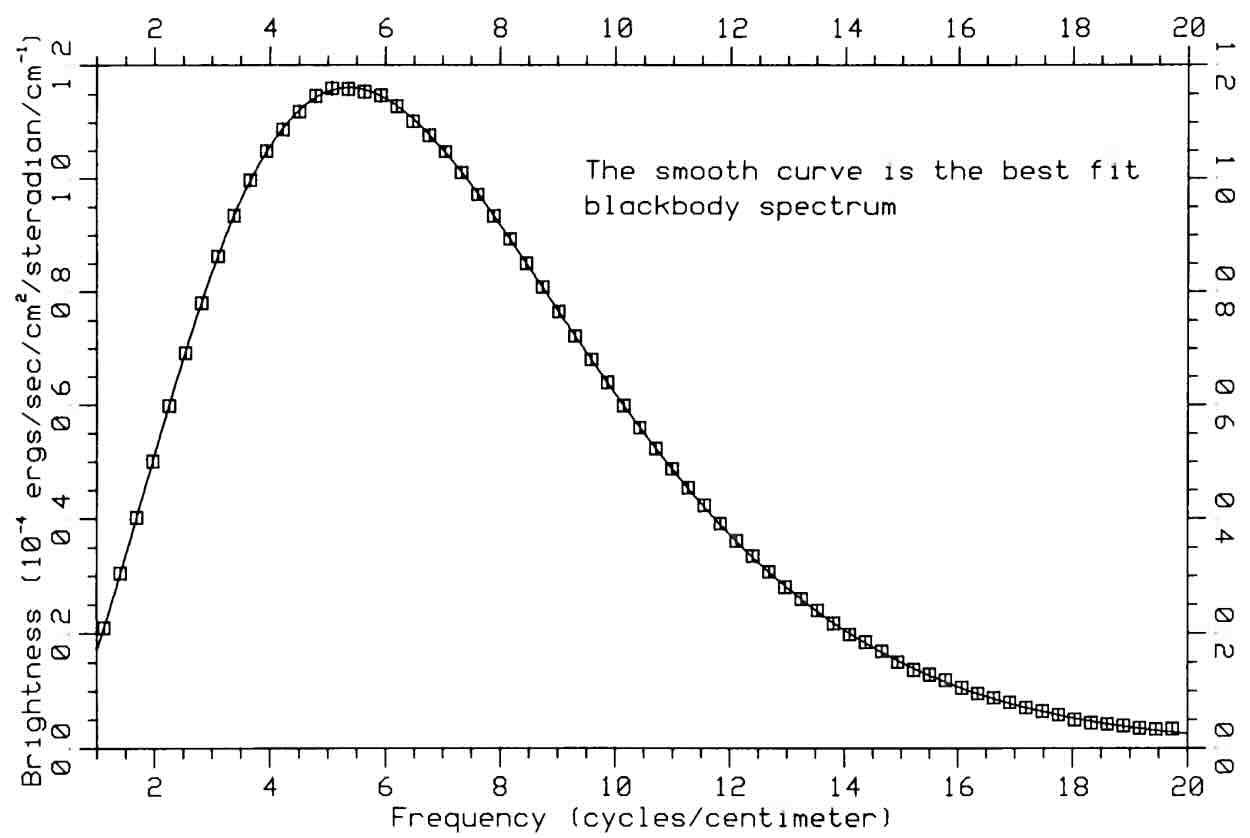} }{\copyright~1990 AAS. Reproduced with permission.}
    \caption[Blackbody radiation spectrum of the CMB measured by \COBE]{The
    \COBE satellite's Far-Infrared Absolute Spectrophotometer (FIRAS) measured
    CMB spectral radiance data (error bars show 1\% observational uncertainty) plotted
    against a 2.735\,K blackbody radiation spectrum fit (solid line). The figure
    is adapted from ref.~\cite{mather1990preliminary} with a permission from the
    authors and the publisher.}
    \label{fig:cobe_spectrum}
\end{figure}

NASA's Cosmic Background Explorer (\COBE) satellite, launched in 1989, provided
definitive evidence for the Big Bang theory through precise measurements of the
CMB spectrum and temperature. As shown in Figure \cref{fig:cobe_spectrum}, the
observed CMB spectral radiance perfectly matched the predicted blackbody
radiation spectrum \cite{mather1990preliminary}.

\section{Temperature anisotropies}
\subsection{Observation by space missions}
The early universe, according to Big Bang theory, existed as a hot, dense plasma.
In this state, frequent light-matter interactions created a fog-like condition,
making electromagnetic observation by telescopes impossible from present. The universe's
expansion and cooling eventually reached a point where these interactions ceased,
allowing light to travel freely for the first time. This crucial moment, called
the recombination epoch, marked the release of the CMB --- the oldest observable
light in the universe. The CMB filled the cosmos instantaneously as
recombination occurred nearly simultaneously throughout the universe. Today,
viewing looking at distant regions from Earth, we observe the CMB emitted for a surface
called the Last Scattering Surface~(LSS), which represents the recombination
epoch occurring approximately 380,000 years after the Big Bang (redshift $z \simeq
1100$).

\begin{figure}[h]
    \centering
    \begin{minipage}{\textwidth}
        \copyrightbox{ \includegraphics[width=1\columnwidth]{ 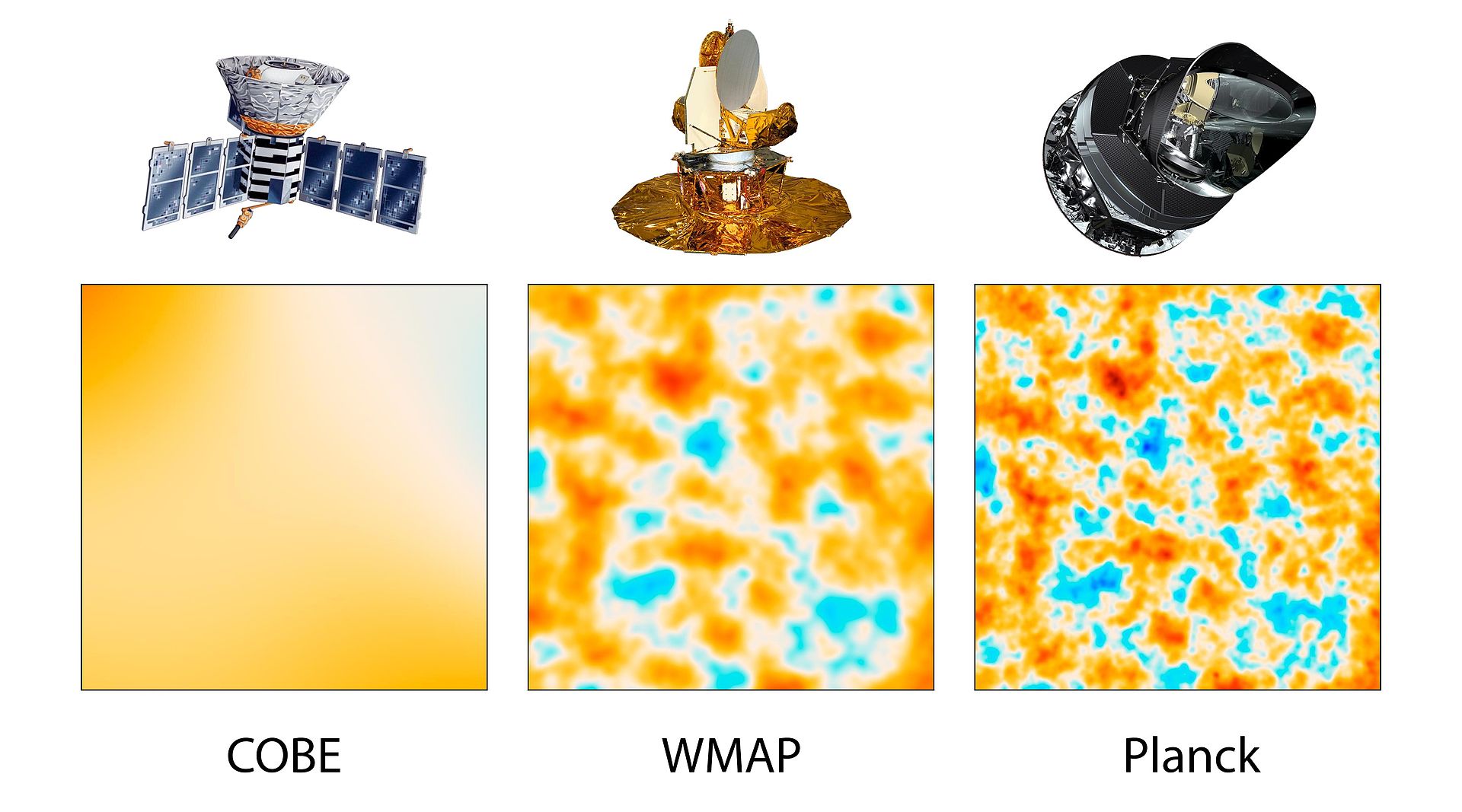 }}{\copyright~2013 NASA/JPL-Caltech/ESA}
        \caption[Overview of past CMB observation satellites]{(top panels)
        Juxtaposition of the \COBE, \WMAP, and \Planck satellite flight
        configurations, accompanied by their respective full-sky observations rendered
        in Mollweide projection, incorporating dipole anisotropy maps (explained
        in \cref{apd:CMB_dipole}). (bottom panels) Comparative visualization of each
        satellite's contribution to CMB map resolution within a $10^{\circ}\times
        10^{\circ}$ sky segment. The angular resolution capabilities evolved
        significantly: \COBE achieved $7^{\circ}$ \cite{bennett1996four}, \WMAP
        attained $0.3^{\circ}$ \cite{tegmark2003high}, and \Planck reached an unprecedented
        $5^{\prime}$ \cite{aghanim2020planck_overview}. Image credit: NASA/JPL-Caltech/ESA.\footnotemark}
        \label{fig:resolution_history}
    \end{minipage}
\end{figure}

\COBE's detailed observations revealed temperature anisotropies on the LSS, detecting
10\,\si{\mu K} variations. These anisotropies, which stem from early universe
quantum fluctuations, offer glimpses into the universe's state beyond the LSS.
Three major space missions -- \COBE, \WMAP, and \Planck have mapped these variations.
\Cref{fig:resolution_history} presents a comparison of their full-sky CMB
temperature anisotropy maps. \WMAP improved upon \COBE's observations,
determining the universe's age as $13.7 \pm 0 .2$ billion years
\cite{peiris2003first}. The subsequent \Planck mission refined this to $13. 797 \pm
0.023$ billion years and conducted comprehensive measurements of galactic foreground
emissions and CMB across frequencies from 26\,GHz to 1139\,GHz, providing
detailed insights into emission components
\cite{aghanim2020planck_cosmo,aghanim2020planck_overview}.

Since \COBE confirmed the CMB's perfect blackbody radiation nature, we can express
the relationship between CMB intensity $I_{\rm CMB}$ and temperature $T_{\rm CMB}$
using the Planck distribution from \cref{eq:Planck}:
\begin{align}
    I_{\rm{CMB}}(\nu) = B_{\nu}(T_{\rm{CMB}}).
\end{align}
The CMB temperature fluctuation $\Delta I(\nu)$ can be written as
\begin{align}
    \Delta I(\nu) = \left. \pdv{B_\nu(T)}{T}\right|_{T=T_{\rm{CMB}}}\Delta T.
\end{align}
In the Rayleigh-Jeans limit ($h\nu \ll k_{B} T$), the relative fluctuation becomes
\begin{align}
    \frac{\Delta I(\nu)}{I_{\rm{CMB}}}= \frac{\Delta T}{T_{\rm{CMB}}},
\end{align}
which remains frequency-independent. While $\Delta I$ is measured in \si{Jy.sr^{-1}},
CMB studies typically use $\Delta T$ in \si{K_{CMB}} (or \si{\mu K_{CMB}}), commonly
abbreviated as \si{K}. This measurement is referred to as the thermodynamic temperature.

\subsection{Quantification of anisotropies}
\footnotetext{\url{https://lambda.gsfc.nasa.gov/education/graphic_history/microwaves.html}}

To estimate cosmological parameters, including the age of the universe, from full-sky
CMB temperature anisotropy maps, observations require quantification of surface
fluctuations through spherical harmonic decomposition. Consider a spherical coordinate
system centered on the observer, where the line-of-sight direction unit vector
$\hat{n}$ is defined as:
\begin{equation}
    \hat{n}= (\sin \theta \cos \phi, \sin \theta \sin \phi, \cos \theta).
\end{equation}
For a temperature distribution $T(\hat{n})$ on the celestial sphere, the mean
temperature $\overline{T}$ is:
\begin{equation}
    \overline{T}= \int \frac{\odif{\Omega}}{4 \pi}T(\hat{n}) = \int_{-1}^{1}\frac{\odif{(\cos \theta)}}{2}
    \int_{0}^{2\pi}\frac{\odif{\phi}}{2\pi}T(\hat{n}).
\end{equation}

\begin{figure}[htbp]
    \centering
    \includegraphics[width=1\columnwidth]{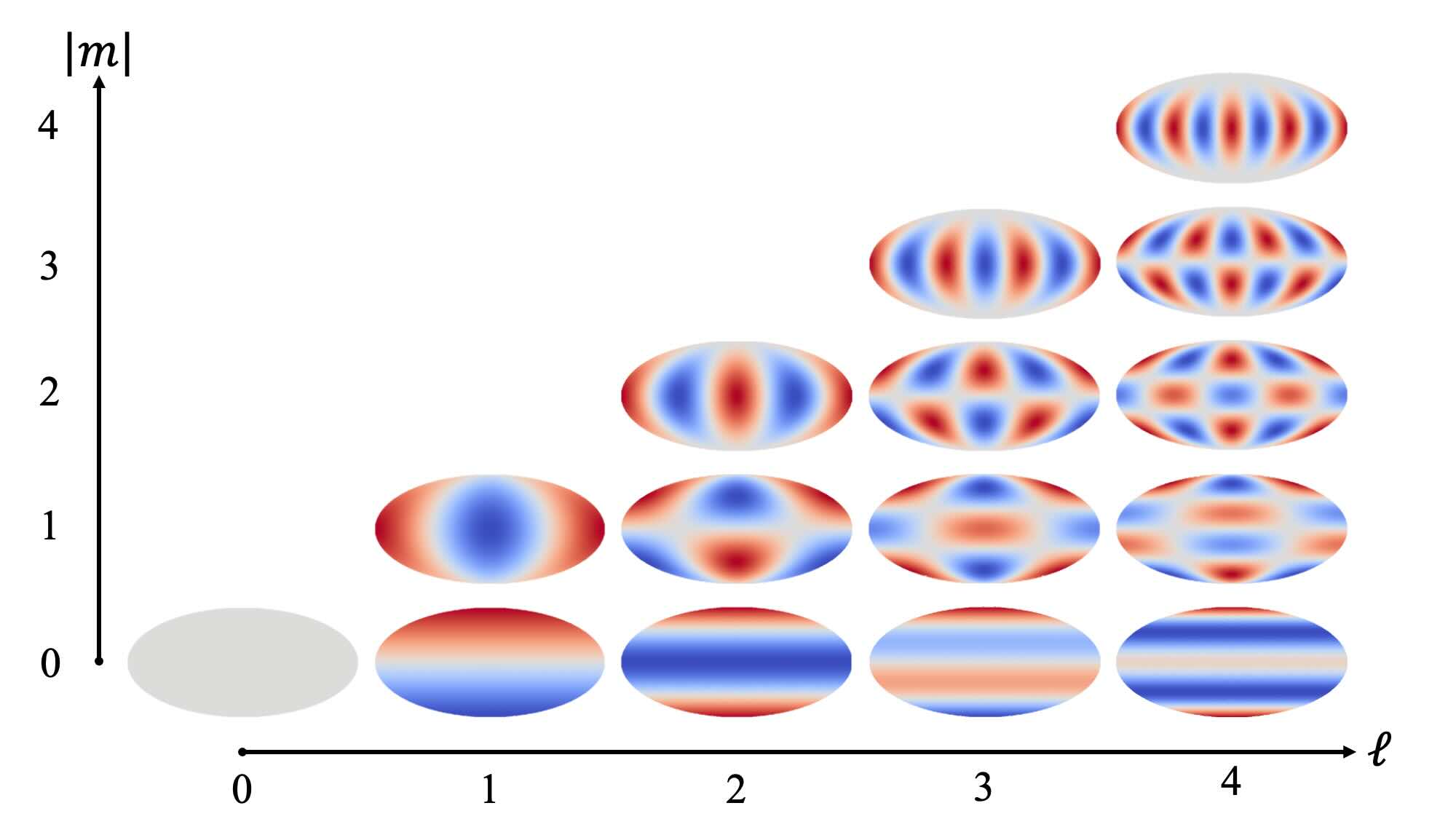}
    \caption{Mollweide projection of spherical harmonics for $\ell\leq4$ and
    $\ab|m|$.}
    \label{fig:sh_map}
\end{figure}

Temperature anisotropy $\Delta T$ is defined as the deviation from this mean: $\Delta
T = T - \overline{T}$. CMB temperature anisotropies arise from two sources: the Doppler
effect due to observer motion and inherent anisotropies from cosmic matter distribution
inhomogeneities. For mathematical analysis, temperature anisotropies can be expanded
in spherical harmonics $Y_{\ell m}(\hat{n})$:
\begin{equation}
    \Delta T (\hat{n}) = \sum_{\ell=1}^{\infty}\sum_{m=-\ell}^{\ell}a_{\ell m}Y_{\ell
    m}(\hat{n}), \label{eq:spherical}
\end{equation}
where $a_{\ell m}$ are harmonic expansion coefficients which can be obtained by
integrating over $\odif[order=2]{\hat{n}}$:
\begin{equation}
    a_{\ell m}= \int \odif[order=2]{\hat{n}}\Delta T(\hat{n})Y_{\ell m}^{*}(\hat{n}
    ).
\end{equation}
Here, $\ell=0$ represents the monopole component (uniform component), $\ell=1$ the
dipole component, and $\ell=2$ the quadrupole component. Higher $\ell$ values correspond
to finer angular structures, with angular scale $\delta \theta\simeq \pi/\ell$. Large-angular
structures are termed low-$\ell$ components, while small-angle structures are
high-$\ell$ components. While the expansion coefficients $a_{\ell m}$ depend on coordinate
origin, their squared sum $\sum_{m=-\ell}^{\ell}a_{\ell m}a^{*}_{\ell m}$
remains rotationally invariant. The spherical harmonics take the form:
\begin{equation}
    Y_{\ell m}(\hat{n}) = (-1)^{m}\sqrt{ \frac{2\ell+1}{4\pi} \frac{(\ell-m)!}{(\ell+m)!}}
    P_{\ell m}(\cos\theta)\exp(im\phi). \label{eq:spherical_harmonics}
\end{equation}
\Cref{fig:sh_map} illustrates spherical harmonics up to $\ell=4$ in Mollweide
projection, showing only $\ab|m|$ due to symmetry between $\pm m$.

\begin{figure}[h]
    \centering
    \copyrightbox{ \includegraphics[width=1\columnwidth]{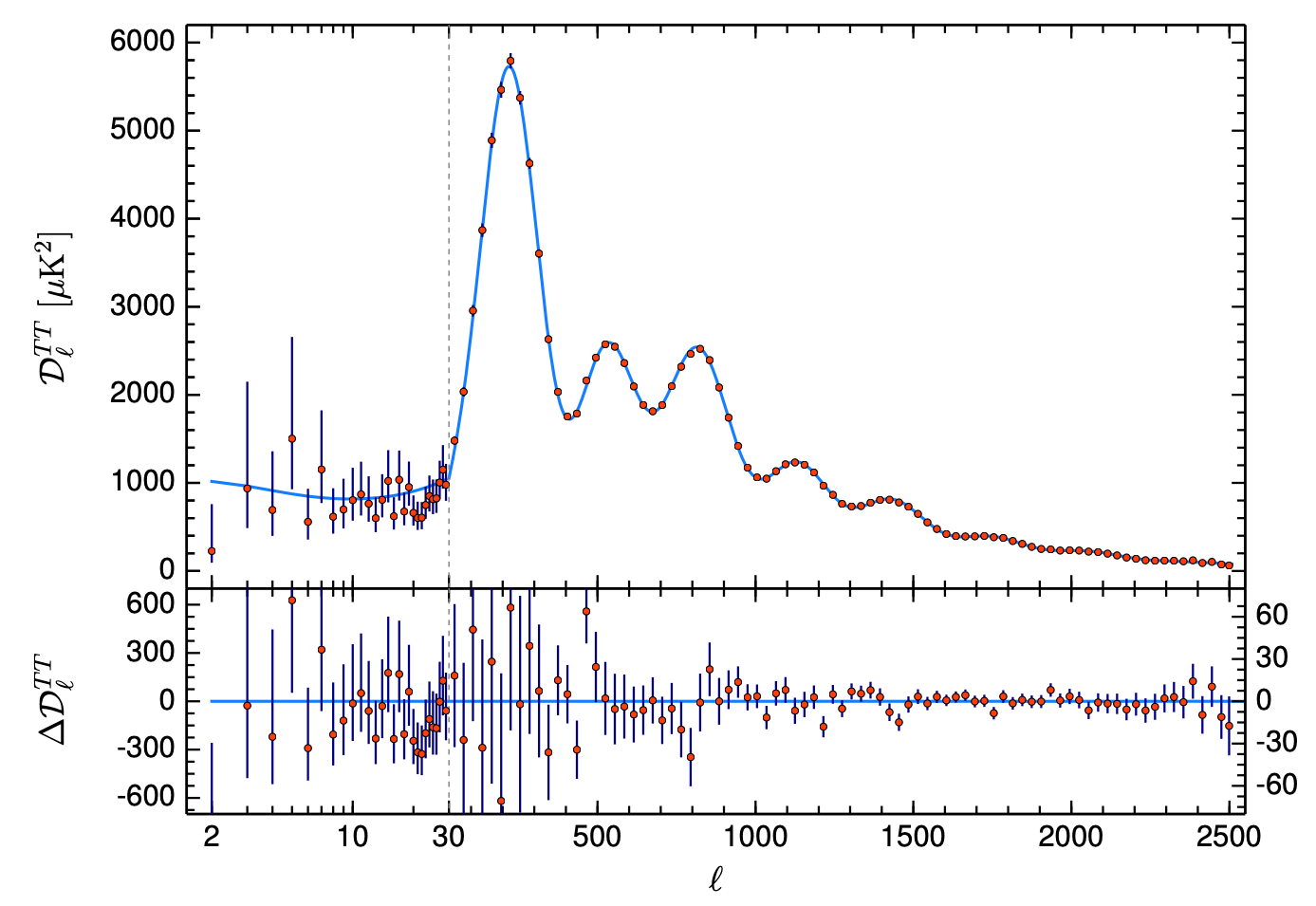} }{\copyright~2020 A\&A}
    \caption[The angular power spectrum of CMB temperature anisotropies measured
    by \Planck.]{CMB temperature anisotropy power spectrum measured by \Planck. The
    horizontal axis shows multipole moment $\ell$, with higher values
    corresponding to smaller angular scales. The vertical axis shows $D_{\ell}$ in
    units of \si{\mu K^2}. Red points indicate measurements, while the blue line
    shows the best-fit $\Lambda$CDM model. The figure is adapted from ref.~\cite{aghanim2020planck_cosmo}
    with a permission from the publisher.}
    \label{fig:planck_Cell}
\end{figure}

The angular power spectrum $C_{\ell}$ is defined as:
\begin{equation}
    C_{\ell}= \frac{1}{2\ell+1}\sum_{m=-\ell}^{\ell}a_{\ell m}a_{\ell m}^{*}. \label{eq:power_spectrum}
\end{equation}
To account for the Sachs-Wolfe effect \cite{SW_effect1967}, CMB studies commonly
use the scaled power spectrum:
\begin{equation}
    D_{\ell}= \frac{\ell(\ell+1)}{2\pi}C_{\ell}.
\end{equation}
\Cref{fig:planck_Cell} shows the CMB temperature anisotropy power spectrum
measured by \Planck. The larger error bars at low-$\ell$ reflect cosmic variance
--- a fundamental limitation arising from having only one observable universe,
even with perfect measurements.

\section{Polarization anisotropies}
\subsection{Stokes parameters}
The Stokes parameters provide a complete description of electromagnetic wave
states and serve as observational quantities for characterizing polarization
distributions across the celestial sphere. Consider measuring the electric field
oscillation direction of incoming light along a line of sight. Treating the
surrounding sky region as a plane, we define a Cartesian coordinate system $(x,y)$.
If we denote the electric field components in the $x$ and $y$ directions as
$E_{x}^{2}$ and $E_{y}^{2}$, the electromagnetic wave intensities can be expressed
as:
\begin{align}
    I_{x}\propto E_{x}^{2}, \\
    I_{y}\propto E_{y}^{2}.
\end{align}

The Stokes parameter $T$, representing total intensity (or temperature anisotropy),
also known as the unpolarized component, is given by:
\begin{align}
    T=I_{x}+ I_{y}.
\end{align}
To characterize polarization, we define the Stokes parameter $Q$ as the
difference between the squared amplitudes of electric fields oscillating in the
$x$ and $y$ directions:
\begin{align}
    Q \propto{E_x}^{2}-{E_y}^{2}.
\end{align}
The Stokes parameter $U$ is similarly defined in a coordinate system $(x',y')$ rotated
by $45$ degrees:
\begin{align}
    U \propto{E}_{x'}^{2}-{E}_{y'}^{2}.
\end{align}
These polarization components $Q$ and $U$ are coordinate-dependent. Under a coordinate
rotation by angle $\varphi$, they transform as:
\begin{align}
    \ab(\mqty{Q' \\ U'}) = \ab(\mqty{\cos2\varphi & \sin2\varphi \\ -\sin2\varphi & \cos2\varphi}) \ab(\mqty{Q \\ U}).
\end{align}
Using complex notation, we can express this transformation more concisely:
\begin{align}
    Q' \pm iU' = \exp(\mp 2i \varphi)(Q \pm iU). \label{eq:stokes_rot}
\end{align}

The factor of 2 in the exponential represents the \spin of the transformation, reflecting
how the Stokes parameters return to their original values under a $180^{\circ}$ rotation.
While a fourth Stokes parameter $V$ exists to describe circular polarization, we
omit its discussion as CMB polarization is known to be purely linear.

\subsection{\texorpdfstring{$E$}{E}-mode and \texorpdfstring{$B$}{B}-mode
polarization}

While Stokes parameters $Q$ and $U$ can describe polarization, their coordinate-dependent
nature can lead to confusion in quantitative polarization analysis. To address
this, we introduce coordinate-independent representations: $E$-mode and $B$-mode
polarization
\cite{seljak1997polarization,zaldarriaga1997allsky,kamionkowski1997statistics}. Consider
a small sky region around an arbitrary line of sight with a 2D Cartesian coordinate
system. Let the position vector from the center be:
\begin{align}
    \bm{\theta}= (x, y) = (\theta\cos\phi, \theta\sin\phi),
\end{align}
For Stokes parameters at position $\bm{\theta}$ on the celestial sphere, we can
express their 2D Fourier expansion using wave vector
$\bm{\ell}=(\ell\cos{\phi_\ell}, \ell \sin{\phi_\ell})$:
\begin{align}
    Q(\bm{\theta}) + iU(\bm{\theta}) = \int\frac{\odif[order=2]{\ell}}{(2\pi)^{2}}a_{\bm{\ell}}\exp(i\bm{\ell}\cdot\bm{\theta}),\label{eq:Q+iU}
\end{align}
Since $Q+iU$ transforms under coordinate rotation, the Fourier coefficients
$a_{\bm{\ell}}$ also change. To compensate for the $\exp(-2i\varphi)$ factor from
\cref{eq:stokes_rot}, we define:
\begin{align}
    a_{\bm{\ell}}= -{}_{2}a_{\bm{\ell}}\exp(2i\phi_{\ell})
\end{align}
Rewriting \cref{eq:Q+iU} yields:
\begin{align}
    Q(\bm{\theta}) \pm iU(\bm{\theta}) = -\int\frac{\odif[order=2]{\ell}}{(2\pi)^{2}}{}_{\pm2}a_{\bm{\ell}}\exp(\pm2i\phi_{\ell}+ i \bm{\ell}\cdot\bm{\theta}),\label{eq:Q+iU_new}
\end{align}
We introduce new quantities $E_{\bm{\ell}}$ and $B_{\bm{\ell}}$ defined as:
\begin{align}
    {}_{\pm2}a_{\bm{\ell}}\equiv -(E_{\bm{\ell}}\pm iB_{\bm{\ell}}),
\end{align}
This transforms \cref{eq:Q+iU_new} into:
\begin{align}
    Q(\bm{\theta}) \pm iU(\bm{\theta}) = \int\frac{\odif[order=2]{\ell}}{(2\pi)^{2}}\ab(E_{\bm{\ell}}\pm iB_{\bm{\ell}}) \exp\ab(\pm 2i \phi_{\ell}+ i \bm{\ell}\cdot \bm{\theta}),\label{eq:Q+iU_new2}
\end{align}
The relationship with original coefficients is:
\begin{align}
    E_{\bm{\ell}} & = -\frac{1}{2}\ab({}_{2}a_{\bm{\ell}}+{}_{-2}a_{\bm{\ell}}), \\
    B_{\bm{\ell}} & = \frac{1}{2}i\ab({}_{2}a_{\bm{\ell}}-{}_{-2}a_{\bm{\ell}}),
\end{align}
with complex conjugates $E_{\bm{\ell}}^{*}= E_{-\bm{\ell}}$ and $B_{\bm{\ell}}^{*}
= B_{-\bm{\ell}}$. The inverse transform is:
\begin{align}
    E_{\bm{\ell}}\pm iB_{\bm{\ell}}= \int \odif[order=2] \theta \ab(Q+iU)(\bm{\theta})\exp\ab(\mp i \phi_{\ell}- i \bm{\ell}\cdot \bm{\theta}),
\end{align}

\begin{figure}[htbp]
    \centering
    \includegraphics[width=0.8\columnwidth]{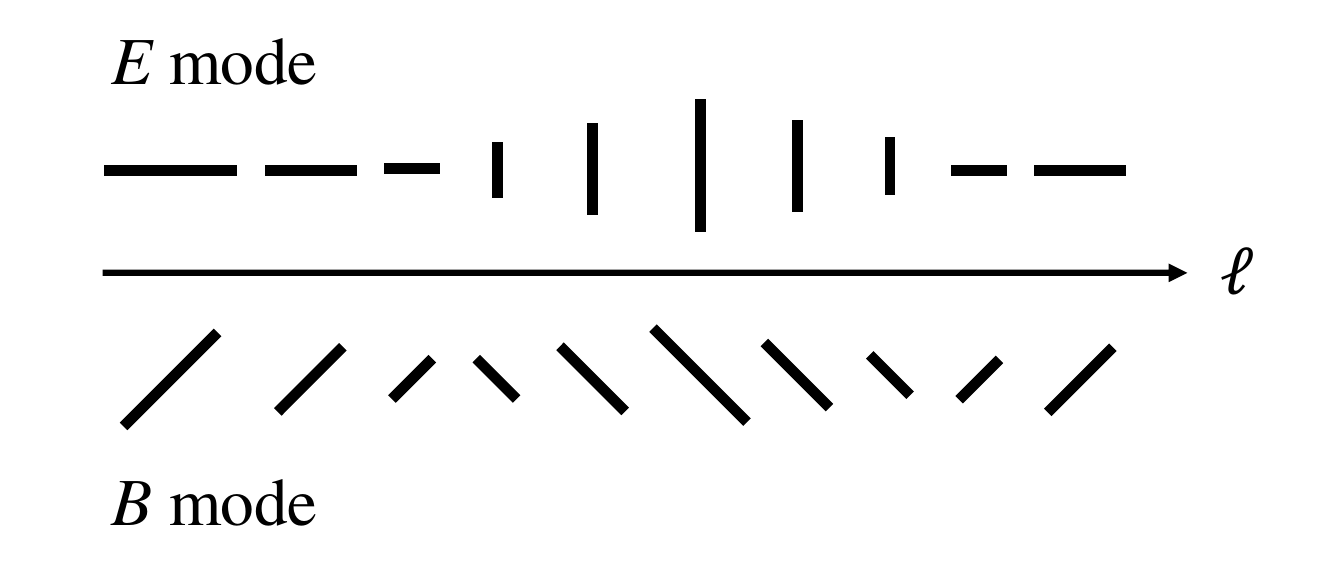}
    \caption[$E$- and $B$-mode polarization patterns]{Visualization of $E$- and $B$-mode
    polarization patterns. Line lengths represent Stokes parameter magnitudes. $\ell$-axis
    is the wavenumber vector direction. $E$-mode polarization is parallel or
    perpendicular to $\ell$, while $B$-mode polarization is rotated by $45^{\circ}$.}
    \label{fig:EandB}
\end{figure}

As shown in \cref{fig:EandB}, $E_{\bm{\ell}}$ represents polarization parallel or
perpendicular to the wave vector $\bm{\ell}$, while $B_{\bm{\ell}}$ represents
polarization rotated by $45^{\circ}$. These are termed $E$- and $B$-mode
polarization, respectively. While coordinate rotation affects both polarization
and $\bm{\ell}$ directions, their relative orientation (parallel, perpendicular,
or $45^{\circ}$) remains invariant, making $E$ and $B$ modes coordinate-independent
quantities. $E$ and $B$ modes can be viewed as Stokes parameters $Q$ and $U$ defined
with $\bm{\ell}$ as the $x$-axis. As evident in \cref{fig:EandB}, $E$-mode polarization
is parity-invariant, while $B$-mode polarization changes sign under parity
transformation, providing a clear distinction between these modes.

To define $E$- and $B$-mode polarization globally across the celestial sphere,
rather than just locally on a plane, we employ spherical harmonics instead of 2D
Fourier transforms. While standard spherical harmonics are invariant under
$\phi \rightarrow \phi+2\pi$ rotations, polarization, being a \spin-2 quantity, is
invariant under $\phi \to \phi+\pi$ transformations. We define basis functions
satisfying this transformation in 2D as:
\begin{align}
    {}_{\pm2}Y(\bm{\ell}) \equiv \frac{1}{\ell^{2}}\ab(\pdv{}{x}\pm i\pdv{}{y})^{2}\exp(i\bm{\ell}\cdot\bm{\theta}) = -\exp\ab(\pm2 i\phi_{\ell}+ i\bm{\ell}\cdot\bm{\theta}),\label{eq:harmonics}
\end{align}
These are called \spin-2 harmonic functions, derived from second-order
derivatives of the standard harmonic functions
$\exp(i\bm{\ell}\cdot \bm{\theta})$ used for Fourier transforms of \spin-0
quantities like temperature anisotropies. Using these, \cref{eq:Q+iU_new2}
becomes:
\begin{align}
    Q(\bm{\theta}) \pm iU(\bm{\theta}) = \int \frac{\odif[order=2]{\ell}}{(2\pi)^{2}}\ab( E_{\bm{\ell}}\pm iB_{\bm{\ell}}){}_{\pm2}a_{\bm{\ell}}{}_{\pm2}Y(\bm{\ell}),\label{eq:Y_ell}
\end{align}

We can generalize the definition of Stokes parameters $Q$ and $U$ by considering
arbitrary orthogonal basis vectors $\bm{e}_{1}, \bm{e}_{2}$ instead of fixed $x$
and $y$ directions. Defining complex basis vectors:
\begin{align}
    \bm{e}_{\pm}\equiv \frac{1}{\sqrt{2}}\ab(\bm{e}_{1}\pm \bm{e}_{2}),
\end{align}
\cref{eq:harmonics} can be rewritten as:
\begin{align}
    {}_{\pm2}Y(\bm{\ell}) = \frac{2}{\ell^{2}}\sum_{i,j}e_{\pm i}e_{\pm j}\tilde{\grad}_{i}\tilde{\grad}_{j}\exp(i\bm{\ell}\cdot\bm{\theta}),
\end{align}
where $\tilde{\grad}$ represents derivatives perpendicular to the line-of-sight direction
$\hat{n}$ on the celestial sphere. In spherical coordinates centered on the
observer, with $\bm{e}_{1}$ along $\theta$ and $\bm{e}_{2}$ along $\phi$, the
relationship between spherical and Cartesian Stokes parameters becomes:
\begin{align}
    (Q+iU)_{\rm spherical}= \exp(-2i\phi)(Q+iU)_{\rm cartesian},
\end{align}
Using \spin-$\pm2$ spherical harmonics $_{\pm2}Y_{\ell m}$, \cref{eq:Y_ell}
becomes \cite{zaldarriaga1997allsky}:
\begin{align}
    (Q \pm iU)(\hat{n}) = \sum_{\ell=2}^{\infty}\sum_{m=-\ell}^{\ell}{}_{\pm2}a_{\ell m}{}_{\pm2}Y_{\ell m}(\hat{n}).
\end{align}
The \spin-2 spherical harmonics can be expressed through second derivatives of standard
spherical harmonics:
\begin{align}
    {}_{\pm2}Y_{\ell m}= 2\sqrt{\frac{(\ell-2)!}{(\ell+2)!}}\sum_{i,j}e_{\pm i}e_{\pm j}\tilde{\grad}_{i}\tilde{\grad}_{j}Y_{\ell m}(\hat{n}),
\end{align}
Defining $_{\pm2}a_{\ell m}\equiv -(E_{\ell m}\pm i B_{\ell m})$, the full-sky $E$-
and $B$-mode polarization components are:
\begin{align}
    E_{\ell m} & = -\frac{1}{2}\ab({}_{2}a_{\ell m}+{}_{-2}a_{\ell m}), \\
    B_{\ell m} & = \frac{1}{2}i\ab({}_{2}a_{\ell m}-{}_{-2}a_{\ell m}),
\end{align}

The power spectra for $E$ and $B$ modes are \cite{kamionkowski1997statistics}:
\begin{align}
    \langle E_{\ell m}E_{\ell' m'}^{*}\rangle = C_{\ell}^{EE}\delta_{\ell\ell'}\delta_{mm'}, \\
    \langle B_{\ell m}B_{\ell' m'}^{*}\rangle = C_{\ell}^{BB}\delta_{\ell\ell'}\delta_{mm'},
\end{align}
The temperature-polarization correlation can be expressed as:
\begin{align}
    C^{XY}_{\ell}= \frac{1}{2\ell+1}\sum_{m}{{}_{s}a_{\ell m}^X}{{}_{s}a_{\ell m}^{Y*}},\label{eq:C_XX}
\end{align}
where $X,Y$ can be $T$, $E$, or $B$, and $s$ denotes the \spin. While
$C^{TT}_{\ell}$, $C^{TE}_{\ell}$, $C^{EE}_{\ell}$, and $C^{BB}_{\ell}$ have even
parity, $C^{TB}_{\ell}$ and $C^{EB}_{\ell}$ have odd parity and vanish due to
the absence of correlation between temperature anisotropies and $B$-mode
polarization, and between $E$- and $B$-mode polarization.

\subsection{Generation of polarization anisotropies}
\begin{figure}[htbp]
    \centering
    \includegraphics[width=1\columnwidth]{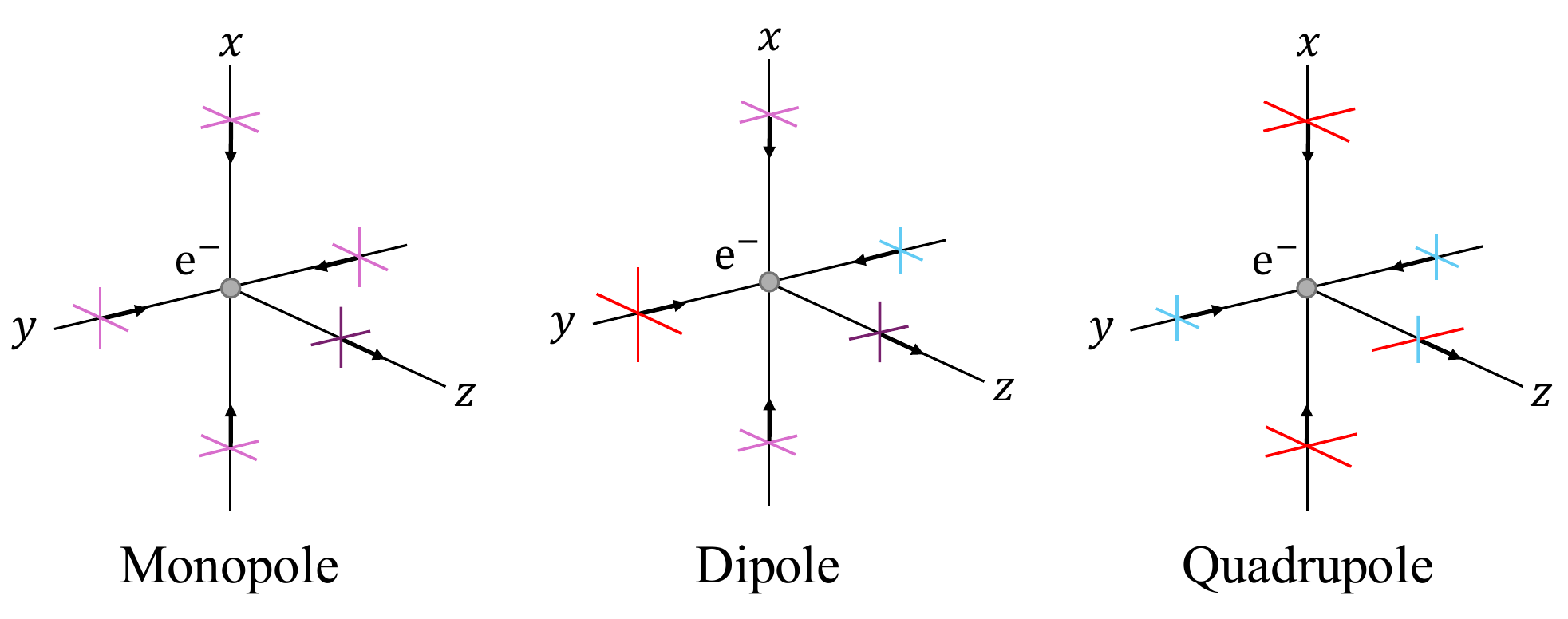}
    \caption[Generation of CMB polarization anisotropies from quadrupole
    radiation fields.]{Generation of CMB polarization anisotropies. The figure represents
    an electron at the origin of a Cartesian coordinate system. (left) Monopole
    radiation field with perfect spherical symmetry. (middle) Dipole radiation field
    with no net polarization along the $z$-axis. (right) Quadrupole radiation
    field producing polarized light along the $z$-axis.}
    \label{fig:pol_generation}
\end{figure}

The CMB polarization originates through Thomson scattering interactions between
photons and electrons \cite{polnarev1985polarization,hu1997polprimer}, and it was
discovered by DASI experiment in 2002 \cite{kovac2002poldetection}. To
understand this mechanism, consider a scenario depicted in \cref{fig:pol_generation},
where an electron is positioned at the origin of a three-dimensional Cartesian
coordinate system. When the surrounding radiation field exhibits monopole symmetry
(left), no net polarization is observed along the $z$-axis due to the perfect
spherical symmetry of the distribution. Similarly, a dipole distribution (middle)
produces no polarized emission along the observation direction. However, when
the radiation field possesses quadrupole anisotropy (right), Thomson scattering preferentially
generates polarized light along the $z$-axis. This fundamental process underlies
the generation of CMB polarization anisotropies.

In viscous fluids, anisotropic stress --- a drag-like force --- acts on the
fluid. Prior to the last scattering epoch, photons interacted with baryons
through Thomson scattering, which suppressed anisotropic stress. However, as recombination
occurred and the coupling between baryons and the photon fluid weakened,
anisotropic stress increased significantly, enabling the formation of anisotropies
\cite{ma1995cosmological}. The non-isotropic stress responsible for generating
quadrupole anisotropies originates from two distinct sources: acoustic waves (scalar
perturbations) and gravitational waves (tensor perturbations).

\subsection{Polarization from scalar perturbations}

\begin{figure}[htbp]
    \centering
    \copyrightbox{ \includegraphics[width=0.49\columnwidth]{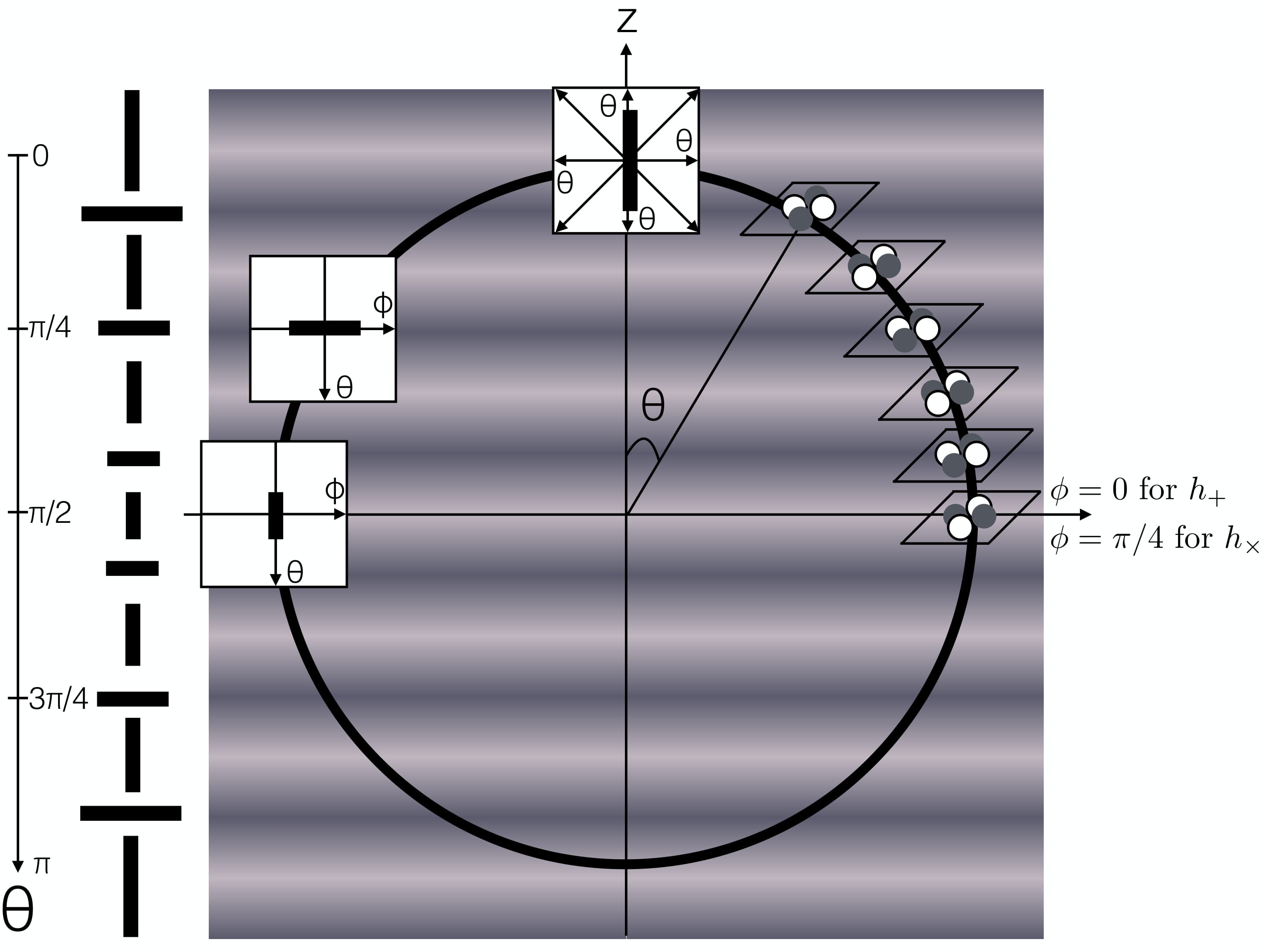} \includegraphics[width=0.49\columnwidth]{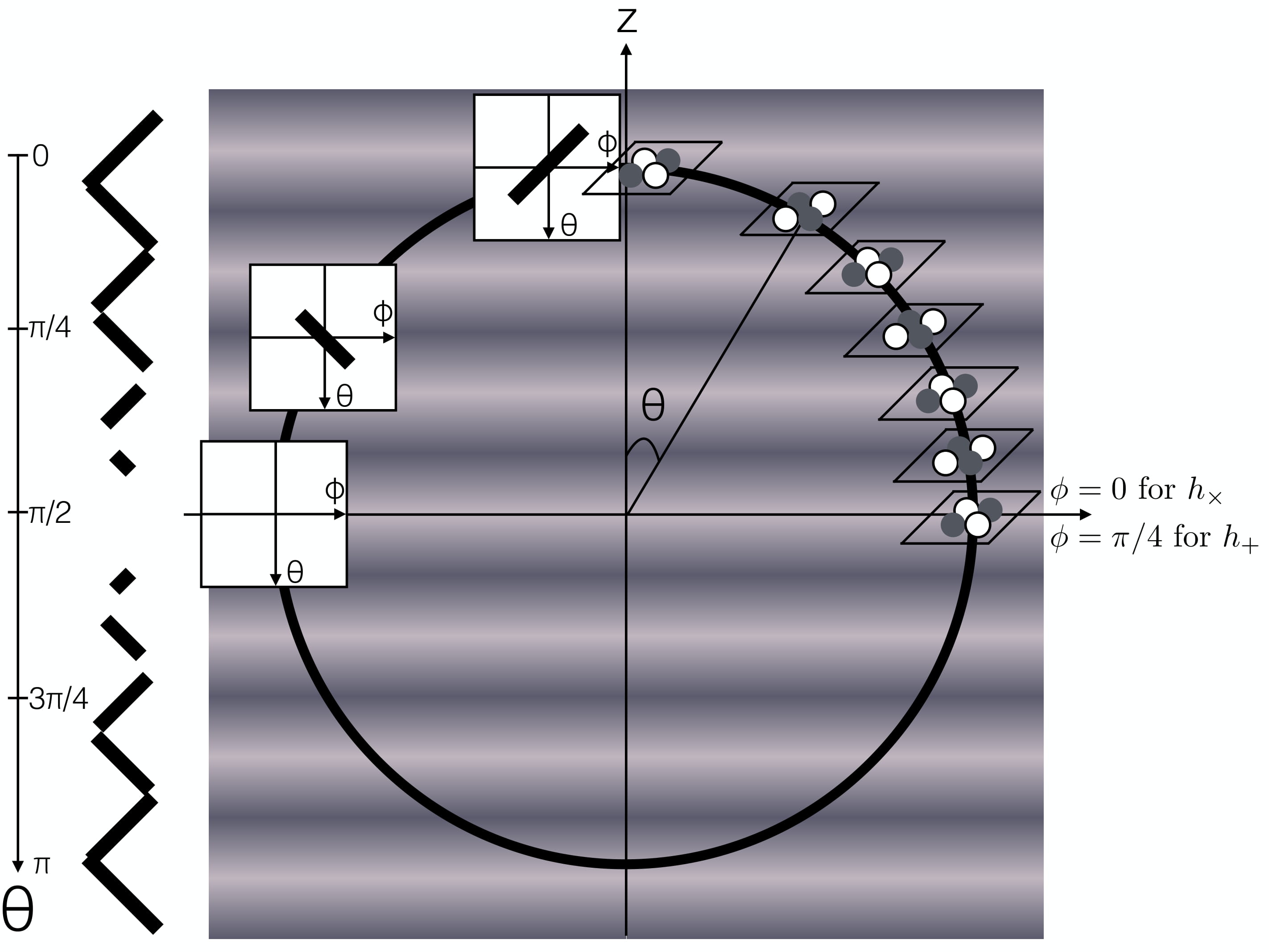} }{\copyright~2022 Springer Nature}
    \caption[ Generation of $E$- and $B$-mode polarization from scalar and
    tensor perturbations. ]{(left) $E$-mode polarization generated by scalar
    perturbations. (right) $B$-mode polarization generated by tensor perturbations.
    The left diagram shows the acoustic wave's effect on the photon fluid, with alternating
    black and gray shading representing regions of varying density. The right
    diagram depicts quadrupole temperature anisotropy distributions generated by
    gravitational waves. The figures are adapted from ref.~\cite{komatsu2022newphysics}
    with a permission fomr the publisher.}
    \label{fig:EB_pol}
\end{figure}

The scalar perturbations, i.e., the acoustic waves in fluids generate scalar anisotropic
stress. \Cref{fig:EB_pol} (left) illustrates how a single acoustic wave with
Fourier wavenumber $\bm{q}$ propagating along the $z$-axis generates $E$-mode
polarization.
The photon fluid develops regions of varying density along the $z$-axis following
the acoustic wave compressions and rarefactions, depicted by alternating black
and gray shading. The four circles in the right hemisphere represent temperature
quadrupole anisotropies as seen by electrons. Each circle contains an electron
at its center, with white circles indicating directions of higher temperature and
dark gray circles showing directions of lower temperature. Consider an observer
at the center of the celestial sphere. As the elevation angle $\theta$ increases
according to the left diagram in \cref{fig:EB_pol} (left), the observed
polarization direction and intensity vary as shown in the right diagram. The length
of the polarization bars is proportional to $\sin^{2}{\theta}$. In a coordinate
system where the Fourier wavevector $\bm{q}$ aligns with the $z$-axis, $Q>0$ and
$U=0$, reflecting the axial symmetry of scalar perturbations which are
independent of azimuthal angle $\phi$. This analysis reveals that under anisotropic
stress from acoustic waves, the polarization direction remains either
perpendicular or parallel to the direction of elevation angle $\theta$ variation,
conclusively identifying it as $E$-mode polarization.

\subsection{Polarization from tensor perturbations}

Under tensor-type anisotropic stress from gravitational waves, axial symmetry is
broken, introducing $\cos2\phi$ and $\sin2\phi$ dependencies in the azimuthal
angle. This generates non-zero $U$ polarization, corresponding to $B$-mode
polarization \cite{seljak1997signature}.

\begin{figure}[htbp]
    \centering
    \copyrightbox{ \includegraphics[width=0.49\columnwidth]{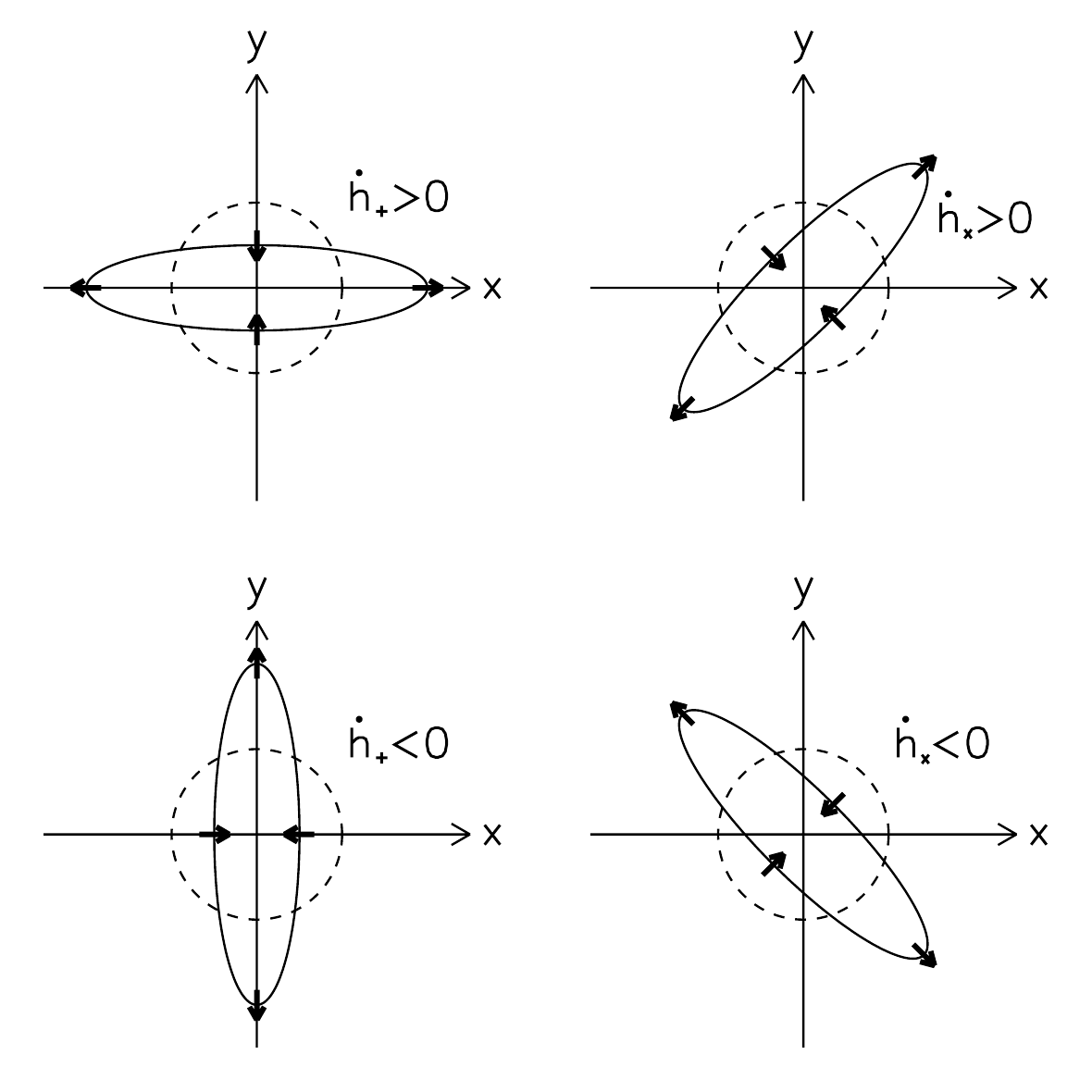} }{\copyright~2022 Springer Nature}
    \caption[Deformation modes of gravitational waves.]{Deformation patterns of
    gravitational waves. The left diagram shows the plus ($+$) mode, with space
    stretching along the $x^{1}$ direction and contracting along the $x^{2}$
    direction. The right diagram illustrates the cross ($\times$) mode, with
    space stretching along the $45^{\circ}$ direction in the $x^{1}$-$x^{2}$
    plane. The figure adapted from ref.~\cite{komatsu2022newphysics} with a
    permission.}
    \label{fig:GW_modes}
\end{figure}

In a three-dimensional space defined by coordinates $x^{1}x^{2}x^{3}$, let $D_{ij}$
represent gravitational waves. With wave propagation along the $x^{3}$ direction,
$D_{ij}$ has non-zero components in the $x^{1}$-$x^{2}$ plane:
\begin{align}
    D_{ij}= \ab(\mqty{h_+ & h_\times & 0 \\ h_\times & -h_+ & 0 \\ 0 & 0& 0}).
\end{align}
As illustrated in \cref{fig:GW_modes}, consider the effect of these gravitational
waves on test particles arranged in a circle in the $x^{1}$-$x^{2}$ plane.
When $h_{+}$ increases, space stretches along the $x^{1}$ direction, increasing
the distance between particles along this axis. Since gravitational waves preserve
area, the distance between particles along the $x^{2}$ direction decreases
correspondingly. The opposite occurs when $h_{+}$ decreases. Conversely, when $h_{\times}$
increases, space stretches along the $45^{\circ}$ direction in the $x^{1}$-$x^{2}$
plane while contracting in the perpendicular direction. These deformation patterns
define the plus ($+$) mode and cross ($\times$) mode, associated with $h_{+}$
and $h_{\times}$ variations respectively.

\begin{figure}[htbp]
    \centering

    \copyrightbox{ \includegraphics[width=0.49\columnwidth]{ 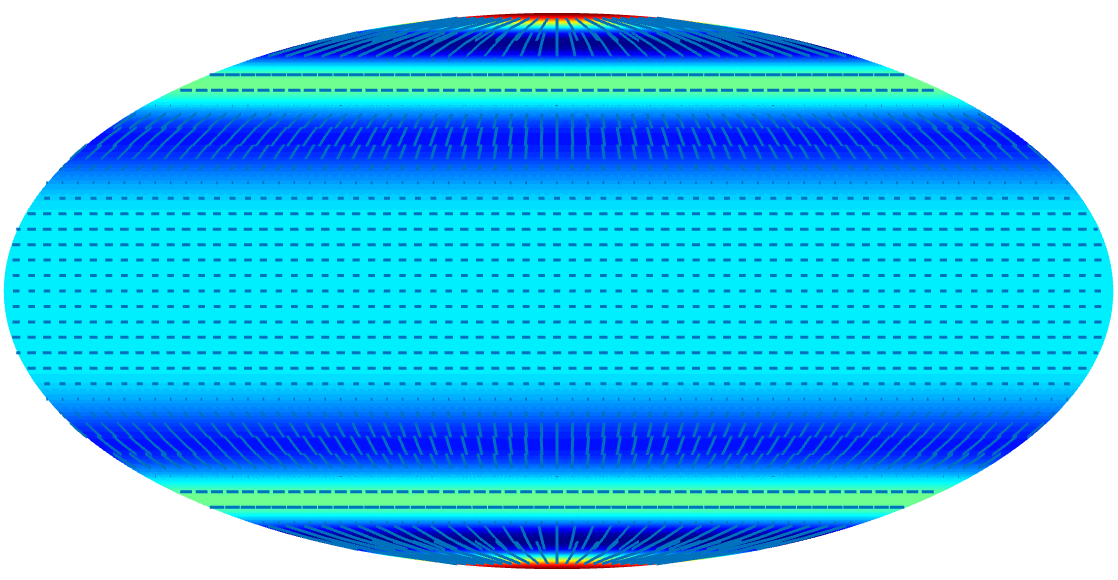 } \includegraphics[width=0.49\columnwidth]{ 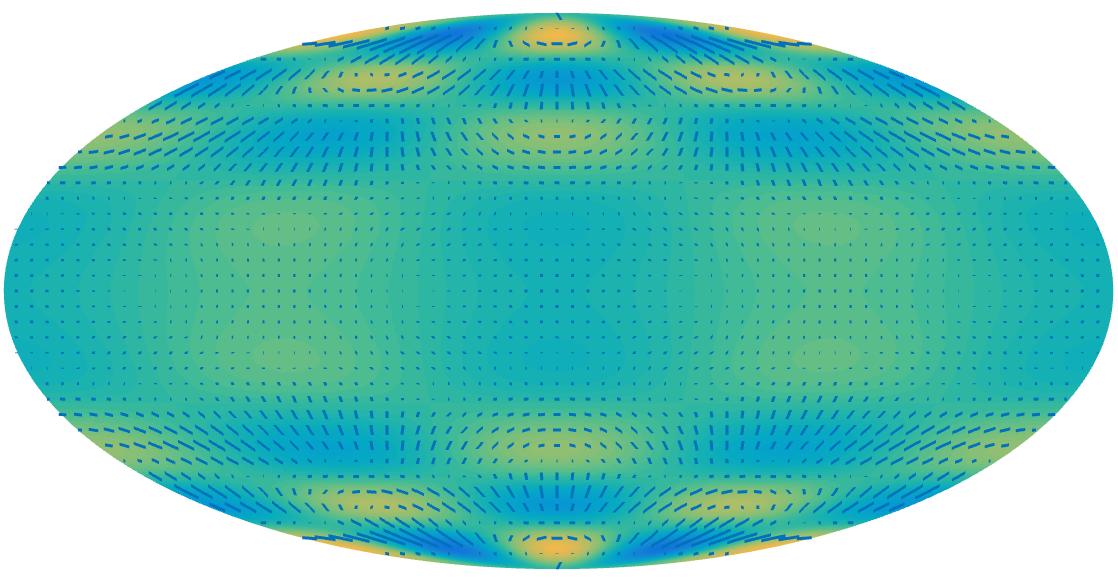 }}{\copyright~2016 Annual Reviews}
    \caption[Full-sky projections of $E$- and $B$-mode polarization]{ (left) Full-sky
    projection illustrating $E$-mode polarization patterns generated by a single
    Fourier mode of the density perturbation field. (right) Full-sky projection
    depicting $B$-mode polarization patterns induced by a single gravitational
    wave. The figures are adapted from ref.~\cite{kamionkowski2016quest} with a
    permission from the publisher.}
    \label{fig:EB_projection}
\end{figure}

For gravitational waves propagating along the $z$-axis, \cref{fig:EB_pol} (right)
shows the quadrupole distribution generated by ${\dot{h}_\times}$ with the horizontal
axis at $\phi=0^{\circ}$, or by ${\dot{h}_+}$ at $\phi=45^{\circ}$. Thomson scattering
of these quadrupole temperature anisotropies by electrons at the LSS produces the
polarization patterns and intensities illustrated in each figure. The observed
polarization distribution represents a superposition of these contributions. Notably,
in \cref{fig:EB_pol} (right), the polarization direction maintains a $45^{\circ}$
angle relative to the direction of polar angle $\theta$ variation, definitively
identifying it as $B$-mode polarization. \Cref{fig:EB_projection} illustrates the
full-sky projection of $E$- and $B$-mode polarization patterns, analogous to the
geometrical representation shown in \cref{fig:EB_pol}, providing a comprehensive
visualization of their distinct characteristics.

\section{Primordial gravitational waves from cosmic inflation}
\begin{figure}[htbp]
    \centering
    \copyrightbox{ \includegraphics[width=1\columnwidth]{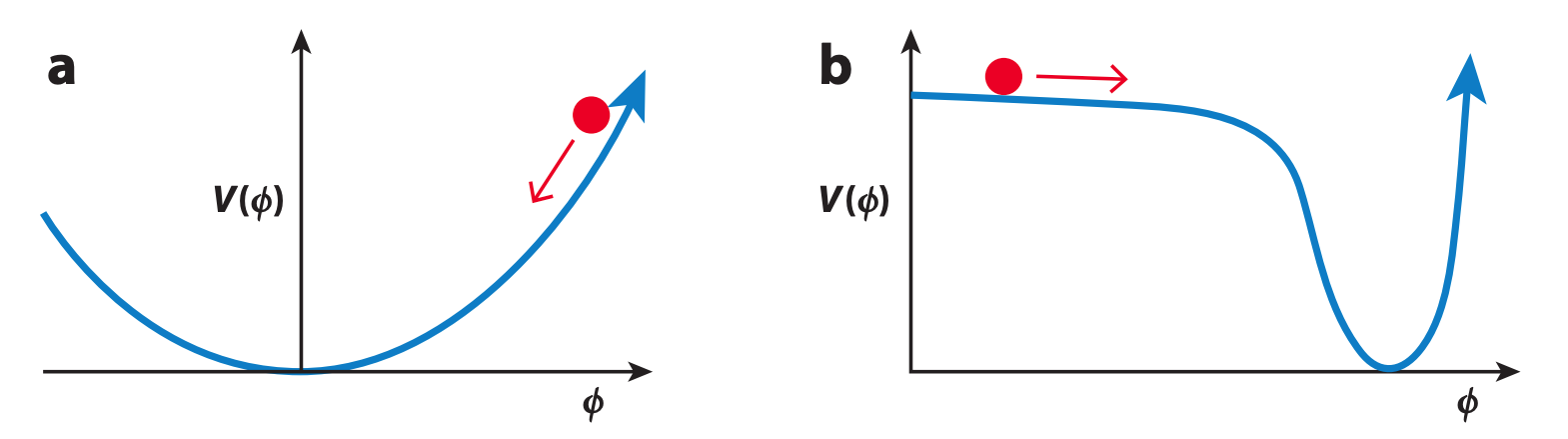} }{{\copyright~2016 Annual Reviews}}
    \caption[Typical potentials in the slow-roll inflation model]{Typical
    potentials in the slow-roll inflation model: quadratic potential (left) and hilltop
    potential (right). The figure is adapted from ref.~\cite{kamionkowski2016quest},
    with permission of the publisher.}
    \label{fig:slow_roll}
\end{figure}

The cosmic inflation provides crucial initial conditions for the Big Bang theory
and offers solutions to several problems associated with it, e.g., the horizon
and flatness problem
\cite{sato1981firstorder,guth1981inflationary,kazanas1980dynamics}. According to
this hypothesis, the universe underwent exponential spatial expansion in its
early stages, stretching quantum fluctuations over a brief period. The field
responsible for driving inflation is called the inflaton field, denoted by $\phi$.
Its time evolution is governed by the equation of motion and the Friedmann
equation \cite{friedmann1924freadmanneq}:
\begin{align}
    \ddot{\phi}+ 3H\dot{\phi}+ V'(\phi) = 0,                                      \\
    H^{2}= \frac{1}{3m_{pl}^{2}}\left[ V(\phi) + \frac{\dot{\phi}^{2}}{2}\right],
\end{align}
where $H$ is the Hubble parameter, $m_{pl}$ is the Planck mass defined as
$m_{pl}= (8\pi G)^{-1/2}$ with $G$ being the gravitational constant, and $V$ is the
potential of the inflaton field. Various inflation models are characterized by
the form of $V$. A notable example is the slow-roll inflation model for a single
scalar field \cite{lyth1999particle}. The slow-roll parameters $\epsilon_{V}$
and $\eta_{V}$ are defined as:
\begin{align}
    \epsilon_{V} & \simeq \frac{m_{pl}^{2}}{2}\left( \frac{V'}{V}\right)^{2}, \\
    \eta_{V}     & \simeq \frac{m_{pl}^{2}V''}{V}.
\end{align}
When $\epsilon_{V}\ll 1$ and $\ab|\eta_{V}| \ll 1$, the inflaton field changes
slowly over time, a process referred to as slow-rolling, analogous to a particle
gently rolling down a slope. Examples of potential shapes are shown in ref.~\cref{fig:slow_roll}
(left) representing a quadratic potential and (right) a hilltop potential. The quadratic
model is also known as the large-field model, while the hilltop potential is a
small-field model. In the large-field model group, there are chaotic
inflationary models, natural inflation models \cite{linder2003chaotic,freese1990natural}.
In the small-field model group, there is the Starobinsky model as a representative
model \cite{starobinsky1980newtype}. Their power spectra are given by:
\begin{align}
    \mathcal{P}_{R}(k) & = A_{s}\left( \frac{k}{k_{*}}\right)^{n_s - 1 + \frac{1}{2} \frac{dn_{s}}{d\ln k} \ln(k/k_*) + \ldots}, \\
    \mathcal{P}_{t}(k) & = A_{t}\left( \frac{k}{k_{*}}\right)^{n_t + \frac{1}{2} \frac{dn_{t}}{d\ln k} \ln(k/k_*) + \ldots},
\end{align}
where $k$ is the wavenumber, $A_{s}$ and $A_{t}$ are the amplitudes of scalar
and tensor modes, respectively, and $n_{s}$ and $n_{t}$ are their spectral
indices. The value of the inflaton field when the mode with wavenumber $k_{*}= a_{*}
H_{*}$ first crosses the Hubble radius is denoted by $\phi_{*}$. The pivot scale
$k_{*}= 0.05$ \si{Mpc^{-1}} is commonly used. The coefficients are related to
the slow-roll parameters as:
\begin{align}
    A_{s}    & \approx \frac{V}{24\pi^{2}m_{pl}^{4}\epsilon_{V}}, \\
    A_{t}    & \approx \frac{2V}{3\pi^{2}m_{pl}^{4}},             \\
    n_{s}- 1 & \approx 2\eta_{V}- 6\epsilon_{V}, \label{eq:ns}    \\
    n_{t}    & \approx -2\epsilon_{V}.
\end{align}
During the inflationary epoch, the potential maintains a notably flat
configuration, characterized by the conditions $\epsilon_{V}\ll 1$ and $\ab|\eta_{V}
| \ll 1$, though these parameters do not precisely vanish. As a consequence, the
scalar spectral index $n_{s}$ exhibits a slight deviation from unity, as evident
from \cref{eq:ns}. This departure from perfect scale invariance, known as the primordial
tilt, stands as a fundamental prediction of inflationary theory, definitively establishing
that $n_{s}\neq 1$. Historically, it was estimated that $n_{s}=0.9646 \pm 0.0098$
by 9-years \WMAP results which implies a non-zero tilt in the primordial spectrum
(i.e., $n_{s}< 1$) at $3.6\sigma$ \cite{hinshaw2013nineyear}, and later by
\Planck 2013 results, they detected $n_{s}= 0.9600 \pm 0.0072$, a $5.5\sigma$
departure from $n_{s}= 1$ \cite{planck2013cosmological}.

\subsection{Primordial gravitational waves and tensor-to-scalar ratio}
The inflation hypothesis, which posits exponential spatial expansion in the
early universe, also predicts the stretching of quantum fluctuations, leading to
the formation of gravitational waves \cite{grishchuk1974amplification}. These primordial
gravitational waves, generated during the universe's inception, are expected to exist
on ultra-long wavelengths spanning billions of light-years \cite{guzzetti2016gravitational}.
The $B$-mode polarization produced by these waves on the LSS is specifically
termed primordial $B$-mode polarization. The parameter representing the strength
of primordial gravitational waves is the tensor-to-scalar ratio, $r$, defined as:
\begin{align}
    r = \frac{\mathcal{P}_{t}(k_{*})}{\mathcal{P}_{R}(k_{*})}\approx 16\epsilon_{V}\approx -8n_{t}.
\end{align}
Accurate measurement of the $B$-mode polarization power spectrum allows for the
determination of $r$. Numerous CMB polarization experiments worldwide aim to detect
this signature, but the primordial $B$-mode polarization remains undetected due
to its extremely weak intensity, and $r$ has not yet been discovered. Currently,
various experiments provide upper limits on $r$. Before the results from \WMAP, the
chaotic inflation model with a potential proportional to $\phi^{4}$ was predominant
\cite{linder2003chaotic}. However, the \WMAP 3-years data released in 2006 ruled
out this model, leading to increasingly stringent constraints on $r$. Consequently,
many quadratic potential models, including those proportional to $\phi^{3}, \phi^{2}
, \phi$ and $\phi^{2/3}$ (classified under the left panel of
\cref{fig:slow_roll}), were also ruled out. This shift in understanding brought attention
to inflation models capable of explaining smaller values of $r$, with the
Starobinsky model \cite{starobinsky1980newtype} (classified under the right panel
of \cref{fig:slow_roll}) being a representative example.

The upper limit on $r$ at the pivot scale $0.002$ from the \Planck 2018 data and
the ground-based \textit{BICEP2/Keck} --- \textit{BK15} dataset is estimated to
be \cite{2021BICEP}:
\begin{align}
    r_{0.002}< 0.036 \quad \text{(95\% confidence level)}.
\end{align}
The value of $r$ is directly related to the energy scale of inflation $V^{1/4}$ \cite{achucarro2022inflation}:
\begin{align}
    V^{1/4}= 1.04\times 10^{16}\si{GeV}\ab(\frac{r}{0.01})^{1/4}.
\end{align}
Detecting a non-zero $r$ from primordial $B$-mode polarization observations
would not only validate the inflation hypothesis but also provide insights into
the inflation mechanism, contributing to the development of fundamental theories
such as quantum gravity and grand unified theories.

\section{Implications of angular power spectrum}
Having established precise definitions for $E$- and $B$-mode polarization, \cref{eq:C_XX}
enables computation of the auto-correlation power spectrum $D_{\ell}$,
illustrated in \cref{fig:power_spectrum}. The CMB power spectrum encodes a wealth
of cosmological information, and through sophisticated analysis techniques such
as maximum likelihood estimation, researchers have successfully constrained fundamental
cosmological parameters \cite{bond1997theoretical}. These calculations were
executed using \texttt{CAMB}, a sophisticated cosmological parameter-based power
spectrum software \cite{antony2000efficient,antony2013efficient,howlett2012cmb}.\footnote{\url{https://camb.readthedocs.io/en/latest/}}
Among these components, the `Primordial $C_{\ell}^{BB}$', representing the power
spectrum of $B$-mode polarization originating from primordial gravitational
waves, remains the sole unmeasured element. The distinct peaks, troughs, and
high-$\ell$ region attenuation observed in these power spectra each arise from
unique physical phenomena.

\subsection{Temperature anisotropies power spectrum}
The power spectrum of temperature anisotropies, denoted by $C_{\ell}^{TT}$,
exhibits remarkable characteristics across different angular scales. At large angular
scales ($\ell \lesssim 20$), the spectrum maintains near-constant values due to
the Sachs-Wolfe effect, independent of the $\ell(\ell+1)/2\pi$ scaling \cite{SW_effect1967}.
A prominent acoustic peak emerges around $\ell \lesssim 200$, which is measured by
BOOMERanG and MAXIMA-1 experiment \cite{bernardis2000boomerang,hanany2000maxima}.

The existence of acoustic oscillations in the pre-recombination universe was theoretically
established in \cite{peebles1970primeval}. The horizon, defined as the causal region
where information propagates at light speed, is approximately determined by the
inverse of the Hubble parameter, commonly known as the Hubble length \cite{bond1987statistics}.
An analogous concept applies to acoustic waves within the baryon-photon fluid
constrained by Thomson scattering, termed the sound horizon. During inflation, fluctuations
remain frozen outside the horizon due to exponential expansion. Upon inflation's
conclusion, these fluctuations re-enter the horizon, establishing density
distributions in the baryon-photon fluid and generating sound waves.

The series of peaks observed around $\ell \lesssim 1000$ correspond to the
fundamental mode and subsequent overtones of these sound waves. For smaller angular
scales ($\ell \gtrsim 1000$), the acoustic oscillations experience significant
attenuation due to the Silk damping effect \cite{silk1968cosmic}. This damping occurs
as the universe transitions to transparency and the baryon-photon fluid coupling
weakens, causing exponential suppression of short-wavelength sound waves.

\subsection{\texorpdfstring{$E$}{E}-mode power spectrum}
The power spectrum $C_{\ell}^{EE}$ of $E$-mode polarization exhibits peaks and
troughs that are anti-phased with those of $C_{\ell}^{TT}$. This phase opposition
stems from the dominant role of scalar anisotropic stress, originating from
acoustic waves, in generating $E$-mode polarization \cite{hu1996acoustic}.

The general solution to the acoustic wave equation can be expressed as a linear combination
of sine and cosine terms with constant coefficients. For adiabatic perturbations,
the cosine term dominates, while for Corld Dark Matter (CDM) curvature
perturbations, the sine term prevails. Observational data of the temperature
power spectrum aligns well with the cosine-model solution characteristic of
adiabatic perturbations. The velocity field potential solution must follow a sine-model
pattern due to the connection between anisotropic stress and energy conservation.

Since scalar perturbations generate anisotropic stress through velocity field gradients,
the resulting $E$-mode power spectrum naturally creates with a phase opposite to
that of the temperature anisotropies power spectrum, as shown in \cref{fig:power_spectrum}.
A comprehensive derivation from the Boltzmann equations can be found in
\cite{hu1997cmbanisotropies}, which provides a detailed theoretical framework for
understanding these phase relationships.

Before recombination, Thomson scattering suppressed anisotropic stress in the
baryon-photon fluid, preventing polarization. However, after recombination,
while acoustic waves began to experience Silk damping, polarization started to form.
Thus, for $\ell \lesssim 200$, $C_{\ell}^{TT}$ begins to decrease, whereas $C_{\ell}
^{EE}$ increases. At higher $\ell$, the $E$-mode power spectrum also diminishes due
to Silk damping.

A notable feature of $E$-mode polarization is the peak at $\ell \lesssim 20$.
This peak arises from the reionization of hydrogen atoms, which were ionized by ultraviolet
radiation from newly formed stars after recombination. This event, known as the cosmic
reionization, occurred around redshift $z \lesssim 20$. The scattering probability
of photons with intergalactic gas, including hydrogen, is described by the optical
depth $\tau$, and the polarization intensity depends on $\tau$.

Temperature anisotropies and polarization with wavelengths shorter than the
Hubble length at redshift $z$ are smoothed by Thomson scattering, decaying as $\exp
(-\tau)$. In contrast, fluctuations with wavelengths comparable to the Hubble length
are less affected by Thomson scattering, and their polarization intensity is
proportional to $\tau$. Since cosmic reionization occurred at $z \lesssim 20$,
new polarization is generated on relatively large angular scales ($\ell \lesssim
10$).

Currently, $\tau$ remains the least precisely determined parameter in standard cosmology,
limiting our ability to compare CMB anisotropies with matter distribution
clustering, particularly for measuring neutrino mass sums \cite{allison2015towards,divalentino2016cosmological,
giusarma2016improvement,boyle2018deconstructing,archidiacono2017physical}.

\subsection{Lensing \texorpdfstring{$B$}{B}-mode power spectrum}
\label{sec:lensing}

The lensing $C_{\ell}^{BB}$ represents the power spectrum of $B$-mode
polarization generated by the gravitational lensing \cite{einstein1911einfluss}.
While $B$-mode polarization was initially thought to be produced solely by tensor
anisotropic stress from gravitational waves, gravitational lensing can convert
$E$-mode polarization into $B$-mode polarization (and vice versa, though this
effect is minor). This lensing effect mixes power spectra of different $\ell$
values. The mixing effect depends on the variance of the angle difference of light
bent by gravitational lensing at two points on the celestial sphere. This can be
expressed as a convolution integral with a Gaussian factor involving the
variance and $\ell$ \cite{seljak1996gravitational}.

As a result, the lensing $C_{\ell}^{BB}$ appears as a smoothed version of the $E$-mode
polarization. This $B$-mode polarization from gravitational lensing was first
detected by the \textit{BICEP2} experiment in 2016 \cite{ade2016bicep2}.
Gravitational lensing affects not only polarization but also the temperature
power spectrum, causing a smoothing effect. This phenomenon was first observed
by ref.~\cite{keisler2011measurement}, predating the detection of lensing $B$-mode
polarization.

\subsection{Primordial \texorpdfstring{$B$}{B}-mode power spectrum}
\label{sec:Primordial_spectrum}

The Primordial $C_{\ell}^{BB}$ represents the power spectrum of $B$-mode polarization
generated by primordial gravitational waves. The peaks at $\ell \sim 2$ and
$\ell \sim 80$ are known as the reionization and recombination bumps, respectively.
While the latter peak is historically termed the recombination bump, it's worth noting
that there was no prior period of proton-electron combination before the last scattering
epoch. At large $\ell$, the power spectrum's attenuation stems from the redshift-induced
amplitude decay of gravitational waves during cosmic expansion, exhibiting a
power-law decay with $\ell$. Unlike acoustic waves affected by Silk damping,
gravitational waves, being transverse waves, experience no viscous dissipation. The
oscillations observed at $\ell \gtrsim 100$ differ from acoustic oscillations;
each $\ell$ corresponds to fluctuations entering the horizon at different times,
with oscillation phases determined by the number of gravitational wave cycles between
horizon entry and last scattering.

\begin{figure}[htbp]
    \centering
    \includegraphics[width=1\columnwidth]{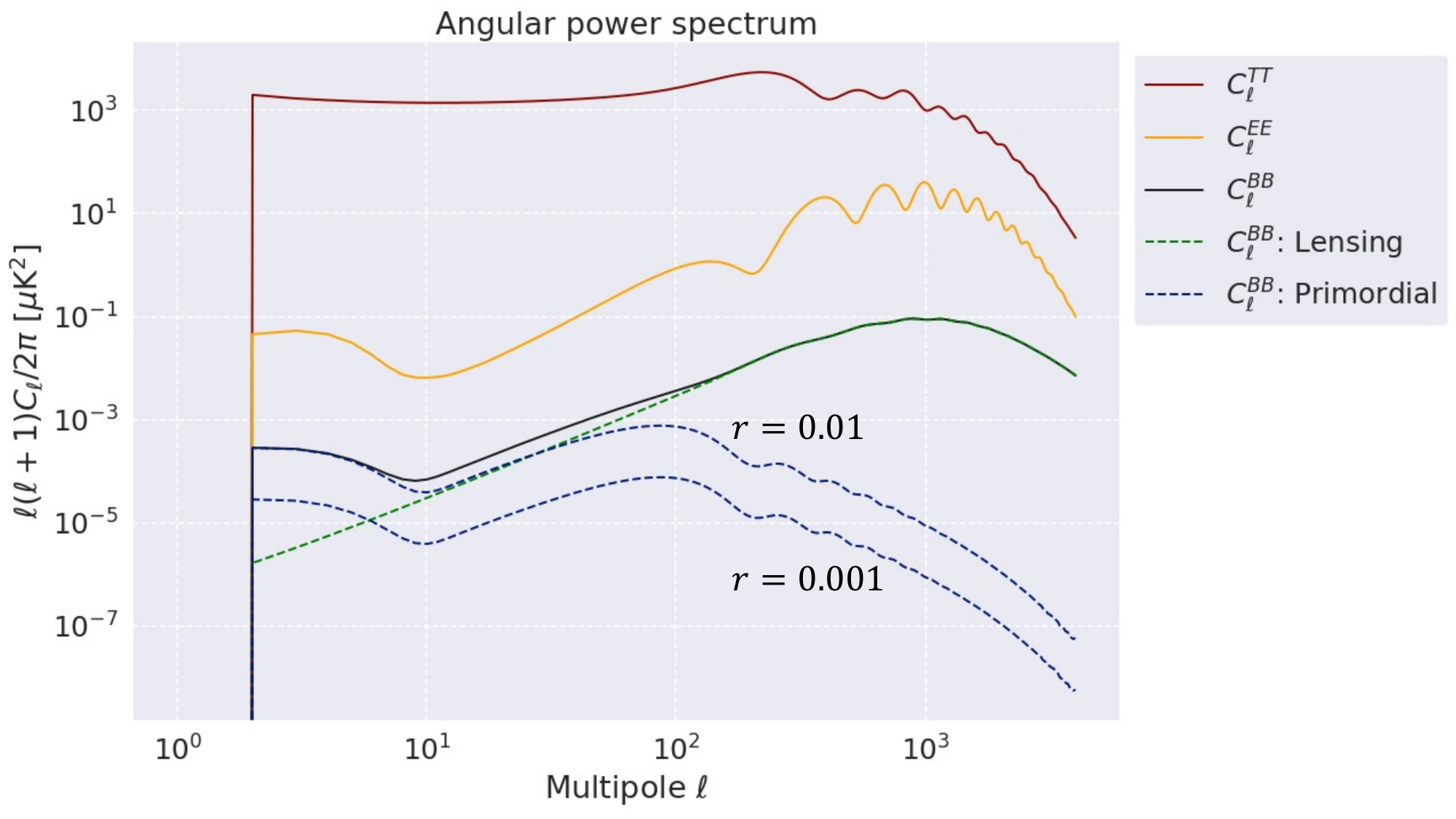}
    \caption[ Power spectra $C^{TT,EE,BB}_{\ell}$ calculated using \texttt{CAMB} ]{Power
    spectra $C^{TT,EE,BB}_{\ell}$ calculated using \texttt{CAMB}. The
    $C^{BB}_{\ell}$ components are shown separately as lensing (green dashed line)
    and primordial (blue dashed line). The observed $C^{BB}_{\ell}$ would be the
    sum of these components; for instance, if the true value is $r = 0.01$, the
    measured $C^{BB}_{\ell}$ would follow the black line.}
    \label{fig:power_spectrum}
\end{figure}

The primordial gravitational wave amplitude, and consequently its power spectrum,
scales proportionally with the tensor-to-scalar ratio $r$.
\Cref{fig:power_spectrum} displays primordial $C_{\ell}^{BB}$ for $r = 0.01$ and
$0.001$. The observed $B$-mode power spectrum $C_{\ell}^{BB}$ comprises both lensing
and primordial components. Since representative single-field slow-roll inflation
models predict $r \geq 0.01$ \cite{lyth1999particle}, we illustrate the total $C_{\ell}
^{BB}$ as the sum of lensing and primordial components assuming $r = 0.01$.

Determining $r$ requires precise measurement of the reionization and recombination
bump amplitudes. Particularly, detecting smaller $r$ values (e.g., $r = 0.001$) necessitates
accurate measurement of the reionization bump. Since this feature manifests at
low $\ell$, corresponding to large angular scales on the celestial sphere,
observations covering wide sky areas become crucial.

In 2014, the \textit{BICEP2} experiment conducted unprecedented large-angular
scale $B$-mode polarization measurements from the ground, initially reporting $r
= 0.20$ \cite{PhysRevLett.112.241101}. However, this single-frequency (150 \si{GHz})
measurement could not adequately account for galactic dust foreground
contamination, and subsequent joint analysis with \Planck data led to its retraction.
This experience highlighted the importance of multi-frequency observations for
proper foreground removal, in addition to large-angular scale coverage.

As of 2022, the \textit{BICEP2} collaboration, incorporating data from \WMAP, \Planck,
and the ground-based \text{Keck} Array, has established an upper limit of $r_{0.05}
< 0.0 36$ at 95\% confidence level \cite{2021BICEP}. This analysis achieved an unprecedented
precision with an uncertainty of $\sigma(r) = 0.009$, representing the most
stringent constraints on $r$ to date.

    \chapter{Scanning strategies of CMB space missions}
\label{chap:CMB_space_missions} \minitoc

\chapabstract{ This chapter delves into the scanning strategies utilized by various CMB space missions, from the pioneering \COBE's Sun Synchronous Orbit to the advanced Lagrange point 2-based missions. We follow the evolution of these strategies, focusing on their unique methods for achieving full-sky coverage. Key technical parameters, such as spin and precession axis angles, and rotation periods, are thoroughly examined. The chapter also introduces the \LiteBIRD mission, which inspired this study, highlighting its use of HWP. Finally, we summarize the scanning parameters across different CMB missions, underscoring the critical role of optimized scanning strategies in attaining high-precision CMB polarization measurements. }

\section{Scanning strategy of past CMB space missions}

\subsection{\COBE}
\COBE was deployed in a Sun Synchronous Orbit (SSO), which is a specialized
polar orbit designed for CMB spectrum and full-sky temperature anisotropy
measurements \cite{boggess1992cobe}. The SSO's key advantage lies in maintaining
a constant angle between the orbital plane and the Sun, ensuring stable solar
radiation exposure throughout the mission. For SSO configurations with inclination
angles above $95^{\circ}$, the Earth's oblate shape induces orbital plane
rotations completing one cycle annually, facilitating comprehensive sky coverage.

The satellite spins at 0.8\,rpm, a rate carefully selected to reduce noise and
systematic effects from radiometer gain and offset variations. This pioneering research
on the relationship between scanning strategies and noise characteristics was thoroughly
examined in \cite{wright1996scanning}. Three primary instruments were aboard: FIRAS,
DMR (Differential Microwave Radiometers), and DIRBE (Diffuse Infrared Background
Experiment). FIRAS, aligned with the spin axis, measured sky frequency spectra
through a $7^{\circ}$ field of view (FoV). The DMR system comprised three receiver
pairs, positioned $120^{\circ}$ apart around the dewar's aperture plane. Each radiometer
measured differential sky signals between horn pairs with $7^{\circ}$ FoV separated
by $60^{\circ}$, positioned $30^{\circ}$ from the spin axis.

Both DMR and DIRBE instruments traced epicyclic patterns, enabling daily scans of
half the sky and facilitating comprehensive multipole measurements for each sky pixel.
\begin{figure}[h]
    \centering
    \copyrightbox{ \includegraphics[width=0.19\columnwidth]{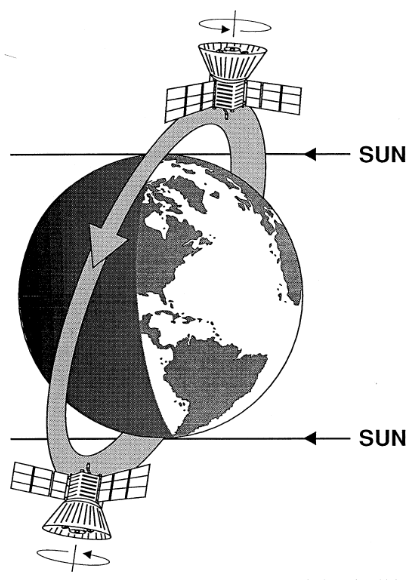} \includegraphics[width=0.80\columnwidth]{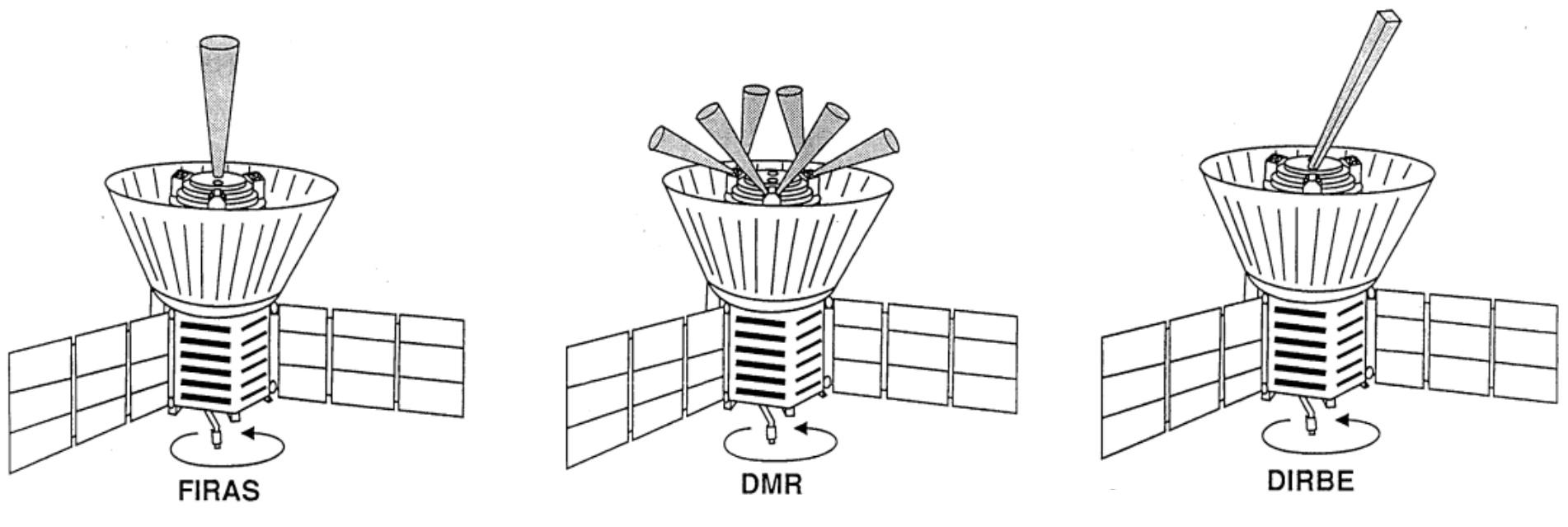}}
    {\copyright~1992 AAS. Reproduced with permission.}
    \caption[\COBE's scanning strategy and instrument layout]{\COBE's SSO
    scanning strategy (leftmost) and instrument layout: FIRAS (center-left), DMR
    (center-right), and DIRBE (rightmost). Rotation schematic shows satellite's
    spin. The figure is adapted from ref.~\cite{boggess1992cobe} with a
    permission from the authors and the publisher.}
    \label{fig:cobe_scan}
\end{figure}

\subsection{\WMAP}

\begin{figure}[htbp]
    \centering
    \copyrightbox{ \includegraphics[width=0.8\columnwidth]{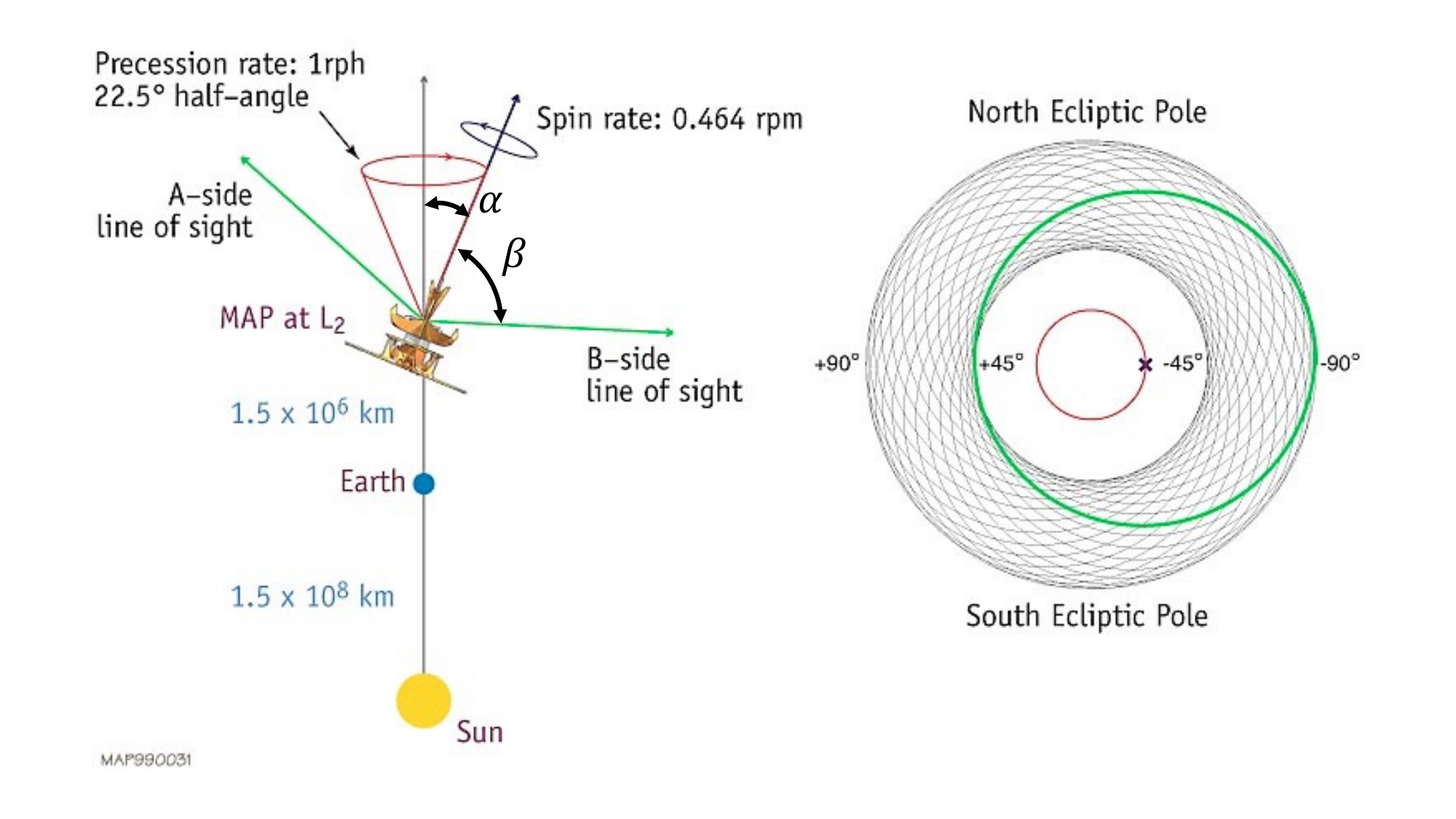} }{\copyright~2010 NASA/WMAP Science Team}
    \caption[\WMAP scanning strategy]{\WMAP scanning strategy (left) and
    telescope scanning trajectory over one precession period of 1\,hour (right).
    Figure credit: NASA/WMAP Science Team\footnotemark}
    \label{fig:wmap_scan}
\end{figure}

\WMAP aimed to observe the full-sky temperature anisotropy previously measured
by \COBE at higher angular resolution, with the goal of estimating cosmological parameters
from the power spectrum \cite{Bennett2003}. As the first CMB satellite deployed
at Lagrange point 2 (\Ltwo), it combined precession, spin, and solar orbit rotations
for comprehensive sky coverage. This three-rotation scanning strategy offers
high flexibility and has become a fundamental concept for \Ltwo-based space
mission, allowing optimization for specific mission objectives. In this scanning
approach, $\alpha$ defines the angle between the Sun-\Ltwo vector and precession
axis, while $\beta$ represents the angle between the spin axis and telescope
pointing direction (\cref{fig:wmap_scan}).

\WMAP's scanning strategy, configured with the parameters shown in \cref{fig:wmap_scan},
was designed to minimize systematic errors through several key considerations: rapid
scanning of most of the sky to minimize $1/f$ noise effects; uniform scanning
angles across sky pixels to suppress systematic effects; multiple observations of
each pixel at different times; maintaining instruments in Earth's shadow for
optimal passive cooling and avoiding radiation from the Sun, Earth, and Moon; constant
angle between the Sun and solar panel plane for thermal and power stability.

With $\alpha+\beta\leq90^{\circ}$, the satellite could observe the full sky
within six months. Additionally, two telescopes were positioned at mirror-image
angles, enabling common-mode noise reduction through differential measurements.
The mission conducted full-sky CMB observations for 9 years and 2 months
following its launch on June 30, 2001.

\footnotetext{\url{https://wmap.gsfc.nasa.gov/mission/observatory_scan.html}}

\subsection{\Planck}

\Planck was designed to conduct high-resolution, high-sensitivity observations of
CMB temperature anisotropies and galactic frequency spectra. The mission
employed a scanning strategy that enabled multiple observations of the same sky
pixel within short time intervals, significantly enhancing temperature mapping precision
\cite{planck_scanning_strategy,delabrouille1998scanning,maris2006flexible}.
Unlike \WMAP, \Planck featured a single telescope configuration rather than a pair,
maximizing optical system size to achieve superior angular resolution.

\begin{figure}[htbp]
    \centering
    \copyrightbox{ \includegraphics[width=0.8\columnwidth]{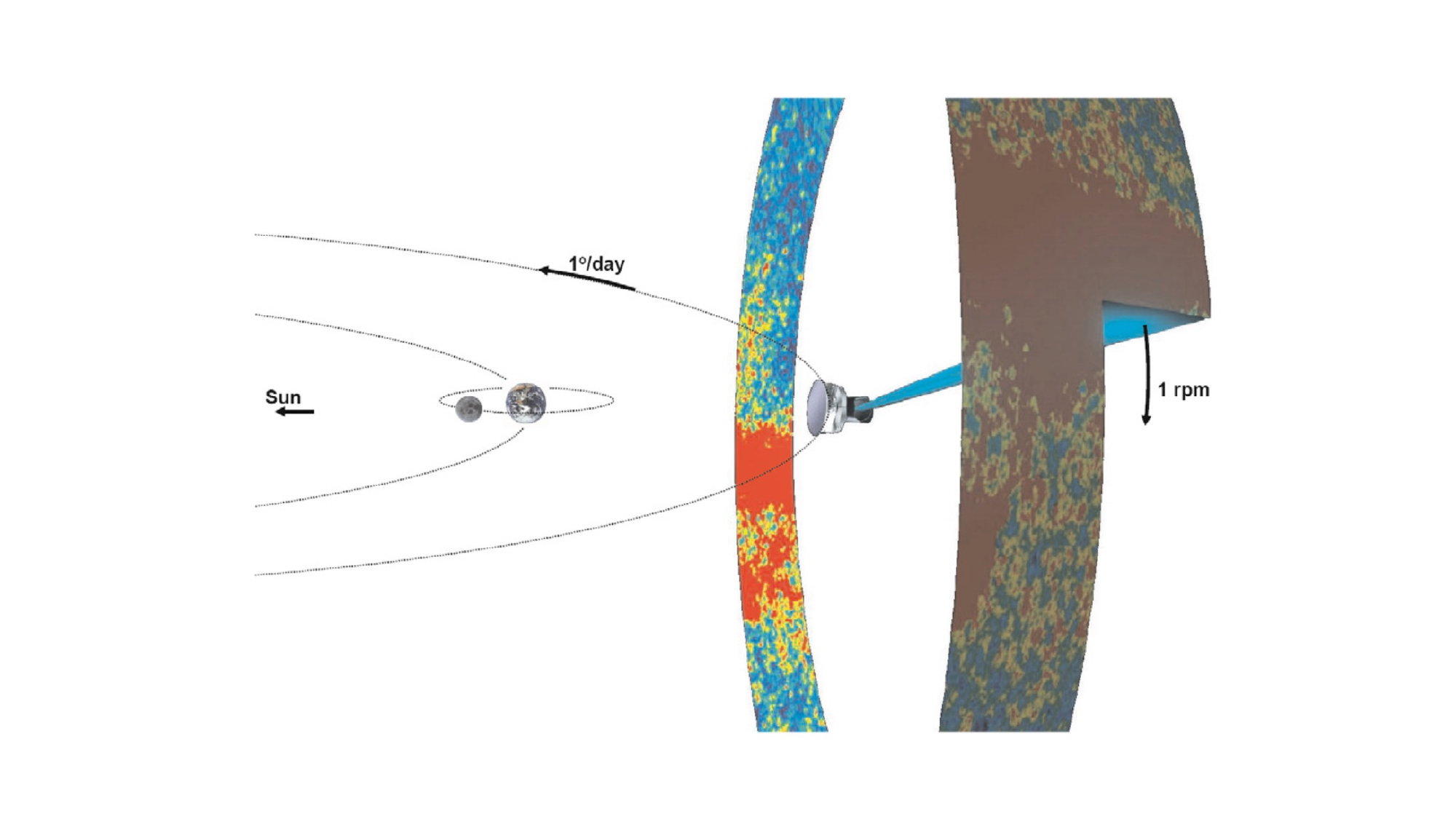} }{\copyright~2010 A\&A}
    \caption[\Planck scanning strategy]{\Planck scanning strategy, characterized
    by its 1\,rpm spin rate and slow 6-month precession period. The figure
    adopted from ref.~\cite{planck_prelaunch} with a permission from the publisher.}
    \label{fig:planck_scan}
\end{figure}

\Planck operated with a 1\,rpm spin rate combined with a slow 6-month precession
period (\cref{fig:planck_scan}). This precession period, the longest among all
CMB space missions considered to date, successfully achieved uniform full-sky coverage.
However, this strategy provided a valuable lesson for future missions: while
effective for temperature measurements, it resulted in redundant polarization angles
relative to sky pixels, thereby reducing polarization sensitivity. Nevertheless,
the slow precession rate offered exceptional satellite control stability and
maneuverability, enabling flexible in-flight operations.

This operational flexibility proved particularly valuable for a calibration using
planets. For instance, Jupiter, being the brightest, served as a crucial calibrator
for gain and pointing measurements. While \Planck's spin axis typically rotated
approximately $1^{\circ}$ per day, the satellite could switch to a `deep scan
mode' when Jupiter was in view. This mode involved slower spin axis rotation, allowing
extended `repointing' maneuvers of Jupiter and improving beam pattern sampling by
a factor of 3.4 \cite{Planck_LFI_beam}.

\section{Summary of scanning parameters across CMB space missions}
This section summarizes the scanning strategies of CMB space missions deployed
at \Ltwo (excluding \COBE due to its different orbit). The scanning strategies are
characterized by four key parameters:
\begin{itemize}
    \item Precession axis angle ($\alpha$): angle between Sun-\Ltwo vector and
        precession axis

    \item Spin axis angle ($\beta$): angle between spin axis and telescope

    \item Precession period ($T_{\alpha}$): time for one complete spin axis
        rotation

    \item Spin period ($T_{\beta}$): period of spacecraft rotation around spin
        axis
\end{itemize}
\Cref{tab:sumarry_scanning_parameters} presents these parameters for \WMAP, \Planck,
\EPIC, \CORE, \PICO, and \LiteBIRD \cite{Bennett2003,planck2014planck_mission,Bock_epic,finelli2018core,PICO2019,takase2024scan}.
Among these, \EPIC and \PICO were NASA proposals, \CORE was an ESA proposal, and
\LiteBIRD is a JAXA mission. All these missions focus on CMB polarization, specifically
targeting the primordial $B$-mode signal from the inflationary epoch.

\begin{table}[htbp]
    \centering
    \begin{tabular}{lllll}
        \hline
        {}        & $\alpha$      & $\beta$        & $T_{\alpha}$ & $T_{\beta}$ \\
        \hline
        \WMAP     & $22.5^{\circ}$  & $70^{\circ}$ & 1\,hr        & 129\,s      \\
        \hline
        \Planck   & $7.5^{\circ}$ & $85^{\circ}$   & 6\,month     & 1\,min      \\
        \hline
        \EPIC     & $45^{\circ}$  & $55^{\circ}$   & 3.2\,hr      & 1\,min      \\
        \hline
        \CORE     & $30^{\circ}$  & $65^{\circ}$   & 4\,day       & 2\,min      \\
        \hline
        \PICO     & $26^{\circ}$  & $69^{\circ}$   & 10\,hr       & 1\,min      \\
        \hline
        \LiteBIRD & $45^{\circ}$  & $50^{\circ}$   & 3.2058\,hr   & 20\,min     \\
        \hline
    \end{tabular}
    \caption[ Scanning strategy parameters for CMB space missions. ]{Scanning
    strategy parameters for CMB space missions deployed at $\rm L_{2}$. Among these
    missions, only \Planck and \WMAP have been completed, while the others are either
    in development or were proposed missions.}
    \label{tab:sumarry_scanning_parameters}
\end{table}

\section{\LiteBIRD space mission}
\label{sec:litebird} 
\begin{figure}[htbp]
    \centering
    \copyrightbox{ \includegraphics[width=1\columnwidth]{ 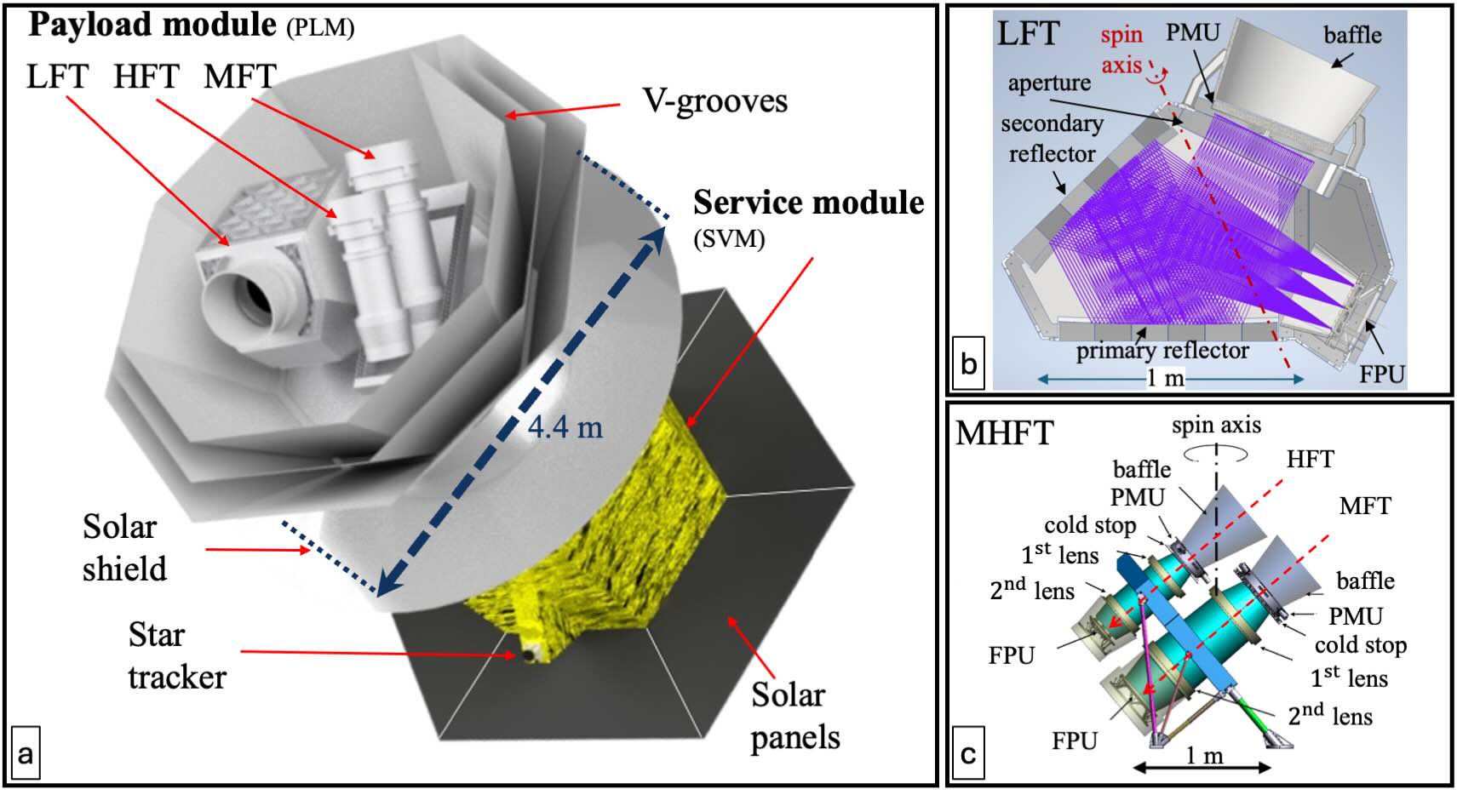}}{\copyright~2024 SPIE}
    \caption[Schematic of \LiteBIRD's spacecraft and payloads]{Architectural
    overview of \LiteBIRD showing: (a) integrated spacecraft design comprising PLM
    and SVM, (b) reflective LFT, and (c) refractive MHFT. The figure is adapted
    from ref.~\cite{ghigna2024litebird}, with permission of the publisher and
    authors.}
    \label{fig:litebird_spacecraft}
\end{figure}

\begin{figure}[htbp]
    \centering
    \copyrightbox{ \includegraphics[width=0.9\columnwidth]{ 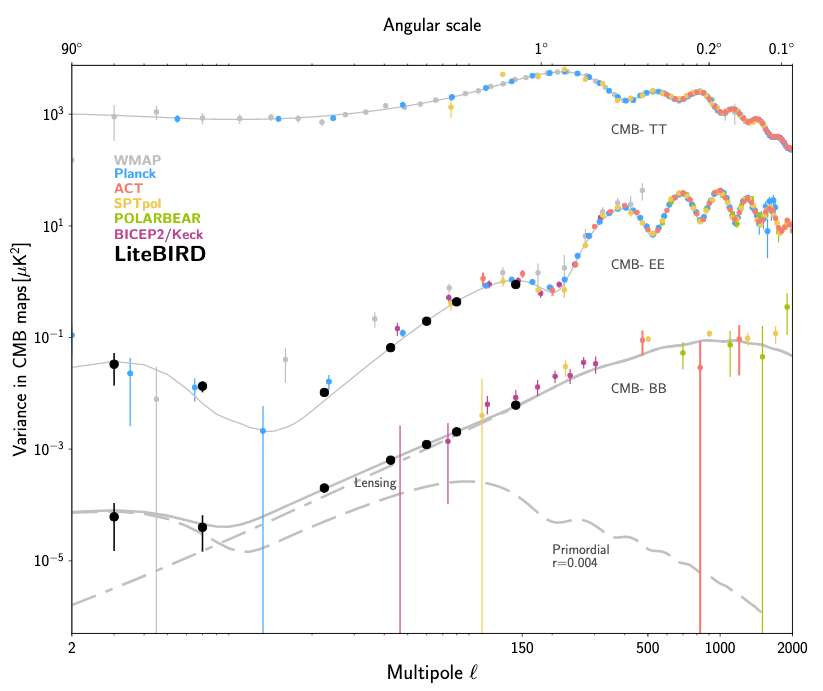 }}
    {\copyright~2022 Oxford University Press}
    \caption[Expected sensitivity of \LiteBIRD for CMB power spectra]{CMB
    angular power spectra components: temperature anisotropy (top), $E$-mode (middle),
    and $B$-mode polarization (bottom). Solid lines represent the $\Lambda$CDM model's
    best-fit power spectra including scale-invariant tensor perturbations ($r = 0
    .004$). The thin dashed line shows the $B$-mode contribution from scale-invariant
    tensor perturbations ($r = 0.004$). Colored points indicate existing CMB measurements
    from various experiments \cite{aghanim2020planck_overview,hinshaw2013nineyear,ade2021improved,adachi2016measurement,sayre2020measurements,bicep2_2018,benett2013finalmaps,polarbear2017measurement,henning2018measurements,planck2018power_spectra,choi2020atacama},
    while black points represent \LiteBIRD's projected polarization sensitivity.
    The figure is adopted from ref.~\cite{PTEP2023} based on Creative Commons CC
    BY license.}
    \label{fig:litebird_cl}
\end{figure}

\LB, scheduled for launch in 2032 as a JAXA L-class mission,
will conduct three years of CMB polarization observations \cite{PTEP2023}. As illustrated
in \cref{fig:litebird_spacecraft}, the spacecraft architecture integrates a
payload module (PLM) and service module (SVM). The observatory's multi-frequency
coverage is achieved through three specialized telescopes: the Low Frequency
Telescope (LFT) \cite{sekimoto2020concept}, Medium Frequency Telescope (MFT),
and High Frequency Telescope (HFT) \cite{montier2018mhft}.
The mission's primary scientific objective is to detect primordial $B$-mode polarization
signatures from cosmic inflation, with anticipated sensitivity shown in
\cref{fig:litebird_cl}. The mission aims to achieve a high-precision measurement
of the tensor-to-scalar ratio, $r$ with its error $\delta r < 0.001$.

To achieve this unprecedented sensitivity, \LB implements two key technological innovations:
a continuously rotating HWP for polarization modulation and an optimized
scanning strategy for comprehensive full-sky polarization measurements. These design
features make \LB an exemplary case study for analyzing polarization-optimized scanning
strategies.

\subsection{Polarization modulation with half-wave plates}
\LiteBIRD employs the continuously rotating HWPs as Polarization Modulator Units
(PMUs) in all three telescopes \cite{sakurai2020breadboard,columbro2020polarization}.
HWPs are manufactured from birefringent materials with different refractive
indices along two orthogonal optical axes. The LFT uses A-cut sapphire, while
the MHFT adopts metamaterial-based metal mesh HWPs. When polarized light enters a
HWP, the difference in refractive indices along each axis creates an optical
path difference, altering the electromagnetic wave's phase. Specifically, an HWP
introduces a half-wavelength phase retardation; other common wave plates include
quarter-wave plates. For HWPs, the polarization direction of incident light is
rotated symmetrically relative to the optical axis. Consequently, rotating an
HWP by an angle $\phi$ rotates the transmitted polarization direction by $2\phi$
relative to the incident polarization. When the HWP continuously rotates with angular
frequency $\phi=\omega t$, the transmitted polarization direction rotates at
$2\omega t$. Since electromagnetic wave intensity is proportional to the square of
the electric field, the polarization signal is ultimately modulated at
$4\omega t$ \cite{galitzki2018simons}.

    \chapter{Novel map-making approach in \spin space}
\label{chap:formalism}

\minitoc

\chapabstract{In this chapter, we introduce map-making methods in a \spin space representation, which transforms angular domain information into Fourier domain. We first establish the formalism for observations without a HWP, showing how the detected signal can be decomposed into \spin components. We then extend this to include HWP modulation, which adds complexity but maintains similar mathematical structure. Finally, we demonstrate how this formalism generalizes to multiple detectors, including orthogonal detector pairs. This \spin space approach enables faster map-making compared to traditional time-domain methods while preserving all relevant information.}

\section{Description of signal and map-making in \spin space}
\subsection{The case without HWP}
\label{sec:formalism_wo_HWP}

We can describe the signal detected by a detector within a sky pixel of
spherical coordinates $\Omega=(\theta, \phi)$ as a function of the detector's
crossing angle $\psi$ as
\begin{equation}
    S^{d}(\Omega,\psi)=h(\Omega,\psi)S(\Omega,\psi), \label{eq:Sd}
\end{equation}
where $S$ is the signal field, which describes the signal detected by the detector
at each visit of the sky pixel, and the real space scan field, $h$, describes
the observation by a detector in each sky pixel under a specific scanning
strategy. In this thesis, we refer a quantity that is function of $\Omega$ and
$\psi$ to a `field' which can be recognized like a state function in the quantum
mechanics. As introduced in ref.~\cite{mapbased}, the real space scan field can be
expressed as
\begin{equation}
    h(\Omega,\psi)= \frac{2 \pi}{\Nhits(\Omega)}\sum_{j}\delta(\psi-\psi_{j}( \Omega
    )),
\end{equation}
where $\delta$ is the Dirac delta function, $\Nhits(\Omega)$ is the number of
hits/observations in the sky pixel $\Omega$, and $\psi_{j}(\Omega)$ is the crossing
angle of the $j^{\rm th}$ visit of the sky pixel $\Omega$.

It is then we can decompose the signal field $S$ into the \spin-$n$ space components,
$\St[n]$ by the Fourier transform $\psi \to n$ as
\begin{align}
    S^{d}(\Omega,\psi) & = \sum_{n}\Sd[n](\Omega)e^{i n\psi},                                                       \\
    \Sd[n](\Omega)     & = \sum_{n'=-\infty}^{\infty}\h[n-n'](\Omega) \St[n'](\Omega),\label{eq:coupling_eq_wo_HWP}
\end{align}
where $\h[n-n']$ is referred to as the orientation function defined as
\begin{equation}
    \begin{split}
        \h[n](\Omega)&= \frac{1}{2\pi}\int d\psi h(\Omega,\psi)e^{-i n\psi}\\&= \frac{1}{\Nhits(\Omega)}
        \sum_{j}e^{-i n\psi_j}(\Omega).
    \end{split}
\end{equation}
Note that when we discuss the \spin moment, we describe it in italics to distinguish
it from the spacecraft's about the maximum inertial axis, which is described in
normal font. In this thesis, we distinguish quantities $x$ and $\tilde{x}$ as
the real (i.e.\ angular) space and \spin (i.e.\ Fourier) space, respectively.
These quantities in \spin space satisfy the relations:
\begin{align}
    \St[n] & = \St[-n]^{*}, \\
    \Sd[n] & = \Sd[-n]^{*}, \\
    \h[n]  & = \h[-n]^{*}.
\end{align}

Once, a signal is given by the bolometric equation
\begin{align}
    S(\Omega,\psi) & = I(\Omega) + Q(\Omega)\cos(2\psi) + U(\Omega)\sin(2\psi) + \mathscr{N}\nonumber                \\
                   & = I(\Omega) + \frac{1}{2}P(\Omega)e^{2i\psi}+ \frac{1}{2}P^{*}(\Omega)e^{-2i\psi}+ \mathscr{N},
\end{align}
where we introduce $P=Q+iU$ and its complex conjugate $P^{*}$. The $\mathscr{N}$
represents the probability density function (PDF) of the noise. Here, we define
a signal that is measured by $j^{\rm th}$ observation at a sky pixel $\Omega$
as
\begin{align}
    d_{j} & = \ab (\mqty{1 & \frac{1}{2}e^{2i\psi_j} & \frac{1}{2}e^{-2i\psi_j}}) \ab (\mqty{I \\ P \\ P^*}) + n_{j}\nonumber \\
          & = \mathbf{w}_{j}\cdot \mathbf{s}+ n_{j},\label{eq:dj}
\end{align}
where $n_{j}$ represents the $j^{\rm th}$ sample of noise given by $\mathscr{N}$,
$\mathbf{w}_{j}$ is a basis vector, and $\mathbf{s}$ is the Stokes vector. In order
to reconstruct the Stokes vector from observations, we minimize:
\begin{align}
    \chi^{2}= \sum_{i,j}\ab(d_{i}-\mathbf{w}_{i}\cdot \mathbf{s})\ab(N^{-1})_{ij}\ab(d_{j}-\mathbf{w}_{j}\cdot \mathbf{s}),
\end{align}
where $N_{ij}$ is the noise covariance matrix. After minimizing $\chi^{2}$, we obtain
the equation of linear regression to estimate Stokes vector as
\begin{align}
    \hat{\mathbf{s}} & = \ab(\sum_{i,j}\mathbf{w}_{i}^{\dagger}(N^{-1})_{ij}\mathbf{w}_{j})^{-1}\ab(\sum_{i,j}\mathbf{w}_{i}^{\dagger}(N^{-1})_{ij}d_{j}), \label{eq:map-making_eq_noise}
\end{align}
where $\hat{\mathbf{s}}$ represents the estimated Stokes vector and $\dagger$ represents
the Hermitian transpose. Let us assume the noise is given by Gaussian
distribution, i.e., white noise which does not have a correlation between the $i^{\rm
th}$ and the $j^{\rm th}$ measurement. Furthermore, we define the symbol for the
average of a quantity $x_{j}$ as $\ab< x_{j}> = \frac{1}{N}\sum_{j}^{N}x_{j}$,
then, \cref{eq:map-making_eq_noise} can be expressed as
\begin{align}
    \hat{\mathbf{s}} & = \ab(\sum_{j}\mathbf{w}_{j}^{\dagger}\mathbf{w}_{j})^{-1}\ab(\sum_{j}\mathbf{w}_{j}^{\dagger}d_{j}) \nonumber                                                                                                                                                                                               \\
                     & = \ab(\mqty{ 1 & \frac{1}{2}\h[2] & \frac{1}{2}\h[-2] \\ \frac{1}{2}\h[-2] & \frac{1}{4} & \frac{1}{4}\h[-4] \\ \frac{1}{2}\h[2] & \frac{1}{4}\h[4] & \frac{1}{4}})^{-1}\ab(\mqty{ \ab< d_j > \\ \frac{1}{2}\ab< d_j e^{2i\psi_j}> \\ \frac{1}{2}\ab< d_j e^{-2i\psi_j}>}). \label{eq:map-making_TOD_wo_HWP}
\end{align}
This equation corresponds to the simple binning map-making approach (e.g.\ ref.~\cite{binning_brawn}),
and the following relation
\begin{align}
    \ab(\mqty{ \ab< d_j > \\ \frac{1}{2}\ab< d_j e^{2i\psi_j}> \\ \frac{1}{2}\ab< d_j e^{-2i\psi_j}>}) = \ab(\mqty{ \Sd[0] \\ \frac{1}{2}\Sd[2] \\ \frac{1}{2}\Sd[-2]}),
\end{align}
where the right part can be obtain by \cref{eq:coupling_eq_wo_HWP}. Of these,
the map-maker in \spin space is given by
\begin{align}
    \hat{\mathbf{s}} & = \ab(\mqty{ 1 & \frac{1}{2}\h[2] & \frac{1}{2}\h[-2] \\ \frac{1}{2}\h[-2] & \frac{1}{4} & \frac{1}{4}\h[-4] \\ \frac{1}{2}\h[2] & \frac{1}{4}\h[4] & \frac{1}{4}})^{-1}\ab(\mqty{ \Sd[0] \\ \frac{1}{2}\Sd[2] \\ \frac{1}{2}\Sd[-2]}). \label{eq:map-making_spin_wo_hwp}
\end{align}
This formalization is introduced in ref.~\cite{mapbased} which allows us to simulate
the map-making faster than TOD-based approach.

\subsection{The case with HWP}
\label{sec:formalism_w_HWP}

Basic formalism of the map-making in \spin space with or without HWP is the same,
but the case with HWP has additional complexity due to the HWP modulation. First,
the signal field $S$ is defined as a function of the detector's crossing angle $\psi$
and the HWP angle $\phi$. The real space scan field $h$ is also a function of
$\Omega$, $\psi$, and $\phi$. The signal detected by a detector within a sky pixel
of spherical coordinates $\Omega=(\theta, \varphi)$ is given by
\begin{equation}
    S^{d}(\Omega,\psi,\phi)=h(\Omega,\psi,\phi)S(\Omega,\psi,\phi).
\end{equation}
Since the signal field is expanded to a two dimensional field given by $\psi$ and
$\phi$, we consider corresponding scan field $h$ as
\begin{equation}
    h(\Omega,\psi, \phi)= \frac{4 \pi^{2}}{N_{\rm hits}(\Omega)}\sum_{j}\delta(\psi
    -\psi_{j})\delta(\phi-\phi_{j}).
\end{equation}
Now we consider Fourier transform to bring the signal field to \spin space. Defining
$n$ and $m$ as the \spin moment that is the variable conjugate to the angle $\psi$
and $\phi$, the transformation $(\psi,\phi)\to(n,m)$ is given by
\begin{align}
    S^{d}(\Omega,\psi,\phi) & = \sum_{n,m}\Sd[n,m](\Omega)e^{i n\psi}e^{i m\phi},                                                                     \\
    \Sd[n,m](\Omega)        & = \sum_{n'=-\infty}^{\infty}\sum_{m'=-\infty}^{\infty}\h[\Delta n, \Delta m](\Omega) \St[n',m'](\Omega), \label{eq:kSd}
\end{align}
where we introduce $\Delta n = n-n'$ and $\Delta m = m-m'$, and define the two dimensional
orientation function, $\h[\Delta n,\Delta m]$ by Fourier transform of the real
space scan field as
\begin{equation}
    \begin{split}
        \h[n,m](\Omega)&= \frac{1}{4\pi^{2}}\int d\psi \int d\phi h(\Omega,\psi,\phi
        )e^{-i n\psi}e^{-i m\phi}\\&= \frac{1}{N_{\rm hits}}\sum_{j}e^{-i(n\psi_j
        + m \phi_j)}.\label{eq:exp_crosslink}
    \end{split}
\end{equation}
Now, we define a signal field given by a bolometer with a HWP as
\begin{align}
    S(\Omega,\psi,\phi) & = I(\Omega) + Q(\Omega)\cos(4\phi-2\psi) + U(\Omega)\sin(4\phi-2\psi) + \mathscr{N}\nonumber                    \\
                        & = I(\Omega) + \frac{1}{2}P(\Omega)e^{-i(4\phi-2\psi)}+ \frac{1}{2}P^{*}(\Omega)e^{i(4\phi-2\psi)}+ \mathscr{N}.
\end{align}
A signal detected by the $j^{\rm th}$ observation at a sky pixel $\Omega$ is given
by
\begin{align}
    d_{j} & = \ab( \mqty{ 1 & \frac{1}{2}e^{-i{(4\phi_j-2\psi_j)}} & \frac{1}{2}e^{i{(4\phi_j-2\psi_j)}} }) \ab(\mqty{ I \\ P \\ P^{*} }) + n_{j}\nonumber \\
          & = \mathbf{w}_{j}\cdot \mathbf{s}+ n_{j}.
\end{align}
By the same procedure as the case without HWP, we can obtain the map-making
equation
\begin{align}
    \hat{\mathbf{s}} & = \ab(\mqty{ 1 & \frac{1}{2}\h[-2,4] & \frac{1}{2}\h[2,-4] \\ \frac{1}{2}\h[2,-4] & \frac{1}{4} & \frac{1}{4}\h[4,-8] \\ \frac{1}{2}\h[-2,4] & \frac{1}{4}\h[-4,8] & \frac{1}{4} })^{-1}\ab(\mqty{ \ab< d_j > \\ \frac{1}{2}\ab< d_j e^{i(4\phi_j-2\psi_j)}> \\ \frac{1}{2}\ab< d_j e^{-i(4\phi_j-2\psi_j)}> }) \label{eq:map-making_TOD_w_HWP} \\
                     & =\ab(\mqty{ 1 & \frac{1}{2}\h[-2,4] & \frac{1}{2}\h[2,-4] \\ \frac{1}{2}\h[2,-4] & \frac{1}{4} & \frac{1}{4}\h[4,-8] \\ \frac{1}{2}\h[-2,4] & \frac{1}{4}\h[-4,8] & \frac{1}{4} })^{-1}\ab(\mqty{ \Sd[0,0] \\ \frac{1}{2}\Sd[2,-4] \\ \frac{1}{2}\Sd[-2,4] }), \label{eq:map-making_spin_w_hwp}
\end{align}
where \cref{eq:map-making_TOD_w_HWP,eq:map-making_spin_w_hwp} correspond to the
time domain and the \spin domain map-making approach, respectively.

\section{Observation with multiple detectors}
\label{sec:multi_detector}

Nowadays, the CMB experiment usually has multiple detectors about $10^{3}$ to $10
^{4}$ to take a statistics. Here we consider the implementation of multiple
detectors in the map-making procedure. This can be simply described by modifying
several quantities in the previous section. We introduce the detector index
$\mu$ and total number of detectors $\Ndets$, then the total number of hits per pixel
$\Ntot$ is given by
\begin{equation}
    \Ntot(\Omega) = \sum_{\mu}\Nhits\pmu(\Omega),
\end{equation}
where $\Nhits\pmu$ is the number of hits of the $\mu^{\rm th}$ detector in the sky
pixel $\Omega$. The orientation function given by total number of observations, $\htot
[n,m]$, is
\begin{equation}
    \htot[n,m](\Omega) = \frac{1}{\Ntot(\Omega)}\sum_{\mu}\h[n,m]\pmu(\Omega)\Nhits
    \pmu(\Omega).
\end{equation}

Here, we define orthogonal pair detector which is named as $\texttt{T}$ and
$\texttt{B}$ that stands for `Top' and `Bottom' detectors. These detector observes
the same direction though different crossing angle $\psi$, let us denote the
crossing angle of the $\texttt{T}$ and $\texttt{B}$ detectors as
$\psi^{\texttt{T}}$ and $\psi^{\texttt{B}}$, respectively. Then, the orientation
function of the $\texttt{T}$ can be exchanged to that of the $\texttt{B}$ by the
following relation
\begin{equation}
    \h[n,m]^{(\texttt{B})}= \h[n,m]^{(\texttt{T})}e^{i n \frac{\pi}{2}}.
\end{equation}

The detected signal in \spin space per detector is given by
\begin{align}
    \Sd[n,m]\pmu(\Omega) & = \sum_{n'=-\infty}^{\infty}\sum_{m'=-\infty}^{\infty}\h[n-n',m-m']\pmu(\Omega) \St[n',m']\pmu(\Omega). \label{eq:Sd_mu}
\end{align}
It is then the total detected signal given by the multiple detectors is
\begin{align}
    \Sd[n,m]^{\rm tot}(\Omega) & = \frac{1}{\Ntot(\Omega)}\sum_{\mu}\sum_{n'=-\infty}^{\infty}\sum_{m'=-\infty}^{\infty}\Nhits\pmu(\Omega) \h[n-n',m-m']\pmu(\Omega) \St[n',m']\pmu(\Omega) \nonumber \\
                               & = \frac{1}{\Ntot(\Omega)}\sum_{\mu}\Nhits\pmu(\Omega) \Sd[n,m]\pmu(\Omega).\label{eq:Sd_tot}
\end{align}
By using these quantities, the map-making equation for the multiple detectors is
given by
\begin{align}
    \hat{\mathbf{s}} & = \ab (\mqty{ 1 & \frac{1}{2}\htot[-2,4] & \frac{1}{2}\htot[2,-4] \\ \frac{1}{2}\htot[2,-4] & \frac{1}{4} & \frac{1}{4}\htot[4,-8] \\ \frac{1}{2}\htot[-2,4] & \frac{1}{4}\htot[-4,8] & \frac{1}{4} })^{-1}\ab (\mqty{ \Sd[0,0]^{\rm tot} \\ \frac{1}{2}\Sd[2,-4]^{\rm tot} \\ \frac{1}{2}\Sd[-2,4]^{\rm tot} }).\label{eq:mapmaking_tot}
\end{align}
In this section we assumed HWP-aware observation though in the case without HWP can
be obtained by same procedure.

    \chapter{Optimization of full-sky scanning strategy}
\label{chap:scanning_strategy_optimization}

\chapabstract{This chapter addresses the optimization of scanning strategies for next-generation CMB space missions, with particular focus on \LiteBIRD. We begin by establishing the parameter space of scanning strategies and analyzing the constraints on both geometric and kinematic parameters. Through detailed simulations and defined evaluation metrics, we explore optimal scanning configurations. Our analysis yields three promising solutions for CMB polarimetry. Based on \LiteBIRD's perspective, we identify and recommend the most suitable configuration. Finally, we benchmark this optimal strategy against those employed by \Planck and proposed for \PICO, evaluating their relative merits for calibration procedures and null-tests. The contents in this chapter are based on the our work published in ref.~\cite{takase2024scan}.}

\minitoc
\section{The parameter space of scanning strategies}

At the second Lagrange point ($\mathrm{L_2}$) of the Sun-Earth system,
pioneering CMB space missions like \WMAP and \Planck executed their observations.
These missions achieved comprehensive celestial coverage through an intricate combination
of spacecraft dynamics: rotational motion about its axis, precessional movement,
and its heliocentric orbital trajectory. As depicted in the left panel of \cref{fig:standard_config_and_T_beta},
the scanning geometry is characterized by two fundamental angles: $\beta$, which
defines the angular separation between the observational line-of-sight and the
spacecraft's principal axis of inertia (serving as the spin axis), and $\alpha$,
which measures the angle between this spin axis and the Sun-spacecraft vector. The
spacecraft undergoes two primary rotational motions: a spin about its principal
axis with period $T_{\beta}$, and a precession of this spin axis around the Sun-spacecraft
vector with period $T_{\alpha}$.

The next generation of CMB experiments introduces an additional complexity through
the implementation of a continuously rotating HWP for polarization modulation.
The HWP's orientation is specified by angle $\phi$ relative to the optical axis,
with a rotation period of $T_{\phi}$. For each rotational component
$j \in \{\alpha,\beta,\phi \}$, the motion can be equivalently characterized by its
period $T_{j}$ or its corresponding frequency:
\begin{align}
    \omega_{j} & = 2\pi/T_{j}~ \rm{[rad/s]}, \\
    f_{j}      & = 1/T_{j}~ \rm{[Hz]},       \\
    \nu_{j}    & = 60/T_{j}~ \rm{[rpm]}.
\end{align}
Throughout this work, we employ these kinetic parameters ($T_{j},\omega_{j},f_{j}$,
or $\nu_{j}$) interchangeably as contextually appropriate. A comprehensive exposition
of the spacecraft's precession and spin dynamics is presented in \cref{apd:scan_motion}.
The scanning strategy incorporates another critical parameter: the sampling rate
$f_{\rm s}$, which quantifies the temporal density of data acquisition. The parameters
$\alpha$ and $\beta$ constitute the 'geometric parameters' of the scanning strategy,
while the temporal measures $T_{j}/\omega_{j}/f_{j}/\nu_{j}$ represent the 'kinematic
parameters'. The scanning strategy's complete characterization requires optimization
within a six-dimensional parameter space $\{\alpha , \beta, T_{\alpha}, T_{\beta}
, \nu_{\phi}, f_{\rm s}\}$. While this optimization presents considerable
complexity, we can systematically reduce its effective dimensionality through mission-specific
constraints and judicious assumptions, particularly for configurations
incorporating a HWP.

\section{Constraints on the parameter space}

\subsection{Constraints on geometric parameters \label{sec:angle-constraint}}

A fundamental geometric constraint requires that
\begin{align}
    \kappa= \alpha+\beta > 90^{\circ}.\label{eq:const_geometric}
\end{align}
This condition, as elaborated in ref.~\cite{OptimalScan}, is imperative for achieving
comprehensive celestial coverage. Given that the effective $\beta$ varies across
detector positions, implementing an elevated $\kappa$ value, typically
$\kappa \sim 95^{\circ}$ \cite{OptimalScan}, is advantageous to accommodate the experiment's
specific Field of View (FoV). This configuration simultaneously optimizes
thermal management by minimizing direct solar radiation exposure. The upper limit
of $\kappa$ is primarily determined by the engineering constraints of the sun-shield
and thermal control systems. Historical precedents include \WMAP and \Planck, both
utilizing $\kappa =92.5^{\circ}$, while \EPIC \cite{Bock_epic} proposed a more
ambitious $\kappa=100^{\circ}$ (achieved with $\alpha =45^{\circ}, \beta=55^{\circ}$;
see \cref{tab:sumarry_scanning_parameters}), representing the most extensive
$\kappa$ value proposed for any CMB space mission. Within a given experimental framework,
$\kappa$ typically remains fixed, establishing a direct relationship between $\alpha$
and $\beta$.

\subsection{Constraints on kinetic parameters}
Having established the geometric parameters, we now examine the temporal aspects
governed by periods $T_{j}$. The incorporation of a HWP in the spacecraft serves
dual critical functions: primarily, it facilitates polarization modulation at frequencies
substantially exceeding the instrument's $1/f$ noise knee frequency $f_{\rm knee}$;
additionally, it enhances the homogeneity of effective crossing angles. The
optimization of HWP revolution frequency in relation to the $1/f$ noise model
lies beyond our present scope.

Given a HWP with revolution frequency $f_{\phi}$ capable of adequately
suppressing $1/f$ noise characterized by a specific $f_{\rm knee}$, we must establish
appropriate constraints for the spacecraft's rotational periods. In contrast to
traditional configurations without a HWP, where incoherent polarimetric receivers
rely exclusively on spacecraft rotation for polarization modulation (necessitating
rapid spin rates as detailed in ref.~\cite{OptimalScan}), the presence of a HWP fundamentally
alters these requirements. The primary consideration shifts to ensuring
sufficient polarization modulation within each sky pixel during the telescope's
transit. Specifically, the HWP's angular velocity must be appropriately scaled
relative to the sky scanning motion. The maximum angular velocity of the pointing
across the celestial sphere can be expressed as
\begin{equation}
    \begin{split}
        \omega_{\rm max}&= \omega_{\alpha}\sin\kappa + \omega_{\beta}\sin{\beta}\\
        &= 2\pi \left( \frac{\sin\kappa}{T_{\alpha}}+ \frac{\sin\beta}{T_{\beta}}
        \right).
    \end{split}
\end{equation}
A comprehensive derivation of this $\omega_{\rm max}$ expression is elaborated in
\cref{apd:sweeping_velocity}.\footnote{This formulation assumes co-directional
precession and spin rotations. For counter-rotating configurations, refer to the
detailed analysis in \cref{apd:rotation_direction}.}

The transit duration $\tau$, representing the time interval during which the pointing
vector traverses an angular distance equivalent to the beam's Full Width at Half
Maximum (FWHM) $\Delta \theta$, is given by
\begin{align}
    \tau = \frac{\Delta \theta}{\omega_{\rm max}}.
\end{align}
For accurate demodulation of the polarization signal modulated by the HWP, the sampling
rate must exceed the Nyquist frequency with an adequate margin. Specifically, since
a HWP rotating at frequency $f_{\phi}$ modulates the polarization signal at $4f_{\phi}$,
the corresponding Nyquist frequency is $8f_{\phi}$. To ensure robust signal recovery,
we introduce a margin factor $\Nmargin$ ($\Nmargin>1$), leading to the following
sampling rate requirement:
\begin{align}
    f_{\rm s}> 8f_{\phi}\Nmargin. \label{eq:sampling_rate}
\end{align}
Furthermore, to ensure adequate polarization modulation within each sky pixel,
the HWP must complete $\Nmod$ ($\Nmod>1$) revolutions while the telescope's line
of sight traverses an angular distance equivalent to the beam's FWHM
$\Delta \theta$, yielding the requirement
\begin{align}
    \Nmod T_{\phi} & < \tau. \label{eq:modulation_condition}
\end{align}
By integrating \cref{eq:sampling_rate,eq:modulation_condition}, we establish a
fundamental constraint:
\begin{align}
    \frac{\Nmod}{\tau}< f_{\phi}< \frac{f_{\rm s}}{8\Nmargin}. \label{eq:total_condition}
\end{align}
This inequality ensures both adequate polarization signal modulation within the beam's
FWHM $\Delta \theta$ and sufficient temporal resolution for signal demodulation.
The resultant constraint on the spacecraft manifests through $\tau$, necessitating
a duration sufficient to satisfy \cref{eq:modulation_condition}. This
effectively imposes an upper threshold on the maximum angular velocity
$\omega_{\rm max}$ at which the instruments can traverse the celestial sphere. To
establish precise constraints on the spacecraft's kinematic parameters,
specifically $T_{\alpha}$ and $T_{\beta}$, we can reformulate \cref{eq:modulation_condition}
into:
\begin{align}
    \frac{\Nmod}{f_{\phi}} & < \frac{\Delta \theta}{2\pi \left( \frac{\sin\kappa}{T_{\alpha}} + \frac{\sin\beta}{T_{\beta}} \right)}. \label{eq:req_for_HWP}
\end{align}
Through algebraic manipulation, we derive a lower bound for $T_{\beta}$:
\begin{align}
    T_{\beta}^{\rm{lower}}\equiv\frac{2\pi \Nmod T_{\alpha}\sin\beta}{\Delta \theta f_{\phi}T_{\alpha}- 2\pi \Nmod \sin\kappa}. \label{eq:T_spin}
\end{align}
While previous studies, notably ref.~\cite{OptimalScan}, considered detector time
constants as a constraint on $T_{\beta}$, this consideration becomes redundant
in configurations incorporating HWP modulation. In such systems, the detector's
time constant primarily influences the HWP rotation frequency rather than
imposing direct constraints on the scanning strategy parameters. By applying the
condition $T_{\beta}<T_{\alpha}$ to \cref{eq:T_spin}, we can establish a lower
bound for $T_{\alpha}$ to maintain stable spacecraft inertial control:
\begin{align}
    T_{\alpha}^{\rm{lower}}\equiv\frac{2\pi \Nmod (\sin\beta + \sin\kappa)}{\Delta \theta f_{\phi}}. \label{eq:T_alpha}
\end{align}
This yields a hierarchical constraint on the rotational periods:
\begin{align}
    \tbl < T_{\beta}< T_{\alpha}.
\end{align}
The precession period $T_{\alpha}$ is further constrained by an upper limit, which
can only be determined through numerical simulations. Prior research
\cite{OptimalScan} indicates that precession periods exceeding one year
compromise the achievement of comprehensive sky coverage.

Through the systematic application of these angular and temporal constraints, we
can effectively reduce the dimensionality of the optimization problem from six to
three parameters. Given that $\kappa$ is typically fixed by instrumental
considerations and incorporating the constraint $\tbl<T_{\beta}$, the parameter space
can be reformulated as:
\begin{align}
    \{\alpha, \beta, T_{\alpha}, T_{\beta}, \nu_{\phi}, f_{\rm s}\}\rightarrow\{\alpha, \kappa-\alpha, T_{\alpha}, \tbl(\alpha,T_{\alpha})<T_{\beta}, \nu_{\phi}(f_{\rm knee}), f_{\rm s}(\nu_{\phi})\}.
\end{align}
The optimization problem ultimately reduces to the exploration of three
essential parameters $\{\alpha , T_{\alpha}, \tbl<T_{\beta}\}$, where $\nu_{\phi}$
and $f_{\rm s}$ are determined by the specified $f_{\rm knee}$ characteristics.

\section{The case of \LB mission}
\label{sec:case_of_LB}

The \LB instrument model employs the following configuration: $\alpha=45^{\circ}$
and $\beta=50^{\circ}$ (yielding $\kappa=95^{\circ}$), aligned with our
discussion in \cref{sec:angle-constraint}. The rotation periods are set to $T_{\alpha}
=3.2058$\,hours (192.348\,minutes) and $T_{\beta}=20$\,minutes, with HWP
rotation rates $\nu_{\phi}=46/39/61$ rpm for LFT/MFT/HFT respectively, and a sampling
rate $f_{\rm s}=19$ Hz\footnote{Precisely defined as
$f_{\rm s}=20\,{\rm MHz}/2^{20}$ \cite{PTEP2023}.}. The telescopes' Fields of
View are $18^{\circ}\times8^{\circ}$ for LFT and $14^{\circ}$ radius for both
MFT and HFT\footnote{MFT and HFT values represent FoV radius.}.

\begin{figure}[ht]
    \centering
    \includegraphics[width=0.49\columnwidth]{
        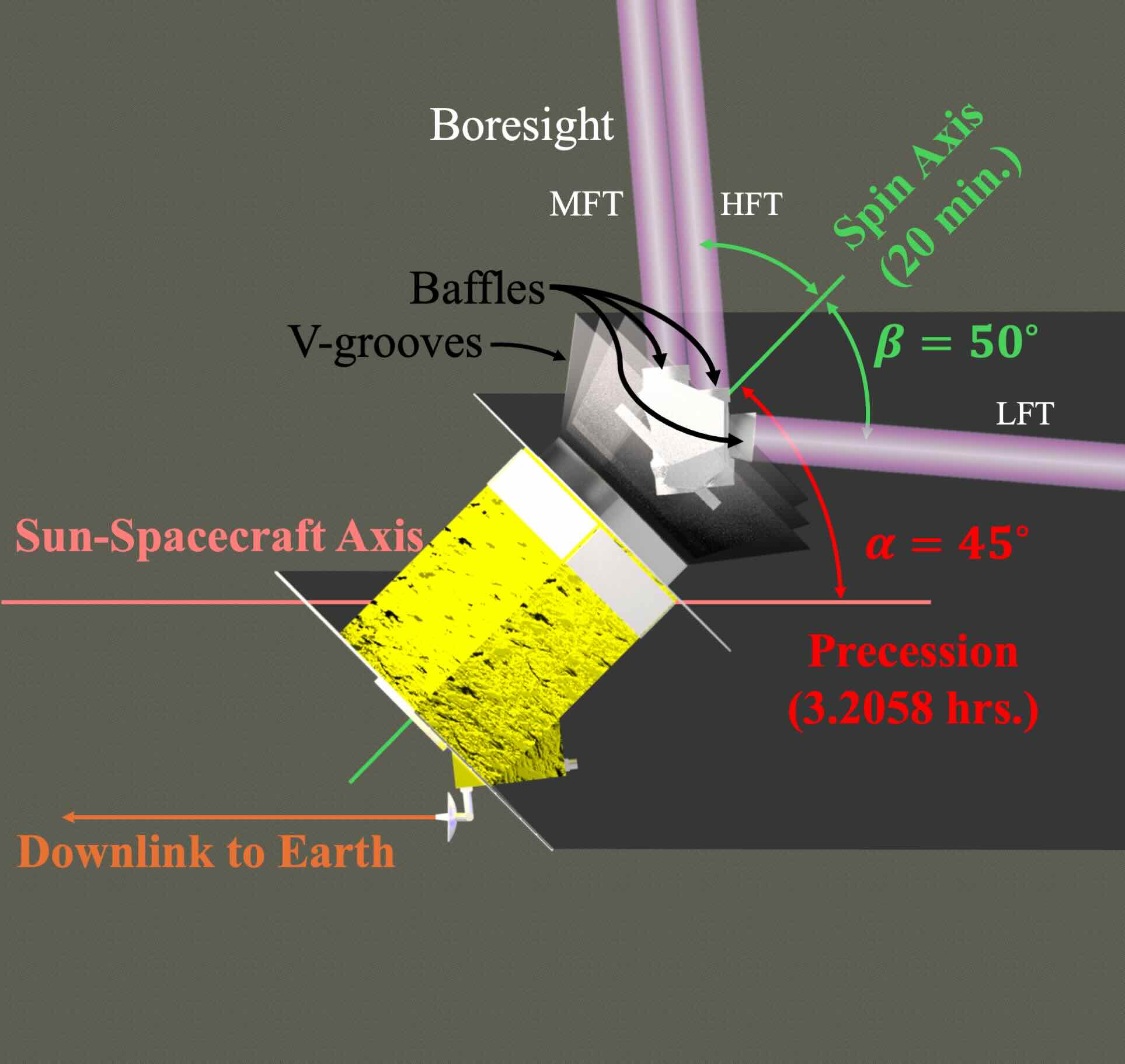
    }
    \includegraphics[width=0.49\columnwidth]{
        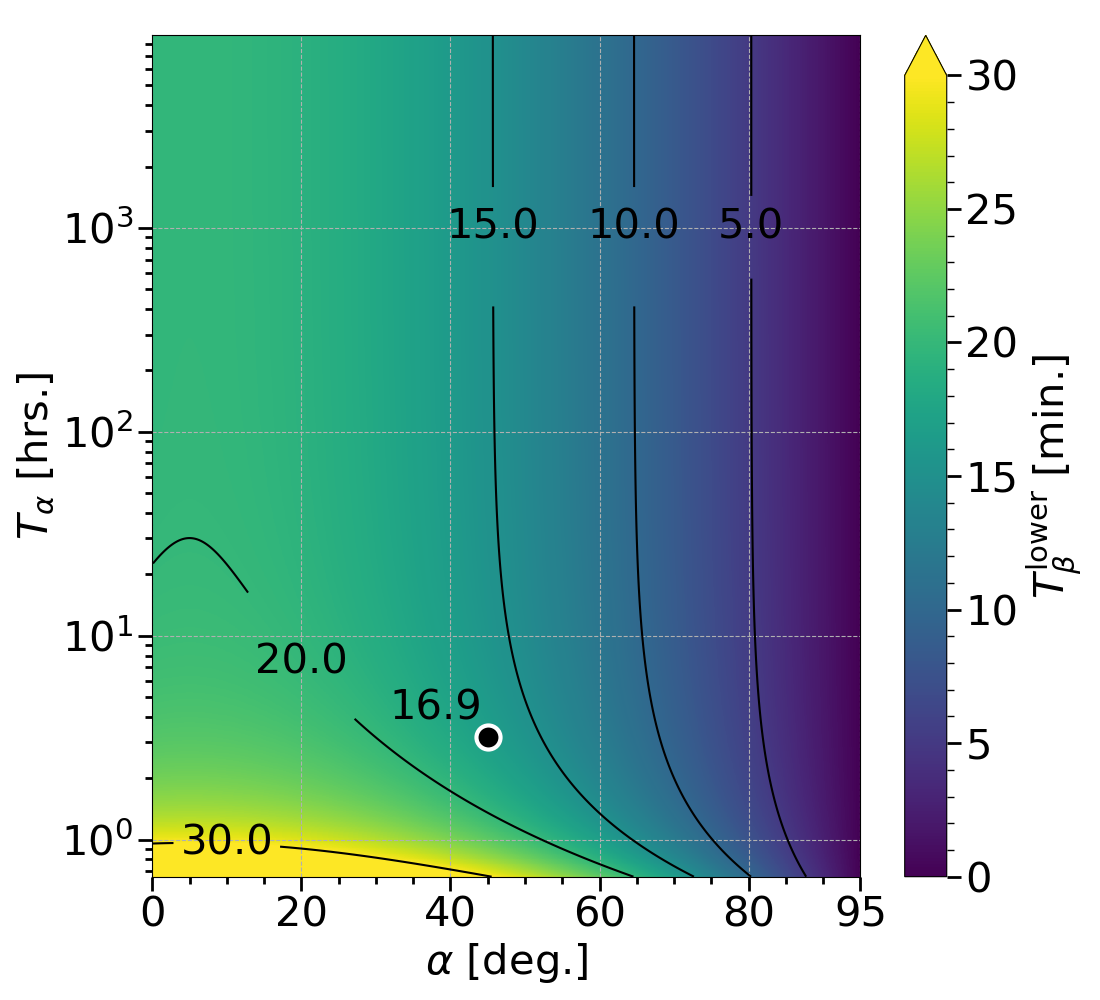
    }
    \caption[The scanning configuration and spin period constraints of \LB.]{(left)
    Illustration of \LB's standard scanning configuration. (right) Lower bound on
    spin period (\Tbetalow) calculated from \cref{eq:T_spin} with $\Nmod=1$,
    using HFT parameters (402 GHz, FWHM=$17.9'$). The black dot indicates the
    \SC parameters
    $(\alpha, T_{\alpha}, T_{\beta})=(45^{\circ}, 192.348\,\rm{min}, 16.9\,\rm{min}
    )$, where $\alpha$ and $T_{\alpha}$ are \LB's nominal values.}
    \label{fig:standard_config_and_T_beta}
\end{figure}

\begin{figure}
    \centering
    \includegraphics[width=0.9\columnwidth]{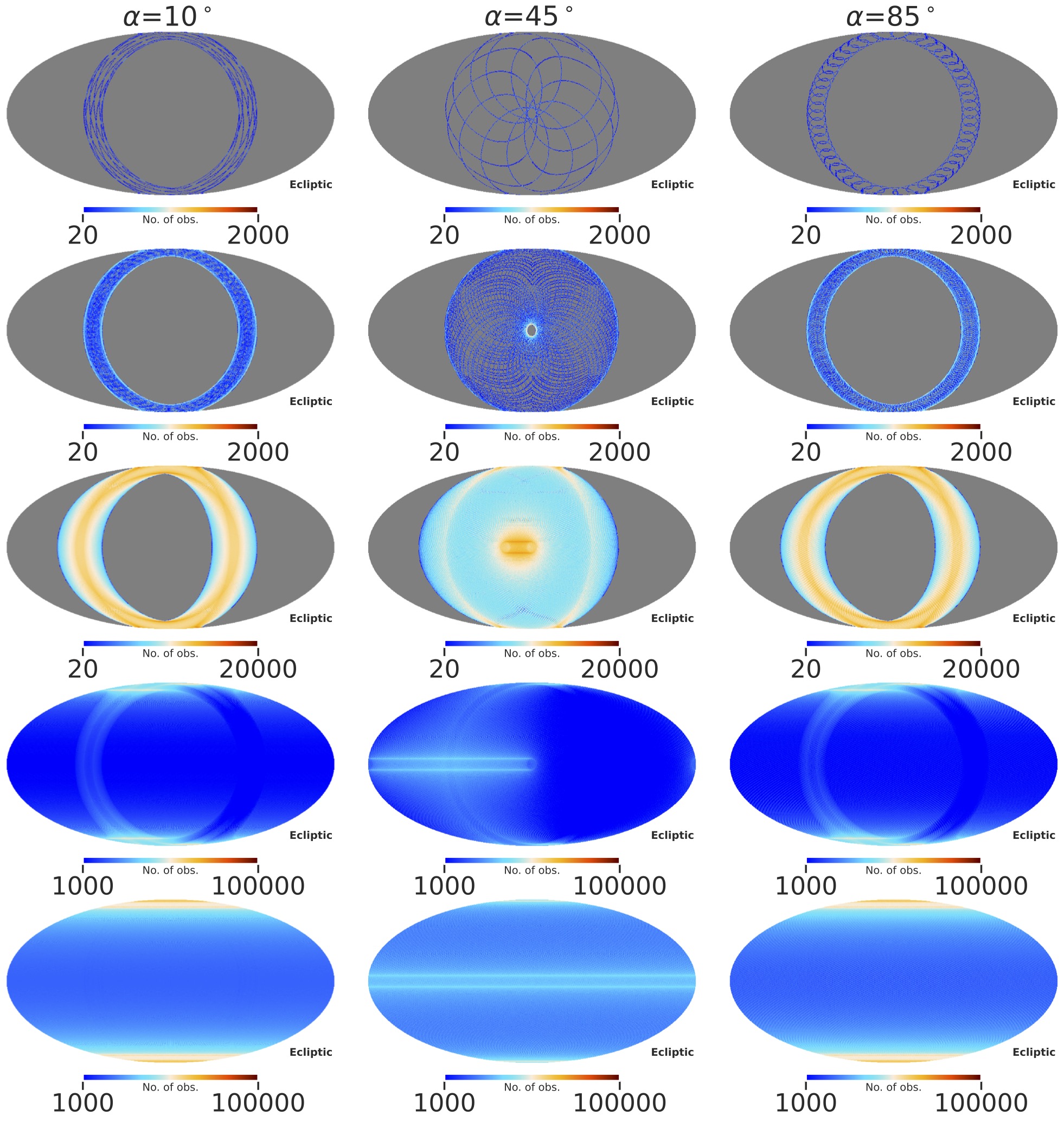}
    \caption[Hit-maps showing observation counts per sky pixel over time for
    different geometric parameters.]{Hit-maps showing observation counts per sky
    pixel over time for different geometric parameters. The central column shows
    the \SC ($\kappa=95^{\circ}$, $T_{\alpha}=192.348$\,minutes). Left and right
    columns show configurations with $\alpha=10^{\circ}$ and $85^{\circ}$ respectively,
    with corresponding $T_{\beta}$ values of 21.9 and 3.8\,minutes calculated
    from \cref{eq:T_spin}. Rows display hit-map evolution from one precession period
    (192.348\,minutes) through 1 day, 1 month, 6 months, to 1 year. Simulations use
    a single boresight detector with 19 Hz sampling rate and $N_{\rm side}=128$ (pixel
    size $\approx 0.46^{\circ}$).}
    \label{fig:hitmaps}
\end{figure}

This configuration, termed the \SC, is depicted in \cref{fig:standard_config_and_T_beta}
(left) and detailed in \cref{tab:LB_standard_config}. \Cref{fig:hitmaps} (middle
column) shows simulated hit-maps counting observations per sky pixel under the
\SC. For comparison, the figure includes hit-maps with $\alpha=10^{\circ}$ (left)
and $\alpha=85^{\circ}$ (right), demonstrating geometric parameter effects on scan
patterns. These simulations maintain the \SC's $T_{\alpha}$ while using
\Tbetalow calculated for each $\alpha$ and $T_{\alpha}$ combination.

This work aims to explore and rigorously justify this configuration choice,
focusing on in-flight calibration and systematic effect suppression. To evaluate
the constraints from \cref{eq:T_spin}, we use the parameters of the HFT, which has
the highest HWP revolution rate and smallest FWHM ($\Delta \theta=17.9'$) at 402
GHz. For simplicity, we apply these parameters across all frequency bands, with
$\Nmod =1$ and $\Nmargin=2$ as recommended in ref.~\cite{PTEP2023}. In the \LB
instrument model, satisfying \cref{eq:T_spin} for the HFT parameters ensures compliance
for LFT and MFT as well. Using these values to calculate \Tbetalow yields the
parameter space shown in \cref{fig:standard_config_and_T_beta} (right), where $\tbl
=16.9$\,minutes at $(\alpha,T_{\alpha})=(45^{\circ},192.348\,\rm{min})$.

\begin{table}[h]
    \centering
    \begin{tabular}{cccccccc}
        \hline
        $\alpha$ & $\beta$ & Precession period & Spin period & \multicolumn{3}{c}{HWP revolution rate} & Sampling rate \\
                 &         &                   &             & LFT                                     & MFT          & HFT &    \\
        {[deg]}  & {[deg]} & {[min]}           & {[min]}     & \multicolumn{3}{c}{[rpm]}               & {[Hz]}        \\
        \hline
        45       & 50      & 192.348           & 20          & 46                                      & 39           & 61  & 19 \\
        \hline
    \end{tabular}
    \caption[Scanning strategy parameters of the \LB mission.]{Parameters of the
    \SC for the \LB mission \cite{PTEP2023}.}
    \label{tab:LB_standard_config}
\end{table}

From the \LB instrument model and the discussion in the previous section, the effective
free parameters of the scanning strategy to optimize are given by
\begin{align}
    \{\alpha, \beta, T_{\alpha}, T_{\beta}, \nu_{\phi}, f_{\rm s}\} \rightarrow \{ \alpha, 95^{\circ}-\alpha, T_{\alpha}, 16.9\,\mathrm{min}< T_{\beta}, 61\,\rm{rpm}, 19\,\rm{Hz}\}. \label{eq:scan_parameter_space}
\end{align}

For the subsequent analysis, we concentrate our simulations on a single detector
positioned at the boresight (the central axis of the telescope's focal plane).
Our comprehensive verification demonstrates that extending the analysis to
detectors situated at the periphery of the FoV yields conclusions consistent
with those derived from the boresight detector, as elaborated in \cref{sec:optimization}.
For an exhaustive treatment of non-boresight detector characteristics, readers are
directed to \cref{apd:other_detector}.

    \section{Metrics for optimization}
\label{sec:metrics}

\subsection{Visibility time of compact sources}

The seminal observations from the \Planck mission underscore the critical significance
of compact astronomical sources, particularly solar system planets and distant
galaxy clusters, in achieving precise beam characterization and pointing calibration
\cite{Planck_HFI_beam, Planck_LFI_beam}. Consequently, an optimal scanning
strategy must strike a delicate balance between maximizing the observation duration
of these calibration sources while simultaneously fulfilling the primary scientific
objectives within the prescribed mission timeline. The calibration precision is
anticipated to correlate directly with the integrated duration during which the
instrument's boresight maintains visibility of these compact sources. Our analysis
seeks to identify optimal parameters in the $\{\alpha, T_{\alpha}\}$ parameter
space that maximize this integrated visibility duration throughout the mission.

\subsection{Forming speed of sky coverage}
A fundamental requirement for any scanning strategy is its ability to efficiently
survey extensive celestial regions. \EPIC is engineered to observe more than 50\%
of the celestial sphere within a single 24\,hours period, achieved through the
careful selection of a 3.2\,hours precession period and a 1\,rpm spin rate.
This configuration enables data acquisition across all necessary angular scales
for comprehensive analysis \cite{bock2008experimental}. To quantify this coverage
efficiency, we introduce the $T_{\rm cover}$ metric, defined as the duration
required to survey a specified fraction of the hemispheric sky. A shorter
$T_{\rm cover}$ indicates more rapid sky coverage, which is instrumental in
mitigating the impact of $1/f$ noise contamination at large angular scales.

\subsection{Hit-map uniformity}

The spatial uniformity of the hit-map distribution across the celestial sphere is
paramount for achieving homogeneous sensitivity. Any significant heterogeneity in
sensitivity across distinct celestial regions can potentially amplify noise variance,
thereby compromising the efficacy of foreground removal.

We define a spherical coordinate system $\Omega_{i}=(\theta_{i},\varphi_{i})$
and employ index $i$ to designate sky pixels in accordance with the \healpix
schema \cite{healpix} (see \cref{apd:healpix}).\footnote{\url{https://healpix.sourceforge.io/}} To
quantitatively assess the degree of sensitivity homogeneity, we employ the
hit-map's standard deviation, $\sigma_{\rm hits}$, mathematically expressed as
\begin{align}
    \sigma_{\rm{hits}}= \sqrt{\frac{1}{N_{\rm pix}-1}\sum_{i=0}^{N_{\rm pix}-1}\left(\ab<\Nhits>-\Nhits(\Omega_{i})\right)^{2}},
\end{align}
where $\Nhits$ represents the aggregate number of observational hits, and $\ab<.>$
denotes the full-sky averaged value. A diminished $\sigma_{\rm hits}$ value
signifies enhanced hit-map uniformity, consequently indicating more consistent
sensitivity across sky pixels. We conduct a comprehensive analysis of the
$\sigma_{\rm hits}$ distribution within the $\{\alpha, T_{\alpha}\}$ parameter space.

\subsection{Cross-link factor}
\label{sec:spin_char}

Building upon the framework established in ref.~\cite{OptimalScan}, we examine how
scanning strategies with diverse crossing angles across sky pixels can effectively
mitigate systematic effects. To quantify the uniformity of these crossing angles,
we introduce the 'cross-link factor', defined mathematically as:
\begin{align}
    \ab|\h[n,m](\Omega)|^{2} & = \ab (\frac{\sum_{j}^{\Nhits}\cos(n\psi_{j}+m\phi_{j})}{\Nhits})^{2}+\ab(\frac{\sum_{j}^{\Nhits}\sin(n\psi_{j}+m\phi_{j})}{\Nhits})^{2}\nonumber \\
                             & = \ab<\cos(n\psi_{j}+m\phi_{j}) >^{2}+ \ab< \sin(n\psi_{j}+m\phi_{j}) >^{2}, \label{eq:crosslink}
\end{align}
where $\psi$ represents the intersection angle between the crossing angle
and the meridian, while $\phi$ denotes the HWP angle.\footnote{The HWP
angle $\phi$ is defined in accordance with standardized polarization conventions,
following either IAU or COSMO (HEALPix) frameworks, see \cref{apd:pol_convention} for detail.}
For our analysis, we maintain the assumption that the crossing angle remains
parallel to the detector polarization angle. The index $j$ corresponds to the
$j^{\rm{th}}$ observation at a given sky pixel, and the pair $(n,m)$ represents
the respective \spin components for the crossing and HWP angles. It is noteworthy
that in circular statistics, the cross-link factor exhibits a direct relationship
with the circular variance $V$, expressed as $V=1-\sqrt{\ab|\h[n,m](\Omega)|^{2}}$.

For an ideally uniform distribution of crossing angles, the cross-link factor
vanishes ($\ab|\h[n,m](\Omega)|^{2}=0$) across all \spin-$(n,m)$ combinations.
Therefore, a minimal cross-link factor indicates optimal cross-linking efficiency.
Previous studies \cite{OptimalScan,mapbased} have demonstrated that minimizing
these factors across the celestial sphere effectively mitigates systematic effects.

This investigation incorporates a continuously rotating HWP into \cref{eq:crosslink}
and analyzes the cross-link factor distribution for each \spin-$(n,m)$ component
within the $\{\alpha, T_{\alpha}\}$ parameter space. Following the methodology
proposed in ref.~\cite{OptimalScan}, we evaluate the sky-averaged cross-link
factor, $\ab<\ab|\h[n,m]|^{2}>$. The theoretical framework developed in
\cref{chap:formalism} establishes the relationship between specific systematic
effects and their corresponding cross-link factors in \spin space, thereby
demonstrating the metric's utility in systematic effects suppression.

    \section{Results}
\label{sec:results}

In this section, we present the optimization metrics results: visibility time of
planets, hit-map uniformity, and cross-link factor. Our analysis focuses on the
special case where $T_{\beta}=\tbl$, allowing us to evaluate all metrics in the $\{
\alpha,T_{\alpha}\}$ space, as shown in \cref{fig:standard_config_and_T_beta} (right).
We demonstrate in \cref{apd:T_beta_scaled} that choosing different values of $T_{\beta}
>\tbl$ merely scales the distribution of metrics in this space while preserving their
optimal values.

\subsection{Visibility time of compact sources}
\label{sec:comp_source_obs}

A planetary observation is recorded when the boresight comes within
$0.5^{\circ}$ of a planet's position. The accumulated duration of such encounters
defines the visibility time. Also, we consider same quantity for Crab Nebula which
is often used as a polarization calibration source. For simulating planetary
motions, we make the following assumptions:
\begin{itemize}
  \item Initial planetary positions are obtained using \texttt{Astropy}
    \cite{astropy} at 2032-04-01T00:00:00 in Barycentric Dynamical Time.\footnote{\url{https://www.astropy.org/}}
    We assume the position of the Crab Nebula does not change during the mission.

  \item Planetary positions are computed at one-second intervals, with positions
    assumed static within each second.

  \item Five outer planets (Mars, Jupiter, Saturn, Uranus, and Neptune) and Crab
    Nebula are used as calibration sources.
\end{itemize}

\begin{figure}[h]
  \centering
  \includegraphics[width=0.24\columnwidth]{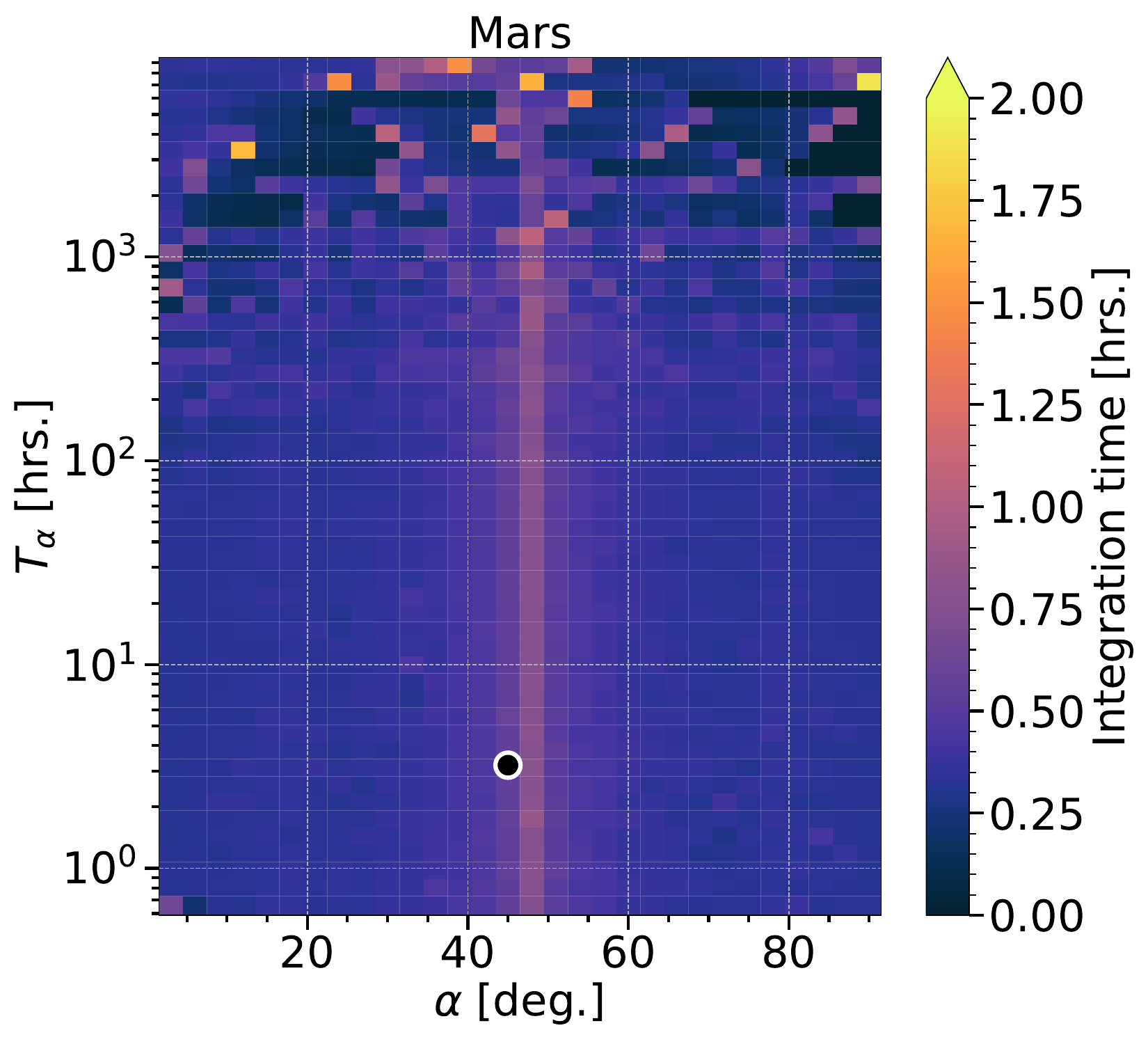}
  \includegraphics[width=0.24\columnwidth]{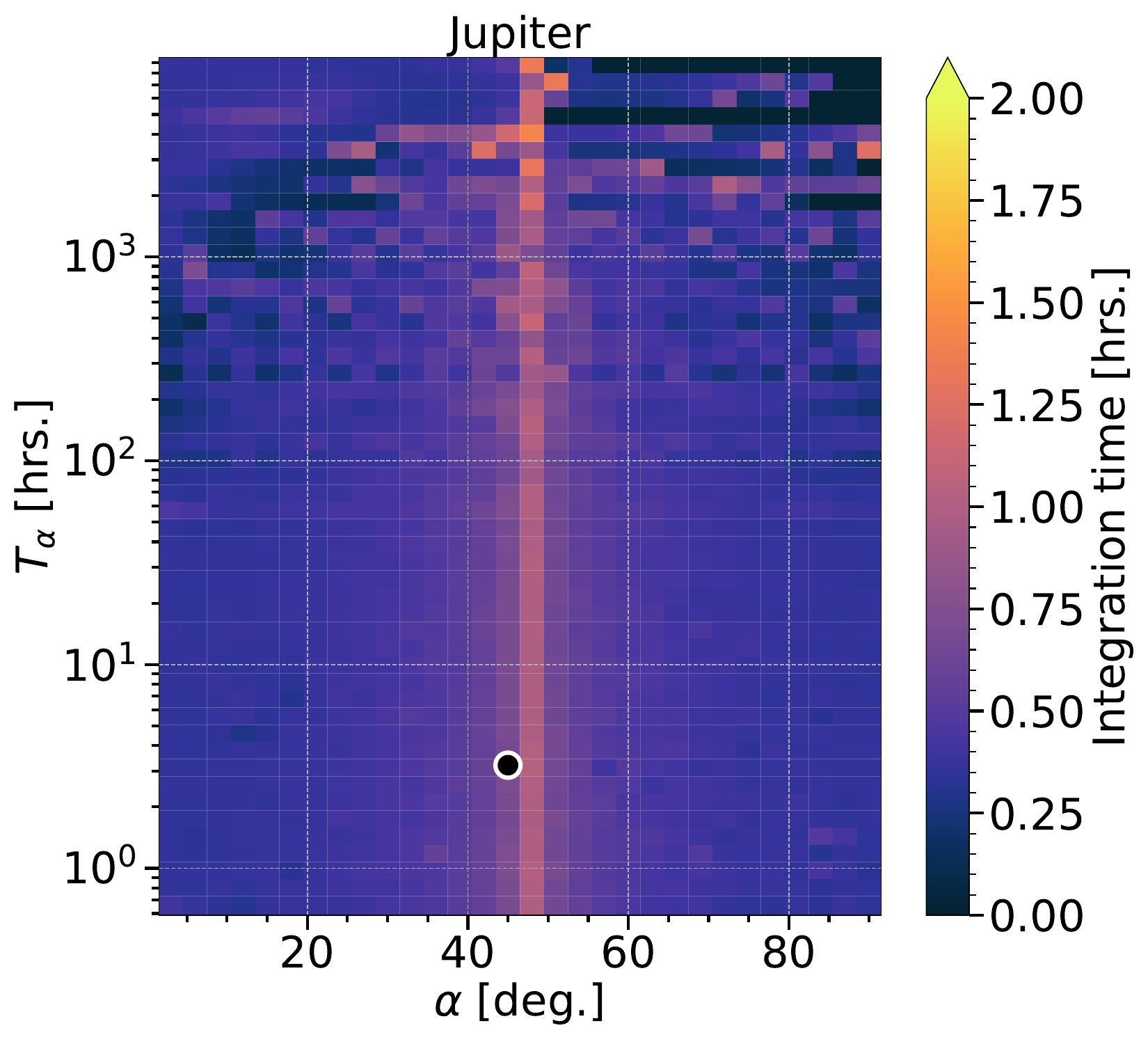}
  \includegraphics[width=0.24\columnwidth]{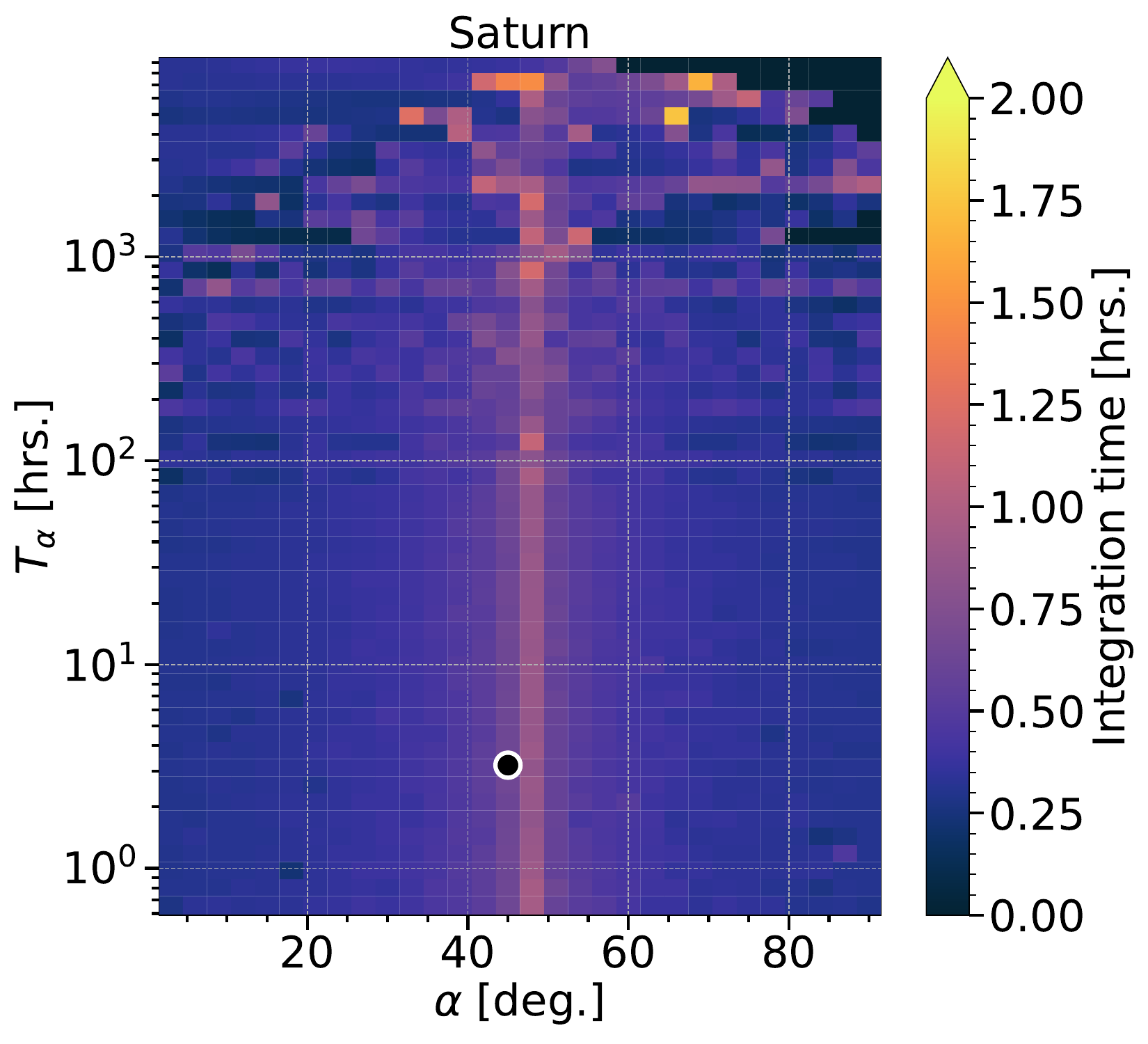}
  \\
  \includegraphics[width=0.24\columnwidth]{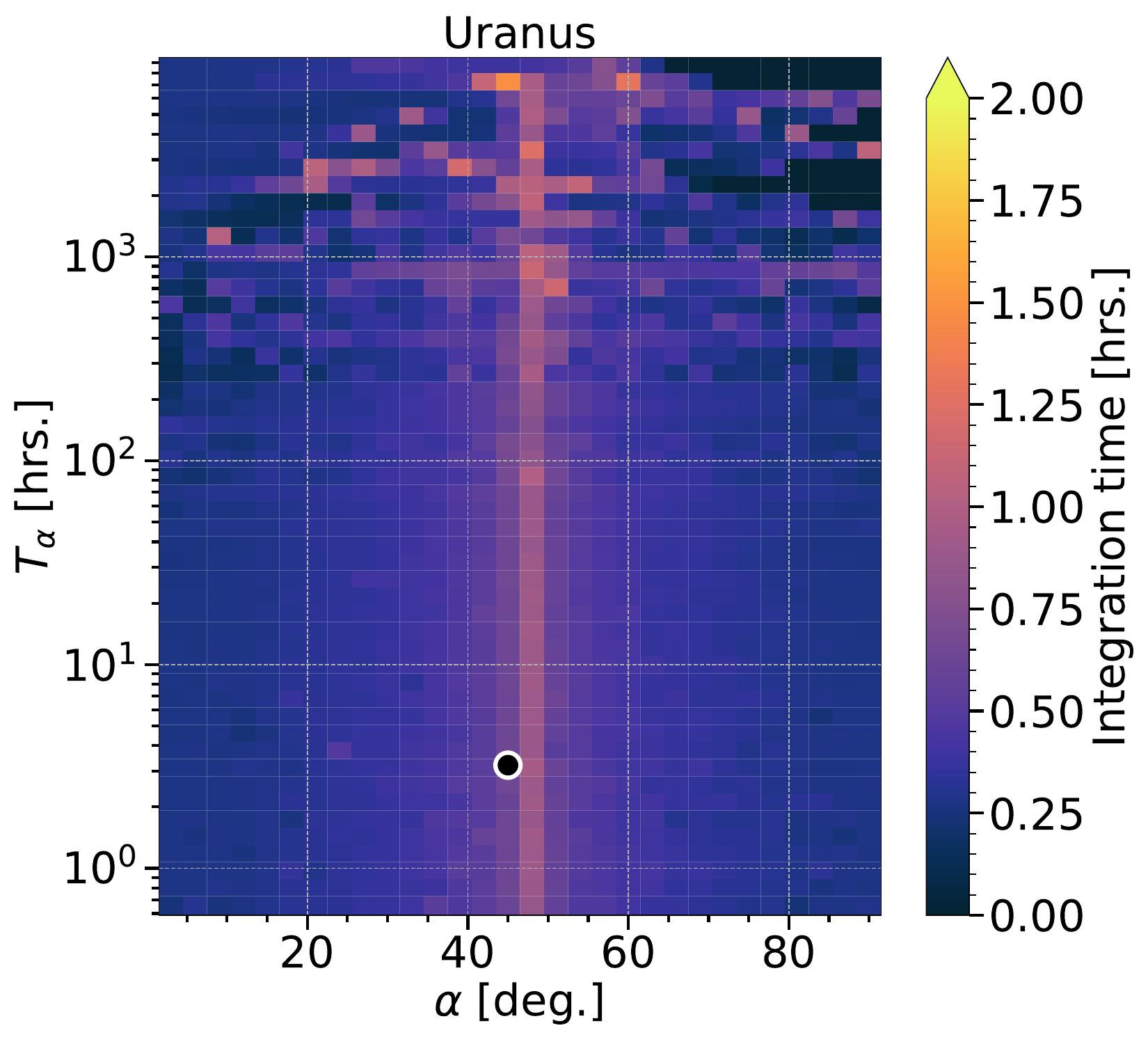}
  \includegraphics[width=0.24\columnwidth]{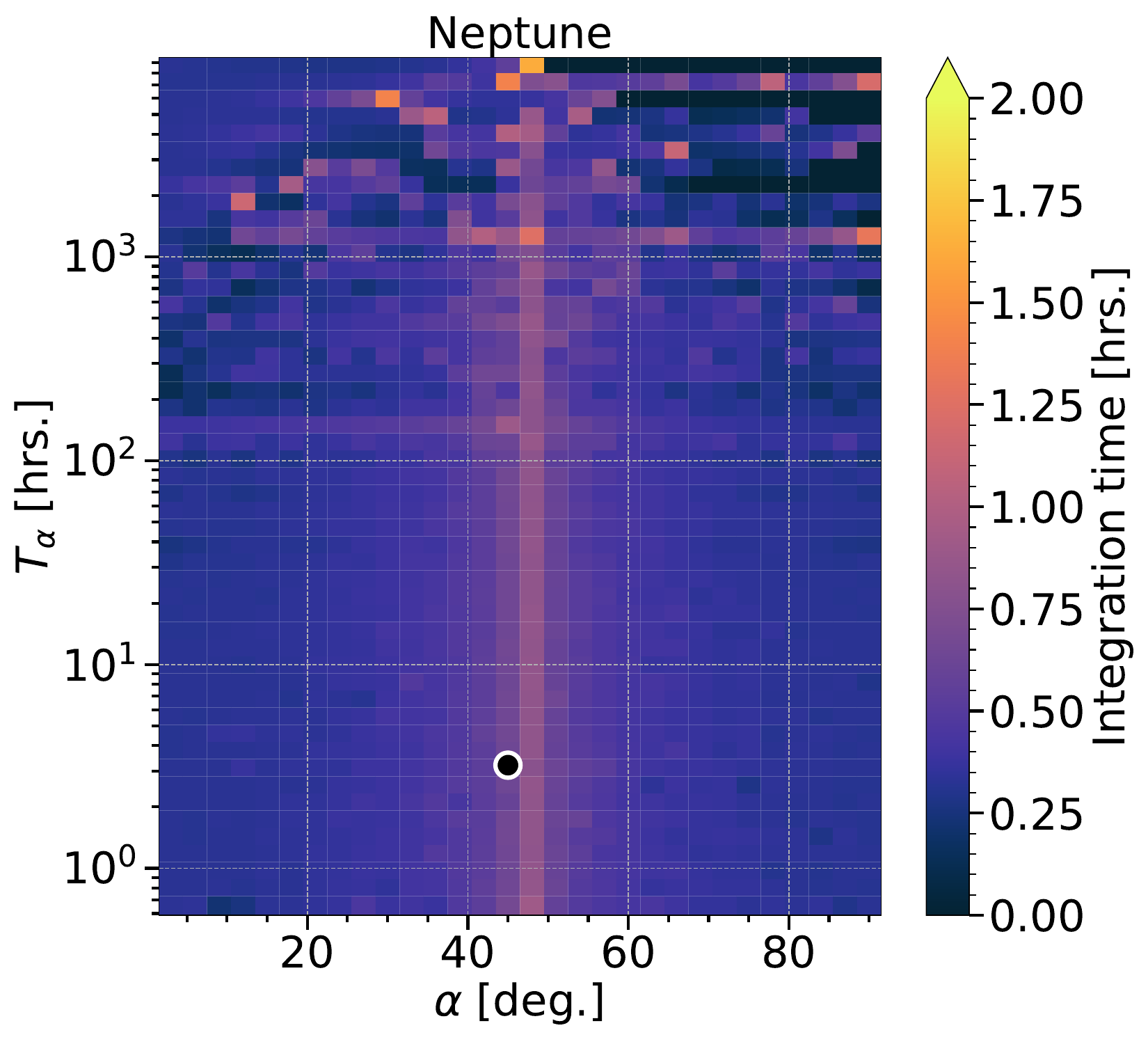}
  \includegraphics[width=0.24\columnwidth]{
    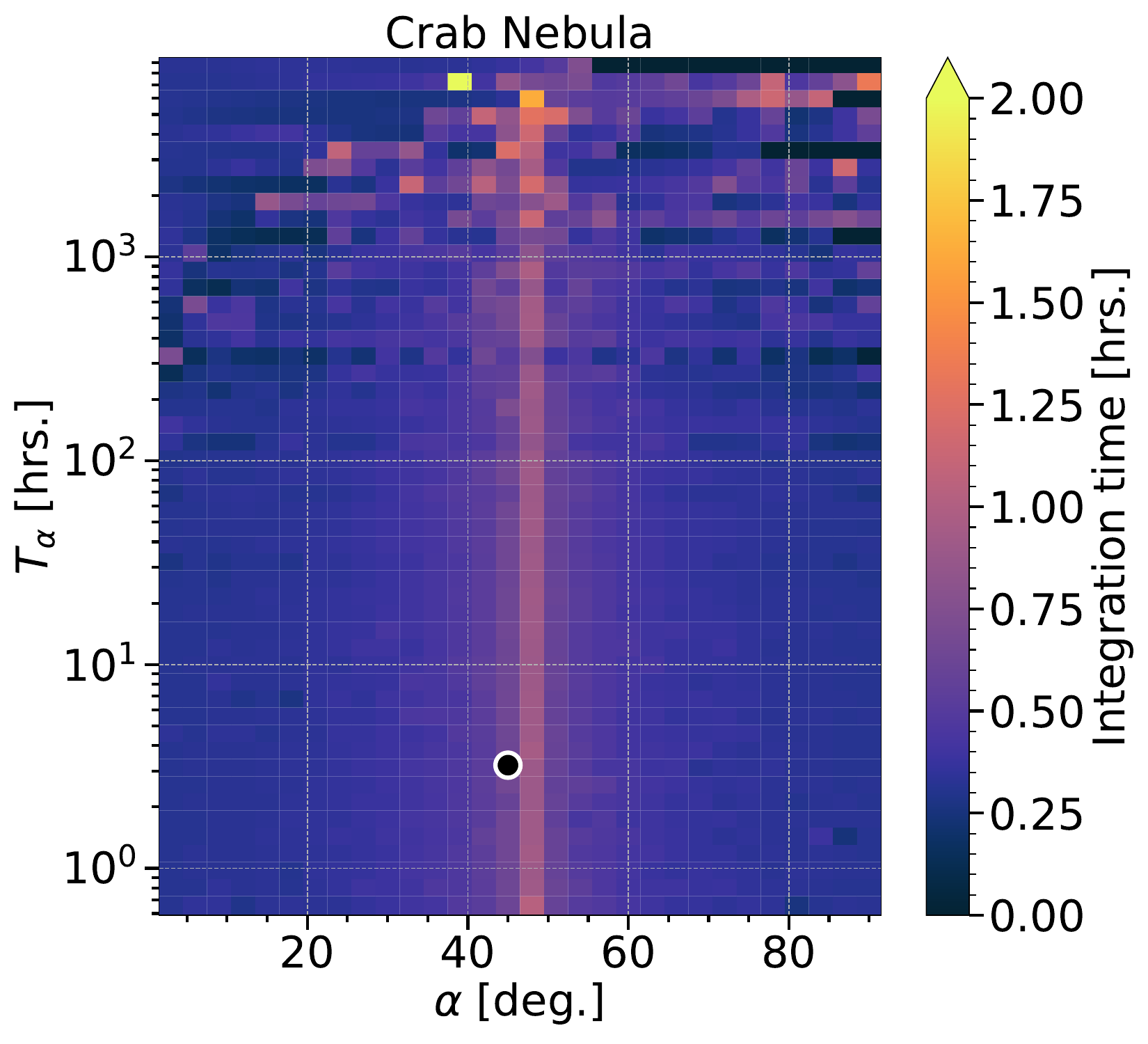
  }
  \includegraphics[width=0.24\columnwidth]{
    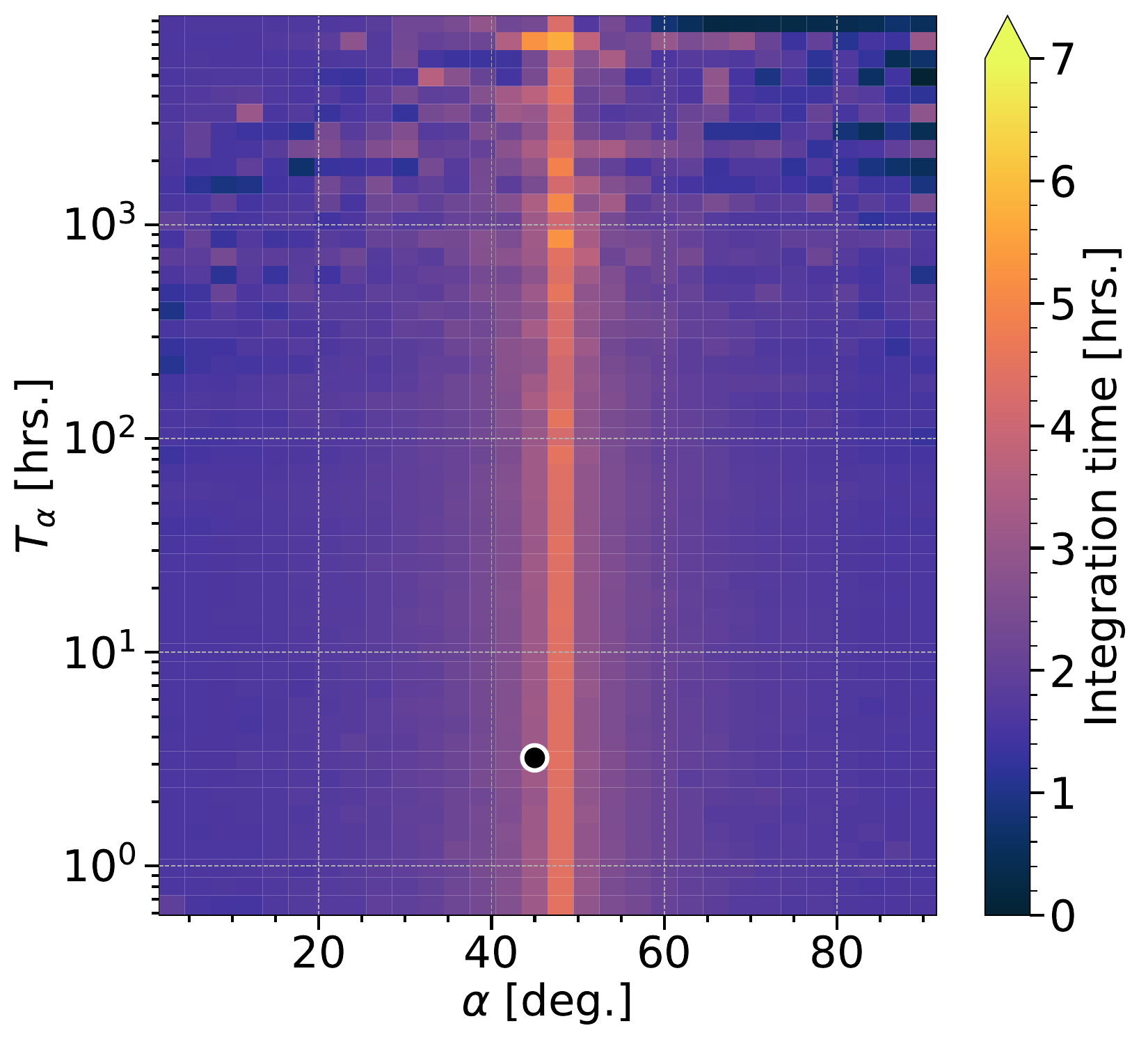
  }
  \caption[Visibility time of planets, Crab Nebula and accumulated visibility time
  for all planets]{Visibility time of planets, Crab Nebula and accumulated visibility
  time for all planets. The order of figure is organized as Mars, Jupiter,
  Saturn, Uranus, Neptune, and Crab Nebula from top left to bottom right. The
  plot on the bottom right shows the accumulated planets visibility time
  distribution.}
  \label{fig:planet_visibility}
\end{figure}

\Cref{fig:planet_visibility} displays the integrated visibility time distribution
in the $\{\alpha, T_{\alpha}\}$ space over the mission's 3-year duration,
accounting for all boresight observations of the considered compact sources. The
distribution pattern is consistent across all planets, with peak visibility
times of 0.77 (Mars), 1.00 (Jupiter), 0.87 (Saturn), and 0.83 (Neptune), 0.90 (Crab
Nebula)\,hours. The accumulated visibility time for all planets is shown in the bottom
right plot of \cref{fig:planet_visibility}, which has 4.4\,hours as the peak.

For $T_{\alpha}\lesssim100$\,hours, the integrated visibility time shows minimal
variation with respect to $T_{\alpha}$. However, the dependence on $\alpha$ is more
pronounced, with the maximum visibility occurring at $\alpha=\beta=47.5^{\circ}$
and displaying symmetrical behavior around this value. This pattern can be understood
by examining the scan trajectories illustrated in \cref{fig:hitmaps}. When
$\alpha=\beta=47.5^{\circ}$, the two parallel scan lines, formed by the orbital
rotation's tangent line shift, intersect at the equator.

The visibility is limited by what we term the 'scan pupil' --- an inner gap where
planets cannot be observed. This pupil is visible in the center of the one-day hit-map
simulation shown in \cref{fig:hitmaps}. Consequently, scanning strategies where $\alpha$
deviates significantly from $47.5^{\circ}$ result in reduced integrated visibility
times.

\subsection{Forming speed of sky coverage}
We simulate the time required to achieve half-sky coverage, $T_{\rm cover}$ in the
$\Nside=128$ map. \Cref{fig:coverage_and_sigma_hit} (left) shows $T_{\rm cover}$,
in the $\{\alpha, T_{\alpha}\}$ space. It is clear as \cref{fig:hitmaps} shown, if
we have too small (or too large) $\alpha$, $T_{\rm cover}$ is large, because the
scan pupil is large. In order to cover the large sky fraction in a short time, we
need to set $3 0^{\circ}\lesssim\alpha\lesssim70^{\circ}$.

\subsection{Hit-map uniformity}
\label{sec:hitmap_dist}

\begin{figure}[htbp]
  \centering
  \includegraphics[width=0.49\columnwidth]{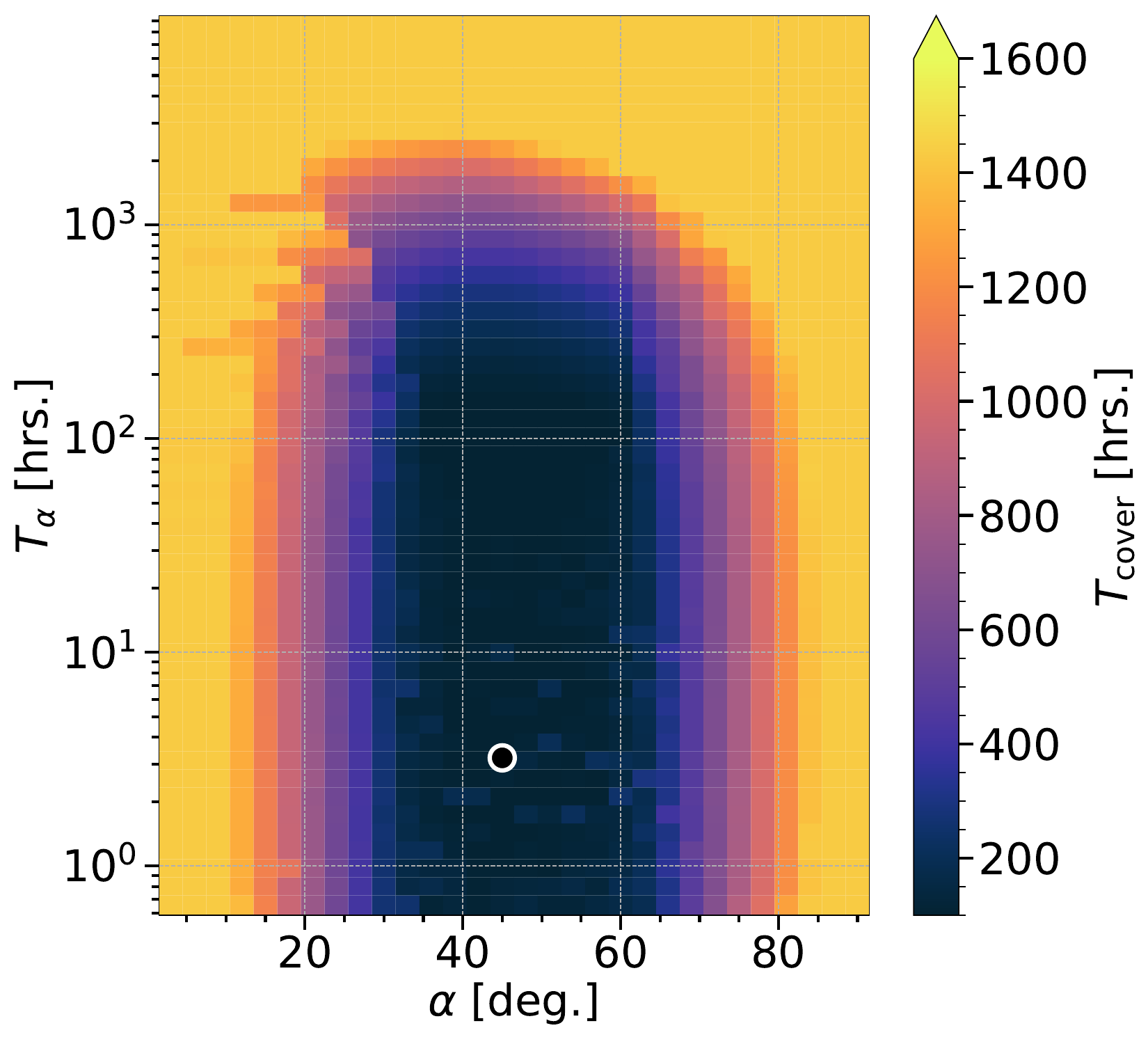}
  \includegraphics[width=0.49\columnwidth]{
    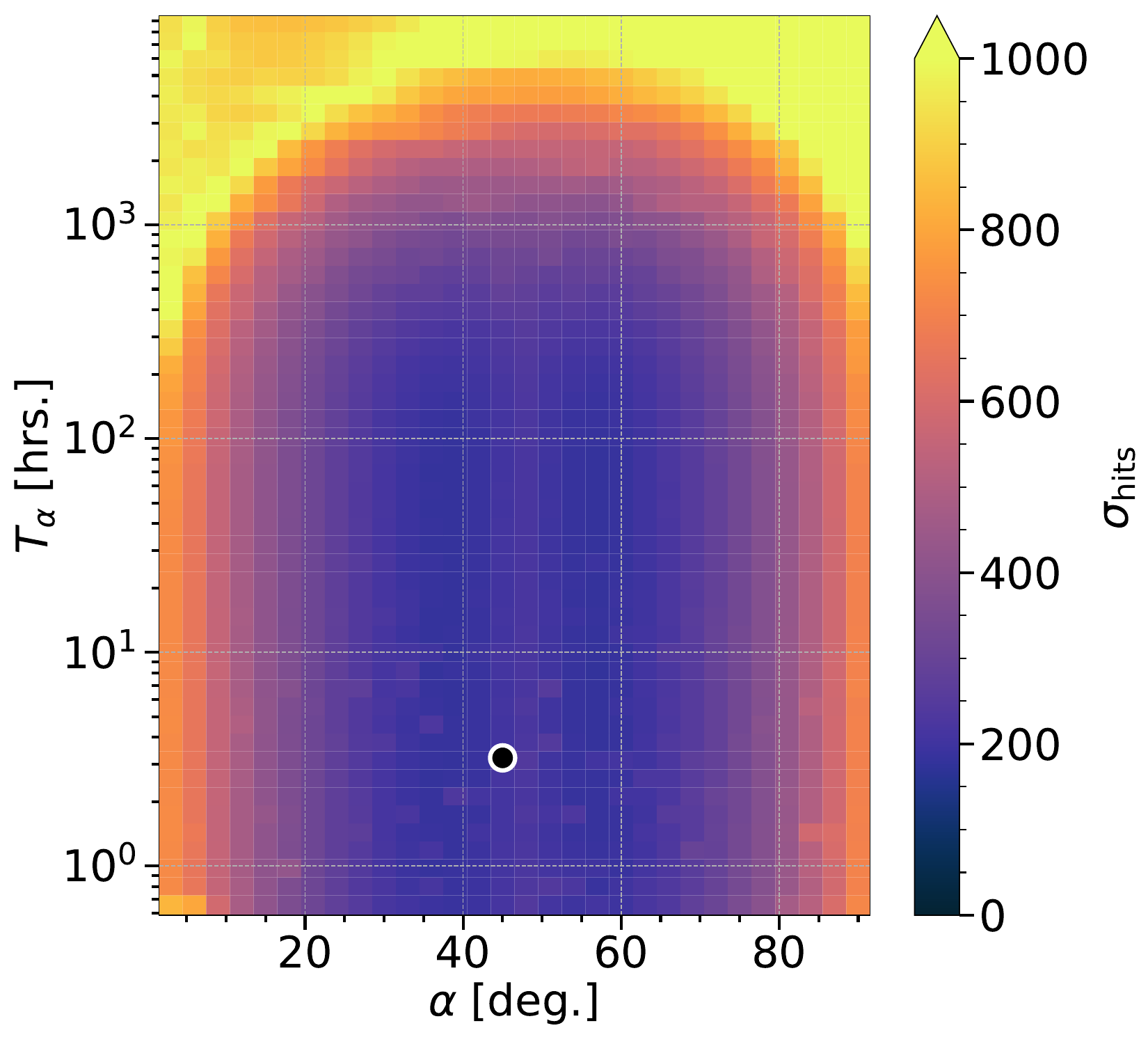
  }
  \caption[Time required to achieve half-sky coverage, $T_{\rm cover}$, and hit-map
  standard deviation, $\sigmahits$]{(left) The time requires to achieve half-sky
  coverage, $T_{\rm cover}$. This is simulated by the $\Nside=128$ map. (right) Hit-map
  standard deviation at $\Nside=256$. Higher $\sigmahits$ values occur at
  extreme $\alpha$ values due to increased polar observations and reduced
  equatorial coverage.}
  \label{fig:coverage_and_sigma_hit}
\end{figure}

The distribution of $\sigmahits$ is shown in \cref{fig:coverage_and_sigma_hit} (right).
High values of $\sigmahits$ occur at extreme values of $\alpha$ (both large and
small), as illustrated in \cref{fig:hitmaps} (left/right). This is because such scanning
strategies result in frequent polar observations but fewer equatorial ones during
the one-year observation period. Similarly, when $T_{\alpha}>100$\,hours, the
limited orbital rotation during precession prevents multiple cycles in the same
sky region, breaking azimuthal symmetry on the hit-map (see \cref{fig:hitmaps}) and
increasing $\sigmahits$.

For $25^{\circ}\lesssim \alpha \lesssim 75^{\circ}$ and $T_{\alpha}\lesssim100$\,hours,
$\sigmahits$ remains relatively stable, with a notable exception at $\alpha=\beta$.
At this special point, trajectory intersection at the equator creates a closed scan
pupil, leading to intensive scanning of poles and equator but reduced coverage
of equatorial regions, thus increasing variance.

This creates an optimization challenge: minimizing $\sigmahits$ conflicts with
maximizing planetary visibility time, which is determined by the scan pupil size.

Notable outliers appear in the region where $T_{\alpha}\lesssim10$\,hours,
showing $\sigmahits$ values that deviate from surrounding pixels. This phenomenon,
caused by spin-precession period resonance, will be examined in \cref{sec:Opt_kinetic}.

\subsection{Cross-link factor}
\label{sec:result_crosslink}

\begin{figure}
  \centering
  \includegraphics[width=0.32\columnwidth]{
    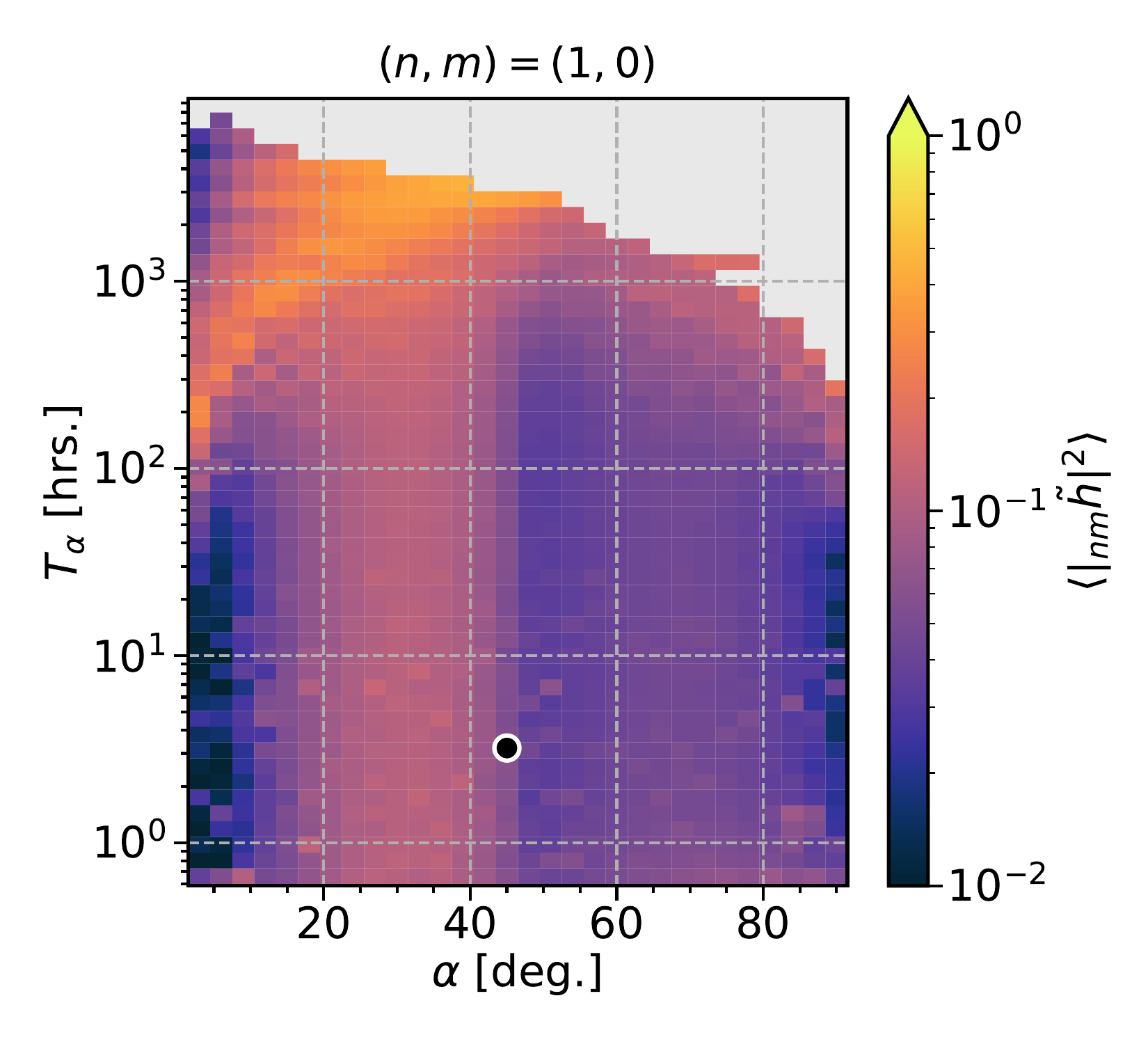
  }
  \includegraphics[width=0.32\columnwidth]{
    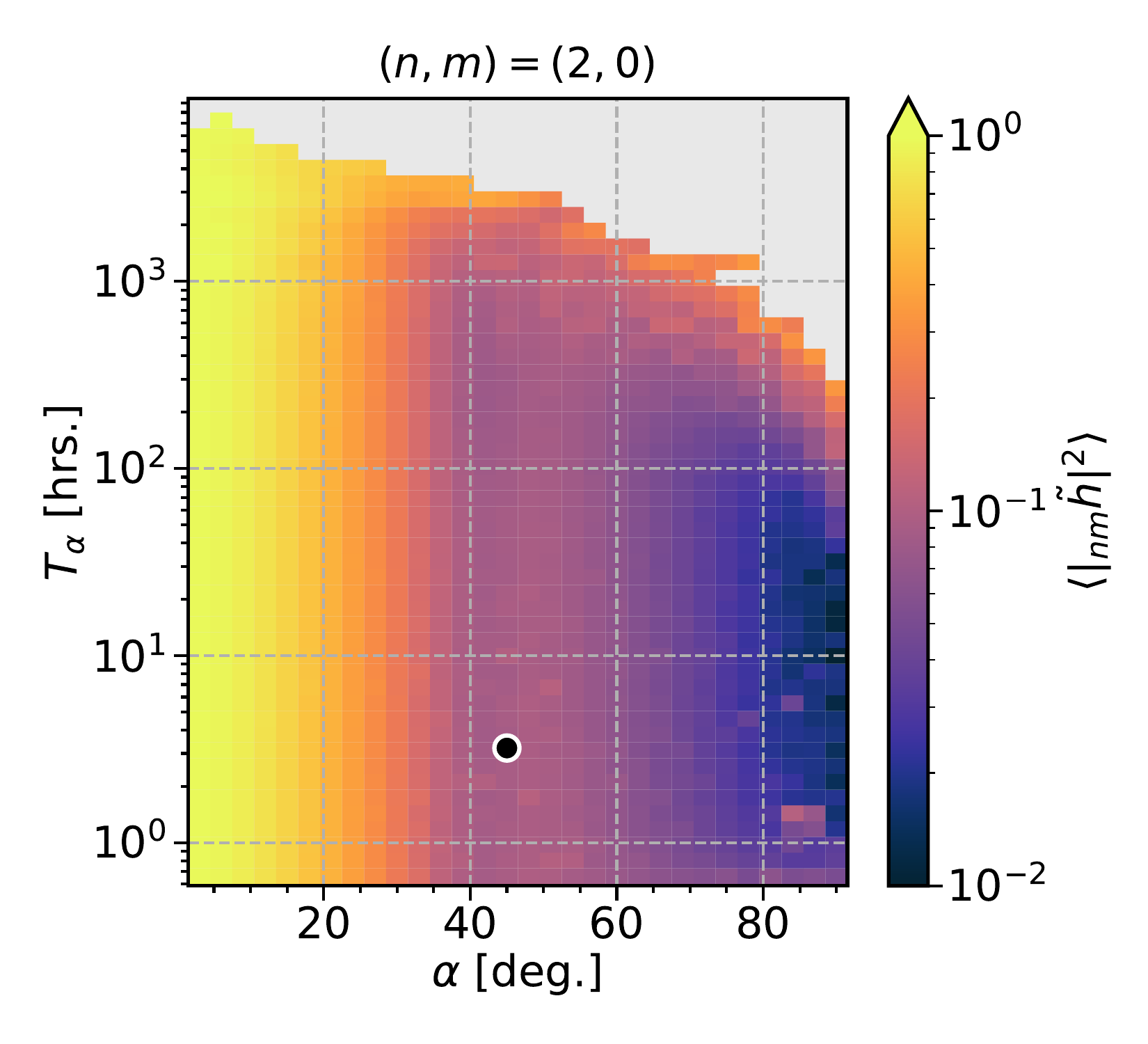
  }
  \includegraphics[width=0.32\columnwidth]{
    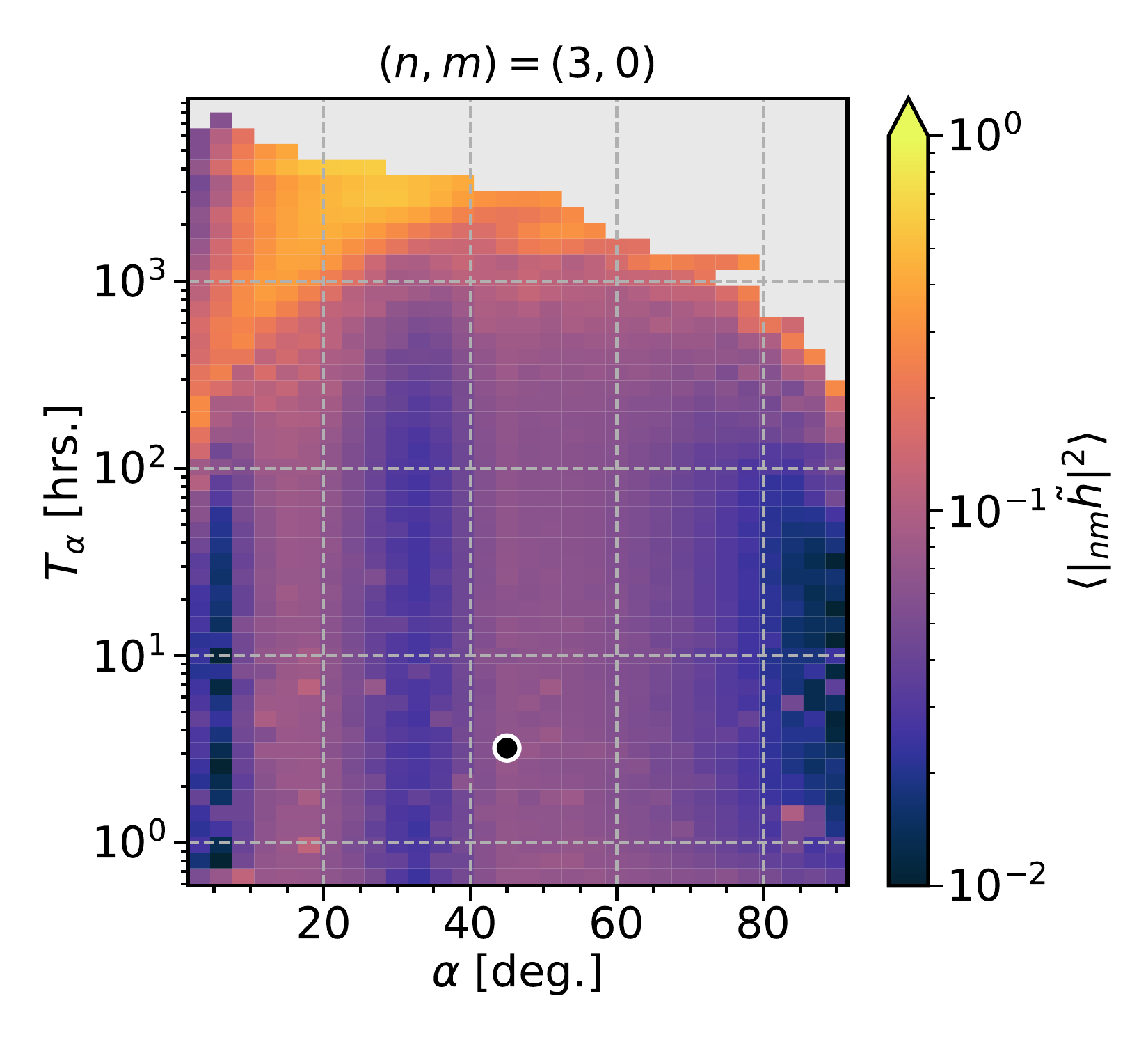
  }
  \\
  \includegraphics[width=0.32\columnwidth]{
    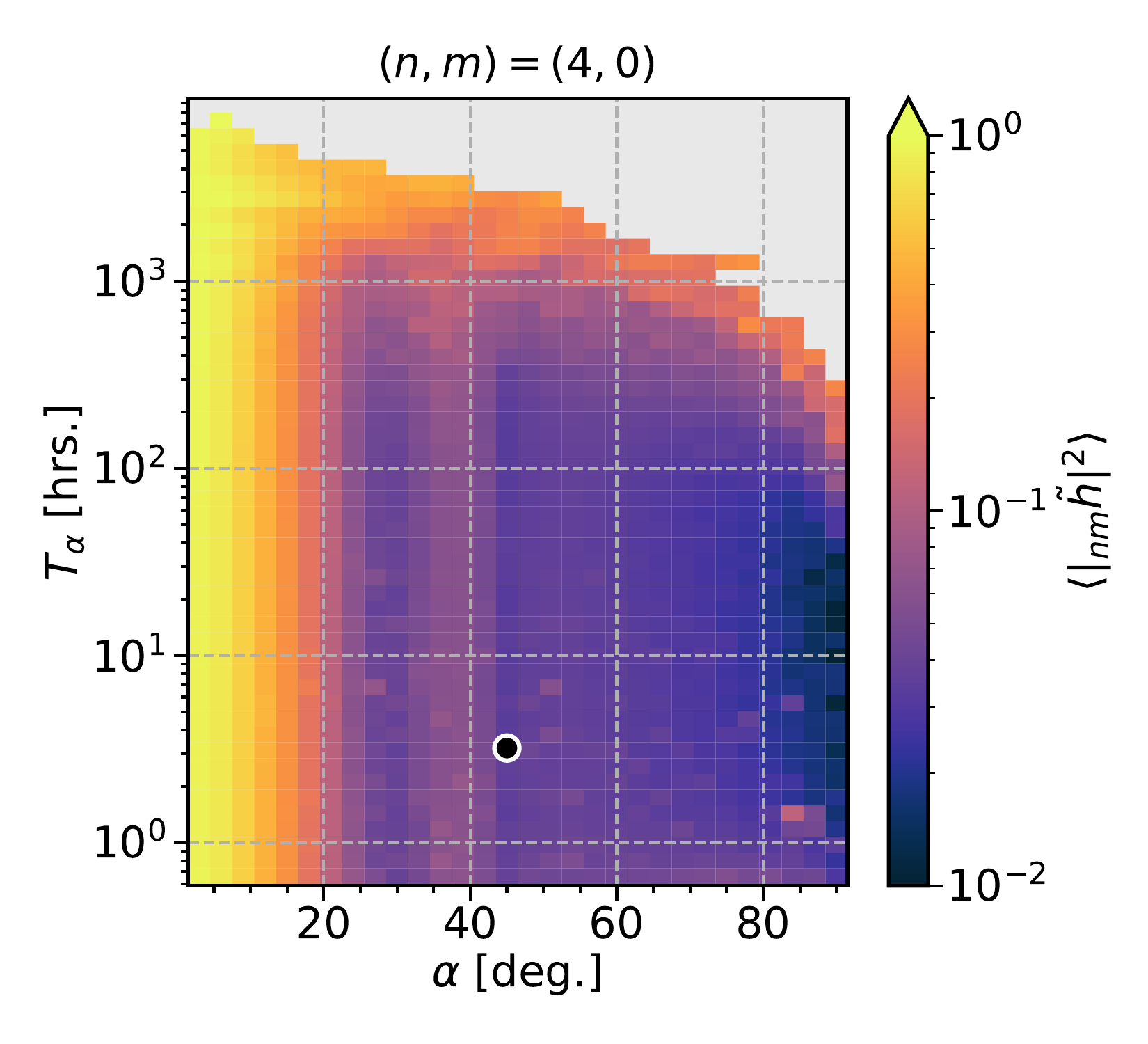
  }
  \includegraphics[width=0.32\columnwidth]{
    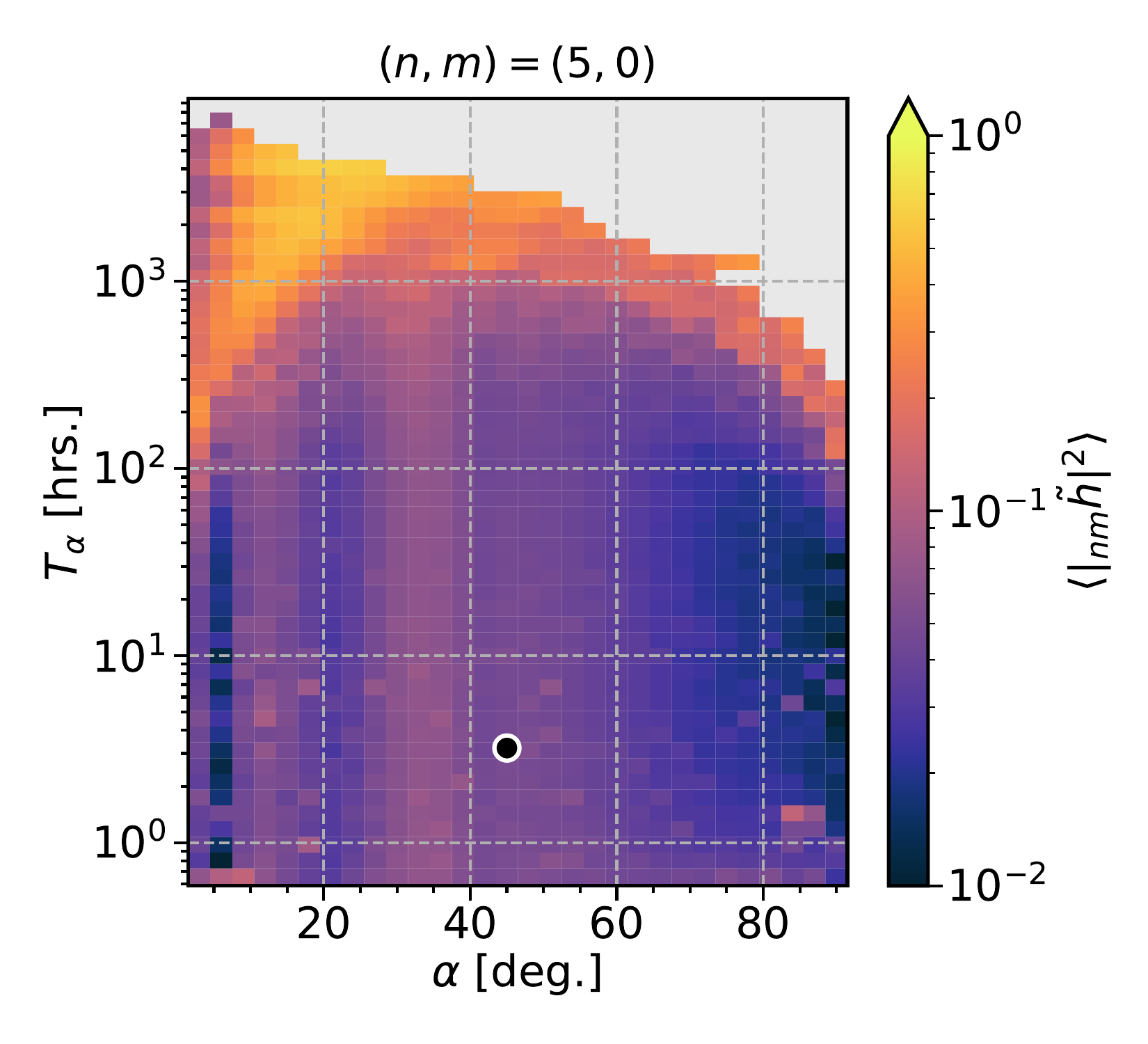
  }
  \includegraphics[width=0.32\columnwidth]{
    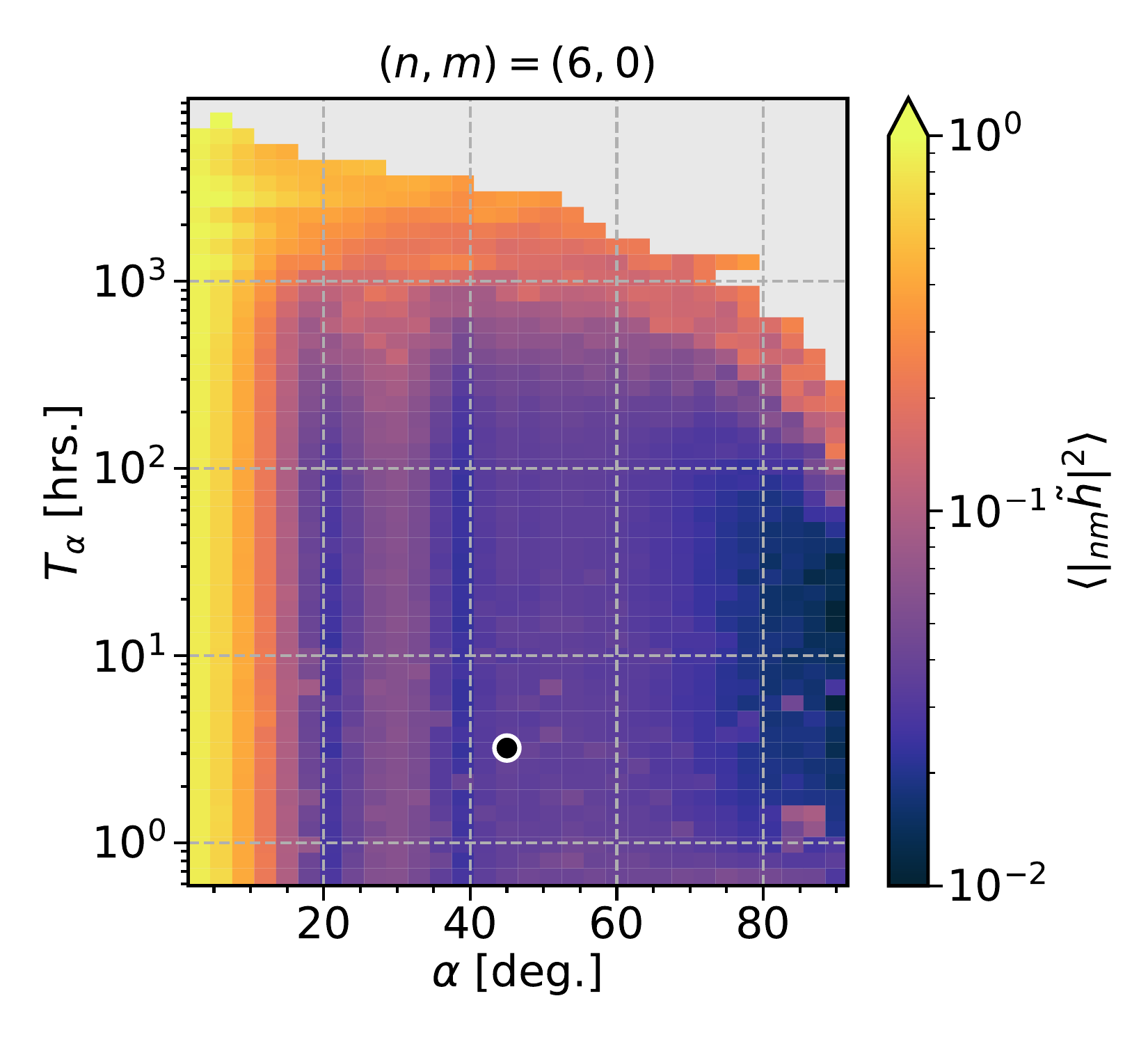
  }
  \\
  \includegraphics[width=0.32\columnwidth]{
    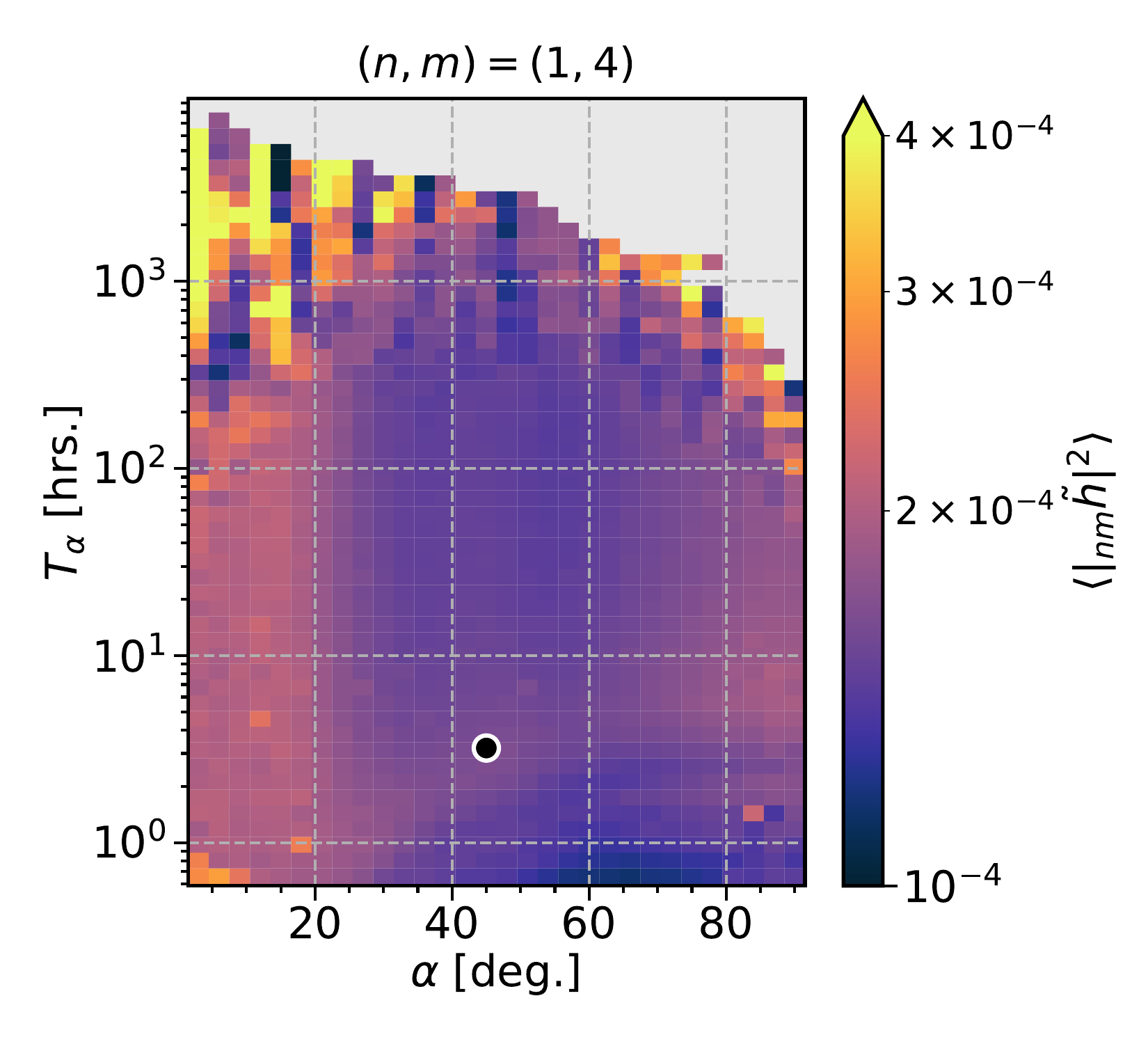
  }
  \includegraphics[width=0.32\columnwidth]{
    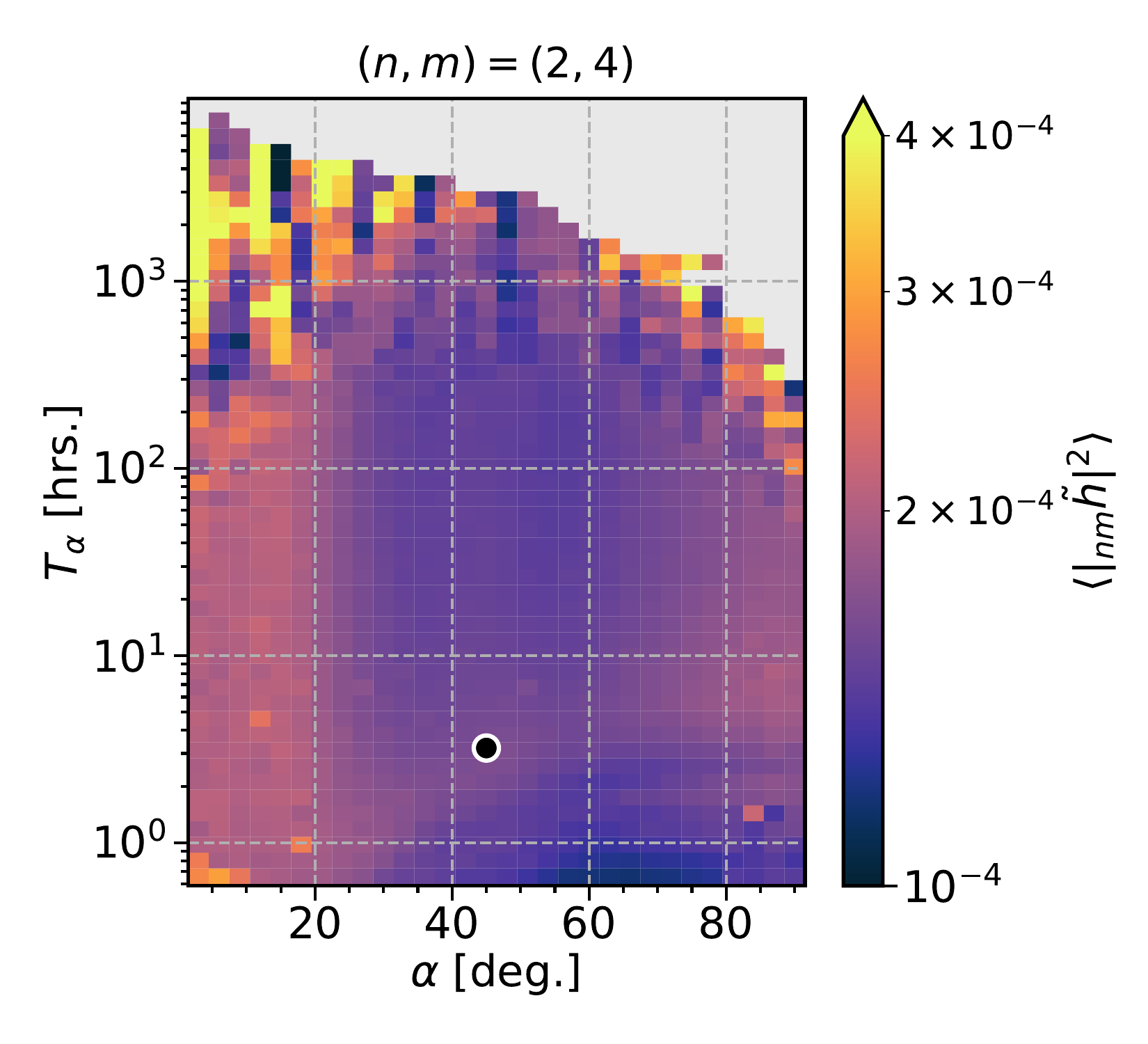
  }
  \includegraphics[width=0.32\columnwidth]{
    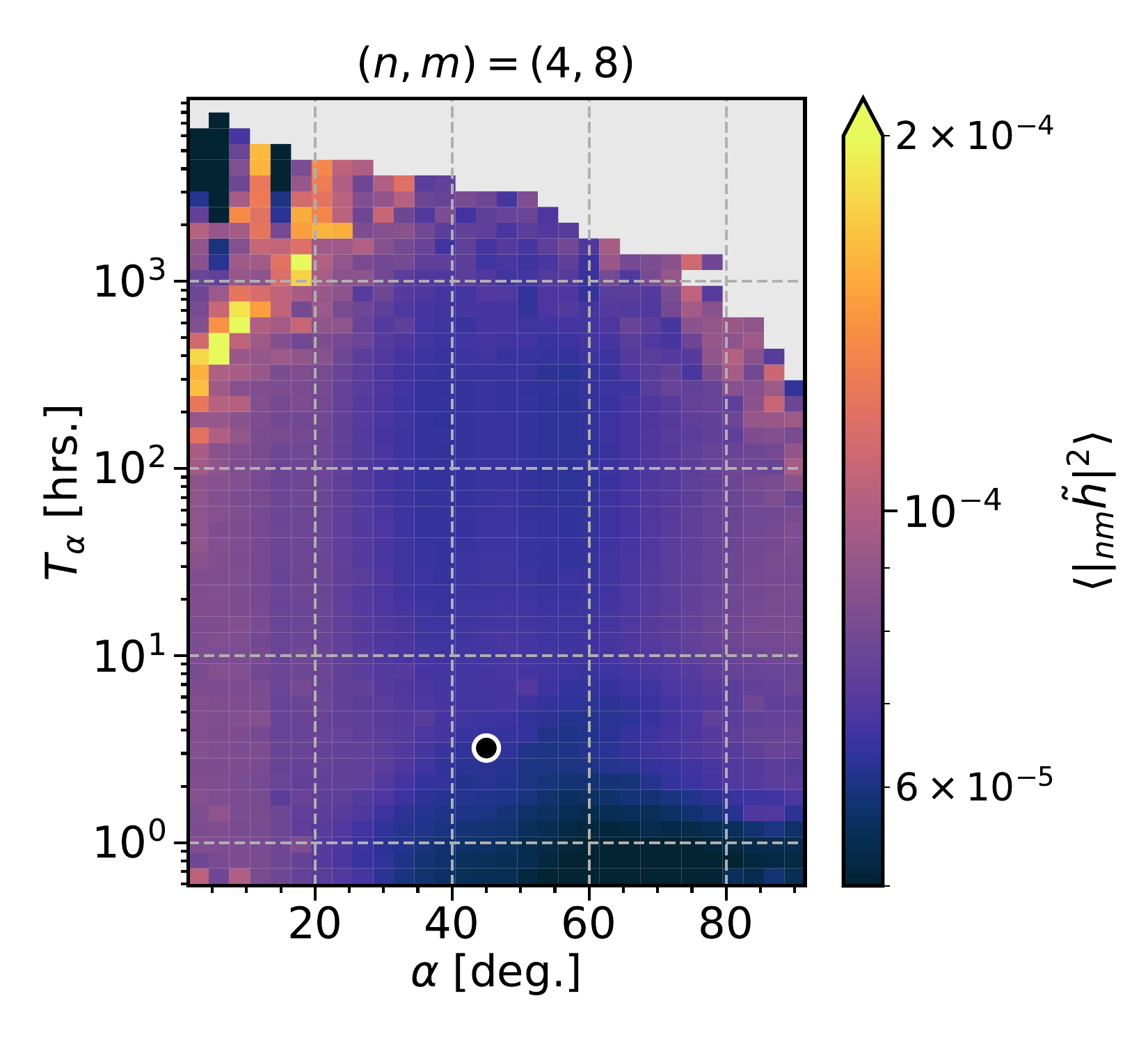
  }
  \caption[Distribution of sky-averaged cross-link factors for different \spin-$(
  n,m)$
  configurations.]{Distribution of sky-averaged cross-link factors for different
  \spin-$( n,m)$ configurations. The top two rows show \spin-$(n,0)$ factors without
  HWP contribution, exhibiting minimal $T_{\alpha}$ dependence but decreasing
  values at larger $\alpha$. The bottom row displays \spin-$(n,m)|_{m=4,8}$ factors
  with HWP contribution, showing uniform distribution across the parameter space
  due to the \cref{eq:T_spin} constraints. The HWP's independent rotation ensures
  consistent cross-link factors regardless of \spin-$n$ when \spin-$m$ is non-zero.}
  \label{fig:cross-links}
\end{figure}

\Cref{fig:cross-links} presents the distribution of cross-link factors for each \spin-$(
n,m)$ values in the $\{\alpha, T_{\alpha}\}$ space. The distributions are
organized in three parts: the top two rows show \spin-$(n,m)|_{m=0}$ factors without
HWP contribution, while the bottom row displays \spin-$(n,m)|_{m=4,8}$ factors
with HWP contribution.

The \spin-$(n,0)$ distributions align closely with findings from ref.~\cite{OptimalScan},
with minor variations attributable to different parameter space sampling. These factors
exhibit smaller values for scanning strategies with large $\alpha$ and small $\beta$.
This reduction occurs because the spin creates smaller rings in this region (visible
in \cref{fig:hitmaps}), resulting in more uniform crossing angles per sky pixel.

In contrast, the \spin-$(n,m)|_{m=4,8}$ factors show remarkably uniform
distribution across the $\{\alpha, T_{\alpha}\}$ space. This uniformity stems from
constraint \cref{eq:T_spin}, where the HWP completes multiple revolutions during
sky pixel transit, generating uniform crossing angles between $0$ and $2\pi$ in a
single observation. This flat behavior extends to all \spin-$(n,m)|_{m=4,8}$
factors where $1\leq n \leq 6$, as demonstrated by the \spin-$(1,4)$ and \spin-$(
2,4)$ examples in \cref{fig:cross-links}.

\subsection{Propagation of cross-link factor to bias}
\label{sec:Propagation}

We use the map-based method employing the \spin formalism from
\cref{chap:formalism} to estimate the measurement bias on $r$, i.e., $\Delta r$,
induced by the pointing offset and HWP non-ideality. This demonstrates how the cross-link
factor value directly relates to $\Delta r$. Detailed estimation methods for
$\Delta r$ are provided in \cref{apd:delta_r}.

The CMB map for this demonstration is generated using \texttt{CAMB} \cite{CAMB}.\footnote{\url{https://camb.readthedocs.io/en/latest/}}
We adopt a 6-parameter $\Lambda\rm{CDM}$ model based on the best fit from \Planck
2018 results: Hubble constant $H_{0}=67.32$, baryon density $\Omega_{b}=0.0494$,
dark matter density $\Omega_{cdm}=0.265$, optical depth to reionization $\tau=0.0
543$, scalar spectral index $n_{s}=0.966$, and amplitude of scalar perturbations
$A_{s}= 2.10\times10^{-9}$ \cite{Planck2018}. The tensor-to-scalar ratio is set
to $r=0$, assuming no primordial $B$ modes.

For the pointing offset simulation, we set the offset parameter
$(\rho,\chi)=(1', 0' )$ in \cref{eq:pointing_offset_field_with_hwp} and use a
CMB-only map as input. The CMB map is smoothed with a symmetric Gaussian beam of
$\rm{FWHM}= 17.9'$, simulating the smallest instrumental beam of \LB, as discussed
in \cref{sec:case_of_LB}. For the instrumental polarization simulation due to
the HWP, we set $\epsilon_{1}=1.0\times10^{-5}$ and $\phi_{QI}=0$ in \cref{eq:HWP_IP_field},
using the CMB solar dipole map as input, as it is a dominant temperature component
\cite{guillaume_HWPIP}. While CMB and foreground anisotropies can contribute as
leakage signals, CMB anisotropies are negligible compared to the dipole. The
complexity of frequency-dependent foregrounds is typically addressed through
specific models, masking, and component separation methods, which is beyond the scope
of this scanning strategy optimization. Note that the magnitude of systematic
effects, i.e., $\rho,\chi$, and $\epsilon_{1}$, are not treated as an issue here;
only the relative penalty of each scanning strategy is identified. Calibration and
mitigation techniques, as discussed in
\cite{planck_pointing_cal,guillaume_HWPIP}, are necessary to address these systematic
effects. The map-based simulations using the \spin formalism in
\cref{chap:formalism} are approximately $10^{4}$ times faster than TOD-based
simulations.

\begin{figure}
  \centering
  \includegraphics[width=0.49\columnwidth]{
    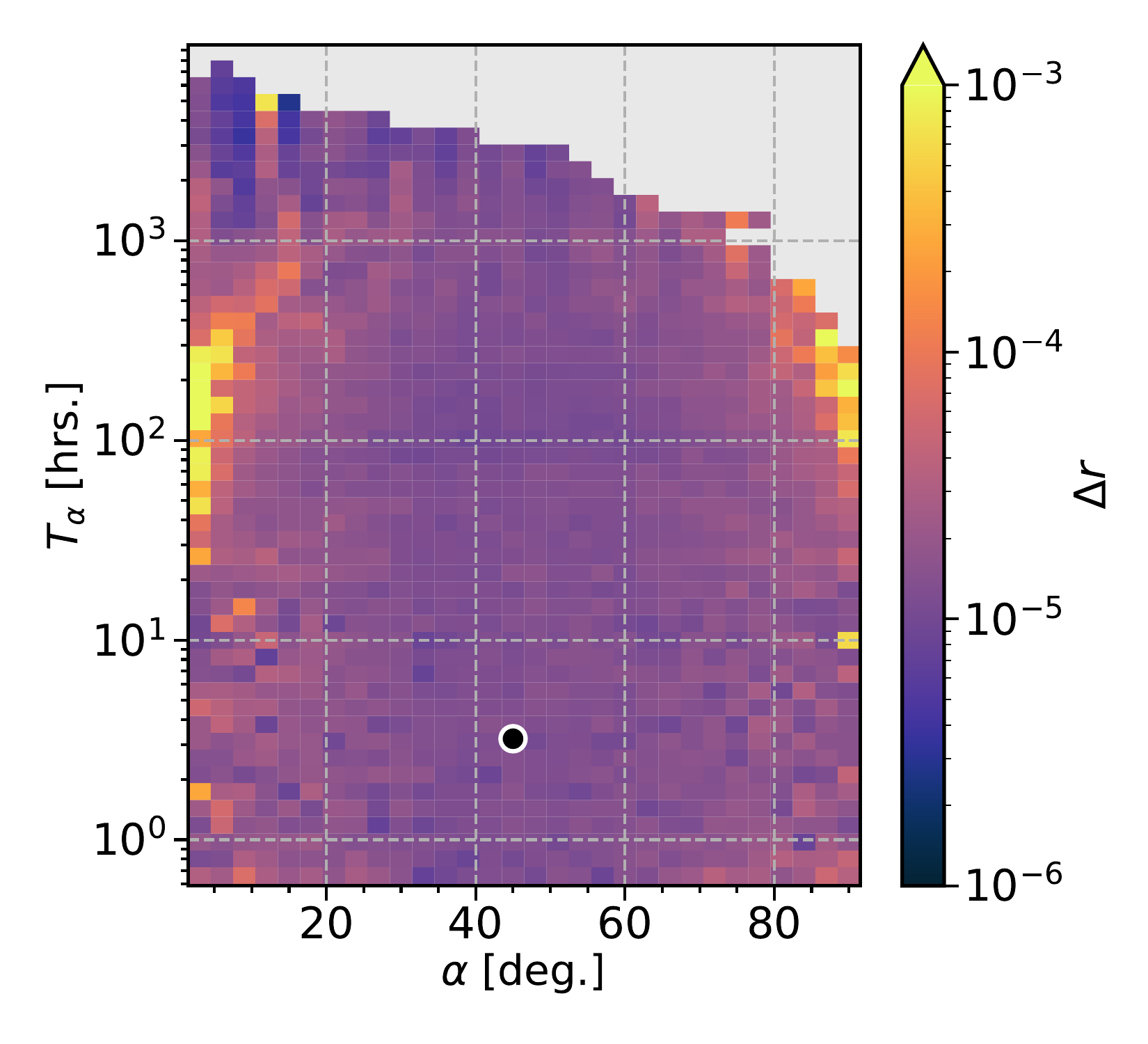
  }
  \includegraphics[width=0.49\columnwidth]{
    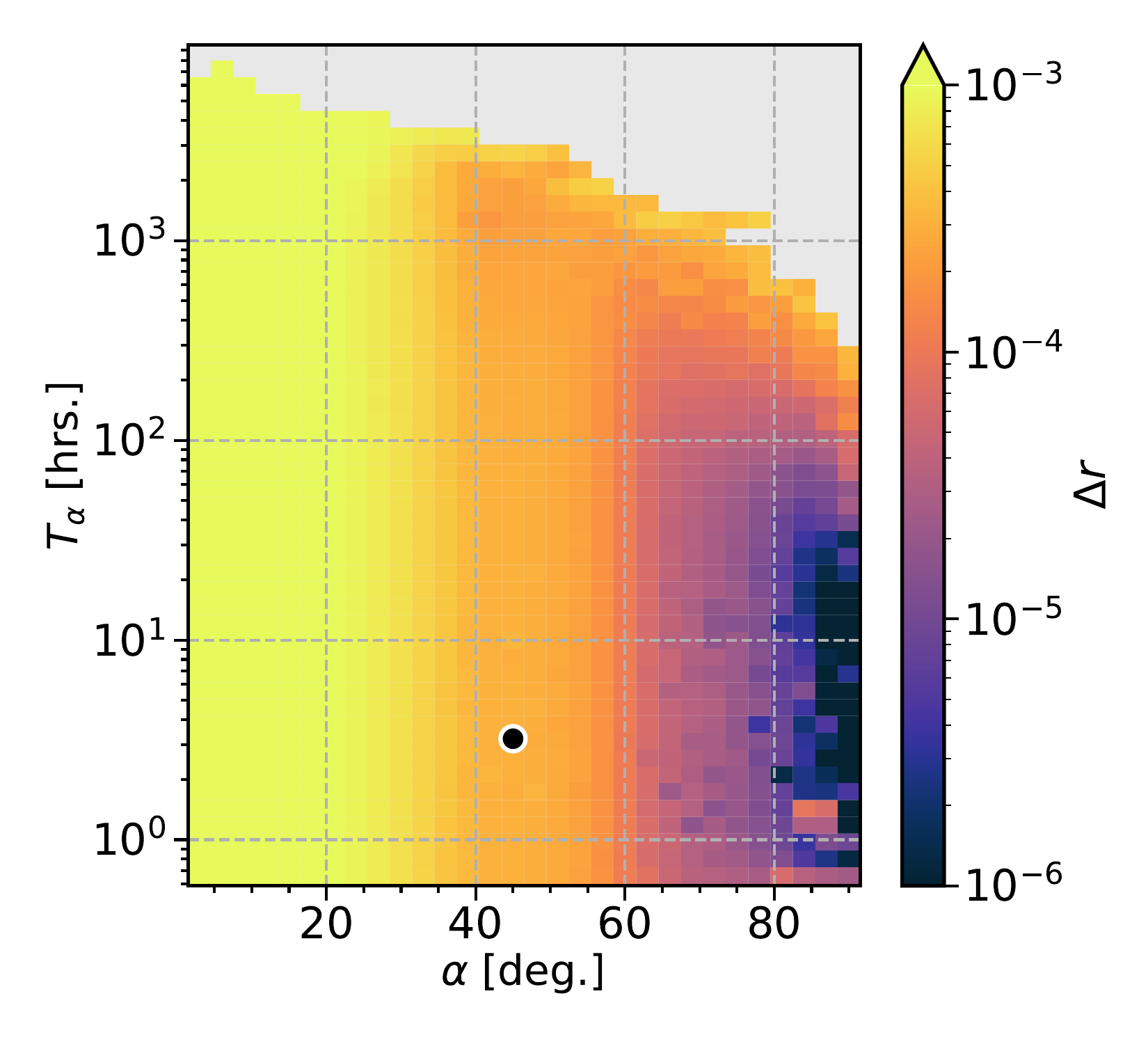
  }
  \caption[Distribution of $\Delta r$ due to pointing offset and instrumental
  polarization with the HWP.]{(left) Distribution of $\Delta r$ due to the
  pointing offset with parameters $(\rho,\chi)=(1',0')$. The input map is the CMB,
  smoothed by a Gaussian beam with $\rm{FWHM}=17.9'$. (right) Distribution of $\Delta
  r$ due to instrumental polarization with the HWP, using
  $\epsilon_{1}=1.0\times10^{-5}$ and $\phi_{QI}=0$. The input map is the CMB solar
  dipole.}
  \label{fig:delta_r}
\end{figure}

\Cref{fig:delta_r} (left) shows the distribution of $\Delta r$ due to the
pointing offset. The overall flat distribution of $\Delta r$, inherited from the
\spin-$(n ,4)$ cross-link factors, indicates that the HWP effectively suppresses
temperature leakage due to the pointing offset. This confirms that the full-sky
average of cross-link factors considering the HWP, as expressed in \cref{eq:crosslink},
is a suitable indicator for the penalty of the scanning strategy for systematic
effects coupled with HWP modulation.

\Cref{fig:delta_r} (right) shows the distribution of $\Delta r$ due to
instrumental polarization with the non-ideal HWP. The distribution structure is similar
to the \spin-$(2 ,0)$ cross-link factor in \cref{fig:cross-links}, indicating
that this systematic effect is not suppressed by the HWP rotation, and only the \spin-$(
2,0 )$ cross-link factor acts as a suppression factor. The justification for the
cross-link factor dependency of HWP non-ideality is found in \cref{apd:HWP_sys}.
These results suggest that systematic effects coupled with the \spin-$(n,4)$ cross-link
factor are suppressed independently of the scanning strategy, and reducing \spin-$(
n,0)$
is crucial even with HWPs.

    \section{Optimization}
\label{sec:optimization}

Based on our previous analysis of how different metrics behave in the $\{\alpha,T
_{\alpha}\}$ parameter space, we confirmed that increasing $T_{\beta}$ from
$\tbl$ does not significantly affect the optimized configuration of cross-link
factors (see \cref{apd:T_beta_scaled}). Our optimization approach will proceed in
two steps. First, we will optimize the geometric parameters, as most metrics
depend primarily on $\alpha$ and remain largely independent of $T_{\alpha}$ for
values up to 100\,hours. Second, with the geometric parameters fixed, we will optimize
the kinematic parameters. This two-step approach will allow us to propose an
effective scanning strategy that balances all considered metrics. In this section,
we begin by examining and comparing different geometric parameter configurations
that have been suggested for next-generation CMB space missions.

\subsection{Optimization of the geometric parameters}
\label{sec:Opt_geometric}

Let us now focus on optimizing the geometric parameters $\alpha$ and $\beta$.
Given the constraints in \cref{eq:const_geometric}, $\beta$ is determined by $\alpha$,
making our optimization effectively single-dimensional. Our key objectives are to
maximize planet visibility time while minimizing both $\sigmahits$ and the cross-link
factors.

The maximum planet visibility time occurs at $\alpha=\beta=47.5^{\circ}$ (the
\BC), yielding 4.4\,hours of integration time. As shown in \cref{fig:planet_visibility}
, planet visibility is symmetrically distributed around this point. However, this
configuration exhibits increased $\sigmahits$, reducing hit-map uniformity. The \SC,
while providing only 3.2\,hours of planet integration time (0.7 times that of \BC),
achieves 1.1 times better hit-map uniformity.

An alternative configuration with $\alpha=50^{\circ}$ that we refer to \FC matches
the \SC's planet integration time and $\sigmahits$, while offering marginally
better cross-link factors (< 5\% improvement). However, comparing the \SC and \FC
respect to the \LB as a model, reveals crucial engineering implications:
\begin{itemize}
  \item The \FC increases solar heat input by 8\% ($\sin50^{\circ}/\cos45^{\circ}
    =1.08$) while reducing solar panel efficiency by 8\%.

  \item Accommodating this increased heat load would require either:
    \begin{itemize}
      \item Optimizing thermal protection mechanism e.g. V-grooves (shown in left
        of \cref{fig:standard_config_and_T_beta}) at the cost of telescope
        volume, or

      \item Maintaining the V-groove design but reducing baffle size, resulting
        in increased sidelobes and systematic effects.
    \end{itemize}
\end{itemize}
While configurations with $60^{\circ}\lesssim \alpha \lesssim70^{\circ}$ show improved
systematic error suppression (see \cref{fig:cross-links,fig:delta_r}),
they further exacerbate heat management issues and reduce solar panel efficiency.
Additionally, the smaller associated $\beta$ values would decrease the CMB solar
dipole signal amplitude, compromising gain calibration.

Given these considerations, and building on the systematic studies in refs.~\cite{PTEP2023,Odagiri_SPIE},
we conclude that the \SC $(\alpha,\beta)=(45^{\circ},50^{\circ})$ represents an
effective choice for the \LiteBIRD mission.

\subsection{Optimization of the kinetic parameters}
\label{sec:Opt_kinetic}

Having established that the \SC geometric parameters $(\alpha,\beta)=(45^{\circ},
50^{\circ})$ provide an effective choice between scientific objectives and spacecraft
design constraints for \LiteBIRD, we now focus on optimizing the kinematic
parameters $T_{\alpha}$ and $T_{\beta}$ while keeping these geometric parameters
fixed.

\subsubsection{Global survey of the kinetic parameter space}

Previous simulations used $T_{\beta}=T_{\beta}^{\rm lower}(\alpha,T_{\alpha})=16.
9$\,min, rather than the \SC's proposed 20\,min. This lower limit on spacecraft spin
period corresponds to an upper limit on spin rate, $\nu_{\beta}^{\rm upper}$, constrained
by \cref{eq:req_for_HWP}. With the HWP handling $1/f$ noise suppression, we can operate
below $\nu_{\beta}^{\rm upper}$ for easier attitude control of the spacecraft.
However, slower spin rates reduce number of spins per precession, degrading
crossing angle uniformity and increasing \spin-$(n,0)$ cross-link factors. We examine
this trade-off between spin rate and cross-link factors.

\begin{figure}
  \centering
  \includegraphics[width=0.32\columnwidth]{
    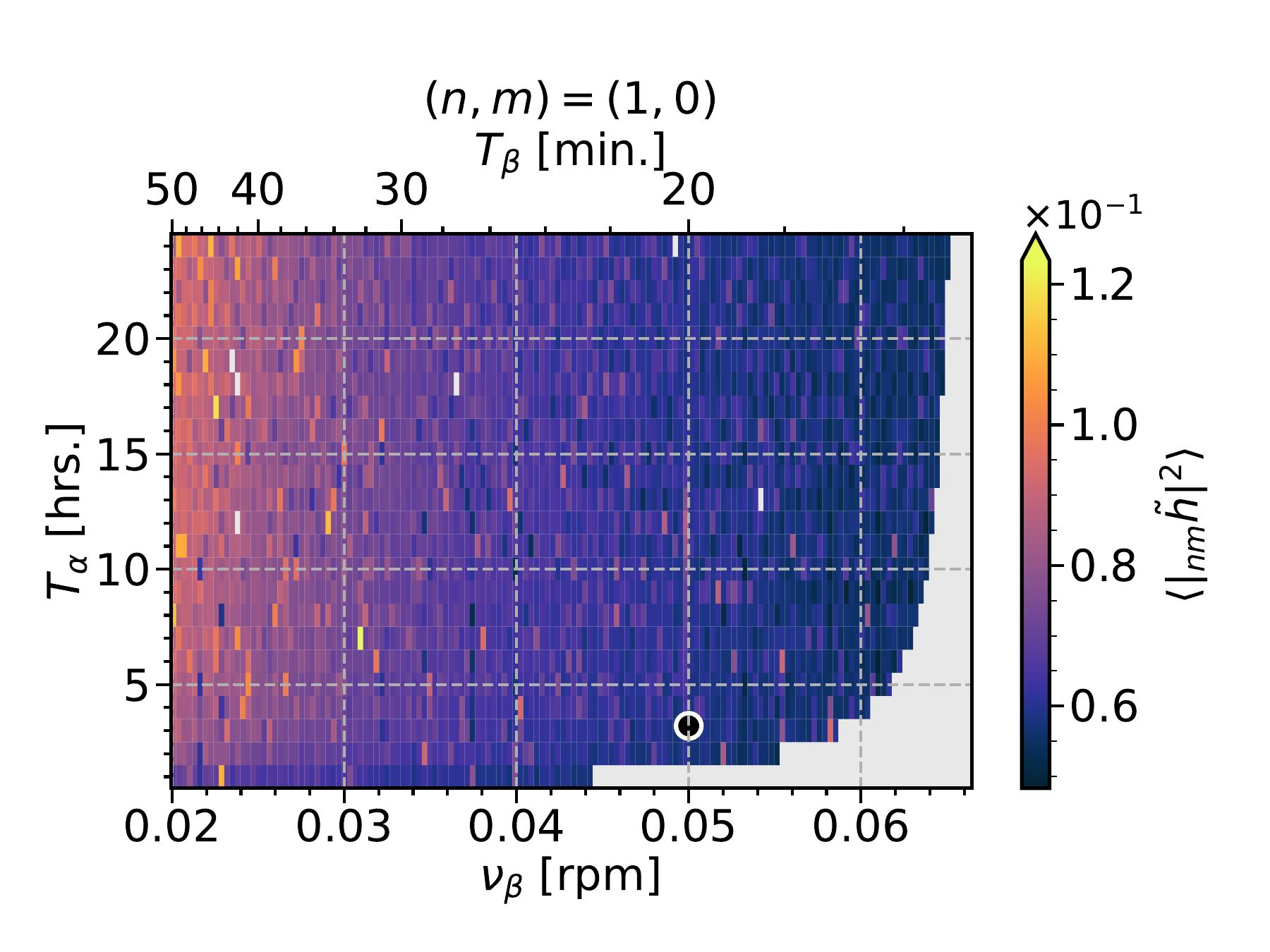
  }
  \includegraphics[width=0.32\columnwidth]{
    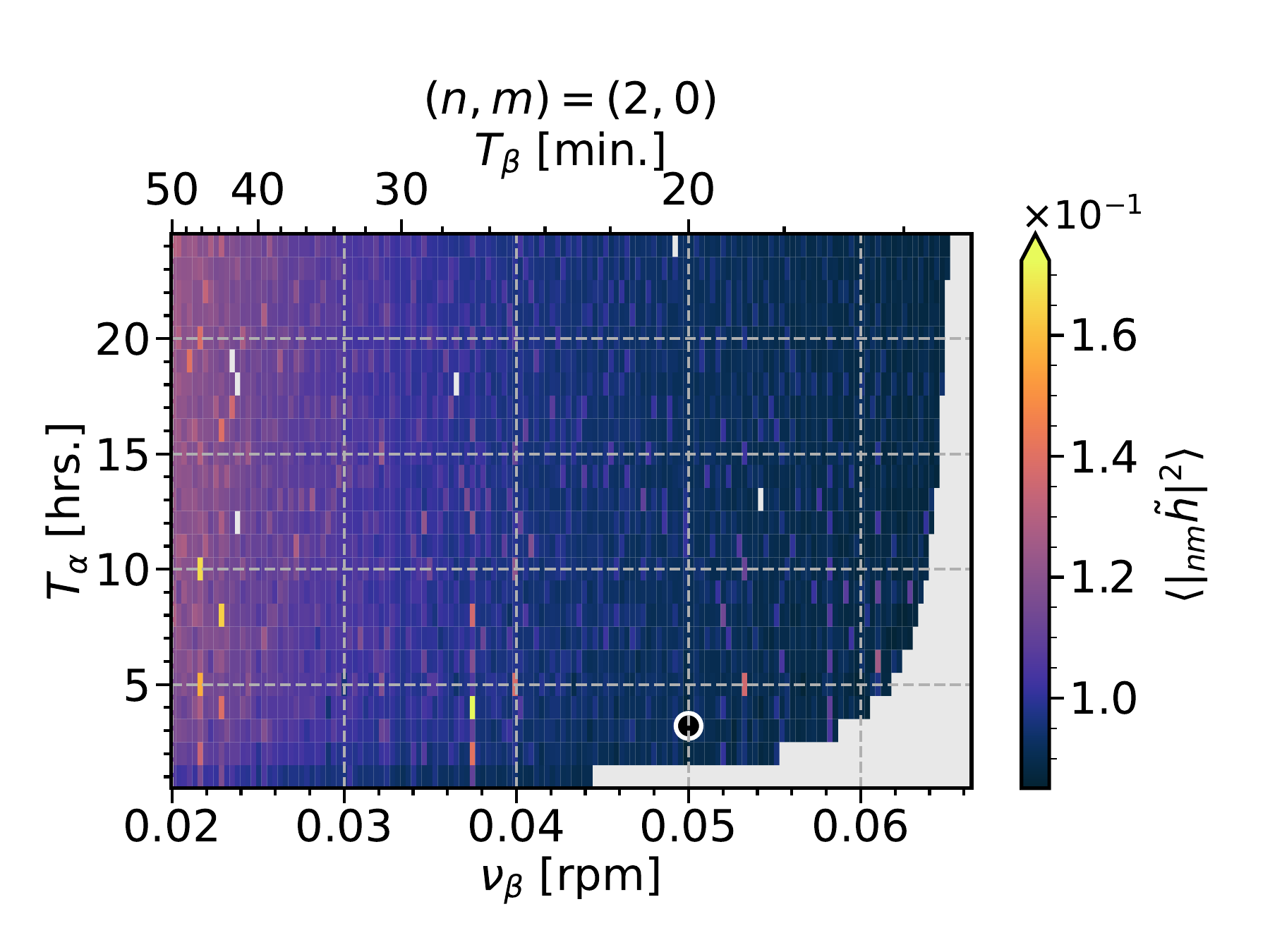
  }
  \includegraphics[width=0.32\columnwidth]{
    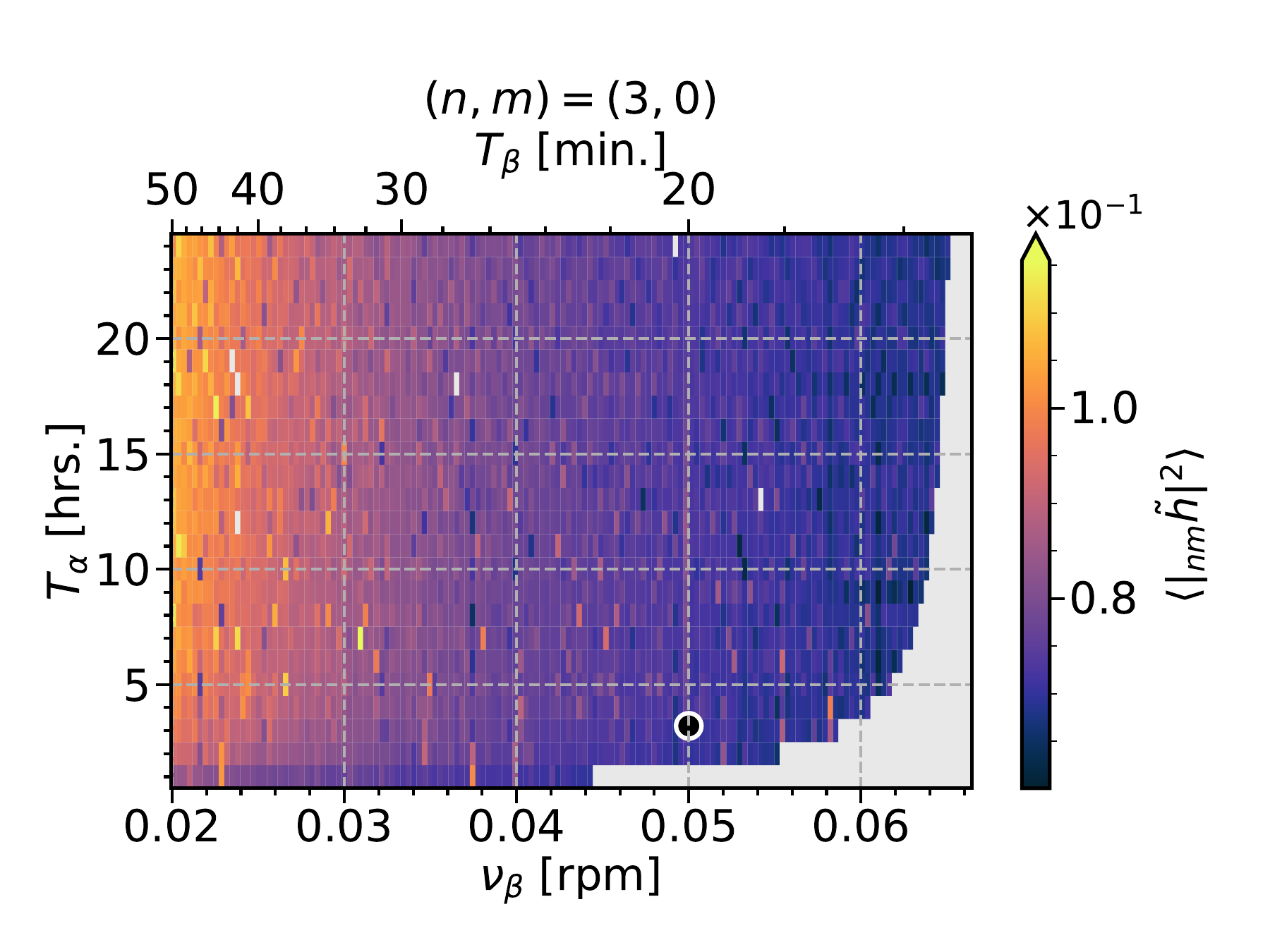
  }
  \\
  \includegraphics[width=0.32\columnwidth]{
    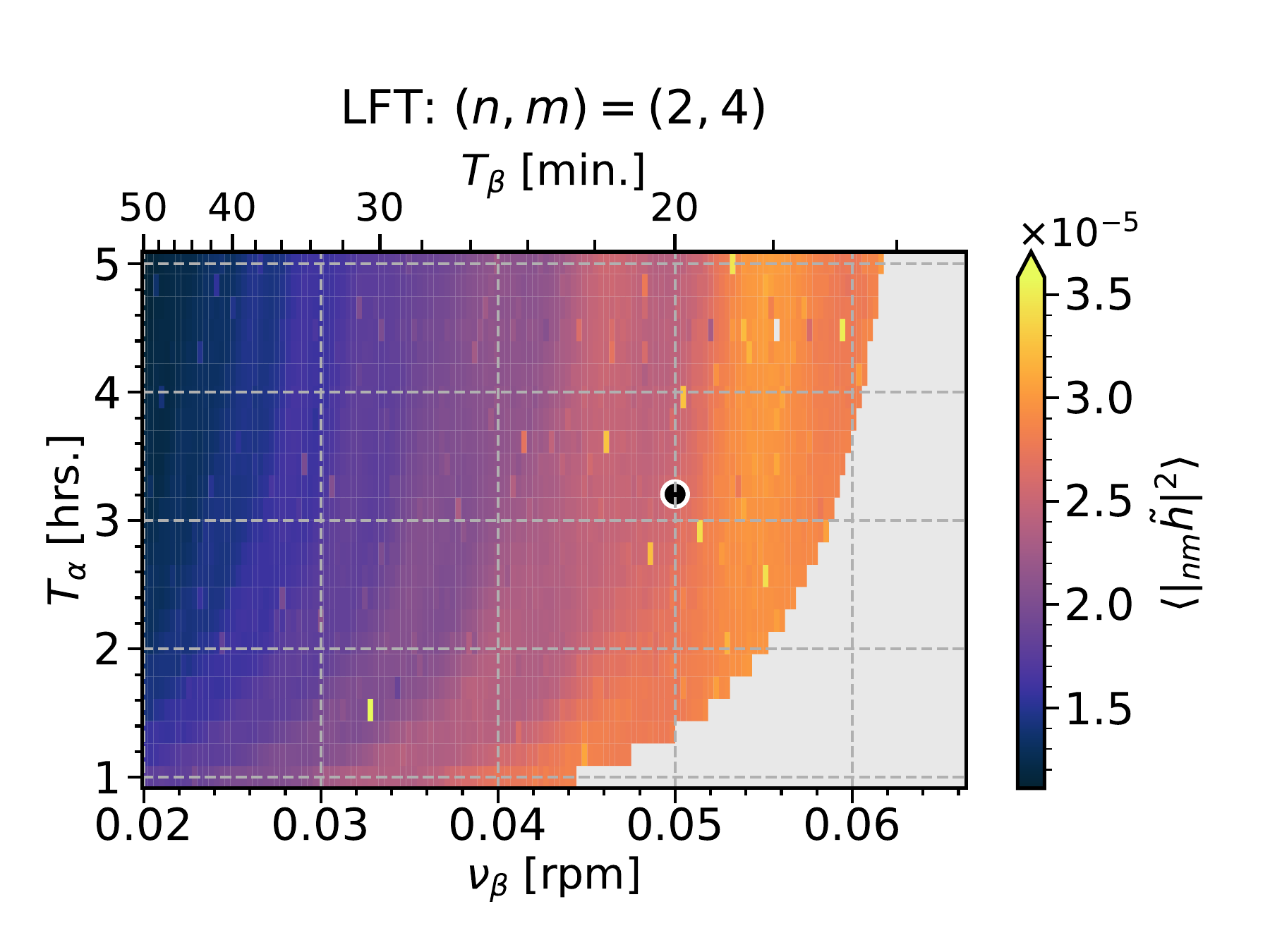
  }
  \includegraphics[width=0.32\columnwidth]{
    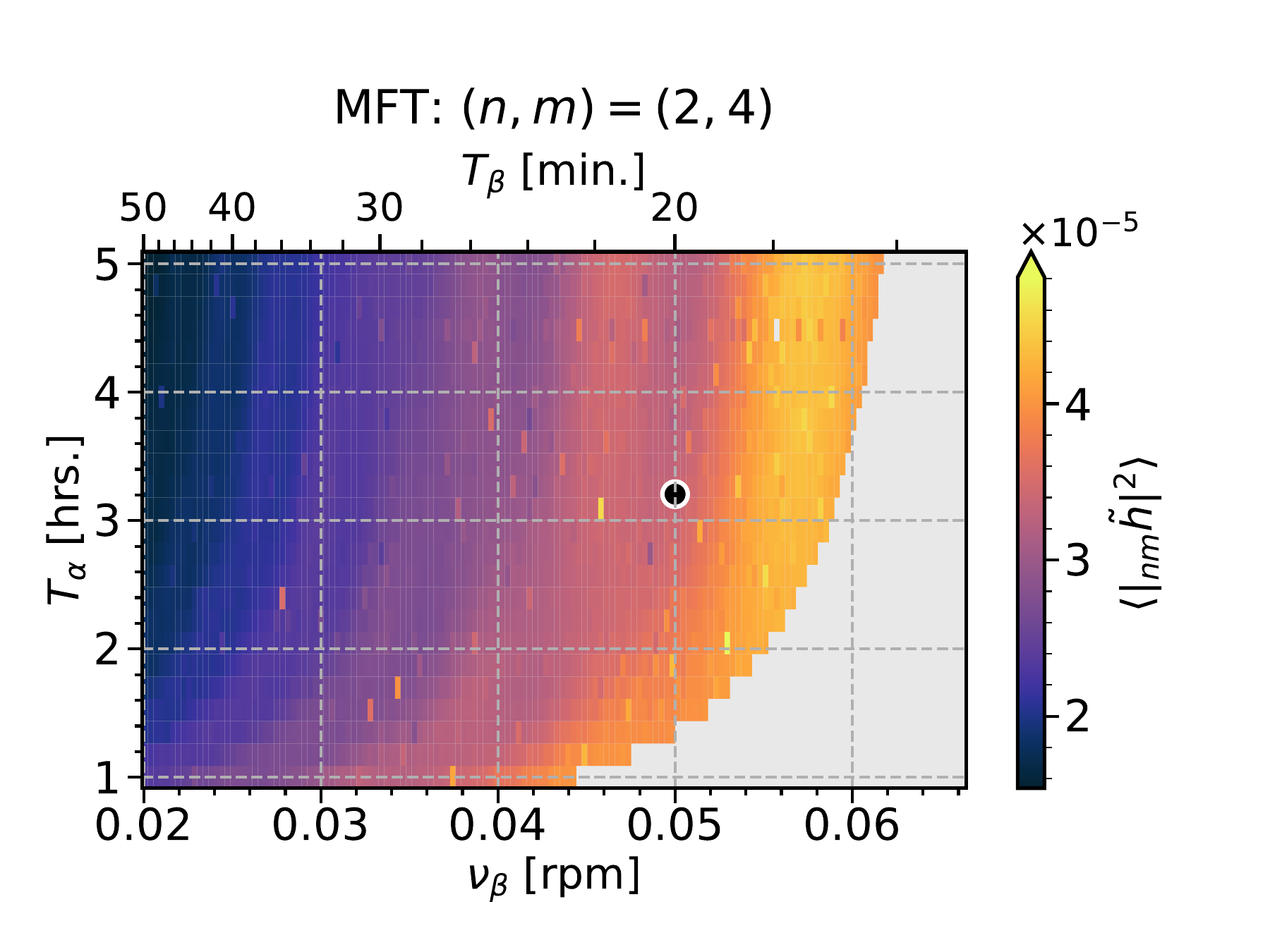
  }
  \includegraphics[width=0.32\columnwidth]{
    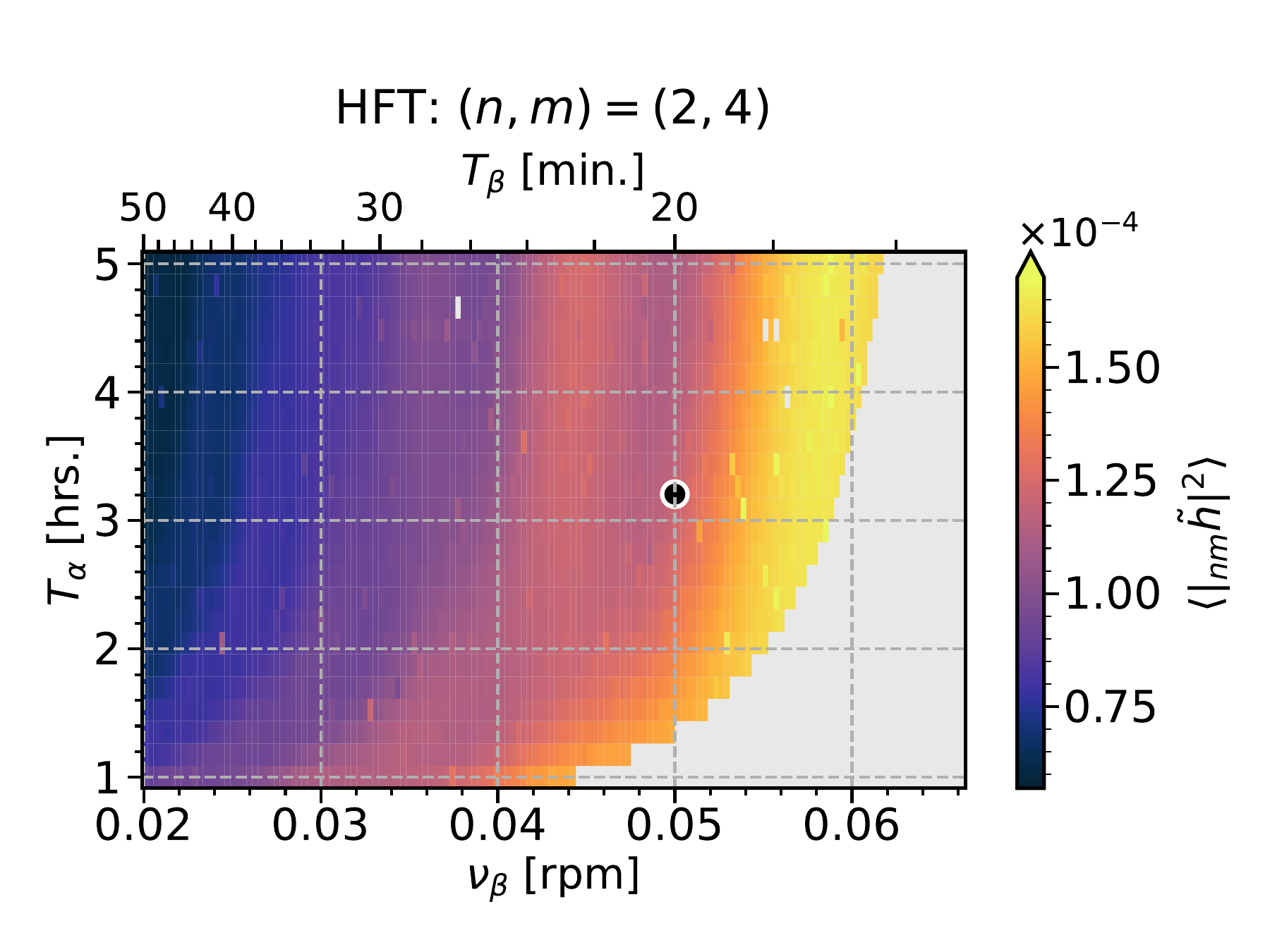
  }
  \caption[Cross-link factors for different \spin-$(n,0)$ configurations and
  \spin-$( 2,4)$ cross-link factors for LFT, MFT and HFT.]{(top panels) Cross-link
  factors for each \spin-$(n,0)$ configuration. (bottom panels) \spin-$(2,4)$ cross-link
  factors shown for LFT (left), MFT (middle) and HFT (right) telescopes. The gray
  shaded region at higher $\nu_{\beta}$ indicates where $\nu_{\beta}<\nu_{\beta}^{\rm{upper}}$
  (\cref{eq:T_spin}) is violated. Missing values within the allowed region are
  due to spin-precession resonance, discussed further in \cref{sec:fine-tuned}.
  While these configurations still achieve nearly full sky coverage, certain pixels
  may be unobserved due to redundant trajectories caused by spin-precession synchronization.}
  \label{fig:rot_period_opt}
\end{figure}

\Cref{fig:rot_period_opt} shows the dependence of cross-link factors on $T_{\alpha}$
and $T_{\beta}$. The top panels display \spin-$(n,0)$ factors, while bottom panels
show \spin-$(2,4)$ factors for each telescope. Bottom panels focus on a narrower,
higher-resolution parameter space with $T_{\alpha}\leq5$\,hours. For LFT/MFT/HFT,
we use HWP rates of 46/39/61\,rpm and $N_{\rm side}$ of 128/128/256, matching
their respective beam FWHMs of 23.7/28.0/17.9 arcmin.\footnote{A \texttt{HEALPix}
pixel spans $\sqrt{4\pi/N_{\rm pix}}$, giving 27.5/13.7\,arcmin for
$N_{\rm side}$=128/256.} Other parameters match the \SC.

The \spin-$(n,0)$ cross-link factors generally decrease with increasing
$\nu_{\beta}$ and remain nearly constant for $T_{\alpha}>2$\,hours.\footnote{This
trend extends to higher \spin-$(n,0)$ factors ($n=4,5,6$) and persists up to
$T_{\alpha}=100$\,hours.} Map outliers show \moire patterns from spin-precession
resonance, which can be eliminated through fine-tuning as discussed in the next section.

Cross-link factors remain stable until $\nu_{\beta}=0.05$\,rpm, showing only a 5\%
increase between 0.04-0.05\,rpm. HWP-related factors increase with $\nu_{\beta}$
but oscillate due to spin-HWP synchronization. We select $\nu_{\beta}=0.05$\,rpm
($T_{\beta}=20$\,min) as the first local minimum below $\nu_{\beta}^{\rm upper}$,
balancing attitude control and systematic error suppression.

Regarding precession period optimization, faster rates benefit CMB solar dipole calibration
by reducing $1/f$ noise impact. At $\nu_{\beta}=0.05$\,rpm, \spin-$(n,4)$ factors
decrease 15\% as $T_{\alpha}$ increases from $T_{\alpha}^{\rm lower}$ to 3.2\,hours,
then stabilize. Therefore, $T_{\alpha}=3.2$\,hours (192\,min) offers an effective
choice between precession speed and cross-linking performance.

\subsubsection{Fine-tuned study of the precession period}
\label{sec:fine-tuned} With our geometric parameters now established, we can
fine-tune the rotation periods to eliminate resonances. We introduce the ratio $\eta$
between precession and spin periods:
\begin{align}
  \eta = \frac{T_{\alpha}}{T_{\beta}}.
\end{align}
Our initial value of $\eta=192~[\rm{min}]/20~[\rm{min}]=9.6$ is rational, causing
the spin to synchronize with precession every 9.6 cycles. This synchronization
creates undesirable linear patterns in the hit-map's azimuthal direction as the
trajectory intersections remain fixed during solar revolution.

Ref.~\cite{hoang2017bandpass} proposed using irrational numbers' decimal parts
to avoid this synchronization. Since irrational numbers cannot be exactly represented
computationally, they must be approximated using rational numbers. Diophantine approximation
theory suggests using truncated continued-fraction expansions of irrational numbers
for this purpose. While ref.~\cite{hoang2017bandpass} recommended the golden ratio
for its optimal convergence properties (also supported by dynamical systems
analysis in ref.~\cite{berry1978}), practical spacecraft operation introduces unavoidable
rotational disturbances. We therefore analyze $\sigmahits$ within
$T_{\alpha}\in [192, 193]$\,min at 0.1\% resolution to identify regions robust against
operational perturbations.

\begin{figure}[h]
  \centering
  \includegraphics[width=1\columnwidth]{
    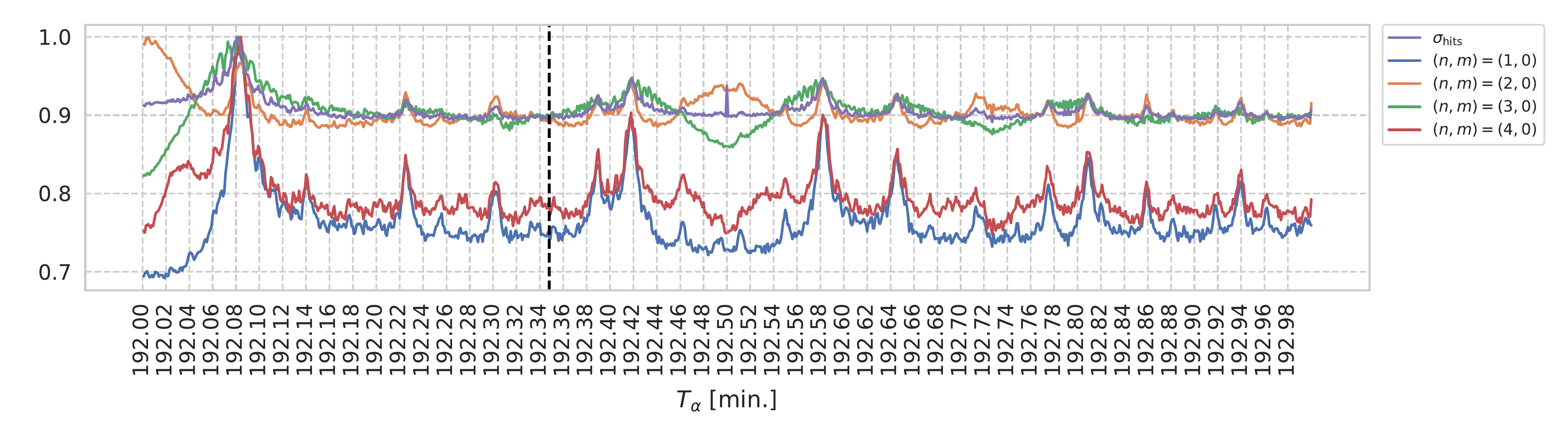
  }
  \caption[Dependence of normalized $\sigma_{\rm hits}$ and \spin-$(n,0)$ cross-link
  factors on $T_{\alpha}$.]{Dependence of normalized $\sigma_{\rm hits}$ and
  \spin-$(n,0)$ cross-link factors on $T_{\alpha}$. The dashed black line indicates
  our chosen value of $T_{\alpha}=192.348$\,min for the \SC. Peaks correspond to
  spin-precession resonances, which produce \moire patterns in the maps as demonstrated
  in \cref{fig:prec_tuning_maps}.}
  \label{fig:prec_tuning}
\end{figure}

\begin{figure}[ht]
  \centering
  \includegraphics[width=0.32\columnwidth]{
    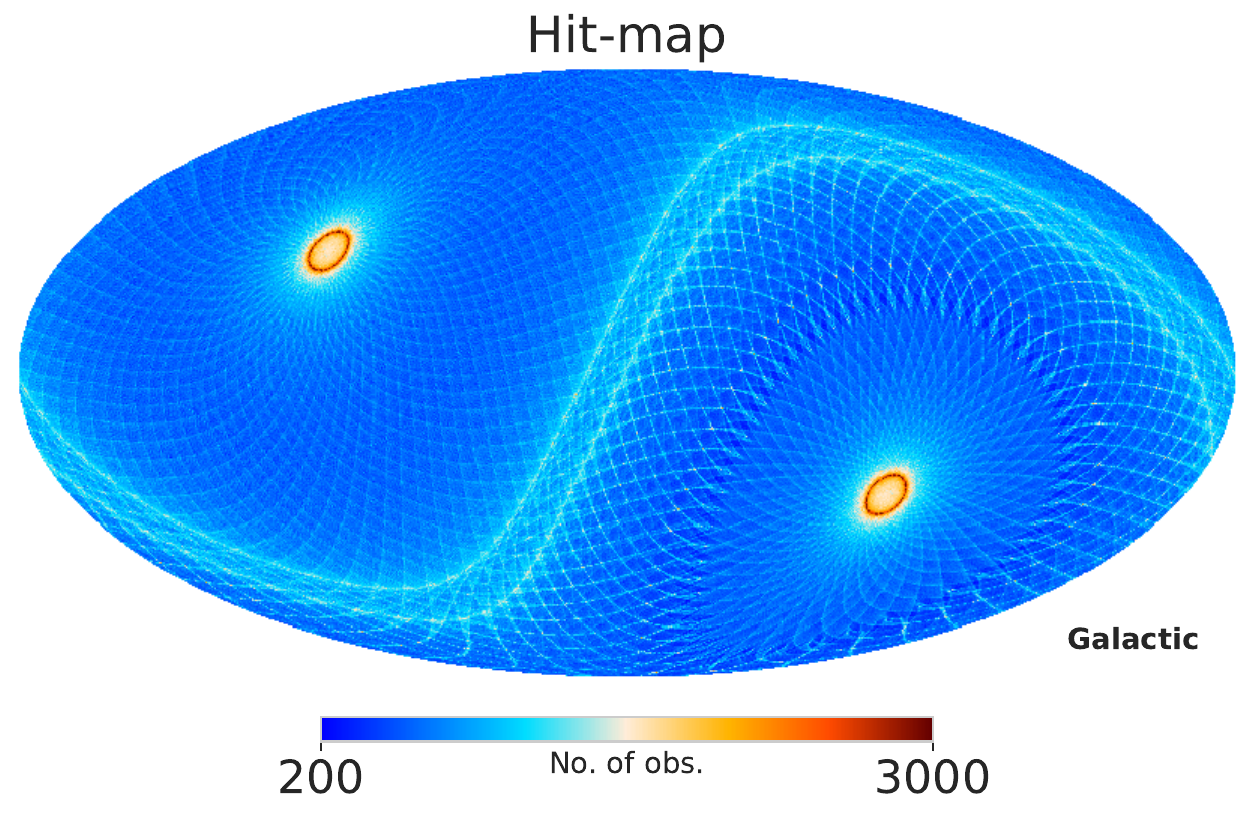
  }
  \includegraphics[width=0.32\columnwidth]{
    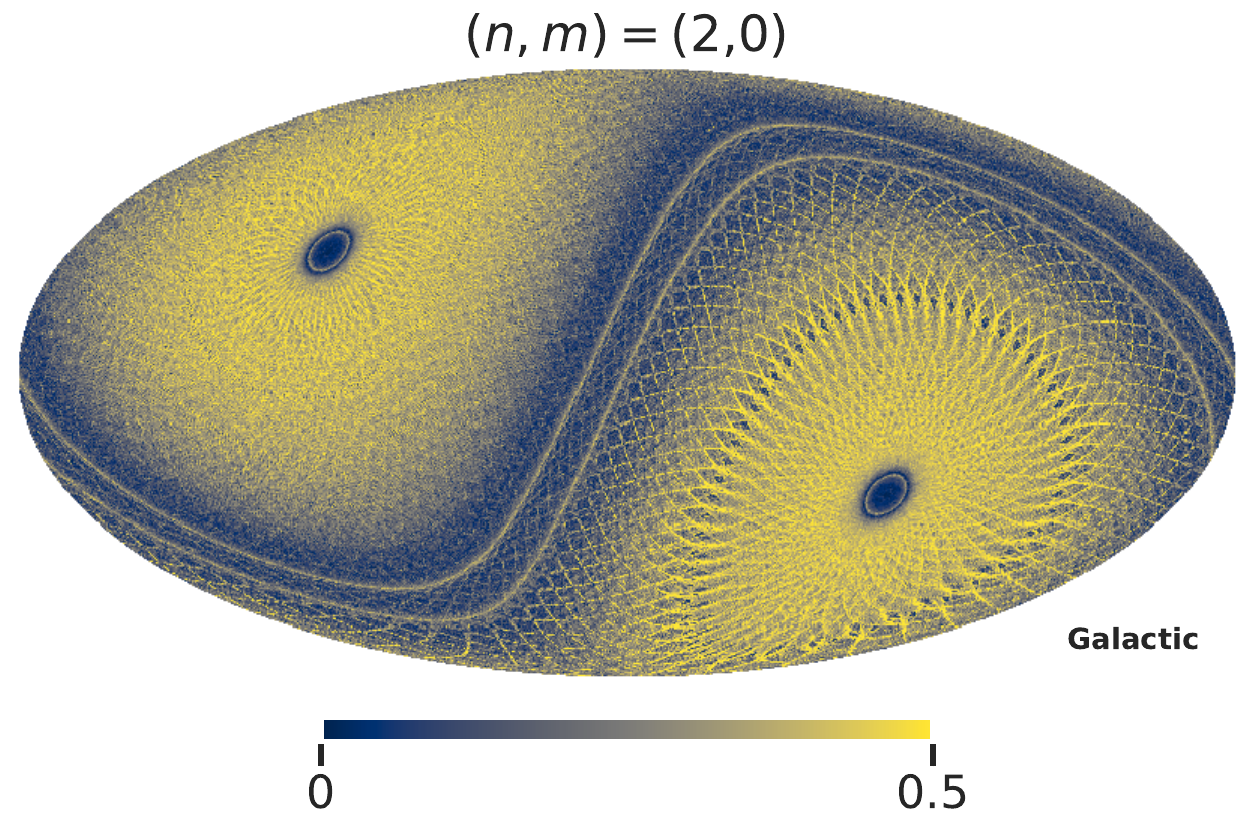
  }
  \includegraphics[width=0.32\columnwidth]{
    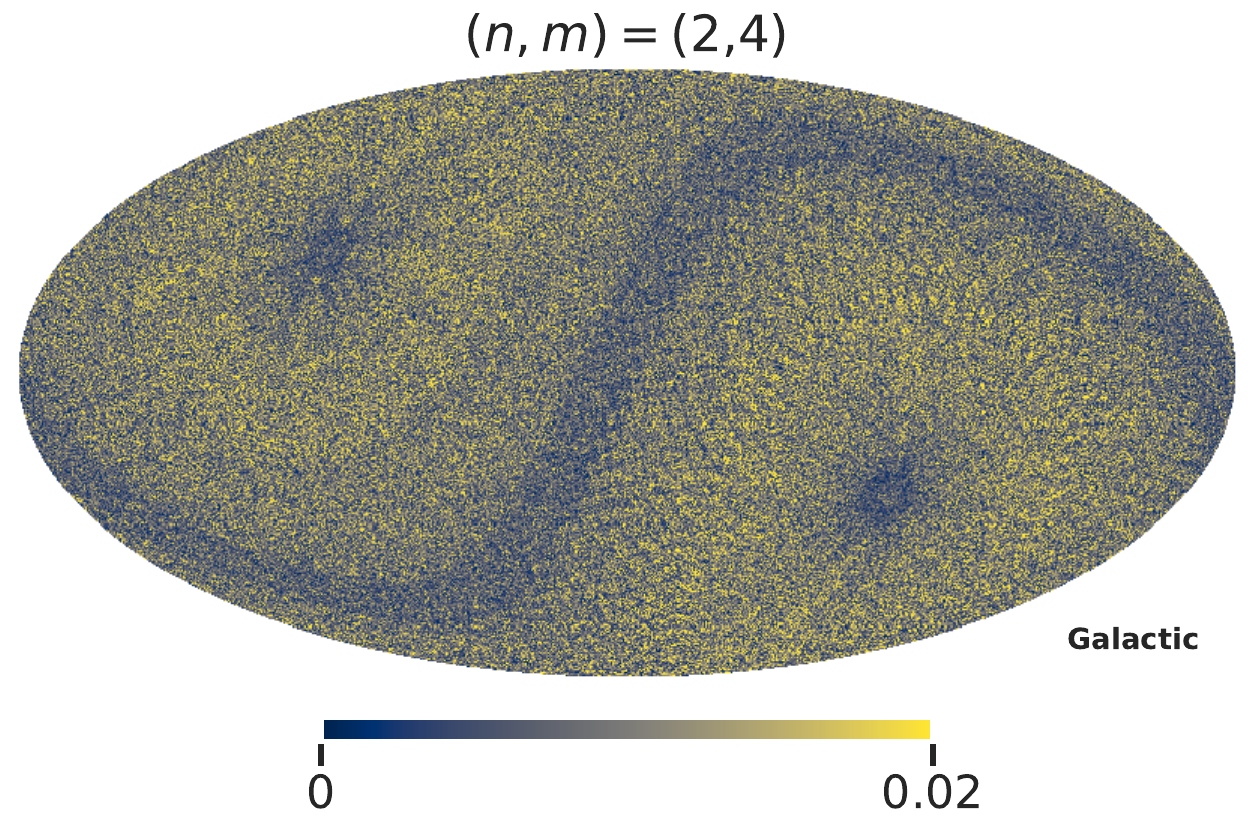
  }
  \\
  \includegraphics[width=0.32\columnwidth]{
    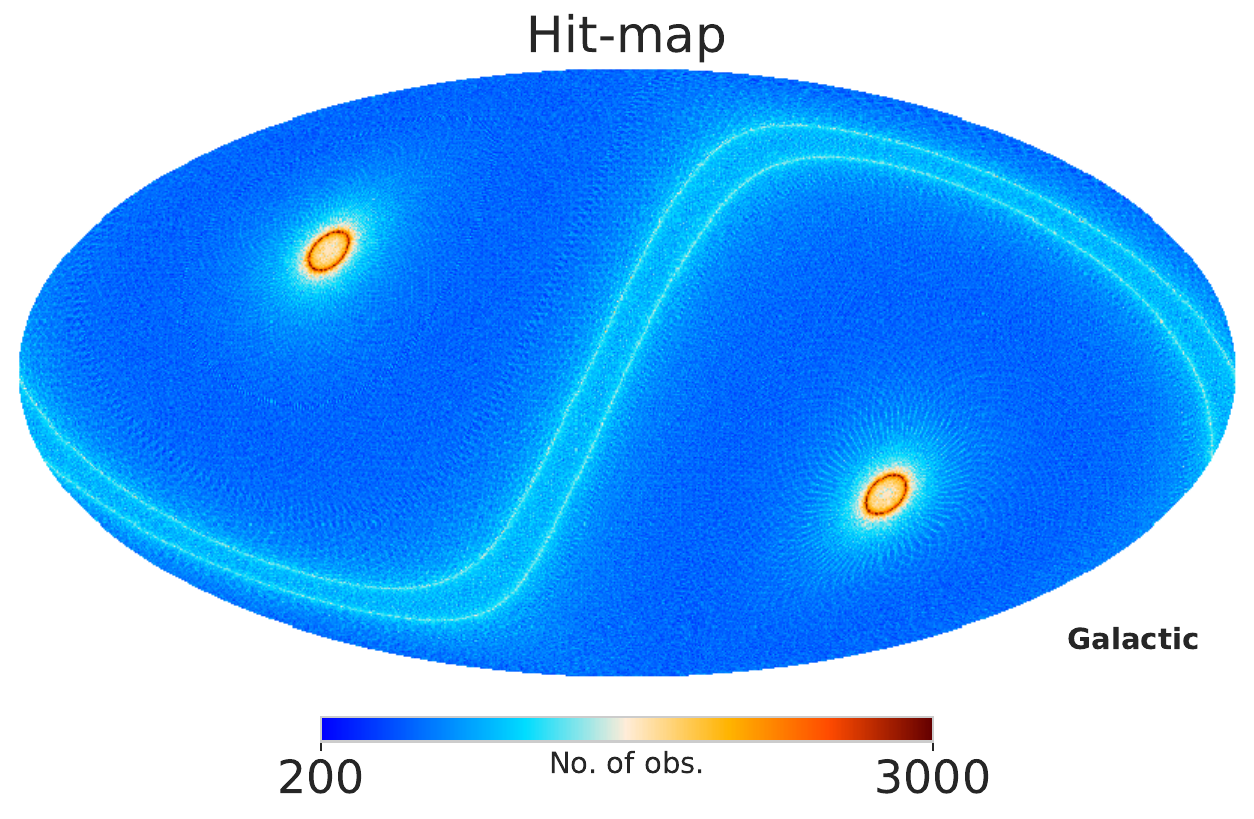
  }
  \includegraphics[width=0.32\columnwidth]{
    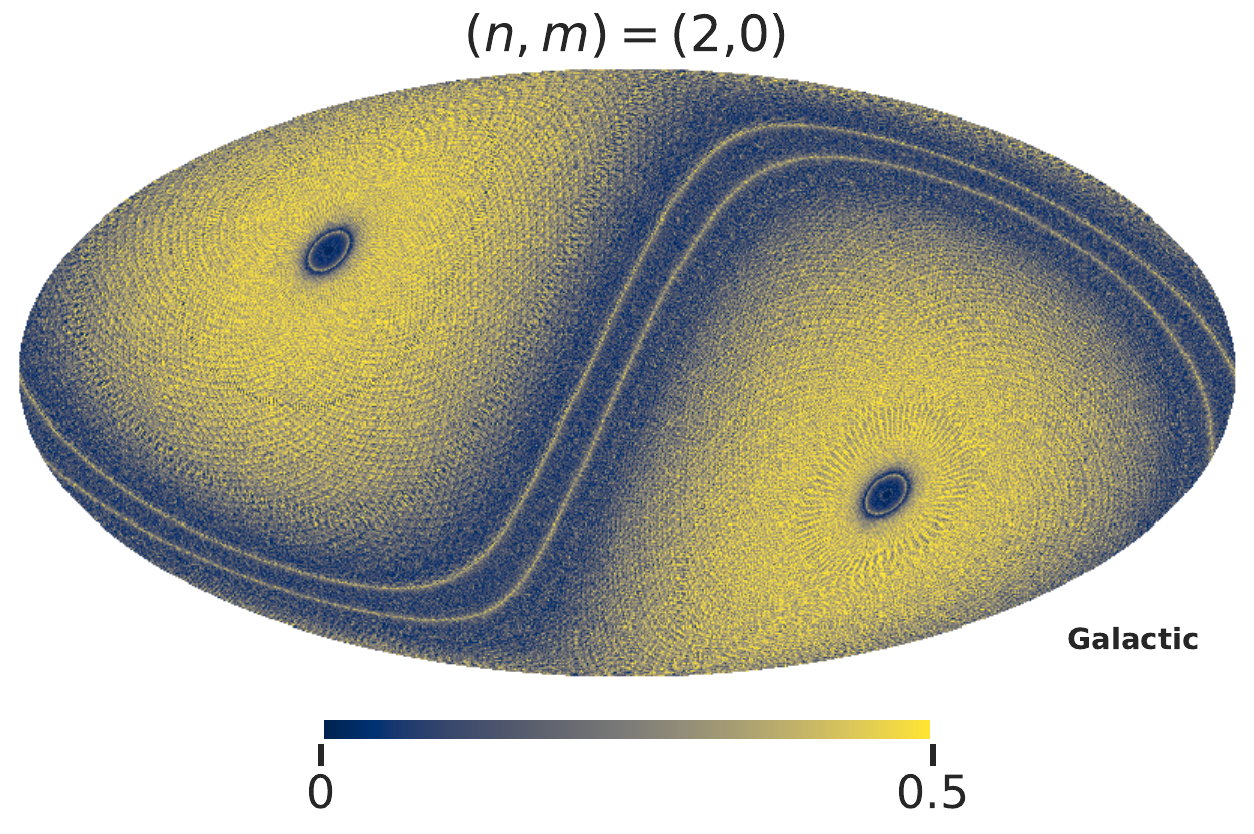
  }
  \includegraphics[width=0.32\columnwidth]{
    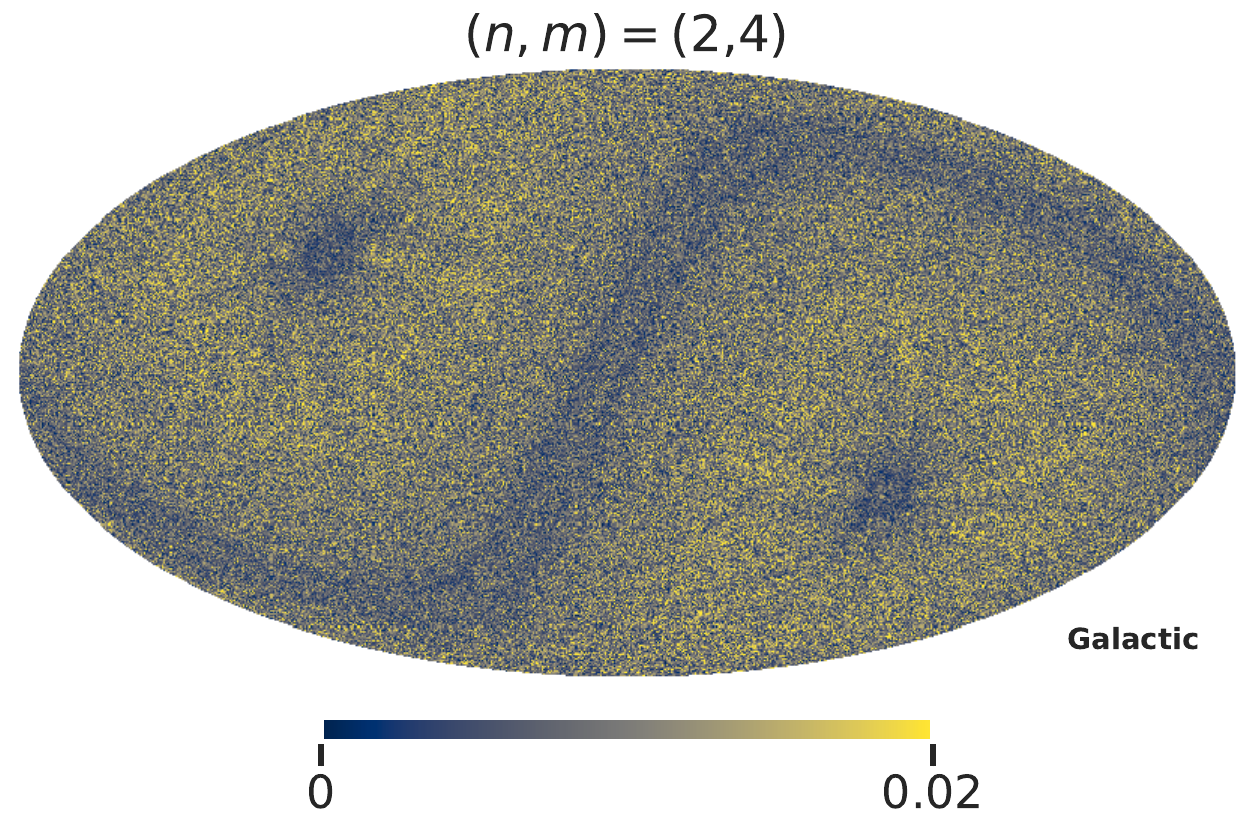
  }
  \caption[Maps showing hit distribution and cross-link factors in Galactic
  coordinates.]{Maps showing hit distribution and cross-link factors in
  Galactic coordinates. Top row ($T_{\alpha}=192.08$\,min): hit-map (left), \spin-$(
  2,0)$ (middle) and \spin-$(2,4)$ (right) cross-link factors. Bottom row: same quantities
  for the \SC's optimized $T_{\alpha}=192.348$\,min. The strong \moire patterns
  visible in the top panels due to spin-precession resonance are eliminated in
  the bottom panels through fine-tuning of the precession period.}
  \label{fig:prec_tuning_maps}
\end{figure}

\Cref{fig:prec_tuning} plots normalized $\sigma_{\rm hits}$ and \spin-$(n,0)$ cross-link
factors versus $T_{\alpha}$. We identify an optimal interval of
$192.3 20 \,\mathrm{min}\leq T_{\alpha}\leq192.370\,\mathrm{min}$ where metrics
remain near minimal with low local variation.\footnote{Analysis of \spin-$(n,m)|_{m=4,8}$
cross-link factors shows minimal resonance effects due to the HWP's relatively
high revolution rate compared to spacecraft motions.} Within this range, we select
$T_{\alpha}=192.348$\,min for its minimal cross-link factors.
\Cref{fig:prec_tuning_maps} demonstrates the improvement: strong \moire patterns
visible at $T_{\alpha}=192.08$\,min (top panels) disappear at our selected value
(bottom panels).

While this analysis assumes perfect rotational stability, practical
implementation will require further study. Future work should model spacecraft inertia
and in-flight stabilization to determine achievable $T_{\alpha}$ precision under
real conditions.

    \section{Implications}
\label{sec:implications}

This section examines crucial considerations beyond previously discussed metrics
when designing scanning strategies. We analyze four key aspects: First,
\cref{sec:beam_reconstruction} explores how scanning strategy impacts beam shape
reconstruction through scanning beam angle analysis. Second, \cref{sec:dipole}
investigates the scanning strategy's effect on CMB solar dipole amplitude, which
is vital for instrumental gain calibration. Third, \cref{sec:skypix} evaluates sky
pixel visit/revisit times across different CMB missions, essential for null-tests.
Finally, \cref{sec:visit} assesses planet observation opportunities for
calibration during mission duration.

\begin{figure}[ht]
  \centering
  \includegraphics[width=1\columnwidth]{
    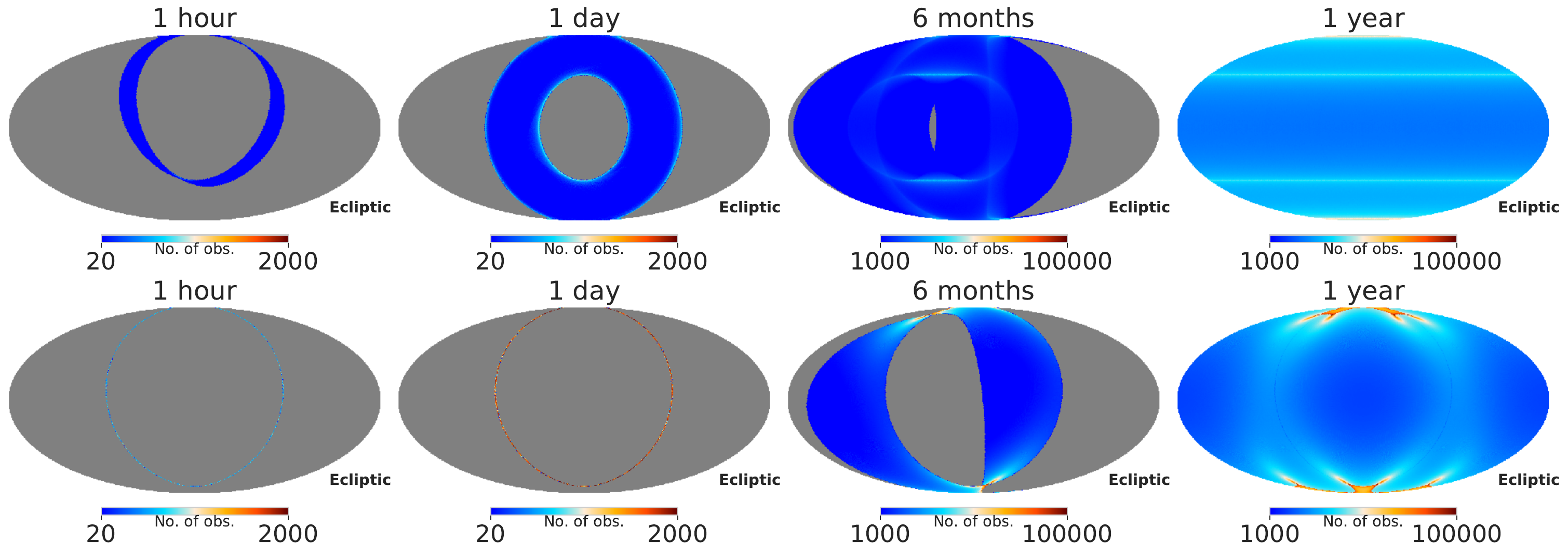
  }
  \caption[Hit-map evolution over time using scanning strategies from \PICO and
  \Planck.]{Hit-map evolution over time using scanning strategies from \PICO (top)
  and \Planck (bottom). From left to right: observations after 1\,hour, 1\,day,
  6\,months, and 1\,year. Simulations use 19\,Hz sampling rate with $\Nside=128$.
  Both missions demonstrate continuous sky mapping with overlapping scan rings that
  shift through slow precession, contrasting with \LiteBIRD's \SC pattern shown
  in \cref{fig:hitmaps}.}
  \label{fig:hitmap_pico_planck}
\end{figure}

For comparison, we examine the \Planck mission and planned \PICO mission.
\Cref{tab:pico_and_planck} details their scanning strategies, while \cref{fig:hitmap_pico_planck}
illustrates the time evolution of their simulated hit-maps.

\begin{table}[t]
  \centering
  \begin{tabular}{lllll}
    \hline
    {}        & $\alpha$      & $\beta$      & $T_{\alpha}$ & $T_{\beta}$ \\
    \hline
    \LiteBIRD & $45^{\circ}$  & $50^{\circ}$ & 3.2058\,hr   & 20\,min     \\
    \hline
    \PICO     & $26^{\circ}$  & $69^{\circ}$ & 10\,hr       & 1\,min      \\
    \hline
    \Planck   & $7.5^{\circ}$ & $85^{\circ}$ & 6\,month     & 1\,min      \\
    \hline
  \end{tabular}
  \caption[Geometric/kinetic parameters of \LiteBIRD, \PICO and \Planck.]{Geometric/kinetic
  parameters of \LiteBIRD, \PICO and \Planck \cite{PICO2019, planck_prelaunch2010}.}
  \label{tab:pico_and_planck}
\end{table}

\subsection{Beam reconstruction systematics}
\label{sec:beam_reconstruction}

Planet observations enable beam shape reconstruction, making this process crucial
for mitigating systematic effects, with scanning strategy playing a vital role
\cite{far_sidelobe}. According to ref.~\cite{Planck_HFI_beam}, systematic effects
in the scan direction (like detector time constant and pointing systematics) can
create degeneracies with sidelobe shapes, hampering reconstruction capabilities.
Despite observing beam shape and detector time constant degeneracy, \Planck
improved Jupiter-based beam reconstruction accuracy using a `deep scan mode'
that reduced spin axis shift when Jupiter was observable \cite{Planck_LFI_beam}.
This demonstrates how appropriate scanning strategy choices can enhance beam measurements.

\begin{figure}[h]
  \centering
  \includegraphics[width=0.45\columnwidth]{
    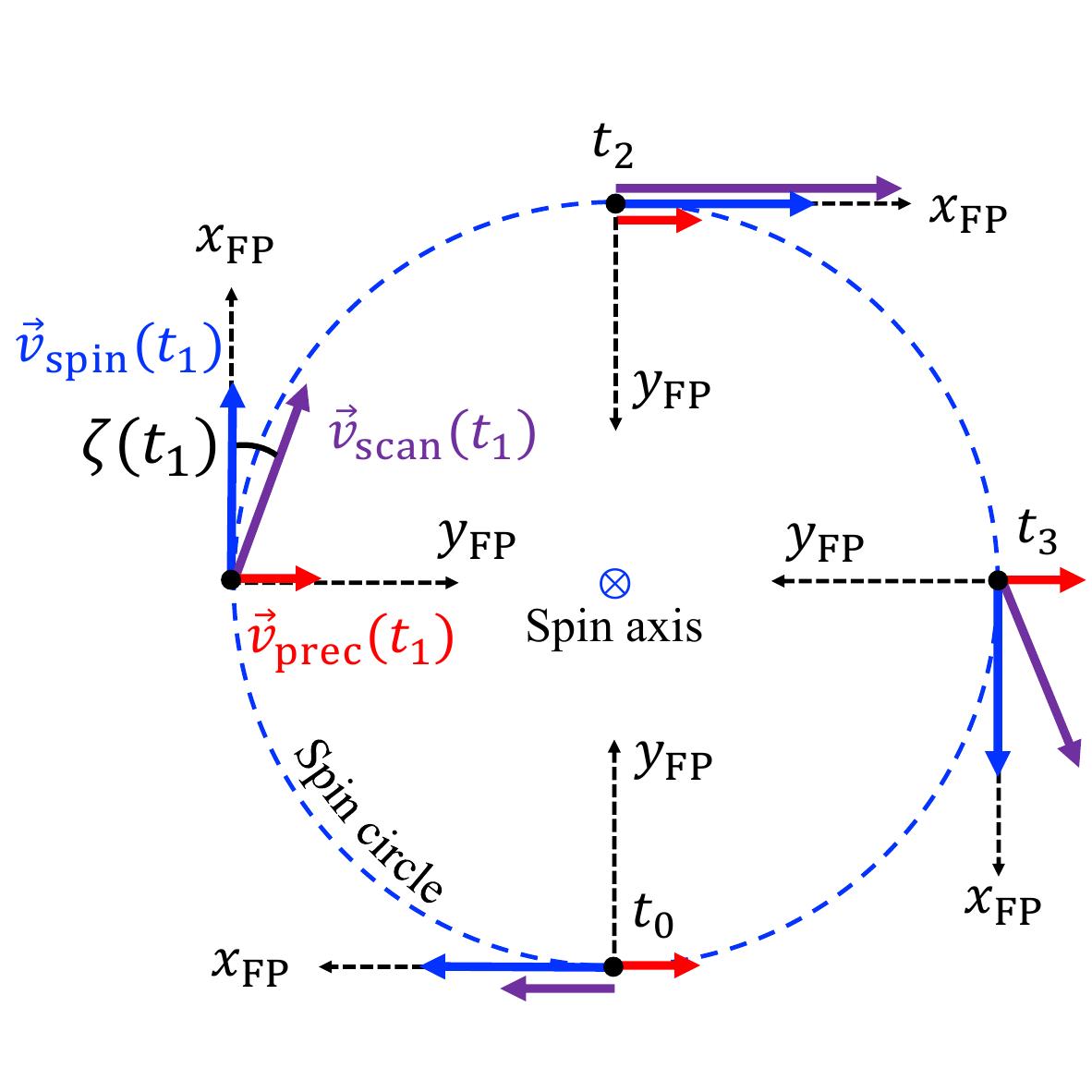
  }
  \includegraphics[width=0.45\columnwidth]{
    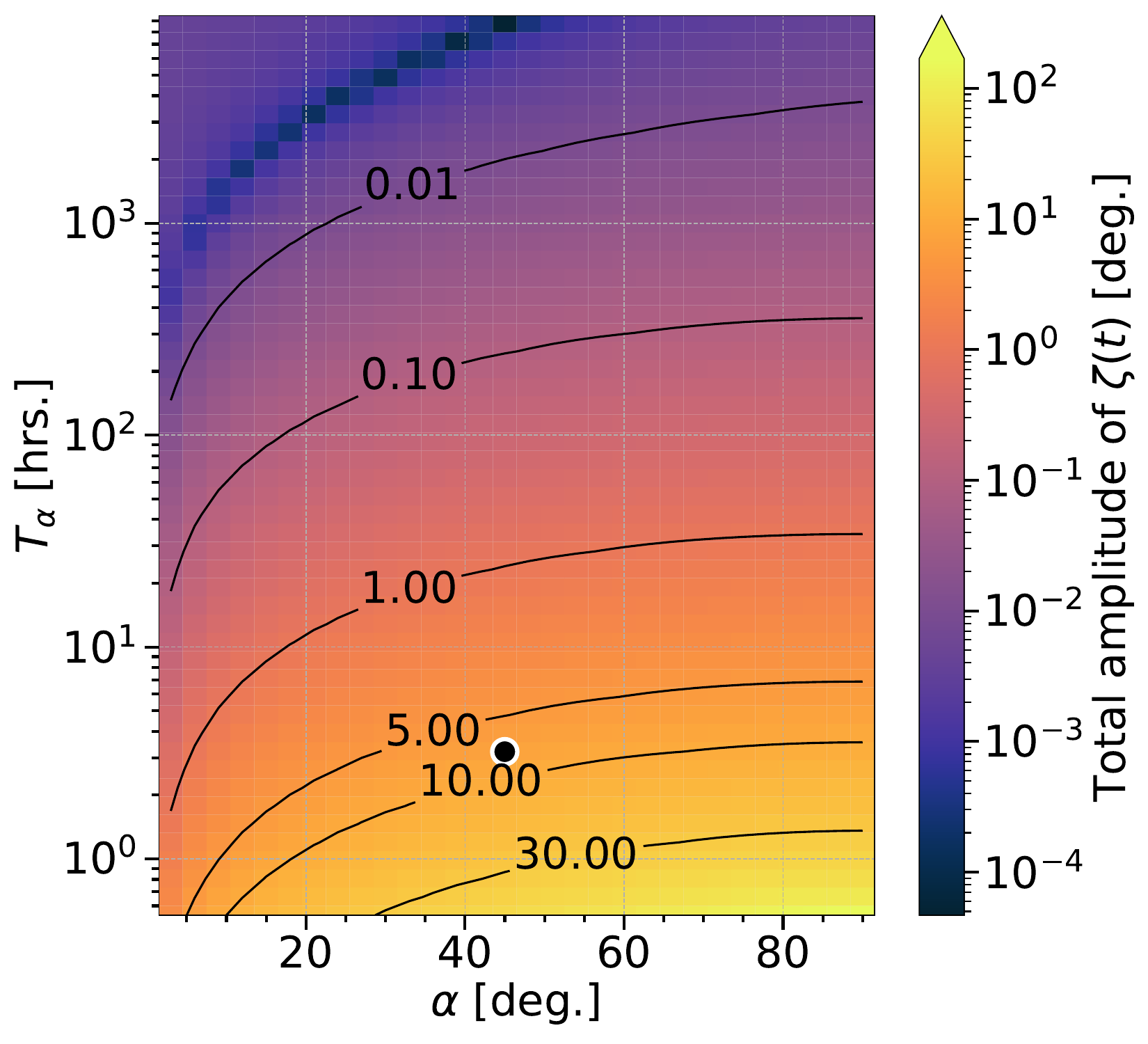
  }
  \caption[Scanning beam angle analysis during one spin cycle.]{(left) Spin
  cycle velocity vectors viewed from spin axis. $x_{\rm FP}$ and $y_{\rm FP}$ show
  detector reference frame coordinates, while $\vec{v}_{\rm prec}$ and
  $\vec{v}_{\rm spin}$ represent precession and spin velocities. Precession pushes
  the spin axis rightward, causing $\vec{v}_{\rm scan}$ (combining $\vec{v}_{\rm
  prec}$ and $\vec{v}_{\rm spin}$) to vary across $x_{\rm FP}$. Angle $\zeta$ measures
  between $\vec{v}_{\rm scan}$ and $x_{\rm FP}$. (right) Total $\zeta(t)$
  amplitude during one spin cycle across $\{\alpha,T_{\alpha}\}$ space.}
  \label{fig:scanbeam}
\end{figure}

\Cref{fig:scanbeam} (left) illustrates velocity vectors during one spin cycle.
We define the scanning beam angle $\zeta$ as the angle between focal plane
coordinate $x$-axis ($x_{\rm FP}$) and scan velocity vector $\vec{v}_{\rm scan}$,
which combines spin ($\vec{v}_{\rm spin}$) and precession ($\vec{v}_{\rm prec}$)
velocities. During scanning, $\vec{v}_{\rm scan}$ direction changes relative to $x
_{\rm FP}$, causing $\zeta(t)$ to oscillate over time. \Cref{fig:scanbeam} (right)
shows the total amplitude of $\zeta(t)$, representing the scanning direction
angle variation range in the detector frame during one spin cycle. Larger amplitudes
facilitate both resolving $\vec{v}_{\rm scan}$-related systematic effect degeneracies
and improving beam-shape/sidelobe characterization.

\LiteBIRD's short $T_{\alpha}$ produces a $\zeta$ amplitude of approximately $5^{\circ}$,
significantly larger than \Planck's 0.5\,arcsec. This larger variation aids sidelobe
reconstruction and helps identify unexpected systematics like extended detector time
constants or transfer function effects. However, \Planck and \PICO strategies,
with larger $T_{\alpha}$, better support frequent short-term compact source observations.

\subsection{Amplitude of CMB solar dipole}
\label{sec:dipole} During sky scanning, the CMB dipole manifests as a 3.3\,mK sinusoidal
signal in the time-ordered data~(TOD) \cite{dipole_SRoll}. This prominent signal
acts as a natural photometric calibrator for detector gain calibration, where larger
signal amplitudes facilitate more precise calibration. However, the frequency spectrum
of this signal plays an equally critical role in calibration quality.

Various instrumental effects can compromise detector gain stability. For
instance, thermal fluctuations typically produce $1/f$-like gain variation spectra.
When these gain fluctuation spectra overlap with dipole signal frequencies, they
can significantly degrade calibration accuracy. Therefore, scanning strategies that
generate dipole signals with higher-frequency spectral peaks are advantageous for
robust calibration.

\begin{figure}[h]
  \centering
  \includegraphics[width=1\columnwidth]{
    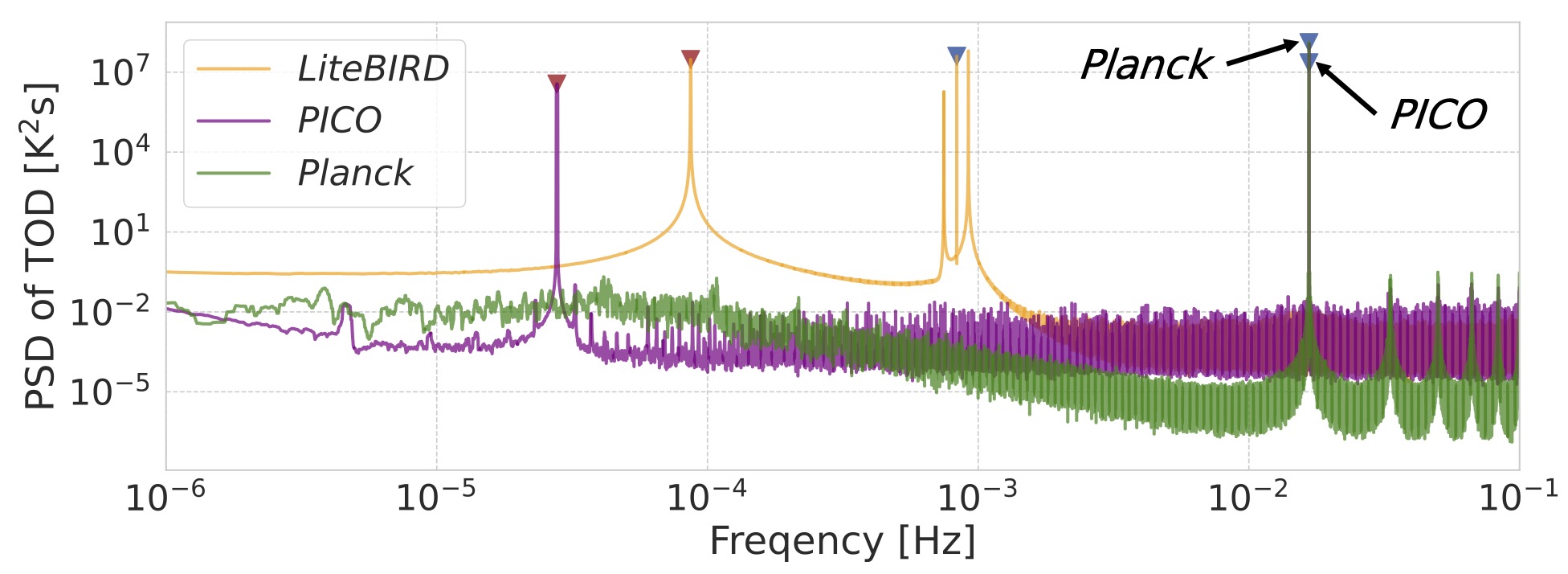
  }
  \caption[ Power spectra of TOD from different space missions scanning a sky containing
  only CMB dipole. ]{Power spectra of TOD from different space missions scanning
  a sky containing only CMB dipole. The red $\blacktriangledown$ and blue $\blacktriangledown$
  markers indicate precession and spin frequencies, respectively. }
  \label{fig:dipole_spectra}
\end{figure}

\Cref{fig:dipole_spectra} presents power spectra analyses of TOD from different space
missions scanning a sky containing only the CMB dipole signal. The spectral
peaks correspond to spacecraft precession (marked by red $\blacktriangledown$)
and spin (marked by blue $\blacktriangledown$) frequencies. Notably, \LiteBIRD
exhibits the highest frequency peak by precession, potentially allowing better separation
between dipole signals and gain fluctuation spectra, thus enabling more precise
calibration. Since spacecraft thermal variations typically occur at the spin frequency
due to continuous solar exposure on one side of spacecraft during each spin cycle,
this separation is particularly valuable. A distinctive feature of \LiteBIRD's spectrum
is the presence of two additional peaks around the spin frequency, located at
$f_{\rm spin}\pm f_{\rm prec}$. These satellite peaks provide additional leverage
for calibration and null-tests by helping distinguish between genuine dipole signals
and systematic gain drifts caused by spin-synchronized disturbances.

\subsection{Sky pixel visit/revisit times}
\label{sec:skypix} To detect unknown time-dependent systematic effects, data is commonly
split into different time periods for null-test analysis through differencing.
The distribution of pixel observation times indicates null-test effectiveness. Uniform
observation of pixels across the mission duration helps detect long-term systematic
effects like gain drift. The pixel revisit time, defined as
\begin{align}
  t^{\rm re}_{j}= t_{j+1}- t_{j},
\end{align}
where $j$ denotes the $j^{\rm th}$ measurement, is another key indicator. A uniform
distribution of revisit times enables analysis across multiple timescales. We
analyze three characteristic points in ecliptic coordinates by \healpix: $(\theta
,\varphi)=(0^{\circ},0^{\circ}),(45^{\circ},180^{\circ})$, and
$(90^{\circ},180^{\circ})$. All simulations use 19\,Hz sampling rate with $N_{\rm
side}=64$ pixelization.

\begin{figure}[h]
  \centering
  \includegraphics[width=0.95\columnwidth]{
    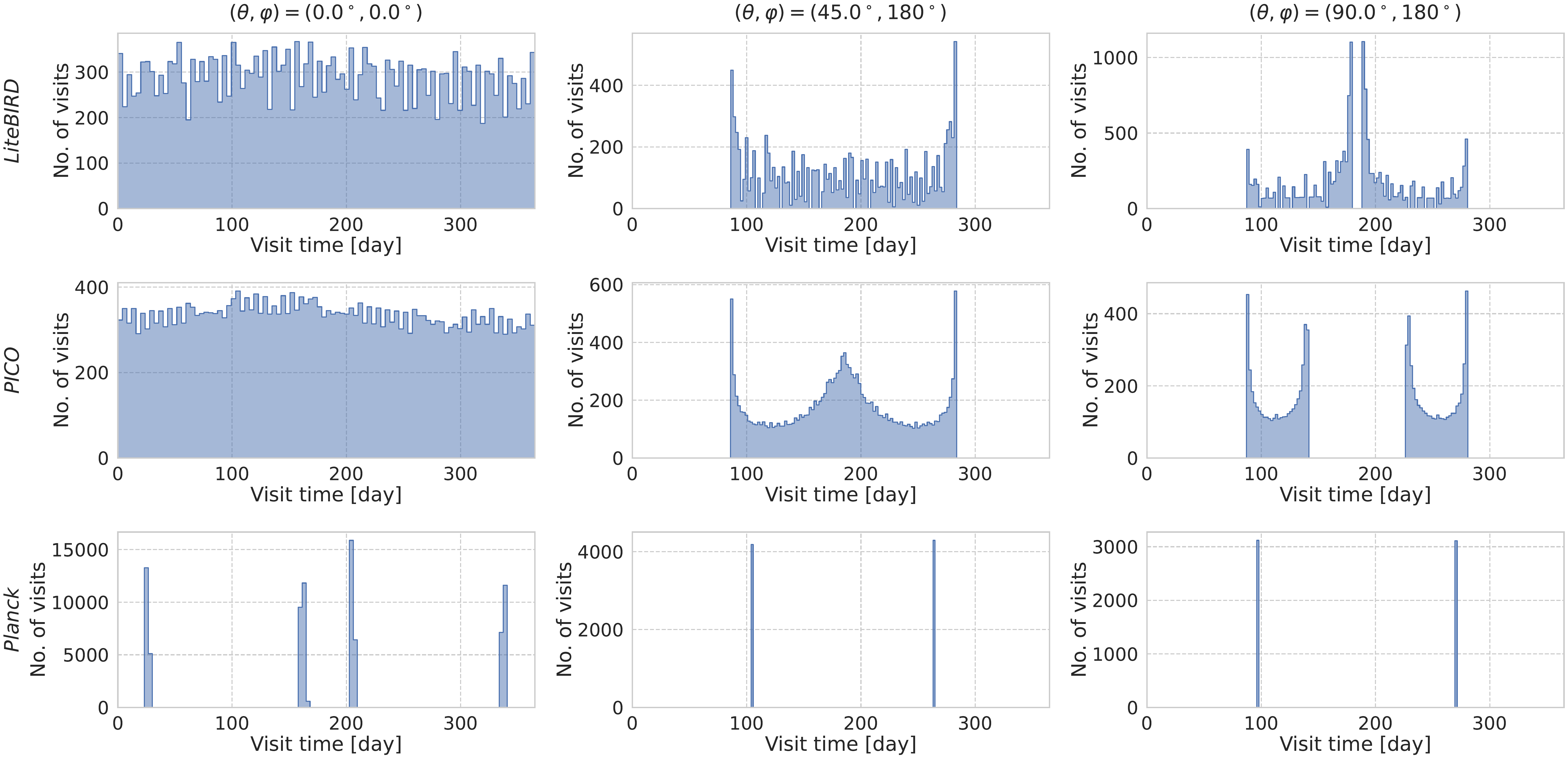
  }
  \caption[Distribution of visit times by spacecraft and sky position. ]{Distribution
  of visit times by spacecraft and sky position. The rows show \LiteBIRD (top),
  \PICO (middle), and \Planck (bottom). Columns display pixel positions in
  ecliptic coordinates: North pole $(0^{\circ},0^{\circ})$, mid-latitude
  $(45^{\circ},180^{\circ})$, and equator $(90^{\circ},180^{\circ})$. All
  simulations use \healpix maps with $\Nside=64$ and 19\,Hz sampling rate. The Sun-$\rm
  {L_2}$ vector at simulation start points to $(\theta,\varphi)=(90^{\circ},0^{\circ}
  )$.}
  \label{fig:visit_hist}
\end{figure}

\Cref{fig:visit_hist} compares visit time distributions across missions. For
polar pixels $(\theta,\varphi)=(0^{\circ},0^{\circ})$, \LiteBIRD and \PICO show consistent
year-round visits, while \Planck exhibits observation gaps. \Planck's large $\beta$
creates wide scan rings, causing intense short-term pixel visits but 6-month
gaps between observations as rings drift. These gaps complicate detection of long-term
effects like gain drift. In contrast, \LiteBIRD and \PICO's larger $\alpha$ and shorter
precession periods enable more frequent pixel visiting. Their main difference
appears in equatorial pixel visit distributions, where gaps reflect scan pupil size
($2|\alpha-\beta|$). Annual unobservable periods occur in regions
$|\alpha-\beta|$ from the equator, determined by scan pupil size and orbital velocity.

\begin{figure}[h]
  \centering
  \includegraphics[width=0.95\columnwidth]{
    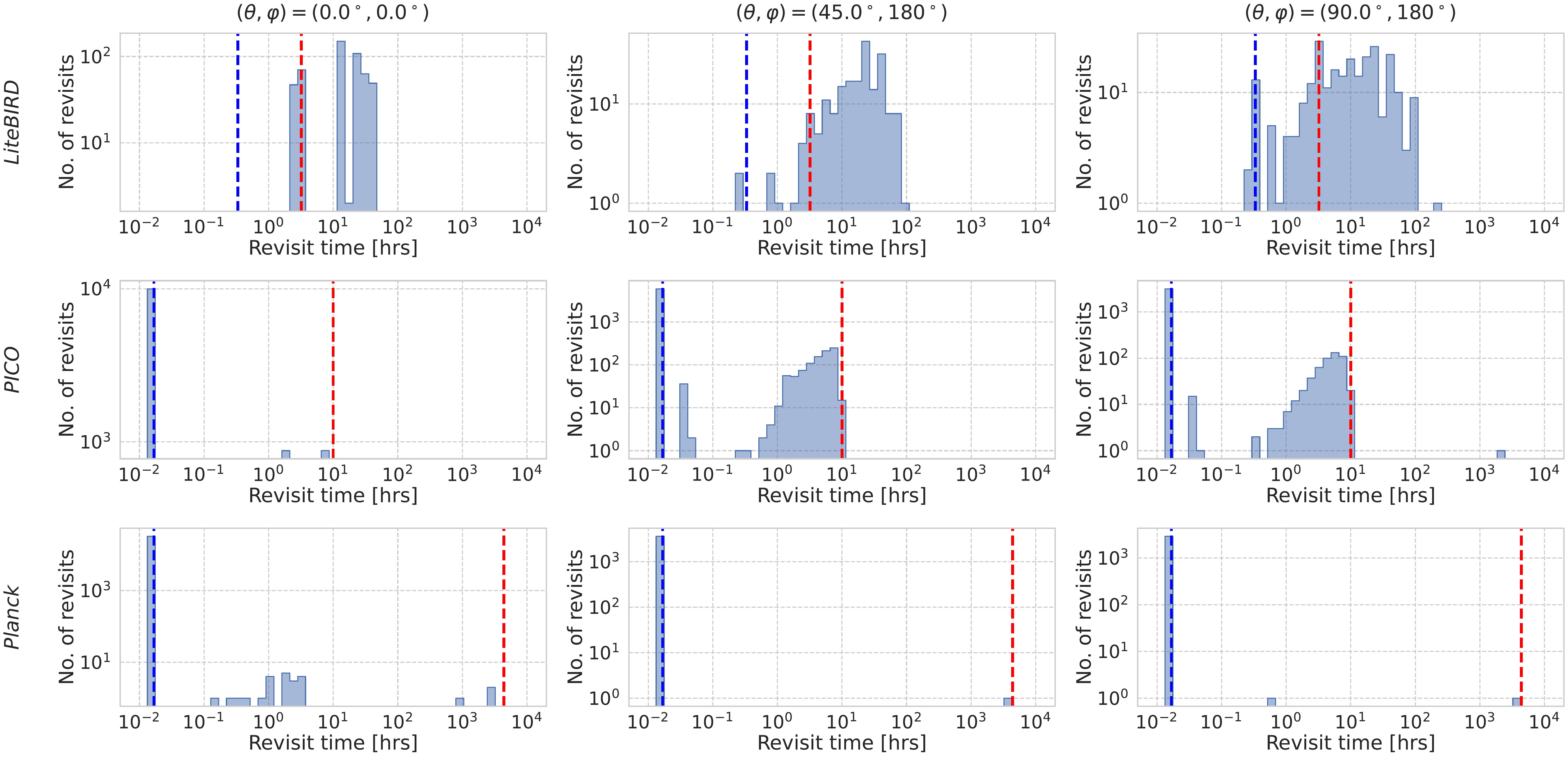
  }
  \caption[Distribution of pixel revisit times across spacecraft and sky
  positions. ]{Distribution of pixel revisit times across spacecraft and sky
  positions, corresponding to panel layout in \cref{fig:visit_hist}. Blue and
  red dashed lines mark spin and precession periods respectively. Note: Some
  revisits occur faster than the spin period because at scan pattern edges, maximum
  precession angular velocity ($\omega_{\rm{max}}$) enables earlier revisits.}
  \label{fig:revisit_hist}
\end{figure}

\Cref{fig:revisit_hist} displays revisit time distributions, with blue and red
dashed lines marking spin and precession periods. \LiteBIRD and \PICO demonstrate
extended pixel visibility and diverse revisit timescales, enabling comprehensive
null-tests. \Planck shows abundant short-term revisits but limited long-term pixel
visibility, constraining temporal null-test capabilities.

These results suggest scanning strategies with small scan pupils and shorter
precession periods ($T_{\alpha}<100$\,hours) optimize time-domain null-tests, aligning
with \cref{sec:optimization} findings.

\clearpage
\subsection{Planet visit/revisit times}
\label{sec:visit} While \cref{sec:comp_source_obs} addressed planet observation integration
time, this section examines two additional crucial aspects: calibration duration
availability throughout the mission and calibration frequency opportunities. We
simulate planet visit and revisit times to understand how scanning strategy
affects planetary visibility, using the same setup as \cref{sec:comp_source_obs}
but analyzing temporal distribution rather than integrated visit times.

\begin{figure}[h]
  \centering
  \includegraphics[width=0.95\columnwidth]{
    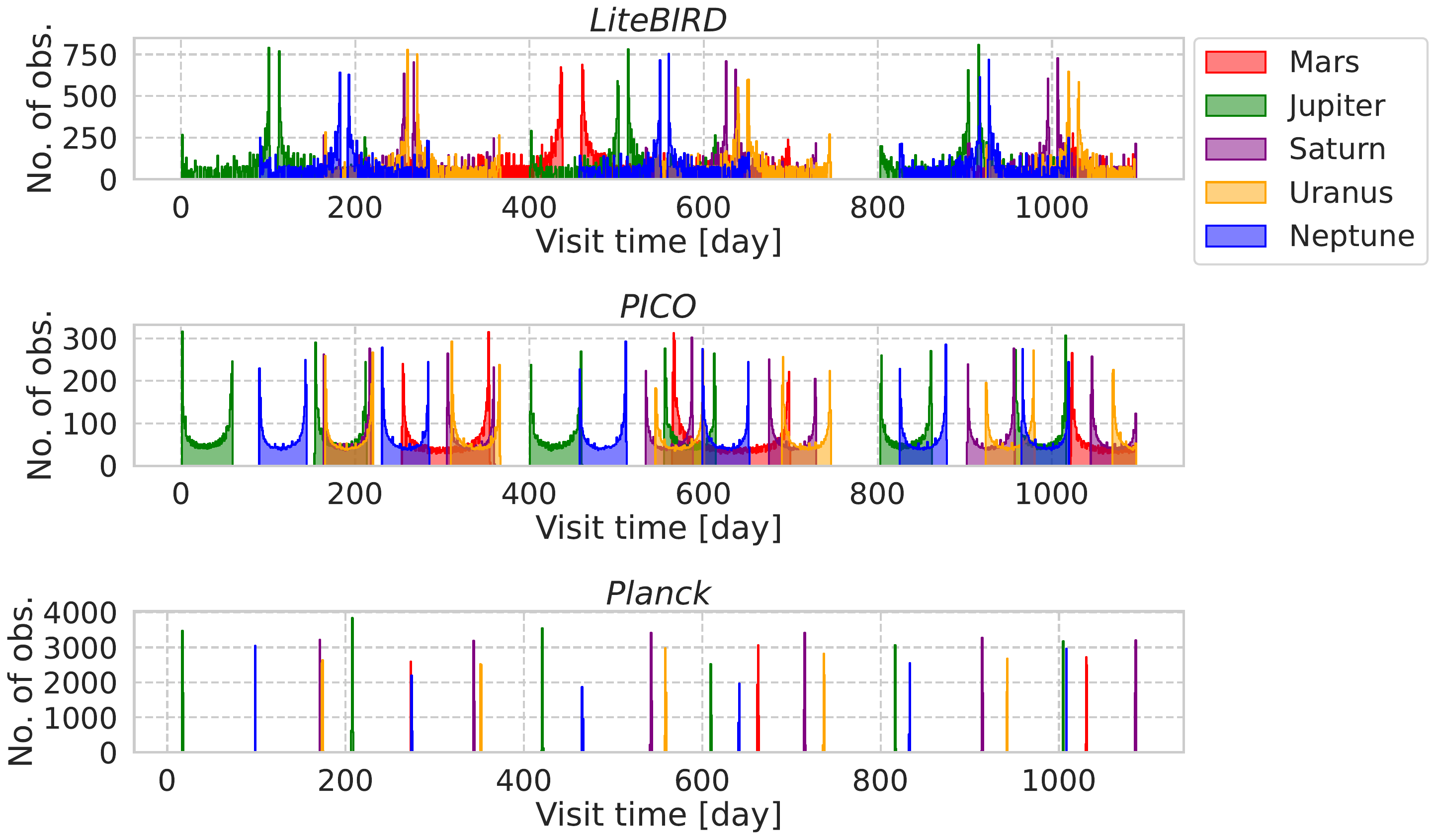
  }
  \caption[Three-year mission planet observation frequency histogram for
  \LiteBIRD, \PICO and \Planck.]{Three-year mission planet observation frequency
  histogram for \LiteBIRD, \PICO and \Planck. Parameters: 1-day bins, simulation
  start 2032-04-01T00:00:00, $0.5^{\circ}$ planet detection threshold, 1-second
  planet position updates.}
  \label{fig:planet_visit_hist}
\end{figure}

\Cref{fig:planet_visit_hist} displays planet visit time distributions across spacecraft.
The distributions mirror those of equatorial sky pixels
$(\theta,\varphi) = (90^{\circ},180^{\circ})$ in \cref{fig:visit_hist}, as planets
orbit near the ecliptic plane. Smaller scan pupils enable longer planetary
observation windows. \PICO's larger scan pupil creates 90-day observation gaps approximately,
though its 1-minute rotation and 10-hour precession periods allow frequent short-term
visibility. \Planck exhibits extreme short-term clustering of observations but with
extensive gaps, complicating long-term calibration. Total planetary visibility times
are: \LiteBIRD (3.1\,hours), \PICO (1.8\,hours), and \Planck (1.6\,hours). \LiteBIRD's
superiority stems from its smaller scan pupil and 20-fold longer spin period,
extending planet transit times.

\begin{figure}[h]
  \centering
  \includegraphics[width=0.95\columnwidth]{
    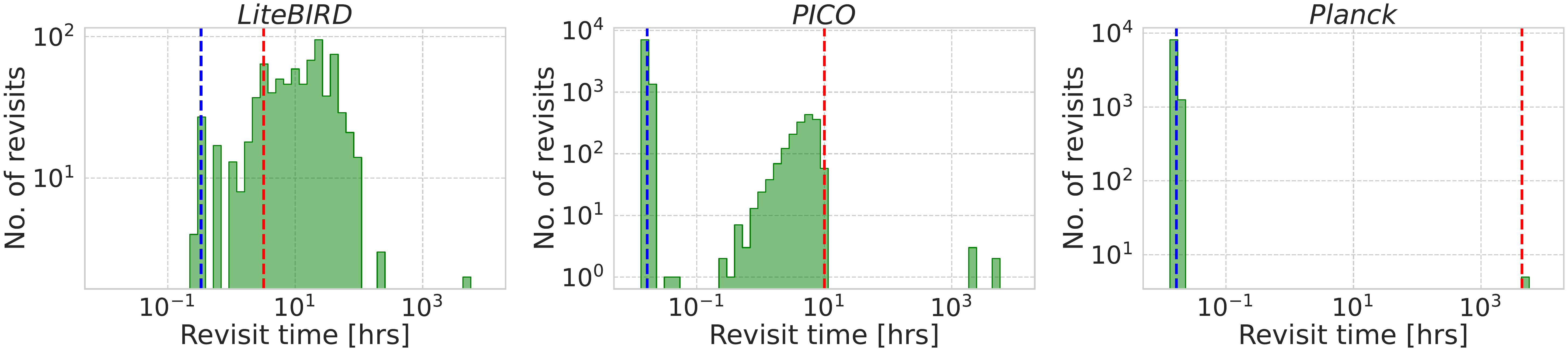
  }
  \caption[Jupiter revisit time distribution.]{Jupiter revisit time
  distribution from \cref{fig:planet_visit_hist}. Blue/red dashed lines indicate
  spin/precession periods. Minimum revisit times match spin periods; maximum 6-month
  intervals reflect coverage gaps. Secondary maxima (scan pupil transit times): \LiteBIRD
  10 days, \PICO 90 days, \Planck 6 months.}
  \label{fig:planet_revisit_hist}
\end{figure}

\Cref{fig:planet_revisit_hist} analyzes Jupiter's revisit time distribution.
\LiteBIRD and \PICO demonstrate diverse revisit timescales, enabling comprehensive
calibration datasets. \Planck favors shorter timescales. Minimizing scan pupil size
maximizes long-term planet visibility and diversifies revisit intervals. Longer
spin periods enhance integrated visibility time.

Spacecraft spin periods determine minimum revisit times. For overlapping scan strategies
like \PICO, precession periods limit maximum revisit times. \LiteBIRD's non-overlapping
spin cycles generate many revisit times exceeding the precession period. All
missions share 6-month maximum revisit intervals, with secondary maxima determined
by scan pupil transit times.

    \chapter{Systematic effects on CMB polarimetry}
\label{chap:systematics} \minitoc

\chapabstract{Building on the \spin formalism established in \cref{chap:formalism}, this chapter systematically analyzes the contaminating effects of systematics on CMB polarization measurements in \LiteBIRD's scanning strategy context. We explore two primary categories of systematic effects: those present in measurements without HWP implementation, and with HWP. Through map-based simulations using \spin formalism, we demonstrate how these systematics manifest in the observed data and quantify their impact on the tensor-to-scalar ratio measurement. Furthermore, we introduce novel mitigation techniques that exploit the \spin properties of systematic effects.
}

This chapter begins by categorizing systematic effects into two primary
classifications: those associated with and without HWP implementation.
Traditional polarization measurements without HWP rely on differential detection,
which involves subtracting signals from orthogonal detector pairs. This
differential approach can introduce several systematic effects, including differential
gain, differential pointing, and differential beam, which are considered typical
systematic challenges \cite{BICEP_syst}.

Although \LiteBIRD implements HWPs, analyzing these differential systematics provides
valuable context for understanding the HWPs' role in systematic mitigation. By examining
\LiteBIRD's hypothetical performance without HWPs, we gain deeper insights into
their benefits. Additionally, we present novel mitigation techniques we have
developed to suppress these systematic effects.

The systematic effects analyzed in this chapter include differential gain and
differential pointing for the case without HWP. For the case with HWP, we
consider absolute pointing offset; pointing disturbance due to HWP rotation; and
instrumental polarization due the HWP non-ideality. By the following sections,
we describe the model of systematic effect in \spin space.

In order to simulate these systematic effects, we developed the simulation framework
\SBM (\Spin-Based Map-making) in \texttt{Python}. This software is distributed in
author's GitHub repository.\footnote{\url{https://github.com/yusuke-takase/SBM}}
To maintain the quality of the code, we wrote unit-tests and applied the `GitHub
Actions' to run the test automatically. The GitHub Actions can run the unit-tests
by a virtual machine in cloud server when the new version of the code is pushed to
the repository.

\section{Systematic effects without HWP}

For modeling differential systematics, we define the signal field for an orthogonal
detector pair \texttt{T}/\texttt{B} as
\begin{align}
    S\tmu & = I + \frac{1}{2}Pe^{2i\psi\tmu}+ \frac{1}{2}P^{*}e^{-2i\psi\tmu}, \\
    S\bmu & = I - \frac{1}{2}Pe^{2i\psi\tmu}- \frac{1}{2}P^{*}e^{-2i\psi\tmu},
\end{align}
where $\psi\bmu=\psi\tmu + \pi/2$. Here, detectors \texttt{T} and \texttt{B} share
the same detector-pixel ID $\mu$ and observe the same sky direction. Without a
HWP, \spin-$m$ is always zero, so we omit it from our formalism.

\subsection{Differential gain}
The signal field with time-independent gain offset can be expressed as
\begin{align}
    S_{\texttt{T}, \mathrm{g}}^{(\mu)} & = (1+g\tmu)S\tmu, \\
    S_{\texttt{B}, \mathrm{g}}^{(\mu)} & = (1+g\bmu)S\bmu,
\end{align}
where $g\tmu$ and $g\bmu$ represent the gain offsets of detectors \texttt{T} and
\texttt{B} respectively. The differential gain field $D\pmu_{\rm g}$ is defined as
\begin{align}
    D\pmu_{\rm g} & = \frac{1}{2}(S_{\texttt{T}, \mathrm{g}}^{(\mu)}- S_{\texttt{B}, \mathrm{g}}^{(\mu)}) \label{eq:diff}                                          \\
                  & = \frac{1}{2}[\Delta g\pmu I + \frac{1}{2}(2+g\tmu+g\bmu)Pe^{2i\psi\tmu}+ \frac{1}{2}(2+g\tmu+g\bmu)P^{*}e^{-2i\psi\tmu}],\label{eq:diff_gain}
\end{align}
where $\Delta g\pmu = g\tmu - g\bmu$. Fourier transforming with respect to
$\psi\tmu$ yields the \spin space form:
\begin{align}
    \Dt[0]\pmu_{\rm g}(\Omega)  & = \frac{1}{2}\Delta g\pmu I(\Omega),       \\
    \Dt[2]\pmu_{\rm g}(\Omega)  & = \frac{1}{4}(2+g\tmu+g\bmu)P(\Omega),     \\
    \Dt[-2]\pmu_{\rm g}(\Omega) & = \frac{1}{4}(2+g\tmu+g\bmu)P^{*}(\Omega).
\end{align}

The scanning strategy's mitigation effect is incorporated by coupling these maps
using \cref{eq:Sd_mu}:
\begin{align}
    \Dd[2]\pmu_{\rm g}(\Omega)  & = \sum_{n'}\h[2-n']\pmu(\Omega) \Dt[n']\pmu_{\rm g}(\Omega) \nonumber                        \\
                                & = \h[4]\pmu \Dt[-2]\pmu_{\rm g}+ \h[2]\pmu \Dt[0]\pmu_{\rm g}+ \h[0]\pmu \Dt[2]\pmu_{\rm g}, \\
    \Dd[-2]\pmu_{\rm g}(\Omega) & ={\Dd[2]\pmu_{\rm g}}^{*}(\Omega).
\end{align}
As we described in \cref{sec:multi_detector}, the total differential signal
$\Dd[ n]^{\rm tot}_{\rm g}$ is obtained through:
\begin{align}
    \Dd[n]^{\rm tot}_{\rm g}(\Omega) & = \frac{1}{\Ntot(\Omega)}\sum_{\mu}\Nhits\pmu(\Omega) \Dd[n]\pmu_{\rm g}(\Omega).\label{eq:Dd_tot}
\end{align}
Since differential detection using orthogonal detector pairs should cancel the temperature
signal, we can exclude the $I$ component estimation from \cref{eq:mapmaking_tot}.
This reduces the matrix dimension from $3\times3$ to $2\times2$, then the map-making
equation by a single pair of detectors can be expressed as
\begin{align}
    \ab (\mqty{ \hat{P}\\ \hat{P^*} }) & =\M[2]^{-1}\ab (\mqty{ \frac{1}{2}\Dd[2]_{\rm g}\pmu \\ \frac{1}{2}\Dd[-2]{\rm g}\pmu }), \label{eq:2x2mapmaking}
\end{align}
where $_{2}M$ is a $2\times2$ matrix defined as
\begin{align}
    \M[2]^{-1}= \ab (\mqty{ \frac{1}{4} & \frac{1}{4}\h[4]\pmu \\ \frac{1}{4}\h[-4]\pmu & \frac{1}{4} })^{-1},
\end{align}
and the map-making equation by the multiple pairs of detectors can be expressed
as
\begin{align}
    \ab (\mqty{ \hat{P}\\ \hat{P^*} }) & = \ab (\mqty{ \frac{1}{4} & \frac{1}{4}\htot[4] \\ \frac{1}{4}\htot[-4] & \frac{1}{4} })^{-1}\ab (\mqty{ \frac{1}{2}\Dd[2]^{\rm tot}_{\rm g} \\ \frac{1}{2}\Dd[-2]^{\rm tot}_{\rm g} }), \label{eq:2x2mapmaking_tot}
\end{align}
where $\hat{P}= \hat{Q}+i\hat{U}$ represents the estimated polarization.

\subsection{Differential pointing}

To describe the pointing offset, we refer the coordinates and formalism
introduced in ref.~\cite{OptimalScan,mapbased}. We use the flat-sky
approximation in the small angular scale and span the sky with a Cartesian
coordinate system. In this reference frame, $x$- and $y$-axis are chosen to be aligned
with east and north, respectively. Assuming that a first-order Taylor expansion around
a point $(x,y)$ is possible when the pointing offset is small, then, the pointing
offset field, $S_{\rm p}$ as
\begin{equation}
    \begin{split}
        S_{\rm p}(\psi) = [1 - (\partial_{x}\Delta x+ \partial_{y}\Delta y)]I&+ \frac{1}{2}
        \left[1-(\partial_{x}\Delta x+ \partial_{y}\Delta y)\right ]Pe^{2i\psi}\\
        &+ \frac{1}{2}\left[1-(\partial_{x}\Delta x+ \partial_{y}\Delta y)\right
        ]P^{*}e^{-2i\psi}.
    \end{split}\label{eq:pointing_offset_field_no_hwp}
\end{equation}
The perturbation term can be defined by using the magnitude of the pointing
offset, $\rho$ and direction of the pointing offset, $\chi$ as
\begin{equation}
    \begin{split}
        \partial_{x}\Delta x+ \partial_{y}\Delta y&= \partial_{x}[\rho \sin( \psi
        +\chi)] + \partial_{y}[\rho \cos(\psi+\chi)]\\&=\frac{\rho}{2}\left [ e^{i(\psi+\chi)}
        \eth + e^{-i(\psi+\chi)}\oeth \right],\label{eq:perturbation}
    \end{split}
\end{equation}
where we introduced the \spin-up (-down) ladder operators,
$\eth=\partial_{y}- i\partial_{x}$, ($\oeth=\partial_{y}+ i\partial_{x}$) as in ref.~\cite{mapbased}.

Now we can describe the signal field of the pointing offset for a detector-\texttt{T}
and \texttt{B} as
\begin{equation}
    \begin{split}
        S_{\texttt{T}, \mathrm{p}}^{(\mu)}(\psi\tmu)&= I - \frac{\rho\tmu}{2}\ab
        [e^{i(\psi\tmu+\chi\tmu)}\eth + e^{-i(\psi\tmu+\chi\tmu)}\oeth]I \\&+ \frac{1}{2}
        \ab[e^{ 2i\psi\tmu}- \frac{\rho\tmu}{2}\ab(e^{i(3\psi\tmu+\chi\tmu)}\eth
        + e^{-i(-\psi\tmu+\chi\tmu)}\oeth)]P\\&+ \frac{1}{2}\ab[e^{-2i\psi\tmu}-
        \frac{\rho\tmu}{2}\ab(e^{i(-\psi\tmu+\chi\tmu)}\eth + e^{-i(3\psi\tmu+\chi\tmu)}
        \oeth)]P^{*},
    \end{split}
\end{equation}
\begin{equation}
    \begin{split}
        S_{\texttt{B}, \mathrm{p}}^{(\mu)}(\psi\bmu)&= I - i\frac{\rho\bmu}{2}\ab
        [e^{i(\psi\tmu+\chi\bmu)}\eth - e^{-i(\psi\tmu+\chi\bmu)}\oeth]I \\&- \frac{1}{2}
        \ab[e^{ 2i\psi\tmu}- i\frac{\rho\bmu}{2}\ab(e^{i(3\psi\tmu+\chi\bmu)}\eth
        - e^{-i(-\psi\tmu+\chi\bmu)}\oeth)]P\\&- \frac{1}{2}\ab[e^{-2i\psi\tmu}-
        i\frac{\rho\bmu}{2}\ab(e^{i(-\psi\tmu+\chi\bmu)}\eth - e^{-i(3\psi\tmu+\chi\bmu)}
        \oeth)]P^{*},
    \end{split}
\end{equation}
where $\rho\tmu$ ($\rho\bmu$) is the magnitude of pointing offset, and $\chi\tmu$
($\chi\bmu$) are the orientation of pointing offset for the detector-\texttt{T}
(-\texttt{B}). By taking the difference of the signal fields as shown in \cref{eq:diff},
we can obtain the differential pointing field as
\begin{equation}
    \begin{split}
        D\pmu_{\rm p}&= -\frac{1}{4}\ab[\zeta\pmu e^{i\psi\tmu}\eth + \zeta^{*\pmu}
        e^{-i\psi\tmu}\oeth]I \\&+ \frac{1}{2}Pe^{ 2i\psi\tmu}- \frac{1}{4}\ab [\ozeta
        \pmu e^{3i\psi\tmu}\eth + \ozeta^{*\pmu}e^{i\psi\tmu}\oeth]P \\&+ \frac{1}{2}
        P^{*}e^{-2i\psi\tmu}- \frac{1}{4}\ab[\ozeta\pmu e^{-i\psi\tmu}\eth + \ozeta
        ^{*\pmu}e^{-3i\psi\tmu}\oeth]P^{*},
    \end{split}
\end{equation}
where we defined $\zeta\pmu$ and $\ozeta\pmu$ as
\begin{align}
    \zeta\pmu  & = \rho\tmu e^{i\chi\tmu}- i\rho\bmu e^{i\chi\bmu},  \\
    \ozeta\pmu & = \rho\tmu e^{i\chi\tmu}+ i \rho\bmu e^{i\chi\bmu}.
\end{align}
By the Fourier transform, we can obtain the \spin space form as
\begin{align}
    \Dt[1]\pmu_{\rm p}(\Omega) & = -\frac{1}{4}\ab (\zeta\pmu\eth I(\Omega) + \ozeta^{*\pmu}\oeth P(\Omega)), \label{eq:t2p} \\
    \Dt[2]_{\rm p}(\Omega)     & = \frac{1}{2}P(\Omega),                                                                     \\
    \Dt[3]\pmu_{\rm p}(\Omega) & = -\frac{1}{4}\ozeta\pmu\eth P(\Omega). \label{eq:p2p}
\end{align}
The coupling with a scan described as follows
\begin{equation}
    \begin{split}
        \Dd[2]\pmu_{\rm p}(\Omega)&= \sum_{n'}\h[2-n']\pmu(\Omega) \Dt[n']\pmu_{\rm
        p}(\Omega ) \\&= \h[5]\pmu \Dt[-3]\pmu_{\rm p}+ \h[4]\pmu \Dt[-2]\pmu_{\rm
        p}+ \h[3]\pmu \Dt[-1]\pmu_{\rm p}\\&+ \h[1]\pmu \Dt[1]\pmu_{\rm p}+ \h[0]
        \pmu \Dt[2]\pmu_{\rm p}+ \h[1]\pmu \Dt[3]\pmu_{\rm p}\\ \Dd[-2]\pmu_{\rm
        p}(\Omega)&={\Dd[2]\pmu_{\rm p}}^{*}(\Omega)
    \end{split}
\end{equation}
Same as the differential gain case, we can perform the map-making by
\cref{eq:2x2mapmaking} after obtaining $\Dd[n]^{\rm tot}_{\rm p}$ by
\cref{eq:Dd_tot}.

\section{Systematic effects with HWP}

\subsection{Absolute pointing offset}
\label{apd:pointing_offset}

Building upon our discussion of differential pointing, we now examine the absolute
pointing offset without considering pair difference that arises when the spacecraft's
actual pointing direction deviates from its intended target. Such deviations can
stem from star tracker calibration errors or misalignment between the star
tracker and telescope mounting. To model this absolute pointing offset, we
extend the formalism developed in \cref{eq:pointing_offset_field_no_hwp} to
incorporate both telescope pointing angle $\psi$ and HWP angle $\phi$, yielding
\begin{equation}
    \begin{split}
        S_{\rm ap}(\psi, \phi) = \ab[1 - (\partial_{x}\Delta x+ \partial_{y}\Delta
        y)]I&+ \frac{1}{2}\ab[1-(\partial_{x}\Delta x+ \partial_{y}\Delta y)]Pe^{-i(4\phi-2\psi)}
        \\&+ \frac{1}{2}\ab[1-(\partial_{x}\Delta x+ \partial_{y}\Delta y)]P^{*}e
        ^{i(4\phi-2\psi)}.
    \end{split}
\end{equation}
The perturbation term defined in \cref{eq:perturbation} can be applied here as well.
Using this definition, we can express the absolute pointing offset field as
\begin{equation}
    \begin{split}
        S_{\rm ap}\pmu(\psi\pmu, \phi)&= I - \frac{\rho\pmu}{2}\ab[e^{i(\psi\pmu+\chi\pmu)}
        \eth + e^{-i(\psi\pmu+\chi\pmu)}\overline{\eth}]I \\&+ \frac{1}{2}\ab[e^{-i(4\phi-2\psi\pmu)}
        - \frac{\rho\pmu}{2}\ab(e^{i(-4\phi+3\psi\pmu+\chi\pmu)}\eth + e^{-i(4\phi-\psi\pmu+\chi\pmu)}
        \overline{\eth}) ]P\\&+ \frac{1}{2}\ab[e^{i(4\phi-2\psi\pmu)}- \frac{\rho\pmu}{2}
        \ab(e^{i(4\phi-\psi\pmu+\chi\pmu)}\eth + e^{-i(-4\phi+3\psi\pmu+\chi\pmu)}
        \overline{\eth}) ]P^{*}.\label{eq:pointing_offset_field_with_hwp}
    \end{split}
\end{equation}
It's important to note that the HWP angle remains coherent across all detectors.
Through Fourier transformation with respect to both angles
$(\psi,\phi) \to (n,m)$, we can represent the signal in spin space as \spin space
as
\begin{align}
    \St[0,0]_{\rm ap}      & = I,                                                                                               \\
    \St[2,-4]_{\rm ap}     & ={\St[-2,4]_{\rm ap}}^{*}= \frac{P}{2},                                                            \\
    \St[1,0]\pmu_{\rm ap}  & ={\St[-1,0]\pmu_{\rm ap}}^{*}= -\frac{\rho\pmu}{2}e^{i\chi\pmu}\eth I, \label{eq:temperature_syst} \\
    \St[1,-4]\pmu_{\rm ap} & ={\St[-1,4]\pmu_{\rm ap}}^{*}= -\frac{\rho\pmu}{4}e^{-i\chi\pmu}\overline{\eth}P,                  \\
    \St[3,-4]\pmu_{\rm ap} & ={\St[-3,4]\pmu_{\rm ap}}^{*}= -\frac{\rho\pmu}{4}e^{i\chi\pmu}\eth P.
\end{align}
The first two components represent pure signals: \spin-$(0,0)$ for temperature
and \spin-$(\pm2,\mp4)$ for polarization $P$ or $P^{*}$. The remaining three terms
characterize systematic effects. Examining the systematic signal in
\cref{eq:temperature_syst}, we observe that the pointing offset's perturbation,
through the action of \spin ladder operators, transforms the original \spin-$(0,0
)$ temperature field into a \spin-$(\pm1,0)$ signal. These ladder operators, representing
the field's gradient, generate spurious odd \spin components absent in the
expected signal. Using \cref{eq:kSd}, we can express the coupling between the
scanning strategy and systematic effects for the pointing offset as
\begin{align}
    {\Sd[0,0]\pmu_{\rm ap}}(\Omega) ={}                                                                                                                                                                                                                                                                                                                                                             & \sum_{n'=-\infty}^{\infty}\sum_{m'=-\infty}^{\infty}\h[0-n',0-m']\pmu(\Omega)\St[n',m']\pmu_{\rm ap}(\Omega)\label{eq:00Sd_abs_pnt}                  \\
    \begin{split}={}&\h[3,-4]\pmu\St[-3,4]\pmu_{\rm ap}+ \h[2,-4]\pmu\St[-2,4]\pmu_{\rm ap}+ \h[1,-4]\pmu\St[-1,4]\pmu_{\rm ap}\\&+ \h[1,0]\pmu \St[-1,0]\pmu_{\rm ap}+ \h[0,0]\pmu \St[0,0]\pmu_{\rm ap}+ \h[-1,0]\pmu\St[1,0]\pmu_{\rm ap}\\&+ \h[-1,4]\pmu\St[1,-4]\pmu_{\rm ap}+ \h[-2,4]\pmu\St[2,-4]\pmu_{\rm ap}+ \h[-3,4]\pmu\St[3,-4]\pmu_{\rm ap}, \nonumber\end{split}                    \\
    {\Sd[2,-4]\pmu_{\rm ap}}(\Omega) ={}                                                                                                                                                                                                                                                                                                                                                            & {\Sd[-2,4]\pmu_{\rm ap}}^{*}(\Omega) = \sum_{n'=-\infty}^{\infty}\sum_{m'=-\infty}^{\infty}\h[2-n',-4-m']\pmu(\Omega)\St[n',m']\pmu_{\rm ap}(\Omega) \\
    \begin{split}={}&\h[5,-8]\pmu\St[-3,4]\pmu_{\rm ap}+ \h[4,-8]\pmu\St[-2,4]\pmu_{\rm ap}+ \h[3,-8]\pmu\St[-1,4]\pmu_{\rm ap}\\&+ \h[3,-4]\pmu\St[-1,0]\pmu_{\rm ap}+ \h[2,-4]\pmu\St[0,0]\pmu_{\rm ap}+ \h[1,-4]\pmu\St[1,0]\pmu_{\rm ap}\\&+ \h[1,0]\pmu \St[1,-4]\pmu_{\rm ap}+ \h[0,0]\pmu \St[2,-4]\pmu_{\rm ap}+ \h[-1,0]\pmu\St[3,-4]\pmu_{\rm ap}. \nonumber\end{split} \label{eq:24Sd_ap}
\end{align}
These equations demonstrate how the scanning strategy's orientation functions
$\h [n,m]$ couple with systematic fields $\St[n,m]$ through multiplication. This
coupling mechanism explicitly reveals how specific \spin-$(n,m)$ systematic effects
can be suppressed through careful design of the scanning strategy.

\subsection{Pointing disturbance due to HWP rotation}
\label{apd:wedge} While the HWP effectively enables polarization measurement through
modulation, its physical imperfections can introduce systematic effects. A notable
example is the non-perfectly flat surfaces of sapphire-based HWPs, which create
a small wedge angle. This imperfection leads to pointing disturbances during HWP
rotation.

Given a HWP wedge angle $w$, refraction causes the pointing direction to deviate
by an angle $\xi$ from its original path, expressed as
\begin{align}
    \xi = (n_{\rm ref}- 1)w,
\end{align}
where $n_{\rm ref}$ represents sapphire's refractive index. During HWP rotation,
the pointing direction traces a circular path with radius $\xi$ around the expected
direction. The pointing disturbance field induced by the HWP wedge angle, denoted
as $S_{\rm w}$, can be formulated as
\begin{equation}
    \begin{split}
        S_{\rm w}\pmu(\psi\pmu, \phi)&= I - \frac{\xi}{2}\ab[e^{i(\psi\pmu+\phi+\chi)}
        \eth + e^{-i(\psi\pmu+\phi+\chi)}\overline{\eth}]I \\&+ \frac{1}{2}\ab[e^{-i(4\phi-2\psi\pmu)}
        - \frac{\xi}{2}\ab(e^{i(-3\phi+3\psi\pmu+\chi)}\eth + e^{-i(5\phi-\psi\pmu+\chi)}
        \overline{\eth}) ]P\\&+ \frac{1}{2}\ab[e^{i(4\phi-2\psi\pmu)}- \frac{\xi}{2}
        \ab(e^{i(5\phi-\psi\pmu+\chi)}\eth + e^{-i(-3\phi+3\psi\pmu+\chi)}\overline
        {\eth}) ]P^{*}.\label{eq:wedge_field}
    \end{split}
\end{equation}

This model is derived by substituting $\chi \to \phi+\chi$ in \cref{eq:pointing_offset_field_with_hwp},
where $\chi$ represents the HWP angle phase. The Fourier transform yields the following
\spin space representation:
\begin{align}
    \St[0,0]_{\rm w}  & = I,                                                               \\
    \St[2,-4]_{\rm w} & = \St[-2,4]_{\rm w}^{*}= \frac{P}{2},                              \\
    \St[1,1]_{\rm w}  & = \St[-1,1]_{\rm w}^{*}= -\frac{\xi}{2}e^{i\chi}\eth I,            \\
    \St[-3,3]_{\rm w} & = \St[3,-3]_{\rm w}^{*}= -\frac{\xi}{4}e^{i\chi}\eth P,            \\
    \St[1,-5]_{\rm w} & = \St[5,-1]_{\rm w}^{*}= -\frac{\xi}{4}e^{-i\chi}\overline{\eth}P.
\end{align}

Since the HWP wedge angle produces a consistent pointing perturbation direction,
the systematic field remains detector-pixel ID ($\mu$) independent. The coupling
between systematic effects and scanning can be expressed as
\begin{align}
    {}_{0,0}{\tilde{S}^d}_{\rm w}(\Omega) ={}                                                                                                                                                                                                                                                                          & \sum_{n'=-\infty}^{\infty}\sum_{m'=-\infty}^{\infty}\h[0-n',0-m'](\Omega)\St[n',m']_{\rm w}(\Omega)\label{eq:00Sd_wedge}      \\
    \begin{split}={}&\h[3,-3]\St[-3,3]_{\rm w}+ \h[2,-4]\St[-2,4]_{\rm w}+ \h[1,-5]\St[-1,5]_{\rm w}\\&+ \h[1,1] \St[-1,-1]_{\rm w}+ \h[0,0] \St[0,0]_{\rm w}+ \h[-1,-1]\St[1,1]_{\rm w}\\&+ \h[-1,5]\St[1,-5]_{\rm w}+ \h[-2,4]\St[2,-4]_{\rm w}+ \h[-3,3]\St[3,-3]_{\rm w}, \nonumber\end{split}                      \\
    {}_{2,-4}{\tilde{S}^d}_{\rm w}={}                                                                                                                                                                                                                                                                                  & {\Sd[-2,4]_{\rm w}}^{*}= \sum_{n'=-\infty}^{\infty}\sum_{m'=-\infty}^{\infty}\h[2-n',-4-m'](\Omega)\St[n',m']_{\rm w}(\Omega) \\
    \begin{split}={}&\h[4,-7]\St[-3,3]_{\rm w}+ \h[4,-8]\St[-2,4]_{\rm w}+ \h[3,-9]\St[-1,5]_{\rm w}\\&+ \h[3,-3]\St[-1,-1]_{\rm w}+ \h[2,-4]\St[0,0]_{\rm w}+ \h[1,-5]\St[1,1]_{\rm w}\\&+ \h[1,1] \St[1,-5]_{\rm w}+ \h[0,0] \St[2,-4]_{\rm w}+ \h[-1,-1]\St[3,-3]_{\rm w}. \label{eq:24Sd_wedge}\nonumber\end{split}
\end{align}

The pointing offset generates spurious \spin-$(\pm1,0)$ systematics from the
\spin-(0,0) temperature field through the \spin ladder operators, which represent
the field's gradient. For pointing systematics due to the HWP wedge angle, the HWP
rotation induces pointing disturbances that sample the gradient fields. This
process generates spurious \spin-$(\pm1,0)$ fields from the original \spin-(0,0)
temperature field. The systematic effect modulates the temperature signal at the
HWP rotation frequency $f_{\phi}$, manifesting as a HWP synchronous systematic
effect at $1f_{\phi}$.

\subsection{Instrumental polarization due to HWP non-ideality}
\label{apd:HWP_sys}

Although HWPs effectively modulate polarization signals, suppress $1/f$ noise, and
reduce polarization measurement uncertainties even with single detectors, their
inherent imperfections introduce systematic effects that require careful
consideration \cite{Giorgio_hwp,Lorenzo_hwp,Essinger_hwp}. Of particular concern
are the non-diagonal terms in the HWP Mueller matrix arising from non-ideality,
which can generate instrumental polarization leading to temperature-to-polarization
leakage which is denoted as $T \to B$.

As discussed in ref.~\cite{guillaume_HWPIP}, this systematic effect
significantly impacts measurements through the CMB solar dipole and Galactic
emission intensity. To quantify the bias on $r$ induced by instrumental polarization,
ref.~\cite{guillaume_HWPIP} introduced the Mueller matrix deviation $\Delta M$ from
an ideal HWP. Considering only the instrumental polarization components at
modulation frequency $4f_{\phi}$ that contribute to $T \to B$ leakage, the systematic
field $\Delta S_{\rm ip}$ can be expressed as
\begin{align}
    \Delta S_{\rm ip}\pmu = \ab[\epsilon_{1}\pmu\cos(4\phi-4\psi\pmu+\phi_{QI})\cos2\psi_{0}\pmu+ \epsilon_{2}\pmu\cos(4\phi-4\psi\pmu+\phi_{UI})\sin2\psi_{0}\pmu]I,
\end{align}
where $\psi_{0}$ represents the detector's polarization angle relative to the focal
plane reference axis, and $\phi_{QI}$ and $\phi_{UI}$ denote the phases
described in ref.~\cite{guillaume_HWPIP}. The parameters $\epsilon_{1}$ and
$\epsilon_{2}$ represent the amplitudes of the Mueller matrix elements at
$4f_{\phi}$, which ideally should be zero but acquire finite values due to HWP non-ideality.

This systematic signal arises from two coupled physical mechanisms: First, for incident
polarization perpendicular to the HWP, the $180^{\circ}$ retardation difference
between polarization states generates a spurious $2f_{\phi}$-signal; Second, the
non-perpendicular s- and p-polarization components flip every $180^{\circ}$ HWP rotation,
producing another $2f_{\phi}$-signal. The coupling between these $2f_{\phi}$-signals
generates spurious $4f_{\phi}$-signals that mimic polarization signals \cite{imada2018instrumentally}.
Notably, this effect persists despite HWP rotation since it originates from the rotation
itself. Under the assumptions $\epsilon_{1}\pmu=\epsilon_{2}\pmu$ and
$\phi_{QI}=\phi_{UI}+\frac{\pi}{2}$, the systematic signal can be expressed in a
more compact form:
\begin{align}
    \Delta S_{\rm ip}\pmu= \frac{\epsilon_{1}\pmu}{2}\ab[e^{i(4\phi-4\psi\pmu+\phi_{QI}-2\psi_{0}\pmu)}+ e^{-i(4\phi-4\psi\pmu+\phi_{QI}+2\psi_{0}\pmu)}]I.\label{eq:HWP_IP_field}
\end{align}
In a coordinate system where $\psi_{0}=0$, Fourier transformation yields the \spin
space representation:
\begin{align}
    {}_{4,-4}\Delta \tilde{S}_{\rm ip}\pmu={}_{-4,4}\Delta \tilde{S}_{\rm ip}^{*(\mu)}= \frac{\epsilon_{1}\pmu}{2}e^{-i\phi_{QI}}I. \label{eq:HWPIP_syst}
\end{align}
The coupling with scanning strategy according to \cref{eq:kSd} results in:
\begin{align}
    {}_{0,0}\Delta \tilde{S}_{\rm ip}^{d(\mu)}  & = \sum_{n'=-\infty}^{\infty}\sum_{m'=-\infty}^{\infty}\h[0-n',0-m']{}_{n',m'}\pmu\Delta\tilde{S}_{\rm ip}\pmu\nonumber                                                \\
                                                & = \h[4,-4]\pmu{}_{-4,4}\Delta \tilde{S}_{\rm ip}\pmu+ \h[-4,4]\pmu{}_{4,-4}\Delta \tilde{S}_{\rm ip}\pmu,                                                             \\
    {}_{2,-4}\Delta \tilde{S}_{\rm ip}^{d(\mu)} & ={}_{-2,4}\Delta \tilde{S}_{\rm ip}^{d*(\mu)}= \sum_{n'=-\infty}^{\infty}\sum_{m'=-\infty}^{\infty}\h[2-n',-4-m']\pmu{}_{n',m'}\Delta\tilde{S}_{\rm ip}\pmu \nonumber \\
                                                & = \h[6,-8]\pmu{}_{-4,4}\Delta \tilde{S}_{\rm ip}\pmu+ \h[-2,0]\pmu{}_{4,-4}\Delta \tilde{S}_{\rm ip}\pmu.
\end{align}
The HWP's contribution significantly suppresses systematic signals coupled to $\h
[\pm4,\mp4]$ and $\h[\pm6,\mp8]$ terms through their real and imaginary parts.
However, $\h[\pm2,0]$ lacks HWP contribution, leaving the associated temperature
leakage only moderated by the scanning strategy. While previous studies have shown
that this bias on $r$ can be compensated \cite{guillaume_HWPIP}, this case
effectively demonstrates how crucial the scanning strategy remains, even with HWP
implementation.

\section{Mitigation techniques to control systematic effects by using \spin}
\label{sec:mitigation}

In this section, we discuss the mitigation of the systematic effects. The map-making
can be recognized as a linear regression problem, and it regresses the signal basis
vector, which corresponds to the $\bm{w}_{j}$ in \cref{eq:dj}. If we reconstruct
not only the Stokes vector but also the systematic effects per \spin-$( n,m)$,
we can suppress the leakage due to the systematic effects to the Stokes vector. Now
we start from a general case that the TOD, $d_{j}$ detected by a single detector
can be written as
\begin{align}
    \begin{split}
    d_{j}&= I + Q\cos(4\phi_{j}-2\psi_{j}) + U\sin(4\phi_{j}-2\psi_{j}) \\&+ \sum_{n\geq 0}\ab[\Z[n,m]^{Q}\cos(n\psi_{j}+m\phi_{j}) + \Z[n,m]^{U}\sin(n\psi_{j}+m\phi_{j})] + n_{j}\\
    &= \ab (\mqty{1 & \cos(4\phi_j-2\psi_j) & \sin(4\phi_j-2\psi_j) & \cdots & \cos(n\psi_j+m\phi_j) & \sin(n\psi_j+m\phi_j)}) \ab (\mqty{I \\ Q \\ U \\ \vdots \\ \Z[n,m]^Q \\ \Z[n,m]^U}) + n_{j}\\&=\bm{w}_{j}\cdot \bm{s}+ n_{j},
    \end{split}
\end{align}
where $\Z[n,m]^{Q}$ and $\Z[n,m]^{U}$ are additional \spin-$(n,m)$ signals that
may be attributed to systematic effects (or sky component, such as $\eth I$ in
the differential pointing systematics) where the superscripts $Q$ and $U$ are in
analogy to the polarization signal \cite{spin_characterisation}. In order to
estimate the signal vector from the measured data samples, we minimize:
\begin{align}
    \chi^{2}= \sum_{i,j}(d_{i}-\bm{w}_{i}\cdot \bm{s})(N^{-1})_{ij}(d_{j}-\bm{w}_{j}\cdot \bm{s}),
\end{align}
where $N_{ij}$ is the noise covariance matrix. After minimization, we can obtain
the equation to estimate the signal vector as
\begin{align}
    \hat{\bm{s}} & = \ab(\sum_{i,j}\bm{w}_{i}^{\dagger}(N^{-1})_{ij}\bm{w}_{j})^{-1}\ab(\sum_{i,j}\bm{w}_{i}^{\dagger}(N^{-1})_{ij}d_{j}),
\end{align}
then, we assume the noise is white noise, i.e., $N_{ij}= \sigma^{2}\delta_{ij}$,
where $\sigma$ is the standard deviation of the noise. The equation can be
simplified as
\begin{align}
    \hat{\bm{s}} & = \ab(\sum_{j}\bm{w}_{j}^{\dagger}\bm{w}_{j})^{-1}\ab(\sum_{j}\bm{w}_{j}^{\dagger}d_{j})\nonumber                                                                        \\
                 & = M^{-1}\ab (\mqty{\ab<d_j> \\ \ab<d_j\cos(4\phi_j-2\psi_j)> \\ \ab<d_j\sin(4\phi_j-2\psi_j)> \\ \vdots \\ \ab<d_j\cos(n\psi_j+m\phi)> \\ \ab<d_j\sin(n\psi_j+m\phi)>}),
\end{align}
where $M$ is the matrix defined as
\begin{align}
    \scalebox{0.75}{$M = \ab( \begin{array}{cccccc}1 & \ab < \cos(4\phi-2\psi_j) > & \ab < \sin(4\phi-2\psi_j) > & \cdots & \ab < \cos(n\psi_j+m\phi) > & \ab < \sin(n\psi_j+m\phi) > \\ \ab < \cos(4\phi-2\psi_j) > & \ab < \cos^2 (4\phi-2\psi_j) > & \ab < \cos(4\phi-2\psi_j) \sin(4\phi-2\psi_j) > & \cdots & \vdots & \vdots \\ \ab < \sin(4\phi-2\psi_j) > & \ab < \cos(4\phi-2\psi_j) \sin(4\phi-2\psi_j) > & \ab < \sin^2(4\phi-2\psi_j) > & \cdots & \vdots & \vdots \\ \vdots & \vdots & \vdots & \ddots & \vdots & \vdots \\ \ab < \cos(n\psi_j+m\phi_j) > & \cdots & \cdots & \cdots & \ab < \cos^2(n\psi_j+m\phi_j) > & \vdots \\ \ab < \sin(n\psi_j+m\phi_j) > & \cdots & \cdots & \cdots & \cdots & \ab < \sin^2(n\psi_j+m\phi_j) >\end{array})$}
    \label{eq:M}.
\end{align}
The mitigation technique with this formalism is introduced by appendix A1.2 of ref.~\cite{spin_characterisation},
and it corresponds the following equation in \spin space as
\begin{align}
    \ab (\mqty{I \\ P \\ P^* \\ \vdots \\ \Z[n,m] \\ \Z[-n,-m]}) = \tilde{M}^{-1}\ab (\mqty{\Sd[0,0] \\ \Sd[2,-4] \\ \Sd[-2,4] \\ \vdots \\ \Sd[n,m] \\ \Sd[-n,-m]}),\label{eq:general_mitigation}
\end{align}
where $\Z[n,m]$ and $\Z[-n,-m]$ are the estimated systematic signals in \spin space,
and $\tilde{M}$ is the matrix defined as
\begin{align}
    \tilde{M}= \ab( \begin{array}{cccccc}1 & \frac{1}{2}\h[-2,4] & \frac{1}{2}\h[2,-4] & \cdots & \frac{1}{2}\h[-n,-m] & \frac{1}{2}\h[n,m] \\ \frac{1}{2}\h[2,-4] & \frac{1}{4} & \frac{1}{4}\h[4,-8] & \cdots & \vdots & \vdots \\ \frac{1}{2}\h[-2,4] & \frac{1}{4}\h[-4,8] & \frac{1}{4} & \cdots & \vdots & \vdots \\ \vdots & \vdots & \vdots & \ddots & \vdots & \vdots \\ \frac{1}{2}\h[n,m] & \cdots & \cdots & \cdots & \frac{1}{4} & \frac{1}{4}\h[2n,2m] \\ \frac{1}{2}\h[-n,-m] & \cdots & \cdots & \cdots & \frac{1}{4}\h[-2n,-2m] & \frac{1}{4}\end{array} ).\label{eq:tildeM}
\end{align}
It allows us to estimate the polarization without contamination due to the
systematic effect given by $\Z[n,m]$.

\section{Results of the systematic effects}
\label{sec:results_syst}

We first consider the case of a single detector at the boresight, so we drop the
superscript $\mu$ which was used to distinguish a detector's pixel ID. The input
map is CMB only which is created by the $\Lambda$CDM model with the \texttt{CAMB},
and the 6-cosmological parameter is set to the \Planck 2018 best-fit values with
no primordial $B$-modes, i.e., $r=0$ as same as we used for
\cref{sec:Propagation}. The scanning strategy is assumed to be the \LiteBIRD's \SC
\cite{takase2024scan}. Finally, the map is smoothed by the Gaussian beam with $\mathrm{FWHM}
=1^{\circ}$.

\subsection{Differential gain}
\begin{figure}[h]
    \centering
    \includegraphics[width=1.0\columnwidth]{
        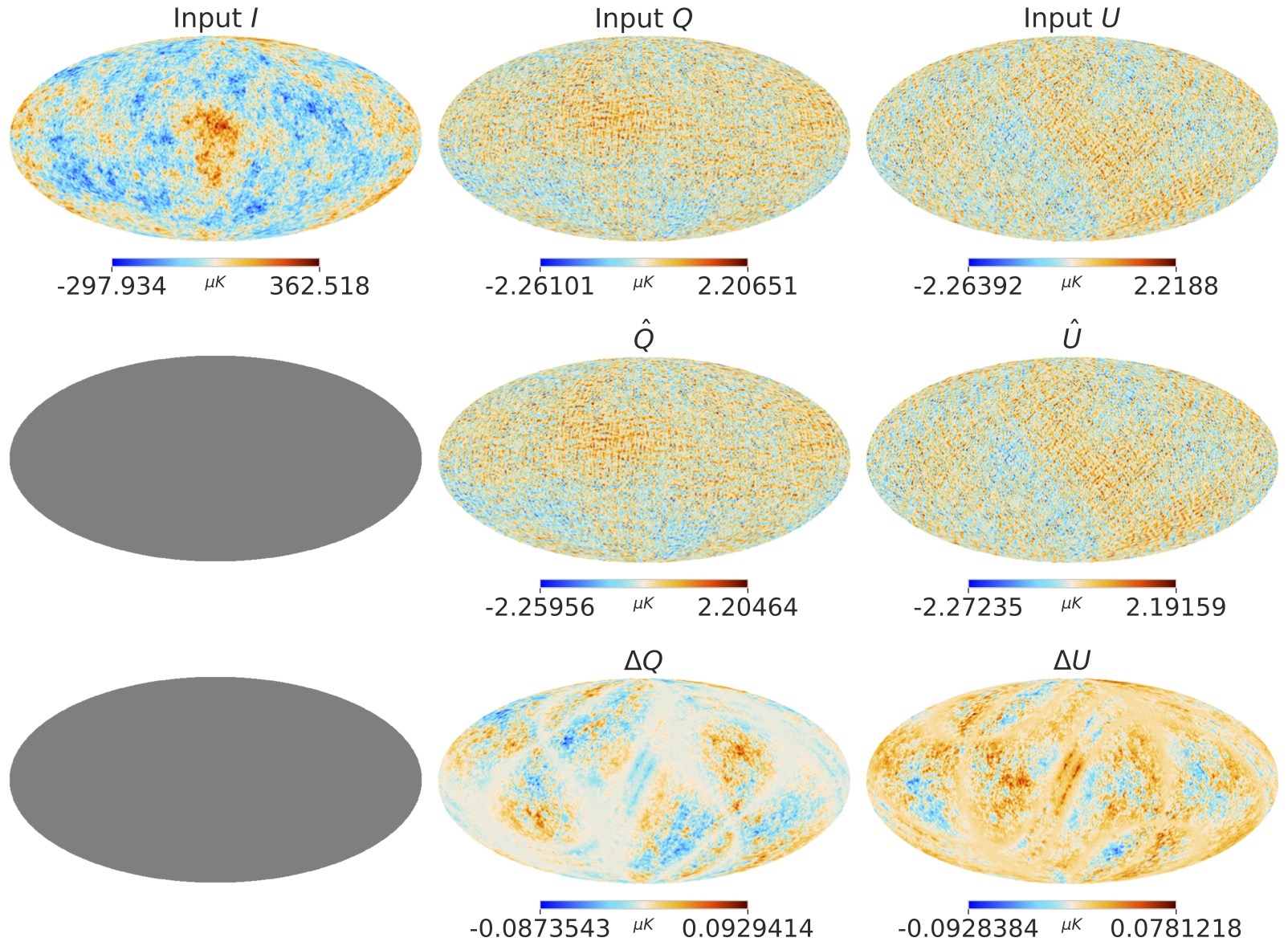
    }
    \caption[Estimated CMB maps and residual maps due to the differential gain
    by the $2\times2$ matrix map-making approach.]{ (top panels) Input CMB maps for
    $I$ (left), $Q$ (middle) and $U$ (right). (middle panels) Estimated CMB maps
    for $\hat{Q}$ (middle) and $\hat{U}$ (right) with the 0.1\% differential
    gain systematics. Due to the differential detection, we do not show the
    temperature map and its residual map as well. (bottom panels) Residual maps given
    by the subtraction between the input map and the estimated map $\Delta Q$ (middle)
    and $\Delta U$ (right). The leakage pattern is given by the structure of
    $\h[2](\Omega)$ and temperature anisotropy. }
    \label{fig:gain_res_map_2x2}
\end{figure}

We assume that a single pair detectors-\texttt{T/B} on the boresight have the gain
offset: $g_{\texttt{T}}=0.001$ and $g_{\texttt{B}}=0$, i.e., $\Delta g=0.001$ (0.1\%)
in \cref{eq:diff_gain}. For the map-making equation, we chose $2\times2$ type
that is shown in \cref{eq:2x2mapmaking}. The input CMB maps ($I$, $Q$ and $U$), estimated
CMB maps ($\hat{Q}$ and $\hat{U}$), and residual maps ($\Delta Q$ and $\Delta U$)
given by the subtraction of the input map from the estimated map are shown in
\cref{fig:gain_res_map_2x2}. Due to the differential gain systematics, we can
see the pattern of temperature-to-polarization leakage, denoted as $T \to P$, in
the residual map. The residual is dominated by the leakage, the structure of the
map is given by the multiplication between the temperature anisotropy and
$\h[2]$ (see \cref{fig:spin-n0_xlink_maps}) so in the case that the real part and
imaginary part of the $\h[2]$ are small, the leakage is suppressed.

\begin{figure}[h]
    \centering
    \includegraphics[width=0.49\columnwidth]{
        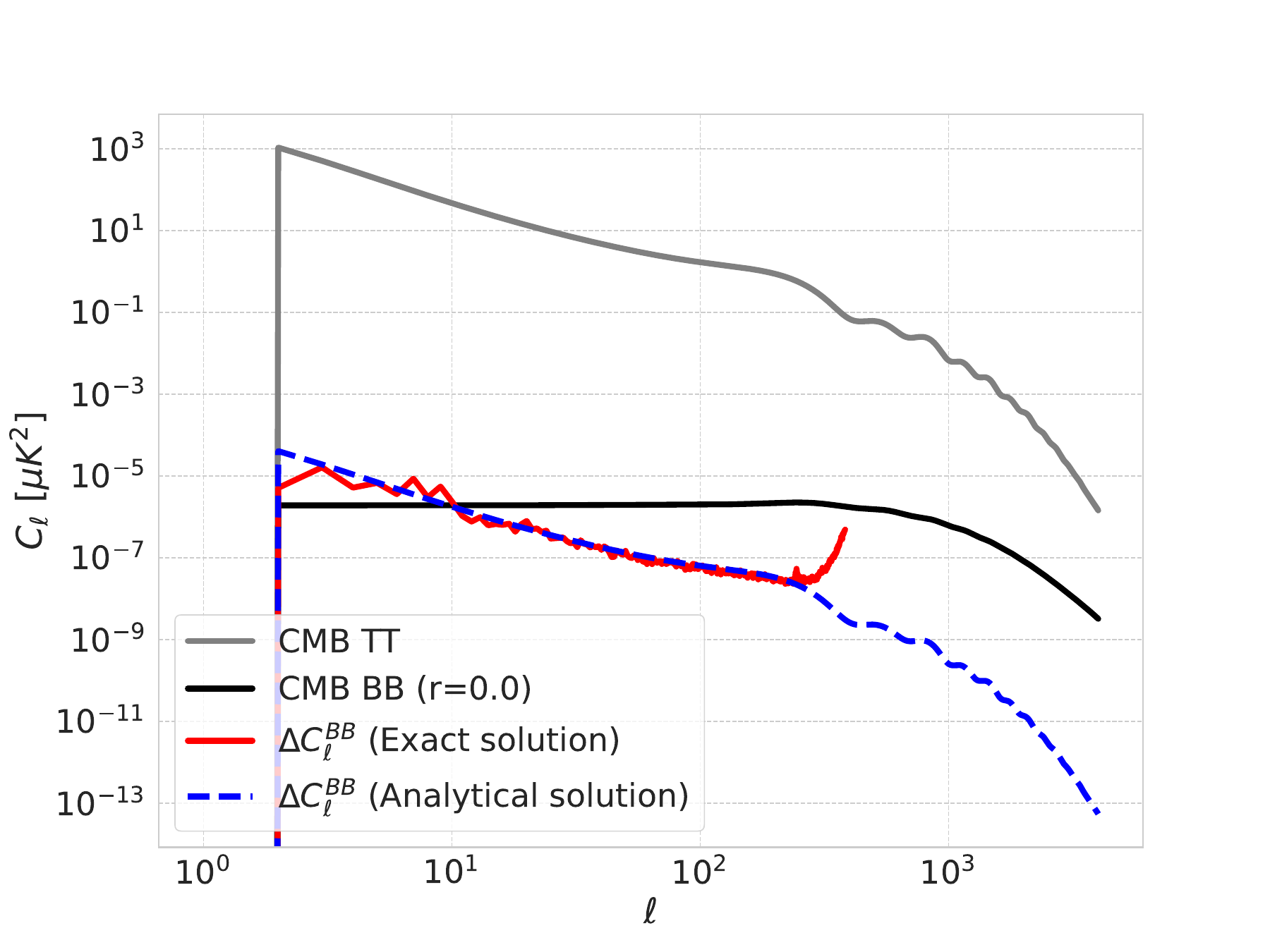
    }
    \includegraphics[width=0.49\columnwidth]{
        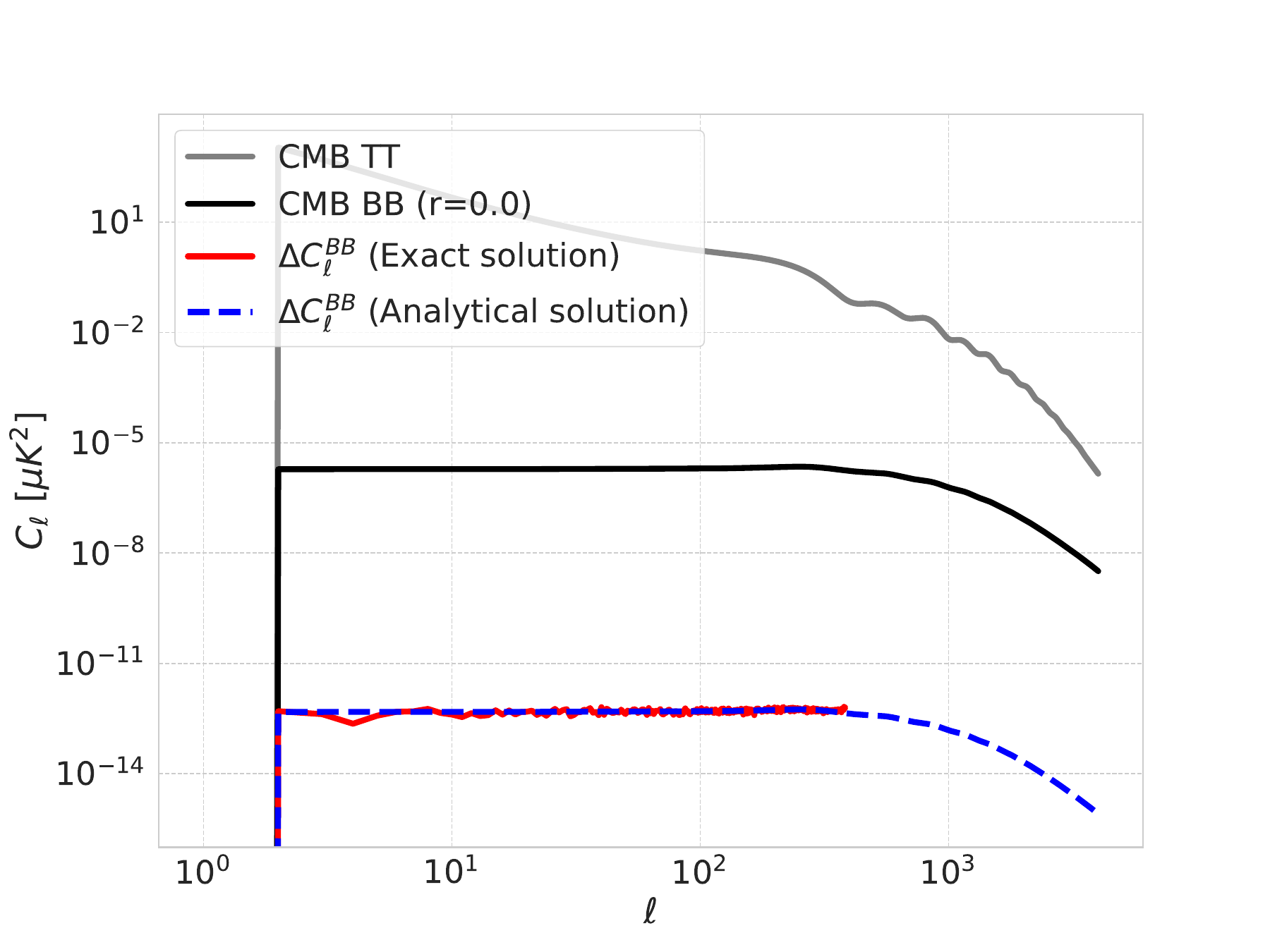
    }
    \caption[Systematic power spectra due to the 0.1\% differential gain systematics.]{Systematic
    power spectra due to the 0.1\% differential gain systematics. (left)
    $\Delta C_{\ell}^{BB}$ by the $2\times2$ matrix map-making approach (\cref{eq:2x2mapmaking}).
    The red solid line is the systematic power spectrum given by \SBM, and the
    blue dashed line is the analytical estimation given by \cref{eq:gain_cl_3x3}.
    The boosting in high-$\ell$ region is due to the deconvolution of the beam's
    transfer function. (middle) $\Delta C_{\ell}^{BB}$ by the $3\times3$ matrix
    map-making approach (\cref{eq:3x3mapmaking_gain}). The red solid line is the
    systematic power spectrum, and the blue dashed line is the analytical
    estimation. The gray/black solid line show the fiducial $C_{\ell}^{TT}$ and $C
    _{\ell}^{TT}$ power spectrum of CMB, respectively. }
    \label{fig:gain_bore_cl}
\end{figure}

The spherical harmonic expansion of the residual maps gives us the systematics power
spectrum, i.e., $\Delta C_{\ell}^{BB}$ which is shown in \cref{fig:gain_bore_cl}
(left) by the solid red line. The boosting where the high-$\ell$ region is made by
the deconvolution of the beam's transfer function. The dashed blue line is given
by the analytical estimation that is given by the following transfer function to
describe $T \to B$ (temperature-to-$B$ mode) leakage \cite{OptimalScan,Hu2003}
\begin{align}
    \Delta C_{\ell}^{BB}= \frac{1}{2}C_{\ell}^{TT}\hmean[2] \Delta g^{2},\label{eq:gain_trans}
\end{align}
where $\hmean[2]$ is the mean value of the $\h[2]$ over the sky. Since the red and
blue line have a good agreement, the simulated result by the \SBM is consistent
with the analytical estimation. We estimated the bias on $r$ i.e.\ $\Delta r$ by
using the likelihood function described in \cref{apd:delta_r}. A maximum multipole
$\ell_{\rm max}$ for the likelihood function is set to $\ell_{\rm max}=191$, and
estimated $\Delta r=0.0023$. The likelihood function that gives this result is shown
in \cref{fig:likelihood_diff_gain}. This value is larger than the \LiteBIRD
mission requirement for total error ($\delta r<0.001$) and is nearly 1000 times
larger than $\Delta r \simeq 10^{-6}$ --- the error budget assigned to
individual systematic effects in ref.~\cite{PTEP2023}, which is a difficult
value to achieve for primordial $B$-mode observations.

\begin{figure}[h]
    \centering
    \includegraphics[width=1.0\columnwidth]{
        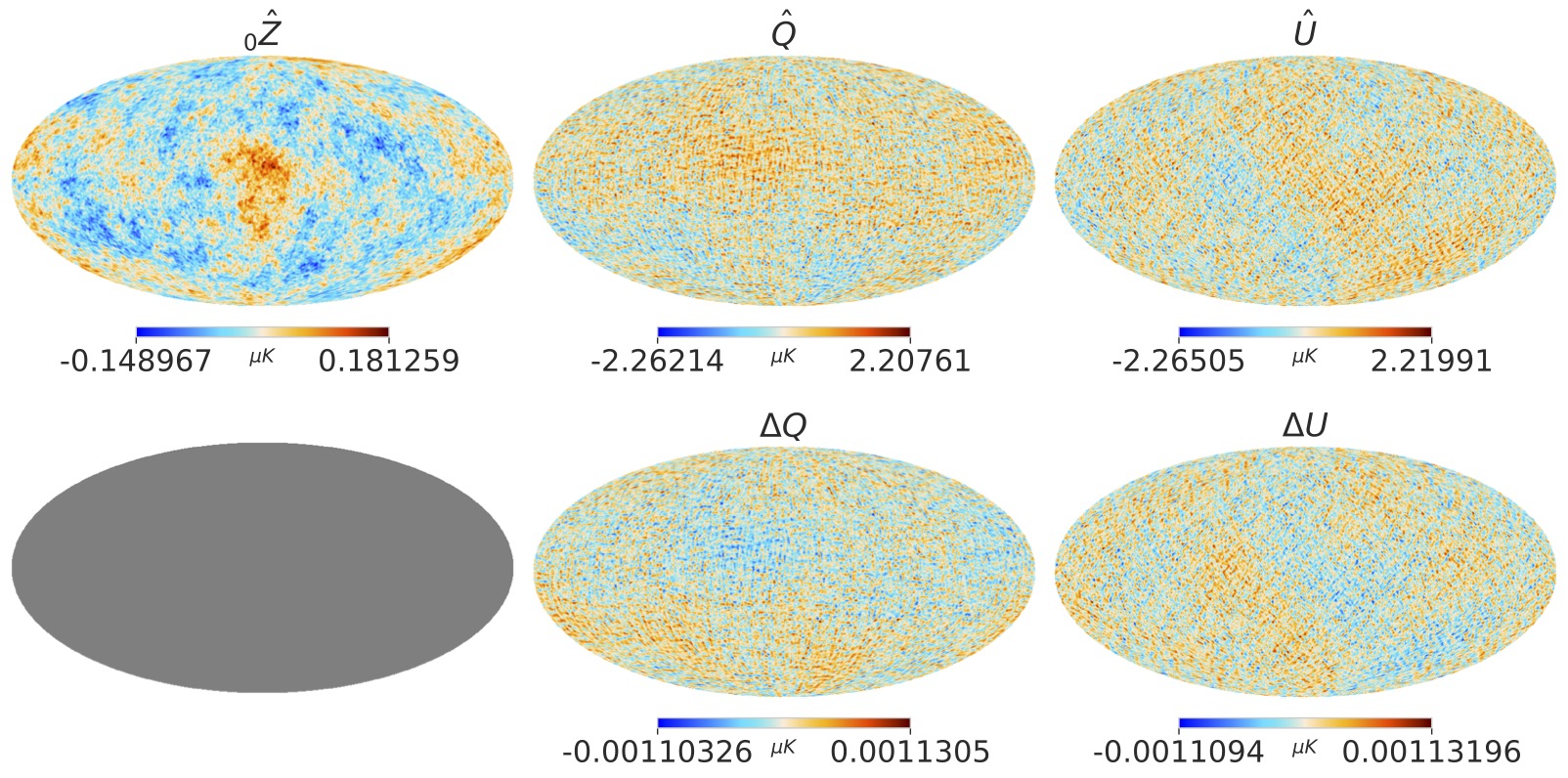
    }
    \caption[Estimated CMB maps and residual maps due to the differential gain
    by the $3\times3$ matrix map-making approach.]{ (top panels) Estimated systematics
    maps and CMB polarizations for ${}_{0}\hat{Z}$ (left), $\hat{Q}$ (middle)
    and $\hat{U}$ by using $3\times 3$ matrix map-making approach in
    \cref{eq:3Mg}. (bottom panels) Residual maps, $\Delta Q$ (middle) and
    $\Delta U$ (right) due to the 0.1\% differential gain systematics.}
    \label{fig:gain_res_map_3x3}
\end{figure}

To manage this huge systematic bias on $r$, we can consider the mitigation by the
$3\times3$ matrix approach as
\begin{align}
    \ab (\mqty{{}_{0}\hat{Z} \\ \hat{P}\\ \hat{P}^*}) ={}_{3}M_{\rm g}^{-1}\ab (\mqty{ \Dd[0]_{\rm g} \\ \frac{1}{2}\Dd[2]_{\rm g} \\ \frac{1}{2}\Dd[-2]_{\rm g} }),\label{eq:3x3mapmaking_gain}        \\
    {}_{3}M_{\rm g}= \ab (\mqty{ 1 & \frac{1}{2}\h[2] & \frac{1}{2}\h[-2] \\ \frac{1}{2}\h[-2] & \frac{1}{4} & \frac{1}{4}\h[-4] \\ \frac{1}{2}\h[2] & \frac{1}{4}\h[4] & \frac{1}{4} }),\label{eq:3Mg}
\end{align}
where ${}_{0}\hat{Z}$ represents the estimated systematic effect with \spin-0 which
corresponds to the $T \to B$ leakage due to the differential gain offset.
Because we estimate the systematic effect, it no longer contaminates the polarization
signal. \Cref{fig:gain_res_map_3x3} shows the estimated CMB maps and the
residual maps by the $3\times3$ map-making approach in \cref{eq:3Mg}. The leakage
pattern that we observed in \cref{fig:gain_res_map_2x2} is vanished in the $\Delta
Q$ and $\Delta U$ residual maps, instead, the leakage is absorbed into the ${}_{0}
\hat{Z}$. Now, the remained systematics effect is only gain offset to the polarization
maps which just scales the magnitude of the maps without changing the pattern.
It gives us the scaled $C_{\ell}^{BB}$ by the given offset as
$\Delta C_{\ell}^{BB}$.

The systematic power spectrum is shown in \cref{fig:gain_bore_cl} (left) in red solid
line. The leakage is suppressed by the mitigation technique, and the systematic power
spectrum is consistent with the analytical estimation which is given by the transfer
function which is scaling $C_{\ell}^{BB}$ and make $B$ mode-to-$B$ mode leakage ($B
\to B$)
\begin{align}
    \Delta C_{\ell}^{BB}= \frac{1}{4}C_{\ell}^{BB}\Delta g^{2}. \label{eq:gain_cl_3x3}
\end{align}
The temperature leakage is captured by the ${}_{0}\hat{Z}$ and it is no longer depends
on scanning strategy as the transfer function does not have cross-linking term.
And estimated $\Delta r$ is less than $10^{-6}$.

\subsection{Differential pointing}

\begin{figure}[h]
    \centering
    \includegraphics[width=0.40\columnwidth]{
        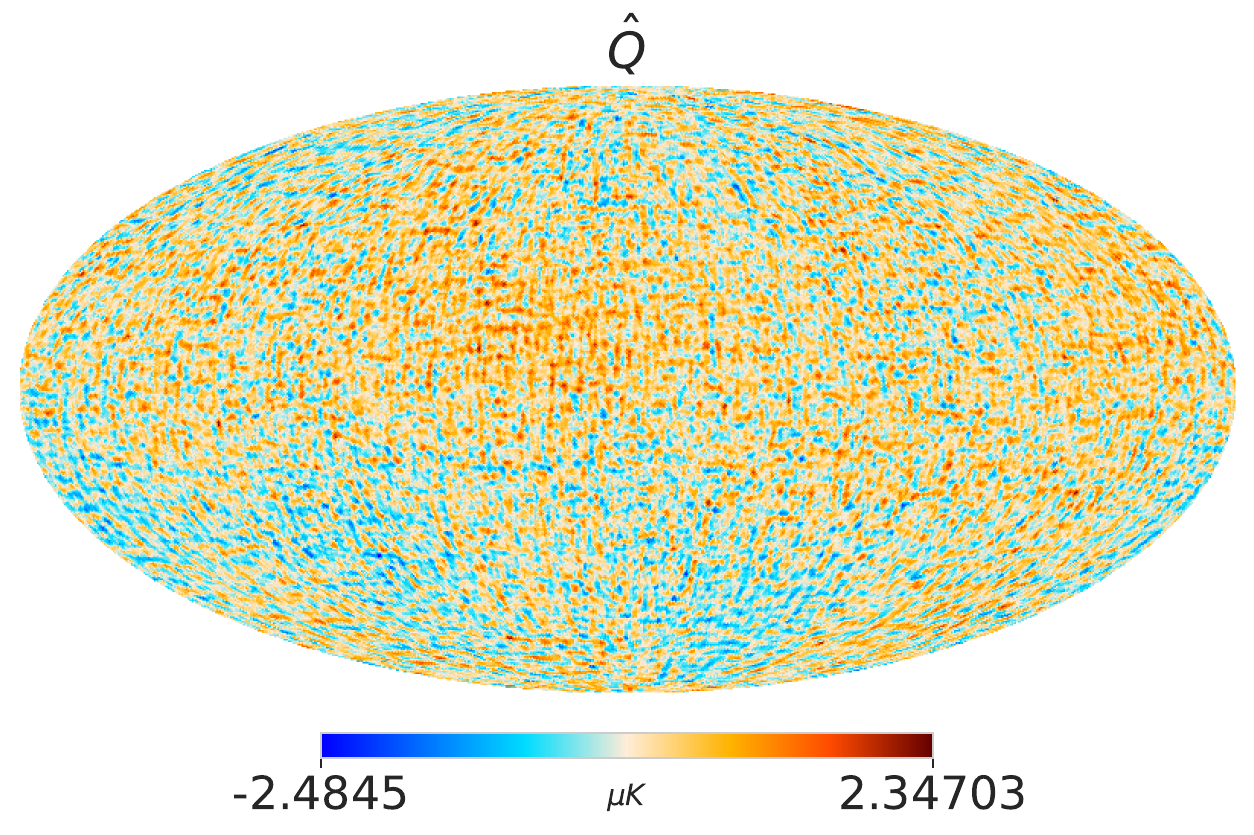
    }
    \hspace{0.05\columnwidth}
    \includegraphics[width=0.40\columnwidth]{
        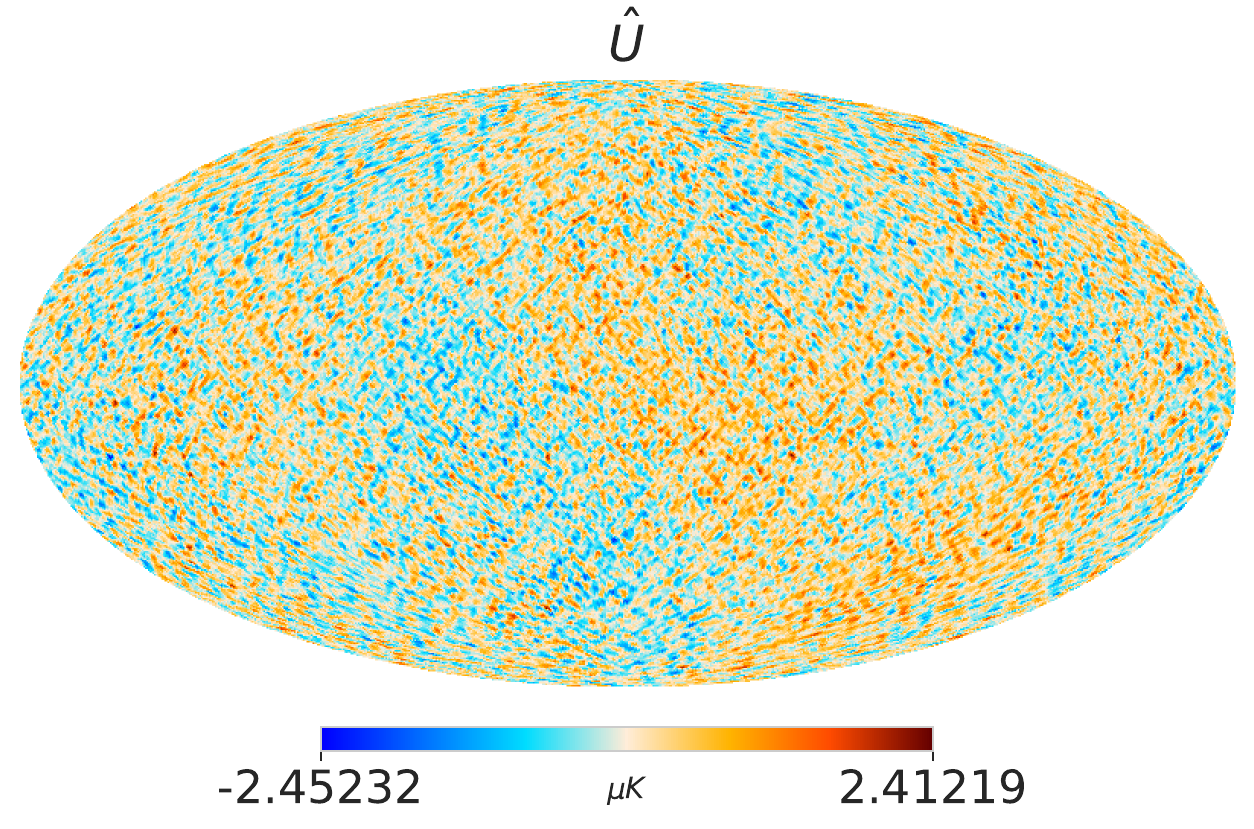
    }
    \\[\baselineskip]
    \includegraphics[width=0.40\columnwidth]{
        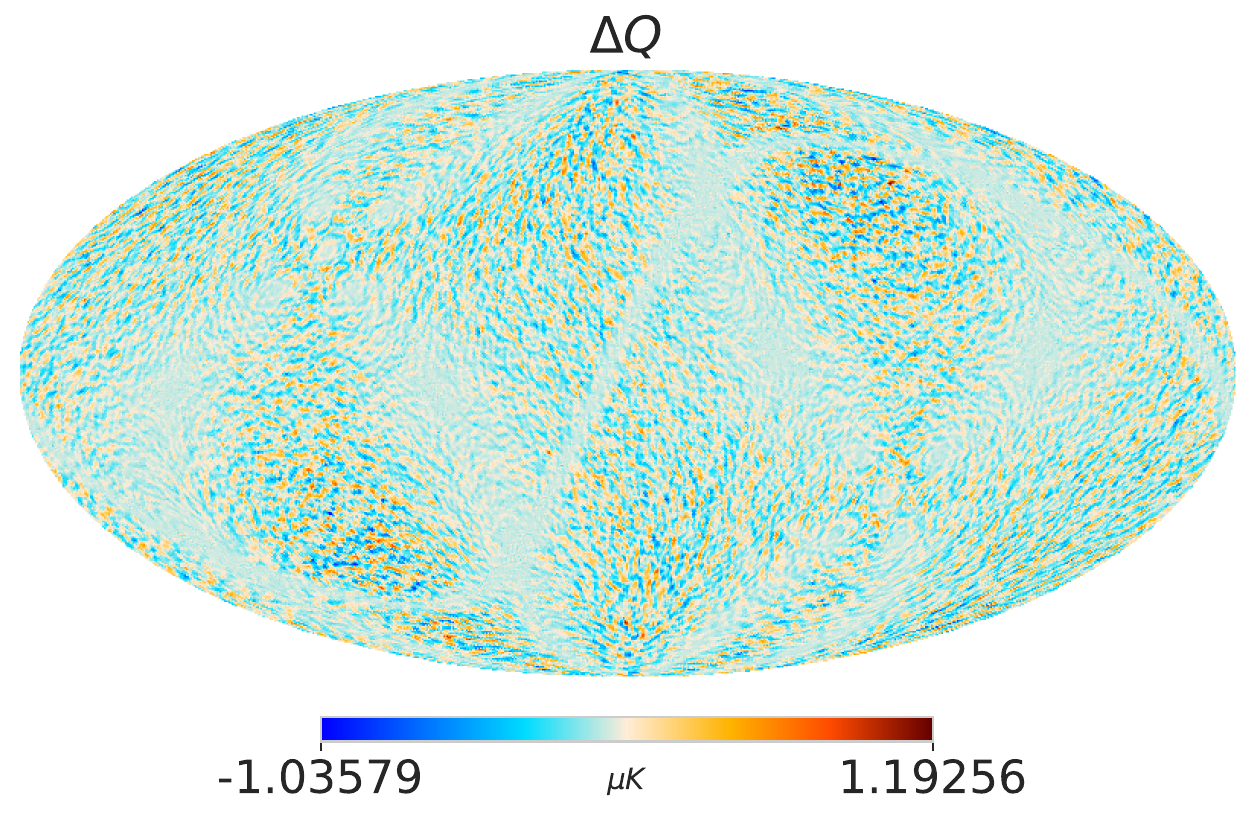
    }
    \hspace{0.05\columnwidth}
    \includegraphics[width=0.40\columnwidth]{
        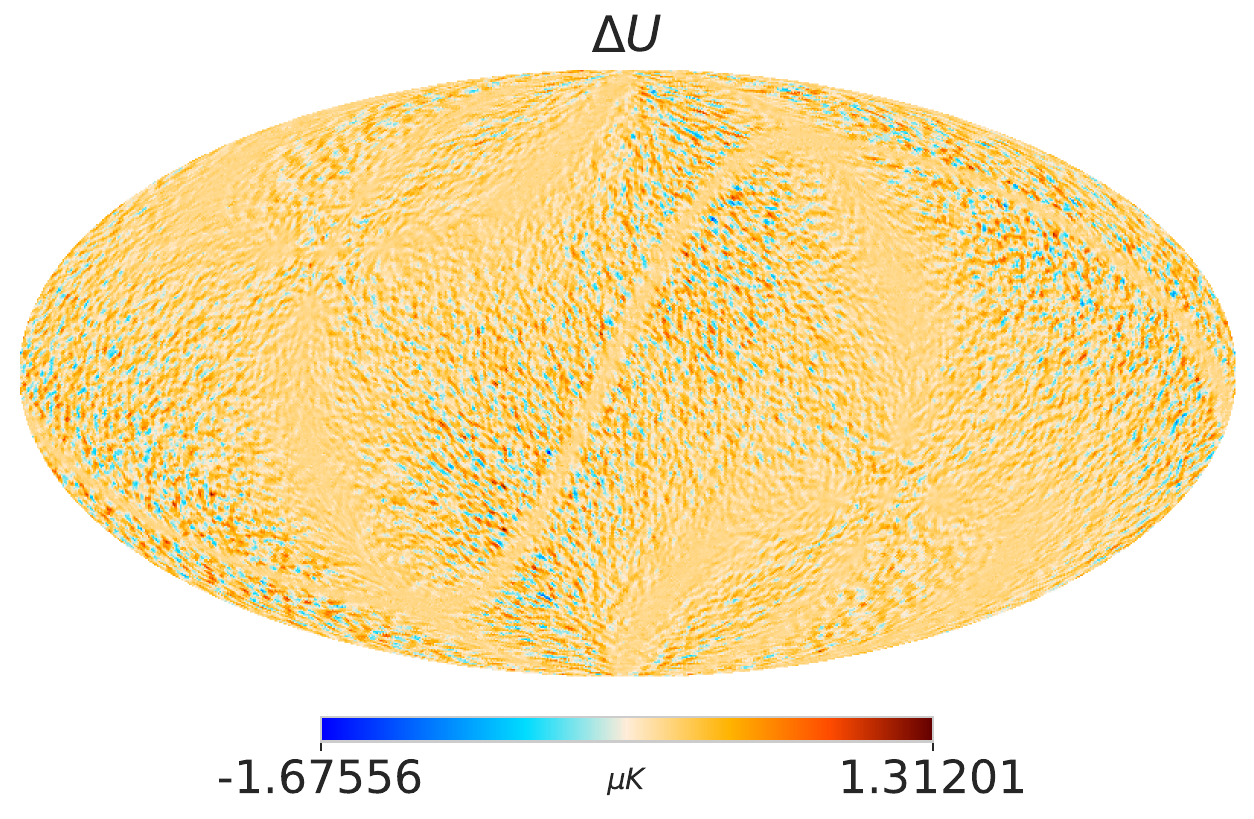
    }
    \caption[Estimated CMB maps and residual maps due to the 1\arcmin differential
    pointing systematics by the $2\times2$ matrix map-making approach.]{Estimated
    $\hat{Q}$ and $\hat{U}$ maps of CMB (top panels), residual maps $\Delta Q$ and
    $\Delta U$ due to the 1\arcmin differential pointing systematics (bottom panels)
    by the $2\times2$ matrix map-making approach. }
    \label{fig:diff_pnt_maps_2x2}
\end{figure}

\begin{figure}[h]
    \centering
    \includegraphics[width=0.32\columnwidth]{
        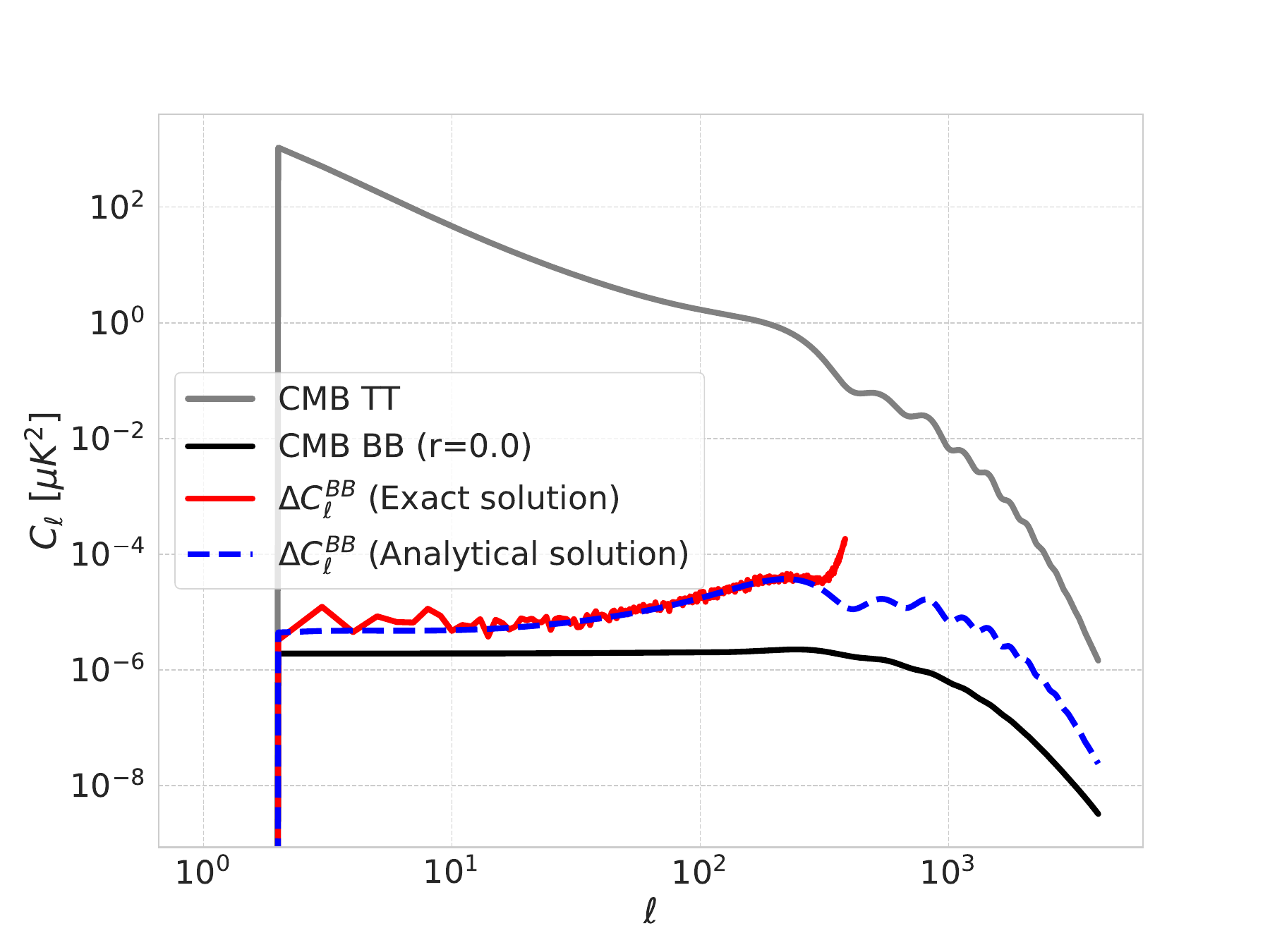
    }
    \includegraphics[width=0.32\columnwidth]{
        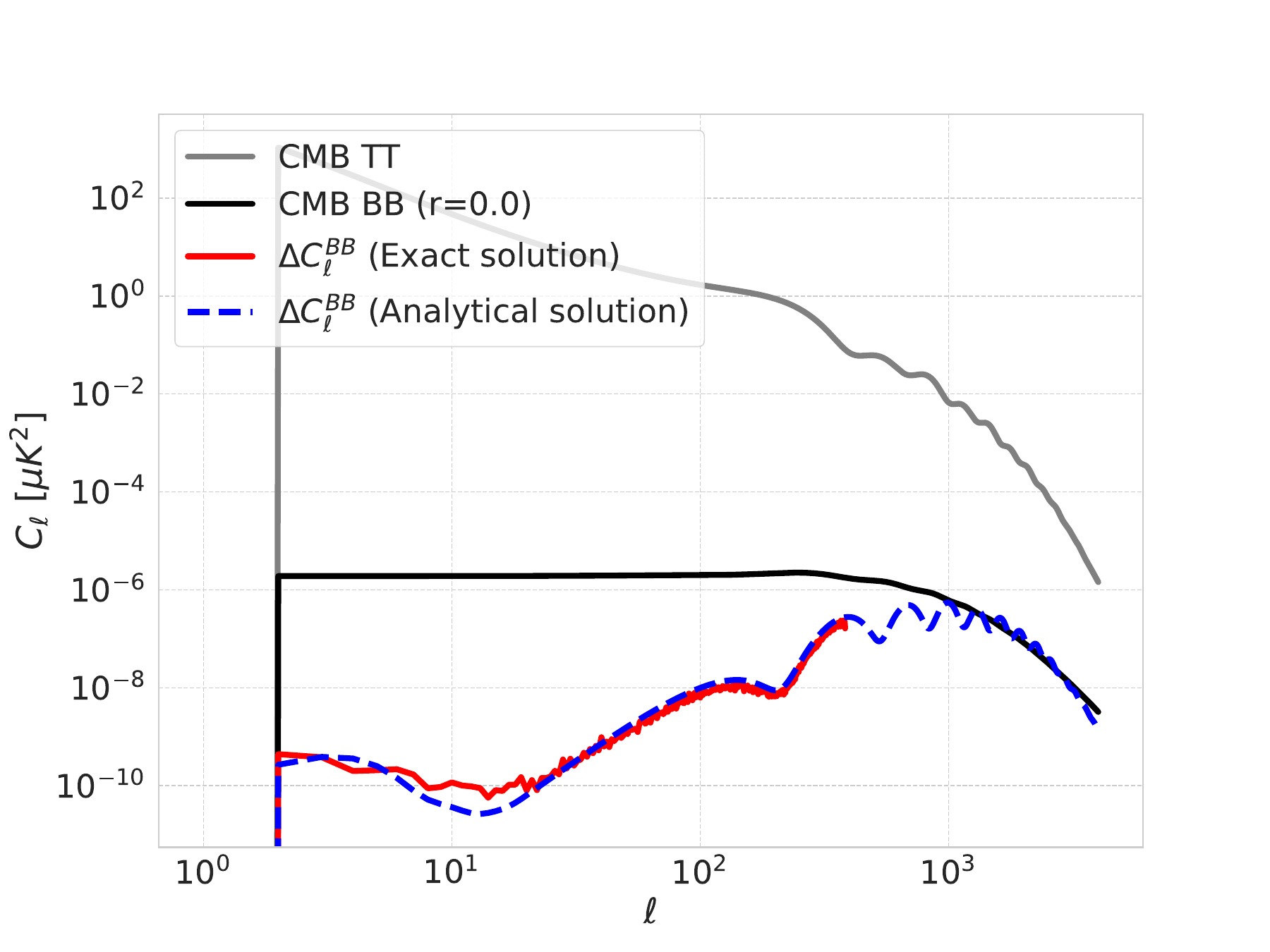
    }
    \includegraphics[width=0.32\columnwidth]{
        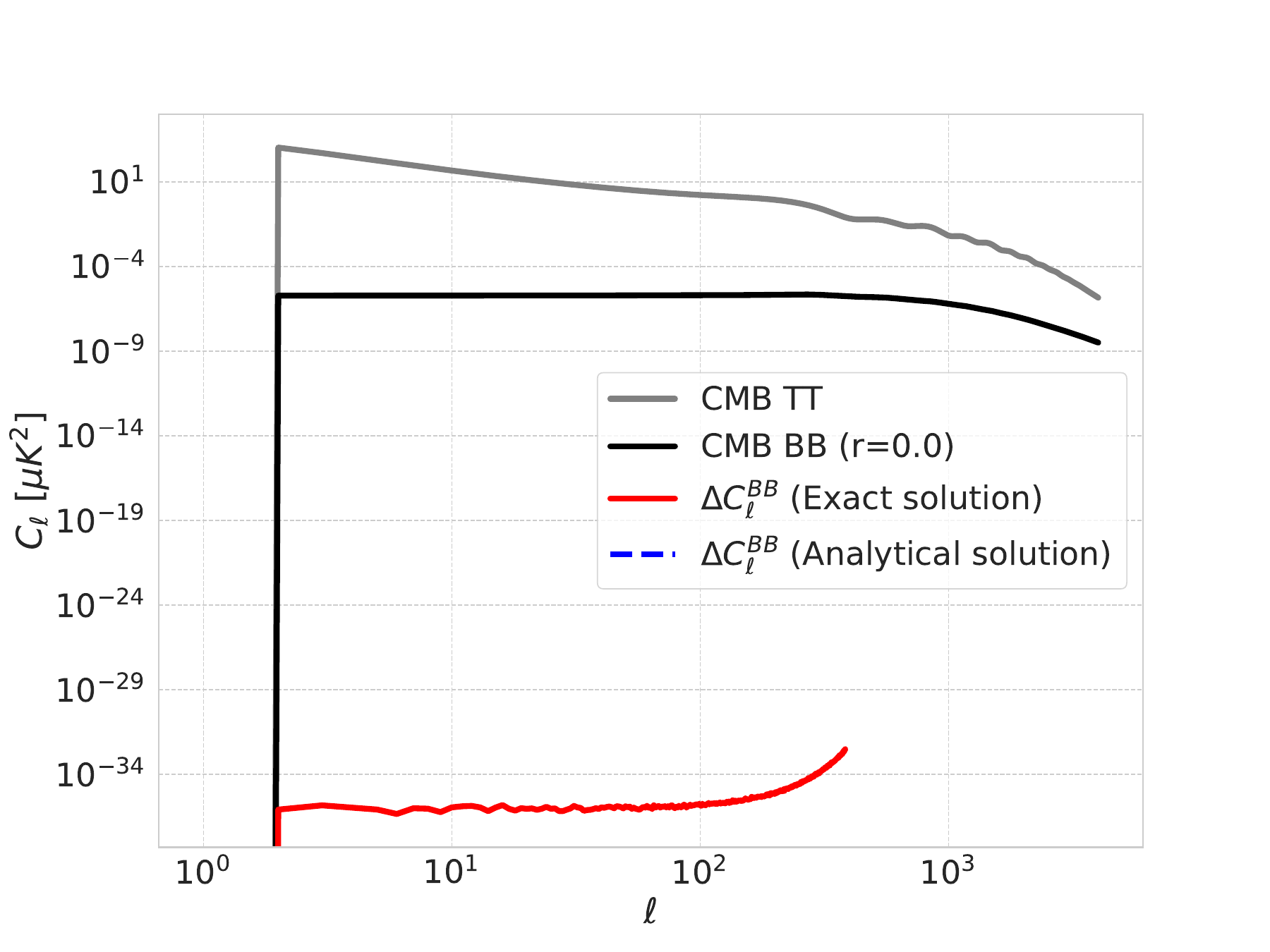
    }
    \caption[Systematic power spectra due to the 1\arcmin differential pointing systematics.
    ]{Systematic power spectra due to the 1\arcmin differential pointing systematics.
    (left) The red solid line is the systematic power spectrum by the $2\times2$
    matrix map-making approach (\cref{eq:2x2mapmaking}). The blue dashed line is
    the analytical estimation given by \cref{eq:pointing_cl_2x2}. (middle) The
    red solid line is the systematic power spectrum by the $4\times4$ matrix map-making
    approach (\cref{eq:4x4mapmaking_pointing}). The blue dashed line is the analytical
    estimation given by \cref{eq:pointing_cl_4x4}. The gray/black solid line
    show the fiducial CMB $TT$/$BB$ power spectrum, respectively. (right) The
    red solid line is the systematic power spectrum by the $6\times6$ matrix map-making
    approach (\cref{eq:4x4mapmaking_pointing}). The approach can fully mitigate differential
    systematics, so the systematic power spectrum is given by only the
    computational noise. }
    \label{fig:pointing_bore_cl}
\end{figure}

We assume that a single pair detectors-\texttt{T/B} has the pointing offset. We
impose its systematics parameter $(\rho, \chi)=(1^{\prime},0^{\prime})$ and
obtained $\hat{Q}$, $\hat{U}$, $\Delta Q$, and $\Delta U$ maps are shown in \cref{fig:diff_pnt_maps_2x2}.
These maps are simulated by the $2\times2$ map-making approach defined by
\cref{eq:2x2mapmaking} and its systematic power spectrum is shown in
\cref{fig:pointing_bore_cl} (left) in red. The blue dashed line shows the analytical
estimation given by the transfer function as \cite{OptimalScan,Hu2003} to describe
the $T \to P$ leakage
\begin{align}
    \Delta C_{\ell}^{BB}= \frac{1}{8}\ab(\hmean[1]+\hmean[3])C_{\ell}^{TT}\rho^{2}\ell^{2}. \label{eq:pointing_cl_2x2}
\end{align}
We estimate $\Delta r$ by the same way with the differential gain case, then we
obtain $\Delta r=2.35$ which is quite huge value.

\begin{figure}[h]
    \centering
    \includegraphics[width=0.32\columnwidth]{
        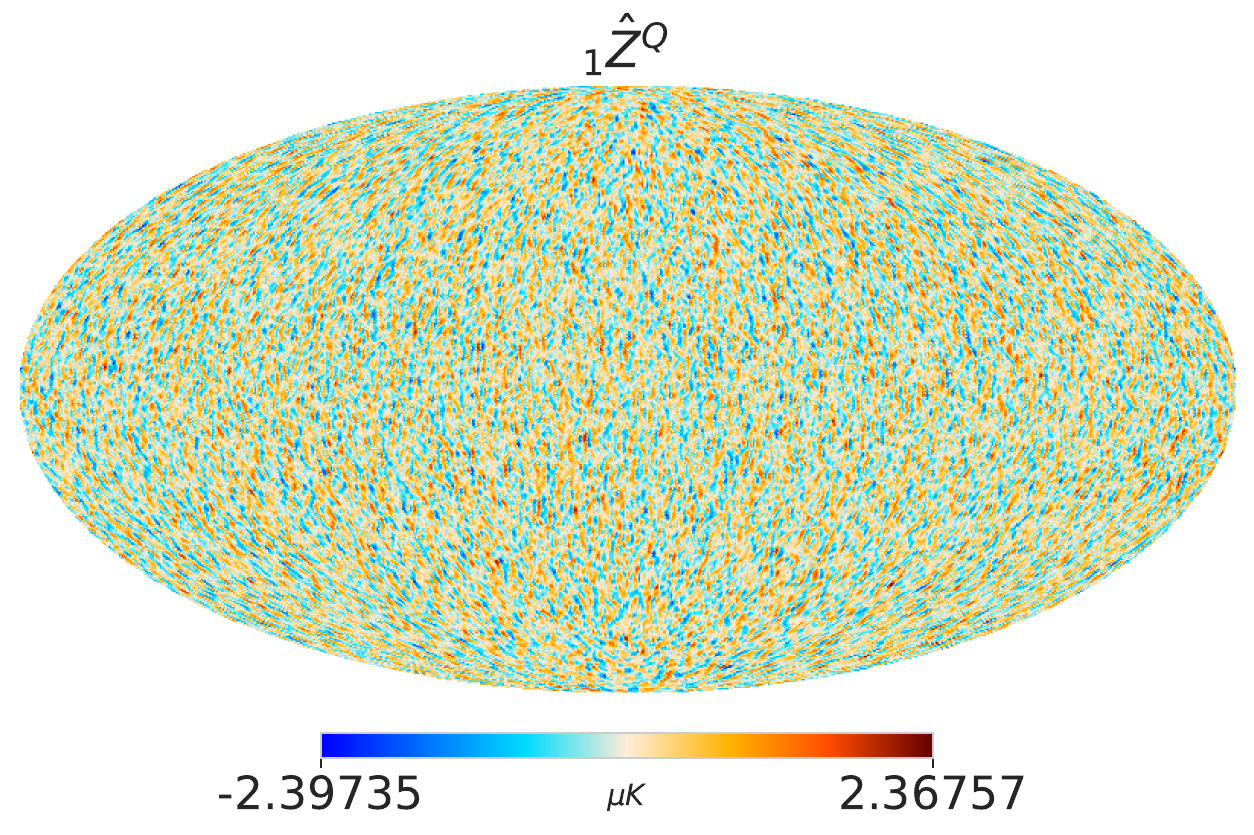
    }
    \includegraphics[width=0.32\columnwidth]{
        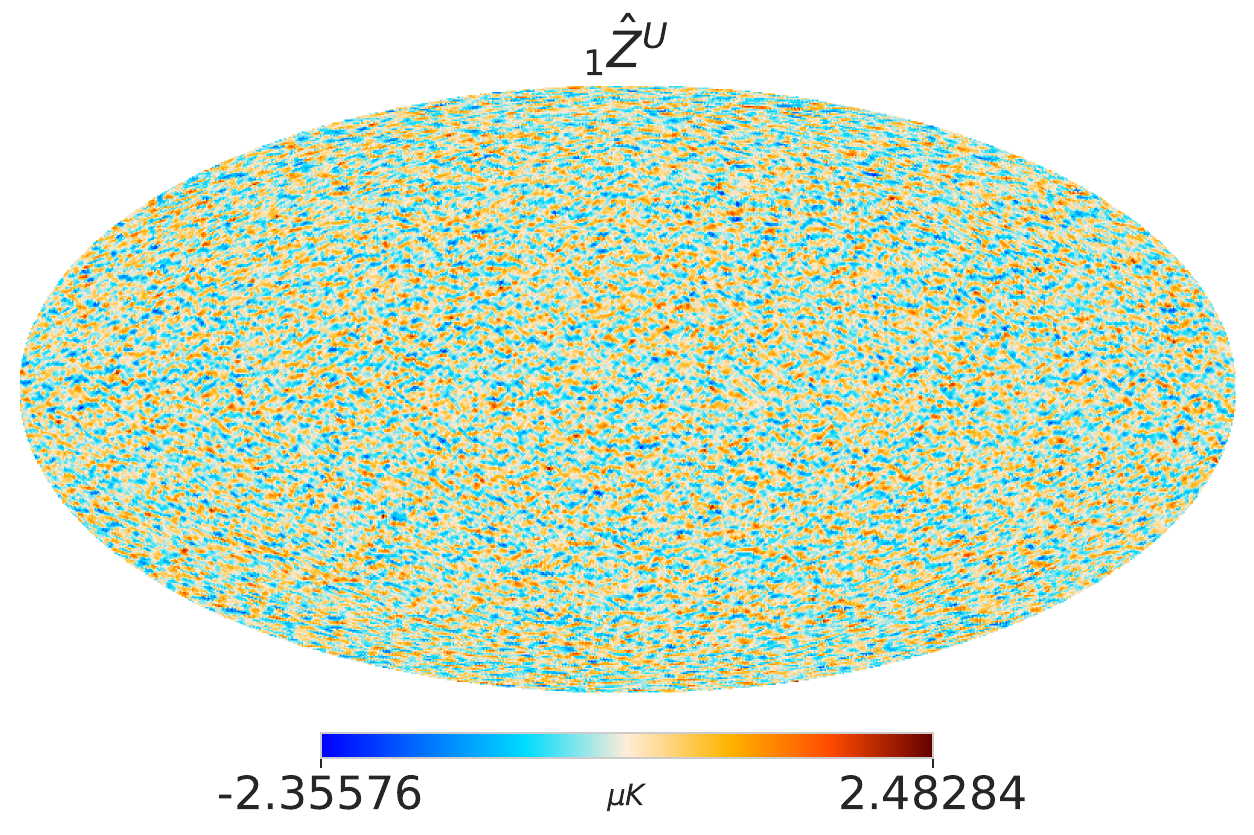
    }
    \includegraphics[width=0.32\columnwidth]{
        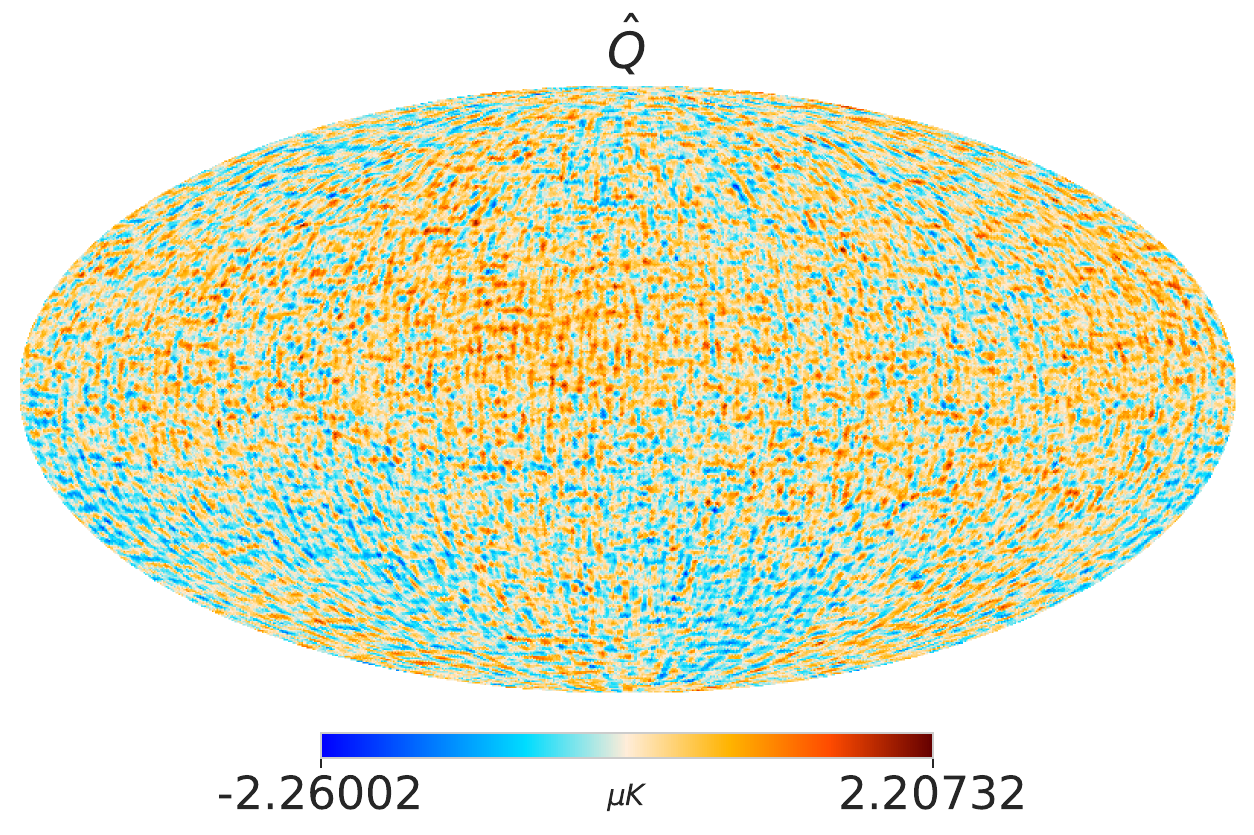
    }
    \\
    \includegraphics[width=0.32\columnwidth]{
        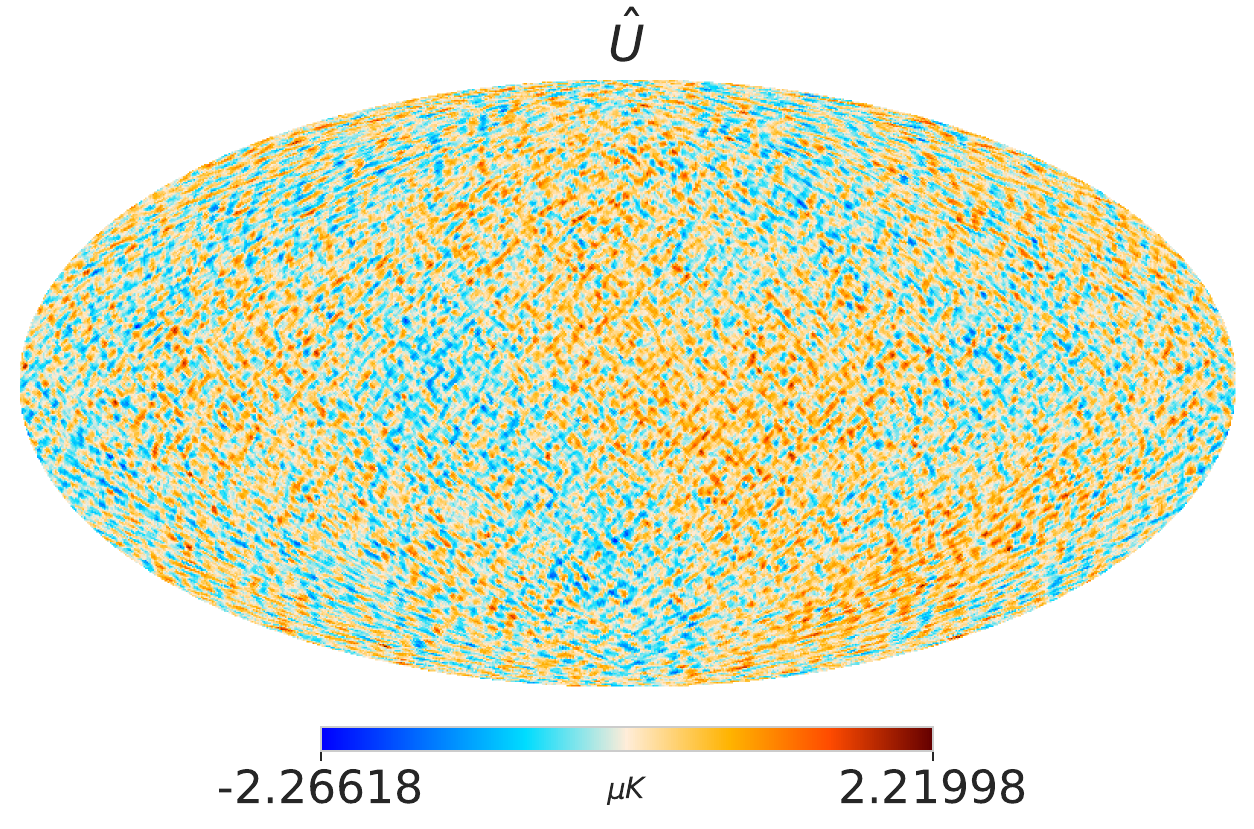
    }
    \includegraphics[width=0.32\columnwidth]{
        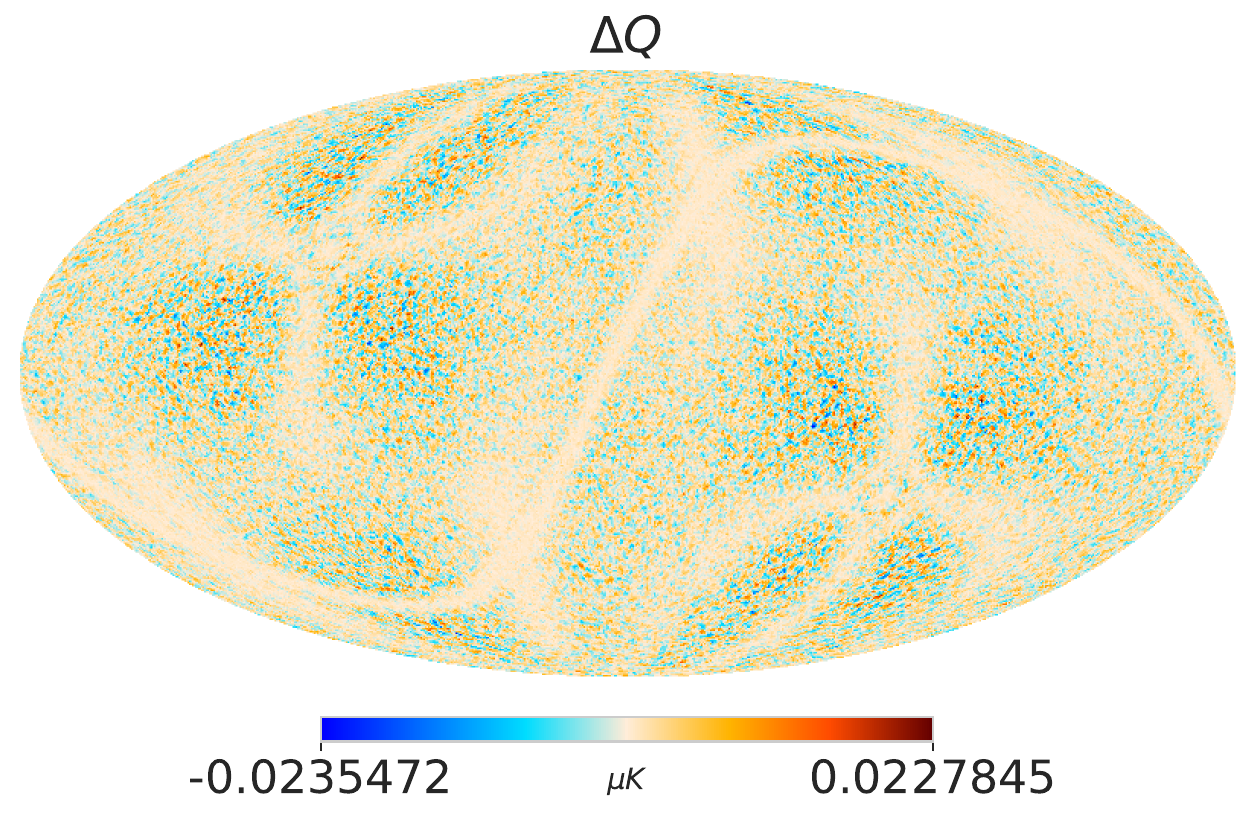
    }
    \includegraphics[width=0.32\columnwidth]{
        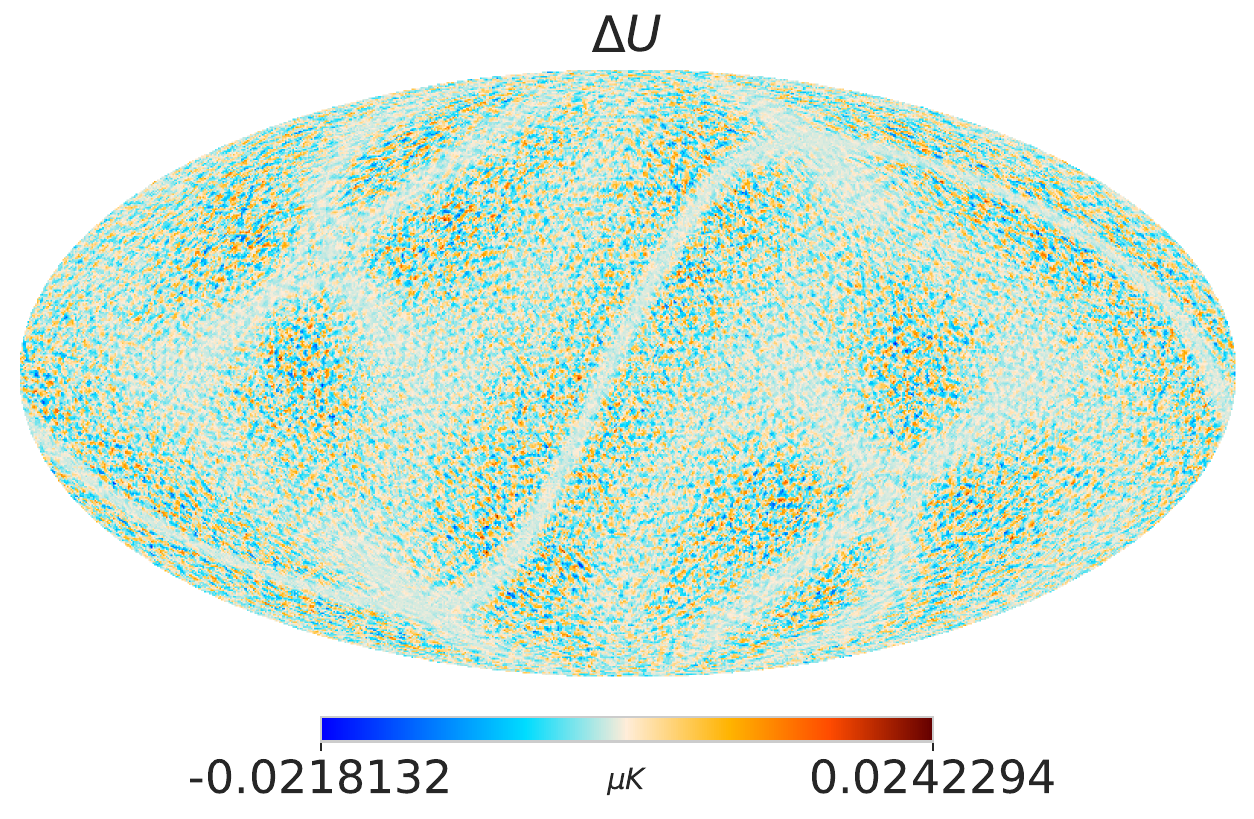
    }
    \caption[Estimated CMB maps and residual maps due to the 1\arcmin differential
    pointing systematics by the $4\times4$ matrix map-making approach.]{Estimated
    CMB maps and residual maps due to the 1\arcmin differential pointing systematics
    by the $4\times4$ matrix map-making approach. It displays $\hZ[1]^{Q}$, $\hZ[
    1]^{U}$, $\hat{Q}$, $\hat{U}$, $\Delta Q$ and $\Delta U$ from top left to
    bottom right. }
    \label{fig:diff_pnt_maps_4x4}
\end{figure}

As we mitigated differential gain systematics by expanding a dimension of the
map-making linear system, we can consider to mitigate the differential pointing systematics
as well. In order to capture temperature gradient-to-polarization leakage which
is expected as a dominant term in \cref{eq:t2p,eq:p2p}, we use following map-maker
\begin{align}
    \ab (\mqty{{}_{1}\hat{Z}\\ {}_{-1}\hat{Z} \\ \hat{P}\\ \hat{P}^* }) ={}_{4}M_{\rm p}^{-1}\ab (\mqty{\frac{1}{2}\Dd[1]_{\rm p} \\ \frac{1}{2}\Dd[-1]_{\rm p} \\ \frac{1}{2}\Dd[2]_{\rm p} \\ \frac{1}{2}\Dd[-2]_{\rm p} }),\label{eq:4x4mapmaking_pointing}                                                                                       \\
    {}_{4}M_{\rm p}= \ab (\mqty{\frac{1}{4} & \frac{1}{2}\h[-1] & \frac{1}{2}\h[2] & \frac{1}{2}\h[-2] \\ \frac{1}{2}\h[1] & \frac{1}{4} & \frac{1}{4}\h[3] & \frac{1}{4}\h[-1] \\ \frac{1}{2}\h[-2] & \frac{1}{4}\h[-3] & \frac{1}{4} & \frac{1}{4}\h[-4] \\ \frac{1}{2}\h[2] & \frac{1}{4}\h[1] & \frac{1}{4}\h[4] & \frac{1}{4}}). \label{eq:4Mp}
\end{align}
\Cref{fig:diff_pnt_maps_4x4} shows $\hat{Q}$ and $\hat{U}$ maps, systematic maps
$\hZ[1]^{Q}$ and $\hZ[1]^{U}$, and residual maps $\Delta Q$ and $\Delta U$ by the
$4\times4$ map-making approach. The leakage pattern is absorbed into the
$\hZ [1]^{Q}$ and $\hZ[1]^{U}$, and the residual maps are free from the temperature
gradient leakage. The systematics power spectrum obtained by
\cref{eq:4x4mapmaking_pointing} is shown in \cref{fig:pointing_bore_cl} (middle).
The blue dashed line is the analytical estimation by the transfer function as
\begin{align}
    \Delta C_{\ell}^{BB}= \frac{1}{4}\ab(\hmean[1]+\hmean[3])C_{\ell}^{EE}\rho^{2}\ell^{2}. \label{eq:pointing_cl_4x4}
\end{align}
Because we capture the systematic field originated by the temperature gradient,
the systematic power spectrum is no longer depends on the temperature and
switches $E$ mode-to-$B$ mode leakage ($E \to B$), as we can see in \cref{eq:pointing_cl_4x4}.
In this case, the $\Delta r$ is estimated as $\Delta r =2.8\times10^{-6}$.

\begin{figure}[h]
    \centering
    \includegraphics[width=0.24\columnwidth]{
        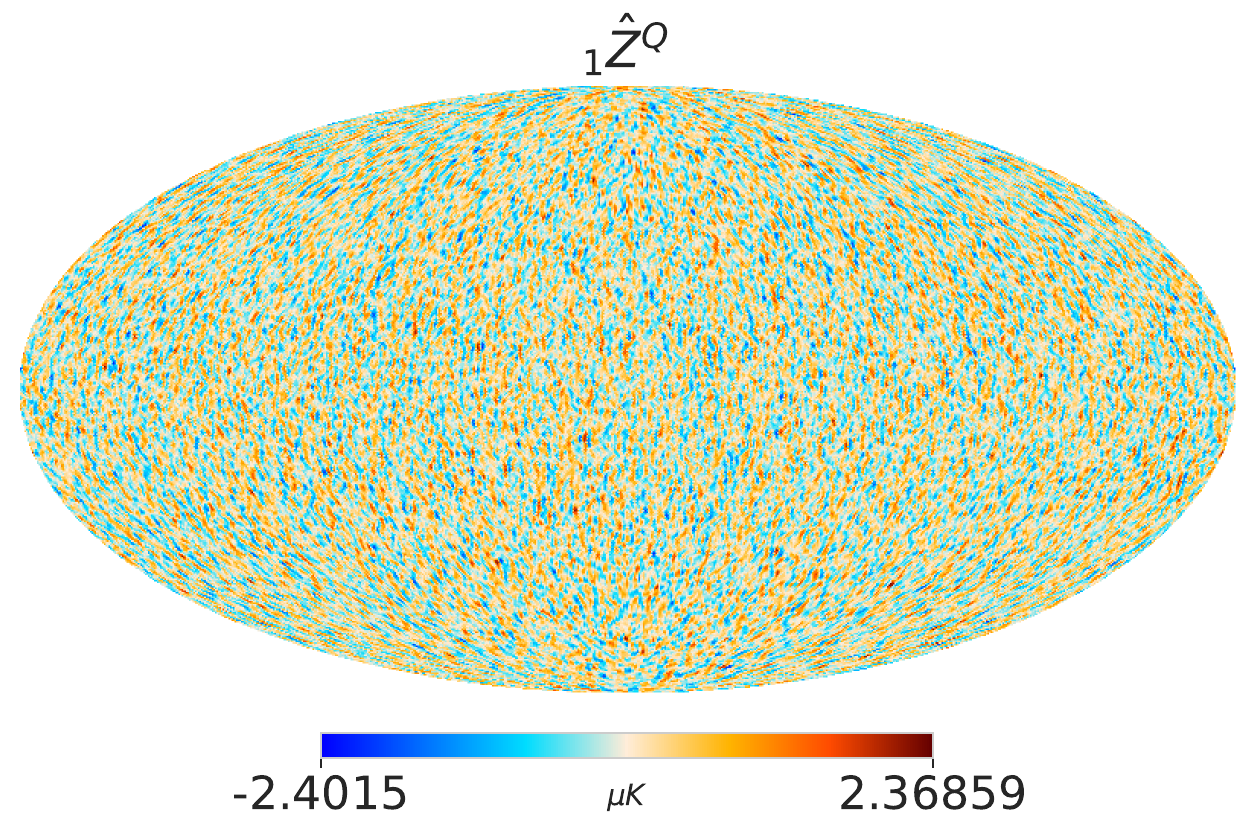
    }
    \includegraphics[width=0.24\columnwidth]{
        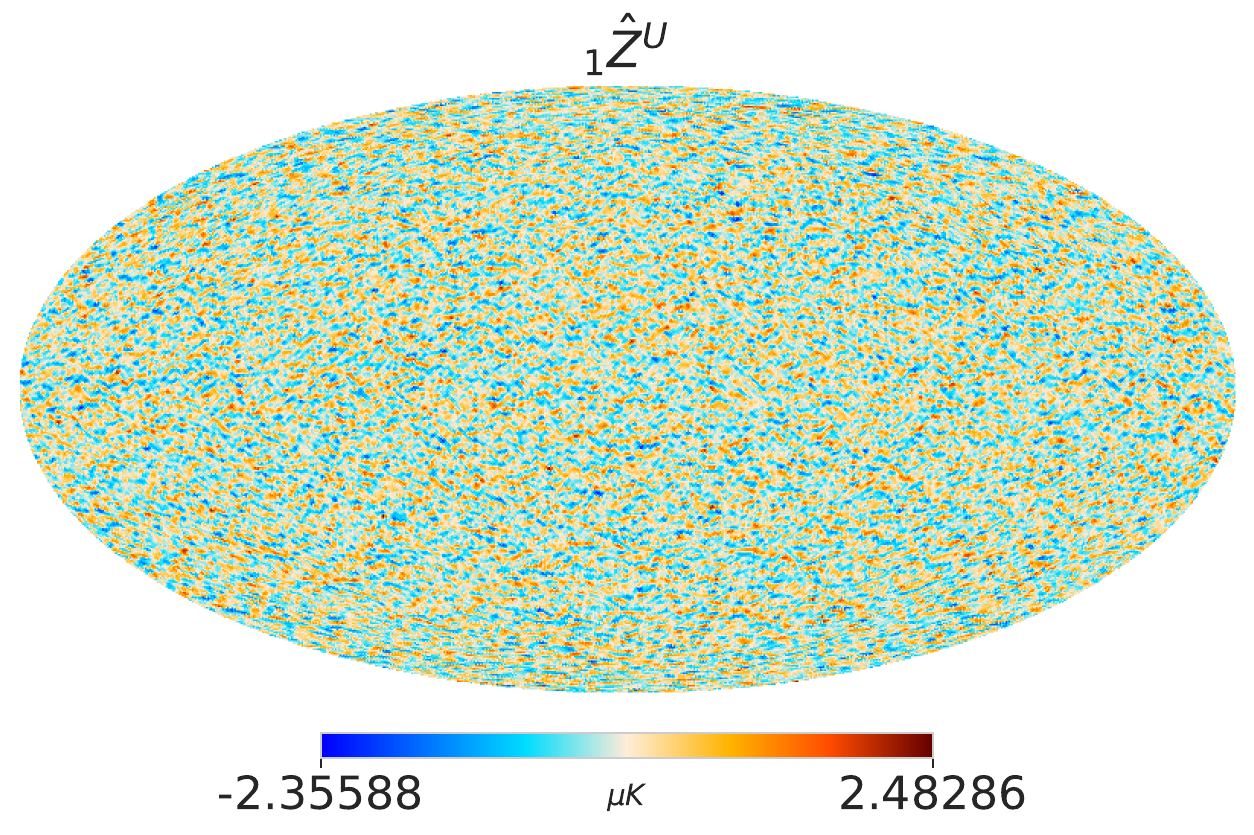
    }
    \includegraphics[width=0.24\columnwidth]{
        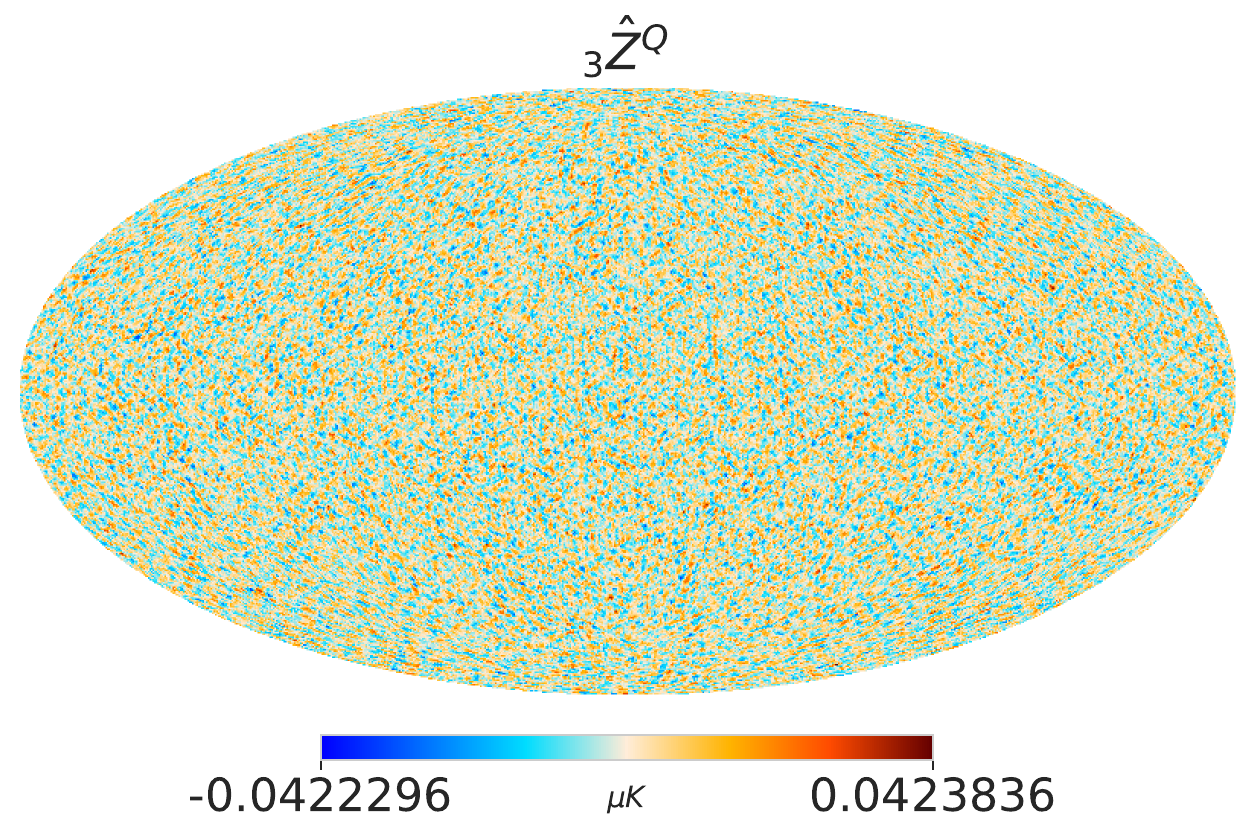
    }
    \includegraphics[width=0.24\columnwidth]{
        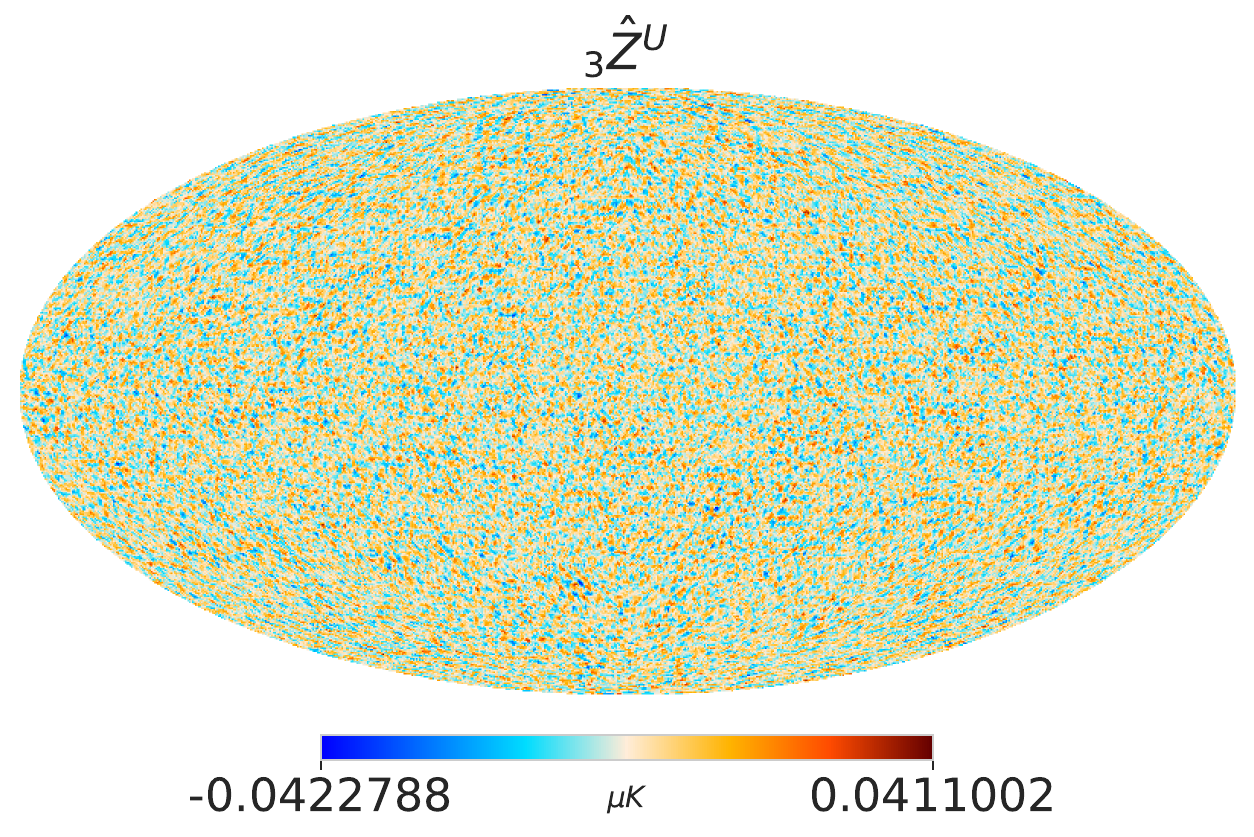
    }
    \\
    \includegraphics[width=0.24\columnwidth]{
        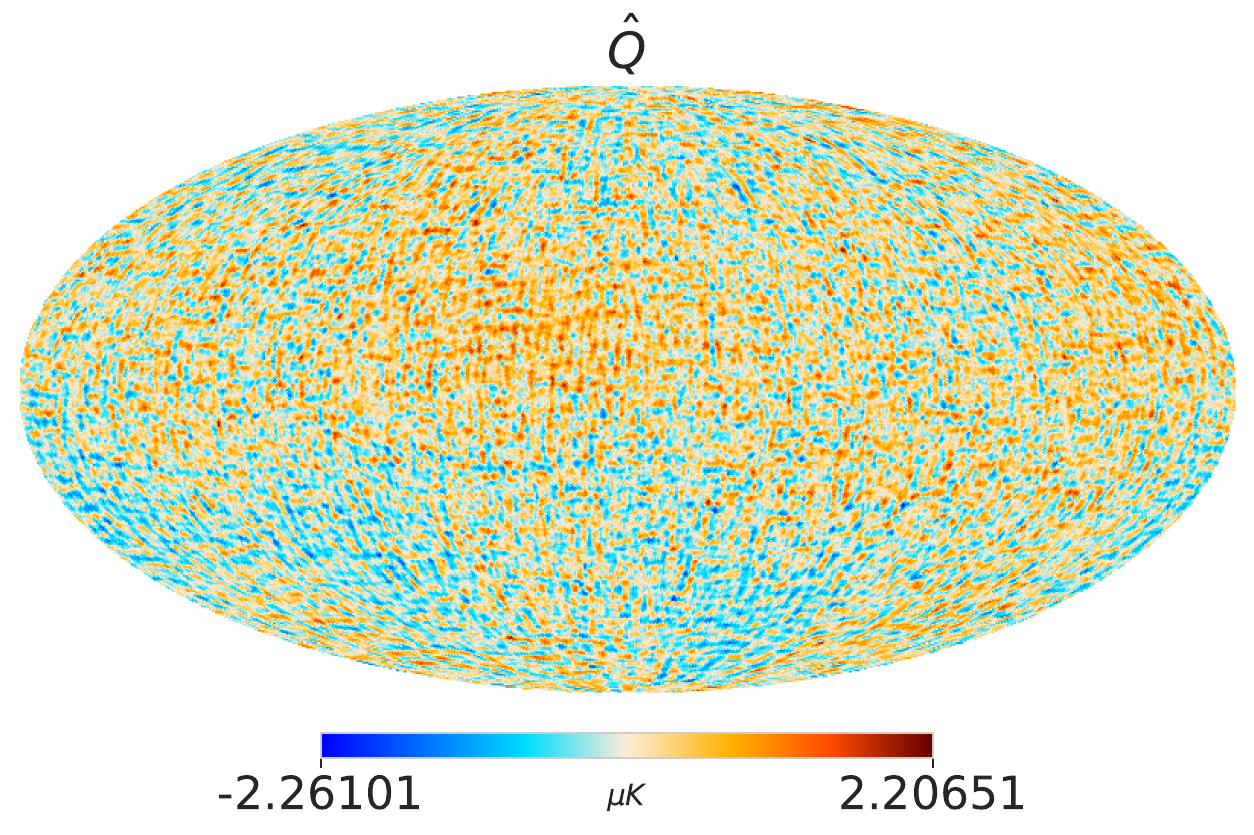
    }
    \includegraphics[width=0.24\columnwidth]{
        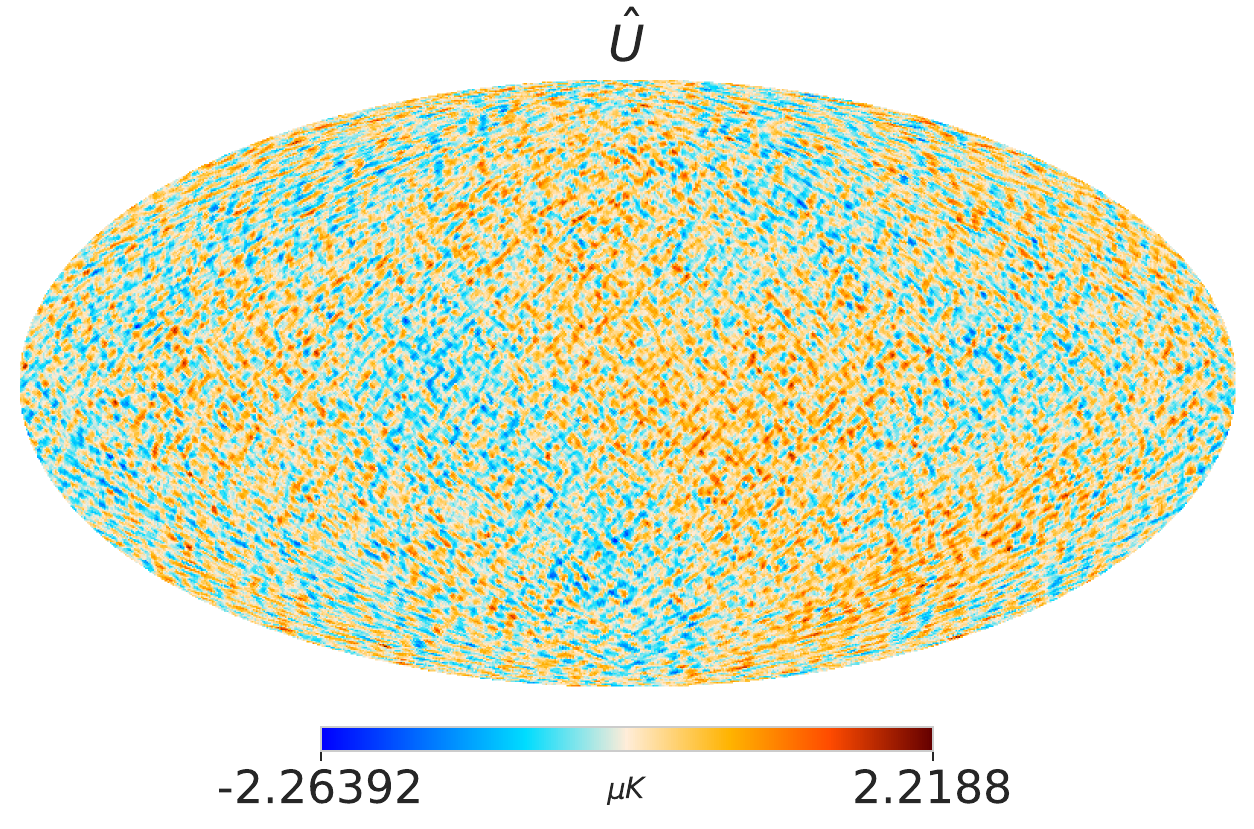
    }
    \includegraphics[width=0.24\columnwidth]{
        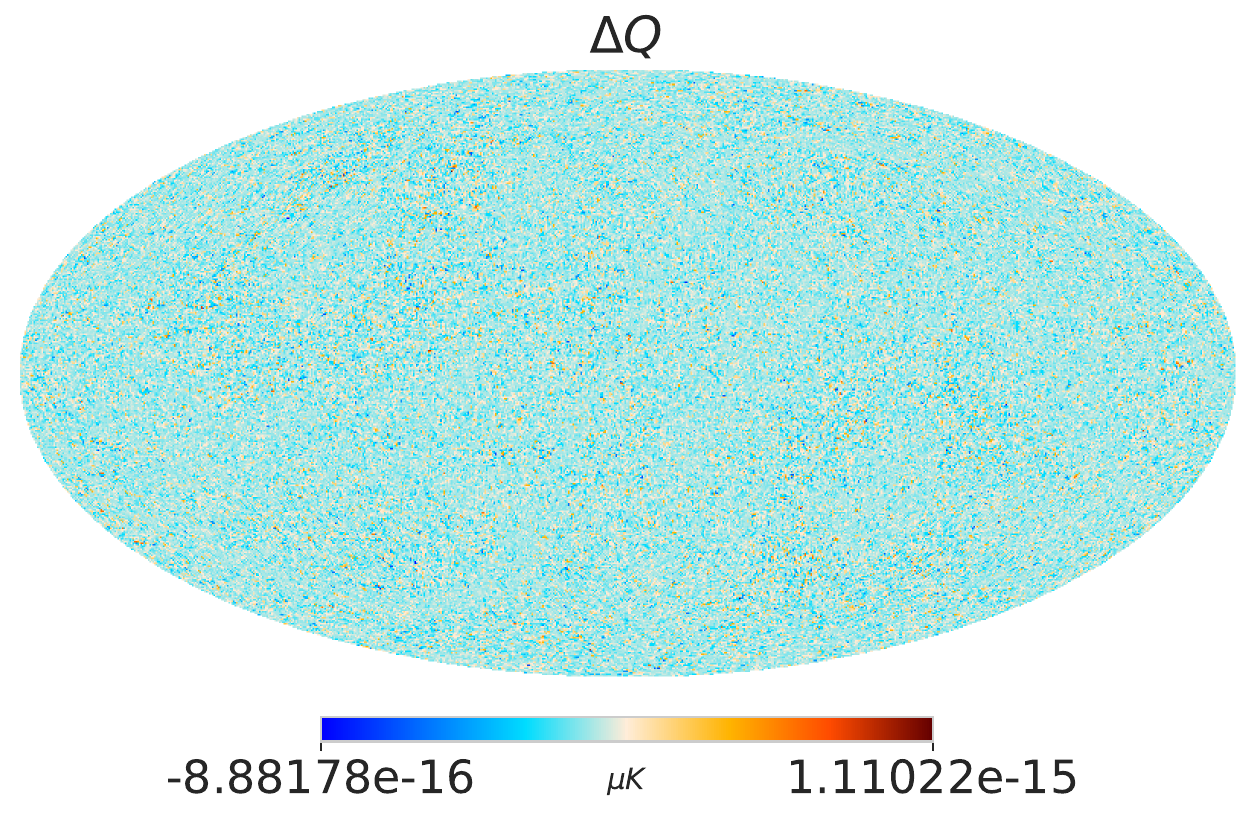
    }
    \includegraphics[width=0.24\columnwidth]{
        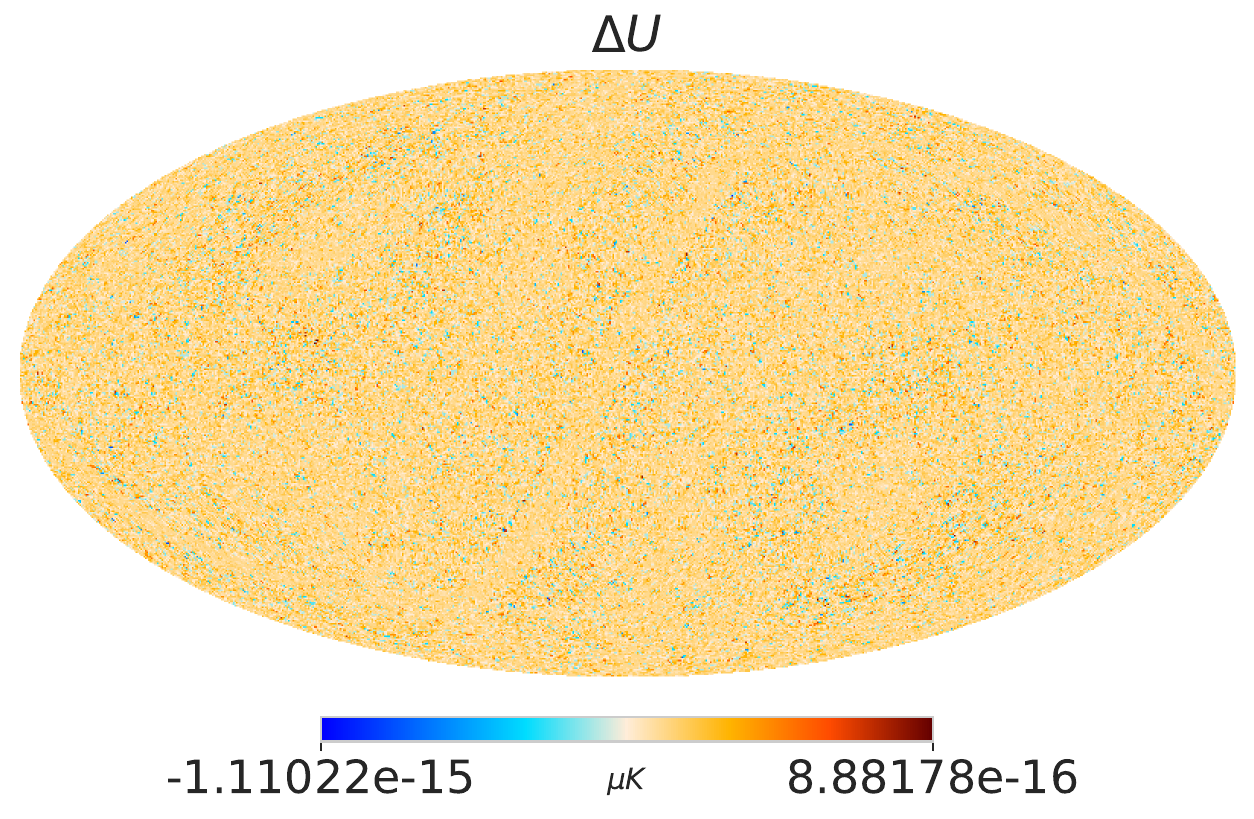
    }
    \caption[Estimated CMB maps and residual maps due to the 1\arcmin differential
    pointing systematics by the $6\times6$ matrix map-making approach.]{Estimated
    CMB maps and residual maps due to the 1\arcmin differential pointing systematics
    by the $6\times6$ matrix map-making approach. It displays $\hZ[1]^{Q}$, $\hZ[
    1]^{U}$, $\hZ[3]^{Q}$, $\hZ[3]^{U}$, $\hat{Q}$, $\hat{U}$, $\Delta Q$ and
    $\Delta U$ from top left to bottom right.}
    \label{fig:diff_pnt_maps_6x6}
\end{figure}

Additionally, if we expand the dimension of the matrix to $6\times6$ to capture polarization
gradient, we can mitigate all the systematics due to the differential pointing
offset. The map-maker is given by
\begin{align}
    \ab (\mqty{{}_{1}\hat{Z}\\ {}_{-1}\hat{Z} \\ \hat{P}\\ \hat{P^*} \\ {}_{3}\hat{Z}\\ {}_{-3}\hat{Z}}) ={}_{6}M_{\rm p}^{-1}\ab (\mqty{\frac{1}{2}\Dd[1]_{\rm p} \\ \frac{1}{2}\Dd[-1]_{\rm p} \\ \frac{1}{2}\Dd[2]_{\rm p} \\ \frac{1}{2}\Dd[-2]_{\rm p} \\ \frac{1}{2}\Dd[3]_{\rm p} \\ \frac{1}{2}\Dd[-3]_{\rm p}}),\label{eq:6x6mapmaking_pointing}                                                                                                                                                                                                                                                                                                                                                                                                                 \\
    {}_{6}M_{\rm p}= \ab (\begin{array}{cccccc}\frac{1}{4} & \frac{1}{2}\h[-1] & \frac{1}{2}\h[2] & \frac{1}{2}\h[-2] & \frac{1}{2}\h[3] & \frac{1}{2}\h[-3] \\ \frac{1}{2}\h[1] & \frac{1}{4} & \frac{1}{4}\h[3] & \frac{1}{4}\h[-1] & \frac{1}{4}\h[4] & \frac{1}{4}\h[-2] \\ \frac{1}{2}\h[-2] & \frac{1}{4}\h[-3] & \frac{1}{4} & \frac{1}{4}\h[-4] & \frac{1}{4}\h[1] & \frac{1}{4}\h[-5] \\ \frac{1}{2}\h[2] & \frac{1}{4}\h[1] & \frac{1}{4}\h[4] & \frac{1}{4} & \frac{1}{4}\h[5] & \frac{1}{4}\h[-1] \\ \frac{1}{2}\h[-3] & \frac{1}{4}\h[-4] & \frac{1}{4}\h[-1] & \frac{1}{4}\h[-5] & \frac{1}{4} & \frac{1}{4}\h[-6] \\ \frac{1}{2}\h[3] & \frac{1}{4}\h[2] & \frac{1}{4}\h[5] & \frac{1}{4}\h[1] & \frac{1}{4}\h[6] & \frac{1}{4}\end{array}).\label{eq:6Mp}
\end{align}
\Cref{fig:diff_pnt_maps_6x6} shows the estimated CMB polarization maps and systematic
maps which is captured by the $6\times6$ matrix map-making approach.
The residual maps are totally free from the systematic effects, and the systematic
power spectrum is shown in \cref{fig:pointing_bore_cl} (right) by red solid line,
though since no systematics is remained in the maps, so the systematic power spectrum
is given by the only computational noise.

\subsection{Absolute pointing offset}
\begin{figure}[h]
    \centering
    \includegraphics[width=0.32\columnwidth]{
        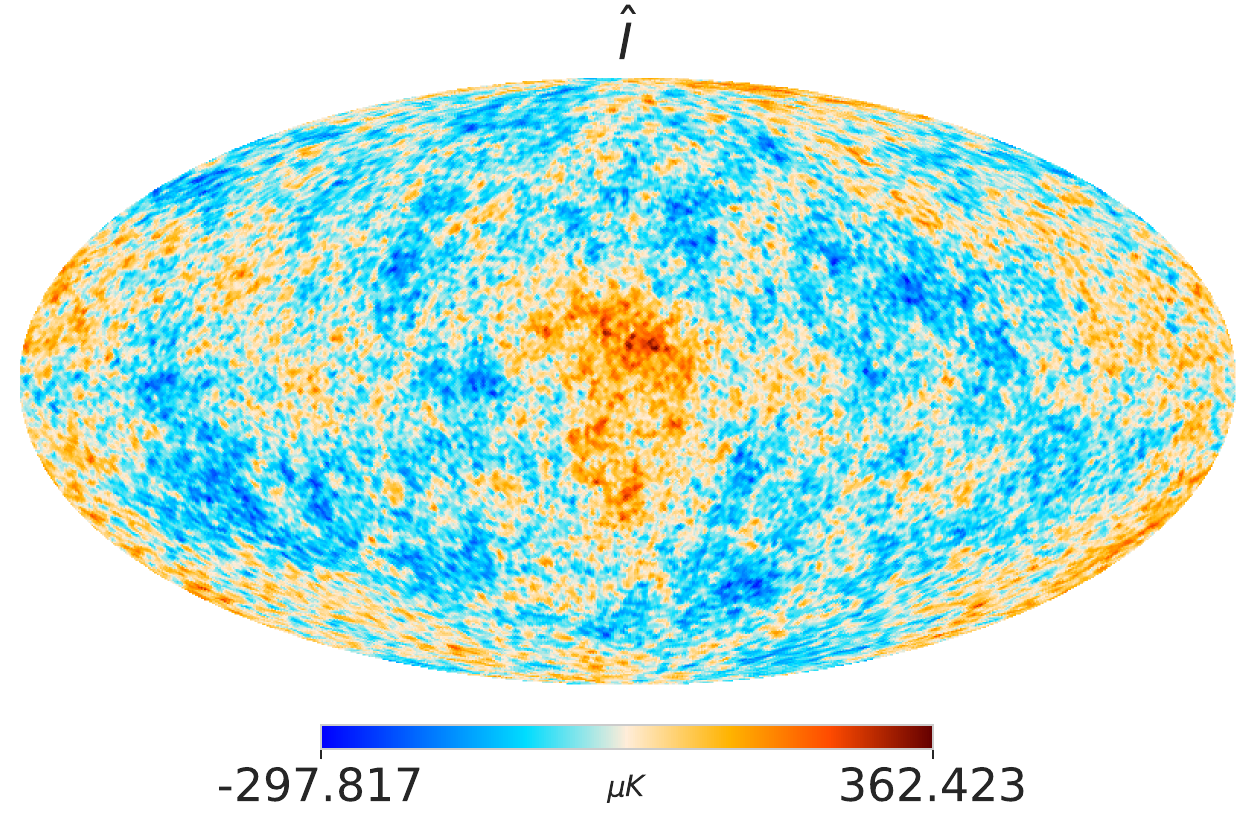
    }
    \includegraphics[width=0.32\columnwidth]{
        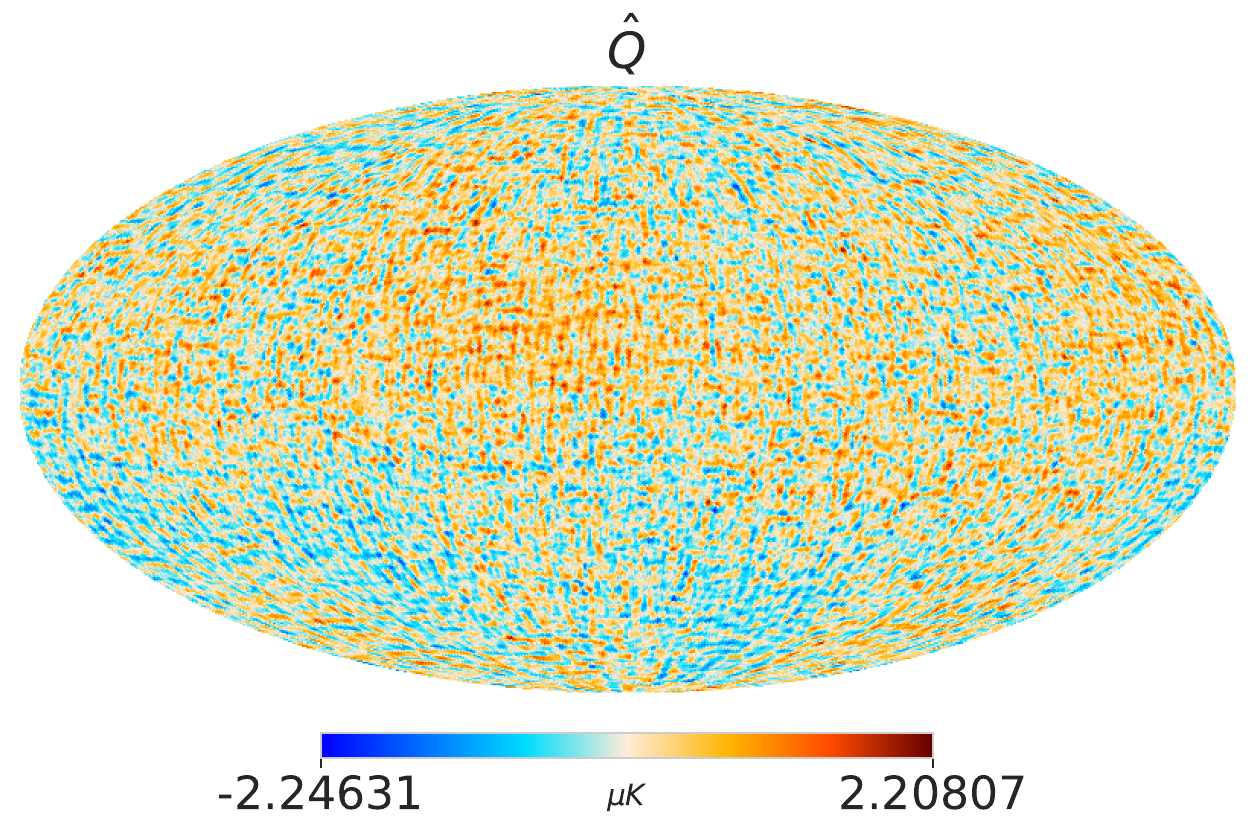
    }
    \includegraphics[width=0.32\columnwidth]{
        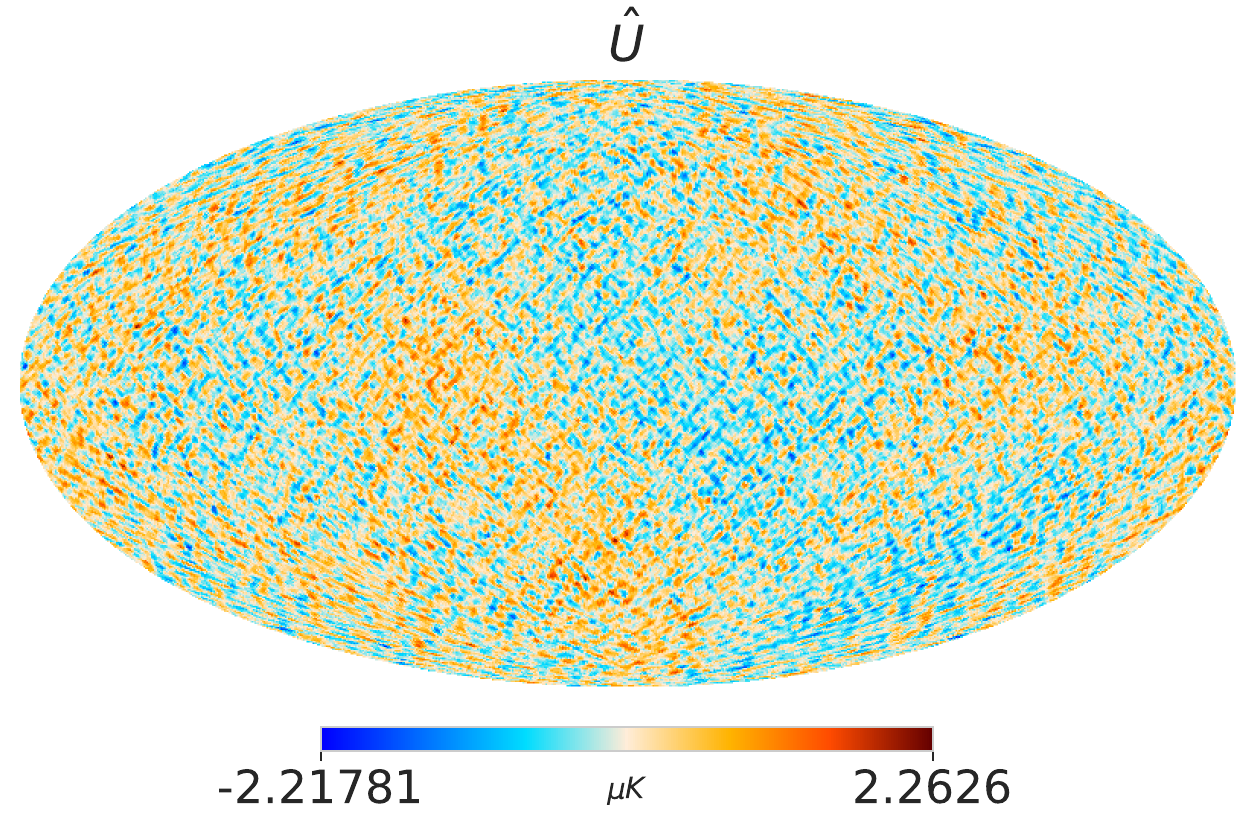
    }
    \\
    \includegraphics[width=0.32\columnwidth]{
        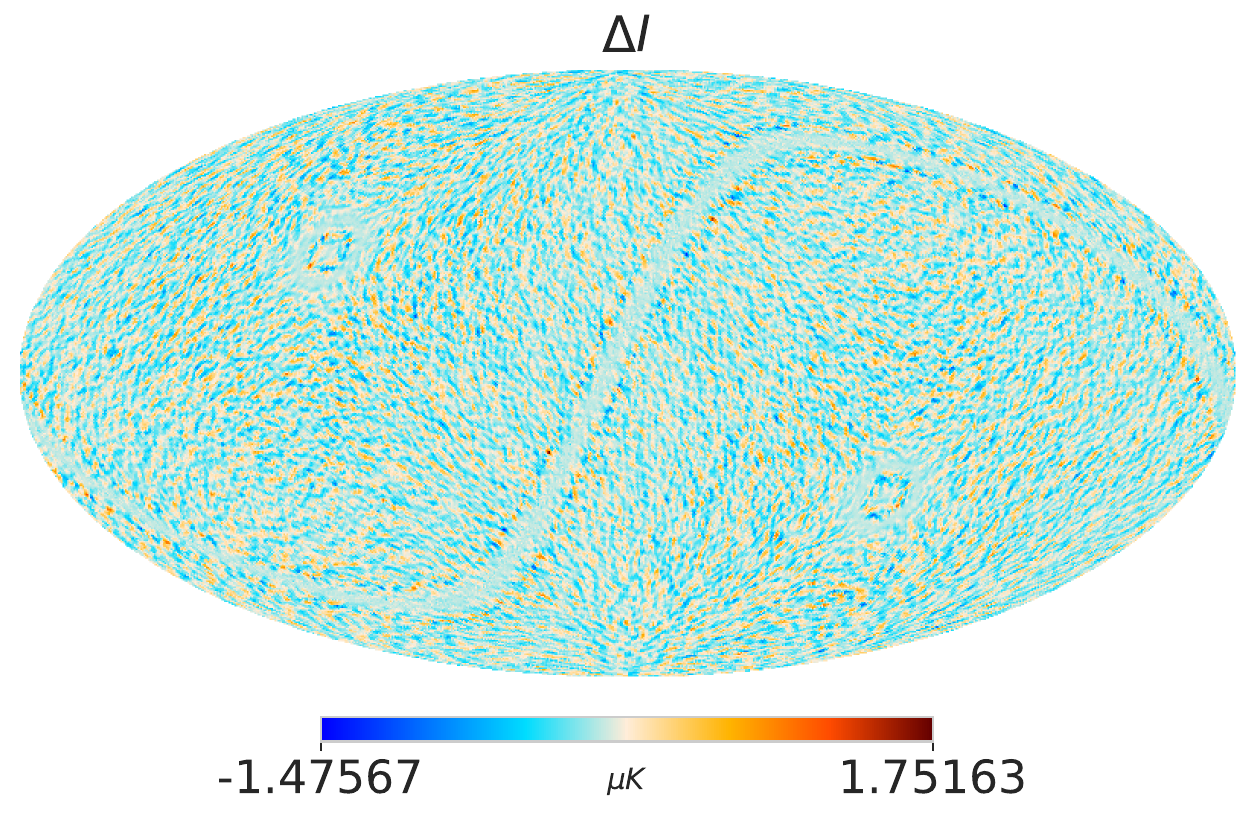
    }
    \includegraphics[width=0.32\columnwidth]{
        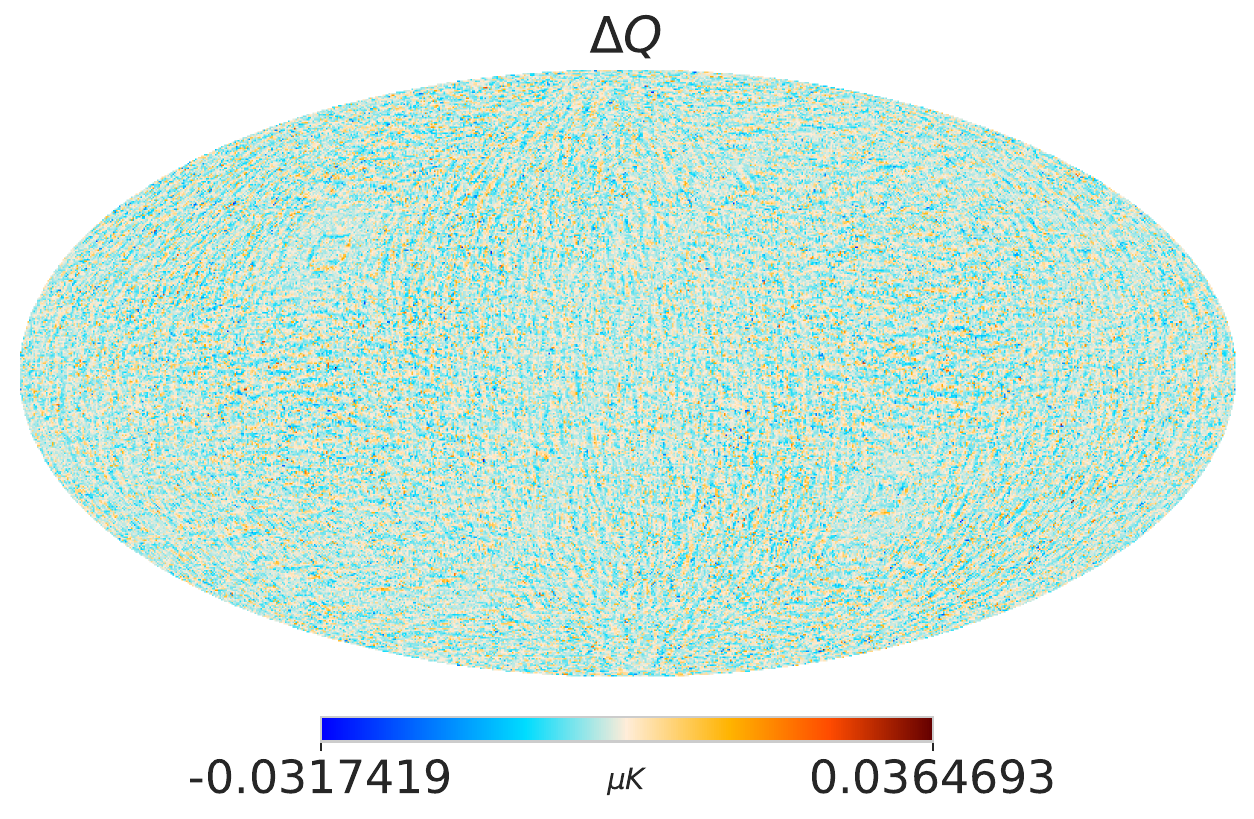
    }
    \includegraphics[width=0.32\columnwidth]{
        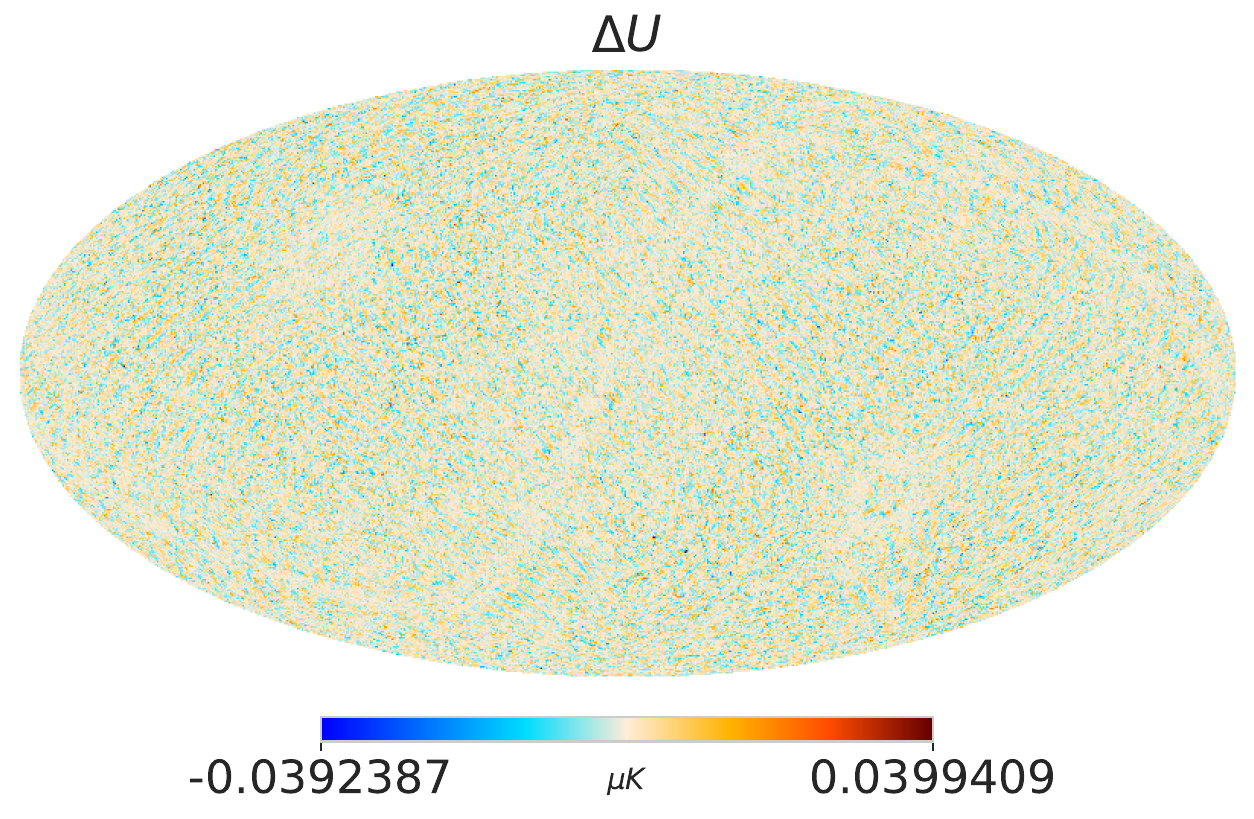
    }
    \caption[Estimated CMB maps and residual maps due to the 1\arcmin absolute pointing
    offset by the $3\times3$ matrix map-making approach with HWP.]{ Estimated CMB
    maps and residual maps due to the absolute pointing offset with HWP,
    $(\rho,\chi)=(1^{\prime},0^{\prime})$ by the $3\times3$ matrix map-making
    approach with HWP. It displays $\hat{I}$, $\hat{Q}$, $\hat{U}$, $\Delta I$, $\Delta
    Q$, and $\Delta U$ from top left to bottom right.}
    \label{fig:abs_pnt_maps_3x3}
\end{figure}

Now we start to discuss the systematic effect with HWP case. We assume that the
pointing systematics parameter $(\rho,\chi)=(1^{\prime},0^{\prime})$ as we set
in \cref{sec:Propagation}. \Cref{fig:abs_pnt_maps_3x3} shows the estimated CMB maps
and residual maps by the $3\times3$ matrix map-making approach shown in \cref{eq:map-making_spin_w_hwp},
we reintroduce it here for the notation that we are using and reader's
convenience:
\begin{align}
    \ab (\mqty{ \hat{I} \\ \hat{P}\\ \hat{P}^* }) & = \M[3]_{\rm ap}^{-1}\ab(\mqty{ \Sd[0,0]_{\rm ap} \\ \frac{1}{2}\Sd[2,-4]_{\rm ap} \\ \frac{1}{2}\Sd[-2,4]_{\rm ap} }), \label{eq:3dim_map-maker_abs_pointing}
\end{align}
where $\M[3]_{\rm ap}$ is given by
\begin{align}
    \M[3]_{\rm ap}= \ab(\mqty{ 1 & \frac{1}{2}\h[-2,4] & \frac{1}{2}\h[2,-4] \\ \frac{1}{2}\h[2,-4] & \frac{1}{4} & \frac{1}{4}\h[4,-8] \\ \frac{1}{2}\h[-2,4] & \frac{1}{4}\h[-4,8] & \frac{1}{4} }).\label{eq:3M_ap}
\end{align}
Even without implementing mitigation techniques, the residual from absolute
pointing offset with HWP shows smaller contamination compared to differential
pointing without HWP (\cref{fig:diff_pnt_maps_2x2}). The HWP contributes to systematic
mitigation through two distinct mechanisms.

First, it reduces systematic contamination through the multiplication of smaller
cross-linking terms. In both differential and absolute pointing cases, the
dominant systematic effect stems from temperature gradient-to-polarization
leakage, corresponding to $\St[\pm1,0]_{\rm ap}$ (or $\St[\pm1]_{\rm p}$) in \cref{eq:temperature_syst,eq:t2p}.
For absolute pointing offset with HWP, $\St[1,0]_{\rm ap}$ couples with $\h[1,-4]$
(and $\St[-1,0]_{\rm ap}$ with $\h[3, -4]$) as shown in \cref{eq:24Sd_ap}. In
contrast, differential pointing without HWP couples $\St[1]_{\rm p}$ with $\h[ 1]$
(and $\St[-1]_{\rm p}$ with $\h[3]$) as in \cref{eq:24Sd_ap}. Since \spin-$(n ,m)$
cross-linking terms with $m>0$ are inherently smaller than \spin-$(n,0)$ terms, the
temperature gradient leakage experiences greater suppression in the HWP-enabled case.
Second, HWP rotation enhances the diagonalization of the covariance matrix in the
map-making equation. The reduced cross-linking terms lead to a more diagonalized
covariance matrix, minimizing the mixing between Stokes parameters during the map-making
process.

\begin{figure}[h]
    \centering
    \includegraphics[width=0.49\columnwidth]{
        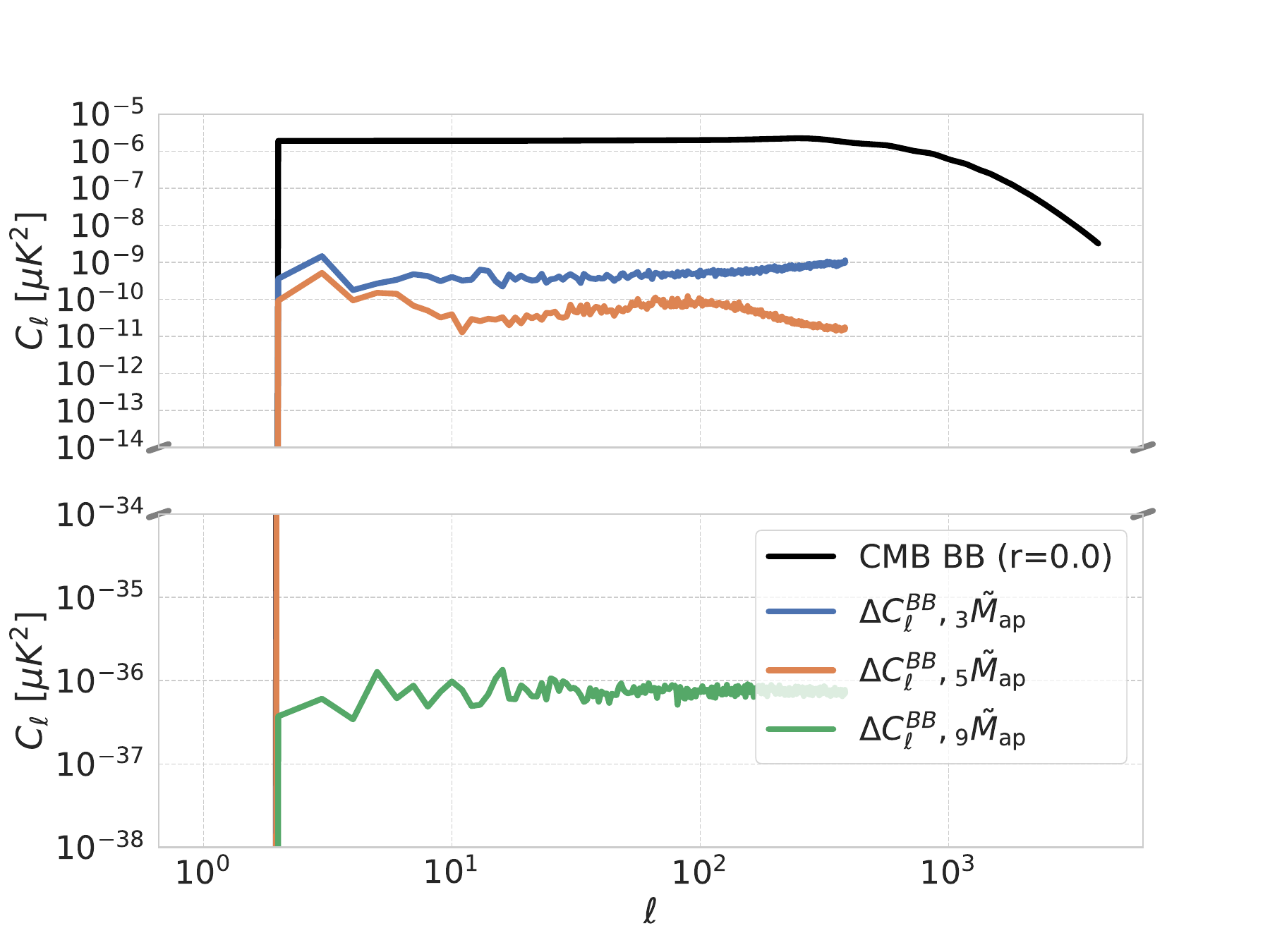
    }
    \includegraphics[width=0.49\columnwidth]{
        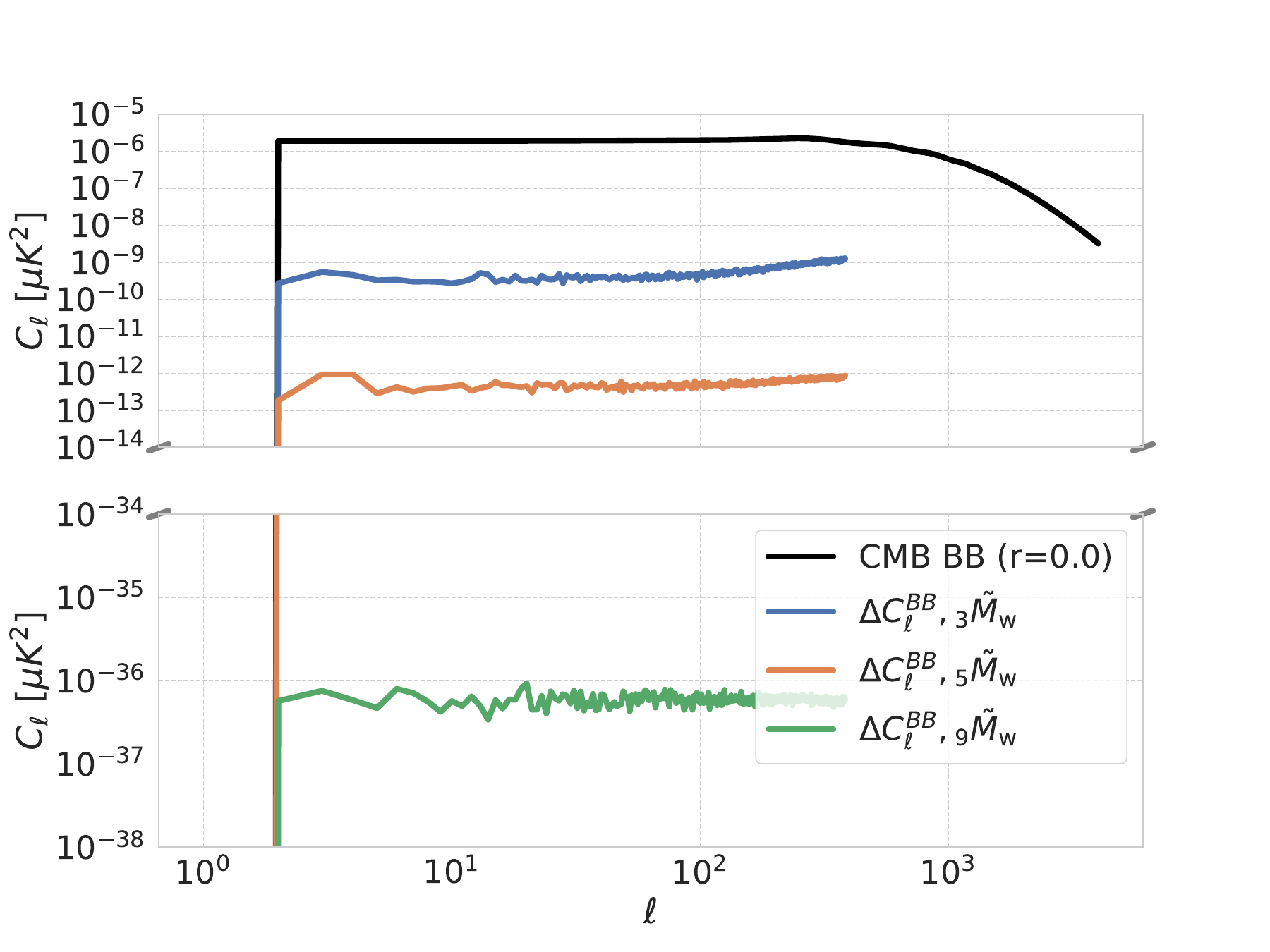
    }
    \caption[Systematic power spectra from absolute pointing offset and HWP-induced
    circular pointing disturbance]{(left) Power spectrum of systematic effects
    arising from absolute pointing offset with HWP, parameterized by $(\rho, \chi
    )=(1^{\prime},0^{\prime})$. Results are shown for three map-making
    approaches: $3 \times 3$ using $\M[3]_{\rm ap}$ (blue), $5\times5$ using
    $\M[5]_{\rm ap}$ (orange), and $9\times9$ using $\M[9]_{\rm ap}$ (green, achieving
    full mitigation). (right) Power spectrum of systematic effects from HWP-induced
    circular pointing disturbance, characterized by pointing perturbation angle
    $\xi=1^{\prime}$ and HWP phase $\chi=0^{\prime}$. Results shown for identical
    map-making approaches as the left panel. The fiducial CMB $B$-mode power
    spectrum is overlaid in black.}
    \label{fig:delta_cl_abs_wedge_pnt}
\end{figure}

The systematic power spectrum derived from $3\times3$ map-making approach is presented
in \cref{fig:delta_cl_abs_wedge_pnt} (left, blue solid line). In contrast to previous
scenarios where analytical estimations of systematic power spectra were feasible
through ensemble-averaged CMB power spectra, the incorporation of HWP adds
complexity via its \spin moment$m$. While it may be feasible to develop a
transfer function mapping signal fields to systematic power spectra by extending
the framework of ref.~\cite{spin_characterisation} to incorporate HWP contributions,
we reserve this analytical development for future investigation.

\begin{figure}[h]
    \centering
    \includegraphics[width=0.32\columnwidth]{
        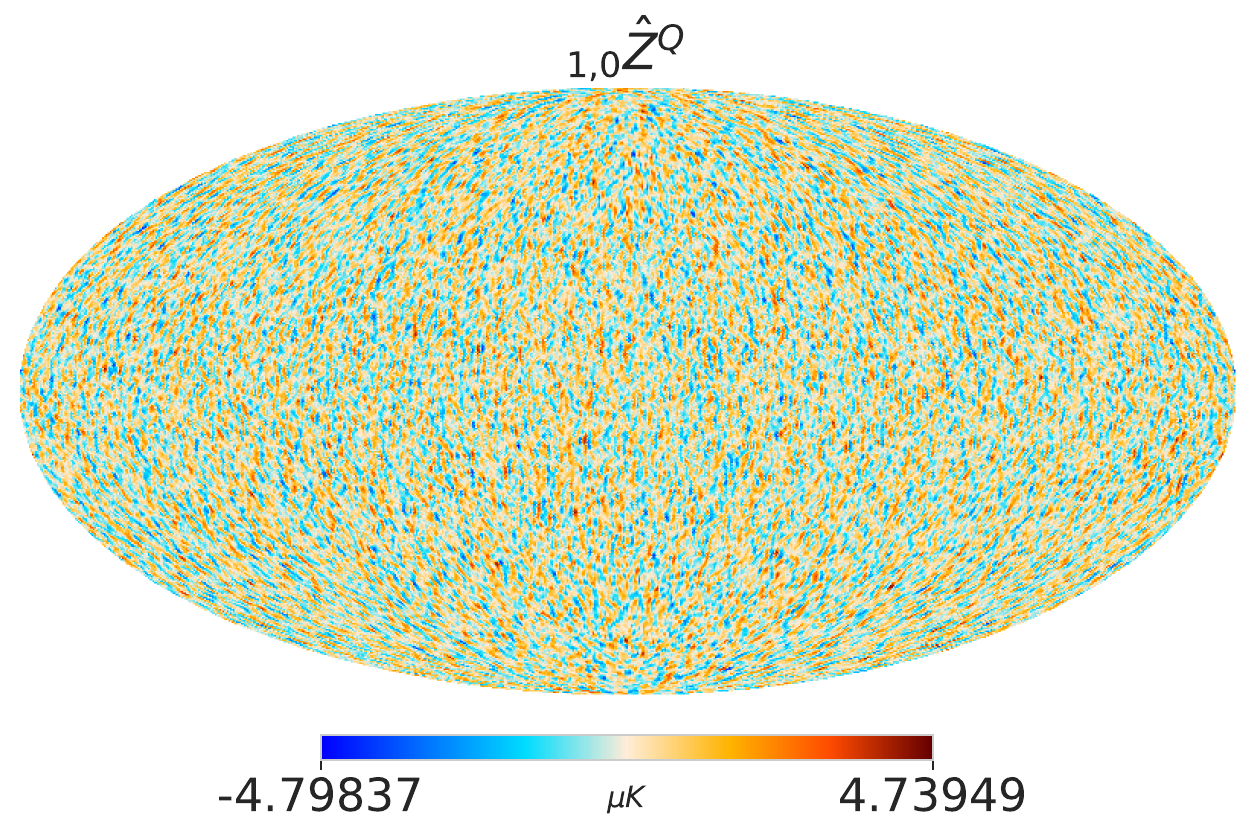
    }
    \includegraphics[width=0.32\columnwidth]{
        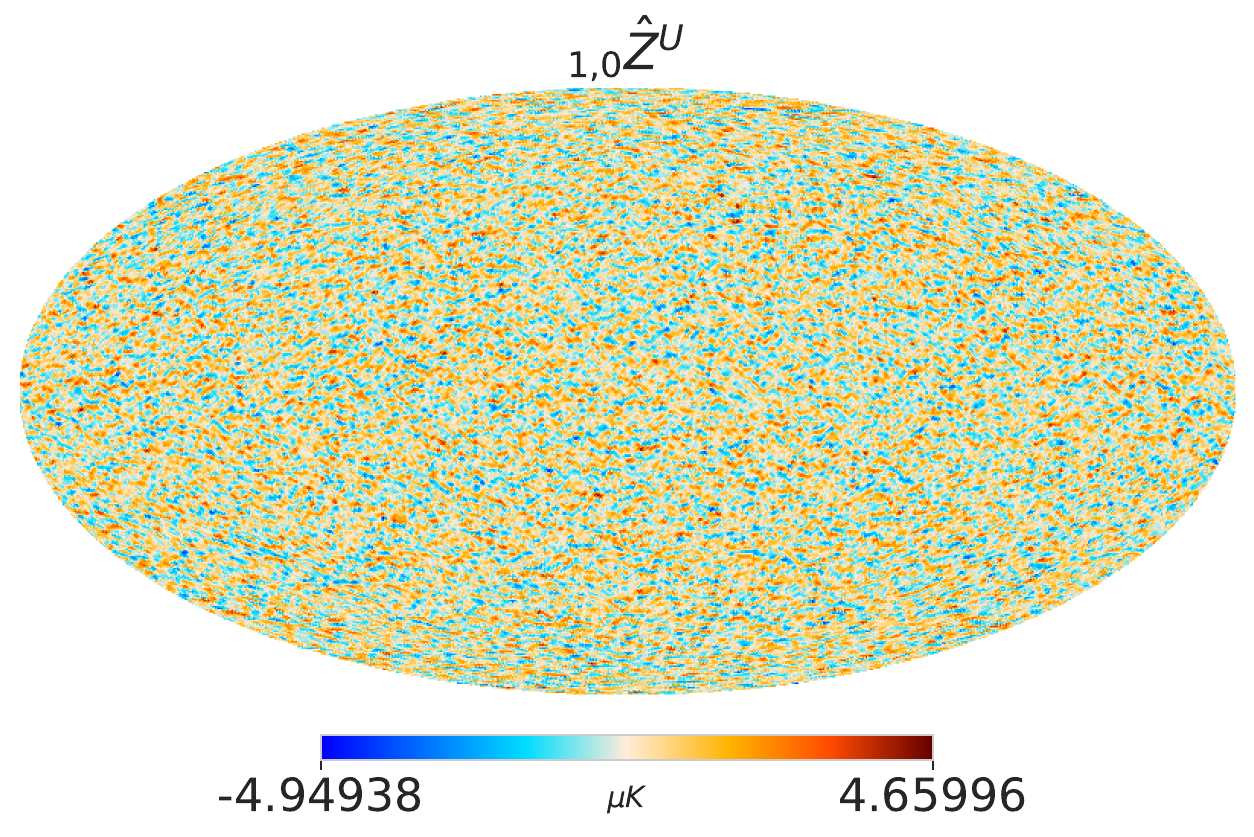
    }
    \\
    \includegraphics[width=0.32\columnwidth]{
        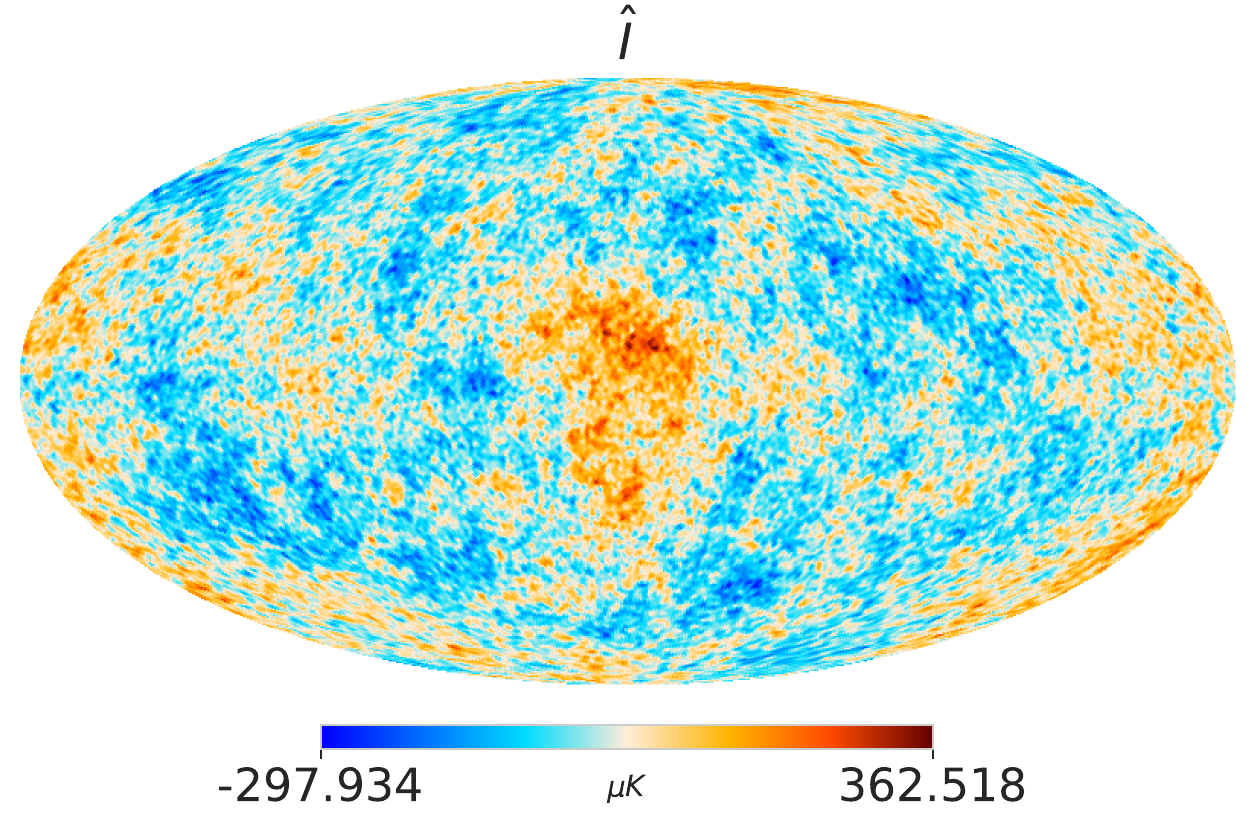
    }
    \includegraphics[width=0.32\columnwidth]{
        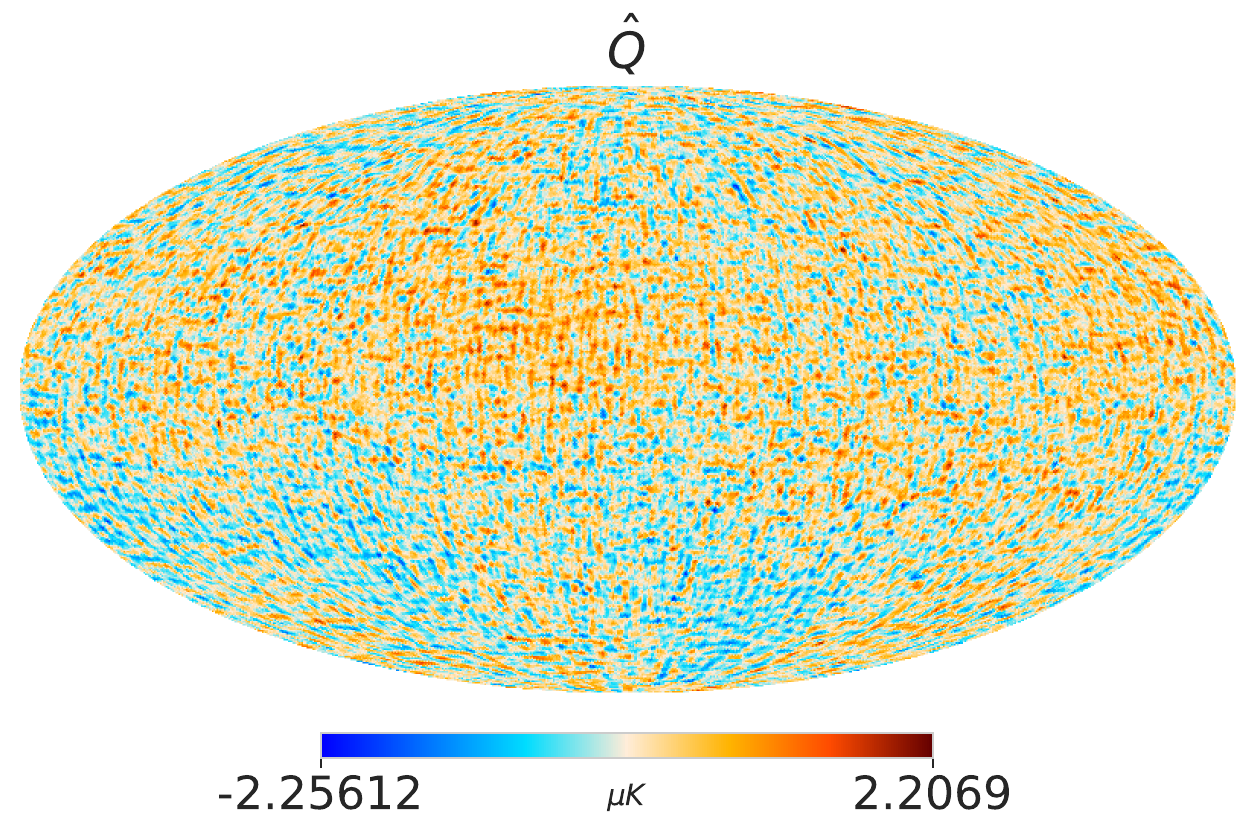
    }
    \includegraphics[width=0.32\columnwidth]{
        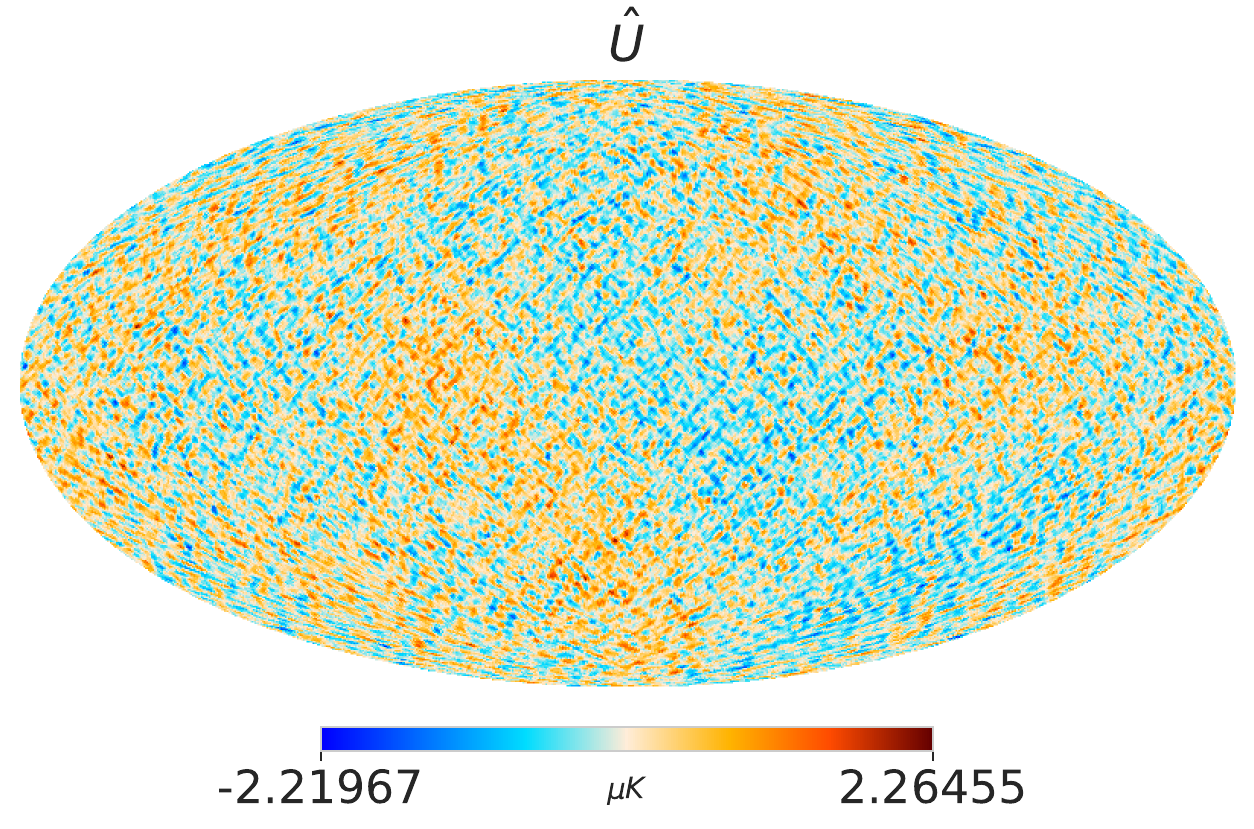
    }
    \\
    \includegraphics[width=0.32\columnwidth]{
        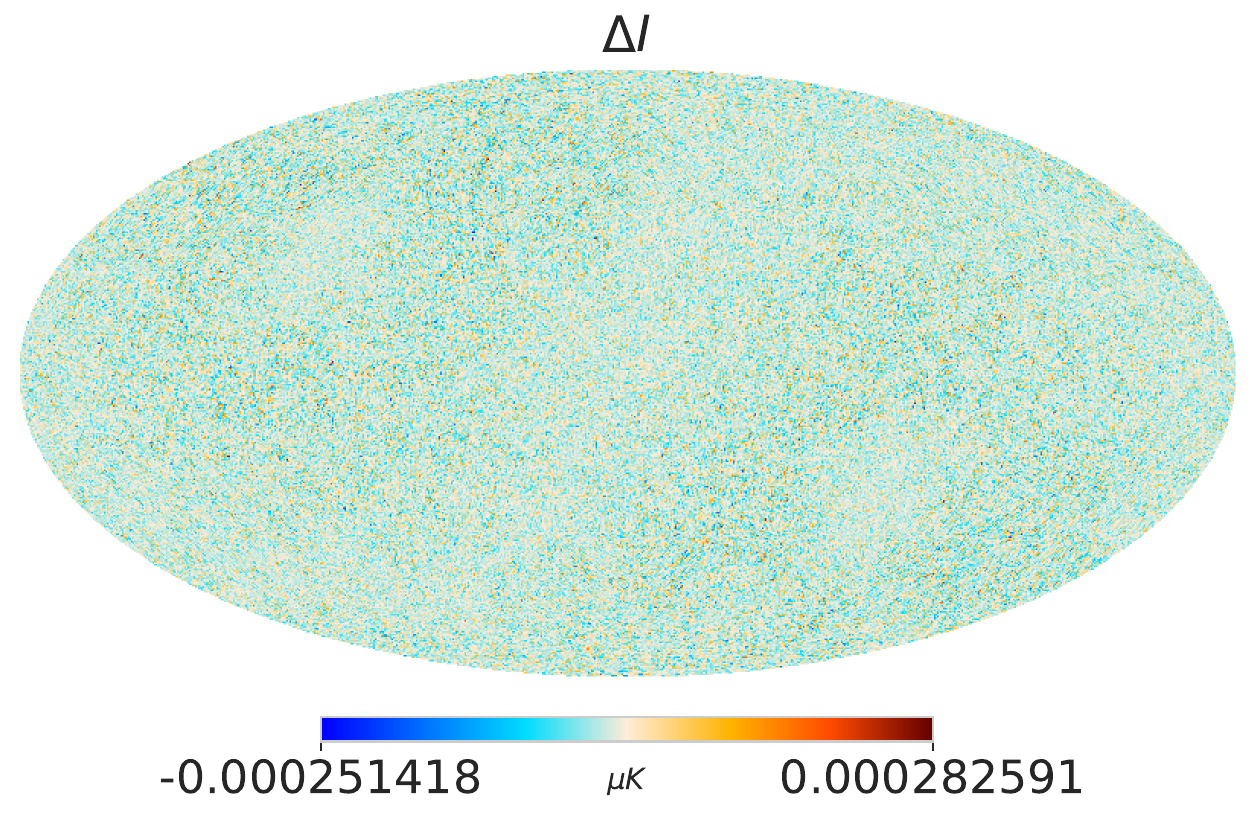
    }
    \includegraphics[width=0.32\columnwidth]{
        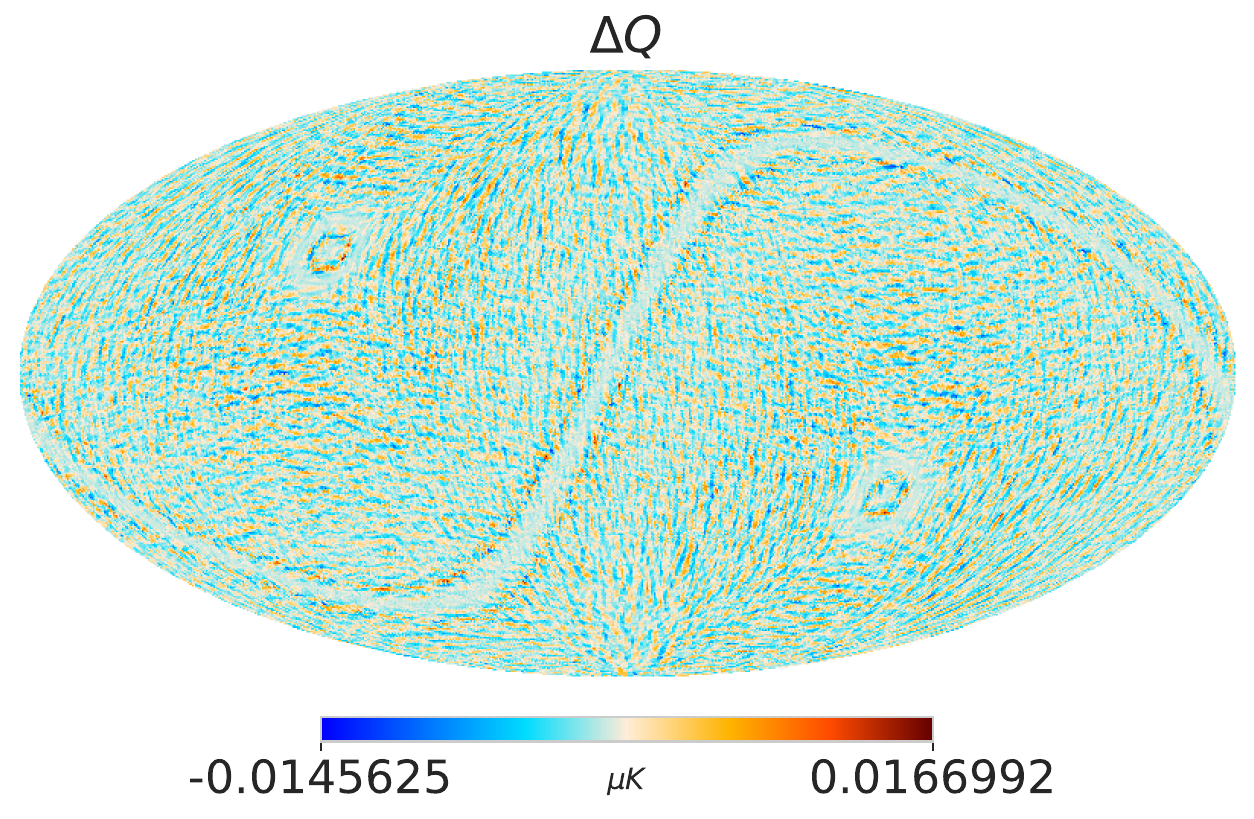
    }
    \includegraphics[width=0.32\columnwidth]{
        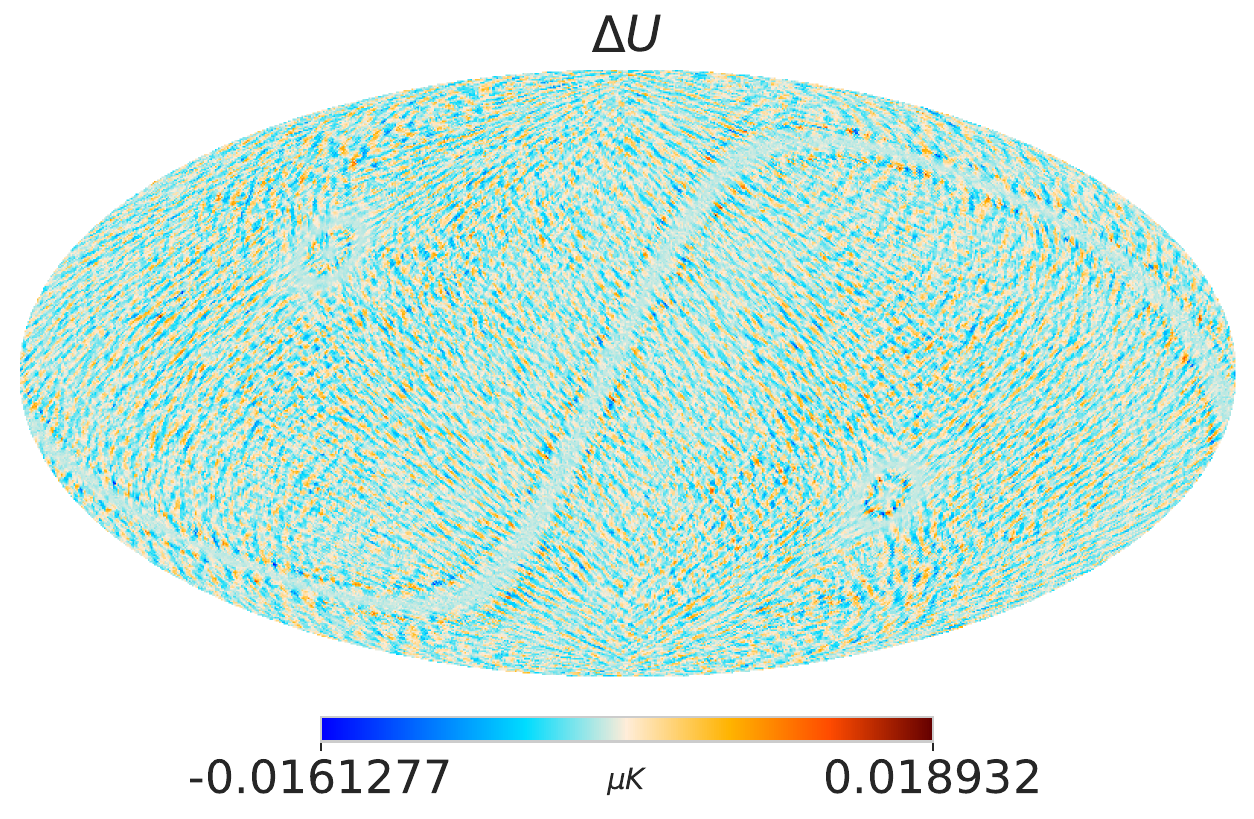
    }
    \caption[Estimated CMB maps and residual maps due to the 1\arcmin absolute pointing
    offset by the $5\times5$ matrix map-making approach with HWP.]{ Estimated CMB
    maps and residual maps due to the absolute pointing offset,
    $(\rho,\chi)=(1^{\prime},0^{\prime})$ by the $5\times5$ matrix map-making
    approach with HWP. Top panels show $\hZ[1,0]^{Q}$ and $\hZ[1,0]^{U}$, middle
    panels show $\hat{I}$, $\hat{Q}$, and $\hat{U}$, and bottom panels show $\Delta
    I$, $\Delta Q$, and $\Delta U$ from left to right. }
    \label{fig:abs_pnt_maps_5x5}
\end{figure}

\begin{figure}[h]
    \centering
    \includegraphics[width=0.32\columnwidth]{
        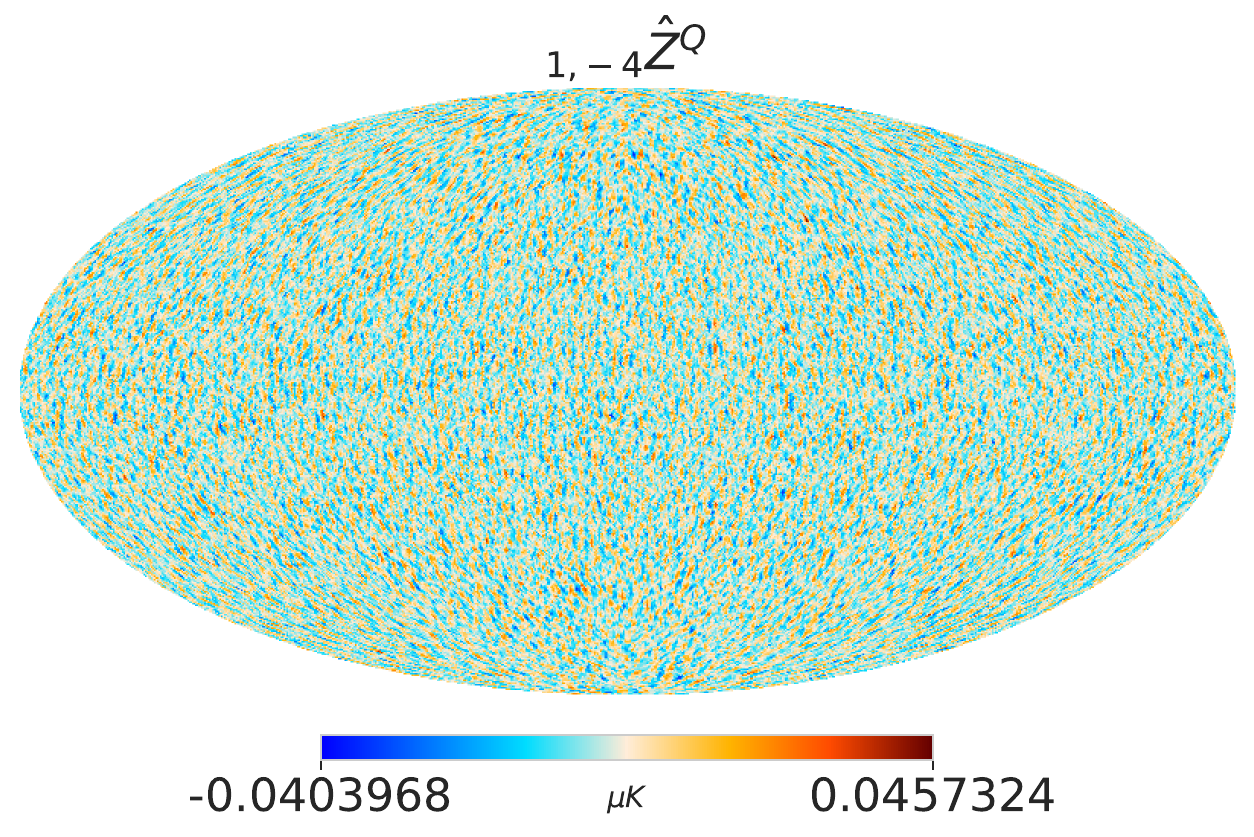
    }
    \includegraphics[width=0.32\columnwidth]{
        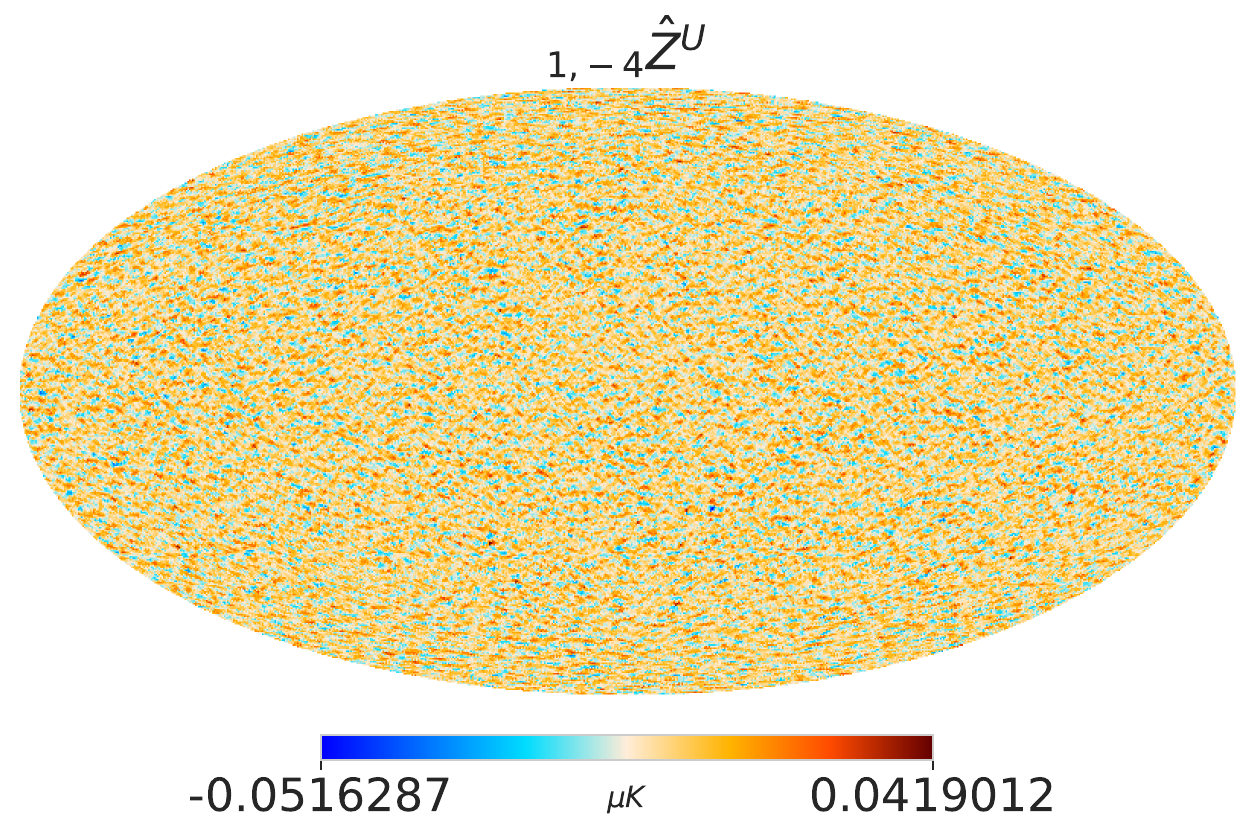
    }
    \includegraphics[width=0.32\columnwidth]{
        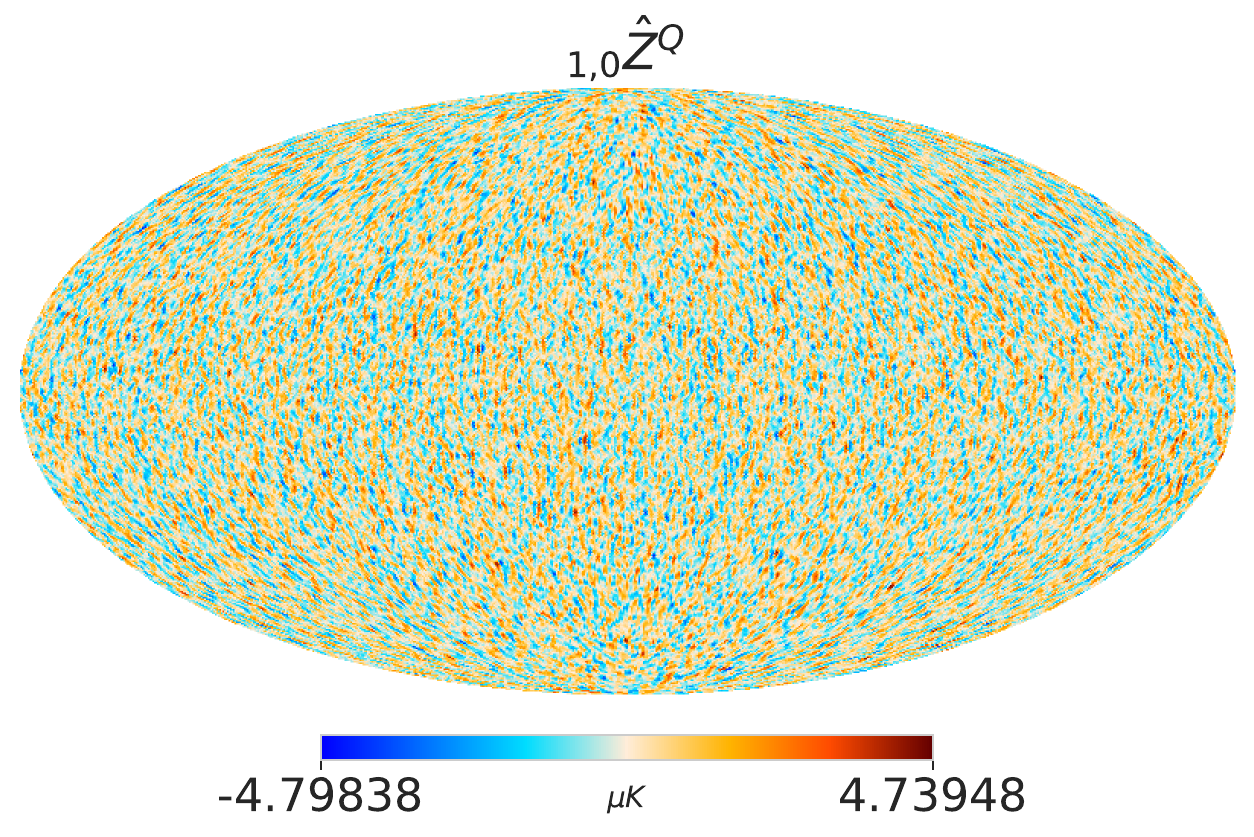
    }
    \\
    \includegraphics[width=0.32\columnwidth]{
        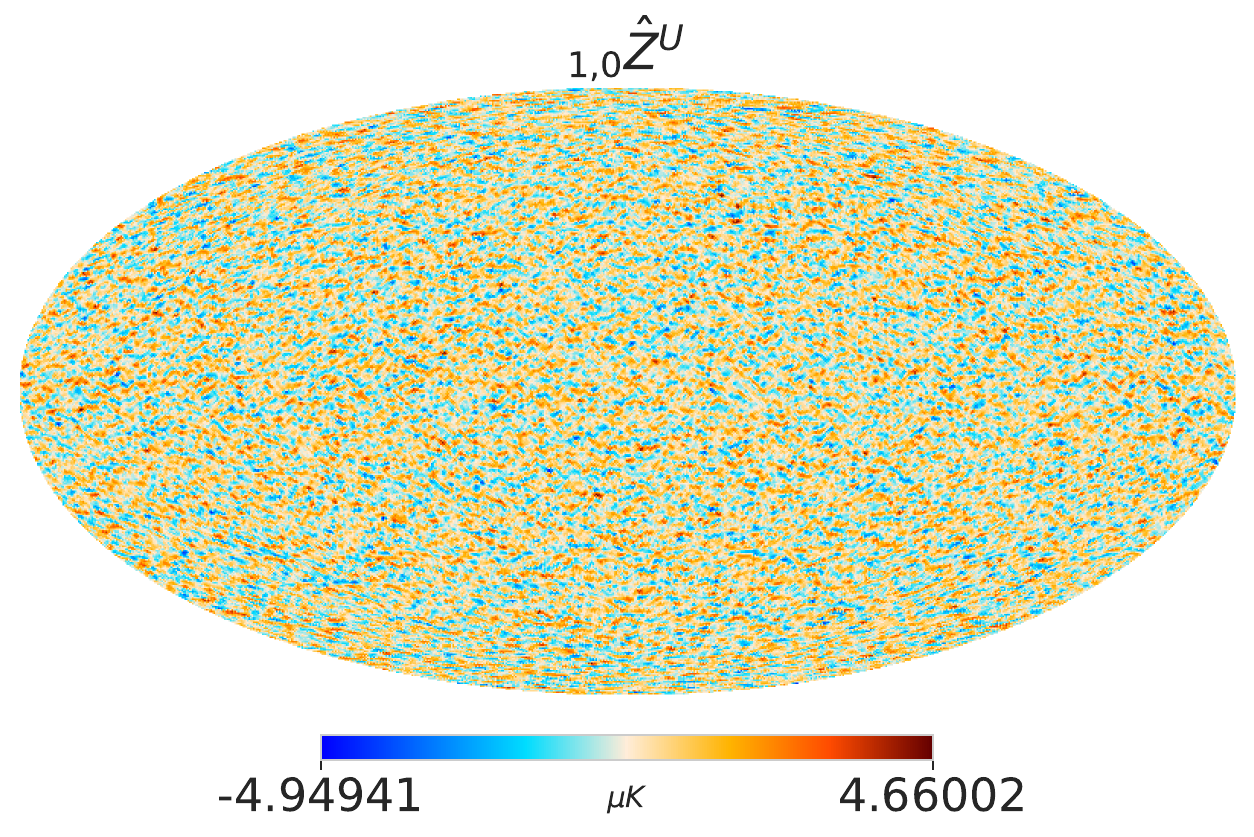
    }
    \includegraphics[width=0.32\columnwidth]{
        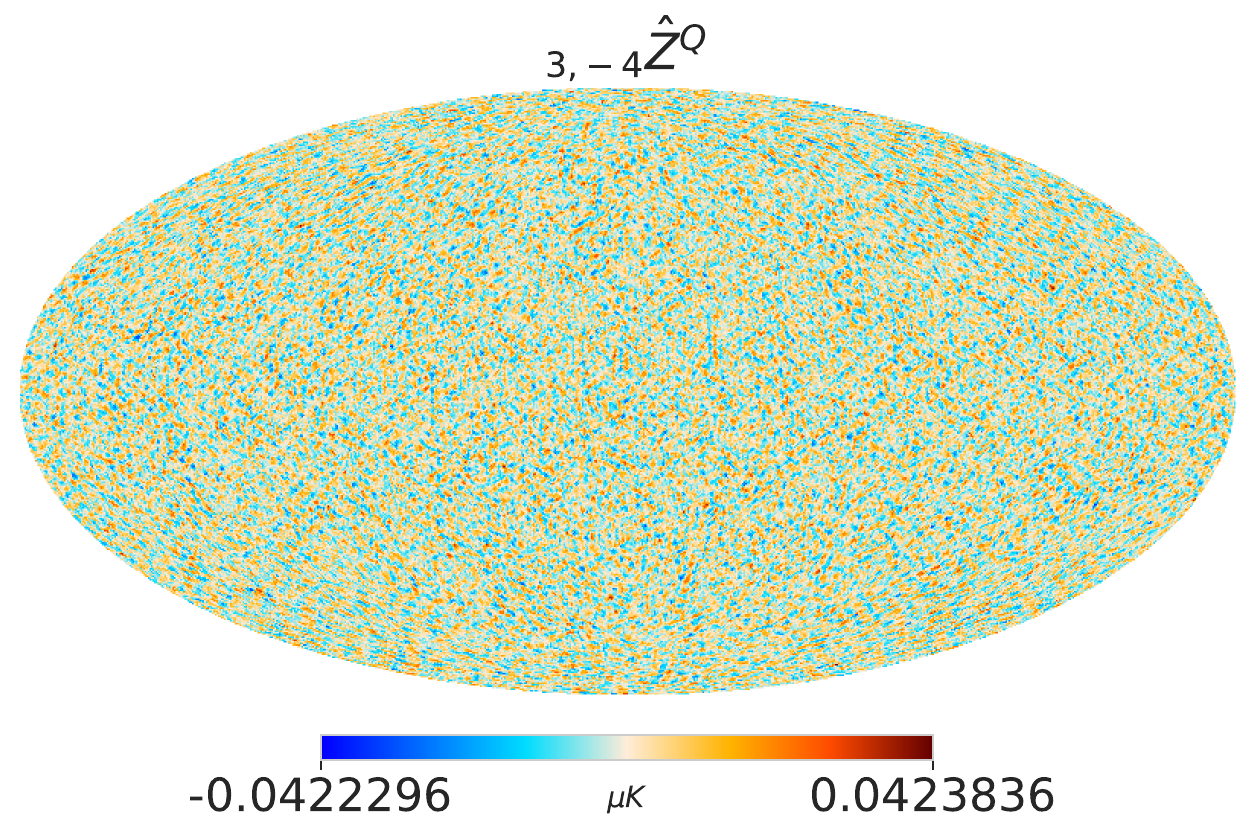
    }
    \includegraphics[width=0.32\columnwidth]{
        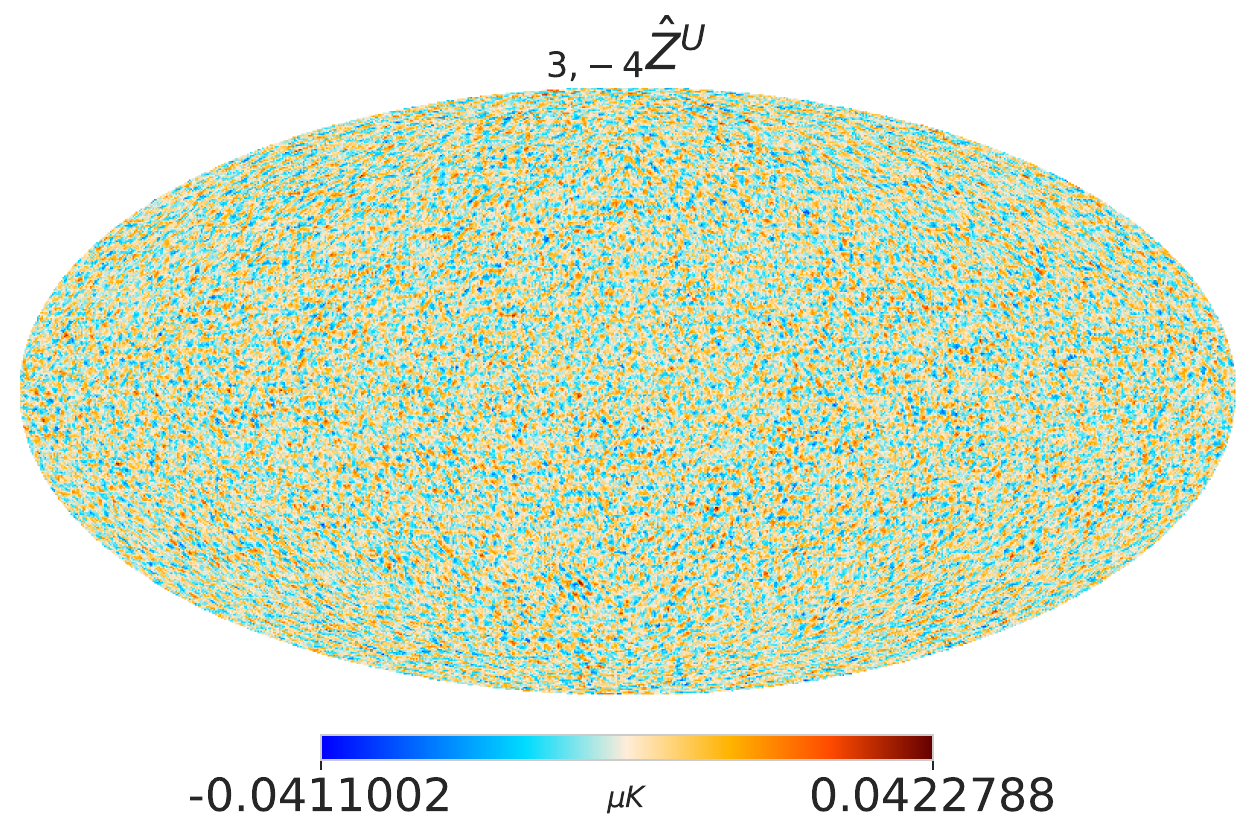
    }
    \\
    \includegraphics[width=0.32\columnwidth]{
        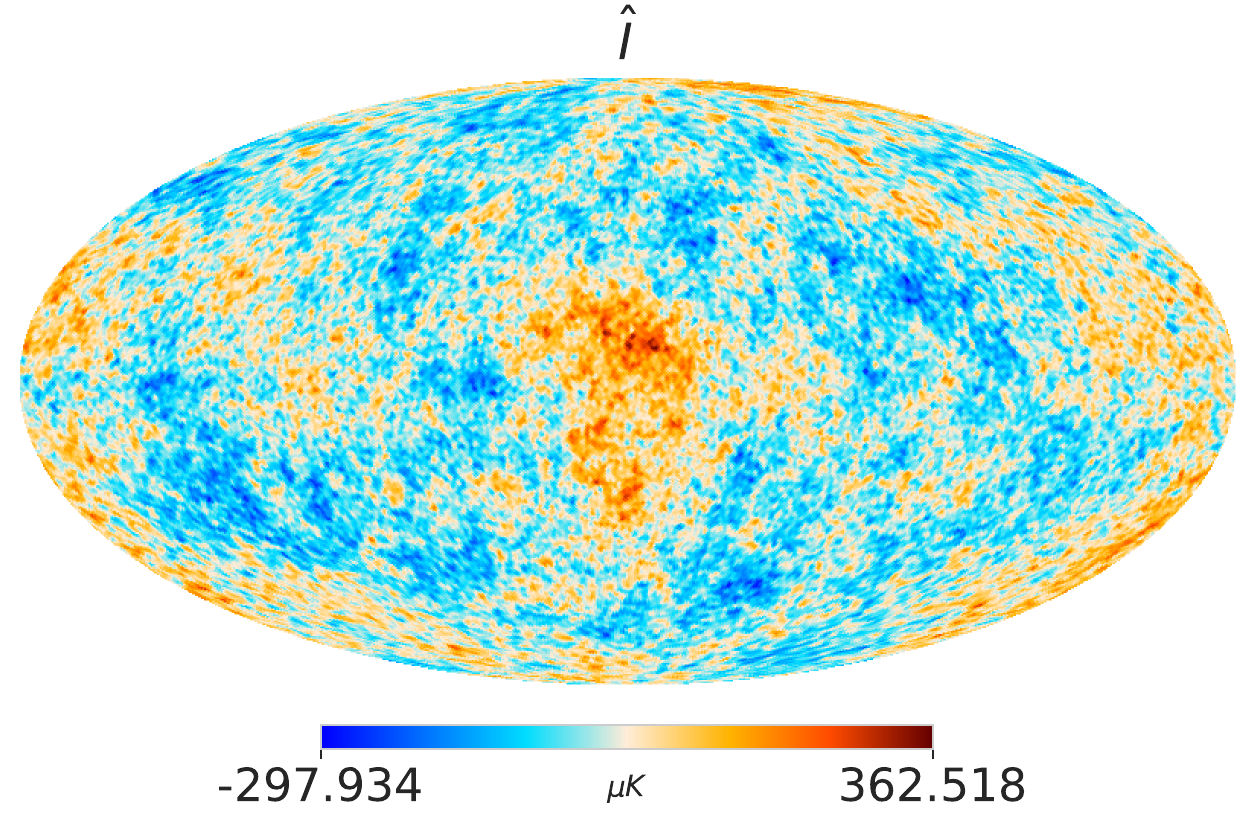
    }
    \includegraphics[width=0.32\columnwidth]{
        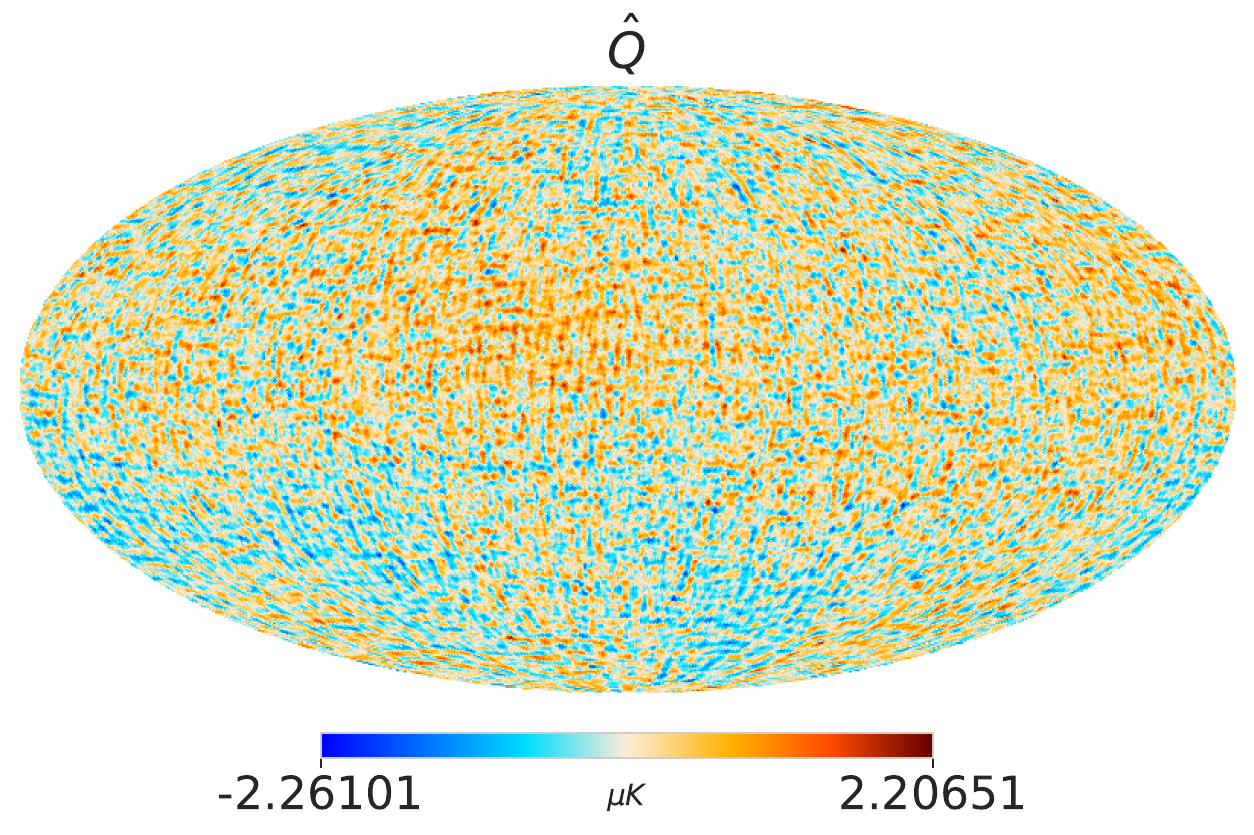
    }
    \includegraphics[width=0.32\columnwidth]{
        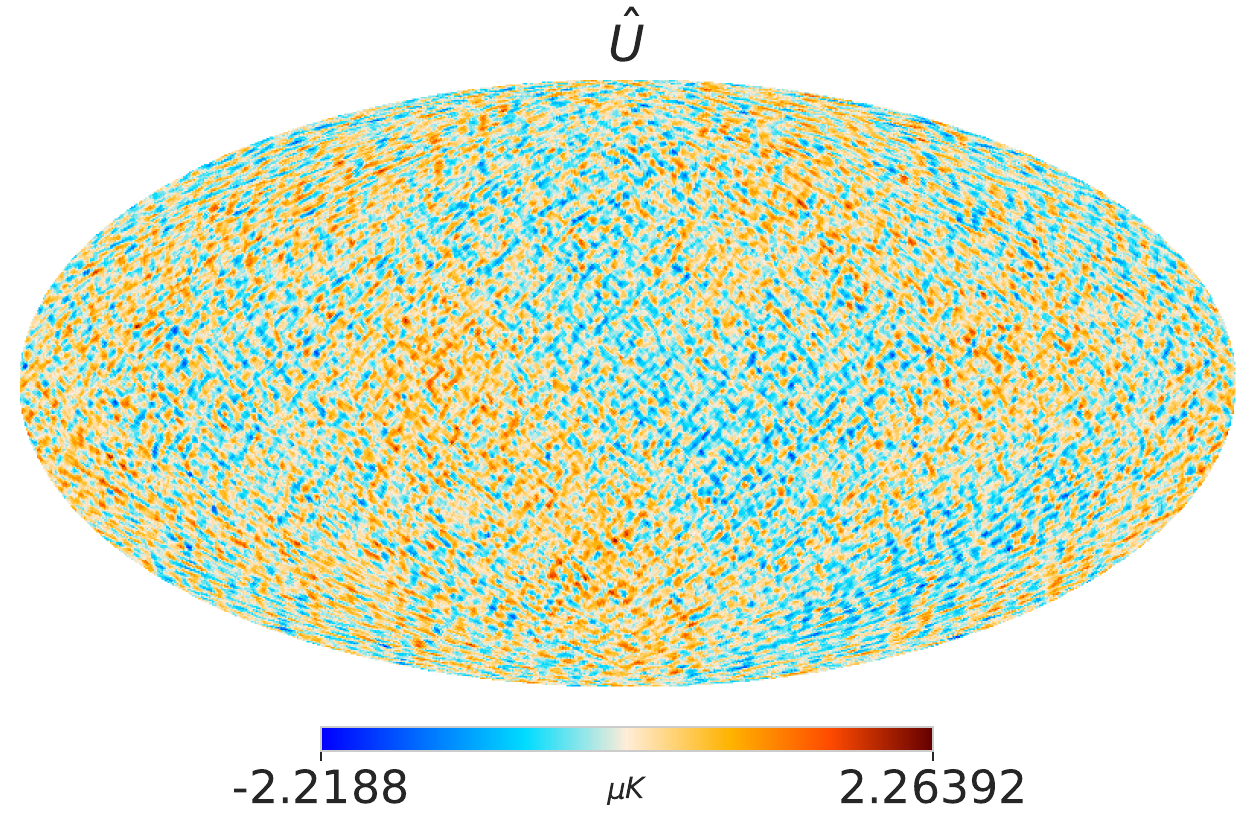
    }
    \\
    \includegraphics[width=0.32\columnwidth]{
        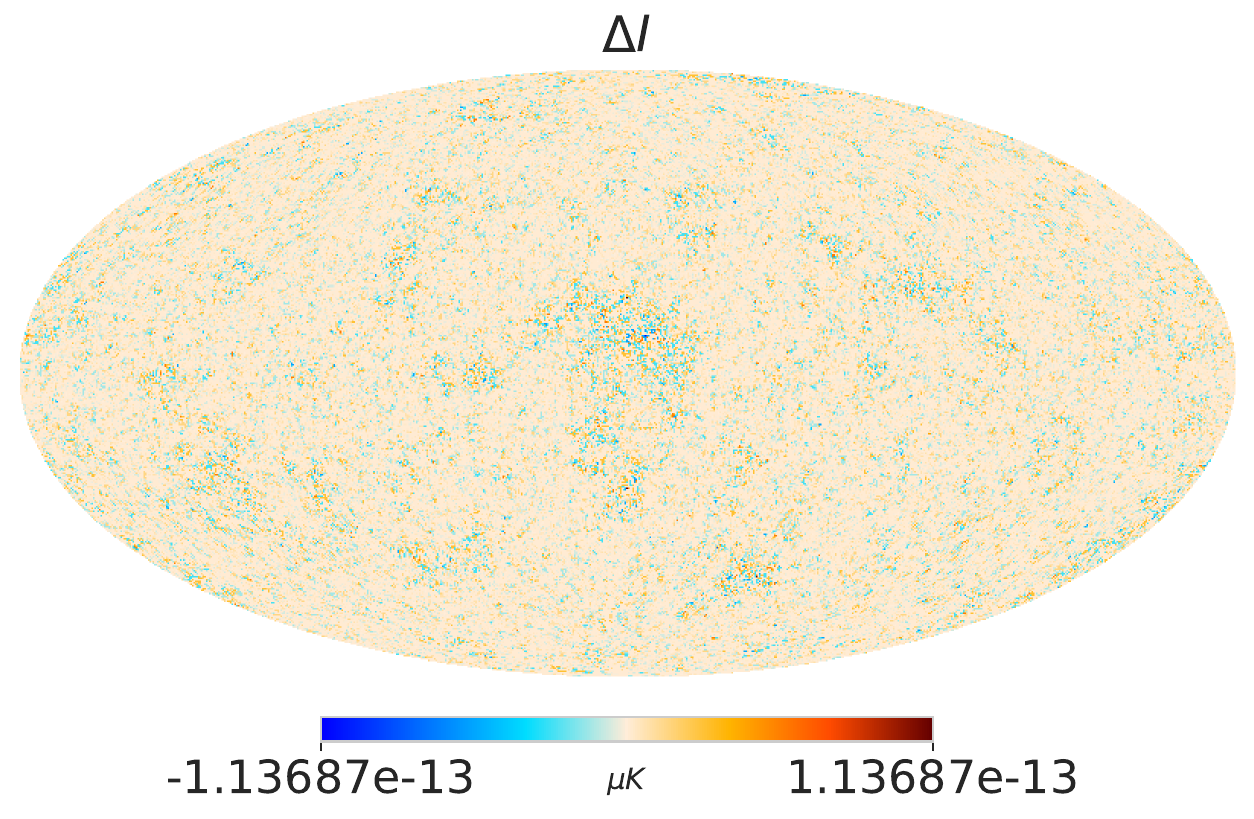
    }
    \includegraphics[width=0.32\columnwidth]{
        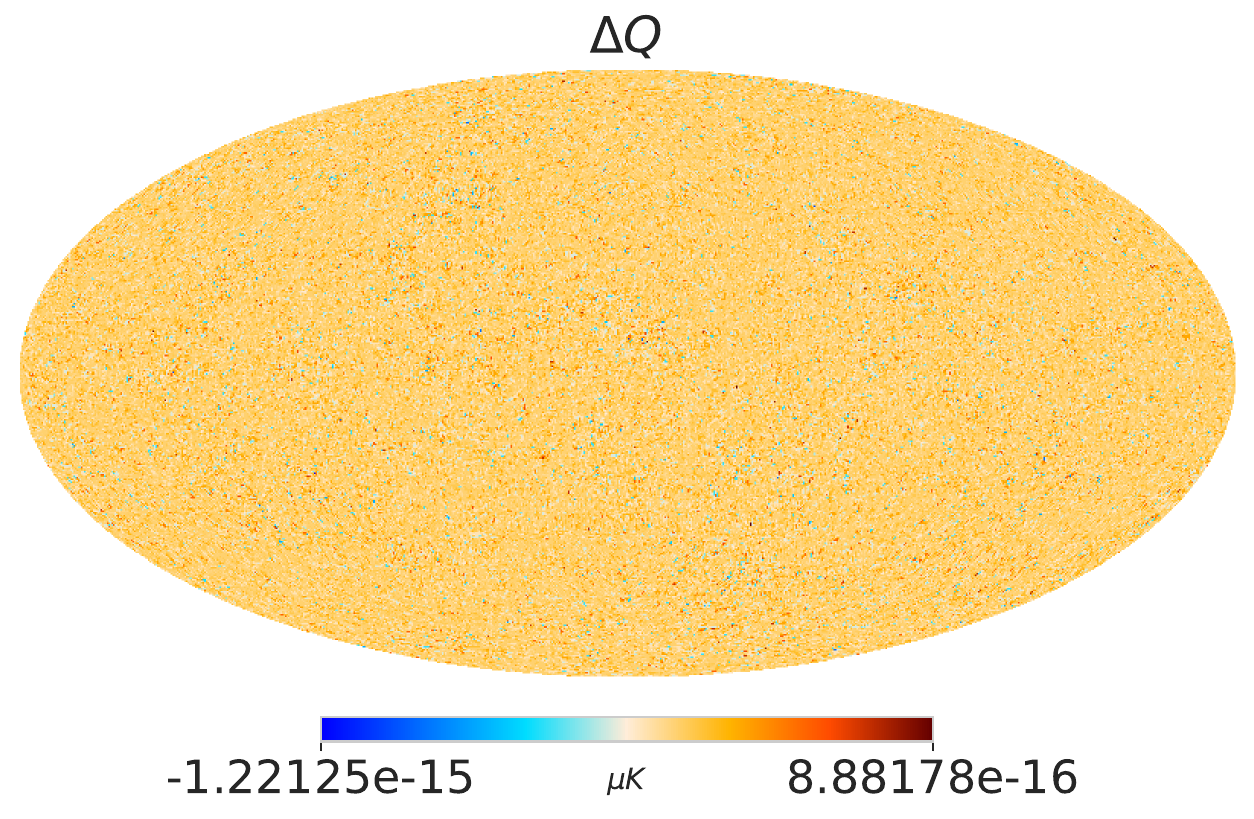
    }
    \includegraphics[width=0.32\columnwidth]{
        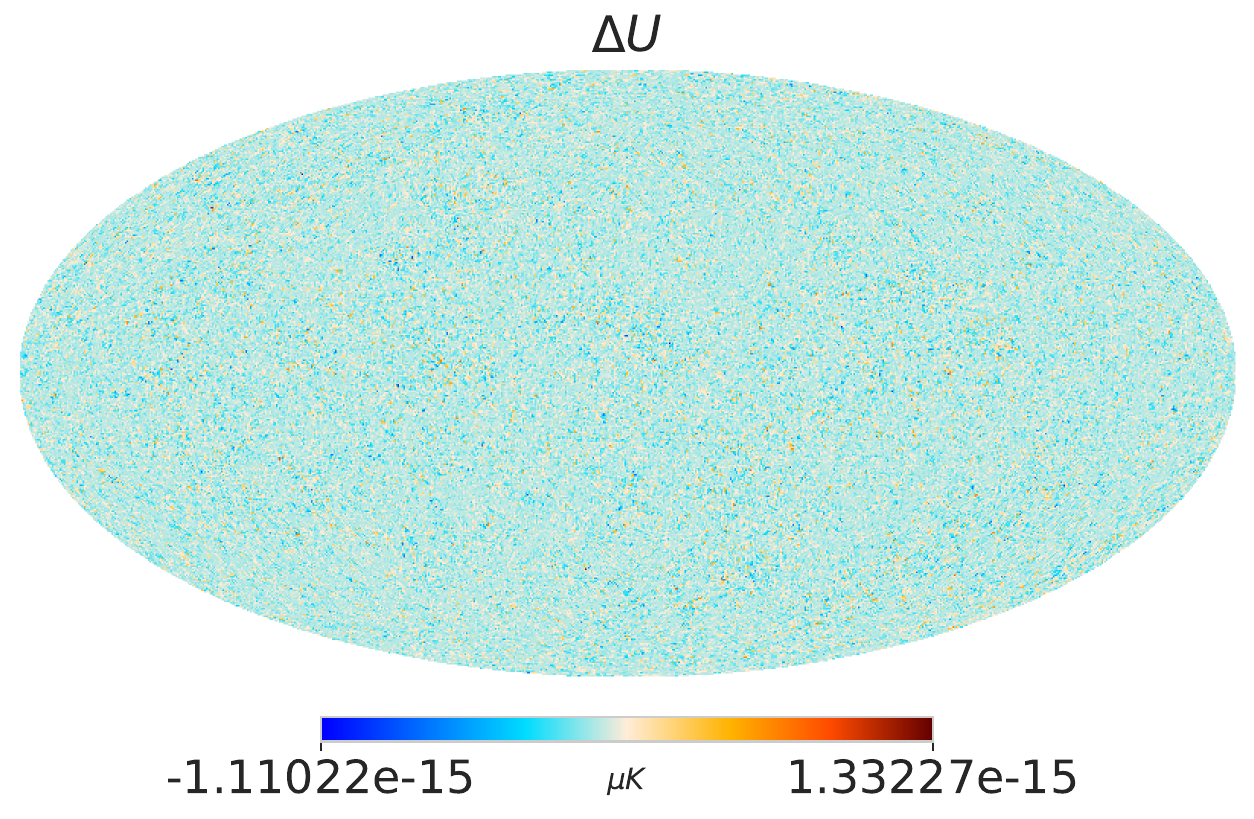
    }
    \caption[Estimated CMB maps and residual maps due to the 1\arcmin absolute pointing
    offset by the $9\times9$ matrix map-making approach with HWP.]{Estimated CMB
    maps and residual maps due to the absolute pointing offset, $(\rho,\chi)=(1^{\prime}
    ,0^{\prime})$ by the $9\times9$ matrix map-making approach with HWP. Top
    panel shows $\hZ[1,-4]^{Q}$, $\hZ[1,-4]^{U}$, and $\hZ[1,0]^{Q}$; second panel
    shows $\hZ[1,0]^{U}$, $\hZ[3,-4]^{Q}$, and $\hZ[3,-4]^{U}$; and third panel
    shows $\hat{I}$, $\hat{Q}$ and $\hat{U}$; and bottom panel shows $\Delta I$,
    $\Delta Q$, and $\Delta U$ from left to right.}
    \label{fig:abs_pnt_maps_9x9}
\end{figure}

Nevertheless, our map-based simulation methodology efficiently generates residual
maps and corresponding systematic power spectra through rapid CMB realizations.\footnote{
A CMB realization involves generating a CMB map by randomizing the phase of CMB spherical
harmonic coefficients $a_{\ell m}$ and computing the ensemble average of the resulting
power spectrum to eliminate cosmic variance.} 
Moreover, this approach surpasses
analytical power spectrum estimation by producing output maps that include
systematic effects, enabling both pixel-space component separation and analysis of
systematic-foreground interactions. Even under use of HWP, we can further improve
systematic mitigation by expanding the map-making matrix to $5\times5$ and $9\times
9$ dimensions. The $5\times5$ matrix map-making approach is defined by
\begin{align}
    \begin{pmatrix}\hat{I}\\{}_{1,0}\hat{Z}\\{}_{-1,0}\hat{Z}\\ \hat{P}\\ \hat{P^*}\end{pmatrix} = \M[5]_{\rm ap}^{-1}\begin{pmatrix}{}_{0,0}{\tilde{S}^d}_{\rm ap}\\{}_{1,0}{\tilde{S}^d}_{\rm ap}\\{}_{-1,0}{\tilde{S}^d}_{\rm ap}\\{}_{2,-4}{\tilde{S}^d}_{\rm ap}\\{}_{-2,4}{\tilde{S}^d}_{\rm ap}\end{pmatrix},
\end{align}
where$\M[5]_{\rm ap}$is given by
\begin{align}
    \M[5]_{\rm ap}= \begin{pmatrix}1&\frac{1}{2}{}_{-1,0}\tilde{h}&\frac{1}{2}{}_{1,0}\tilde{h}&\frac{1}{2}{}_{-2,4}\tilde{h}&\frac{1}{2}{}_{2,-4}\tilde{h}\\ \frac{1}{2}{}_{1,0}\tilde{h}&\frac{1}{4}&\frac{1}{4}{}_{2,0}\tilde{h}&\frac{1}{4}{}_{-1,4}\tilde{h}&\frac{1}{4}{}_{3,-4}\tilde{h}\\ \frac{1}{2}{}_{-1,0}\tilde{h}&\frac{1}{4}{}_{-2,0}\tilde{h}&\frac{1}{4}&\frac{1}{4}{}_{-3,4}\tilde{h}&\frac{1}{4}{}_{1,-4}\tilde{h}\\ \frac{1}{2}{}_{2,-4}\tilde{h}&\frac{1}{4}{}_{1,-4}\tilde{h}&\frac{1}{4}{}_{3,-4}\tilde{h}&\frac{1}{4}&\frac{1}{4}{}_{4,-8}\tilde{h}\\ \frac{1}{2}{}_{-2,4}\tilde{h}&\frac{1}{4}{}_{-3,4}\tilde{h}&\frac{1}{4}{}_{-1,4}\tilde{h}&\frac{1}{4}{}_{-4,8}\tilde{h}&\frac{1}{4}\end{pmatrix}, \label{eq:5M_ap}
\end{align}
and the $9\times9$ matrix map-making approach is defined by
\begin{align}
    \begin{pmatrix}\hat{I}\\{}_{1,0}\hat{Z}\\{}_{-1,0}\hat{Z}\\ \hat{P}\\ \hat{P^*}\\{}_{3,-4}\hat{Z}\\{}_{-3,4}\hat{Z}\\{}_{1,-4}\hat{Z}\\{}_{-1,4}\hat{Z}\end{pmatrix} = \M[9]_{\rm ap}^{-1}\begin{pmatrix}{}_{0,0}{\tilde{S}^d}_{\rm ap}\\{}_{1,0}{\tilde{S}^d}_{\rm ap}\\{}_{-1,0}{\tilde{S}^d}_{\rm ap}\\{}_{2,-4}{\tilde{S}^d}_{\rm ap}\\{}_{-2,4}{\tilde{S}^d}_{\rm ap}\\{}_{3,-4}{\tilde{S}^d}_{\rm ap}\\{}_{-3,4}{\tilde{S}^d}_{\rm ap}\\{}_{1,-4}{\tilde{S}^d}_{\rm ap}\\{}_{-1,4}{\tilde{S}^d}_{\rm ap}\end{pmatrix},
\end{align}
where $\M[9]_{\rm ap}$ is given by
\begin{align}
    \M[9]_{\rm ap}= \begin{pmatrix}1&\frac{1}{2}{}_{-1,0}\tilde{h}&\frac{1}{2}{}_{1,0}\tilde{h}&\frac{1}{2}{}_{-2,4}\tilde{h}&\frac{1}{2}{}_{2,-4}\tilde{h}&\frac{1}{2}{}_{-3,4}\tilde{h}&\frac{1}{2}{}_{3,-4}\tilde{h}&\frac{1}{2}{}_{-1,4}\tilde{h}&\frac{1}{2}{}_{1,-4}\tilde{h}\\ \frac{1}{2}{}_{1,0}\tilde{h}&\frac{1}{4}&\frac{1}{4}{}_{2,0}\tilde{h}&\frac{1}{4}{}_{-1,4}\tilde{h}&\frac{1}{4}{}_{3,-4}\tilde{h}&\frac{1}{4}{}_{-2,4}\tilde{h}&\frac{1}{4}{}_{4,-4}\tilde{h}&\frac{1}{4}{}_{0,4}\tilde{h}&\frac{1}{4}{}_{2,-4}\tilde{h}\\ \frac{1}{2}{}_{-1,0}\tilde{h}&\frac{1}{4}{}_{-2,0}\tilde{h}&\frac{1}{4}&\frac{1}{4}{}_{-3,4}\tilde{h}&\frac{1}{4}{}_{1,-4}\tilde{h}&\frac{1}{4}{}_{-4,4}\tilde{h}&\frac{1}{4}{}_{2,-4}\tilde{h}&\frac{1}{4}{}_{-2,4}\tilde{h}&\frac{1}{4}{}_{0,-4}\tilde{h}\\ \frac{1}{2}{}_{2,-4}\tilde{h}&\frac{1}{4}{}_{1,-4}\tilde{h}&\frac{1}{4}{}_{3,-4}\tilde{h}&\frac{1}{4}&\frac{1}{4}{}_{4,-8}\tilde{h}&\frac{1}{4}{}_{-1,0}\tilde{h}&\frac{1}{4}{}_{5,-8}\tilde{h}&\frac{1}{4}{}_{1,0}\tilde{h}&\frac{1}{4}{}_{3,-8}\tilde{h}\\ \frac{1}{2}{}_{-2,4}\tilde{h}&\frac{1}{4}{}_{-3,4}\tilde{h}&\frac{1}{4}{}_{-1,4}\tilde{h}&\frac{1}{4}{}_{-4,8}\tilde{h}&\frac{1}{4}&\frac{1}{4}{}_{-5,8}\tilde{h}&\frac{1}{4}{}_{1,0}\tilde{h}&\frac{1}{4}{}_{-3,8}\tilde{h}&\frac{1}{4}{}_{-1,0}\tilde{h}\\ \frac{1}{2}{}_{3,-4}\tilde{h}&\frac{1}{4}{}_{2,-4}\tilde{h}&\frac{1}{4}{}_{4,-4}\tilde{h}&\frac{1}{4}{}_{1,0}\tilde{h}&\frac{1}{4}{}_{5,-8}\tilde{h}&\frac{1}{4}&\frac{1}{4}{}_{6,-8}\tilde{h}&\frac{1}{4}{}_{2,0}\tilde{h}&\frac{1}{4}{}_{4,-8}\tilde{h}\\ \frac{1}{2}{}_{-3,4}\tilde{h}&\frac{1}{4}{}_{-4,4}\tilde{h}&\frac{1}{4}{}_{-2,4}\tilde{h}&\frac{1}{4}{}_{-5,8}\tilde{h}&\frac{1}{4}{}_{-1,0}\tilde{h}&\frac{1}{4}{}_{-6,8}\tilde{h}&\frac{1}{4}&\frac{1}{4}{}_{-4,8}\tilde{h}&\frac{1}{4}{}_{-2,0}\tilde{h}\\ \frac{1}{2}{}_{1,-4}\tilde{h}&\frac{1}{4}{}_{0,-4}\tilde{h}&\frac{1}{4}{}_{2,-4}\tilde{h}&\frac{1}{4}{}_{-1,0}\tilde{h}&\frac{1}{4}{}_{3,-8}\tilde{h}&\frac{1}{4}{}_{-2,0}\tilde{h}&\frac{1}{4}{}_{4,-8}\tilde{h}&\frac{1}{4}&\frac{1}{4}{}_{2,-8}\tilde{h}\\ \frac{1}{2}{}_{-1,4}\tilde{h}&\frac{1}{4}{}_{-2,4}\tilde{h}&\frac{1}{4}{}_{0,4}\tilde{h}&\frac{1}{4}{}_{-3,8}\tilde{h}&\frac{1}{4}{}_{1,0}\tilde{h}&\frac{1}{4}{}_{-4,8}\tilde{h}&\frac{1}{4}{}_{2,0}\tilde{h}&\frac{1}{4}{}_{-2,8}\tilde{h}&\frac{1}{4}\end{pmatrix}. \label{eq:9M_ap}
\end{align}
\Cref{fig:abs_pnt_maps_5x5,fig:abs_pnt_maps_9x9} present the estimated CMB and
residual maps derived from the $5\times5$ and $9\times9$ matrix map-making
approaches defined in \cref{eq:5M_ap,eq:9M_ap}, respectively. The $5\times5$
approach effectively isolates the temperature gradient-to-polarization leakage, with
the resulting temperature gradient maps displayed in the top panels of \cref{fig:abs_pnt_maps_5x5}
. This isolation successfully removes temperature leakage from the residual maps.
The corresponding systematic power spectrum, shown in \cref{fig:delta_cl_abs_wedge_pnt}
(left, orange solid line), exhibits a non-flat structure characteristic of CMB $E$-mode
polarization, though its amplitude remains suppressed through cross-linking.
The $9\times9$ approach achieves complete mitigation of systematic effects arising
from absolute pointing offset, as evidenced by the residual maps shown in the
bottom panels of \cref{fig:abs_pnt_maps_9x9}. This comprehensive approach enables
detailed examination of all systematic field components, including temperature-gradient
fields ($\hZ[\pm1,0]$) and polarization-gradient fields ($\hZ[\pm1,\mp4], \hZ[\pm
3,\mp4]$), providing valuable insights into the magnitude of spurious signals generated
by systematic effects associated with specific \spin moments.

Analysis of the tensor-to-scalar ratio bias yielded $\Delta r < 10^{-6}$ for all
scenarios driven by the matrices we used with the absolute pointing offset parameters
$(\rho,\chi)=(1^{\prime},0^{\prime})$.

\subsection{Pointing disturbance due to HWP rotation}
\label{sec:wedge}

\begin{figure}[h]
    \centering
    \includegraphics[width=0.32\columnwidth]{
        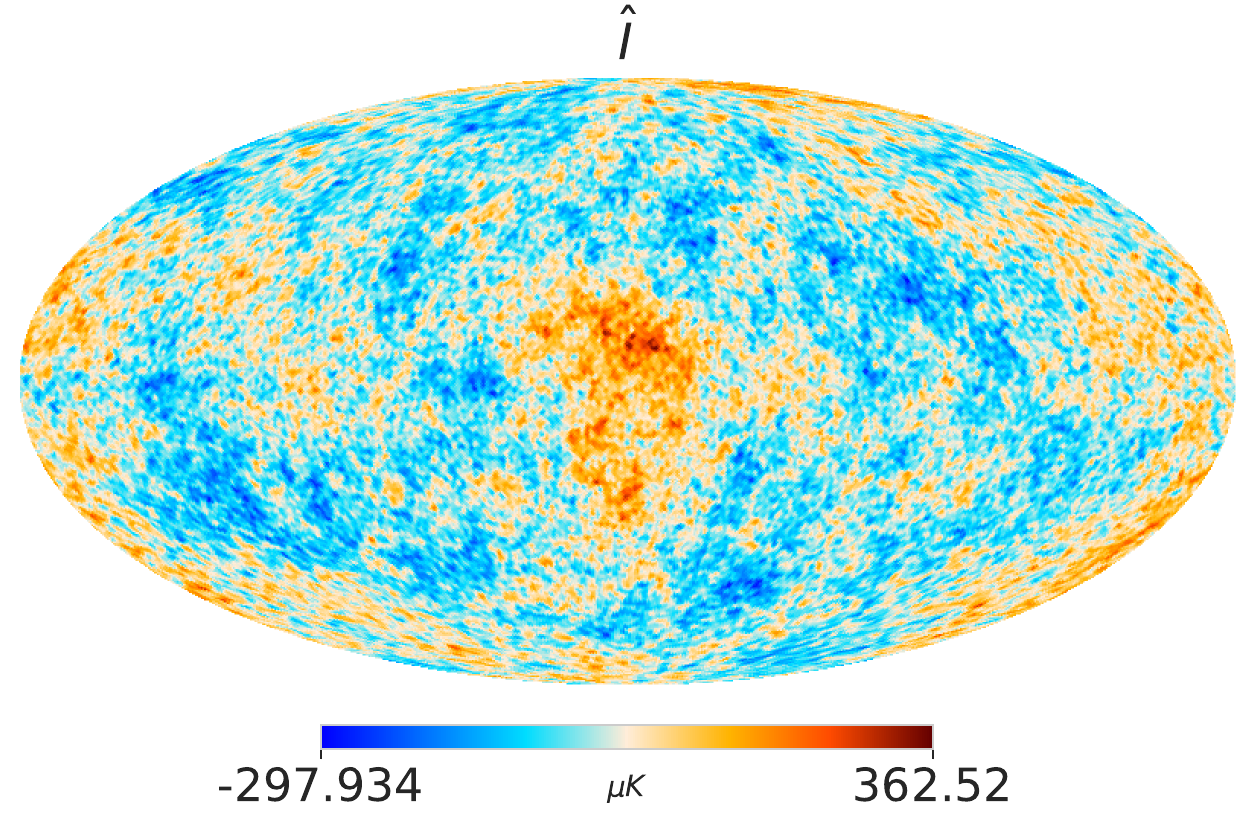
    }
    \includegraphics[width=0.32\columnwidth]{
        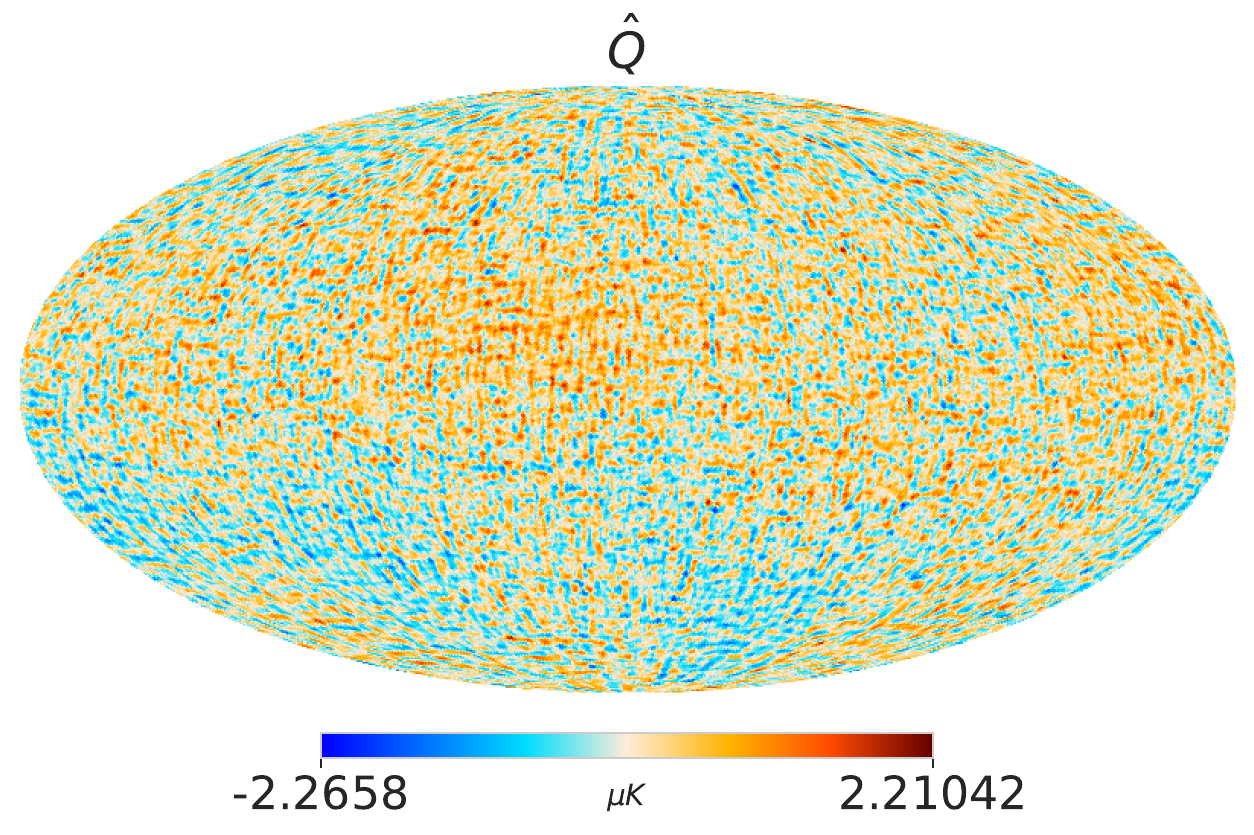
    }
    \includegraphics[width=0.32\columnwidth]{
        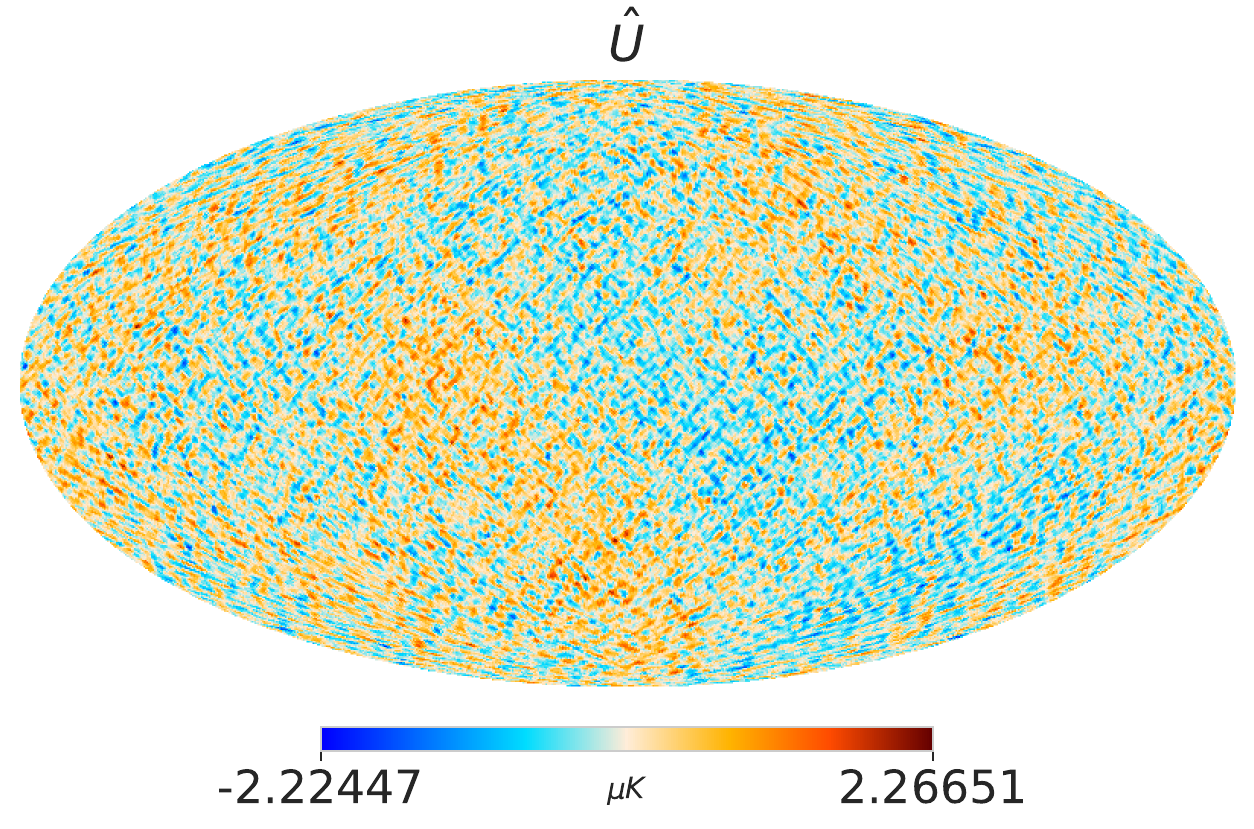
    }
    \\
    \includegraphics[width=0.32\columnwidth]{
        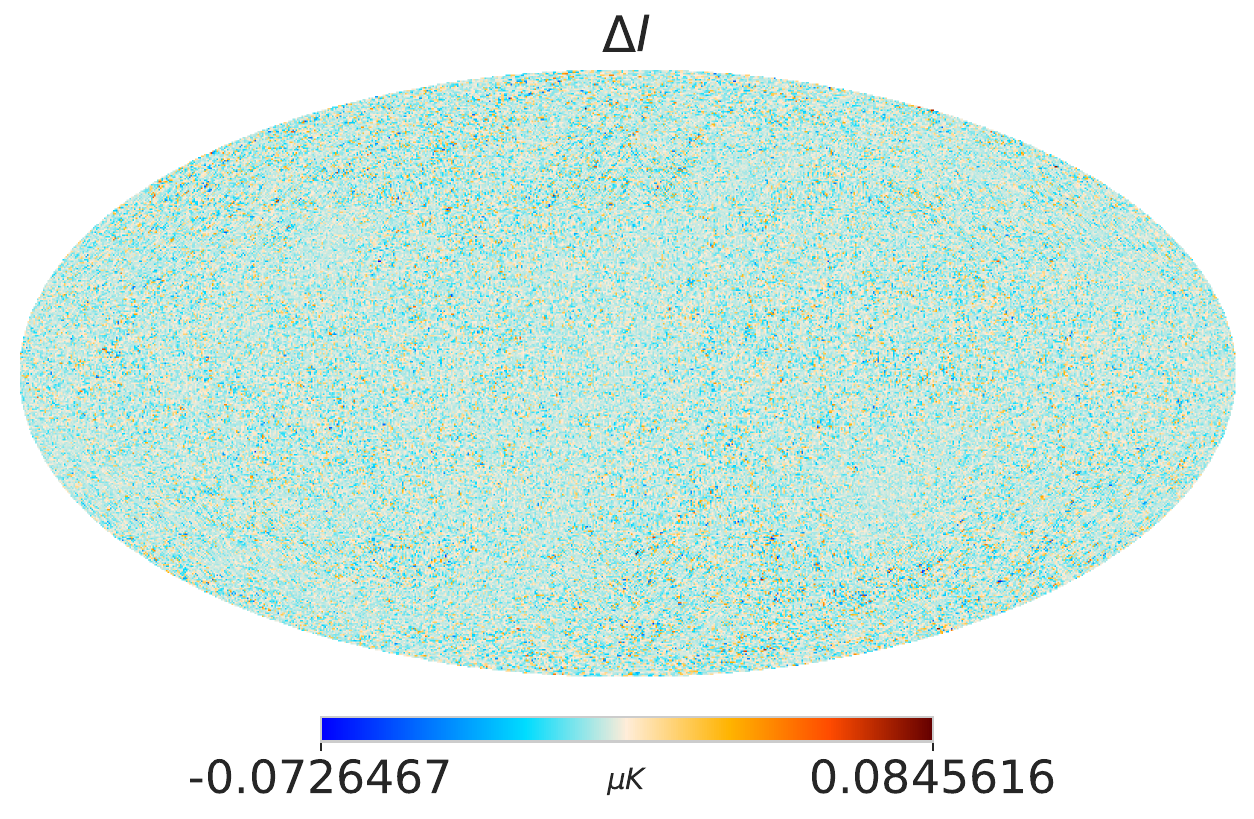
    }
    \includegraphics[width=0.32\columnwidth]{
        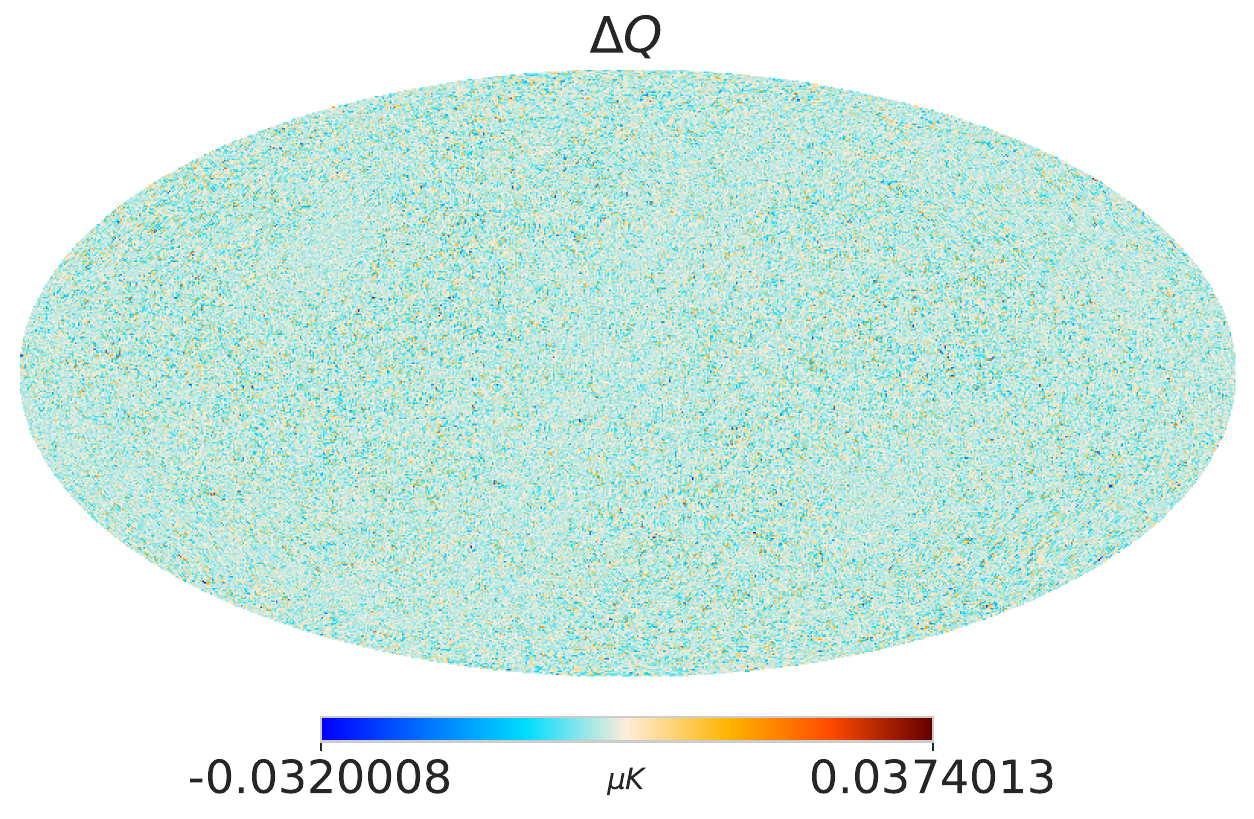
    }
    \includegraphics[width=0.32\columnwidth]{
        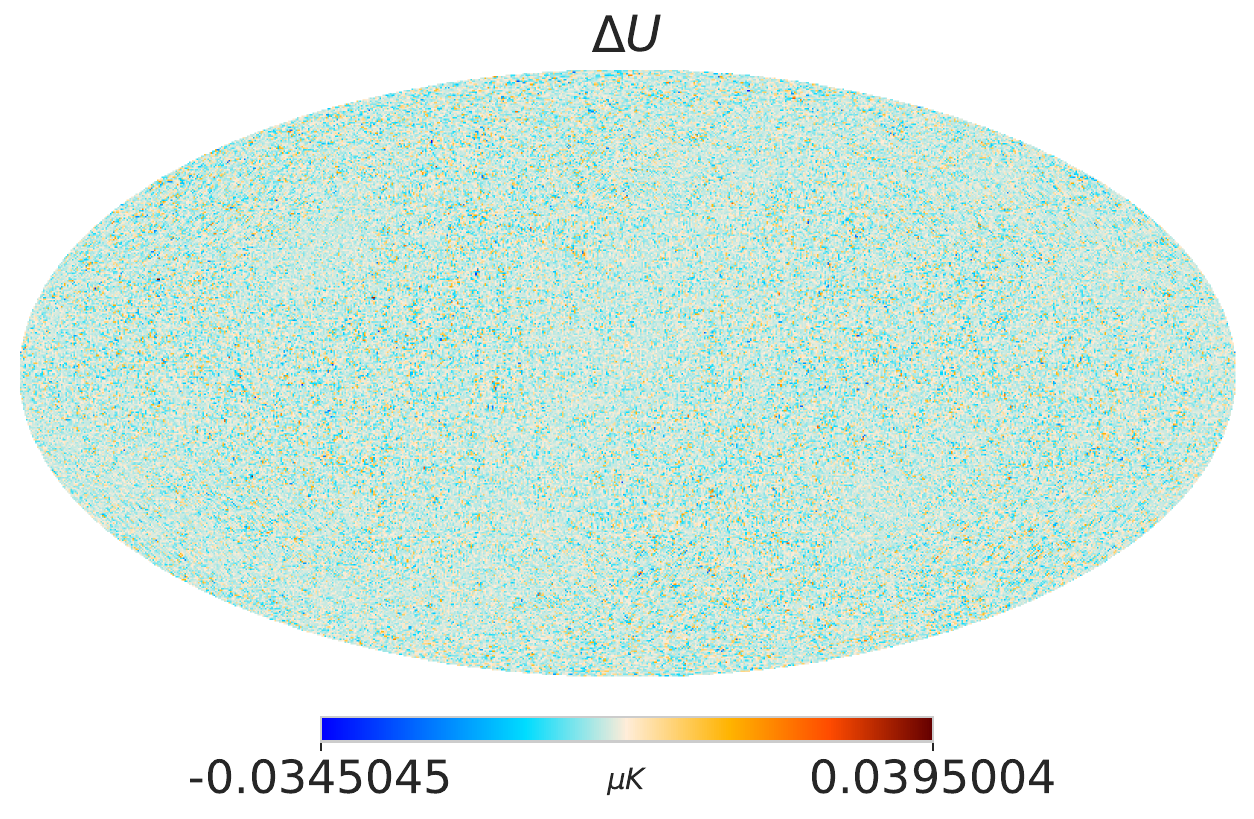
    }
    \caption[ Estimated CMB maps and residual maps due to the HWP wedge effect by
    the $3\times3$ matrix map-making approach with HWP. ]{Estimated CMB maps and
    residual maps due to the HWP wedge effect by the $3\times3$ matrix map-making
    approach with HWP. It displays $\hat{I}$, $\hat{Q}$, $\hat{U}$, $\Delta I$, $\Delta
    Q$, and $\Delta U$ from top left to bottom right. The imposed systematic parameter
    is $(\xi,\chi)=(1^{\prime},0^{\prime})$}
    \label{fig:wedge_maps_3x3}
\end{figure}

The pointing systematic effects arising from the HWP wedge angle exhibit
characteristics analogous to those observed in the absolute pointing offset
scenario. We impose the systematic parameters $(\xi,\chi)=(1^{\prime},0^{\prime})$
which describes the pointing perturbation due to the HWP wedge angle and HWP phase,
respectably. \Cref{fig:wedge_maps_3x3} illustrates the estimated CMB maps and corresponding
residual maps generated through the $3\times3$ matrix map-making framework outlined
in \cref{eq:3M_ap}.

Although the systematic field differs from the absolute pointing offset case, we
can employ the same $3 \times 3$ map-maker, as the underlying mathematical framework
remains consistent. Comparing the residual maps ($\Delta Q$ and $\Delta U$)
between the absolute offset of the pointing (\cref{fig:abs_pnt_maps_3x3}) and
the HWP wedge effect reveals remarkably similar magnitudes and structures, which
aligns with our expectations given that the HWP rotation effectively minimizes cross-linking
in both cases.

In particular, while the temperature residual ($\Delta I$) does not directly
contribute to systematic contamination of the $B$ mode, the HWP wedge effect exhibits
smaller temperature residuals compared to the absolute pointing offset. This
phenomenon arises from the different coupling mechanisms: in the wedge effect of
the HWP, the temperature gradient fields $\St[\mp1,\mp1]$ are coupled with
$\h[\pm1,\pm1]$ without the contribution of the HWP (\cref{eq:00Sd_wedge}), whereas
the absolute pointing offset involves the coupling between $\St[\mp1,0]$ and $\h[
\pm1,0]$ (\cref{eq:00Sd_abs_pnt}). The HWP rotation in the wedge effect
effectively averages signals around sky pixels, resulting in reduced temperature
residuals.

The systematic power spectra originating from the HWP wedge effect are presented
in \cref{fig:delta_cl_abs_wedge_pnt} (right). The solid blue line depicts the
spectrum derived from the $3\times3$ matrix map-making approach, where the systematic
effect manifests itself primarily as leakage $T \to B$, analogous to the case of
absolute pointing offset. The comparable magnitudes of residual maps in both cases
naturally lead to similar levels in their systematic power spectra.

\begin{figure}[h]
    \centering
    \includegraphics[width=0.32\columnwidth]{
        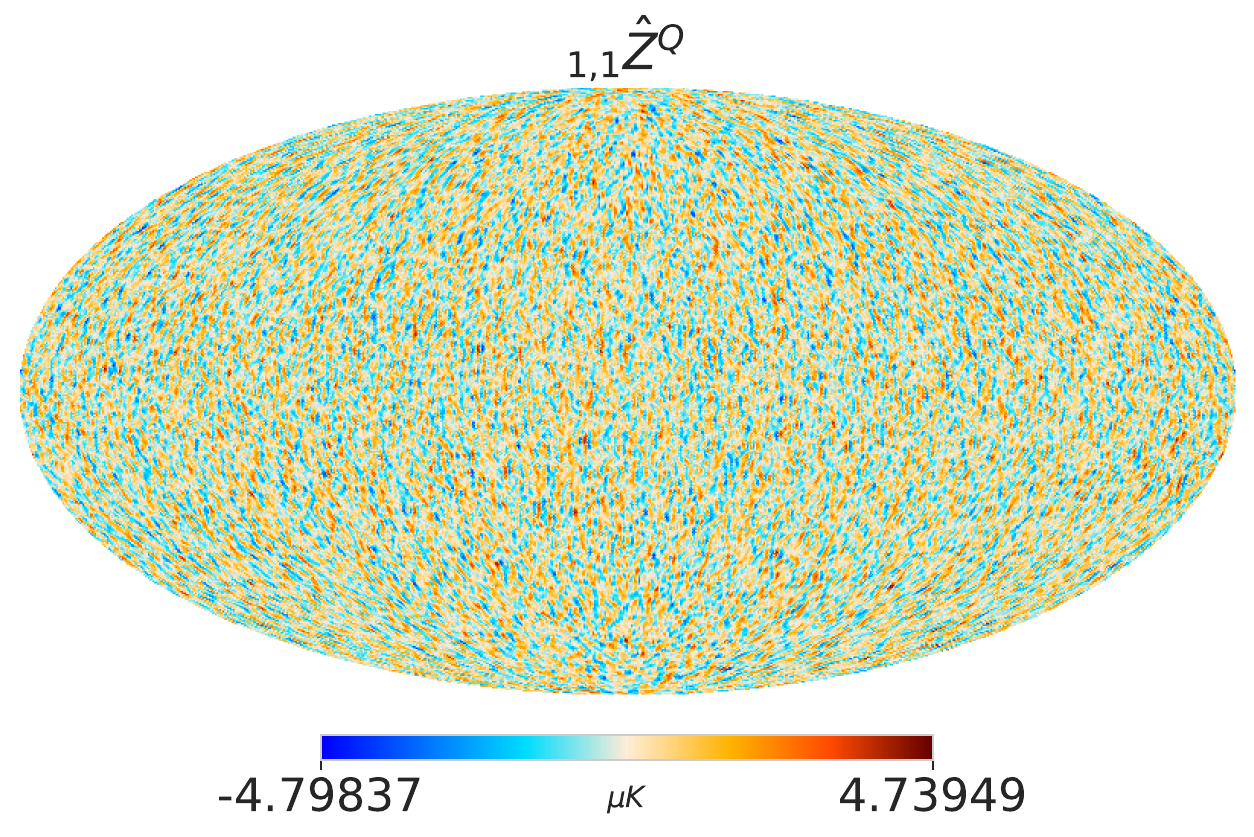
    }
    \includegraphics[width=0.32\columnwidth]{
        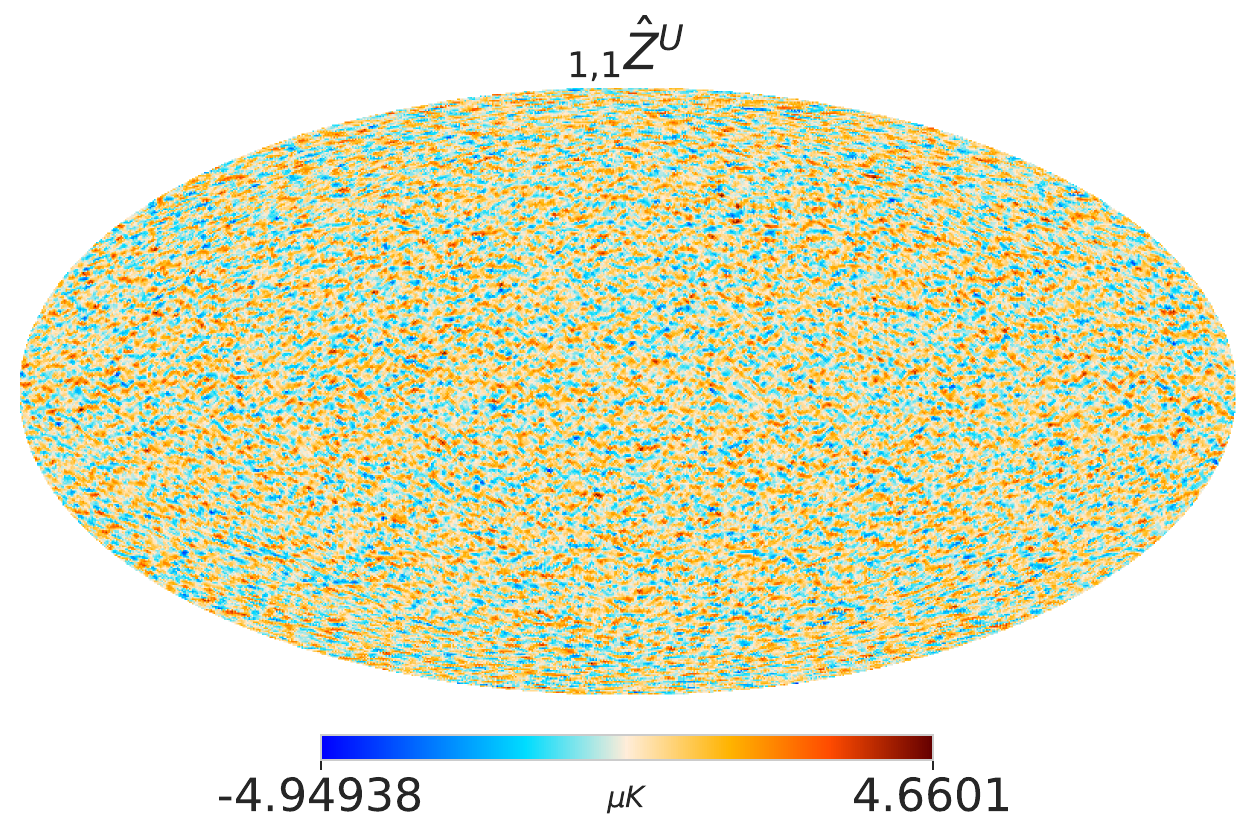
    }
    \\
    \includegraphics[width=0.32\columnwidth]{
        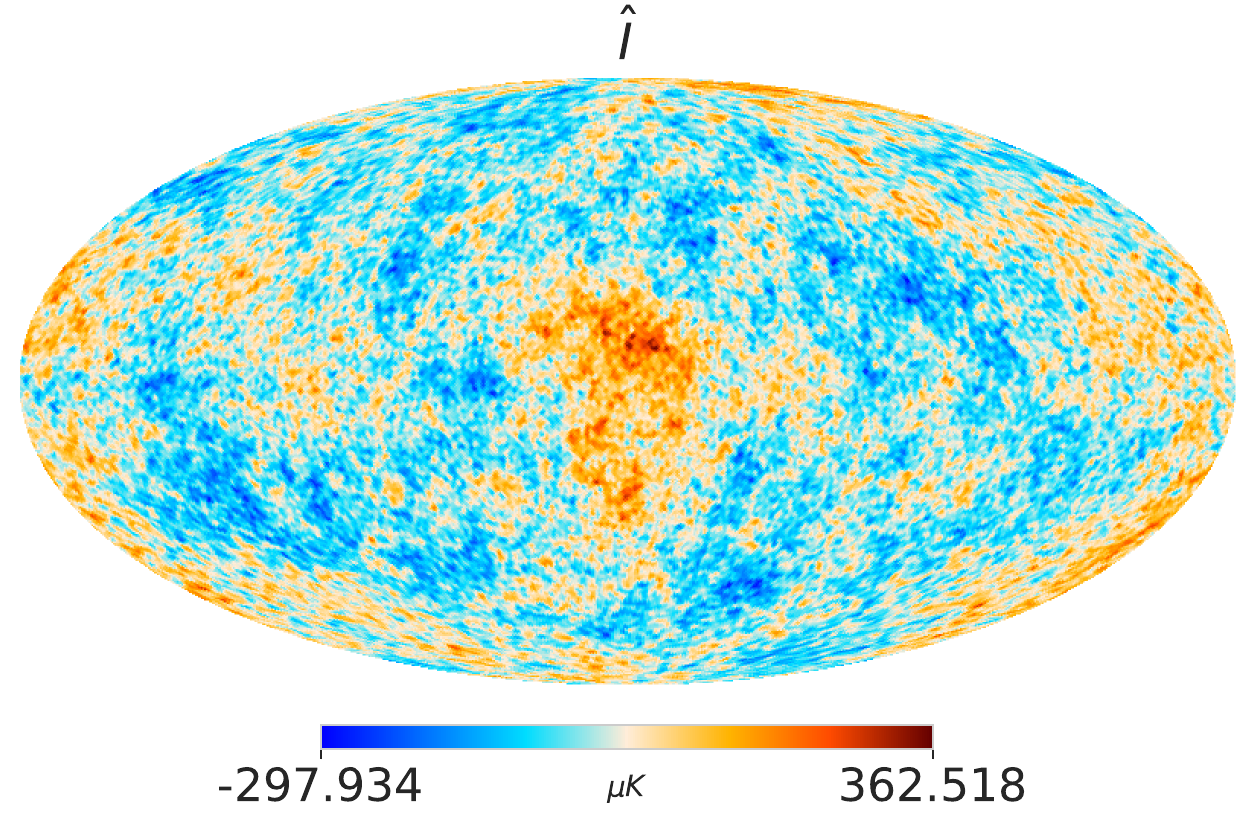
    }
    \includegraphics[width=0.32\columnwidth]{
        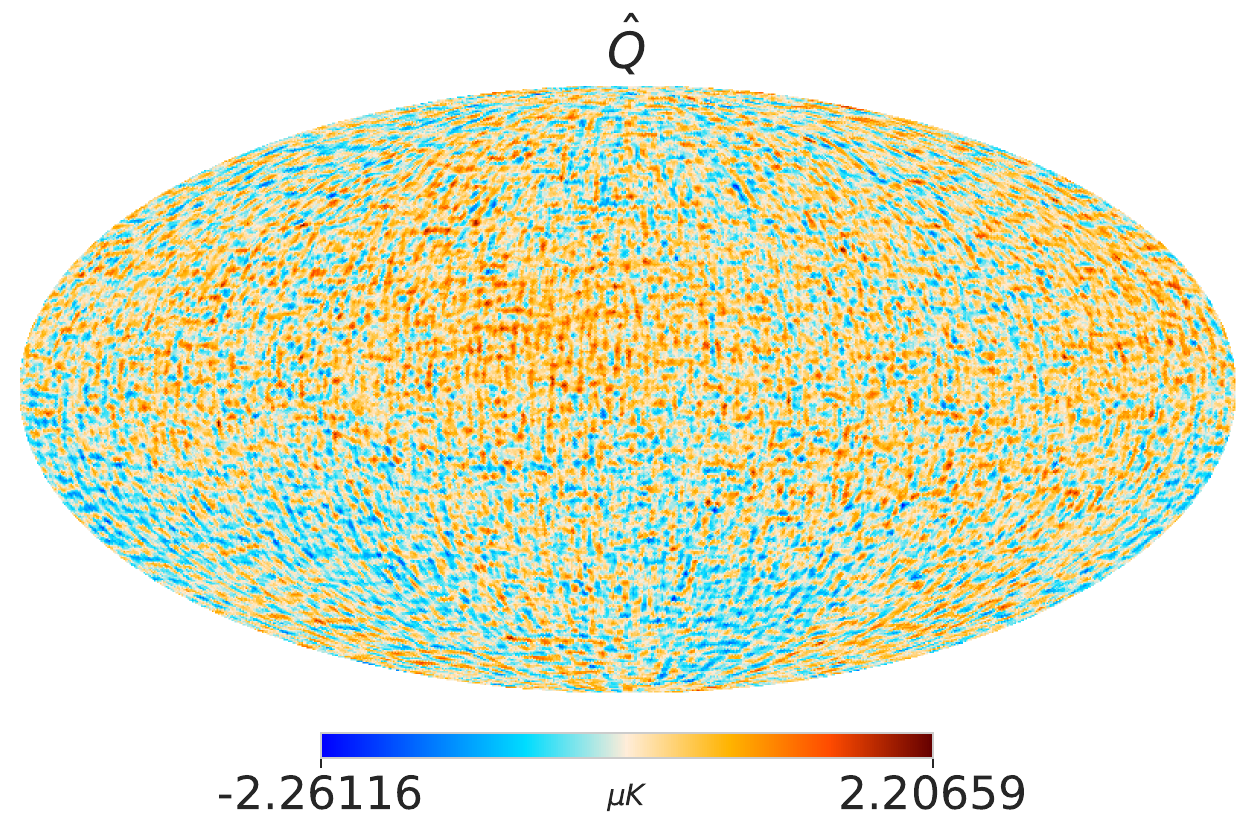
    }
    \includegraphics[width=0.32\columnwidth]{
        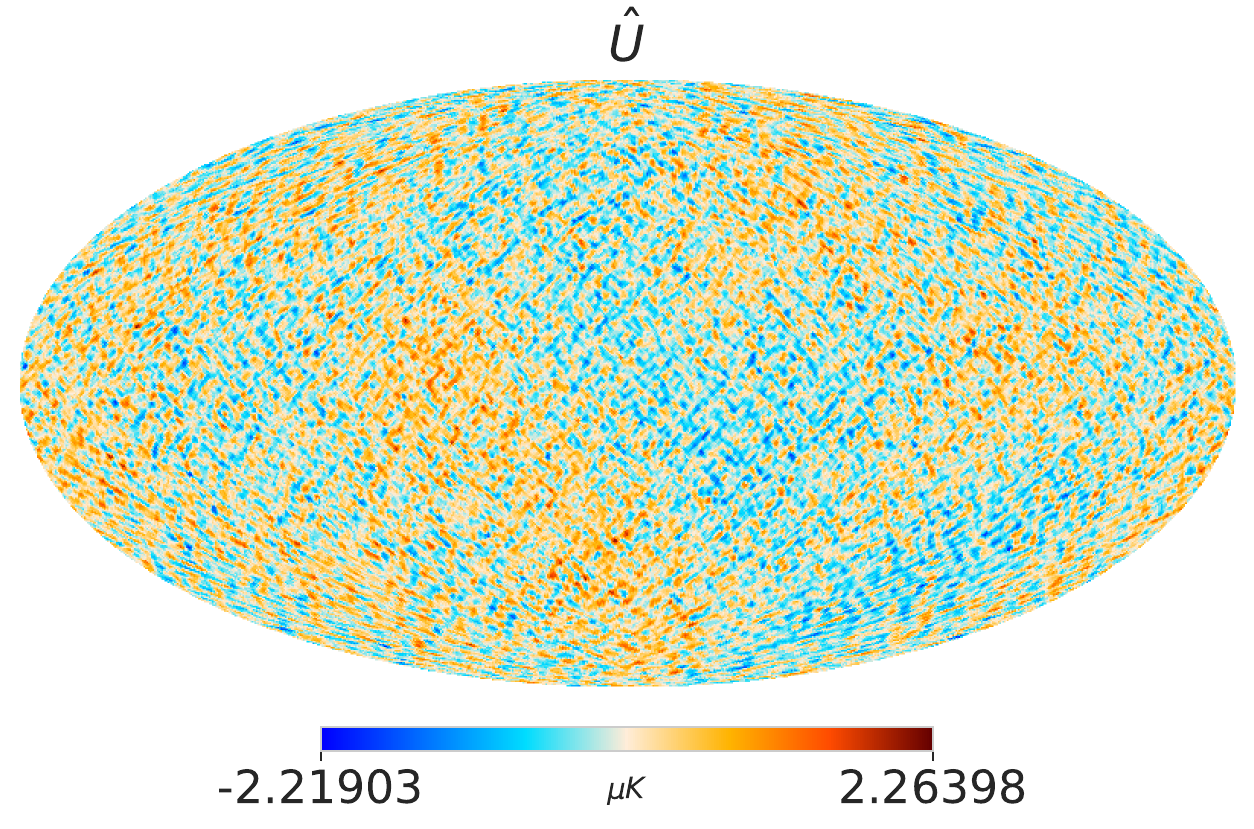
    }
    \\
    \includegraphics[width=0.32\columnwidth]{
        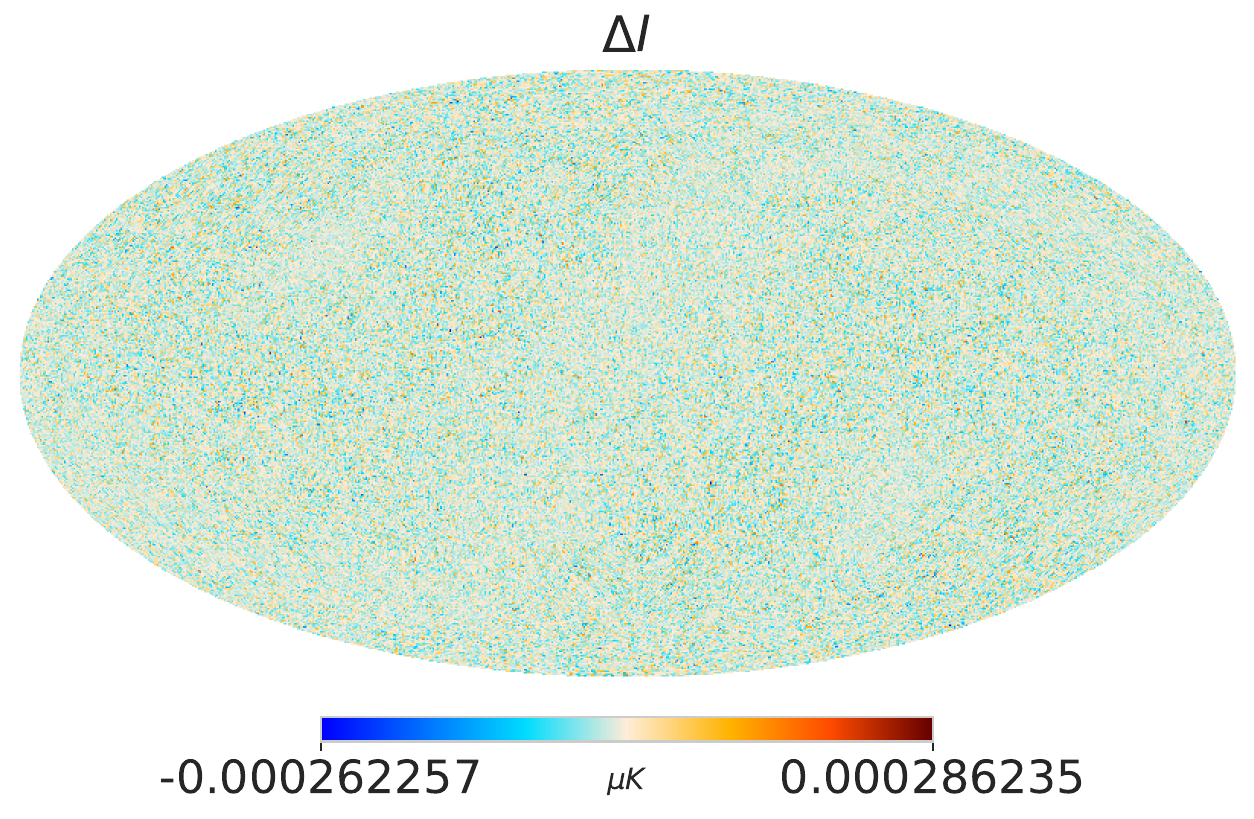
    }
    \includegraphics[width=0.32\columnwidth]{
        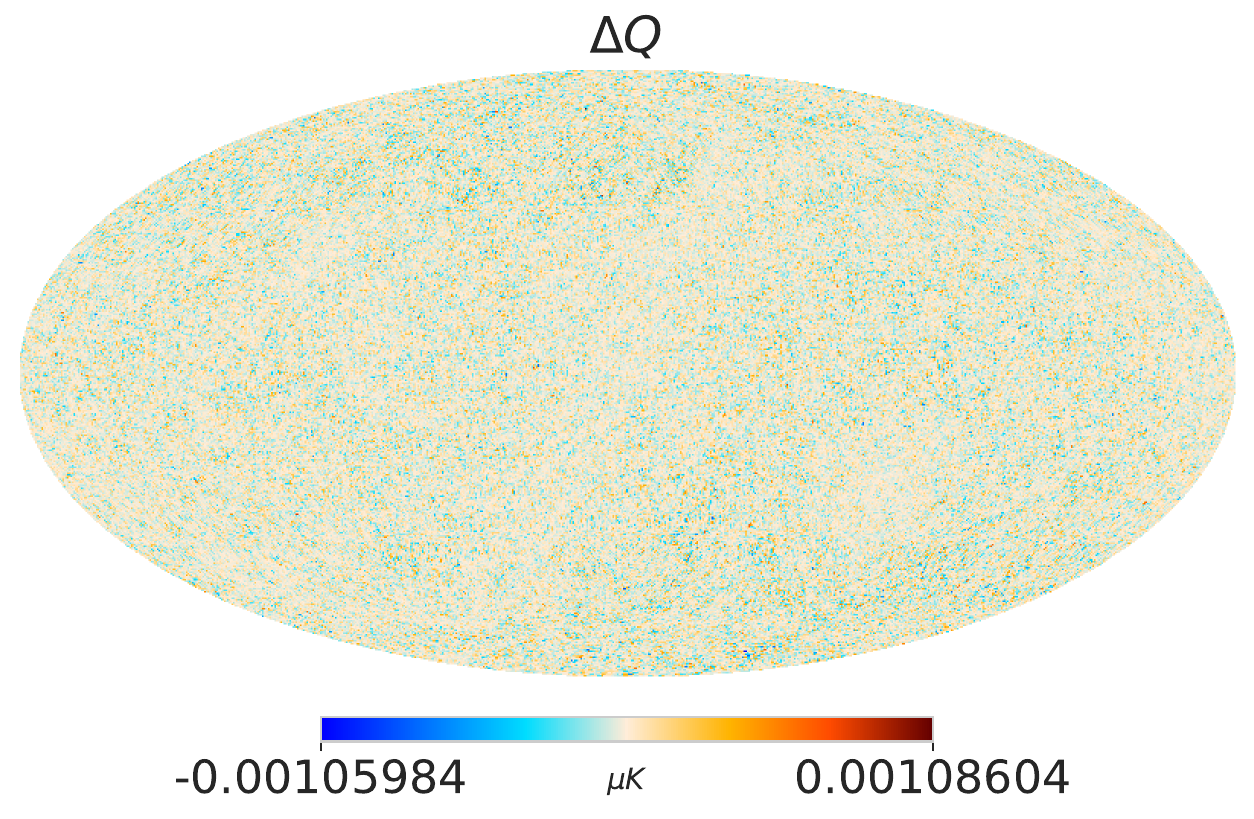
    }
    \includegraphics[width=0.32\columnwidth]{
        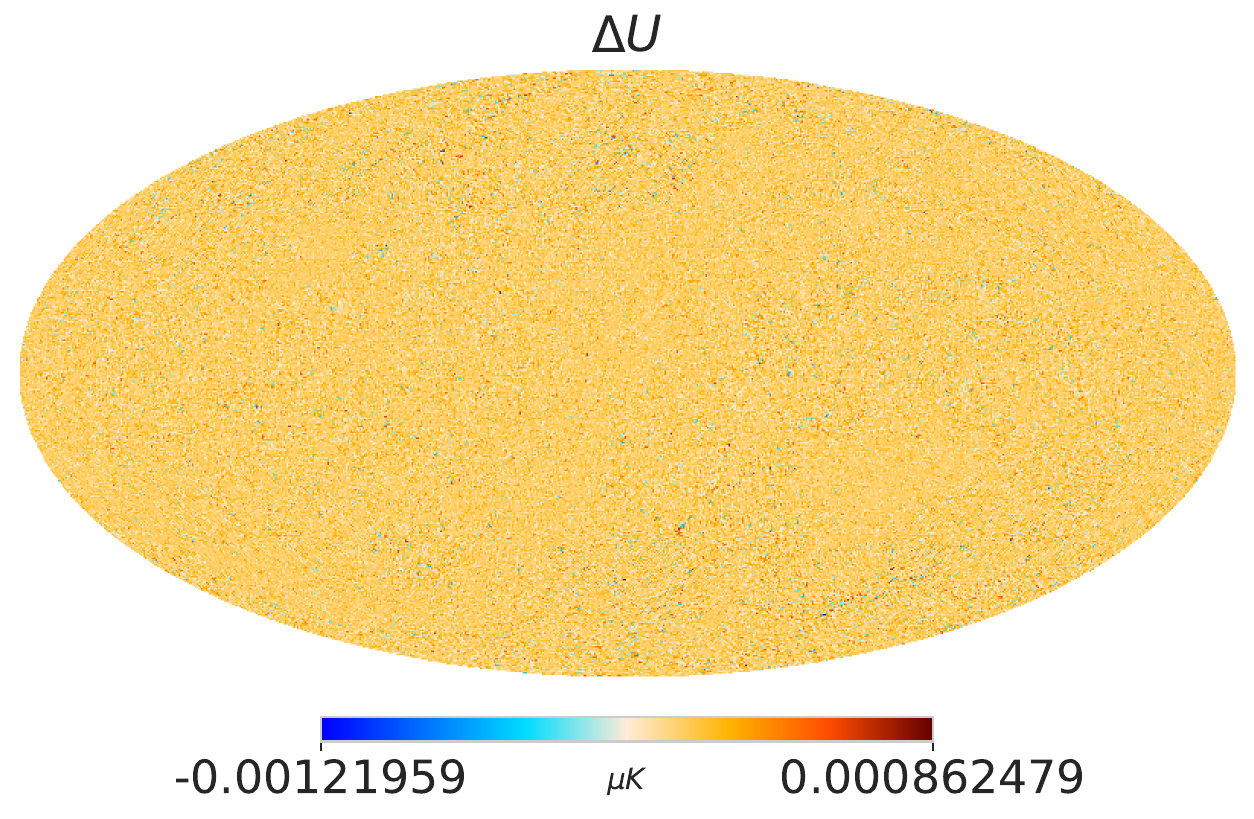
    }
    \caption[Estimated CMB maps and residual maps due to the HWP wedge effect by
    the $5\times5$ matrix map-making approach with HWP.]{Estimated CMB maps and
    residual maps due to the HWP wedge effect by the $5\times5$ matrix map-making
    approach with HWP. It displays $\hZ[1,1]^{Q}$, $\hZ[1,1]^{U}$, $\hat{I}$,
    $\hat{Q}$, $\hat{U}$, $\Delta I$, $\Delta Q$, and $\Delta U$ from top left
    to bottom right. The systematics parameter for the HWP wedge is same as the previous
    $3\times3$ matrix map-making approach case. The imposed systematics
    parameter is $(\xi,\chi)=(1^{\prime},0^{\prime})$.}
    \label{fig:wedge_maps_5x5}
\end{figure}

\Cref{fig:wedge_maps_5x5} presents the estimated CMB maps and residual maps generated
using the $5\times5$ matrix map-making approach defined by
\begin{align}
    \ab(\begin{matrix}\hat{I}\\{}_{1,1}\hat{Z}\\{}_{-1,-1}\hat{Z}\\ \hat{P}\\ \hat{P^*}\end{matrix}) & =\M[5]_{\rm w}^{-1}\ab(\begin{matrix}{}_{0,0}{\tilde{S}^d}_{\rm w}\\{}_{1,1}{\tilde{S}^d}_{\rm w}\\{}_{-1,-1}{\tilde{S}^d}_{\rm w}\\{}_{2,-4}{\tilde{S}^d}_{\rm w}\\{}_{-2,4}{\tilde{S}^d}_{\rm w}\end{matrix}),
\end{align}
where $\M[5]_{\rm w}$ is given by
\begin{align}
    \M[5]_{\rm w}= \ab(\begin{matrix}1&\frac{1}{2}{}_{-1,-1}\tilde{h}&\frac{1}{2}{}_{1,1}\tilde{h}&\frac{1}{2}{}_{-2,4}\tilde{h}&\frac{1}{2}{}_{2,-4}\tilde{h}\\ \frac{1}{2}{}_{1,1}\tilde{h}&\frac{1}{4}&\frac{1}{4}{}_{2,2}\tilde{h}&\frac{1}{4}{}_{-1,5}\tilde{h}&\frac{1}{4}{}_{3,-3}\tilde{h}\\ \frac{1}{2}{}_{-1,-1}\tilde{h}&\frac{1}{4}{}_{-2,-2}\tilde{h}&\frac{1}{4}&\frac{1}{4}{}_{-3,3}\tilde{h}&\frac{1}{4}{}_{1,-5}\tilde{h}\\ \frac{1}{2}{}_{2,-4}\tilde{h}&\frac{1}{4}{}_{1,-5}\tilde{h}&\frac{1}{4}{}_{3,-3}\tilde{h}&\frac{1}{4}&\frac{1}{4}{}_{4,-8}\tilde{h}\\ \frac{1}{2}{}_{-2,4}\tilde{h}&\frac{1}{4}{}_{-3,3}\tilde{h}&\frac{1}{4}{}_{-1,5}\tilde{h}&\frac{1}{4}{}_{-4,8}\tilde{h}&\frac{1}{4}\end{matrix}).\label{eq:5M_w}
\end{align}
This enhanced approach successfully isolates the temperature leakage,
transforming the dominant systematic effect from $T \to B$ to predominantly
$E \to B$ leakage.

The systematic power spectrum obtained through the $5\times5$ matrix map-making
approach is depicted by the orange solid line in
\cref{fig:delta_cl_abs_wedge_pnt} (right). Notably, the $\Delta C_{\ell}^{BB}$
exhibits a completely flat structure, contrasting with the results from the $5\times
5$ matrix approach applied to absolute pointing offset. This distinction arises
from the fundamental differences in their coupling mechanisms between the
systematic fields and cross-link maps. In the absolute pointing offset scenario,
when $\h[n,0]$ couples with the polarization gradient without HWP contribution (see
\cref{eq:24Sd_ap}), the structural integrity of the $\h[n,0]$ map (illustrated
in \cref{fig:spin-n0_xlink_maps}) persists, facilitating phase-coherent leakage
from $E$-mode to $B$-mode polarization.

Conversely, the HWP wedge effect demonstrates no $m=0$ cross-links in \cref{eq:24Sd_wedge},
indicating that all systematic effect components are suppressed by $\h[n,m]$ ($m
\neq 0$) through HWP contribution. These cross-link maps exhibit highly flat structures
due to HWP rotation (as shown in \cref{fig:spin-n4_xlink_maps}), manifesting as
a flat angular power spectrum. When convolved with any sky signal, regardless of
its intrinsic angular structure, this randomization process effectively smooths out
spatial correlations, yielding a flat angular power spectrum. This mechanism parallels
the $E$ mode to $B$ mode conversion induced by gravitational lensing (discussed
in \cref{sec:lensing}): while absolute offset produces phase-coherent $E \to B$
leakage, the HWP wedge effect smooth out its phase information.

\begin{figure}[h]
    \centering
    \includegraphics[width=0.32\columnwidth]{
        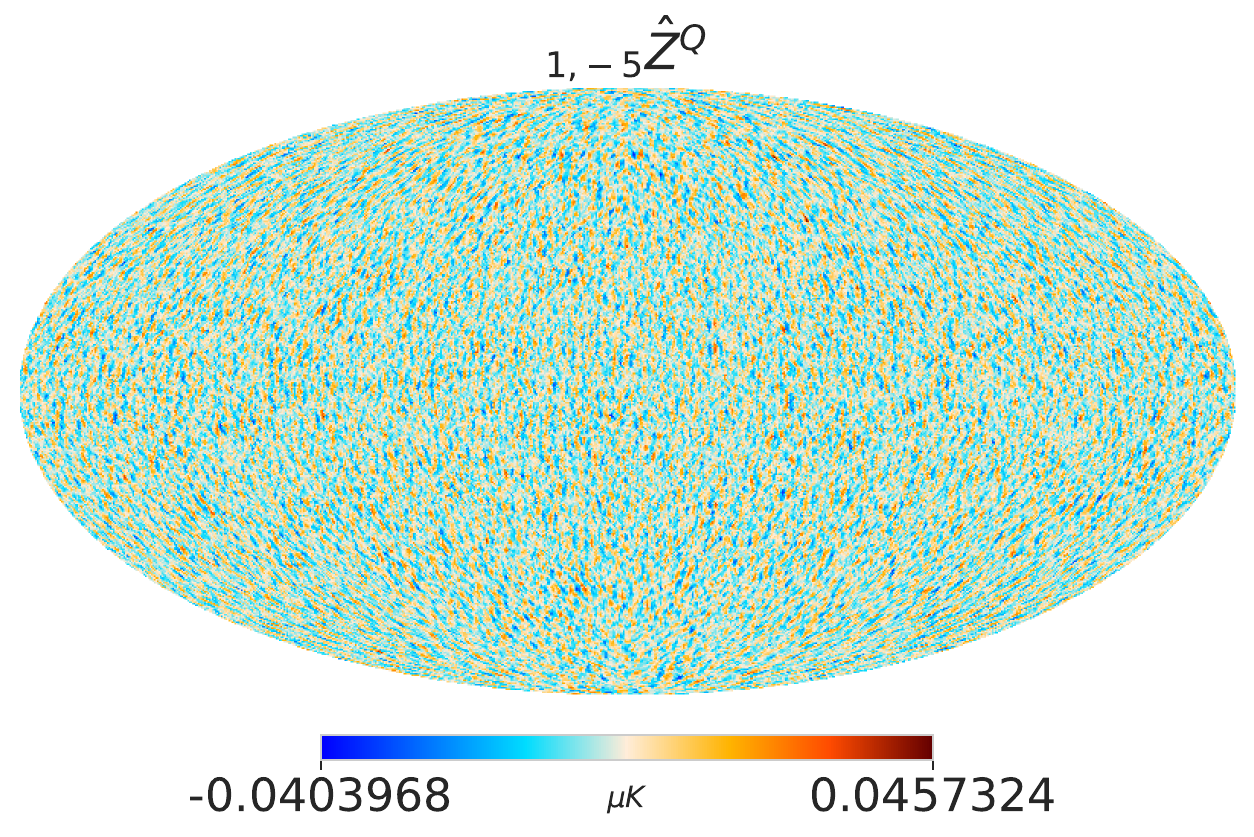
    }
    \includegraphics[width=0.32\columnwidth]{
        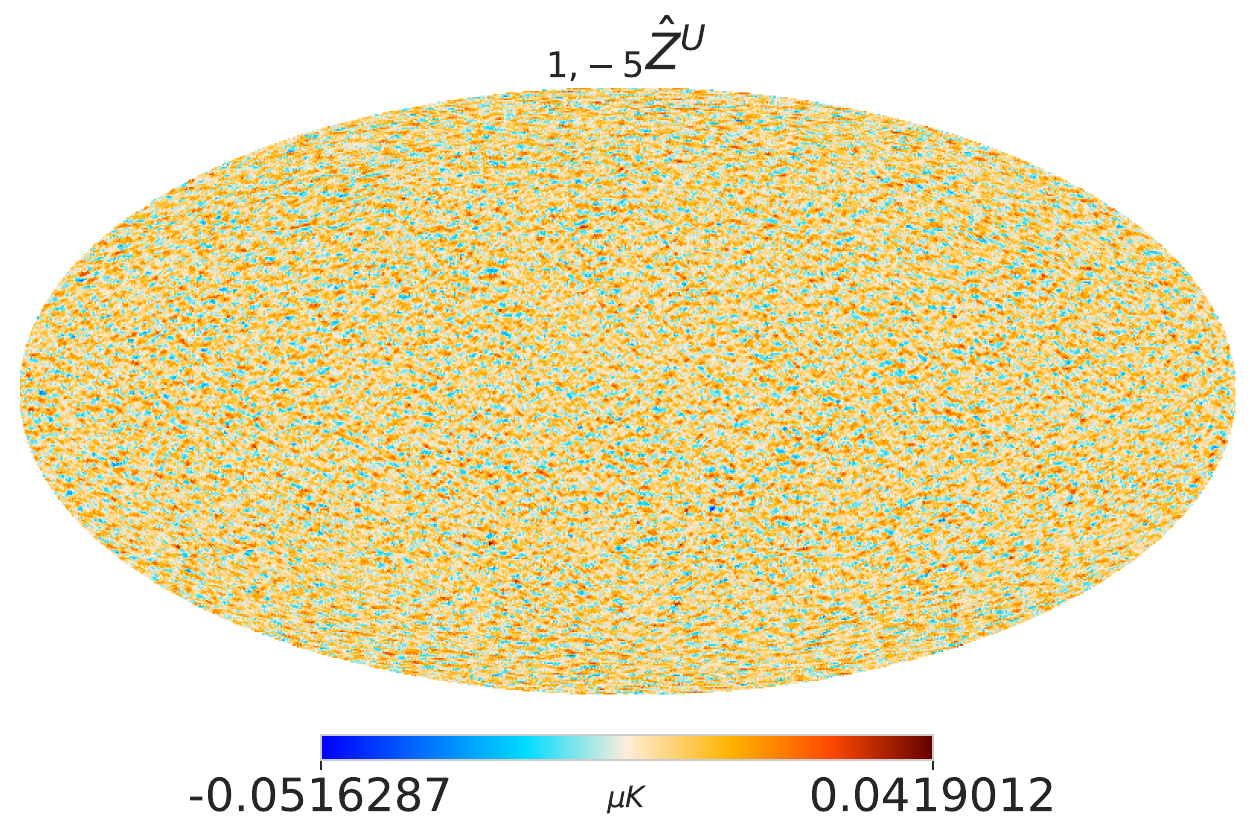
    }
    \includegraphics[width=0.32\columnwidth]{
        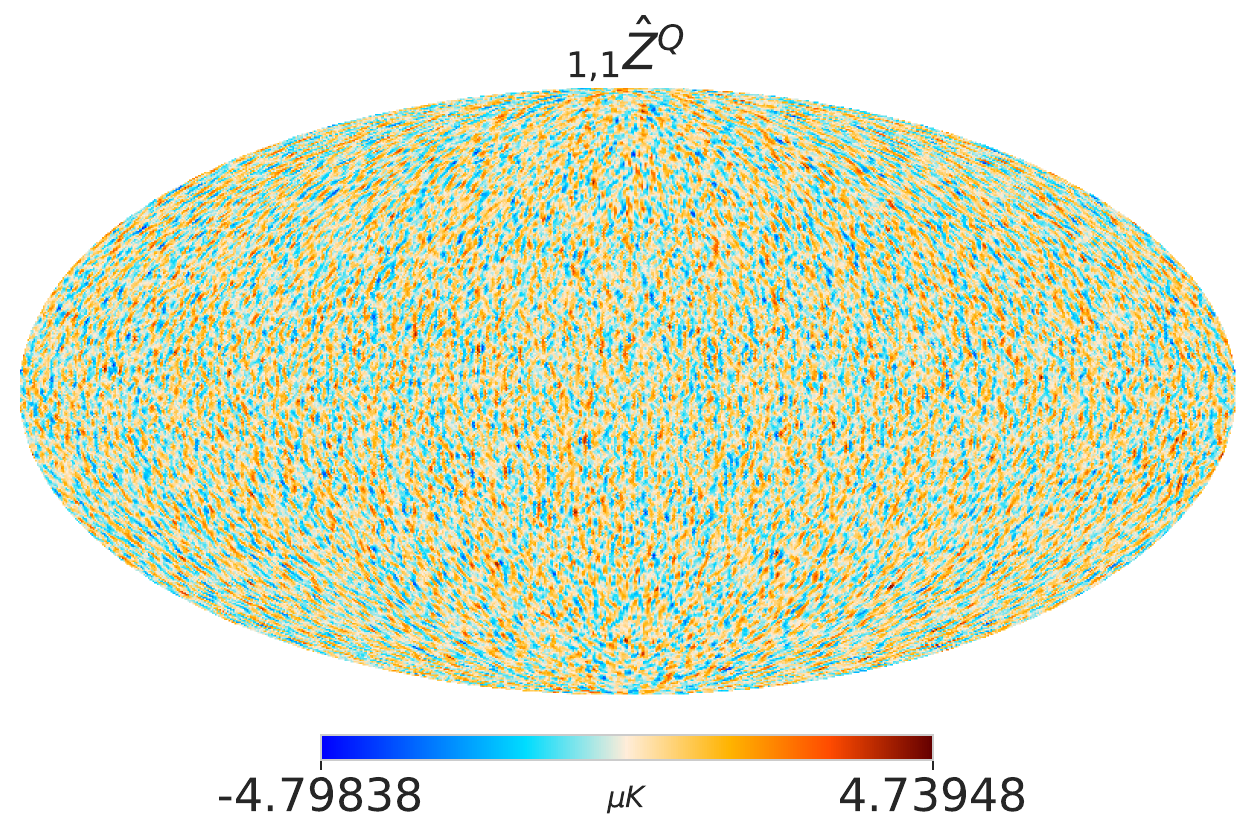
    }
    \\
    \includegraphics[width=0.32\columnwidth]{
        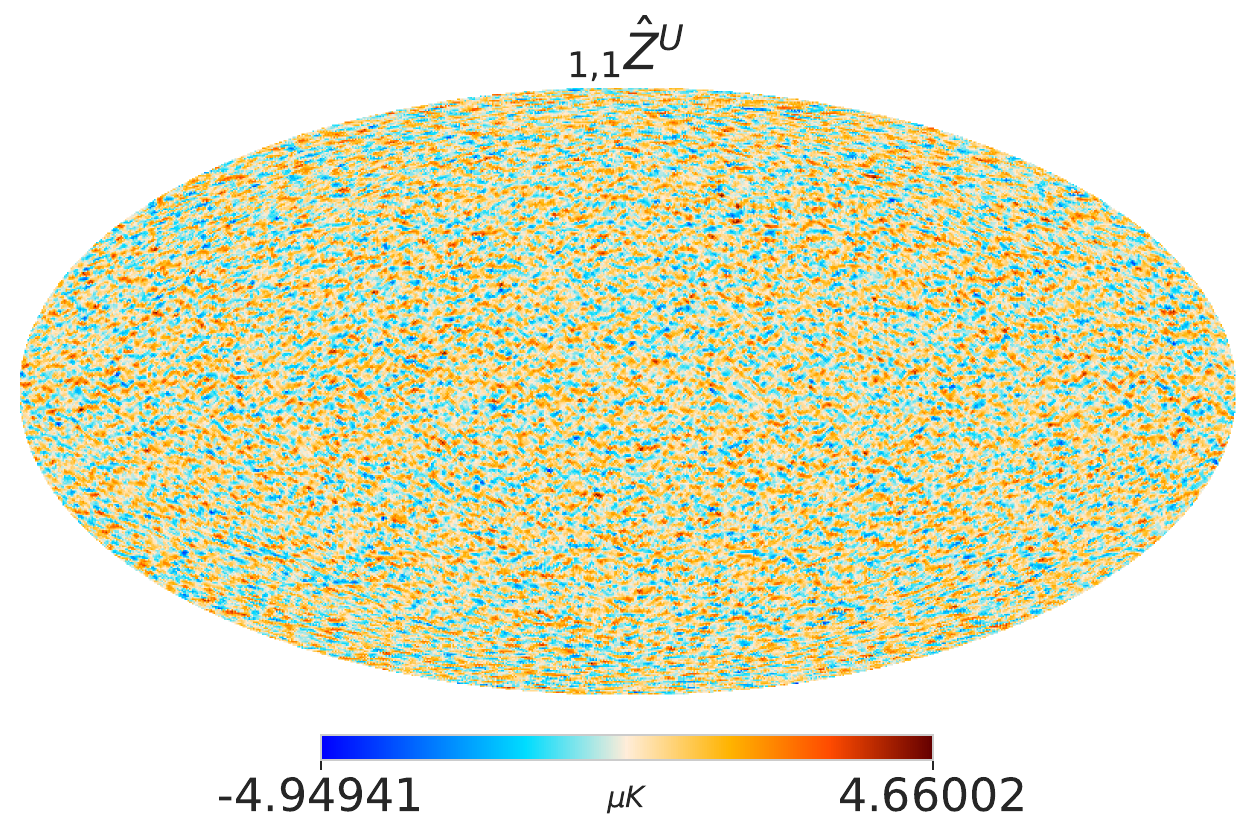
    }
    \includegraphics[width=0.32\columnwidth]{
        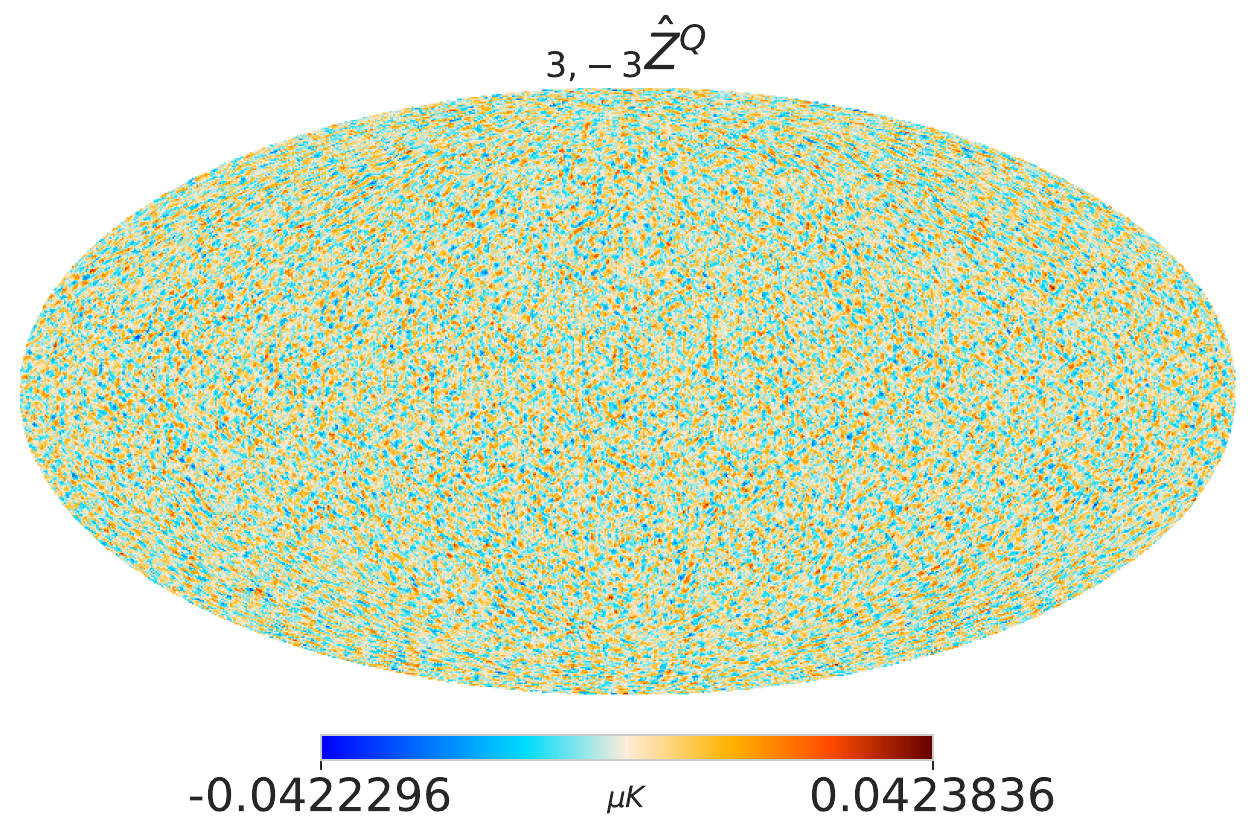
    }
    \includegraphics[width=0.32\columnwidth]{
        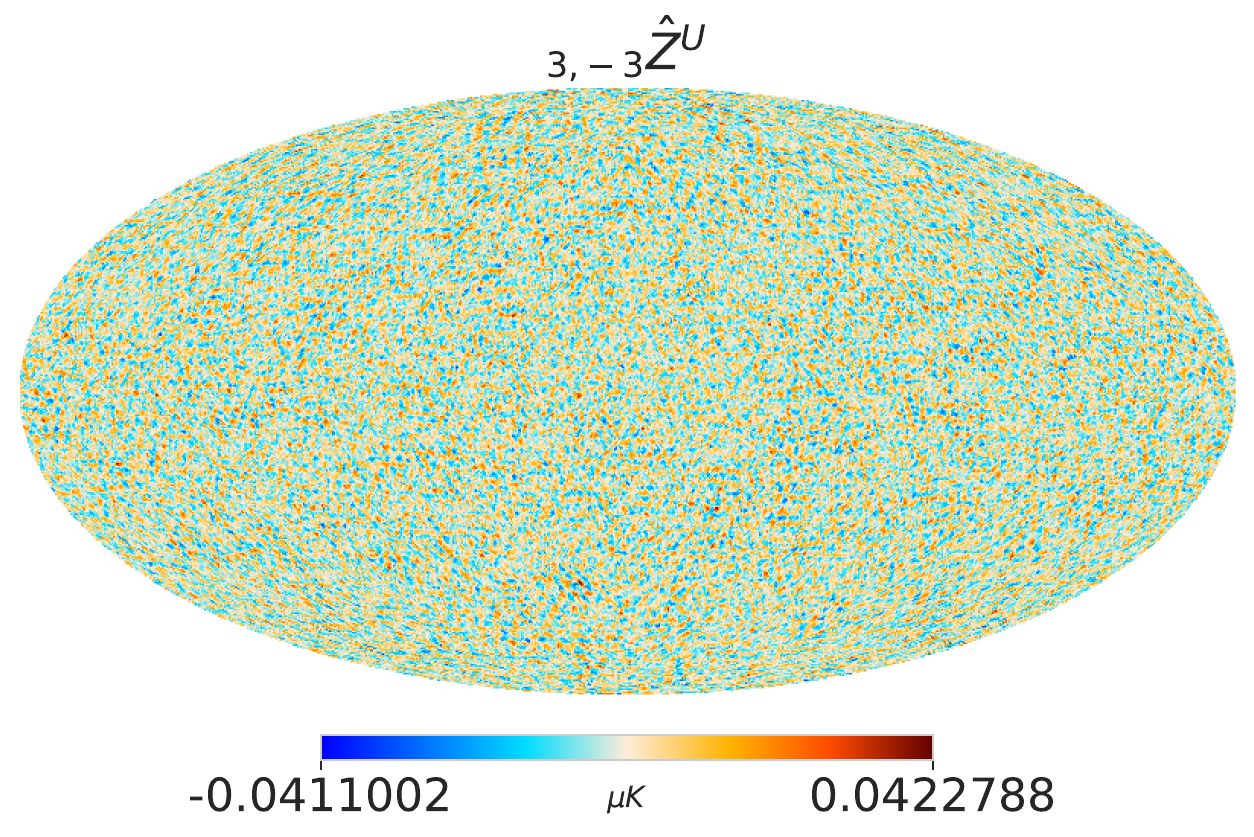
    }
    \\
    \includegraphics[width=0.32\columnwidth]{
        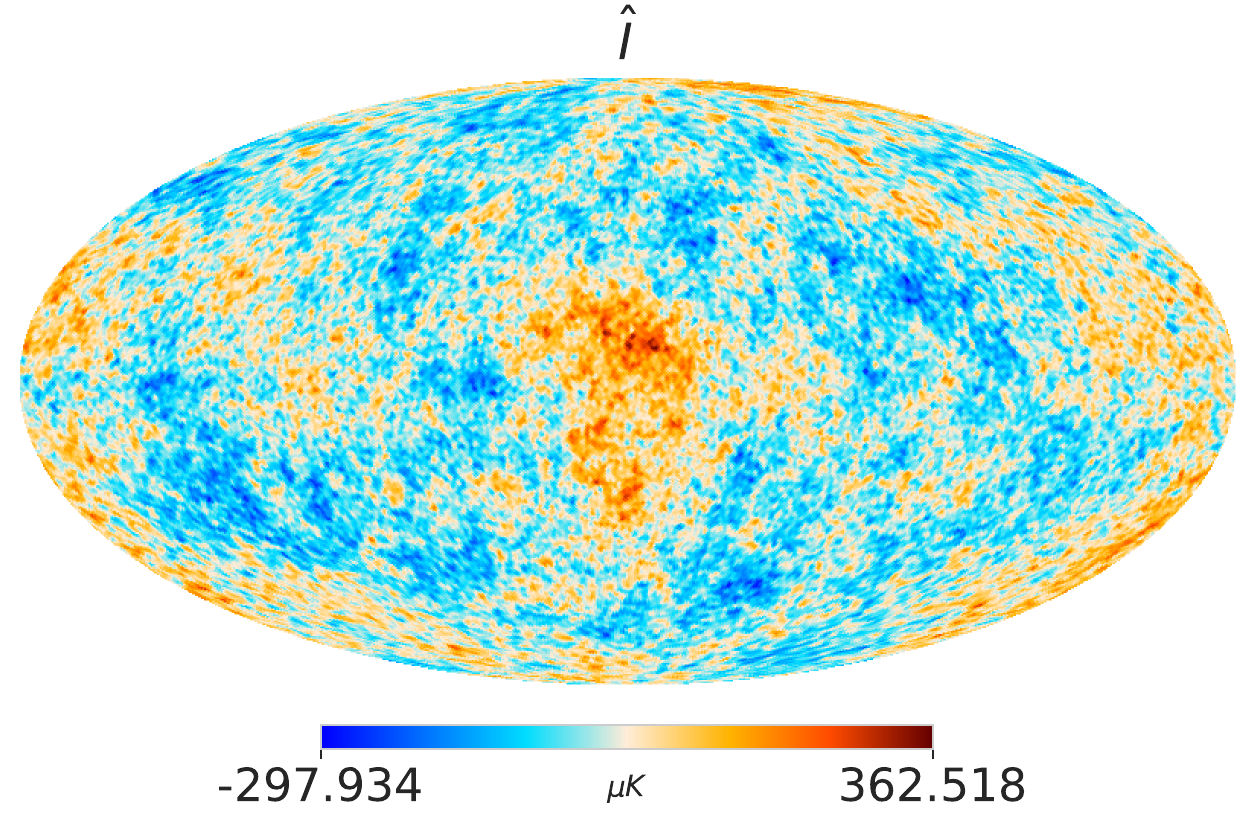
    }
    \includegraphics[width=0.32\columnwidth]{
        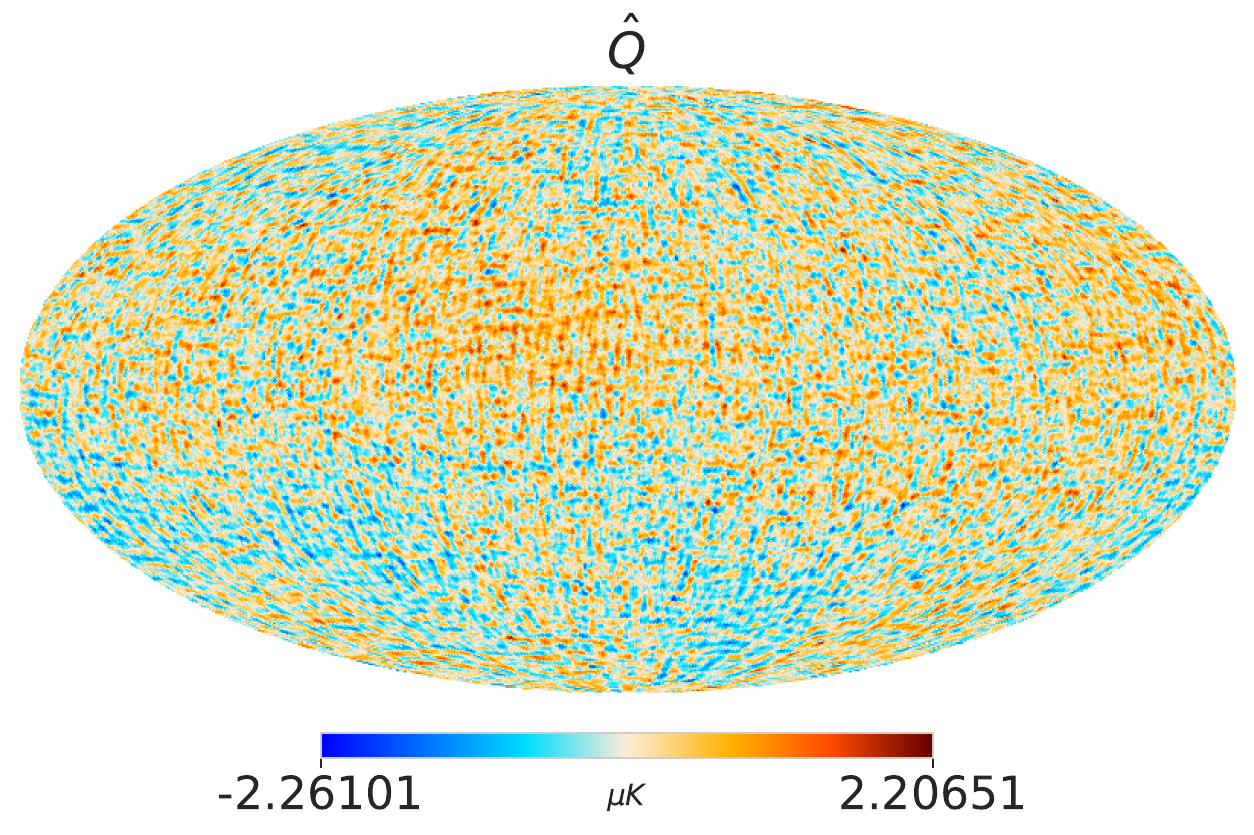
    }
    \includegraphics[width=0.32\columnwidth]{
        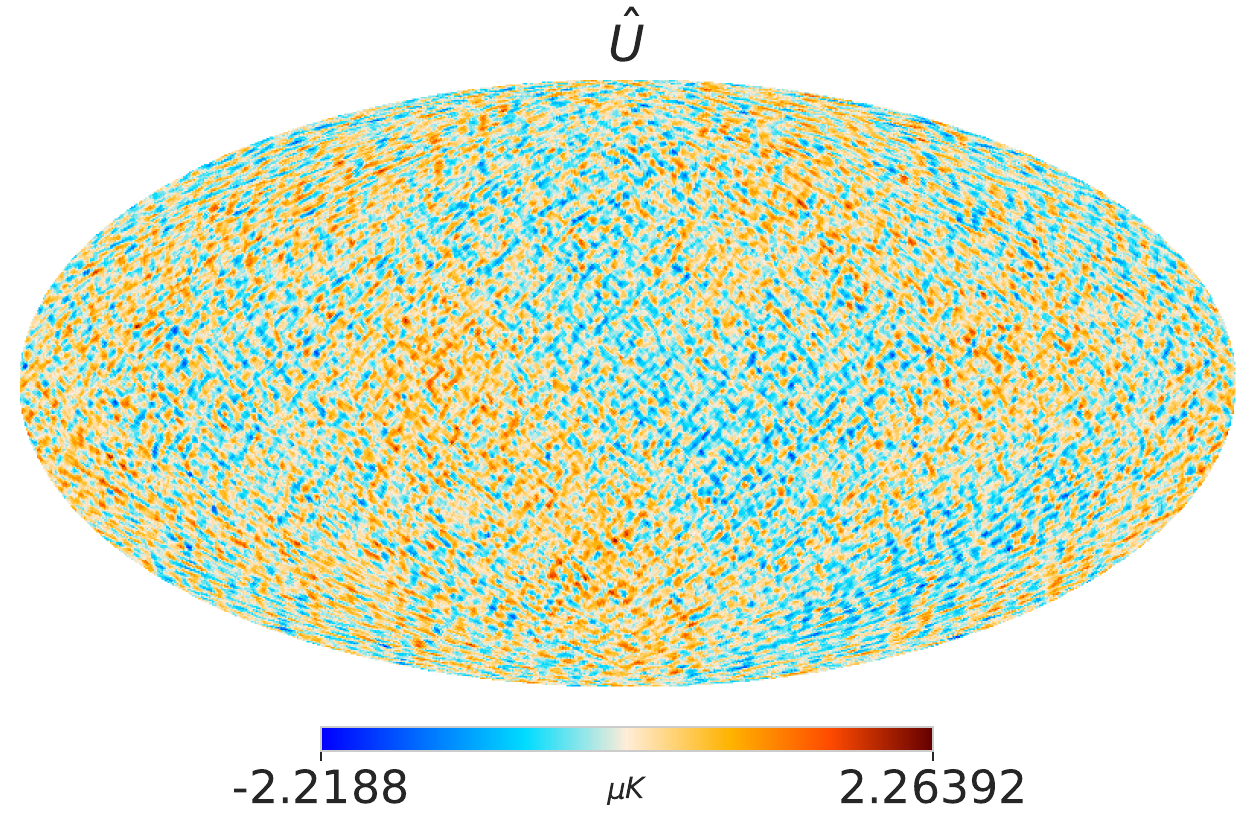
    }
    \\
    \includegraphics[width=0.32\columnwidth]{
        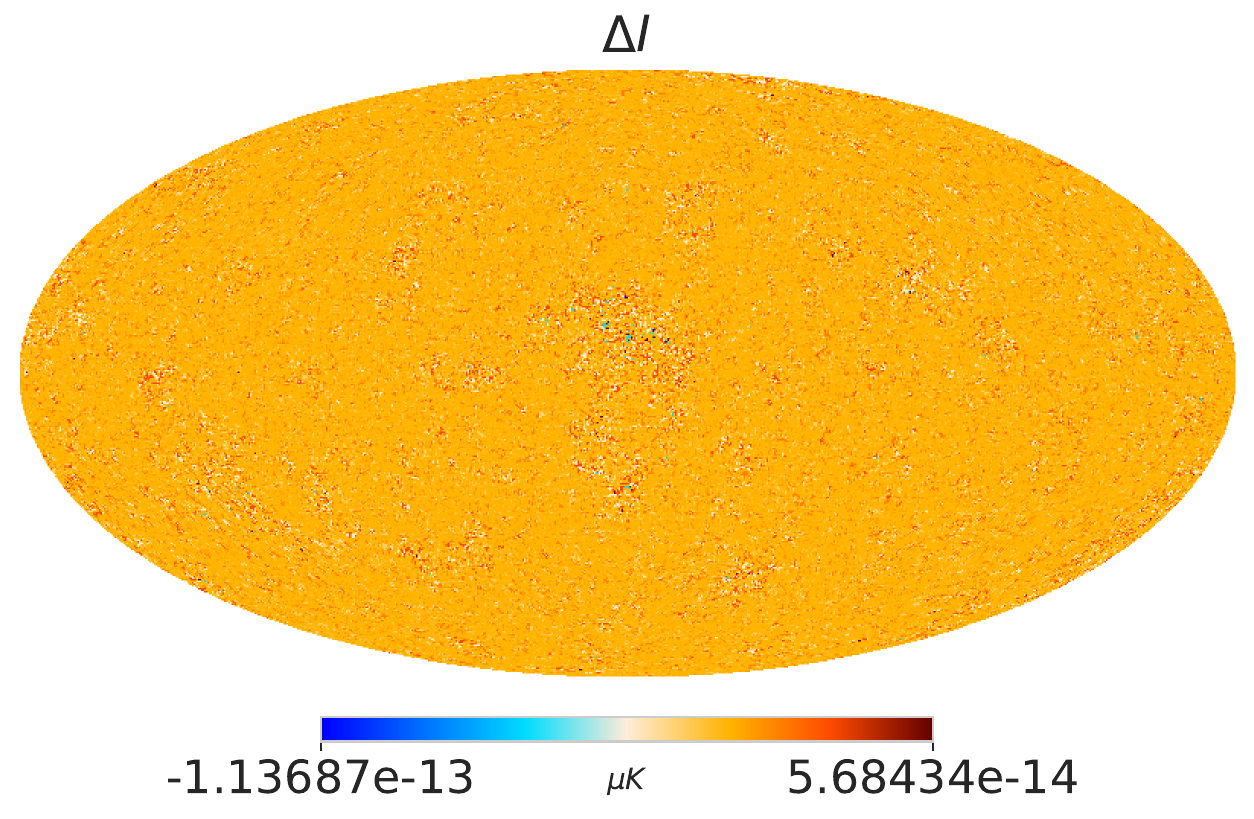
    }
    \includegraphics[width=0.32\columnwidth]{
        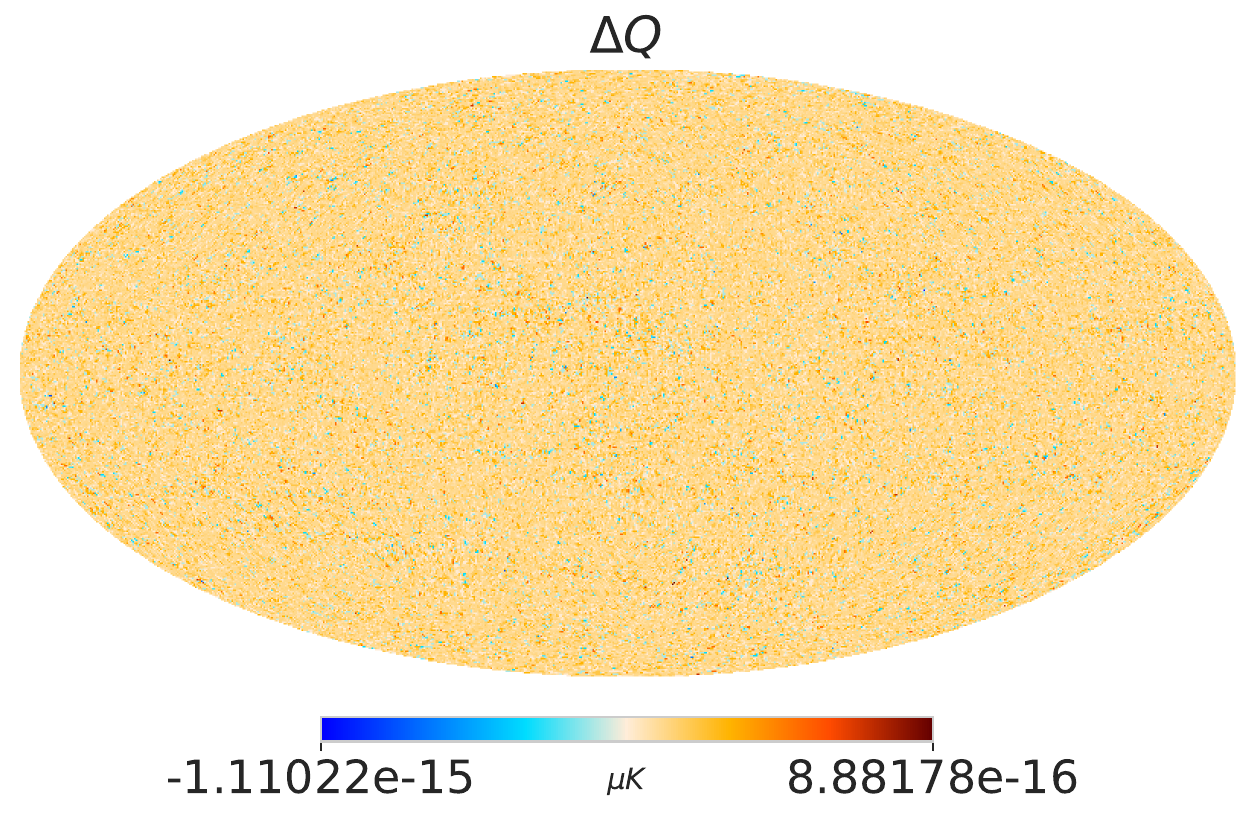
    }
    \includegraphics[width=0.32\columnwidth]{
        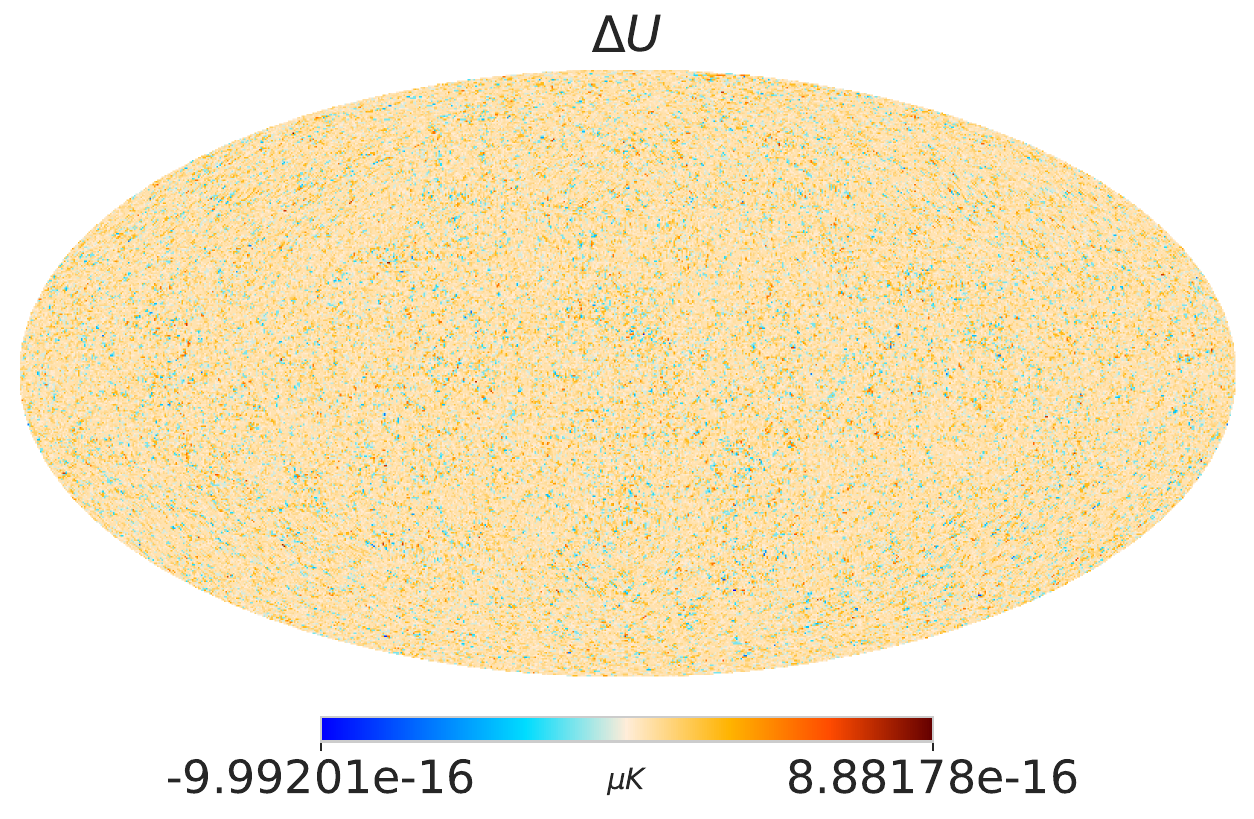
    }
    \caption[ Estimated CMB maps and residual maps due to the HWP wedge effect by
    the $9\times9$ matrix map-making approach with HWP.]{Esimtated CMB maps and
    residual maps due to the HWP wedge effect by the $9\times9$ matrix map-making
    approach with HWP. It displays $\hZ[1,-5]^{Q}$, $\hZ[1,-5]^{U}$, $\hZ[1,1]^{Q}$,
    $\hZ[1,1]^{U}$, $\hZ[3,-3]^{Q}$, $\hZ[3,-3]^{U}$, $\hat{I}$, $\hat{Q}$, $\hat
    {U}$, $\Delta I$, $\Delta Q$, and $\Delta U$ from top left to bottom right. The
    systematics parameter for the HWP wedge is same as the previous $3\times3$ matrix
    map-making approach case.}
    \label{fig:wedge_maps_9x9}
\end{figure}

\Cref{fig:wedge_maps_9x9} presents the CMB maps and residual maps estimated through
the $9\times9$ matrix map-making approach defined by:
\begin{align}
    \ab(\begin{matrix}\hat{I}\\{}_{1,1}\hat{Z}\\{}_{-1,-1}\hat{Z}\\ \hat{P}\\ \hat{P^*}\\{}_{3,-3}\hat{Z}\\{}_{-3,3}\hat{Z}\\{}_{1,-5}\hat{Z}\\{}_{-1,5}\hat{Z}\end{matrix}) & = \M[9]_{\rm w}^{-1}\ab(\begin{matrix}{}_{0,0}{\tilde{S}^d}_{\rm w}\\{}_{1,1}{\tilde{S}^d}_{\rm w}\\{}_{-1,-1}{\tilde{S}^d}_{\rm w}\\{}_{2,-4}{\tilde{S}^d}_{\rm w}\\{}_{-2,4}{\tilde{S}^d}_{\rm w}\\{}_{3,-3}{\tilde{S}^d}_{\rm w}\\{}_{-3,3}{\tilde{S}^d}_{\rm w}\\{}_{1,-5}{\tilde{S}^d}_{\rm w}\\{}_{-1,5}{\tilde{S}^d}_{\rm w}\end{matrix}),\label{eq:9M_w}
\end{align}
where $\M[9]_{\rm w}$ is given by
\begin{align}
\scalebox{0.95}{$
    \M[9]_{\rm w}= \ab(\begin{matrix}1&\frac{1}{2}{}_{-1,-1}\tilde{h}&\frac{1}{2}{}_{1,1}\tilde{h}&\frac{1}{2}{}_{-2,4}\tilde{h}&\frac{1}{2}{}_{2,-4}\tilde{h}&\frac{1}{2}{}_{-3,3}\tilde{h}&\frac{1}{2}{}_{3,-3}\tilde{h}&\frac{1}{2}{}_{-1,5}\tilde{h}&\frac{1}{2}{}_{1,-5}\tilde{h}\\ \frac{1}{2}{}_{1,1}\tilde{h}&\frac{1}{4}&\frac{1}{4}{}_{2,2}\tilde{h}&\frac{1}{4}{}_{-1,5}\tilde{h}&\frac{1}{4}{}_{3,-3}\tilde{h}&\frac{1}{4}{}_{-2,4}\tilde{h}&\frac{1}{4}{}_{4,-2}\tilde{h}&\frac{1}{4}{}_{0,6}\tilde{h}&\frac{1}{4}{}_{2,-4}\tilde{h}\\ \frac{1}{2}{}_{-1,-1}\tilde{h}&\frac{1}{4}{}_{-2,-2}\tilde{h}&\frac{1}{4}&\frac{1}{4}{}_{-3,3}\tilde{h}&\frac{1}{4}{}_{1,-5}\tilde{h}&\frac{1}{4}{}_{-4,2}\tilde{h}&\frac{1}{4}{}_{2,-4}\tilde{h}&\frac{1}{4}{}_{-2,4}\tilde{h}&\frac{1}{4}{}_{0,-6}\tilde{h}\\ \frac{1}{2}{}_{2,-4}\tilde{h}&\frac{1}{4}{}_{1,-5}\tilde{h}&\frac{1}{4}{}_{3,-3}\tilde{h}&\frac{1}{4}&\frac{1}{4}{}_{4,-8}\tilde{h}&\frac{1}{4}{}_{-1,-1}\tilde{h}&\frac{1}{4}{}_{5,-7}\tilde{h}&\frac{1}{4}{}_{1,1}\tilde{h}&\frac{1}{4}{}_{3,-9}\tilde{h}\\ \frac{1}{2}{}_{-2,4}\tilde{h}&\frac{1}{4}{}_{-3,3}\tilde{h}&\frac{1}{4}{}_{-1,5}\tilde{h}&\frac{1}{4}{}_{-4,8}\tilde{h}&\frac{1}{4}&\frac{1}{4}{}_{-5,7}\tilde{h}&\frac{1}{4}{}_{1,1}\tilde{h}&\frac{1}{4}{}_{-3,9}\tilde{h}&\frac{1}{4}{}_{-1,-1}\tilde{h}\\ \frac{1}{2}{}_{3,-3}\tilde{h}&\frac{1}{4}{}_{2,-4}\tilde{h}&\frac{1}{4}{}_{4,-2}\tilde{h}&\frac{1}{4}{}_{1,1}\tilde{h}&\frac{1}{4}{}_{5,-7}\tilde{h}&\frac{1}{4}&\frac{1}{4}{}_{6,-6}\tilde{h}&\frac{1}{4}{}_{2,2}\tilde{h}&\frac{1}{4}{}_{4,-8}\tilde{h}\\ \frac{1}{2}{}_{-3,3}\tilde{h}&\frac{1}{4}{}_{-4,2}\tilde{h}&\frac{1}{4}{}_{-2,4}\tilde{h}&\frac{1}{4}{}_{-5,7}\tilde{h}&\frac{1}{4}{}_{-1,-1}\tilde{h}&\frac{1}{4}{}_{-6,6}\tilde{h}&\frac{1}{4}&\frac{1}{4}{}_{-4,8}\tilde{h}&\frac{1}{4}{}_{-2,-2}\tilde{h}\\ \frac{1}{2}{}_{1,-5}\tilde{h}&\frac{1}{4}{}_{0,-6}\tilde{h}&\frac{1}{4}{}_{2,-4}\tilde{h}&\frac{1}{4}{}_{-1,-1}\tilde{h}&\frac{1}{4}{}_{3,-9}\tilde{h}&\frac{1}{4}{}_{-2,-2}\tilde{h}&\frac{1}{4}{}_{4,-8}\tilde{h}&\frac{1}{4}&\frac{1}{4}{}_{2,-10}\tilde{h}\\ \frac{1}{2}{}_{-1,5}\tilde{h}&\frac{1}{4}{}_{-2,4}\tilde{h}&\frac{1}{4}{}_{0,6}\tilde{h}&\frac{1}{4}{}_{-3,9}\tilde{h}&\frac{1}{4}{}_{1,1}\tilde{h}&\frac{1}{4}{}_{-4,8}\tilde{h}&\frac{1}{4}{}_{2,2}\tilde{h}&\frac{1}{4}{}_{-2,10}\tilde{h}&\frac{1}{4}\end{matrix})$}.
\end{align}
This approach successfully isolates both temperature gradient and polarization
gradient, achieving complete mitigation of systematic effects. The resulting
systematic power spectrum, shown as the green solid line in \cref{fig:delta_cl_abs_wedge_pnt}
(right), demonstrates full suppression of systematic contamination.

When analyzing the impact on the tensor-to-scalar ratio, we found $\Delta r < 10^{-6}$
across all scenarios with pointing offset disturbance parameters $(\xi,\chi)=(1^{\prime}
,0^{\prime})$ induced by the HWP wedge angle.

\subsection{Instrumental polarization due to HWP non-ideality}

To evaluate this systematic effect, we modify the input map and analysis method.
The systematic field is defined by the deviation signal, which includes only the
systematic effect without the fiducial signal we aim to measure, as shown in
\cref{eq:HWP_IP_field}. As discussed in \cref{sec:Propagation}, the primary
source of this systematic effect is the solar dipole. Therefore, we define the
input map as the sum of the CMB temperature anisotropies and the solar dipole
map, excluding any polarization. This approach allows us to focus on the
$T \to B$ leakage induced by the instrumental polarization due to HWP non-ideality.
For the systematic parameters, we set
$(\epsilon_{1},\phi_{QI})=(1.0\times10^{-5}, 0)$ as defined in
\cref{sec:Propagation}.

\begin{figure}[h]
    \centering
    \includegraphics[width=0.32\columnwidth]{
        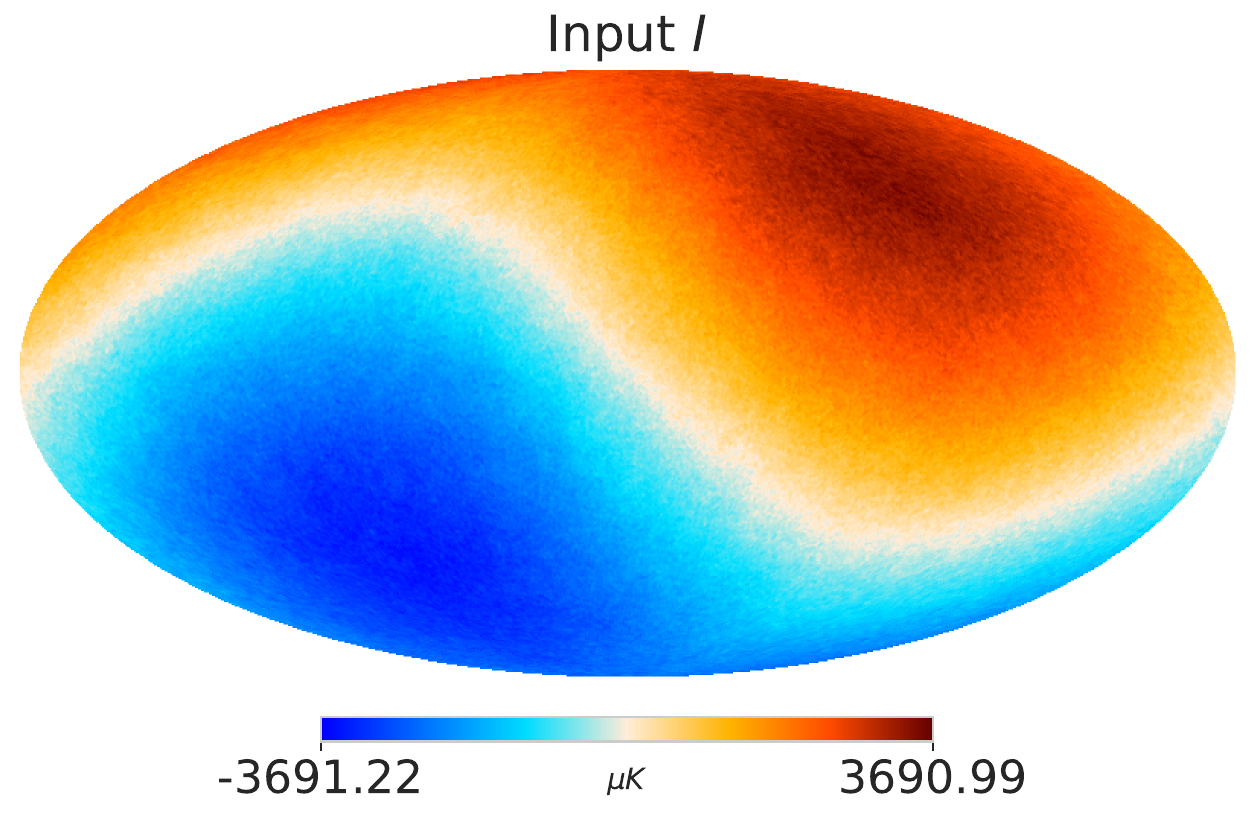
    }
    \includegraphics[width=0.32\columnwidth]{
        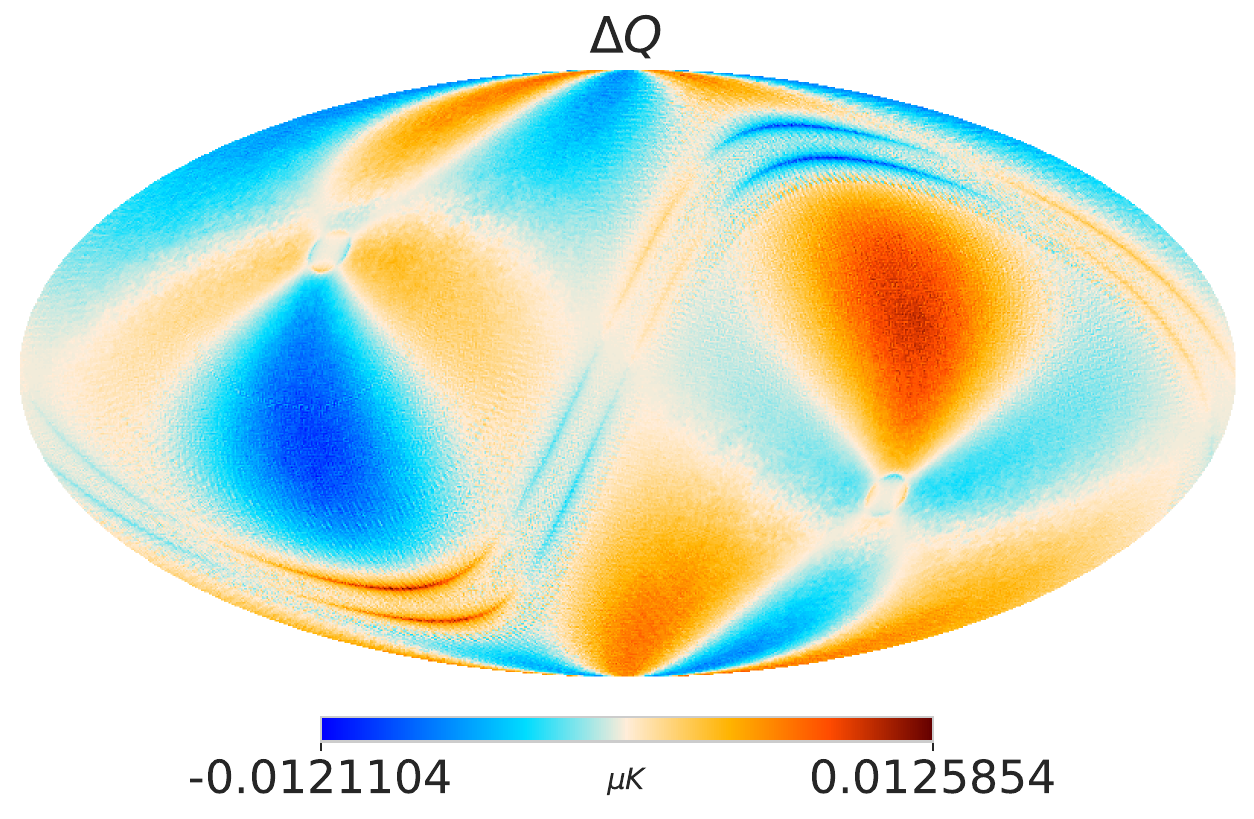
    }
    \includegraphics[width=0.32\columnwidth]{
        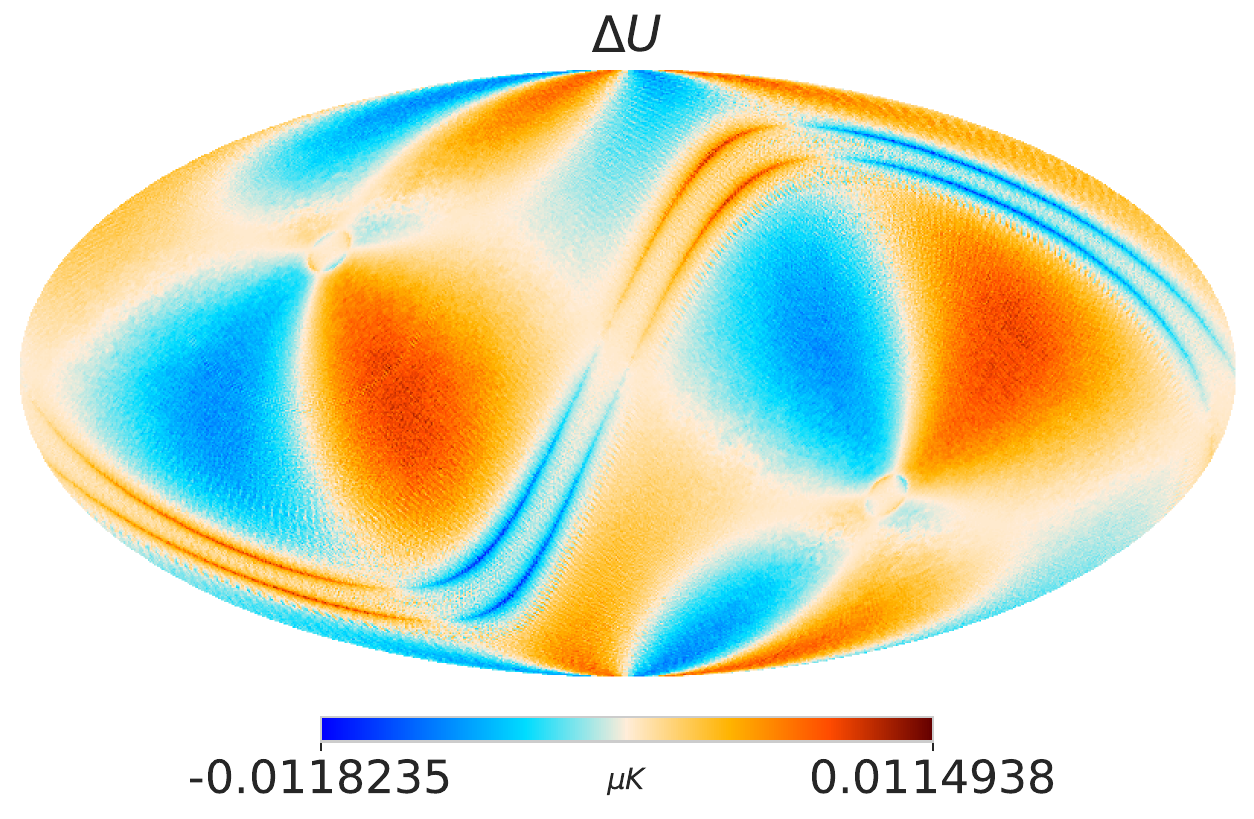
    }
    \caption[Input map and residual maps of instrumental polarization due to HWP
    non-ideality simulation using the $3\times3$ matrix map-making approach.]{(left)
    Input map for instrumental polarization due to HWP non-ideality, comprising the
    sum of CMB temperature anisotropies and the solar dipole map. (middle/right)
    Residual maps $\Delta Q$ and $\Delta U$ resulting from instrumental
    polarization using the $3\times3$ matrix map-making approach with HWP.}
    \label{fig:hwpip_maps_3x3}
\end{figure}

\Cref{fig:hwpip_maps_3x3} shows the input map and the residual maps derived from
the $3\times3$ matrix map-making approach (defined by \cref{eq:3M_ap}). In the residual
maps, the temperature-to-polarization leakage pattern is clearly visible, created
by the convolution between the solar dipole and $\h[\pm2,0]$ cross-link maps.

\begin{figure}[h]
    \centering
    \includegraphics[width=0.32\columnwidth]{
        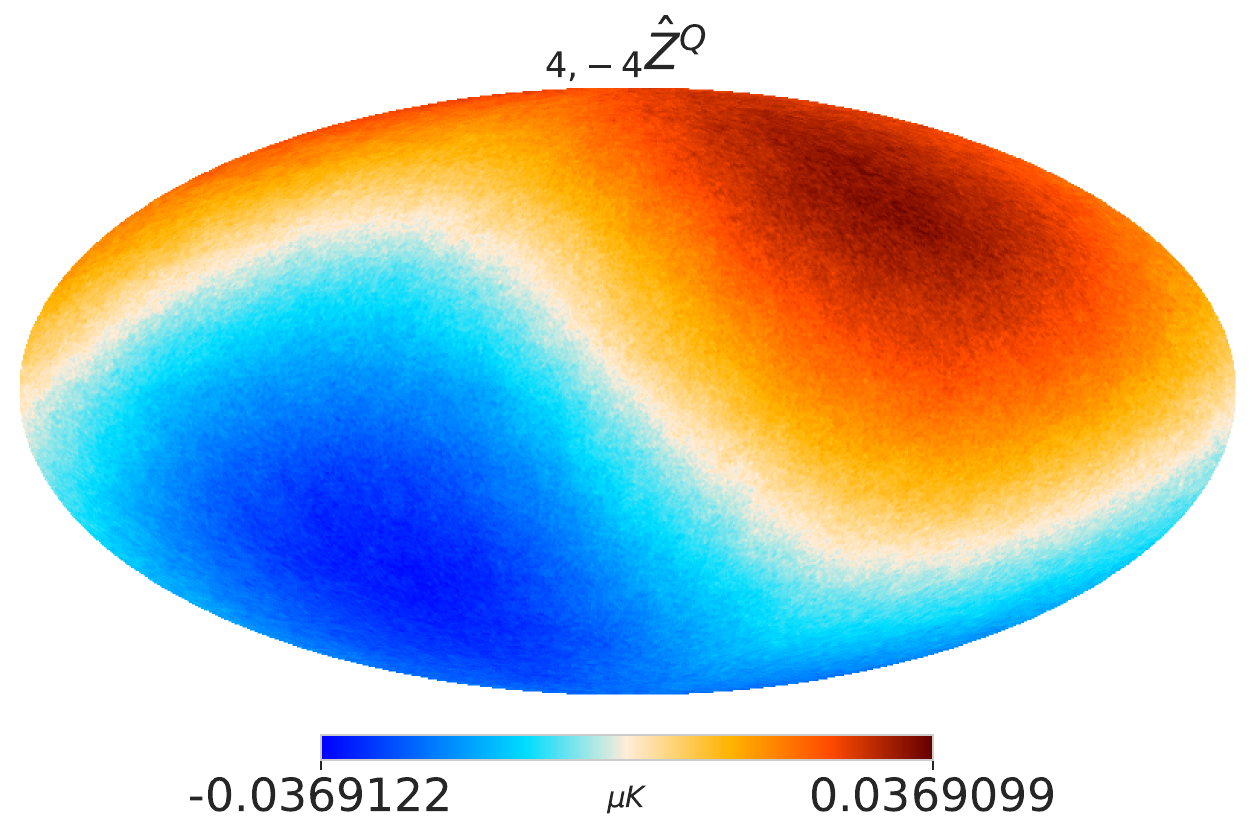
    }
    \includegraphics[width=0.32\columnwidth]{
        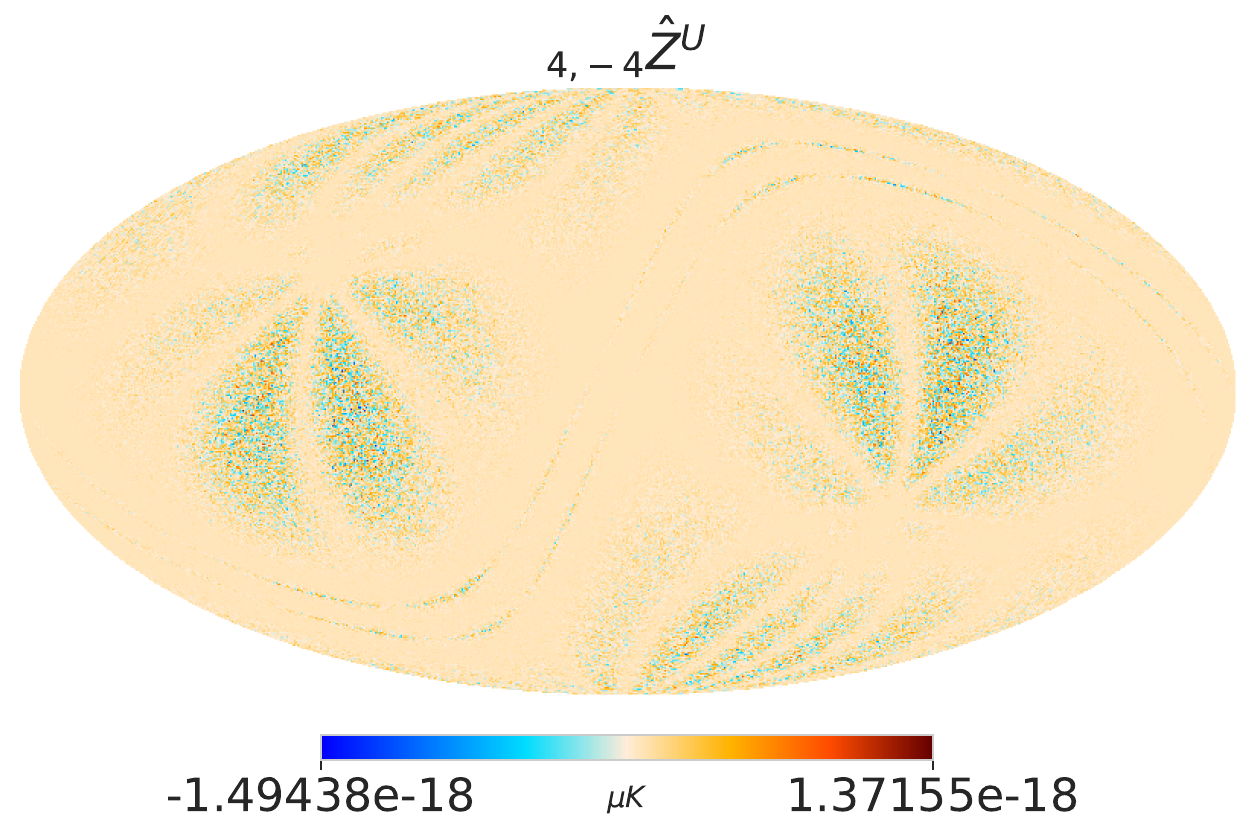
    }
    \\
    \includegraphics[width=0.32\columnwidth]{
        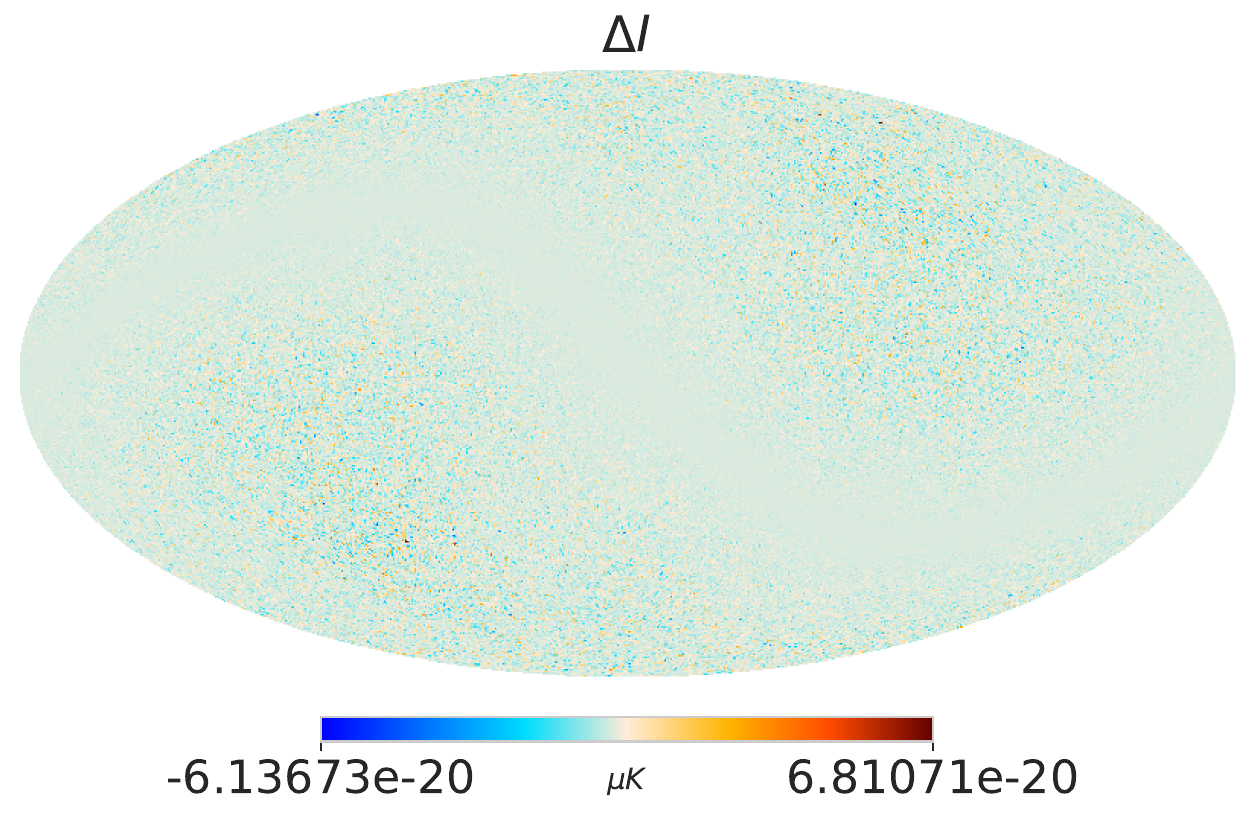
    }
    \includegraphics[width=0.32\columnwidth]{
        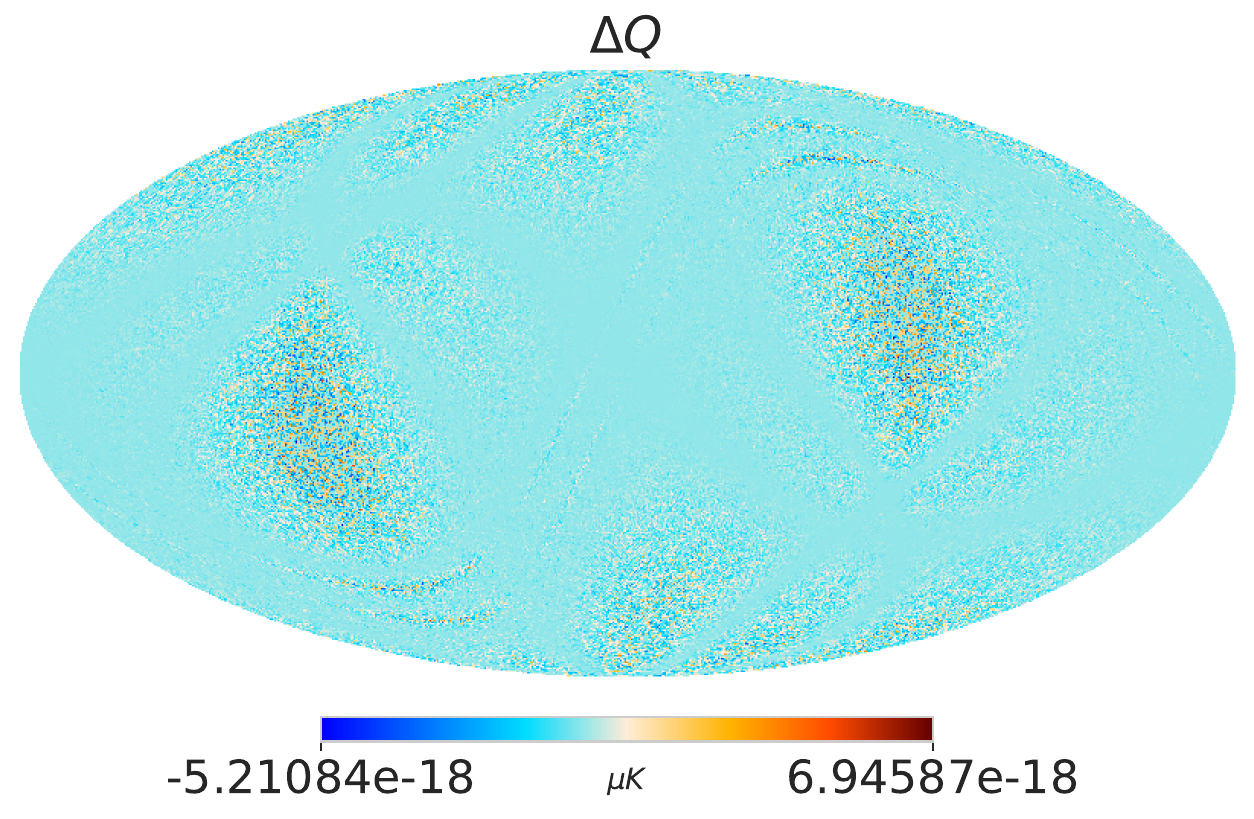
    }
    \includegraphics[width=0.32\columnwidth]{
        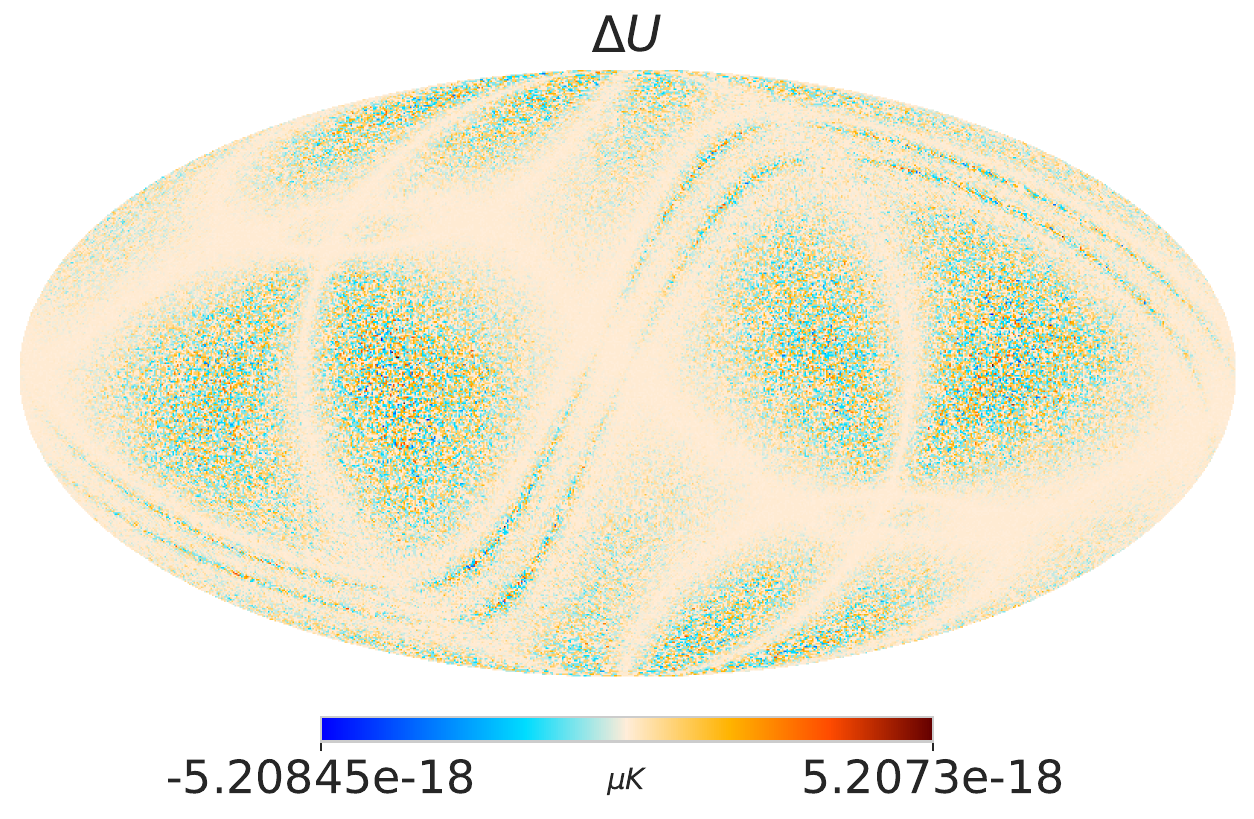
    }
    \caption[Estimated CMB maps and residual maps due to instrumental
    polarization using the $5\times5$ matrix map-making approach with HWP.]{Estimated
    CMB maps and residual maps due to instrumental polarization using the
    $5\times5$ matrix map-making approach with HWP. The top panels show $\hZ[4,-4
    ]^{Q}$ and $\hZ[4,-4]^{U}$, while the middle and bottom panels display
    $\Delta I$, $\Delta Q$, and $\Delta U$. The systematic parameters for instrumental
    polarization are $(\epsilon_{1},\phi_{QI})=(1.0\times10^{-5}, 0)$ as used in
    \cref{sec:Propagation}.}
    \label{fig:hwpip_maps_5x5}
\end{figure}

To effectively mitigate the systematic effect, we capture the leakage term using
the following $5\times 5$ map-making approach:
\begin{align}
    \ab(\begin{matrix}\hat{I}\\ \hat{P}\\ \hat{P^*}\\{}_{4,-4}\hat{Z}\\{}_{-4,4}\hat{Z}\end{matrix}) & = \M[5]_{\rm ip}^{-1}\ab(\begin{matrix}{}_{0,0}{\tilde{S}^d}_{\rm w}\\{}_{2,-4}{\tilde{S}^d}_{\rm w}\\{}_{-2,4}{\tilde{S}^d}_{\rm w}\\{}_{4,-4}{\tilde{S}^d}_{\rm w}\\{}_{-4,4}{\tilde{S}^d}_{\rm w}\end{matrix}),
\end{align}
where $\M[5]_{\rm ip}$ is given by
\begin{align}
    \M[5]_{\rm ip}= \ab(\begin{matrix}1&\frac{1}{2}{}_{-2,4}\tilde{h}&\frac{1}{2}{}_{2,-4}\tilde{h}&\frac{1}{2}{}_{-4,4}\tilde{h}&\frac{1}{2}{}_{4,-4}\tilde{h}\\ \frac{1}{2}{}_{2,-4}\tilde{h}&\frac{1}{4}&\frac{1}{4}{}_{4,-8}\tilde{h}&\frac{1}{4}{}_{-2,0}\tilde{h}&\frac{1}{4}{}_{6,-8}\tilde{h}\\ \frac{1}{2}{}_{-2,4}\tilde{h}&\frac{1}{4}{}_{-4,8}\tilde{h}&\frac{1}{4}&\frac{1}{4}{}_{-6,8}\tilde{h}&\frac{1}{4}{}_{2,0}\tilde{h}\\ \frac{1}{2}{}_{4,-4}\tilde{h}&\frac{1}{4}{}_{2,0}\tilde{h}&\frac{1}{4}{}_{6,-8}\tilde{h}&\frac{1}{4}&\frac{1}{4}{}_{8,-8}\tilde{h}\\ \frac{1}{2}{}_{-4,4}\tilde{h}&\frac{1}{4}{}_{-6,8}\tilde{h}&\frac{1}{4}{}_{-2,0}\tilde{h}&\frac{1}{4}{}_{-8,8}\tilde{h}&\frac{1}{4}\end{matrix})^{-1}.
\end{align}
This map-maker captures the modulated temperature term, $\hZ[\pm4,\mp4]$, and estimates
it as shown in \cref{fig:hwpip_maps_5x5} (top panels). The residual maps' level is
significantly reduced compared to the $3\times3$ matrix map-making approach, as
shown in the middle and bottom panels of \cref{fig:hwpip_maps_5x5}. The estimated
$\Delta r = 0.00046$ in the case of the $3\times3$ matrix map-making approach,
while the $5\times5$ matrix map-making approach yields $\Delta r < 10^{-6}$. This
result implies that some HWP systematics driven by harmonics due to the HWP rotation
can be mitigated using the map-making approach in terms of \spin.

\begin{figure}[h]
    \centering
    \includegraphics[width=0.70\columnwidth]{
        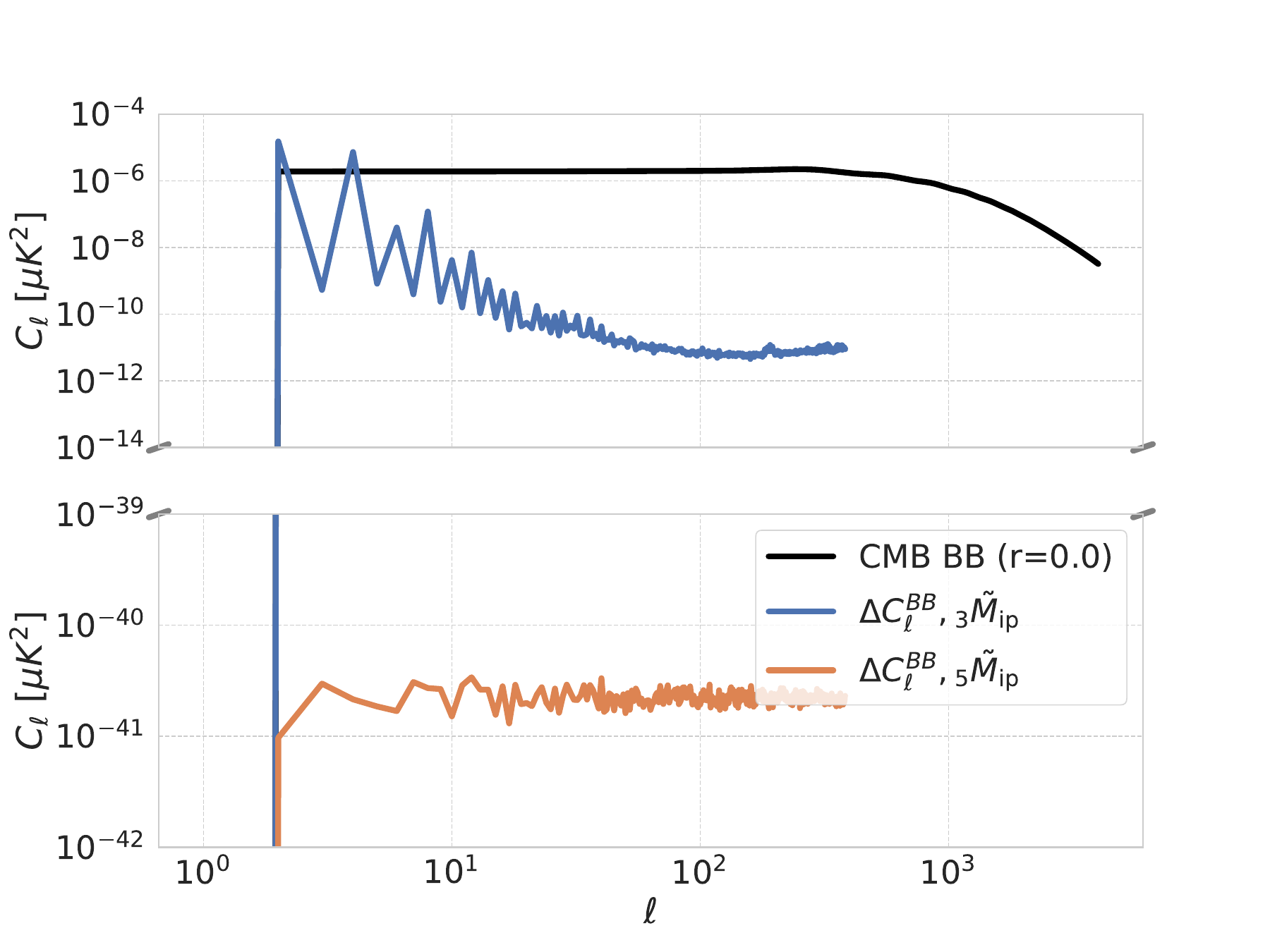
    }
    \caption[Systematic power spectrum $\Delta C_{\ell}^{BB}$ due to
    instrumental polarization using the $3\times3$ and $5\times5$ matrix map-making
    approaches with HWP.]{Systematic power spectrum $\Delta C_{\ell}^{BB}$ due
    to instrumental polarization using the $3\times3$ matrix map-making approach
    (solid blue line) and the $5\times5$ matrix map-making approach (solid orange
    line) with HWP. The systematic parameters for instrumental polarization are $(
    \epsilon_{1},\phi_{QI})=(1.0\times10^{-5}, 0)$ as used in \cref{sec:Propagation}.}
    \label{fig:delta_cl_with_hwpip}
\end{figure}

The systematic power spectrum obtained using this map-making approach is shown in
\cref{fig:delta_cl_with_hwpip}. The solid blue line represents the derived
$\Delta C_{\ell}^{BB}$, highlighting a noticeable bump in the low-$\ell$ region due
to the dipole leakage.

The orange solid line shows the systematic power spectrum derived from the
$5\times 5$ matrix map-making approach, demonstrating a significant reduction in
systematic contamination compared to the $3\times3$ matrix map-making approach.
This result indicates that the $T \to B$ leakage induced by instrumental
polarization due to HWP non-ideality can be effectively mitigated using the
$5\times5$ matrix map-making approach.

    \chapter{Conclusion}
\label{chap:conclusion} \minitoc

\section{Summary of scanning strategy optimization}

This study has investigated the scanning strategy parameter space for spacecraft
missions equipped with a HWP, utilizing $\{\alpha, T_{\alpha}, T_{\beta}\}$ as
the primary variables. Our analysis focused on four fundamental metrics essential
for $B$-mode observation: planet visibility time, forming speed of sky coverage,
hit-map uniformity, and cross-link factor.

Through the application of our custom-developed scan simulator \Falcons, we
conducted a comprehensive analysis of metric distributions across the
$\{\alpha, T_{\alpha}, T_{\beta}\}$ parameter space. Our findings substantiate that
the configuration adopted by \LiteBIRD, as detailed in ref.~\cite{PTEP2023} and known
as the \SC, achieves an optimal balance between instrumental calibration, systematic
effect suppression, and the implementation of robust null-tests for the mission.

From a more comprehensive perspective, optimal metric performance can be achieved
within our defined scanning strategy parameter space (\cref{fig:standard_config_and_T_beta},
right) by maintaining $\alpha\simeq\beta$ and constraining $T_{\alpha}$ to less
than 100\,hours. Our analysis revealed that the considered metrics exhibit small
dependence on $T_{\alpha}$ within this kinetic parameter domain (as evidenced in
\cref{fig:planet_visibility,fig:coverage_and_sigma_hit,fig:cross-links}),
emphasizing the paramount importance of geometric parameters in scanning
strategy optimization. Through rigorous examination detailed in \cref{sec:Opt_geometric},
we systematically eliminated suboptimal configurations and determined that the
geometric configuration of $(\alpha,\beta)=(45^{\circ},50^{\circ})$ achieves an
optimal equilibrium between our metrics and the requirements imposed by
\LiteBIRD's instrumental architecture.

Throughout our analysis, we presented metric distributions within the $\{\alpha,
T_{\alpha}\}$ parameter space while maintaining $T_{\beta}$ at $\tbl$. This
methodological approach was validated by our discovery that, in the regime where
$T_{\beta}<20$\,min, the structural characteristics of these metric
distributions demonstrates invariant scaling properties with respect to
$T_{\beta}$, thereby preserving the spatial configuration of optimal solutions.
A comprehensive investigation of this invariant scaling behavior is elaborated
in \cref{apd:T_beta_scaled}.

This invariant scaling property is further evidenced in
\cref{fig:rot_period_opt}, which illustrates the correlation between cross-link factors
within the $T_{\beta}$ and $T_{\alpha}$ parameter space. Our detailed analysis
in \cref{sec:Opt_kinetic} demonstrates that the configuration of $T_{\beta}=20$\,min
and $T_{\alpha}=192$\,min represents an optimal compromise for minimizing cross-link
factors. These findings were derived under the assumption of \LiteBIRD's current
HWP rotation rate. For alternative experimental configurations with elevated
$f_{\rm knee}$ characteristics, enhanced HWP rotation speeds would be necessary to
achieve efficient modulation. Under such circumstances, the lower boundary of the
spin period, as defined by \cref{eq:T_spin}, would be reduced, permitting more
rapid spin rates. Nevertheless, \cref{fig:rot_period_opt} clearly demonstrates
that reducing the spin period below $T_{\beta}=20$\,min offers no additional
performance benefits.

Extended spin periods are beneficial for spacecraft attitude control, while shortened
precession periods advantageously shift the CMB solar dipole signal used for gain
calibration to higher frequencies, thereby suppressing low-frequency gain
fluctuations (see \cref{fig:dipole_spectra}). As shown in \cref{fig:scanbeam}, a
shortened precession period also allows for also allows for a wider distribution
of scanning beam angles, effectively eliminating potential degeneracies between beam
shape and various systematic effects. To avoid the manifestation of \moire
patterns in both hit-map and cross-link factor distributions-resulting from s in-precession
resonance phenomena (\cref{fig:prec_tuning_maps}) --- a meticulous refinement of
the precession period was performed. This optimization process culminated in the
selection of $T_{\alpha}=192.348$\,min, a value specifically chosen for its
absence of nearby resonance peaks with significant standard deviations (\cref{fig:prec_tuning}).

Beyond the active mitigation of systematic effects through cross-link factor
reduction, scanning strategies must incorporate robust null-test frameworks to evaluate
the effectiveness of in-flight calibration strategies and facilitate the detection
of unforeseen systematic effects. In \cref{sec:implications}, we compared the sky
pixel visit/revisit times and the visit/revisit times to planets—key indicators for
designing calibration strategies and effective null-tests—across the scanning strategies
of \Planck, \PICO, and \LiteBIRD's \SC. Our analysis revealed that \LiteBIRD and
\PICO offer extensive daily coverage, allowing continuous observation of specific
sky pixels and planets over extended periods. The frequent pixel revisitations
at diverse temporal intervals exhibited by these missions facilitate various
calibration strategies and null-test implementations through the strategic
segmentation of observational data across multiple timescales for both sky
pixels and planetary sources. Conversely, \Planck's scanning strategy, although not
specifically optimized for polarization studies, exhibited exceptional
efficiency in attaining rapid deep sensitivity and ensuring stable pixel
observations—features particularly beneficial for temperature measurements. Moreover,
while not explicitly detailed in the manuscript, our investigation encompassed the
directional parameters of spacecraft rotation, including spin and precession
orientations. Our analyses, predicated on counterclockwise spin, precession, and
orbital motion, were subsequently validated through comprehensive simulations examining
all permutations of clockwise and counterclockwise rotations, revealing
negligible directional dependencies. A thorough exposition of this investigation
is presented in \cref{apd:rotation_direction}.

The implementation of \Falcons proved transformative in our analytical approach,
performing grid searches on supercomputing platforms while achieving optimal
memory utilization through thread-parallelization. Furthermore, our elucidation
of the relationship between the scanning strategy parameter space and \LiteBIRD's
\SC, within our defined constraints, provides invaluable insights for the
architectural design of future space-based polarimetric missions. The adaptable framework
of our scan simulator \Falcons extends beyond the specific requirements of the \LiteBIRD
mission, offering broader applicability in the field.

\section{Summary of the systematic effect studies}
The \spin-based map-making methodology developed described in
\cref{chap:formalism} has proven exceptionally effective for both scanning
strategy optimization and systematic effect evaluation. This approach achieved enhanced
versatility through Fourier transformation of HWP rotation angles, enabling
accommodation of more complex experimental configurations. Compared to conventional
TOD-based binning map-making techniques, our method achieved approximately
$10^{4}$-fold computational acceleration by transforming individual pixel
binning operations into efficient map, i.e., image convolutions within Fourier (\spin)
space.

The primary advantage of our methodology extends beyond mere computational efficiency
to provide deeper insights into how systematic effects influence polarimetry.
While TOD-based simulations struggled to elucidate the interplay between
systematic effects and scanning strategies, our approach, leveraging the
inherent rotational symmetry, i.e., \spin of polarization and systematic effects,
enabled analytical understanding of the relationship between scanning strategy's
cross-linking and systematic effects. This framework revealed both the mechanisms
by which specific systematic effects induce $T \to B$ or $E \to B$ leakage and their
suppression through the cross-linking of scanning strategy, as we discussed in
\cref{chap:systematics}.

Furthermore, we developed an enhanced map-maker capable of simultaneous polarization
and systematic effect estimation through extension of the regression estimation linear
system as shown in \cref{sec:mitigation}. This methodology enables separation of
polarization components from systematic effects utilizing \spin properties, resulting
in more precise polarization maps.

Performance evaluation of our optimized scanning strategy demonstrated that with
HWP implementation, in \cref{sec:results_syst}, nearly all systematic effects were
suppressed to $\Delta r<10^{-6}$ (less than $1/10^{3}$ of \LiteBIRD's science
objective). Notably, comparable suppression levels ($\Delta r\lesssim10^{-6}$) were
achievable even without HWP utilization through our developed suppression techniques.

However, the absence of HWP necessitates allocation of sensitivity to systematic
effect estimation, inevitably increasing statistical errors. This presents an
intriguing trade-off between active systematic error suppression through HWP
implementation versus achieving sensitivity through increased detector count or
observation time while employing our methodology without HWP.

\section{Future perspective}

An important remaining challenge lies in the analysis of systematic effects.
While we pioneered the world's first formulation of a map-based formalism using
\spin with HWP, we have yet to develop methodologies for directly calculating systematics
power spectra from systematic fields in with HWP scenarios, as we did for the
differential gain and pointing by \cref{eq:gain_trans,eq:pointing_cl_2x2}. Such development
would enable the calculation of systematic power spectra without CMB ensemble averaging
and cosmic variance effects, potentially leading to substantial computational
efficiency obtains and deeper insights into systematic effects.

In our research, we conducted simulations using CMB-only input maps to elucidate
the impact of systematic effects on CMB maps and $B$-mode power spectra. However,
real observations are complicated by foreground emissions, including synchrotron
and dust radiation from the Milky Way superimposed on the CMB signal. The analysis
of systematic effects incorporating these foreground contributions remains an important
avenue for future investigation.

As mentioned at the conclusion of the previous section, a crucial area for
future research involves resolving the trade-off between active systematic error
suppression through HWP implementation versus achieving enhanced sensitivity
through increased detector count or extended observation time without HWP while
employing our methodology.

Notably, the software tools developed and utilized in this research—\Falcons and
\SBM—are publicly available on the author's GitHub repository, enabling third-party
verification and reproduction of our results.\footnote{\url{https://github.com/yusuke-takase}}
Future work will also encompass maintaining software quality standards and
developing comprehensive documentation.

    \appendix
    \chapter{Additional derivations and frameworks}
\minitoc

\chapabstract{ This appendix provides additional derivations to support the discussion in the manuscript. We first derive the dipole temperature anisotropy of the CMB in. Next, we describe the spacecraft scanning motion in detail. We then discuss the \healpix software framework, which is used for spherical data analysis. Finally, we present a method to estimate the bias on the tensor-to-scalar ratio $r$. }

\section{Fundamentals of cosmology}

\subsection{Dipole temperature anisotropy of the CMB}
\label{apd:CMB_dipole}

Consider a system where the temperature distribution of the CMB on the celestial
sphere is isotropic with a temperature $T_{0}$. This is called the rest frame of
the CMB. Next, consider an observer moving with velocity $\bm{v}$ relative to the
rest frame of the CMB. Due to the Doppler effect of light, the energy of photons
arriving from the direction of motion increases, while the energy of photons
arriving from the opposite direction decreases. Therefore, the temperature
distribution $T(\hat{n})$ observed in the direction of the unit vector $\hat{n}$
is given by
\begin{align}
    T(\hat{n}) = \frac{T_{0}\sqrt{1 - \frac{v^2}{c^2}}}{1 - \frac{\bm{v}\cdot\hat{n}}{c}},
\end{align}
where $c$ is the speed of light. This can be expanded in spherical harmonics,
and if the origin of the celestial coordinates $\theta$ is taken in the
direction of the velocity vector, the expansion coefficients $a_{\ell m}$ become
zero for all values of $m$ except $m=0$. The Earth orbits the Sun at
approximately $30\,\si{km/s}$. From the rest frame of the CMB, the direction of
the Earth's velocity vector changes with the seasons, so the temperature anisotropy
of the CMB observed from Earth varies seasonally. Excluding the anisotropy due
to the Earth's orbital motion, the remaining temperature anisotropy is caused by
the Sun's orbital motion around the galactic center, the mutual interaction
between our galaxy and the Andromeda galaxy, and the influence of the large-scale
structure of galaxies. The temperature anisotropy we observe is the sum of all
these velocity vectors. If the velocity vector of the solar system relative to the
rest frame of the CMB is denoted as $\bm{v}_{\rm{solar~system}}$, it can be
written as
\begin{align}
    \bm{v}_{\rm{solar~system}}=(\bm{v}_{\rm{solar~system}}-\bm{v}_{\rm{galactic}}) + (\bm{v}_{\rm{galactic}}- \bm{v}_{\rm{local~group}}) + \bm{v}_{\rm{local~group}},
\end{align}
The first and second terms on the right-hand side are measured by astronomical
observations, and their magnitudes are approximately $220\,\si{km/s}$ and $80\,\si
{km/s}$, respectively. The third term is difficult to measure by astronomical observations,
but it can be estimated by detailed observations of the temperature anisotropy
of the CMB. The amplitude of CMB dipole is measured as $3361.90\pm0.36$\,\si{\mu K}
by \Planck \cite{dipole_plank2021}.

\section{Scanning motion of spacecraft}
\label{apd:scan_motion}

This section defines the spacecraft's spin and precession motions in detail.
Consider an orthonormal coordinate frame $xyz$ where the spacecraft's spin axis
aligns with the $z$-axis. Let $\beta$ be the angle between the telescope boresight
and spin axis. In the $xyz$ frame, the boresight direction vector is $\bm{n}_{0}=
(\sin\beta, 0, \cos\beta)$. The `spin' motion, with angular velocity $\omega_{\beta}$
around the $z$-axis, transforms $\bm{n}_{0}$ via the time-dependent rotation
matrix $R_{z}$:
\begin{align}
    \bm{n}_{\rm spin}(t) & = R_{z}(\omega_{\beta}t)\bm{n}_{0}\nonumber                                                                                                                       \\
                         & = \ab(\mqty{ \cos\omega_\beta t & -\sin\omega_\beta t & 0 \\ \sin\omega_\beta t & \cos\omega_\beta t & 0 \\ 0 & 0 & 1}) \ab(\mqty{ \sin\beta \\ 0 \\ \cos\beta}),
\end{align}
where $R_{j}~(j\in\{x,y,z\})$ represents rotation matrices around respective axes.
To incorporate spin axis time dependence, we align the $z$-axis with the Sun-spacecraft
axis (see \cref{fig:standard_config_and_T_beta}, left). The spin axis is tilted by
angle $\alpha$ through a $y$-axis rotation:
\begin{align}
    \bm{n}_{\rm spin}'(t) & = R_{y}(\alpha)\bm{n}_{\rm spin}(t) \nonumber                                                                 \\
                          & = \ab(\mqty{ \cos\alpha & 0 & \sin\alpha \\ 0 & 1 & 0 \\ -\sin\alpha & 0 & \cos\alpha}) \bm{n}_{\rm spin}(t),
\end{align}
Here, $\bm{n}_{\rm spin}'(t)$ describes the boresight rotation around the tilted
spin axis. The `precession' motion adds rotation to the spin axis around the $z$-axis
with angular velocity $\omega_{\alpha}$:
\begin{align}
    \bm{n}(t) & = R_{z}(\omega_{\alpha}t)\bm{n}_{\rm spin}'(t)\nonumber                                                                                            \\
              & = \ab(\mqty{ \cos\omega_\alpha t & -\sin\omega_\alpha t & 0 \\ \sin\omega_\alpha t & \cos\omega_\alpha t & 0 \\ 0 & 0 & 1}) \bm{n}_{\rm spin}'(t),
\end{align}

The complete rotational motion combines spin and precession through the matrix
chain:
\begin{align}
    \bm{n}(t) & = R_{z}(\omega_{\alpha}t)R_{y}(\alpha )R_{z}(\omega_{\beta}t)\bm{n}_{0}\nonumber                                                                                                                                                                                                                                                                                                                          \\
              & = \ab(\mqty{ \cos\omega_\alpha t & -\sin\omega_\alpha t & 0 \\ \sin\omega_\alpha t & \cos\omega_\alpha t & 0 \\ 0 & 0 & 1}) \ab(\mqty{ \cos\alpha & 0 & \sin\alpha \\ 0 & 1 & 0 \\ -\sin\alpha & 0 & \cos\alpha}) \ab(\mqty{ \cos\omega_\beta t & -\sin\omega_\beta t & 0 \\ \sin\omega_\beta t & \cos\omega_\beta t & 0 \\ 0 & 0 & 1}) \ab(\mqty{ \sin\beta \\ 0 \\ \cos\beta}). \label{eq:matrix_chain}
\end{align}
By the chain, the motion of spacecraft except for the orbital motion around the
Sun is fully described.

\subsection{Sweep angular velocity on the sky}
\label{apd:sweeping_velocity}

We now derive the maximum angular velocity of the telescope's boresight sweep across
the sky.\footnote{We ignore the orbital rotation around the Sun due to its
negligible contribution to spin and precession.} Using \cref{eq:matrix_chain},
the angle $\Delta A$ traversed by $\bm{n}(t)$ between times $t$ and
$t + \Delta t$ is:
\begin{align}
    \Delta A = \ab|\bm{n}(t+\Delta t) - \bm{n}(t)|.
\end{align}
The sweep angular velocity $\odv{A}{t}$ is:
\begin{align}
    \odv{A}{t}= \ab|\odv{\bm{n}(t)}{t}|.
\end{align}
The maximum sweep velocity $\omega_{\rm max}$ occurs when $\omega_{\beta}t = 0$:
\begin{align}
    \omega_{\rm max} & = \ab|\odv{A}{t}|_{\rm max}\nonumber                                                        \\
                     & = \omega_{\alpha}\sin(\alpha+\beta) + \omega_{\beta}\sin\beta. \label{eq:sweeping_velocity}
\end{align}
Note that this assumes aligned precession and spin rotation directions. For
analysis of reversed rotation directions and their impact on scanning strategy
optimization, see \cref{apd:rotation_direction}.

\section{Method to estimate the bias on \texorpdfstring{$r$}{r}}
\label{apd:delta_r}

We can obtain temperature and polarization maps with specific systematic effects
using the formalism described in this section. The residual maps are created by subtracting
the original input map from the output map containing systematic effects:
\begin{align}
    \Delta I & ={}_{0,0}\tilde{S}^{d}- \St[0,0],   \\
    \Delta P & ={}_{2,-4}\tilde{S}^{d}- \St[2,-4],
\end{align}
where $\Delta P$ decomposes into Stokes parameters as
$\Delta Q=\Re\ab[\Delta P]$ and $\Delta U = \Im\ab[\Delta P]$. For cases where signal
fields are already described solely by systematic signal fields (as with HWP non-ideality
in \cref{apd:HWP_sys}), we can obtain the residuals directly:
\begin{align}
    \Delta I & ={}_{0,0}\Delta \tilde{S}^{d},           \\
    \Delta Q & = \Re\ab[{}_{2,-4}\Delta \tilde{S}^{d}], \\
    \Delta U & = \Im\ab[{}_{2,-4}\Delta \tilde{S}^{d}].
\end{align}

To assess potential systematic bias $\Delta r$ on the tensor-to-scalar ratio $r$,
we define a likelihood function $L(r)$ as:
\begin{align}
    \log{L(r)}=\sum^{\ell_{\rm max}}_{\ell=\ell_{\rm min}}\log{P_\ell(r)},
\end{align}
using multipole range $(\ell_{\rm min},\ell_{\rm max})=(2,191)$ \cite{PTEP2023}.
Here, $P_{\ell}(r)$ is defined as:
\begin{align}
    \log{P_\ell(r)}=-f_{\rm sky}\frac{2\ell+1}{2}\ab[\frac{\hat{C}_{\ell}}{C_{\ell}}+\log{C_\ell}- \frac{2\ell-1}{2\ell+1}\log{\hat{C}_\ell}],
\end{align}
where $\hat{C}_{\ell}$ and $C_{\ell}$ represent measured and modeled $B$-mode power
spectra respectively \cite{likelihood}. We use $f_{\rm sky}=1$ for full-sky
coverage. 
\begin{figure}[h]
    \centering
    \includegraphics[width=0.55\columnwidth]{
        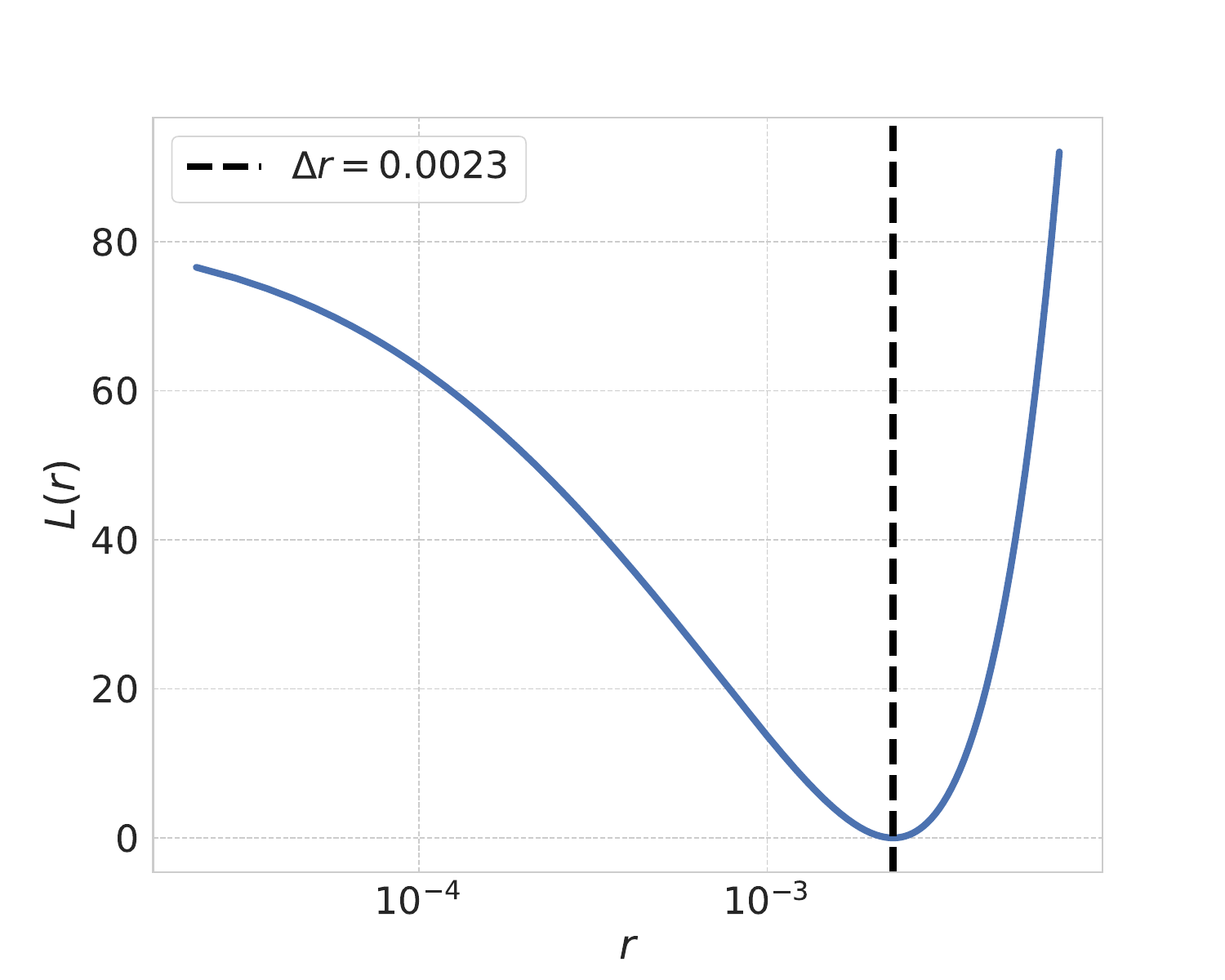
    }
    \caption[Likelihood function to estimate $\Delta r$]{Likelihood function
    $L(r)$ for the 0.1\% differential gain systematics case. }
    \label{fig:likelihood_diff_gain}
\end{figure}
The power spectra are:
\begin{align}
    \hat{C}_{\ell} & = \Delta C_{\ell}+ C_{\ell}^{\rm lens}+N_{\ell},        \\
    C_{\ell}       & = r C_{\ell}^{\rm tens}+ C_{\ell}^{\rm lens}+ N_{\ell},
\end{align}
where $\Delta C_{\ell}$ is the systematic effects power spectrum from $\Delta Q$
and $\Delta U$, $C_{\ell}^{\rm lens}$ is the lensing $B$-mode spectrum,
$C_{\ell}^{\rm tens}$ is the tensor mode with $r = 1$ \cite{PTEP2023}, and $N_{\ell}$
represents noise power spectrum.

The systematic bias $\Delta r$ is defined as the value maximizing $L(r)$:
\begin{align}
    \diff{L(r)}{r}[r=\Delta r] = 0.
\end{align}
Results in \cref{sec:Propagation} use this definition.

\section{The HEALPix software package}
\label{apd:healpix}

\healpix (Hierarchical Equal Area and isoLatitude Pixelisation) is a software
framework that enables systematic handling of spherical data through
pixelization of the celestial sphere.\footnote{\url{https://healpix.sourceforge.io/}}
It has become the standard tool for data analysis in modern astrophysics and
astronomy.

\begin{figure}[h]
    \centering
    \copyrightbox{
    \includegraphics[width=0.45\columnwidth]{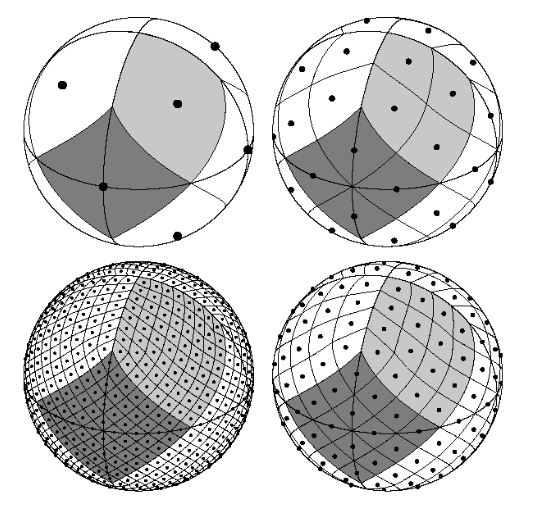}
    }{\copyright~1999 Krzysztof~M.~Gorski et al.}
    \caption[\healpix spherical division]{Orthographic projection of the
    celestial sphere divided into \healpix pixels. Bold lines represent the equator
    and meridians. Light gray regions show polar base-resolution pixels, while dark
    gray regions show equatorial base-resolution pixels, illustrating different
    pixel shapes in these regions. Clockwise from top-left: $\Nside=1,2,4,8$, corresponding
    to total pixel counts $\Npix=12,48,192,76 8$ per \cref{eq:npix}. The figure
    adapted from ref.~\cite{gorski1999healpix}, with permission of the authors.}
    \label{fig:healpix}
\end{figure}

\begin{figure}[h]
    \centering
    \copyrightbox{
    \includegraphics[width=0.45\columnwidth]{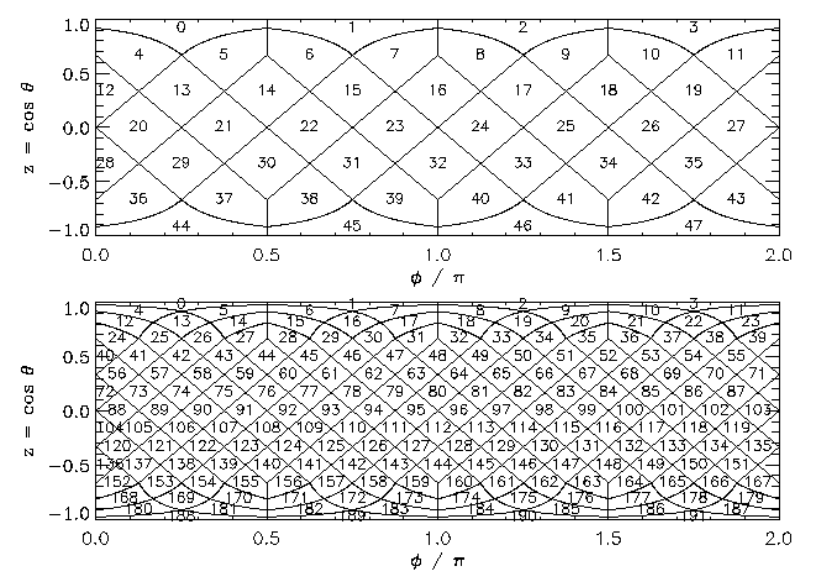}
    }{\copyright~1999 Krzysztof~M.~Gorski et al.}
    \caption[\healpix pixel ordering]{Cylindrical projection of the HEALPix-pixelized
    celestial sphere. The `Ring' ordering scheme numbers pixels from $z=\cos\theta
    = 1$ to $z=\cos\theta=-1$ with increasing $\phi$. The figure adapted from
    ref.~\cite{gorski1999healpix}, with permission of the authors.}
    \label{fig:ring_order}
\end{figure}

The \healpix resolution parameter $\Nside$ is defined as a power of 2:
\begin{align}
    \Nside= 2^{n},
\end{align}
where $n$ is an integer. This scheme divides the sphere into equal-area pixels (see
\cref{fig:healpix}). The total number of pixels $\Npix$ relates to $N_{side}$ as:
\begin{align}
    \Npix= 12 \times \Nside^{2},\label{eq:npix}
\end{align}
In CMB research, the Ring ordering scheme is commonly used for pixel numbering, where
indices spiral from the north pole to the south pole, as shown in
\cref{fig:ring_order}.

\subsection{Polarization convention on the sphere}
\label{apd:pol_convention} 

Two major conventions exist for defining polarization on a
sphere. The IAU convention defines the coordinate system at any tangent plane with positive $z$-axis pointing toward the
sphere's center and positive $x$-axis in the direction of decreasing polar angle
$\theta$. The Stokes parameters $Q$ and $U$ are defined in this frame.

In contrast, the COSMO (HEALPix) convention defines the positive $z$-axis outward
from the sphere's surface and positive $x$-axis in the direction of increasing polar
angle $\theta$. This difference results in a sign flip for the $U$ parameter.
These coordinates are summarized by NASA's LAMBDA (Legacy Archive for Microwave Background Data Analysis) project.\footnote{\url{https://lambda.gsfc.nasa.gov/product/about/pol_convention.html}}

\chapter{Supplementary results and discussions}
\minitoc

\chapabstract{ This appendix presents supplementary analyses and discussions that complement the scanning strategy optimization addressed in the manuscript. We begin by validating our map-based simulation methodology through comparison with TOD-based simulations. Subsequently, we examine how variations in the spin period influence key performance metrics. We then investigate the impact of detector positioning relative to the boresight on optimization parameters. Finally, we analyze how different spacecraft rotation configurations affect the overall scanning performance metrics. }

\section{Comparison between TOD and map-based simulation}
\label{apd:TOD_comparison}

We validate our map-based simulation with \spin formalism against general TOD-based
simulation performed by \cref{eq:map-making_TOD_w_HWP}, using the same input maps
and systematic parameters as \cref{sec:Propagation}. Following systematic effect
injection and map-making, we compute $\Delta Q$, $\Delta U$ maps and
$\Delta C_{\ell}$ according to \cref{apd:delta_r}.

\begin{figure}
    \centering
    \includegraphics[width=0.49\columnwidth]{
        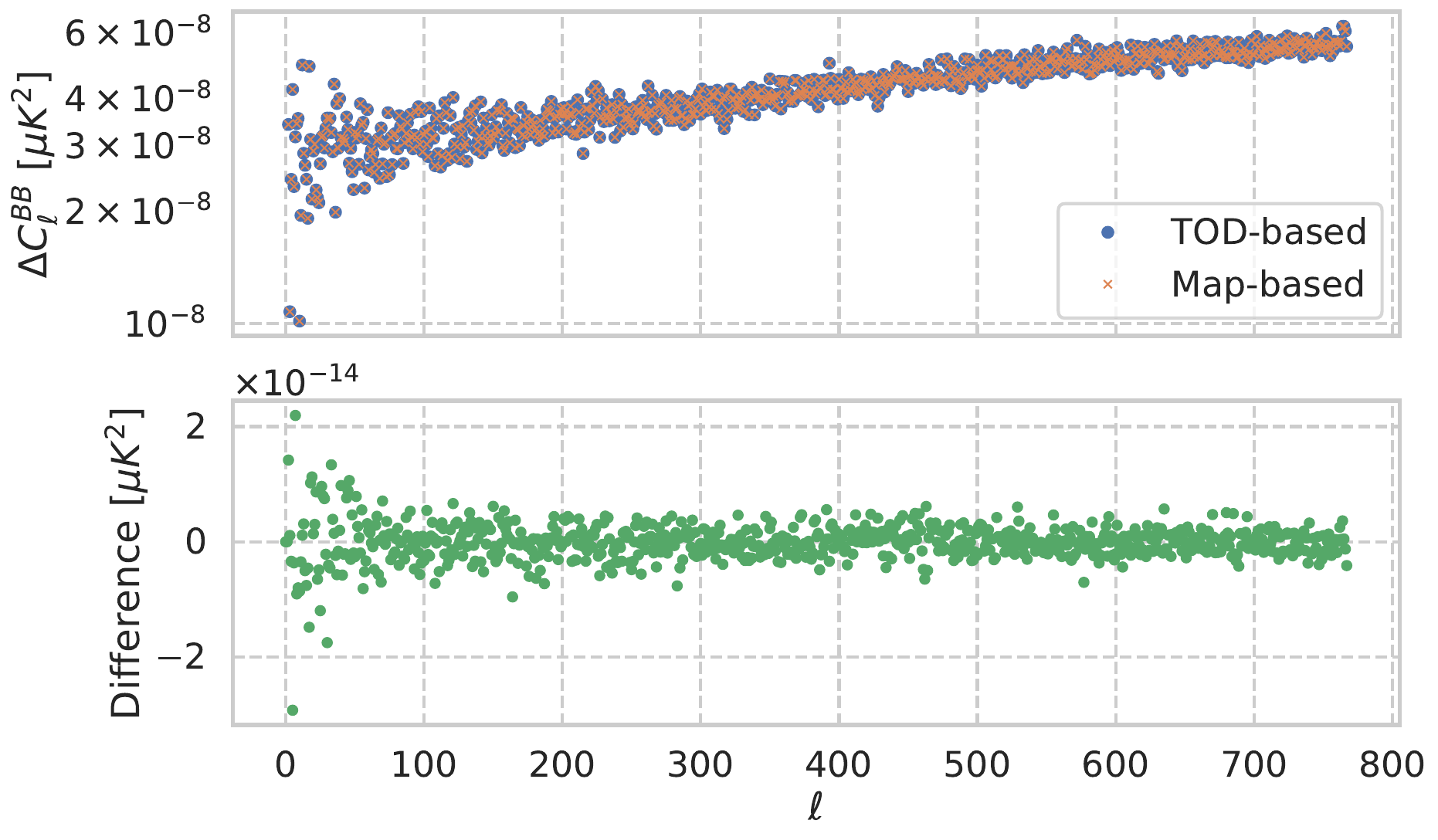
    }
    \includegraphics[width=0.49\columnwidth]{
        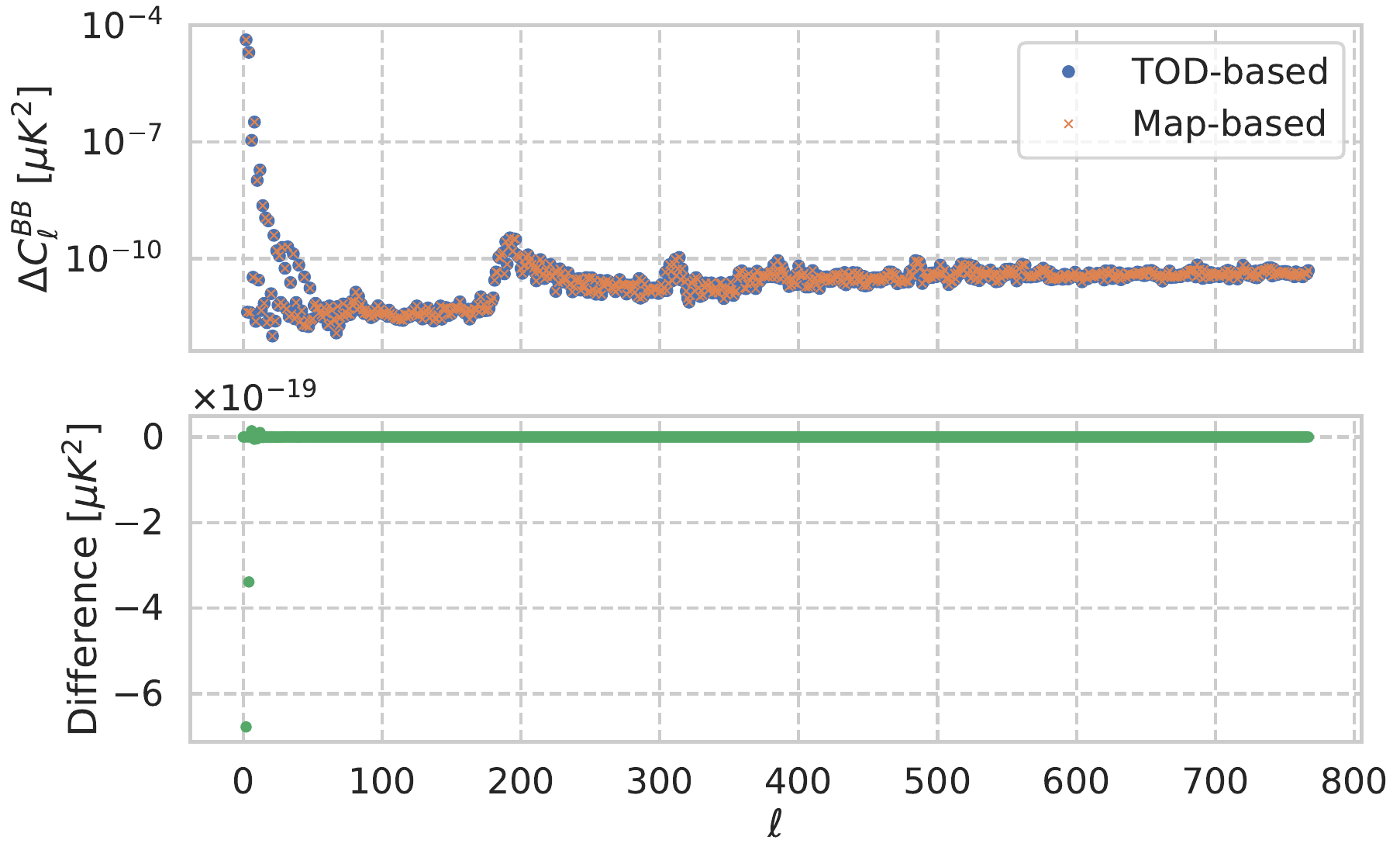
    }
    \caption[Verification of systematic effect $B$-mode power spectra between
    TOD-based and map-based simulations.]{Systematic effect $B$-mode power
    spectra ($\Delta C_{\ell}^{BB}$) comparing pointing offset (left) and HWP
    non-ideality (right). Blue dots show TOD-based simulation results, orange
    crosses show map-based simulation results, and green dots (bottom) show their
    difference. The left panel uses CMB-only input with
    $(\rho,\chi)=(1^{\prime},0^{\prime})$, while the right panel uses solar dipole
    input with $(\epsilon_{1},\phi_{QI})=(1.0\times 10^{-5},0)$.}
    \label{fig:cl}
\end{figure}

\Cref{fig:cl} shows the comparison of $\Delta C_{\ell}$ between TOD-based (blue dots)
and map-based approaches (orange crosses) for two cases:
\begin{itemize}
    \item Left: CMB-only input with pointing offset
        $(\rho,\chi)=(1^{\prime},0^{\prime})$

    \item Right: Solar dipole-only input with HWP non-ideality
        $(\epsilon_{1},\phi_{QI})=(1.0\times10^{-5},0)$
\end{itemize}

The difference between methods (green dots) remains at
$\mathcal{O}(10^{-14})\,\mu K^{2}$, confirming their numerical equivalence. For
the HWP non-ideality case, we observe a jump at $\ell \simeq 200$ arising from the
east-west structure in the hit map (see \cref{fig:hitmaps} middle bottom).

This comparison confirms that our map-based simulations achieve scientific equivalence
with TOD simulations, even when incorporating HWP contributions.

\section{Impact of the value of the spin period on the metrics}
\label{apd:T_beta_scaled}

In this paper, we presented results in the $\{\alpha,T_{\alpha}\}$ space using a
fixed $T_{\beta}=\tbl=16.9$\,min. While $T_{\beta}$ can vary freely above $\tbl$,
creating a three-dimensional parameter space $\{\alpha,T_{\alpha}, \tbl<T_{\beta}
\}$, we demonstrate that varying $T_{\beta}$ merely rescales the metrics without
affecting their optimal values.

\begin{figure}[h]
    \centering
    \includegraphics[width=0.49\columnwidth]{
        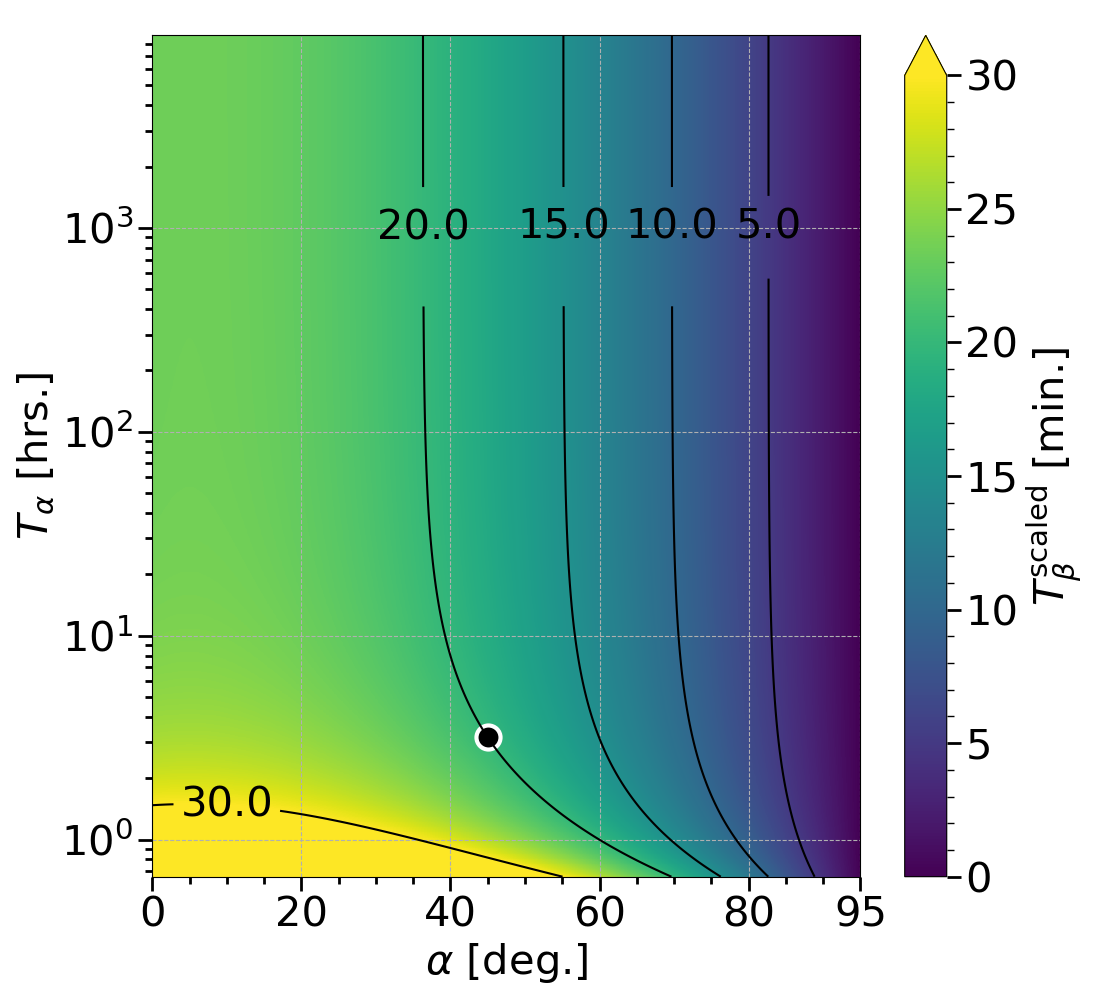
    }
    \includegraphics[width=0.49\columnwidth]{
        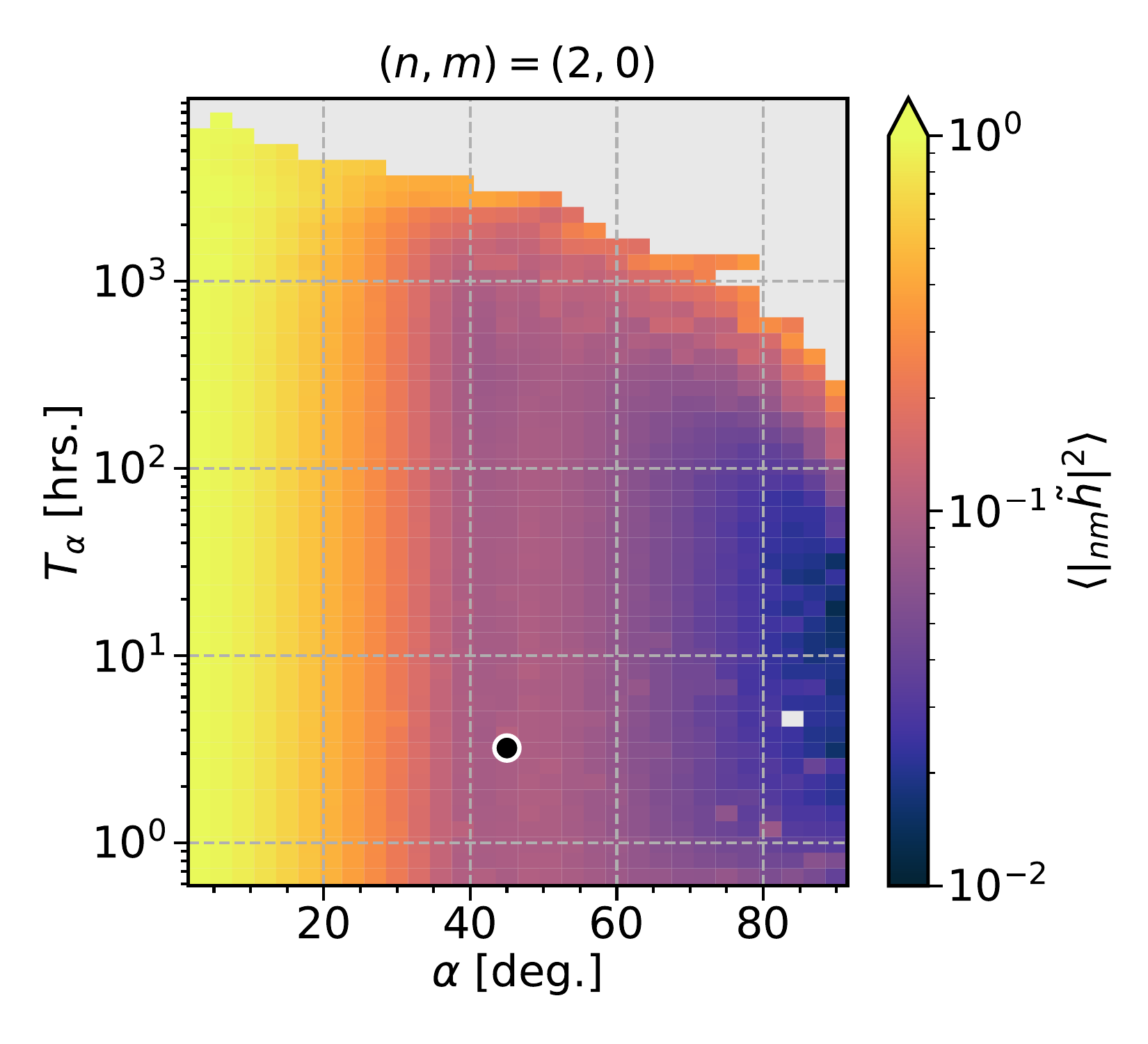
    }
    \\
    \includegraphics[width=0.49\columnwidth]{
        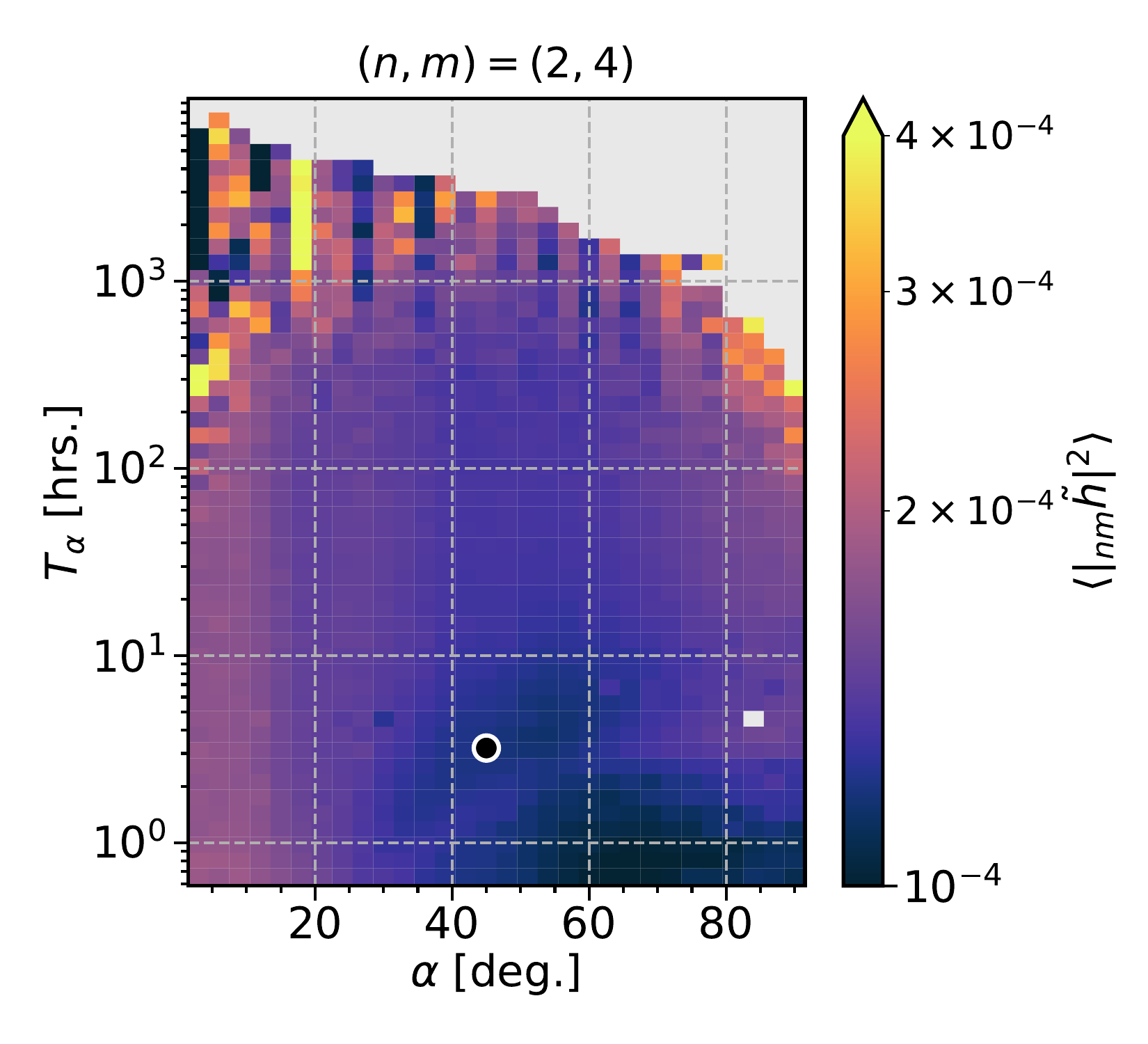
    }
    \includegraphics[width=0.49\columnwidth]{
        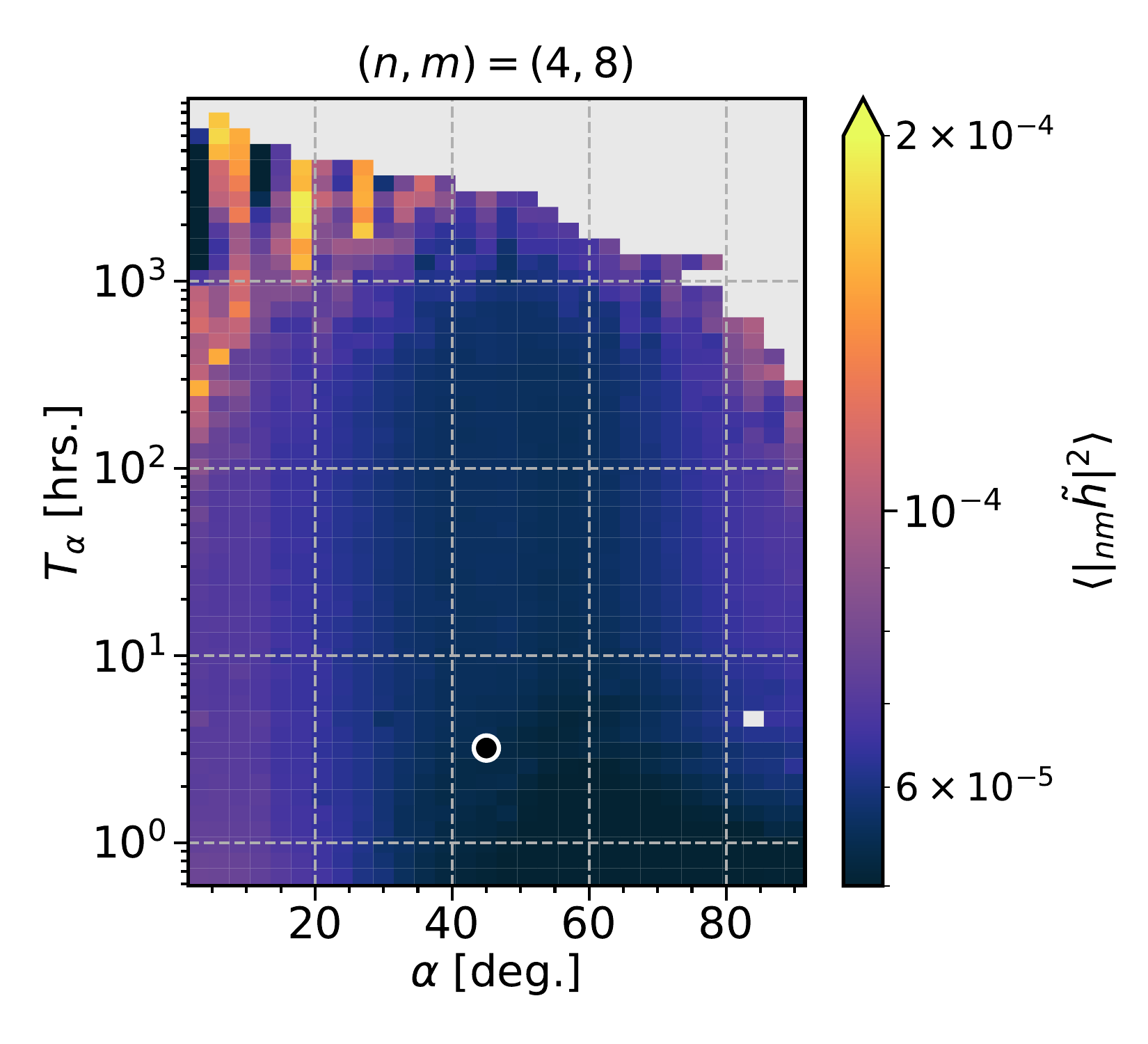
    }
    \caption[Impact of scaling $\tbl$ on the \spin-$m$ cross-link factors.]{(top
    left) $\{\alpha,T_{\alpha},T_{\beta}^{\rm scaled}\}$ space derived by
    scaling $\tbl$ to achieve $T_{\beta}=20$\,min at the \SC point. (top right) \spin-$(
    2,0)$ cross-link factors. (bottom left/right) \spin-$(2,4)$/\spin-$( 4,8)$
    cross-link factors. Non-zero \spin-$m$ cross-link factors show reduced values
    compared to $\tbl$ calculations due to increased HWP rotation time per sky pixel.
    The blank point in the bottom right corner indicates unobserved sky pixels
    from spin-precession resonances, resolvable through minor precession period adjustments.}
    \label{fig:T_beta_scaled}
\end{figure}

To illustrate this, we examine how rescaling $\tbl$ affects the cross-link factor
distribution in the $\{\alpha,T_{\alpha}\}$ space. \Cref{fig:T_beta_scaled} (top
left) shows the $\{\alpha,T_{\alpha}, T_{\beta}^{\rm scaled}\}$ space where we
multiply the \SC point $(\alpha,T_{\alpha})=(45^{\circ}, 192.348\,\rm{min})$ by
a constant to achieve $T_{\beta}=20$\,min.

The resulting cross-link factors are shown in \cref{fig:T_beta_scaled}: \spin-$(2
,0)$
(top right), \spin-$(2,4)$ (bottom left), and \spin-$(4,8)$ (bottom right). The \spin-$(
2,0)$ factor, which excludes HWP contributions, remains nearly identical to results
using $\tbl$. The \spin-$(2,4)$ and \spin-$(4,8)$ factors maintain their flat
distribution pattern but show smaller values due to slower spin allowing more
HWP rotation per sky pixel. While this helps suppress systematic effects,
excessive slowdown can degrade \spin-$(n,0)$ cross-link factors, as discussed in
\cref{sec:Opt_kinetic}. These results confirm that changing rotation periods
only scales values without altering the global structure.

This scaling principle extends to hit-map uniformity and planet visibility
metrics, as these primarily depend on geometric parameters like sky scanning patterns
and scan pupil size.

\section{Metrics for detectors located away from the boresight}
\label{apd:other_detector}

Here we analyze how metrics vary for detectors positioned away from the
boresight (focal plane center) in our \SC. These detectors have different effective
$\beta$ angles ($\beta^{\rm eff}$) relative to the boresight, expressed as:
\begin{align}
    \beta^{\rm eff}_{i}= \beta + \beta_{i},
\end{align}
where $i$ denotes the detector index. Different $\beta_{i}$ values yield unique
scan trajectories and metric values. However, detectors with
$\beta^{\rm eff}_{i}$ close to $\beta$ maintain nearly identical trajectories
and metrics to the boresight. We simulate planets visibility time and cross-link
factors for detectors at extreme $\beta_{i}$ values ($\pm14^{\circ}$) within the
MFT/HFT's $14^{\circ}$ radius field of view.

\begin{figure}[h]
    \centering
    \includegraphics[width=0.32\columnwidth]{
        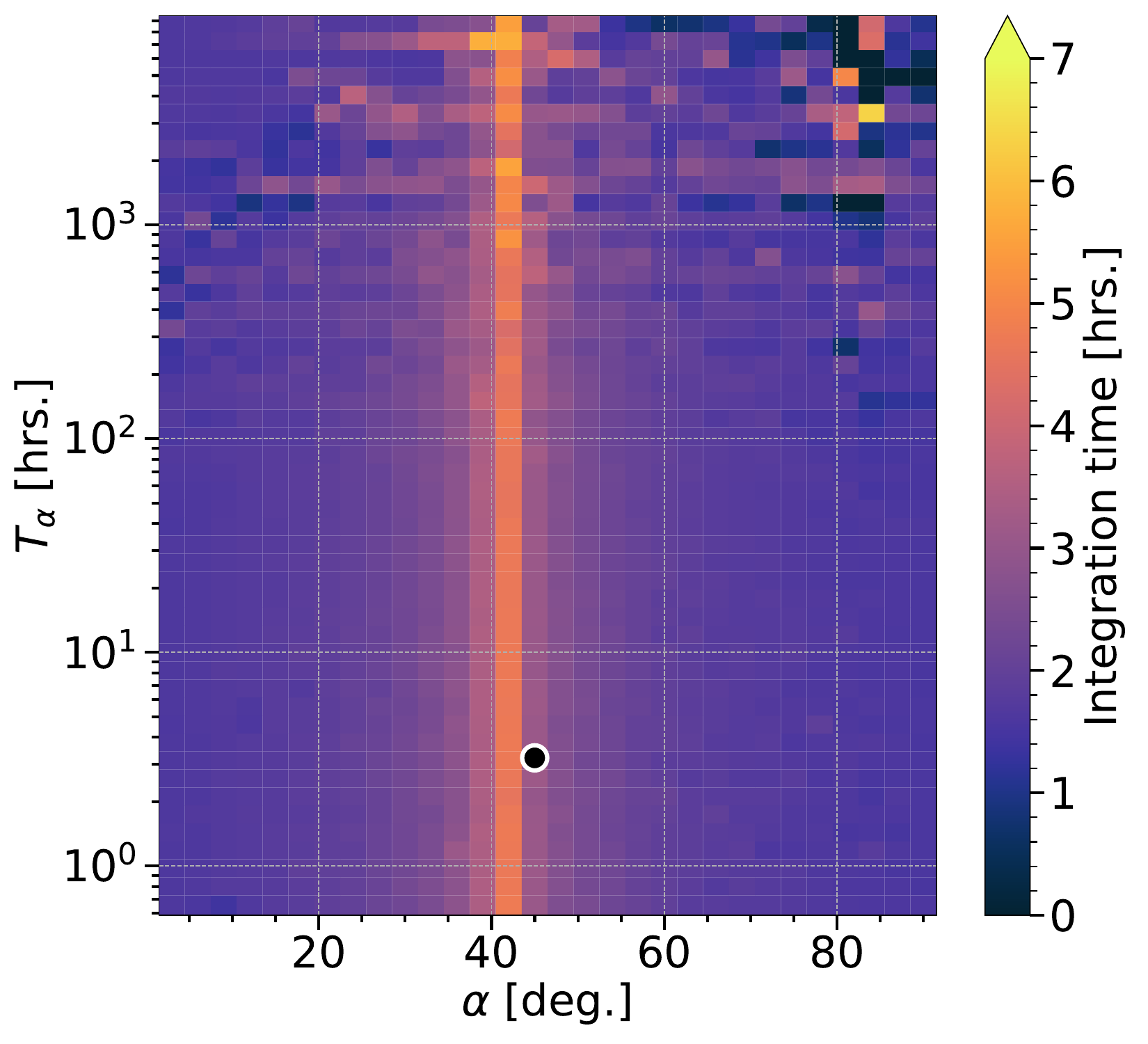
    }
    \includegraphics[width=0.32\columnwidth]{
        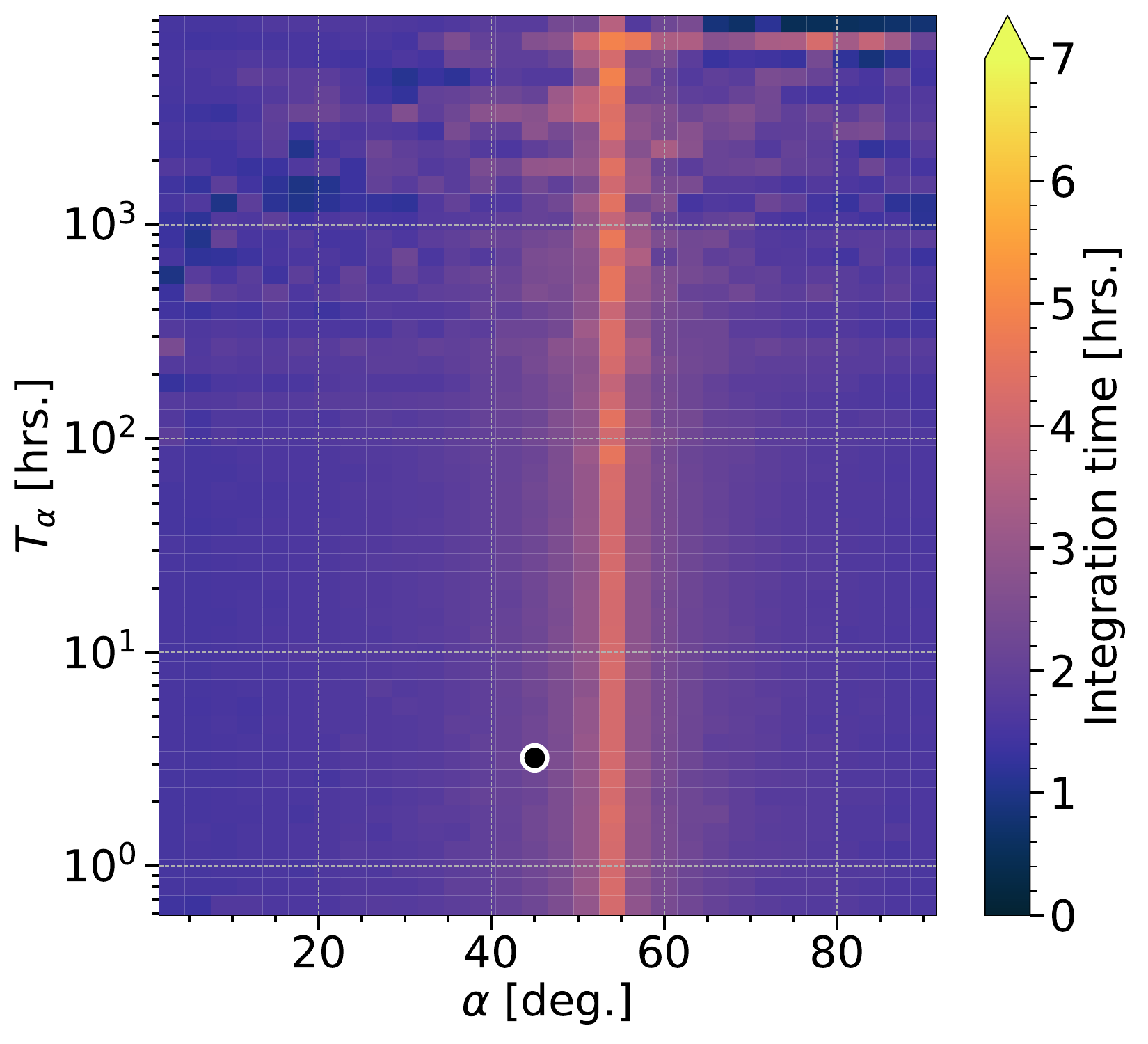
    }
    \includegraphics[width=0.32\columnwidth]{
        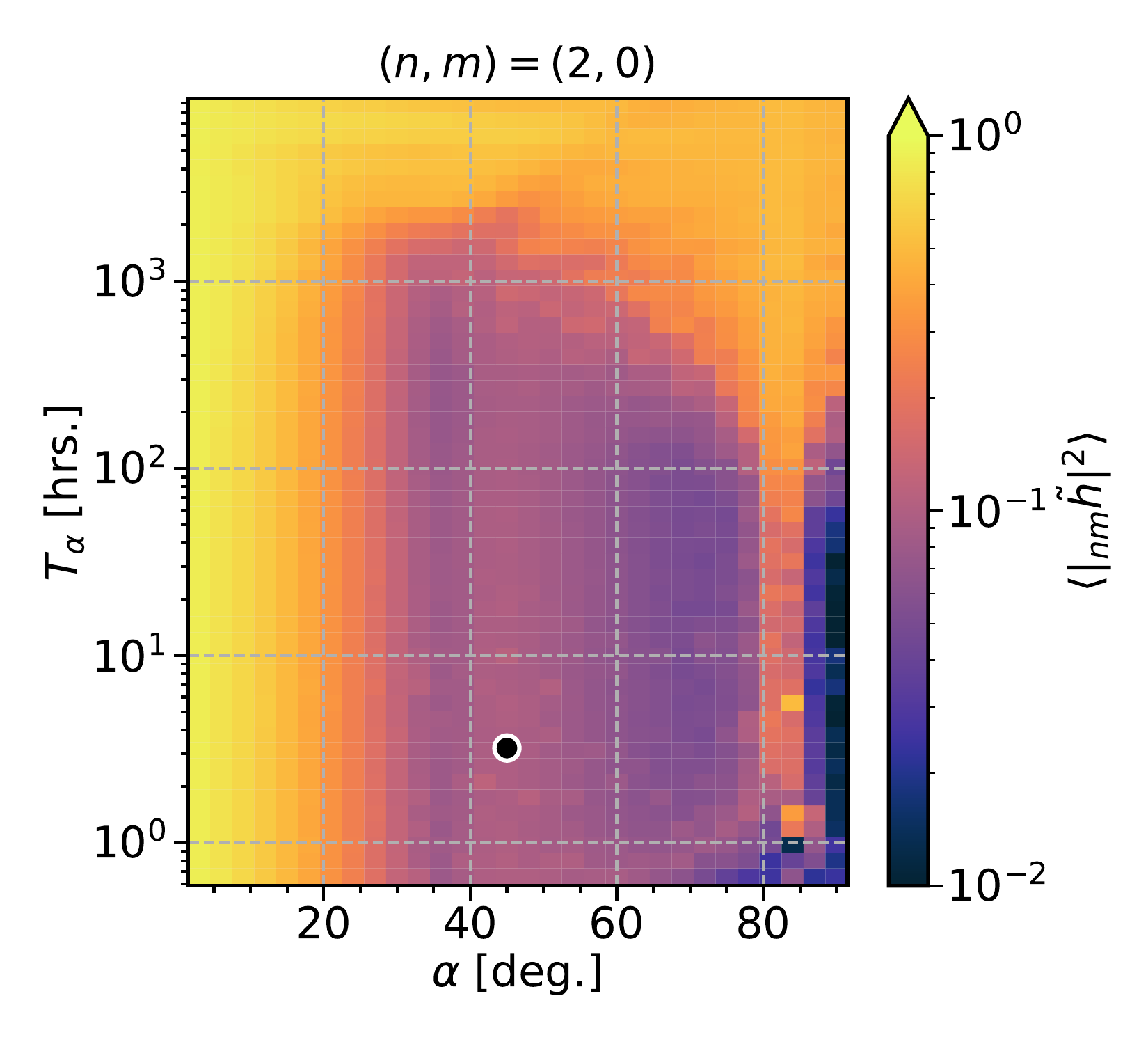
    }
    \\
    \includegraphics[width=0.32\columnwidth]{
        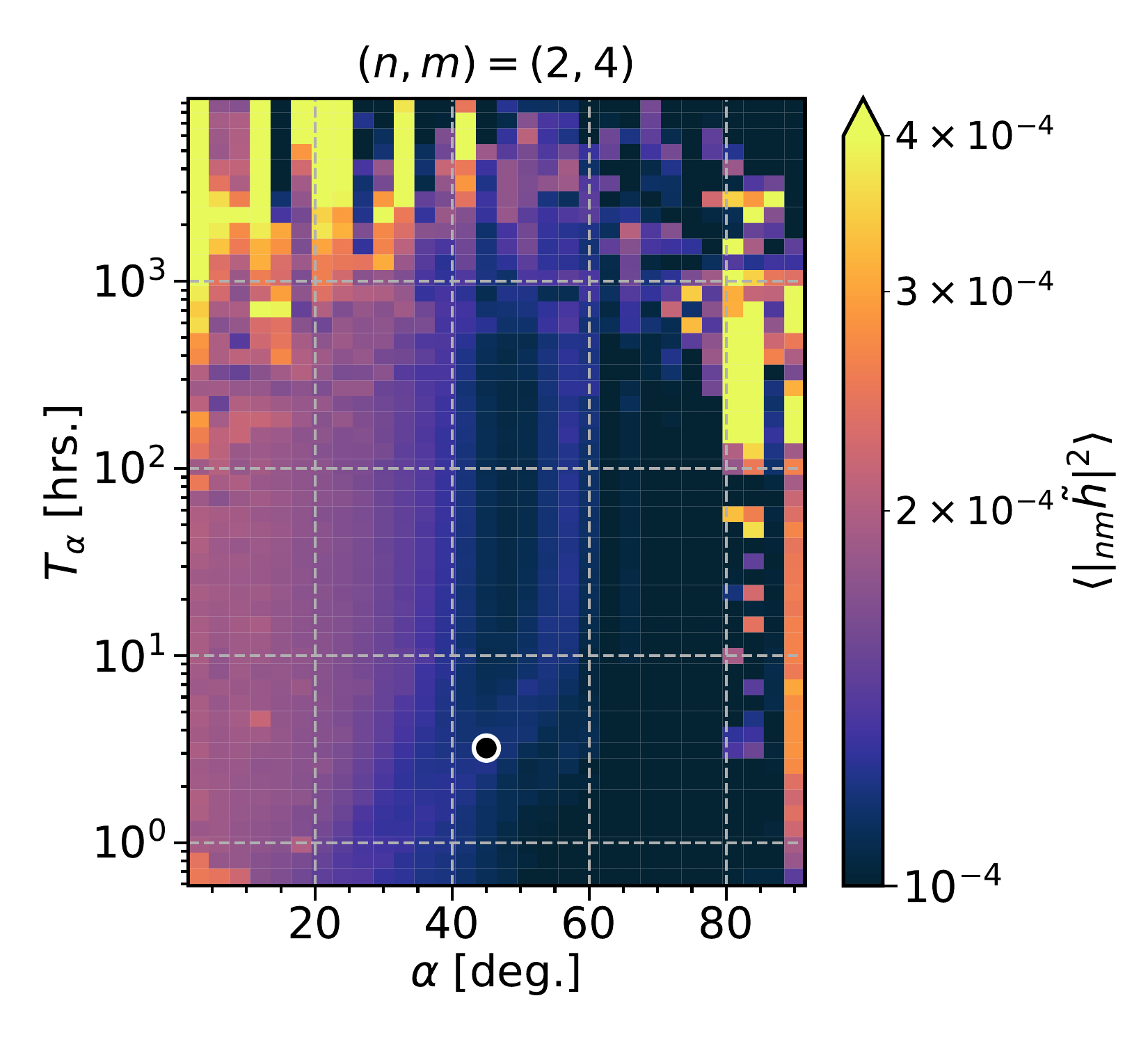
    }
    \includegraphics[width=0.32\columnwidth]{
        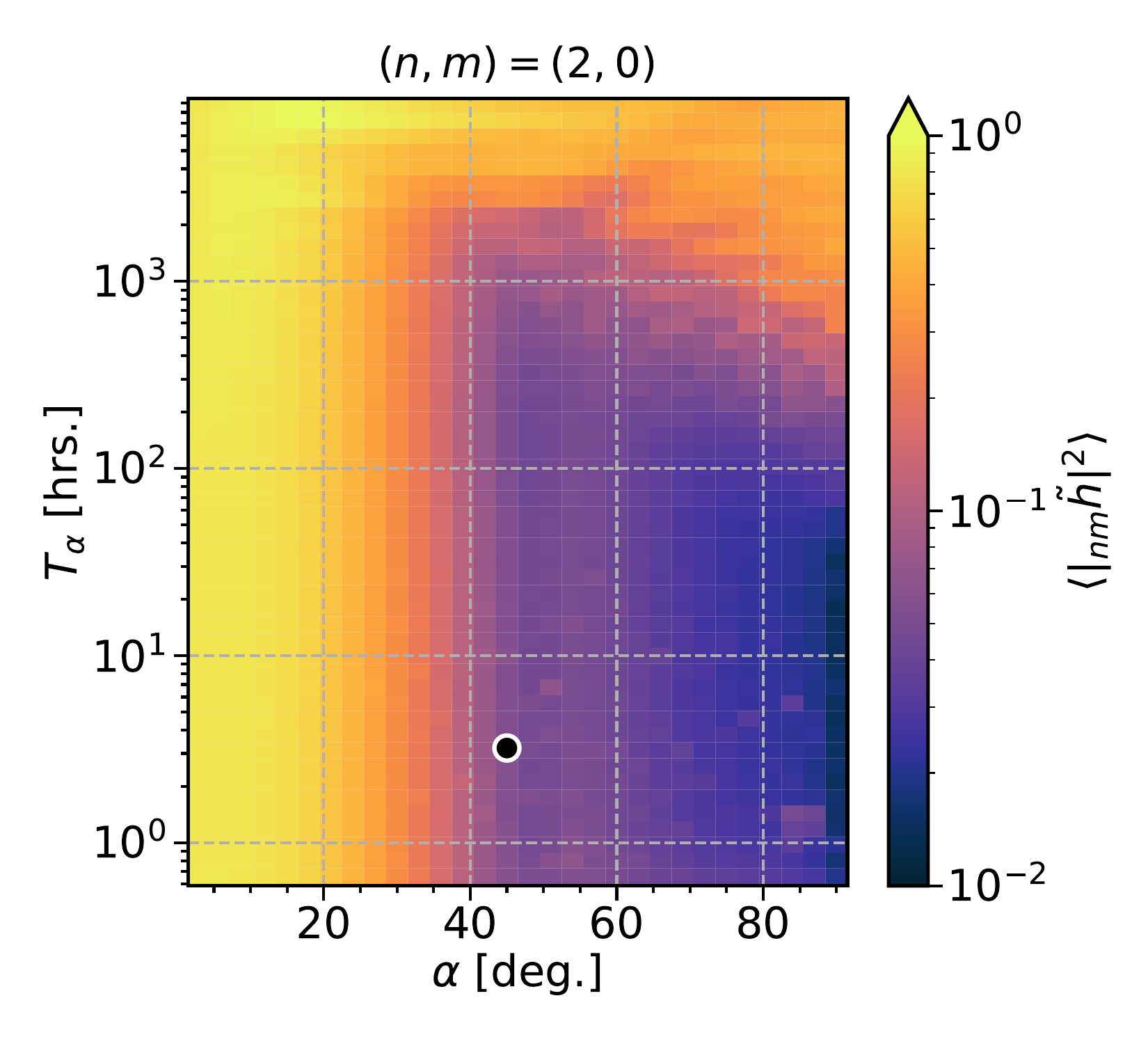
    }
    \includegraphics[width=0.32\columnwidth]{
        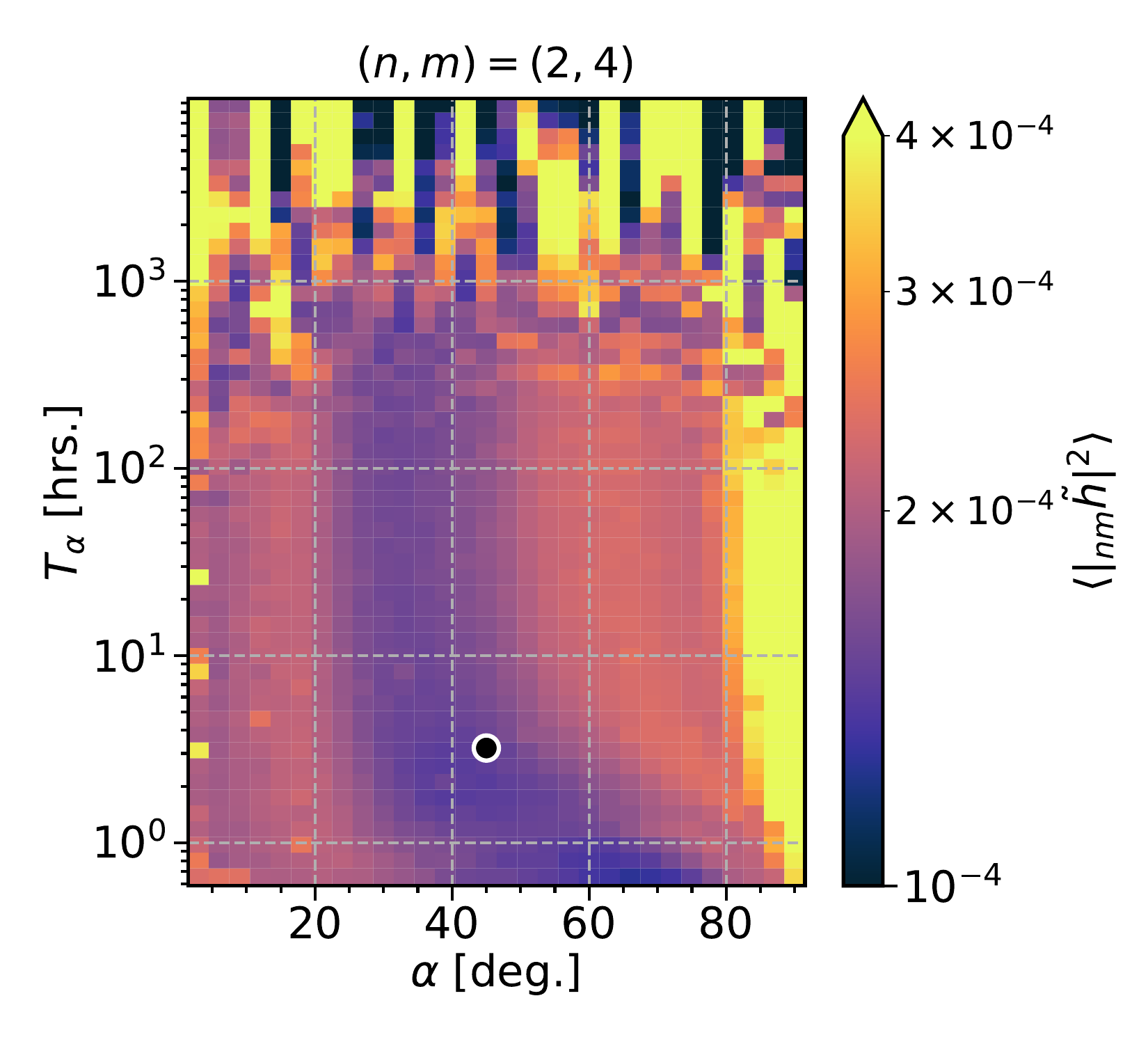
    }
    \caption[Metrics for detectors located away from the boresight.]{(top left/top
    middle) Integrated planet visibility time simulated by a detector which has
    $\beta_{i}=-14^{\circ}/14^{\circ}$. (top right/bottom left) \spin-$(2,0)$ cross-link
    factor simulated by a detector which has $\beta_{i}=-14^{\circ}/14^{\circ}$.
    (bottom middle/bottom right) \spin-$(2,4)$ cross-link factor simulated by a
    detector which has $\beta_{i}=-14^{\circ}/14^{\circ}$. Values for sky pixels
    with diverging cross-link factors are ignored and averaged over the entire sky.}
    \label{fig:metrics_other_det}
\end{figure}

\Cref{fig:metrics_other_det} shows planet visibility integration times (top left/middle:
$\beta_{i}=\mp14^{\circ}$). Like \cref{fig:planet_visibility}, these remain
independent of $T_{\alpha}$, peaking at $\alpha=(\kappa + \beta_{i})/2$. The
results demonstrate that our \SC provides substantial planet visibility for both
boresight and extreme-$\beta^{\rm eff}_{i}$ detectors.

The figure also shows \spin-$(2,0)$ cross-link factors (top right/bottom left: $\beta
_{i}=\mp14^{\circ}$) and \spin-$(2,4)$ factors (bottom middle/right:
$\beta_{i}=\mp14^{\circ}$). For some effective $\beta$ values, detectors cannot satisfy
\cref{eq:const_geometric}, causing unobservable regions near ecliptic poles with
divergent cross-link factors. We present full-sky averages excluding these divergent
pixels.

The $\beta_{i}=-14^{\circ}$ detector's limited sky coverage prevents full-sky observation.
While \spin-$(2,0)$ factors maintain similar structure between
$\beta_{i}=\pm14^{\circ}$, values increase for $\beta_{i}=-14^{\circ}$ due to reduced
crossing angle uniformity in the narrower accessible region. The \spin-$(2,4)$
factors show structural changes with generally lower values for
$\beta_{i}=-14^{\circ}$, as slower scanning (per \cref{eq:sweeping_velocity}) allows
more HWP rotations per sky pixel. Despite these variations, the \SC maintains low
cross-link factors across all detector positions.

\section{Discussion on the rotation direction of the spacecraft}
\label{apd:rotation_direction}

While we defined forward (counterclockwise) rotation in \cref{eq:matrix_chain}, four
possible spin-precession rotation direction combinations exist. Since these
combinations create different trajectories relative to orbital direction, we must
verify our conclusions hold regardless of rotation direction.

We classify the four combinations using rotation matrix signs and define spin-precession
coherence as their product, shown in \cref{tab:rotation_convention}.

\begin{table}[ht]
    \centering
    \begin{tabular}{lllll}
        \hline
        Sign of precession        & $+$ & $+$ & $-$ & $-$ \\
        \hline
        Sign of spin              & $+$ & $-$ & $+$ & $-$ \\
        \hline
        Spin-precession coherence & $+$ & $-$ & $-$ & $+$ \\
        \hline
    \end{tabular}
    \caption{Convention for spin and precession coherence.}
    \label{tab:rotation_convention}
\end{table}

Positive coherence follows \cref{eq:sweeping_velocity} for $\omega_{\rm max}$,
while negative coherence causes one rotation to counteract the other, reducing sweep
velocity. For negative coherence, we transpose the $R_{z}$ rotation matrix in \cref{eq:matrix_chain}:
\begin{align}
    \bm{n}(t) & = R_{z}(\omega_{\alpha}t)R_{y}(\alpha )R_{z}^{\top}(\omega_{\beta}t)\bm{n}_{0},
\end{align}
where $\top$ denotes matrix transposition. The sweep velocity becomes:
\begin{align}
    \omega_{\rm max}^{\pm}= \omega_{\alpha}\sin(\alpha\pm\beta) + \omega_{\beta}\sin\beta,
\end{align}
with $+/-$ indicating positive/negative coherence. This modification affects $T_{\beta}
^{\rm lower}$ and $T_{\alpha}^{\rm lower}$ constraints:
\begin{align}
    T_{\beta}^{\rm{lower},\pm}  & = \frac{2\pi \Nmod T_{\alpha}\sin\beta}{\Delta \theta f_{\phi}T_{\alpha}- 2\pi \Nmod \sin(\alpha\pm\beta)},\label{eq:T_spin_minus} \\
    T_{\alpha}^{\rm{lower},\pm} & = \frac{2\pi \Nmod (\sin\beta + \sin(\alpha\pm\beta))}{\Delta \theta f_{\phi}}.
\end{align}

\begin{figure}
    \centering
    \includegraphics[width=0.49\columnwidth]{
        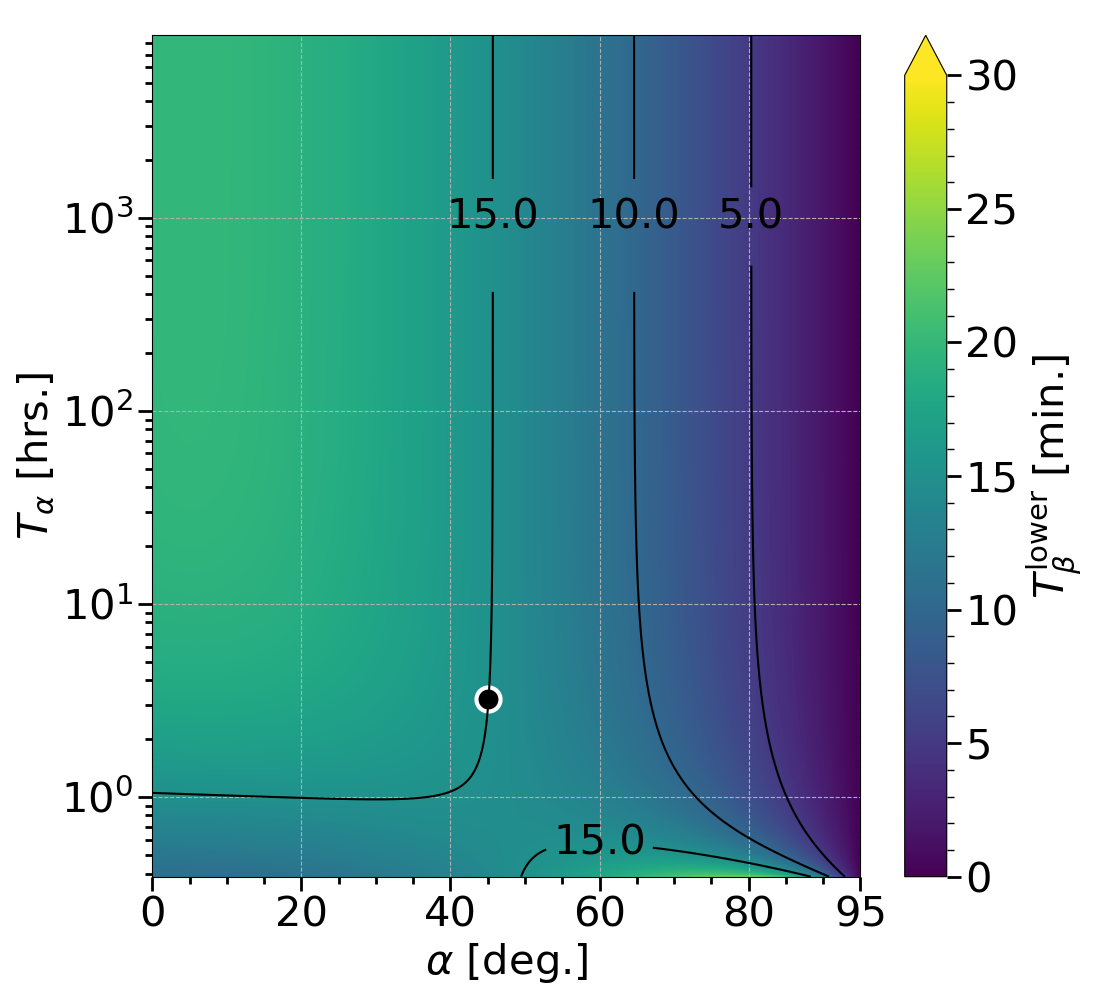
    }
    \includegraphics[width=0.49\columnwidth]{
        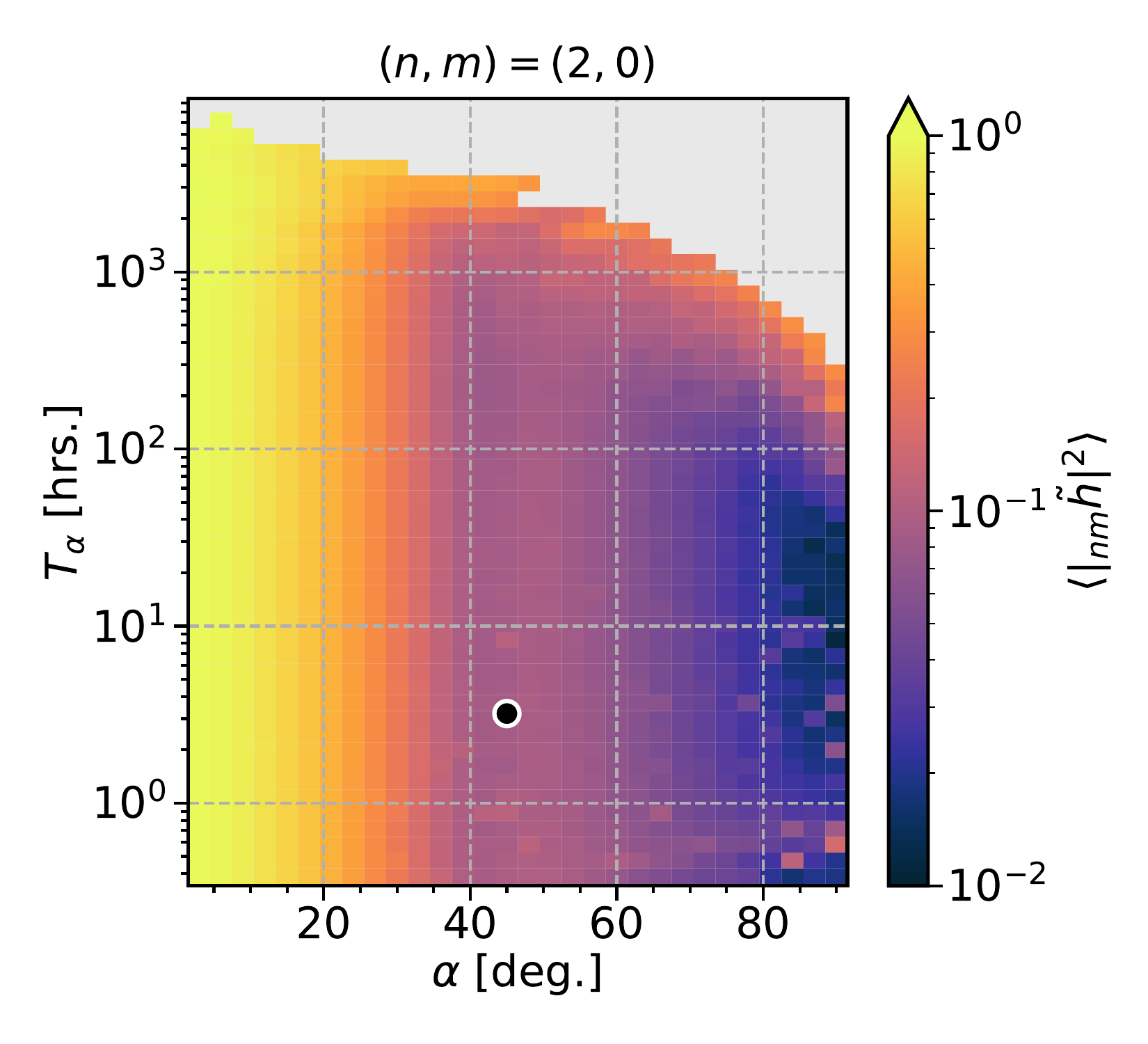
    }
    \\
    \includegraphics[width=0.49\columnwidth]{
        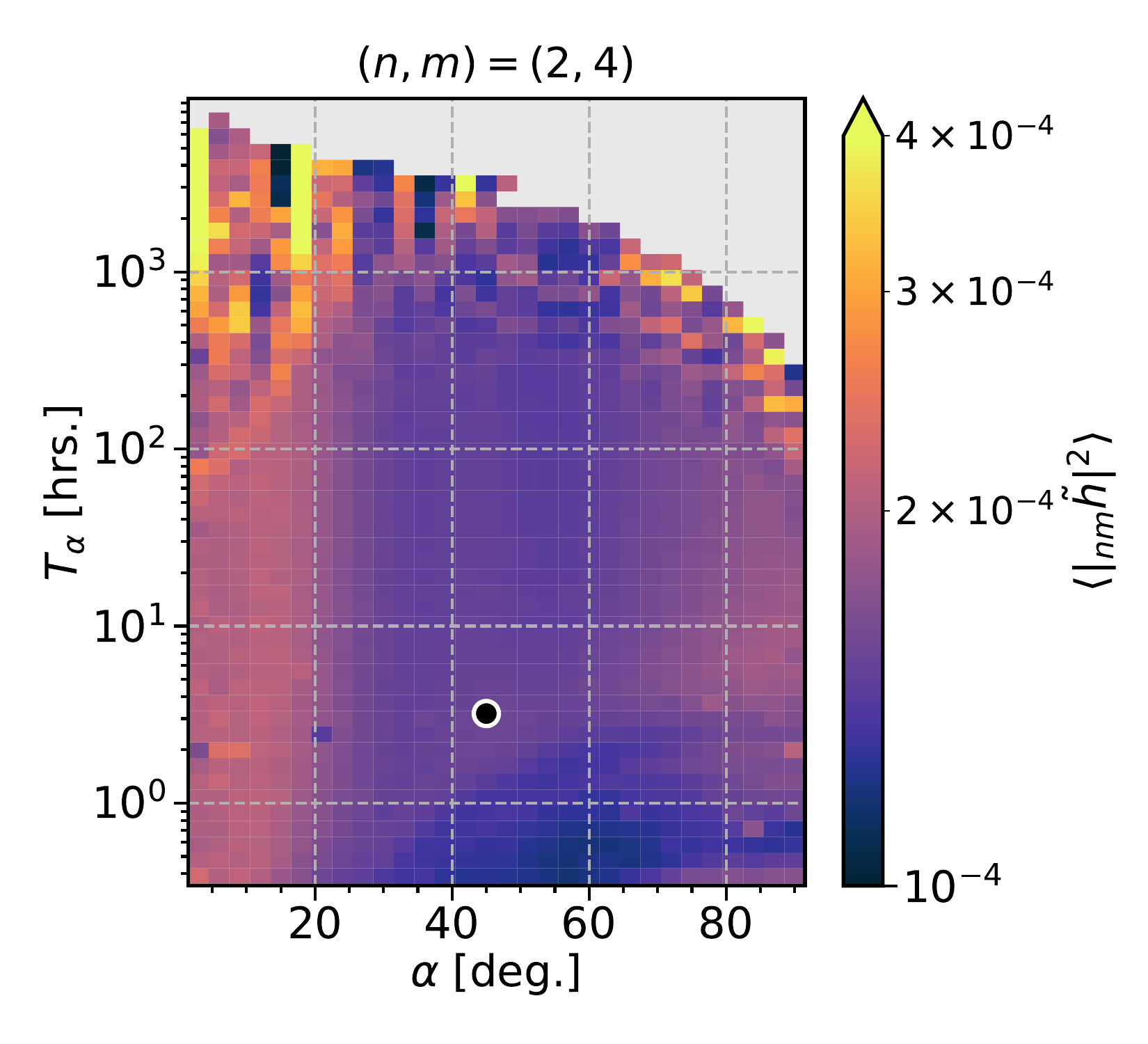
    }
    \includegraphics[width=0.49\columnwidth]{
        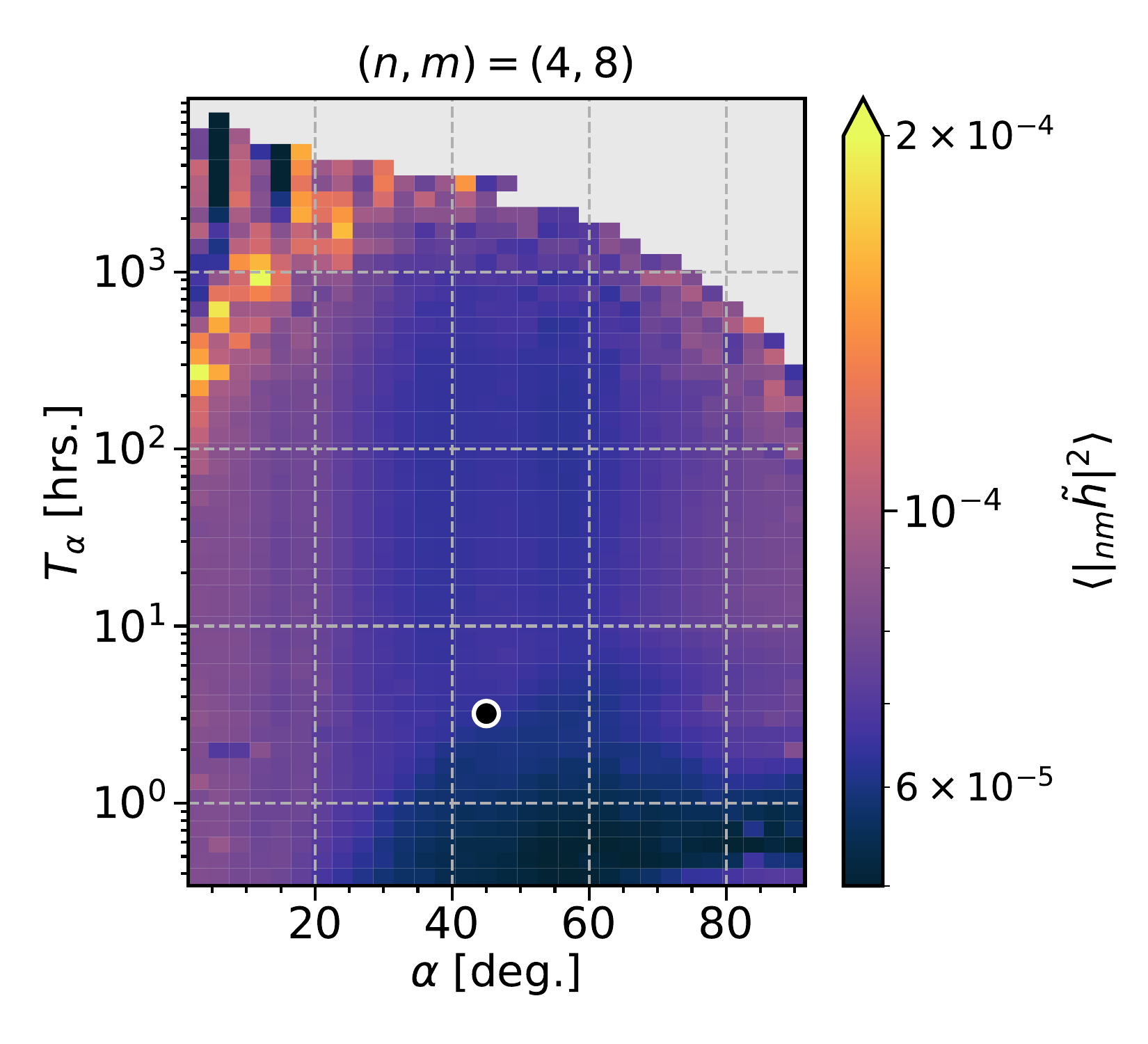
    }
    \caption[Impact of spacecraft rotation direction on \spin-$m$ cross-link
    factors. ]{(top left) $\{\alpha, T_{\alpha}, T_{\beta}^{\rm lower,-}\}$
    space given by \cref{eq:T_spin_minus}. (top right) \spin-$(2,0)$ cross-link factors
    simulated using $T_{\beta}^{\rm lower,-}$. (bottom left/right) \spin-$(2,4)$/\spin-$(
    4 ,8)$
    cross-link factors simulated using $T_{\beta}^{\rm lower,-}$. Maps use $(\rm{prec.}
    ,\rm{spin})=(+,-)$ configuration. The change from
    $\omega_{\rm max}^{+}\rightarrow \omega_{\rm max}^{-}$ reduces sweep speed
    and increases HWP rotations per sky pixel visit, resulting in smaller non-zero
    \spin-$m$ cross-link factors compared to $T_{\beta}^{\rm lower,+}$ case in \cref{sec:result_crosslink}.}
    \label{fig:CCW_figures}
\end{figure}

Negative coherence produces a different $T_{\beta}^{\rm lower}(\alpha,T_{\alpha})$
space than \cref{fig:standard_config_and_T_beta} (right), shown in
\cref{fig:CCW_figures} (top left), allowing shorter $T_{\beta}^{\rm lower}$ due to
$\omega_{\rm max}^{+}\to \omega_{\rm max}^{-}$.
We examined all metrics across all four combinations, finding the \SC maintains its
advantages. The remaining panels in \cref{fig:CCW_figures} show cross-link
factors for $(\rm{prec.},\rm{spin})=(+, -)$, demonstrating minimal changes from \cref{fig:cross-links}.

Notably, negative coherence configurations like $(\rm{prec.},\rm{spin})=(+,-)$
can reduce maximum sweep velocity without parameter changes. For example, the
\SC's $\omega_{\rm max}^{+}=0.26$\,deg/s reduces to $\omega_{\rm max}^{-}=0.23$\,deg/s
with reversed spin, increasing data samples and HWP rotations per sky pixel.

The impact of this velocity reduction on physical results, particularly with
time-correlated noise and HWP-synchronized systematics, requires future end-to-end
simulation study. However, our \SC maintains required scanning capabilities regardless
of rotation directions, providing sufficient information for basic spacecraft
design.

\chapter{Additional figures}
\label{apd:additional_figures}

\minitoc

\chapabstract{ This appendix shows specific cross-link maps for \spin-$(n,m)$ configurations. }

\section{Cross-link maps}
\begin{figure}[htbp]
    \centering
    \includegraphics[width=0.49\columnwidth]{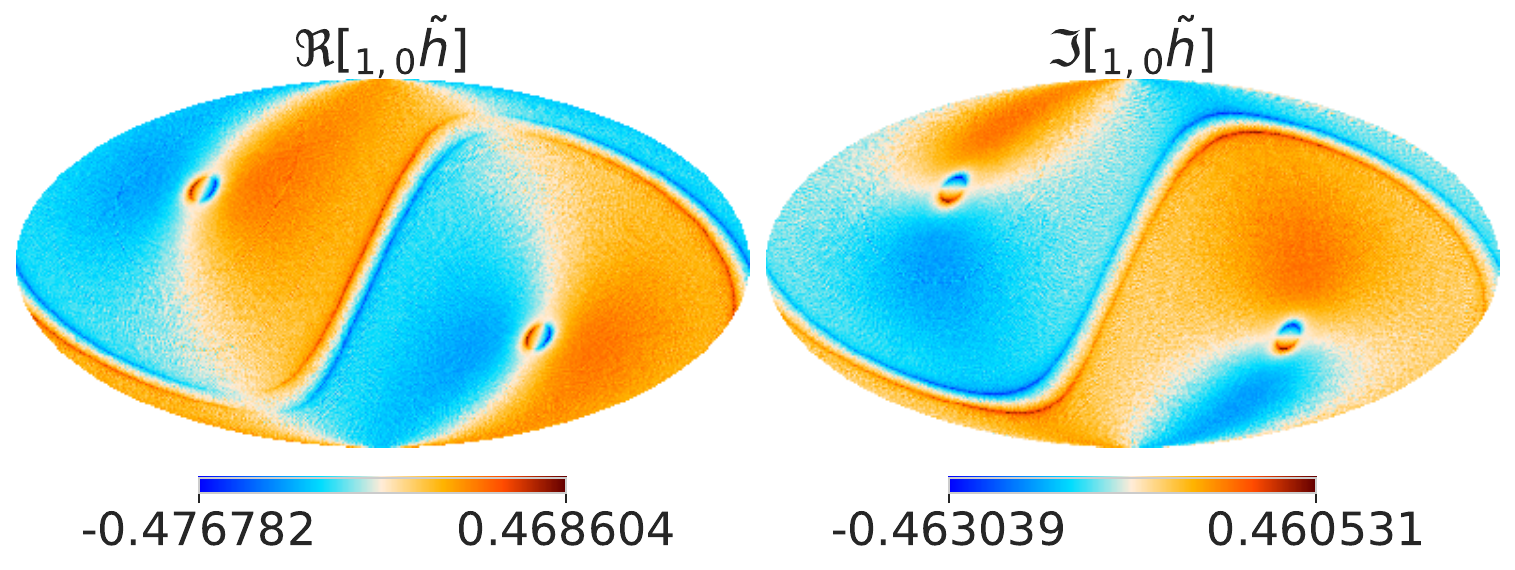}
    \includegraphics[width=0.49\columnwidth]{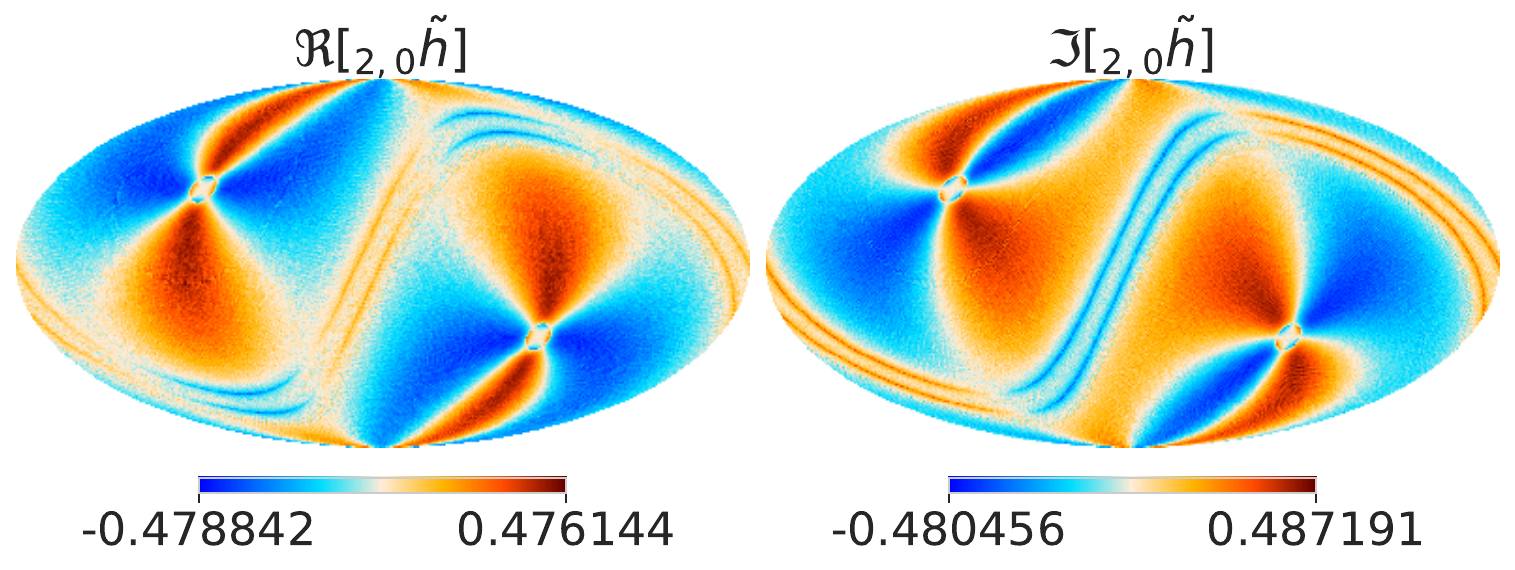}
    \\
    \includegraphics[width=0.49\columnwidth]{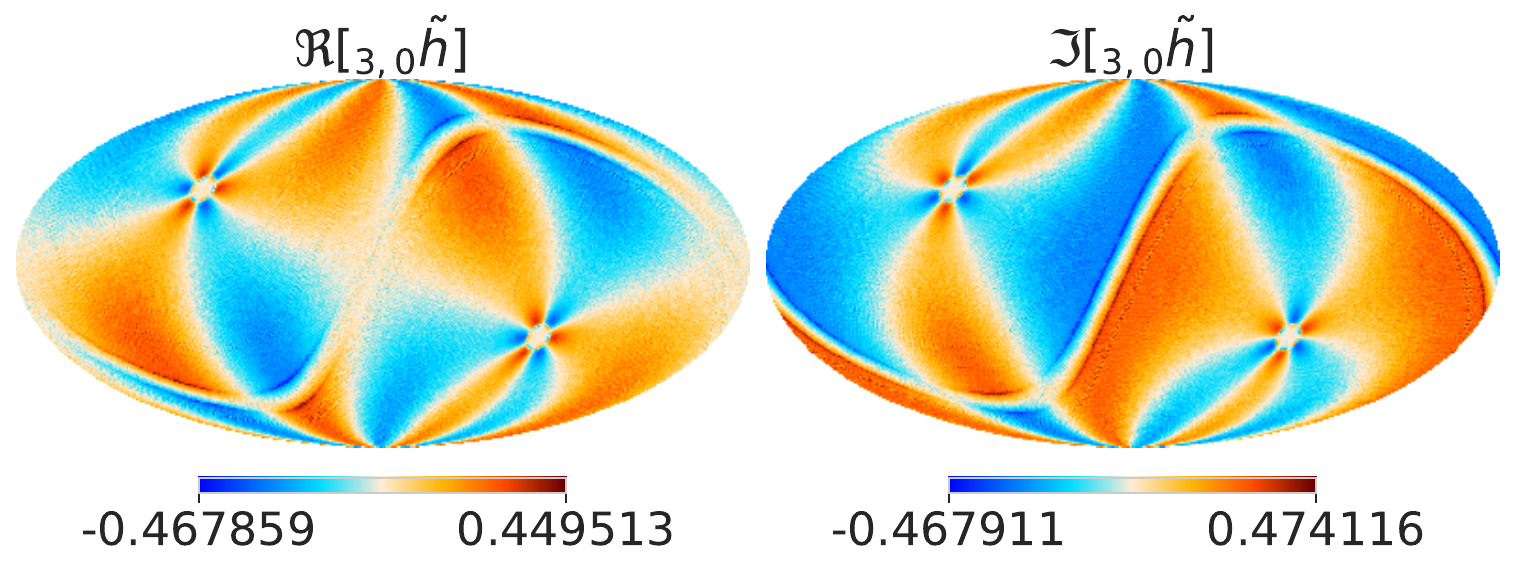}
    \includegraphics[width=0.49\columnwidth]{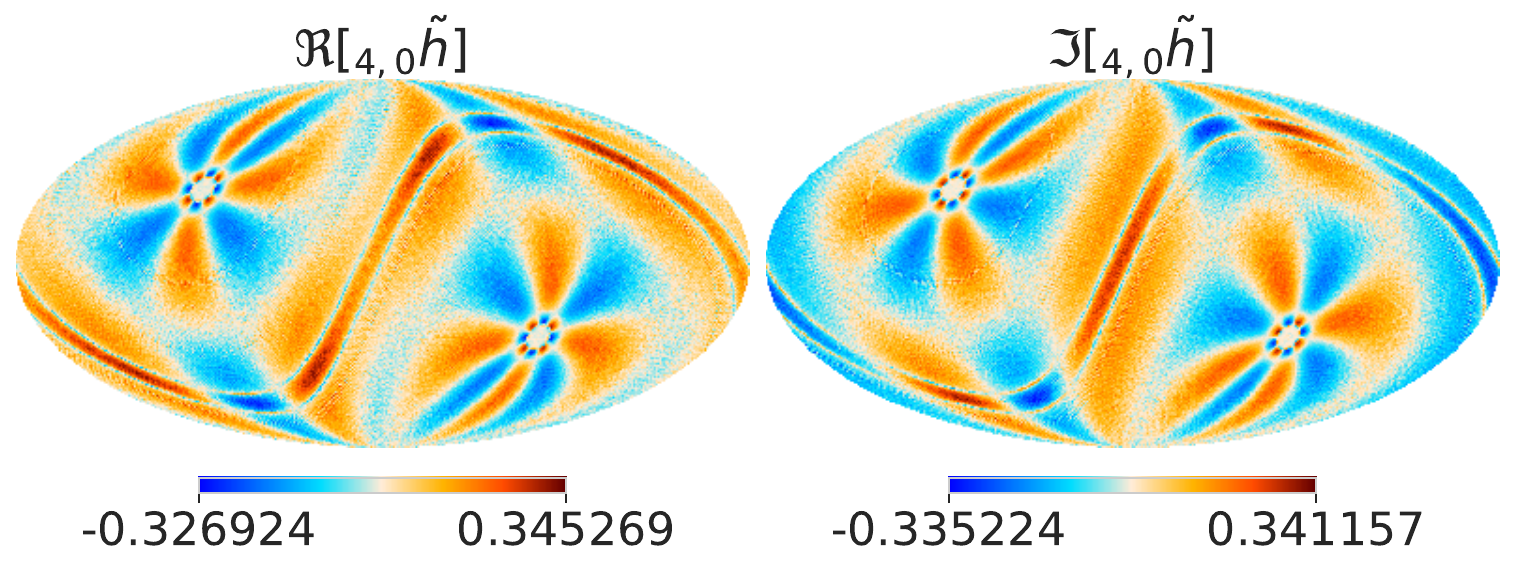}
    \\
    \includegraphics[width=0.49\columnwidth]{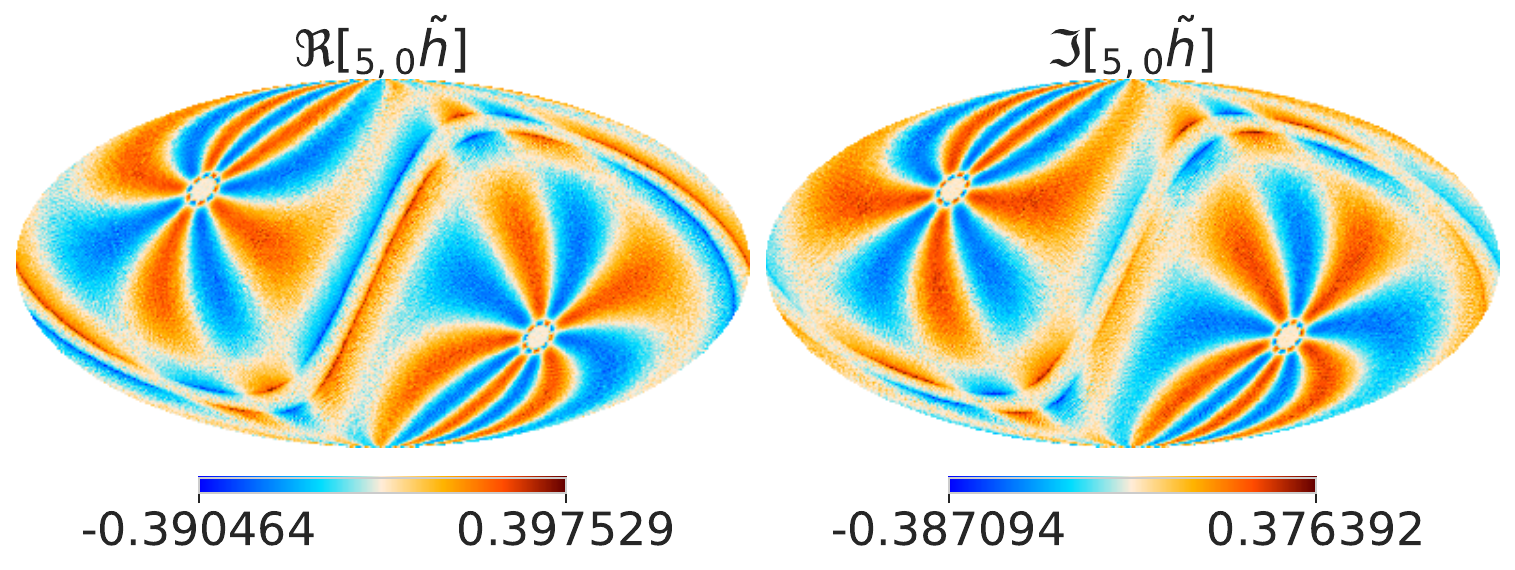}
    \includegraphics[width=0.49\columnwidth]{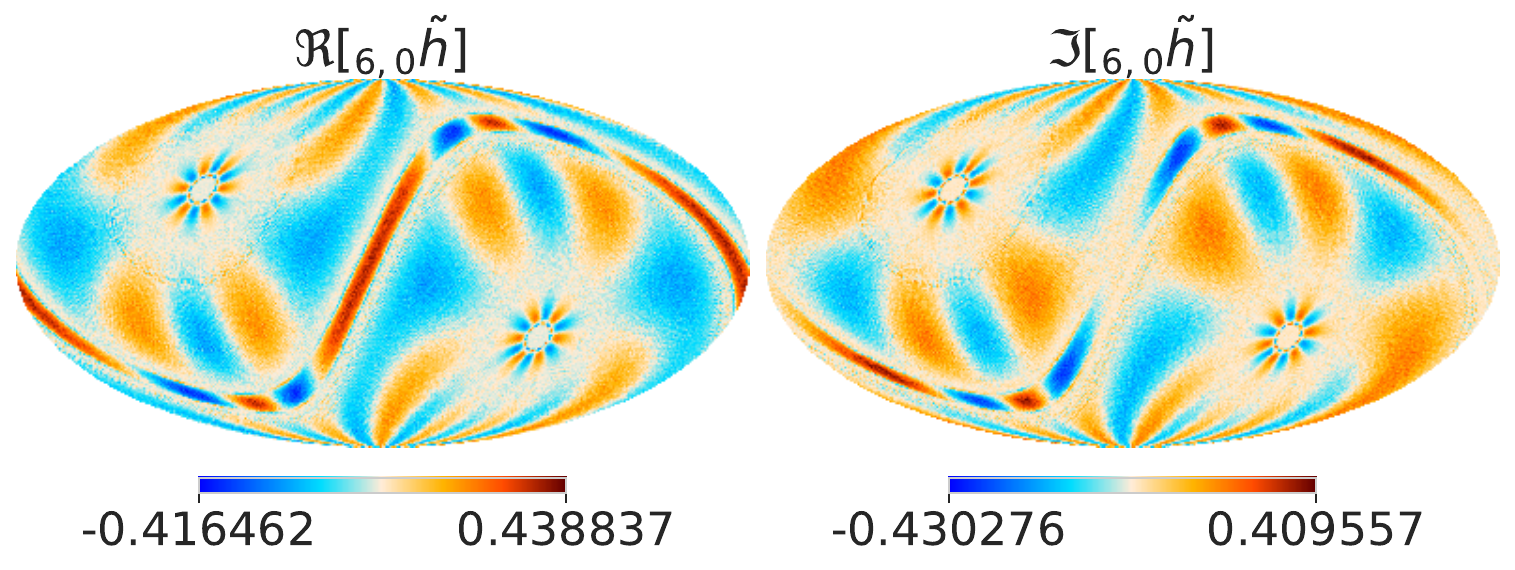}
    \\
    \includegraphics[width=0.49\columnwidth]{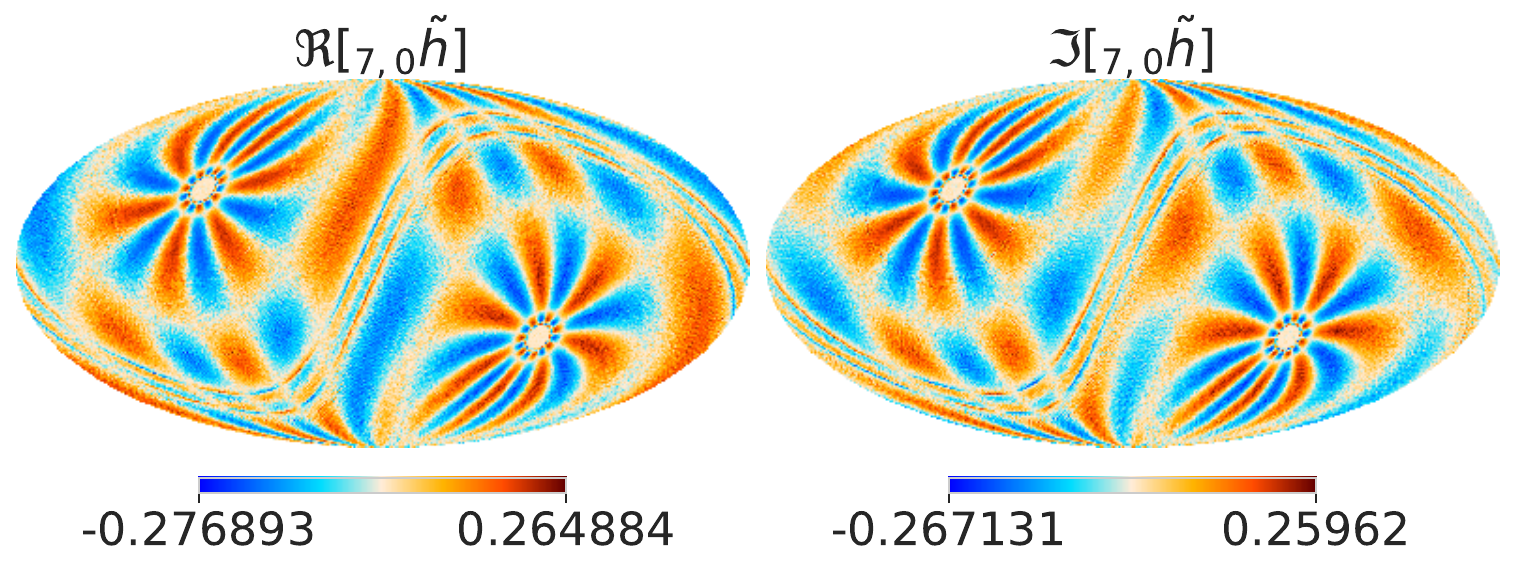}
    \includegraphics[width=0.49\columnwidth]{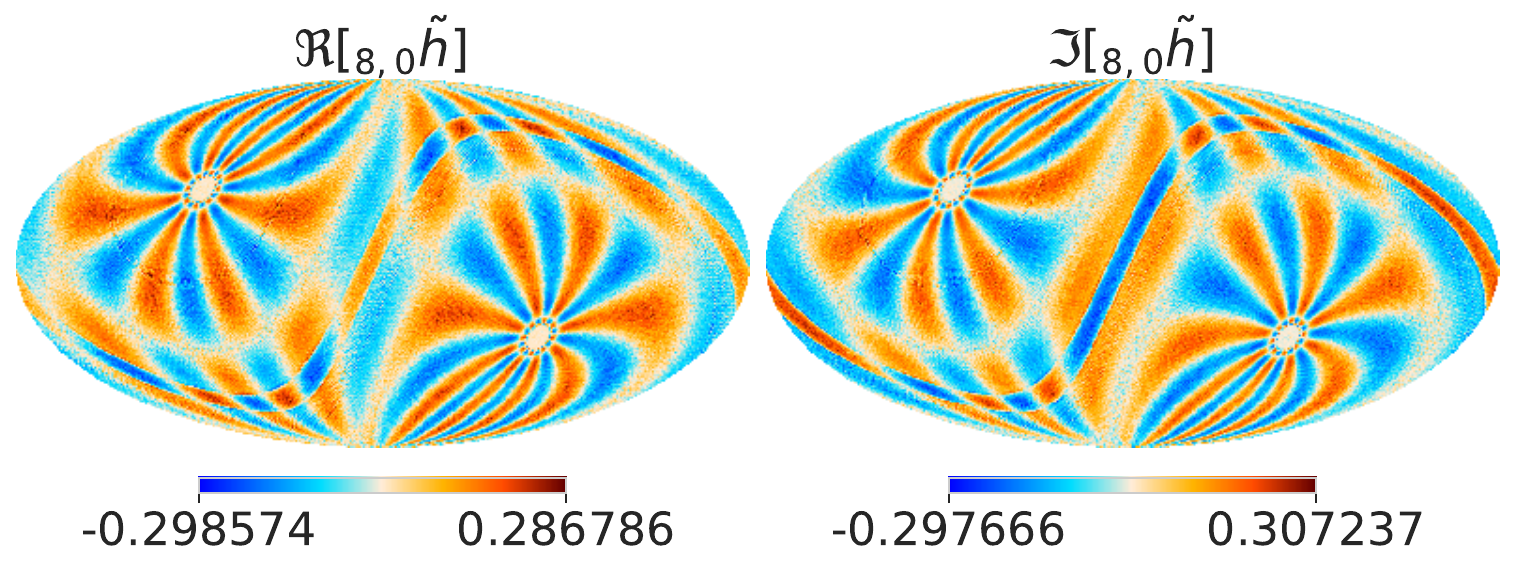}
    \\
    \includegraphics[width=0.49\columnwidth]{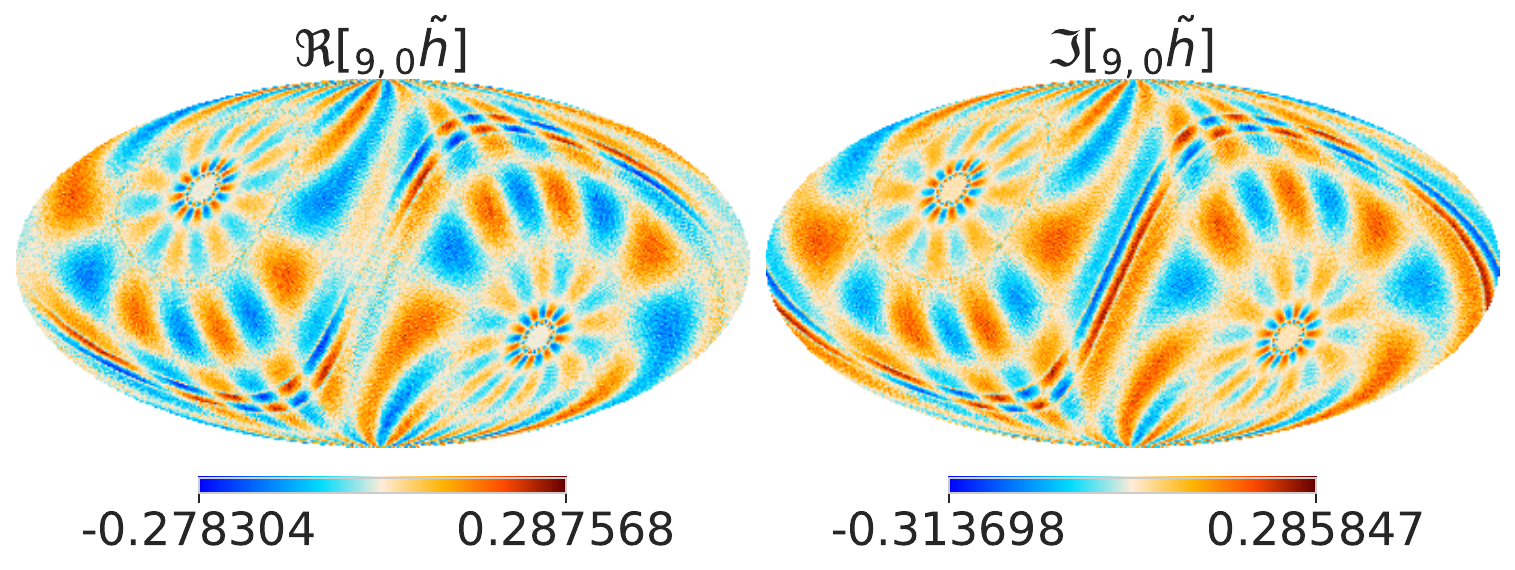}
    \includegraphics[width=0.49\columnwidth]{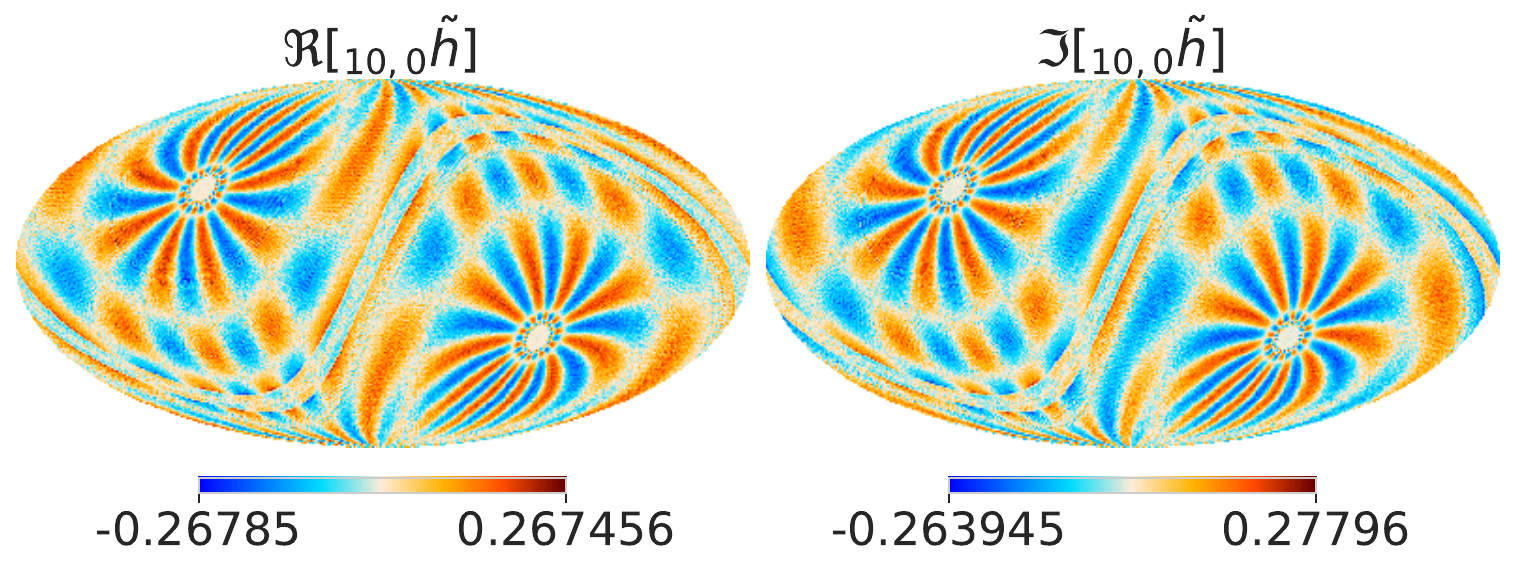}
    \caption[\Spin-$(n,0)$ cross-link factors for $n=1$ to $10$.]{\Spin-$(n,0)$
    cross-link factors for $n=1$ to $10$.}
    \label{fig:spin-n0_xlink_maps}
\end{figure}

\begin{figure}[htbp]
    \centering
    \includegraphics[width=0.49\columnwidth]{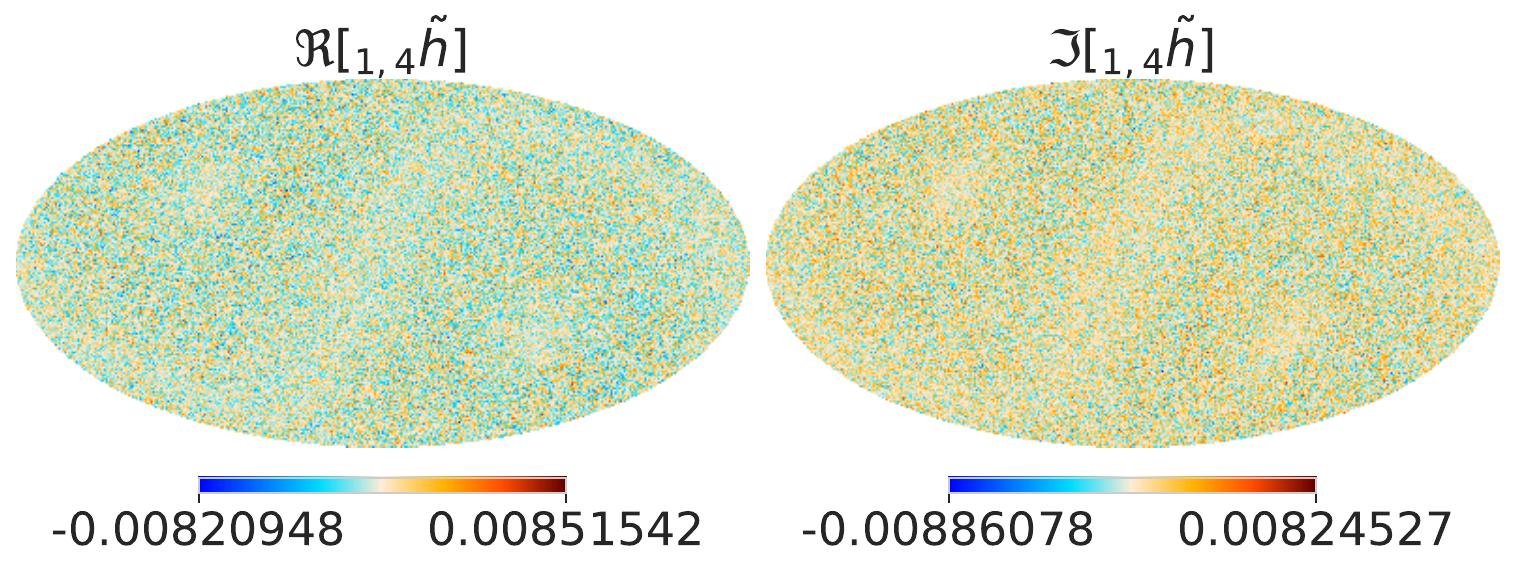}
    \includegraphics[width=0.49\columnwidth]{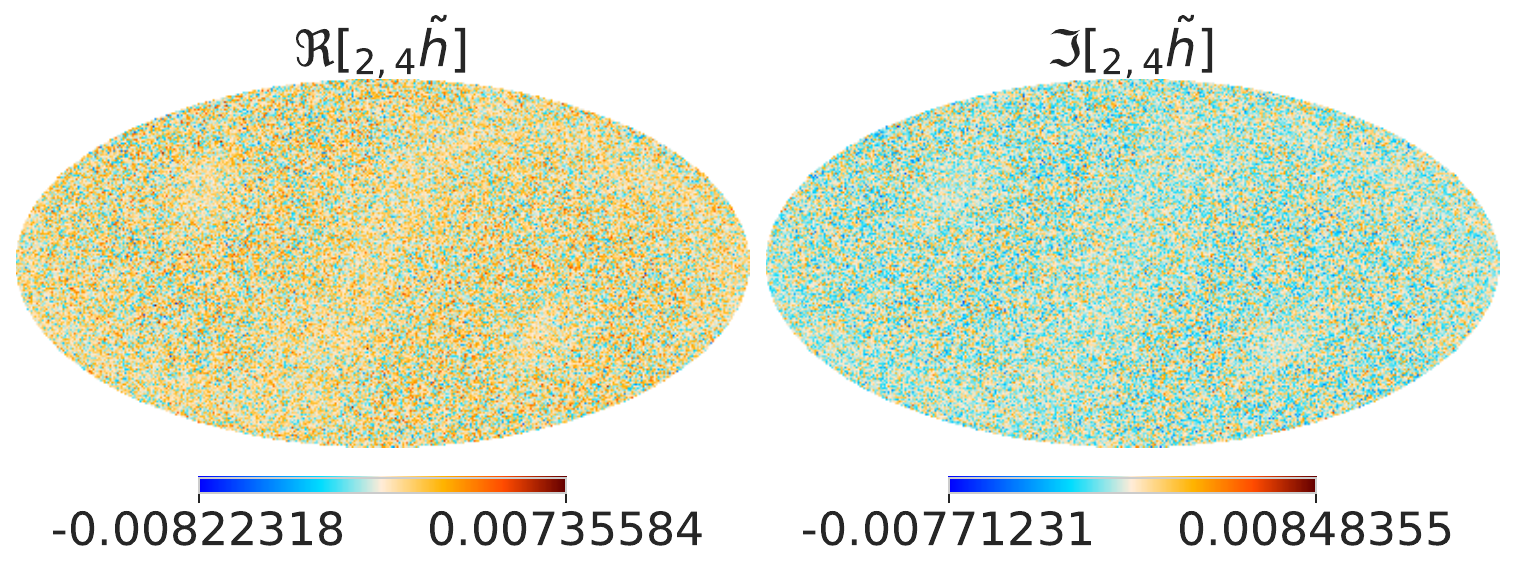}
    \\
    \includegraphics[width=0.49\columnwidth]{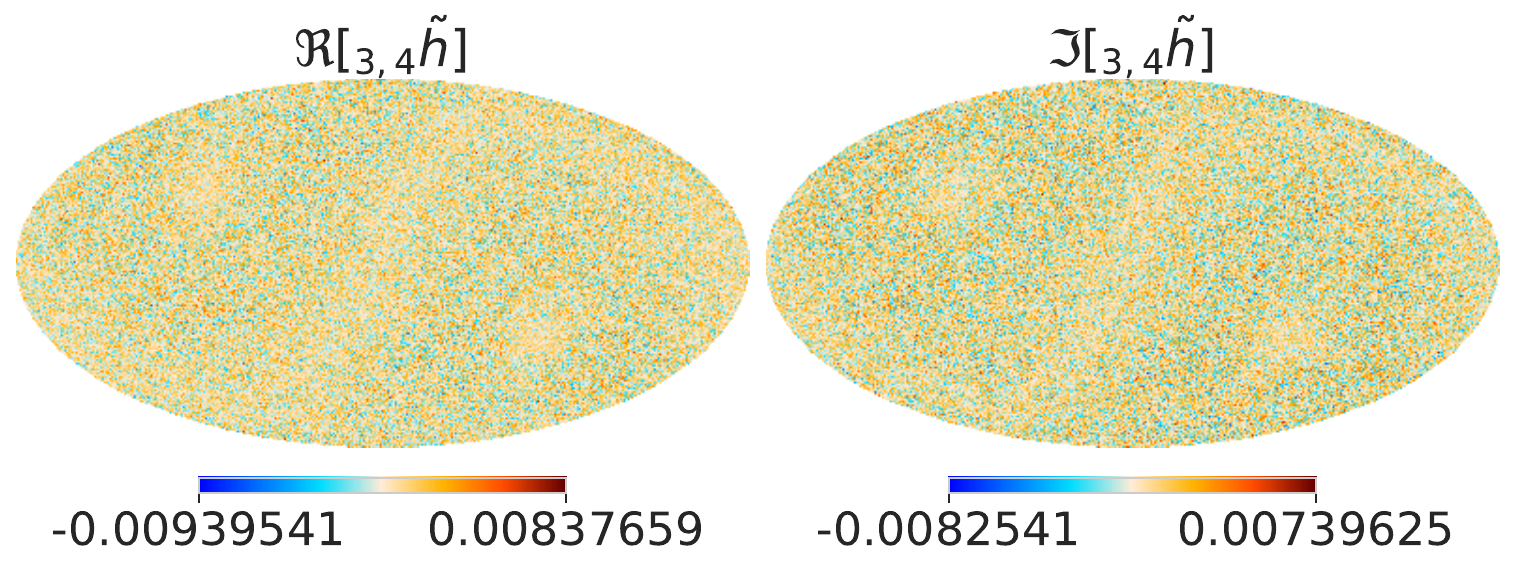}
    \includegraphics[width=0.49\columnwidth]{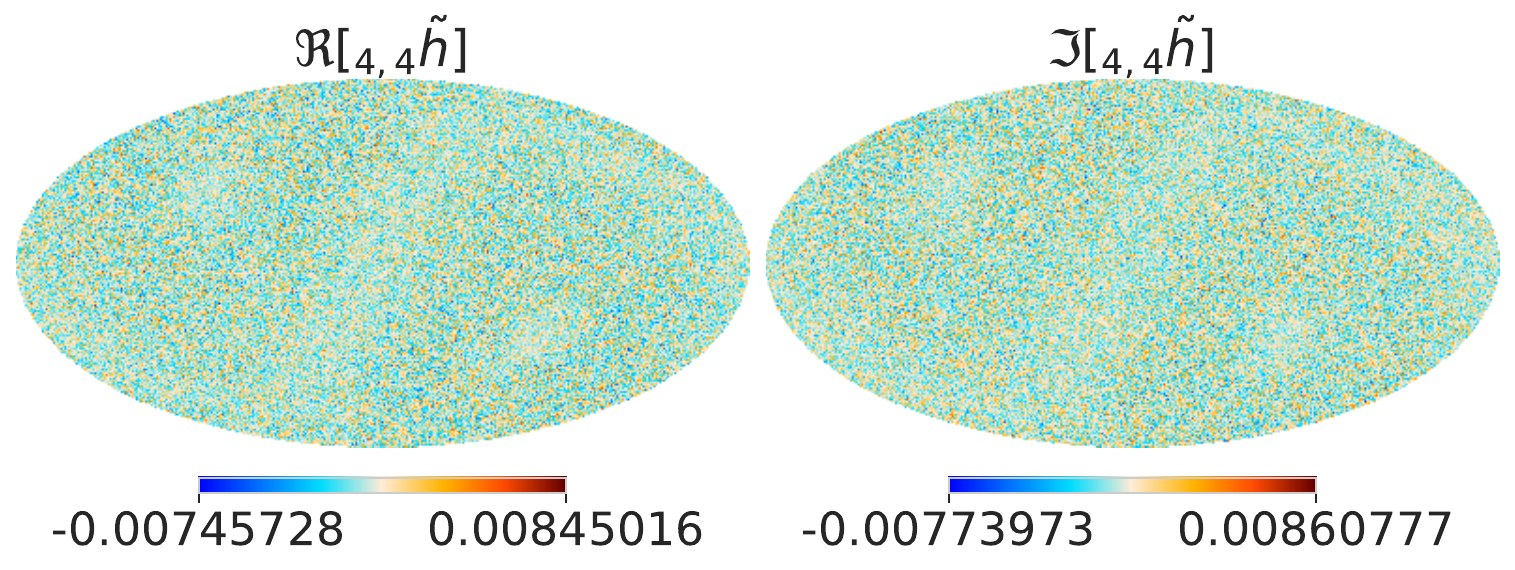}
    \\
    \includegraphics[width=0.49\columnwidth]{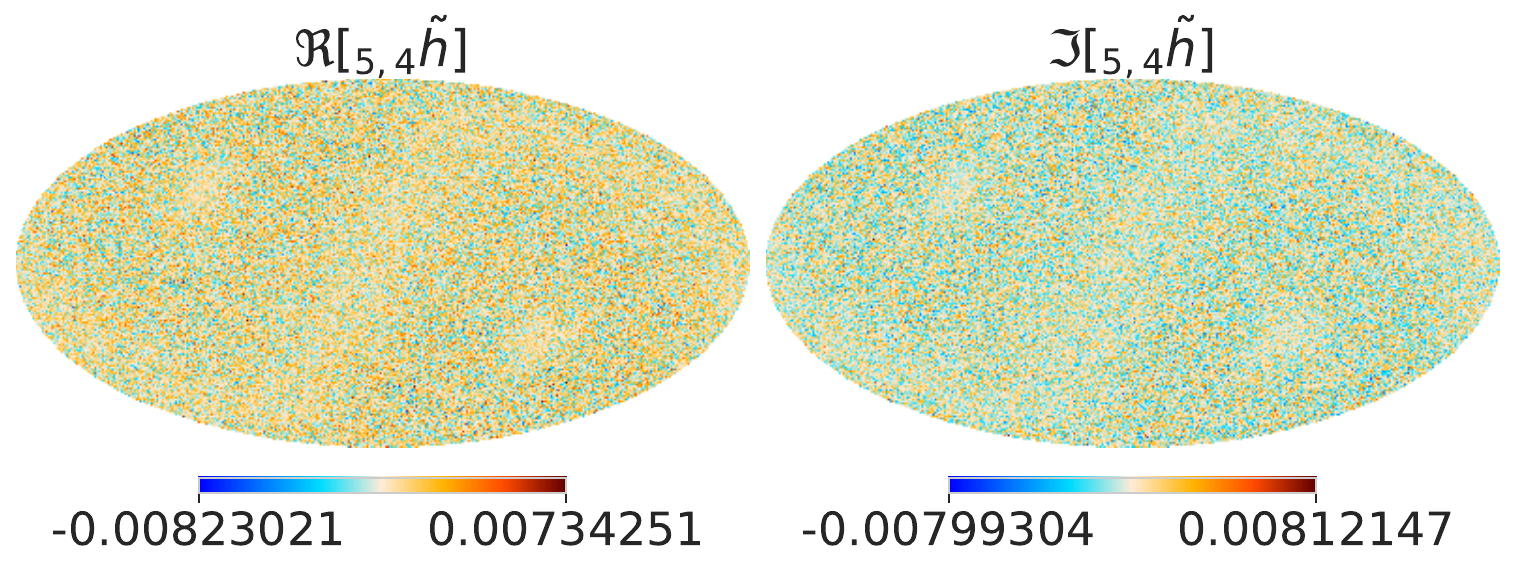}
    \includegraphics[width=0.49\columnwidth]{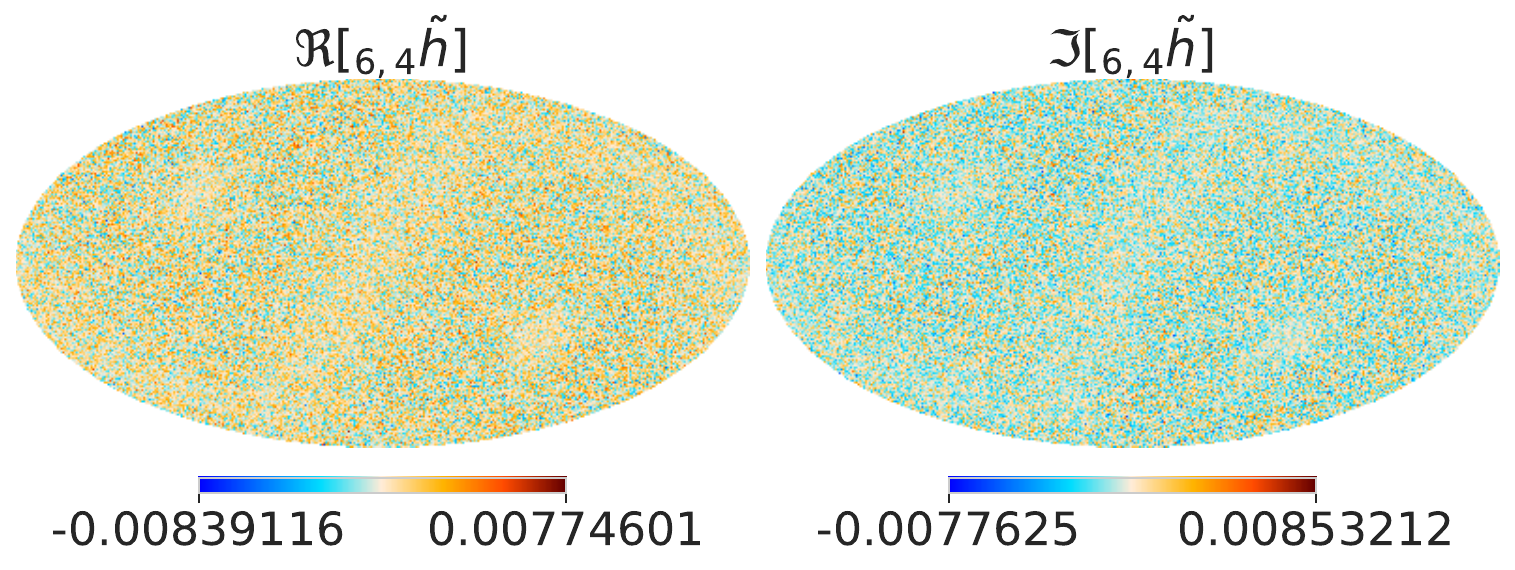}
    \\
    \includegraphics[width=0.49\columnwidth]{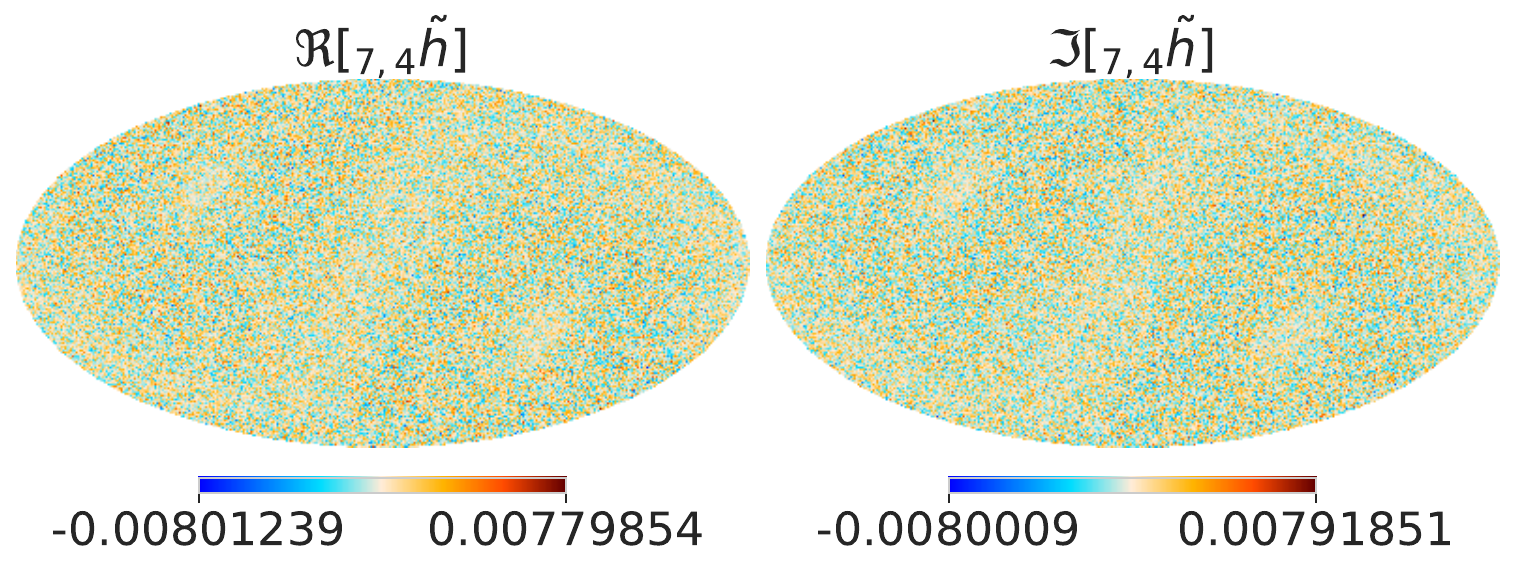}
    \includegraphics[width=0.49\columnwidth]{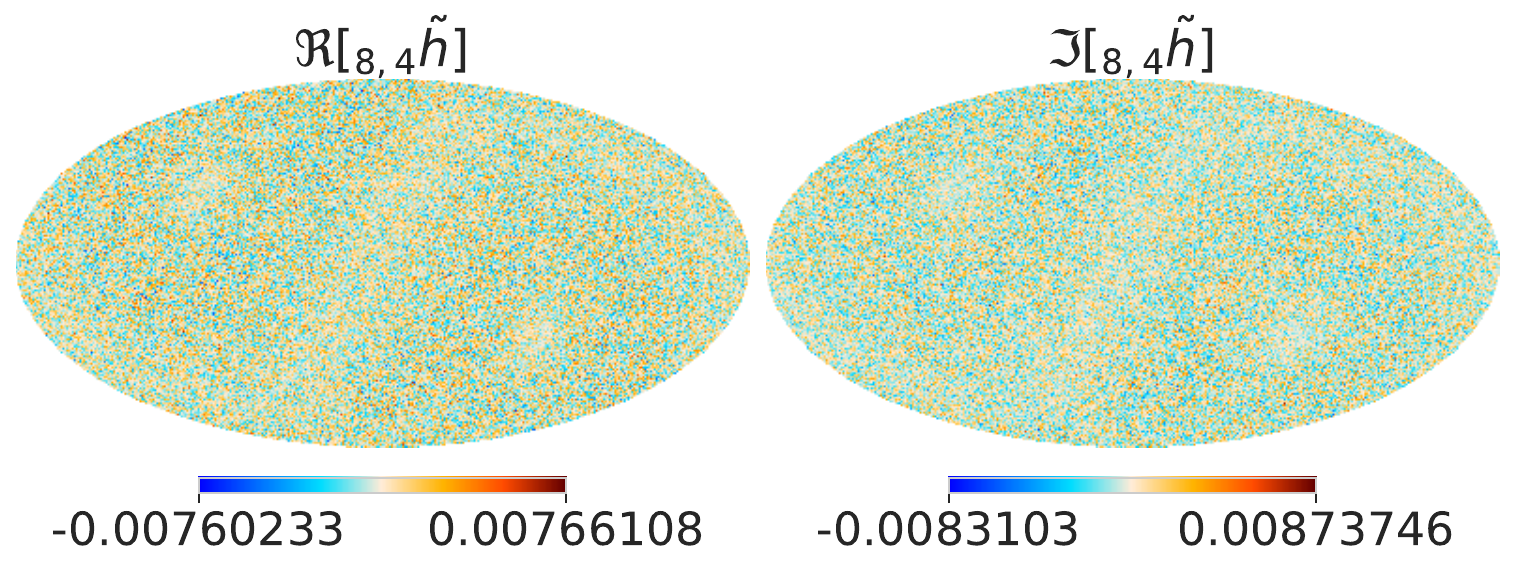}
    \\
    \includegraphics[width=0.49\columnwidth]{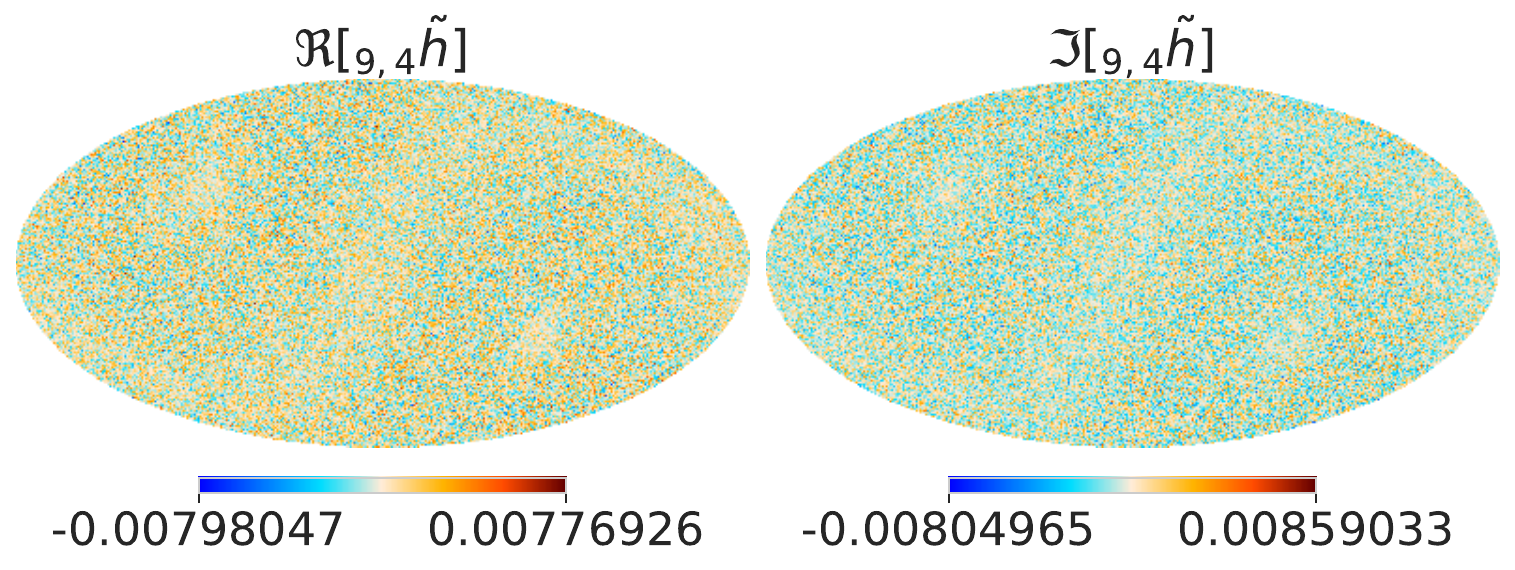}
    \includegraphics[width=0.49\columnwidth]{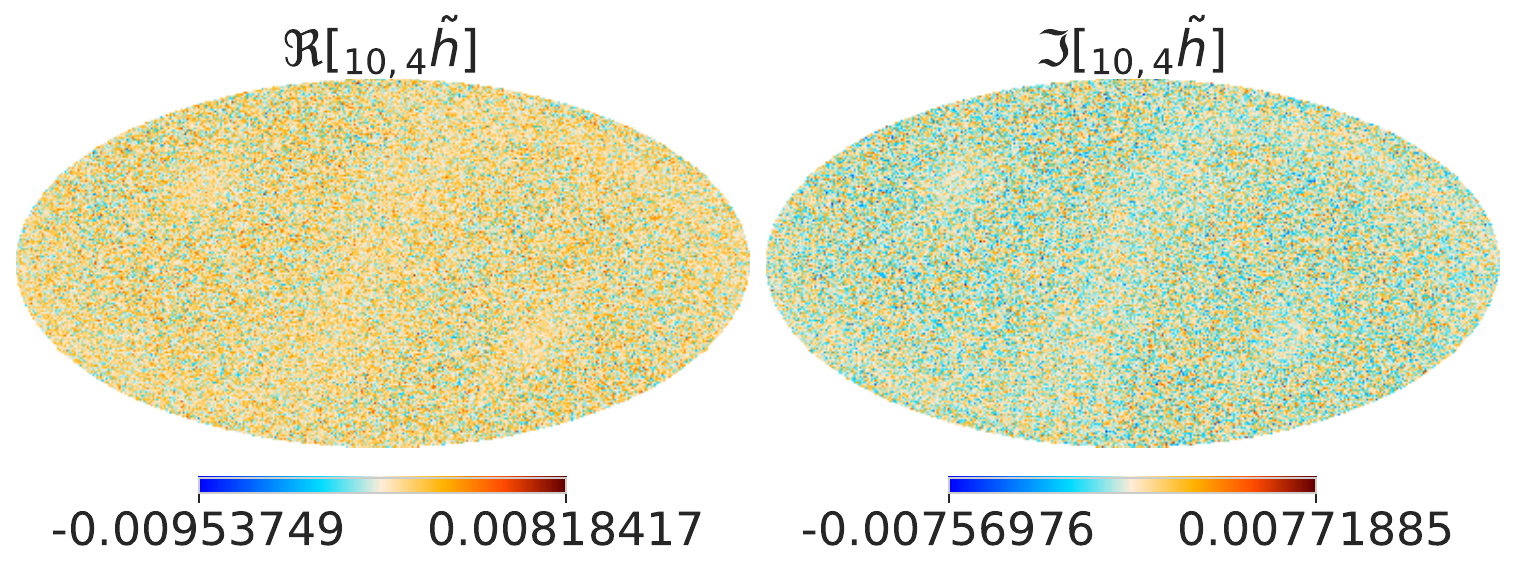}
    \caption[\Spin-$(n,4)$ cross-link factors for $n=1$ to $10$.]{\Spin-$(n,4)$
    cross-link factors for $n=1$ to $10$.}
    \label{fig:spin-n4_xlink_maps}
\end{figure}

We present cross-link maps for \spin-$(n,0)$ for $n=1$ to $10$ in \cref{fig:spin-n0_xlink_maps},
\spin-$(n,4)$ for $n=1$ to $10$ in \cref{fig:spin-n4_xlink_maps} and \spin-$(n,8)$
for $n=1$ to $10$ in \cref{fig:spin-n0_xlink_maps}. The maps are calculated by 3-years
simulation by \LiteBIRD's \SC with $\Nside=128$ configuration.

\begin{figure}[htbp]
    \centering
    \includegraphics[width=0.49\columnwidth]{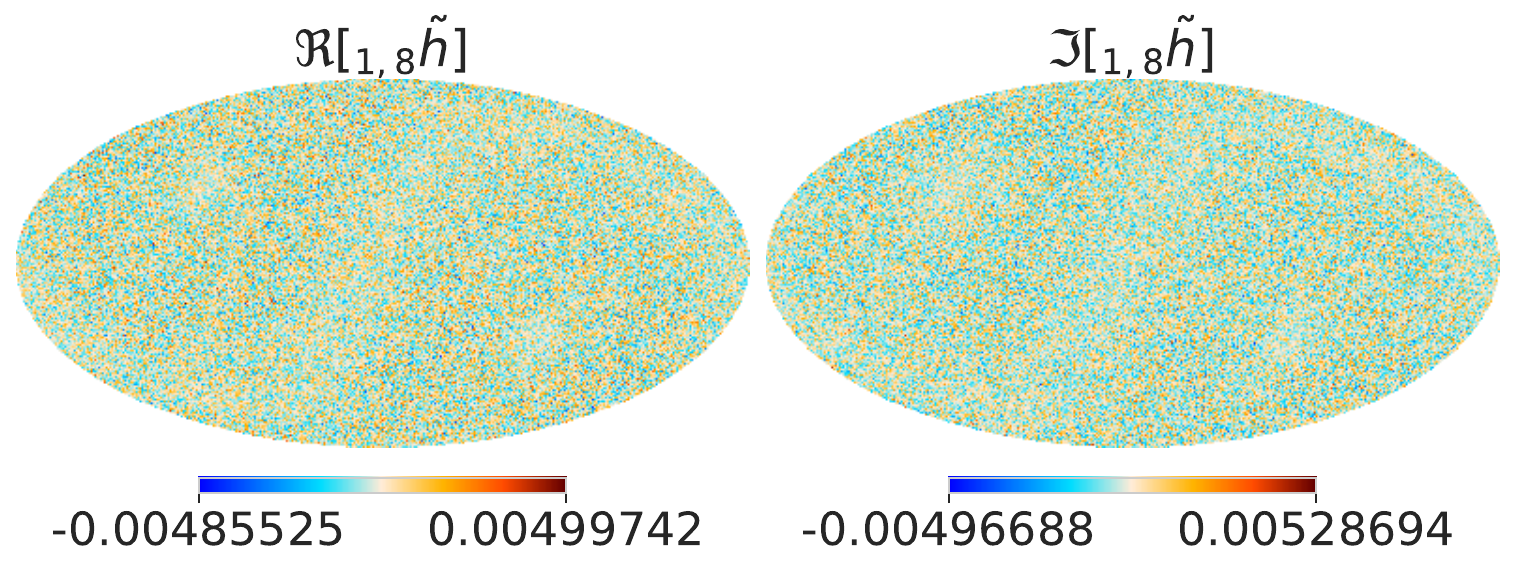}
    \includegraphics[width=0.49\columnwidth]{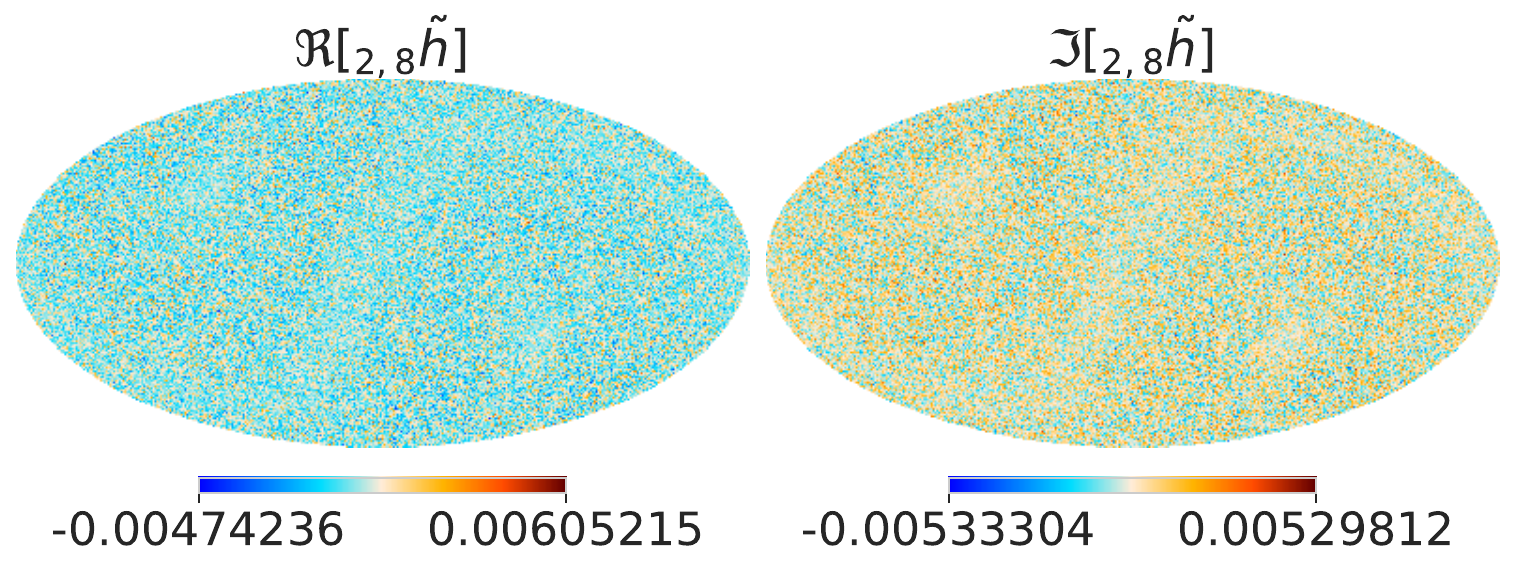}
    \\
    \includegraphics[width=0.49\columnwidth]{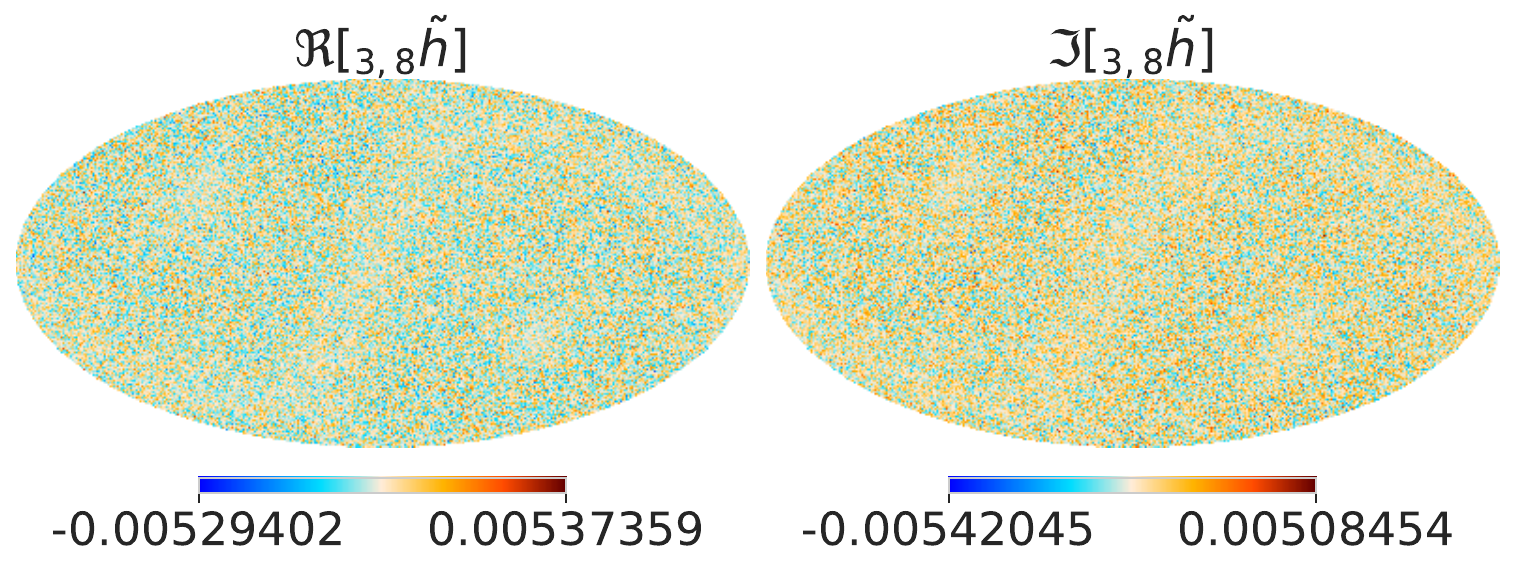}
    \includegraphics[width=0.49\columnwidth]{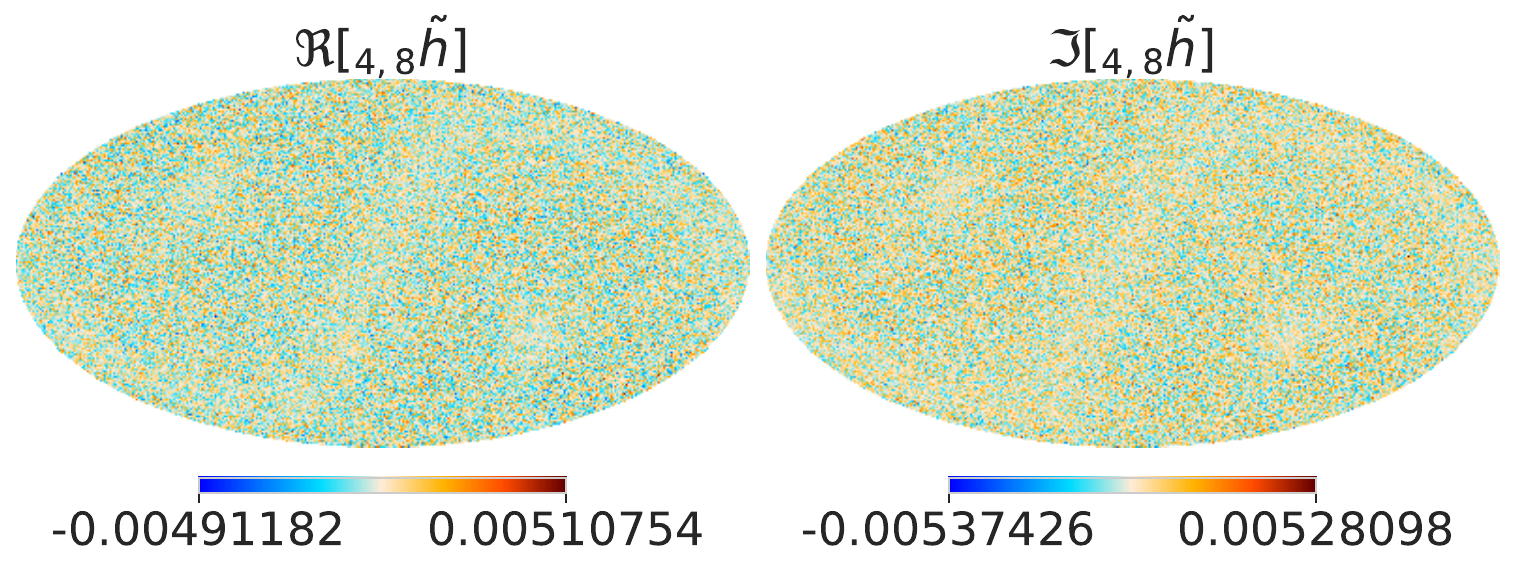}
    \\
    \includegraphics[width=0.49\columnwidth]{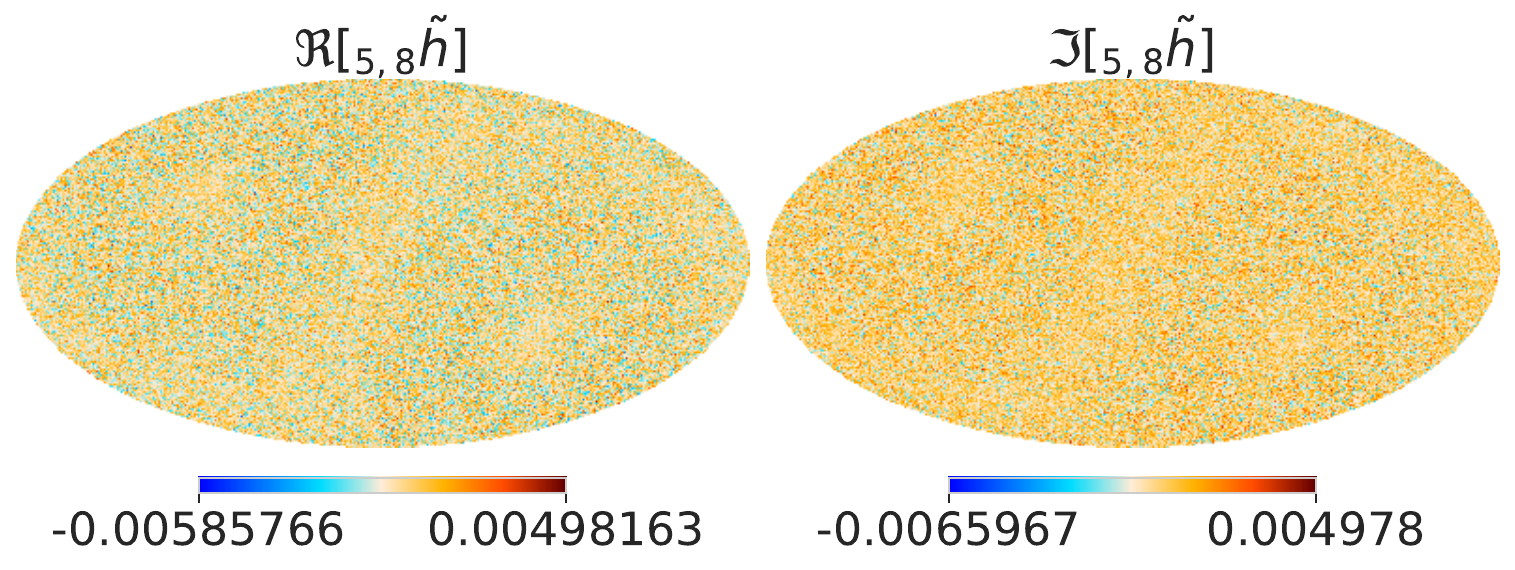}
    \includegraphics[width=0.49\columnwidth]{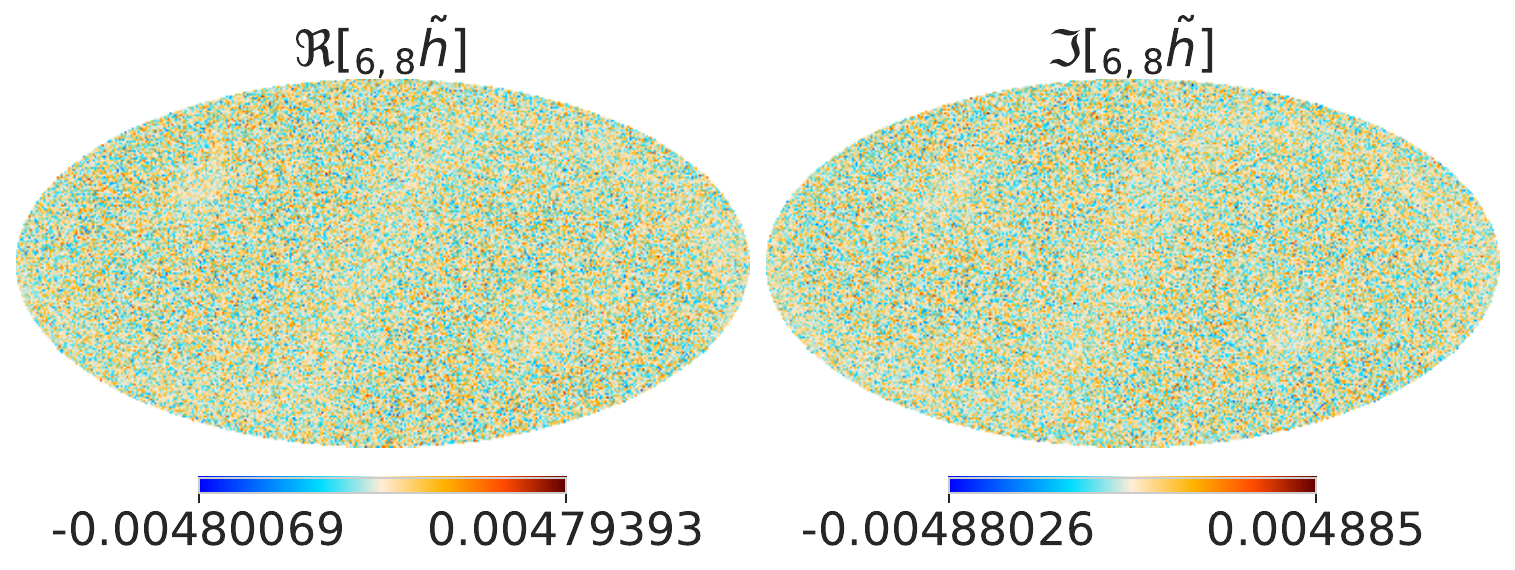}
    \\
    \includegraphics[width=0.49\columnwidth]{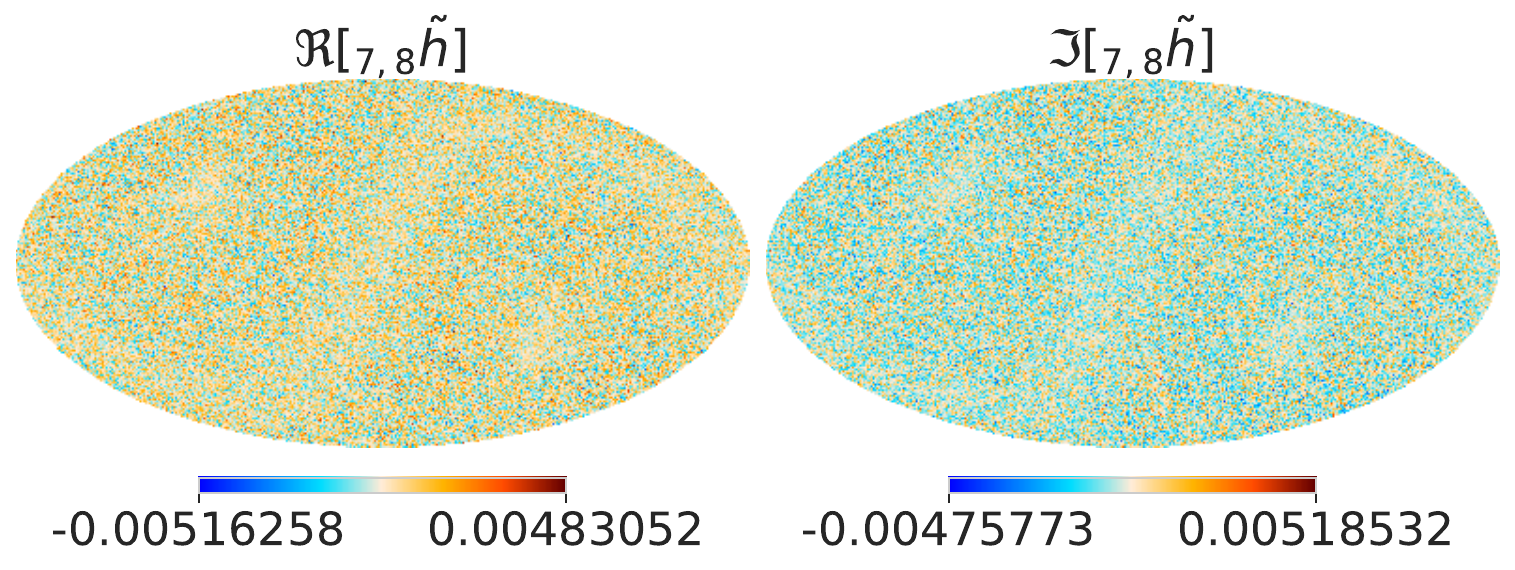}
    \includegraphics[width=0.49\columnwidth]{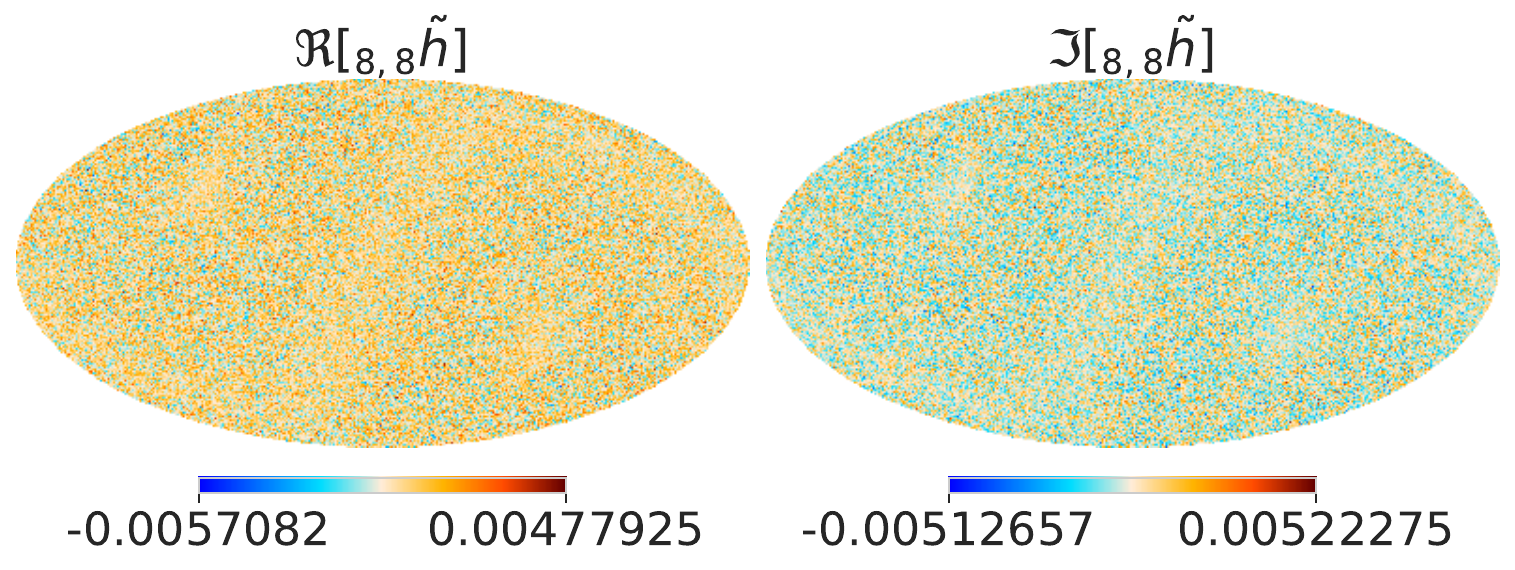}
    \\
    \includegraphics[width=0.49\columnwidth]{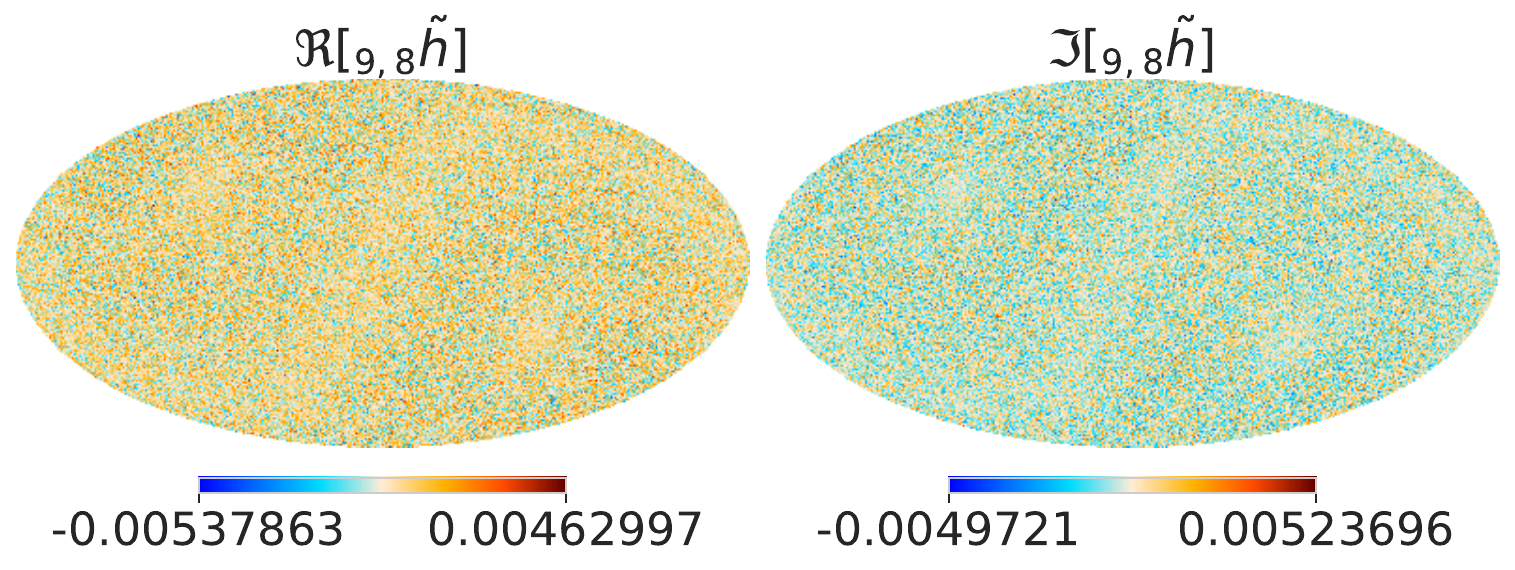}
    \includegraphics[width=0.49\columnwidth]{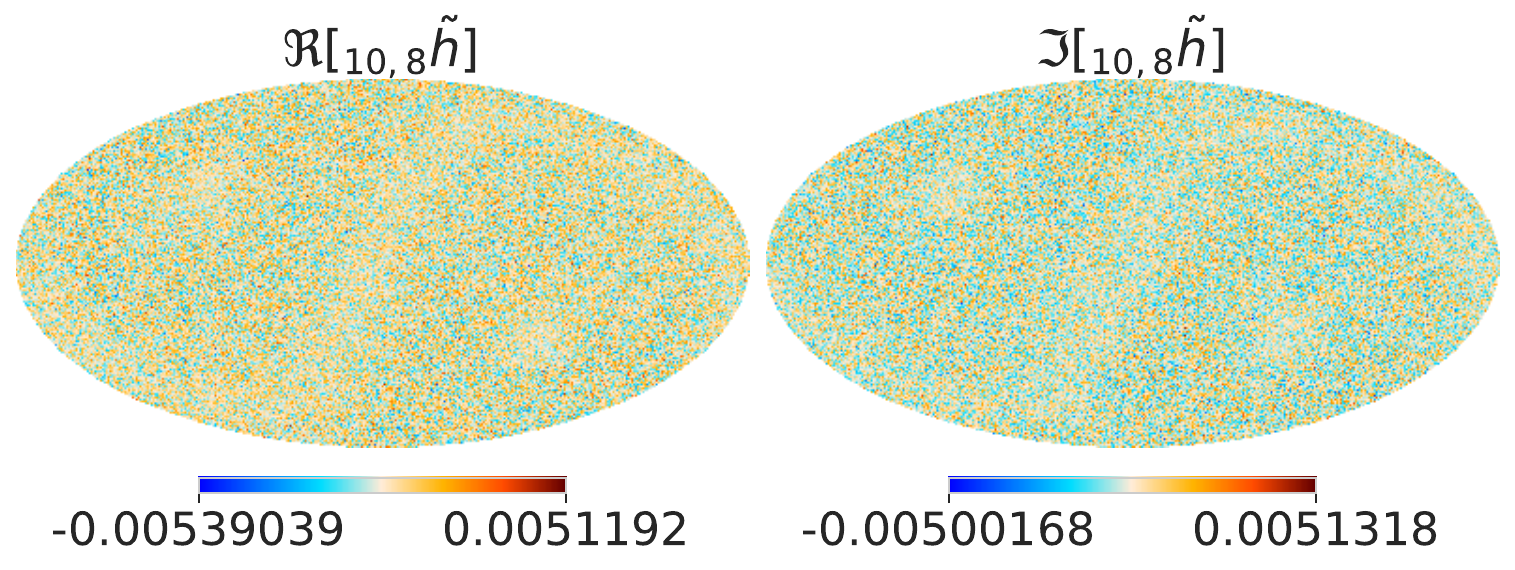}
    \caption[\Spin-$(n,8)$ cross-link factors for $n=1$ to $10$.]{ \Spin-$(n,8)$
    cross-link factors for $n=1$ to $10$. }
    \label{fig:spin-n8_xlink_maps}
\end{figure}

    \bibliographystyle{./class/JHEP}
    
    %\bibliographystyle{apalike} %citeがXX et al.になる
    %\bibliographystyle{unsrt} %citeが番号になる
    
    %--- During a paper writing, it should be on---
    %\bibliography{class/bibliography.bib}
    %----------------------------------------------
    % When we submit it to arxiv, we must comment out:
    % \bibliography{class/bibliography.bib}
    % and generate output.bbl by overleaf's compile which is avilable 
    % [Submit]->[arxiv]->[Download...]
    % change its name output.bbl to bibliography.bbl
    % and put it class/bibliography.bbl
    % then, turn on: 
    
\providecommand{\href}[2]{#2}\begingroup\raggedright\endgroup
 % for arXiv    
    
    \chapter*{Acknowledgments}

I would like to express my heartfelt gratitude to all those who have supported
me in writing this doctoral thesis.

First and foremost, I owe my deepest appreciation to my supervisor, Hirokazu
Ishino. Over the course of six years, spanning my undergraduate, master's, and
doctoral studies, he has guided me and imparted invaluable knowledge. In the
beginning, I struggled immensely with both cosmology and simulations, but thanks
to his supervisions, I was able to complete my doctoral dissertation. His
passionate mentorship and unwavering support have been the cornerstone of my
academic journey.

Next, I extend my gratitude to Yuya Nagano, Léo Vacher, and Samantha Stever, who
have always been by my side throughout my research life. Coding together in the lab
and sharing laughs over jokes are memories I cherish deeply. Without their
companionship, I would not have made it this far.

\begin{figure}[htbp]
    \begin{center}
        \includegraphics[width=1.0\columnwidth]{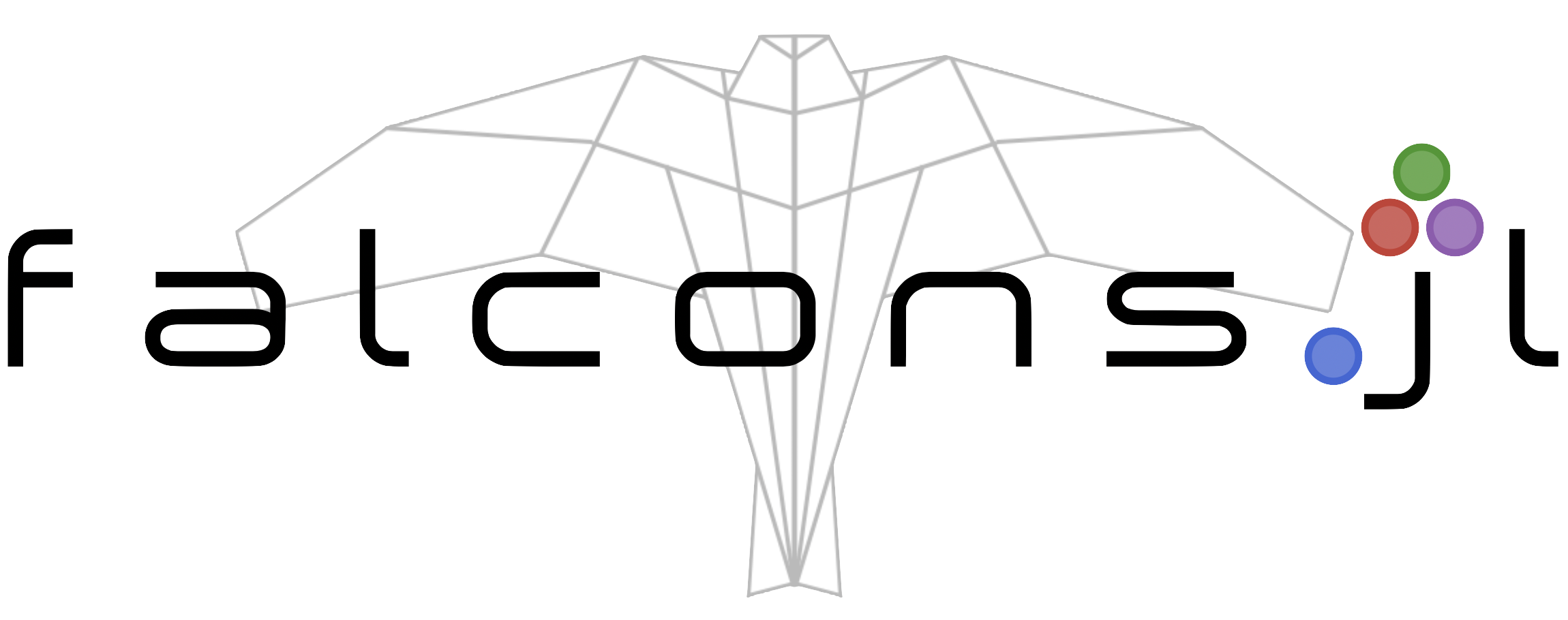}
        \caption[Nice logo of \Falcons]{The \Falcons logo was designed by
        Jonathan Aumont. The motif is based on Japanese origami, and the \texttt{Julia}
        extension --- `\texttt{.jl}' has the same coloring as the four colored
        dots in the logo of the \texttt{Julia} language itself \cite{Julia-2017}.
        The author is very pleased with this sophisticated design.}
        \label{fig:Falcons.jl}
    \end{center}
\end{figure}

I am also grateful to Guillaume Patanchon, Tomotake Matsumura, Yuki Sakurai,
Tommaso Ghigna, Ludovic Montier, Wang Wang, Jonathan Aumont (He kindly created
my \Falcons logo which is shown in \cref{fig:Falcons.jl}), Koki Ishizaka, and Eiichiro
Komatsu for their invaluable advice throughout my research. Meeting them has
been a stroke of luck, and their guidance has been instrumental in advancing my studies.

To my esteemed Italian friends, Marco Bortolami and Nicolò Raffuzzi, I extend my
heartfelt thanks. Our collaborative research has been incredibly stimulating,
and discussing technical topics like pointing systematics and MPI, as well as Italian
culture, has been a delightful experience. Avinash Anand, who helped me when I
suffered from dehydration during an face-to-face meeting on Elba Island, is a lifesaver.
Without him, I might not be here today. I am deeply grateful to him and all the
\LiteBIRD collaborators who supported me during that time.

I also wish to thank my friends in the lab and collaboration, Kunimoto Komatsu, Naoya Doi, Kazuya Kazahaya,
Ryota Uematsu, Serika Tsukatsune, Mami Morinaga, Kiyoshi Ikuma, Ryuji Omae, Mitsuhiro
Higuchi, Shunsuke Okumura, Kento Kakinoki, Kosuke Aizawa and Ryosuke Akizawa. The discussions and enjoyable
times we shared in the lab have been invaluable to me.

The greatest fortune in my research career has been encountering the fascinating
subject of CMB and becoming a member of the \LiteBIRD collaboration. I am deeply
grateful to the \LiteBIRD collaboration for nurturing me into a cosmologist.

Finally, I would like to express my gratitude to my family and my wife Asuka, 
who have always supported and encouraged my research life from afar.

    %-------------------
\end{document}